\documentclass[UKenglish]{scrbook}

\usepackage{ifthen}

\usepackage[labelfont=bf]{caption}

\usepackage[super]{nth}

\usepackage{mathtools}

\usepackage{graphicx}
\usepackage{epstopdf}

\usepackage{float}
\usepackage{scrhack}

\usepackage[shortlabels]{enumitem}

\usepackage{babel}

\usepackage{iflang}

\KOMAoptions{fontsize=11pt, paper=a4, twoside}
\KOMAoptions{listof=totoc}

\KOMAoptions{DIV=12, BCOR=5mm}
\KOMAoptions{headinclude=true, footinclude=false}

\usepackage[headsepline]{scrlayer-scrpage}

\usepackage{amsmath, amssymb}

\usepackage[utf8]{inputenc}

\usepackage{newtxtext}
\usepackage{newtxmath}

\usepackage[autostyle=true]{csquotes}

\DeclareQuoteStyle[mybritish]{myenglish}%
  {\textquotedblleft}{\textquotedblright}%
  {\textquoteleft}{\textquoteright}
\DeclareQuoteAlias[mybritish]{myenglish}{myenglish}
\DeclareQuoteAlias{myenglish}{british}
\DeclareQuoteAlias{myenglish}{UKenglish}

\usepackage{xfrac}

\usepackage{siunitx}
\usepackage{graphicx}
\usepackage[labelformat=simple]{subcaption}

\usepackage{rotating}
\usepackage{array}
\usepackage{booktabs}
\usepackage{dcolumn}
\usepackage[pageshow]{xtab}

\usepackage[version=3]{mhchem}

\usepackage{microtype}
\usepackage[table]{xcolor}
\usepackage{xspace}
\usepackage{setspace}
\usepackage{scrdate,scrtime}
\usepackage{longtable}

\usepackage{tikz}
\usetikzlibrary{shapes,arrows}

\usepackage[acronym, nomain, nosuper, nonumberlist, toc, style=listdotted]{glossaries}
\setacronymstyle{long-sc-short}

\usepackage{tocloft}
\usepackage[backend=biber, backref=true, style=alphabetic, sorting=ynt, block=space, giveninits=true]{biblatex}
\DeclareFieldFormat{labelalpha}{\thefield{entrykey}}
\DeclareFieldFormat{extraalpha}{}

\definecolor{darkred}{rgb}{0.5,0,0}
\definecolor{darkgreen}{rgb}{0,0.5,0}
\definecolor{darkblue}{rgb}{0,0,0.5}
\definecolor{darkyellow}{rgb}{0.5,0.5,0}
\definecolor{darkcyan}{rgb}{0,0.5,0.5}
\definecolor{darkmagenta}{rgb}{0.5,0,0.5}

\PassOptionsToPackage{hyphens}{url}
\usepackage{hyperref}
\hypersetup{colorlinks=true,linkcolor=blue,citecolor=darkmagenta,urlcolor=darkgreen,bookmarksopen}

\usepackage[capitalise]{cleveref}
\crefname{section}{Section}{Sections}
\crefname{subsection}{Section}{Sections}

\def\bfseries{\fontseries\bfdefault\selectfont\boldmath}

\setkomafont{title}{\normalfont\bfseries\huge}
\setkomafont{subtitle}{\normalfont\Large}

\KOMAoptions{numbers=noendperiod}
\KOMAoptions{chapterprefix=true,appendixprefix=true}

\newcommand{\chapquote}[3]{\begin{quotation} \textit{#1} \end{quotation} \begin{flushright} - #2, \textit{#3}\end{flushright} }

\setkomafont{caption}{\normalfont\small}
\setcapindent{0pt}

\newlength{\figwidth}
\deffootnote{1em}{1em}{\textsuperscript{\thefootnotemark}\ }

\KOMAoptions{DIV=last}
\AtEveryBibitem{%
  \clearfield{month}
}

\DeclareFieldFormat[article]{title}{\textbf{#1}\isdot}
\DeclareFieldFormat[misc]{title}{\textbf{#1}\isdot}
\DeclareFieldFormat[online]{title}{\textbf{#1}\isdot}
\DeclareFieldFormat[unpublished]{title}{\textbf{#1}\isdot}
\DeclareFieldFormat[book]{title}{\textbf{#1}\isdot}
\DeclareFieldFormat[inproceedings]{title}{\textbf{#1}\isdot}

\DeclareFieldFormat[article]{journaltitle}{#1\isdot}
\DeclareFieldFormat[article]{journalsubtitle}{#1\isdot}

\DeclareFieldFormat[article]{volume}{\textbf{#1}\isdot}

\AtEveryBibitem{
  \ifentrytype{article}{\clearfield{number}}{}
}

\AtEveryBibitem{%
  \clearlist{location}
}

\DeclareFieldFormat{pages}{\mkfirstpage[{\mkpageprefix[bookpagination]}]{#1}}

\renewcommand*{\newunitpunct}{\addcomma\space}

\renewbibmacro{in:}{}

\DefineBibliographyStrings{UKenglish}{%
  page = {},
  pages = {}
}

\renewbibmacro*{institution+location+date}{%
  \iflistundef{institution}
    {\setunit*{\addcomma\space}}
    {\setunit*{\addcolon\space}}%
  \printlist{institution}%
  \setunit*{\addcomma\space}%
  \usebibmacro{date}%
  \newunit}
\DeclareSIUnit\electronvolt{\text{e\kern-.1em V}}
\DeclareSIUnit\clight{\text{\ensuremath{c}}}
\DeclareSIUnit\gauss{G}
\DeclareSIUnit\PeV{P\eV}
\DeclareSIUnit\EeV{E\eV}
\DeclareSIUnit\erg{erg}
\DeclareSIUnit\nucleon{n}
\DeclareSIUnit\parsec{pc}
\DeclareSIUnit\years{\text{years}}

\DeclareMathOperator{\sign}{sgn}

\newcommand{\diff}{\mathop{}\!\mathrm{d}}
\newcommand{\logten}[1]{\mathop{\log_{10}(#1)}}

\newcommand{\abs}[1]{\left\lvert#1\right\rvert}
\newcommand{\given}[1][]{\:#1\vert\:}
\newcommand{\SIvarOp}[4]{#1~#2~\SI{#3}{#4}}
\newcommand{\SIvarApprox}[3]{#1~$\approx$~\SI{#2}{#3}}
\newcommand{\SIvarEquals}[3]{#1~=~\SI{#2}{#3}}
\newcommand{\SIapprox}[2]{$\approx$~\SI{#1}{#2}}
\newcommand{\SIapproxScientific}[2]{$\approx$~\SI[retain-unity-mantissa=false]{#1}{#2}}
\newcommand{\SIScientific}[2]{\SI[retain-unity-mantissa=false]{#1}{#2}}
\newcommand{\SIvarApproxScientific}[3]{#1~$\approx$~\SI[retain-unity-mantissa=false]{#2}{#3}}

\newlength{\leftstackrelawd}
\newlength{\leftstackrelbwd}
\def\leftstackrel#1#2{\settowidth{\leftstackrelawd}%
{${{}^{#1}}$}\settowidth{\leftstackrelbwd}{$#2$}%
\addtolength{\leftstackrelawd}{-\leftstackrelbwd}%
\leavevmode\ifthenelse{\lengthtest{\leftstackrelawd>0pt}}%
{\kern-.5\leftstackrelawd}{}\mathrel{\mathop{#2}\limits^{#1}}}

\newacronym{AMS}{AMS}{Alpha Magnetic Spectrometer}
\newacronym{LHC}{LHC}{Large Hadron Collider}
\newacronym{ISM}{ISM}{Interstellar Medium}
\newacronym{ACC}{ACC}{Anti-Coincidence Counter}
\newacronym{RICH}{RICH}{Ring-Imaging Cherenkov counter}
\newacronym{ECAL}{ECAL}{Electromagnetic Calorimeter}
\newacronym{TOF}{TOF}{Time Of Flight}
\newacronym{TRD}{TRD}{Transition Radiation Detector}
\newacronym{TAS}{TAS}{Tracker Alignment System}
\newacronym{MDR}{MDR}{Maximum Detectable Rigidity}
\newacronym{MIP}{MIP}{Minimum Ionizing Particle}
\newacronym{BDT}{BDT}{Boosted Decision Tree}
\newacronym{ROC}{ROC}{Receiver Operating Characteristic}
\newacronym{KDE}{KDE}{Kernel Density Estimation}
\newacronym{SAA}{SAA}{South Atlantic Anomaly}
\newacronym{TOI}{TOI}{Top-Of-Instrument}
\newacronym{ISS}{ISS}{International Space Station}
\newacronym{MPV}{MPV}{Most Probable Value}
\newacronym{CMB}{CMB}{Cosmic Microwave Background}
\newacronym{HMF}{HMF}{Heliospheric Magnetic Field}
\newacronym{HCS}{HCS}{Heliospheric Current Sheet}
\newacronym{SN}{SN}{Supernova}
\newacronym{SNRs}{SNRs}{Supernova Remnants}
\newacronym{DSA}{DSA}{Diffusive Shock Acceleration}
\newacronym{WIMPs}{WIMPs}{Weakly Interacting Massive Particles}
\newacronym{CME}{CME}{Coronal Mass Ejections}
\newacronym{TDRS}{TDRS}{Tracking and Data Relay Satellite}
\newacronym{MSFC}{MSFC}{Marshall Space Flight Center}
\newacronym{POCC}{POCC}{Payload Operation Control Center}
\newacronym{HRDL}{HRDL}{High Rate Data Link}

\makeglossaries

\definecolor{darkRed}{rgb}{0.6, 0.0, 0.0}
\definecolor{darkBlue}{rgb}{0.0, 0.0, 0.6}
\definecolor{darkGreen}{rgb}{0.0, 0.6, 0.0}
\definecolor{darkMagenta}{rgb}{0.6, 0.0, 0.6}
\definecolor{lightOrange}{rgb}{1.0, 0.8, 0.0}

\definecolor{elecColor}{rgb}{0.0, 0.0, 0.6}
\newcommand{\elecColorText}{dark blue}

\definecolor{ccElecColor}{rgb}{0.4, 0.4, 1.0}
\newcommand{\ccElecColorText}{light blue}

\definecolor{posiColor}{rgb}{0.0, 0.6, 0.6}

\definecolor{ccPosiColor}{rgb}{0.4, 1.0, 1.0}

\definecolor{protColor}{rgb}{0.6, 0.0, 0.0}
\newcommand{\protColorText}{dark red}

\definecolor{ccProtColor}{rgb}{1.0, 0.4, 0.4}
\newcommand{\ccProtColorText}{light red}

\definecolor{ccColor}{rgb}{1.0, 0.0, 1.0}

\definecolor{fitResultColor}{rgb}{1.0, 0.6, 0.0}
\newcommand{\fitResultColorText}{orange}

\newcommand{\ccProtPeakNovoSymbol}{\mu_{\mathrm{ccprot,~novo}}}
\newcommand{\ccProtPeakNovo}{$\ccProtPeakNovoSymbol$}

\newcommand{\ccProtWidthNovoSymbol}{\sigma_{\mathrm{ccprot,~novo}}}
\newcommand{\ccProtWidthNovo}{$\ccProtWidthNovoSymbol$}

\newcommand{\ccProtTailNovoSymbol}{\tau_{\mathrm{ccprot,~novo}}}
\newcommand{\ccProtTailNovo}{$\ccProtTailNovoSymbol$}

\newcommand{\protPeakNovoSymbol}{\mu_{\mathrm{prot,~novo}}}
\newcommand{\protPeakNovo}{$\protPeakNovoSymbol$}

\newcommand{\protWidthNovoSymbol}{\sigma_{\mathrm{prot,~novo}}}
\newcommand{\protWidthNovo}{$\protWidthNovoSymbol$}

\newcommand{\protTailNovoSymbol}{\tau_{\mathrm{prot,~novo}}}
\newcommand{\protTailNovo}{$\protTailNovoSymbol$}

\newcommand{\protFractionNovoSymbol}{\alpha_{\mathrm{prot,~novo}}}
\newcommand{\protFractionNovo}{$\protFractionNovoSymbol$}

\newcommand{\protPeakGausSymbol}{\mu_{\mathrm{prot,~gaus}}}
\newcommand{\protPeakGaus}{$\protPeakGausSymbol$}

\newcommand{\protWidthGausSymbol}{\sigma_{\mathrm{prot,~gaus}}}

\newcommand{\protWidthGausDeltaSymbol}{{\delta_\sigma}_{\mathrm{prot,~gaus}}}
\newcommand{\protWidthGausDelta}{$\protWidthGausDeltaSymbol$}

\newcommand{\elecPeakNovoSymbol}{\mu_{\mathrm{elec,~novo}}}
\newcommand{\elecPeakNovo}{$\elecPeakNovoSymbol$}

\newcommand{\elecWidthNovoSymbol}{\sigma_{\mathrm{elec,~novo}}}
\newcommand{\elecWidthNovo}{$\elecWidthNovoSymbol$}

\newcommand{\elecTailNovoSymbol}{\tau_{\mathrm{elec,~novo}}}
\newcommand{\elecTailNovo}{$\elecTailNovoSymbol$}

\newcommand{\elecFractionNovoSymbol}{\alpha_{\mathrm{elec,~novo}}}
\newcommand{\elecFractionNovo}{$\elecFractionNovoSymbol$}

\newcommand{\elecPeakGausSymbol}{\mu_{\mathrm{elec,~gaus}}}
\newcommand{\elecPeakGaus}{$\elecPeakGausSymbol$}

\newcommand{\elecWidthGausSymbol}{\sigma_{\mathrm{elec,~gaus}}}
\newcommand{\elecWidthGaus}{$\elecWidthGausSymbol$}

\newcommand{\thesistitle}{Precision measurement of the cosmic-ray electron and positron fluxes as a function of time and energy with the Alpha Magnetic Spectrometer on the International Space Station}
\newcommand{\thesistitleGerman}{Präzisionsmessung der kosmischen Elektron- und Positronflüsse als Funktion der Zeit und Energie mit dem Alpha Magnetic Spectrometer auf der Internationalen Raumstation}
\newcommand*{\thesisauthor}{Diplom-Physiker\\\textbf{Nikolas Zimmermann}}
\newcommand*{\thesistown}{Köln}

\renewcommand*{\pagenumbering}[1]{}

\begin{document}

\frontmatter

\typeout{Document \jobname, Info: Final version of a PhD thesis}
\begin{titlepage}
  \addtolength{\oddsidemargin}{1.0cm}\addtolength{\topmargin}{1.0cm}
  \rmfamily\setlength{\parindent}{0pt}
  \begin{center}
    \begin{onehalfspace}
      \bfseries\huge
      \thesistitle
    \end{onehalfspace}
    \vspace*{16ex}

    Von der Fakultät für Mathematik, Informatik und Naturwissenschaften der RWTH Aachen
    University zur Erlangung des akademischen Grades eines Doktors der Naturwissenschaften
    genehmigte Dissertation

    \vspace*{6ex}
    vorgelegt von \\
    \vspace*{6ex}
    {\large \thesisauthor} \\
    \vspace*{6ex}
    aus \\
    \thesistown \\

    \vspace*{16ex}
    {
      \large
      \begin{tabular}{ll}
        Berichter:    & Universitätsprofessor Dr. rer. nat. Stefan Schael \\
                      & Universitätsprofessor Dr. rer. nat. Christopher Wiebusch \\
      \end{tabular}
    } \\
    \vspace*{4ex}
    { \large Tag der mündlichen Prüfung: 17. Februar 2020 } \\
    \vspace*{10ex}
    Diese Dissertation ist auf den Internetseiten der Universitätsbibliothek verfügbar.
  \end{center}
\end{titlepage}

\begin{titlepage}
  \addtolength{\oddsidemargin}{1.0cm}\addtolength{\topmargin}{1.0cm}
  \rmfamily\setlength{\parindent}{0pt}
  \begin{center}
    {
      \scshape\huge
      \begin{onehalfspace}
        RWTH Aachen University \\
        1. Physikalisches Institut B
      \end{onehalfspace}
    }
    \vspace*{6ex}

    \rule{\linewidth}{2.5pt}
      \begin{onehalfspace}
        \bfseries\huge
        \thesistitle
      \end{onehalfspace}
      \vspace*{-2ex}
    \rule{\linewidth}{2.5pt}

    \vspace*{12ex}

    {\LARGE PhD Thesis} \\
    \vspace*{2ex}
    {\LARGE September 2019} \\
    \vspace*{7ex}
    by \\
    \vspace*{7ex}
    {\Large \thesisauthor}
    \vspace*{20ex}

    {
      \Large
      \begin{tabular}{ll}
        Supervisors:    & Prof. Dr. Stefan Schael \\
                        & Prof. Dr. Christopher Wiebusch
      \end{tabular}
    }
  \end{center}
\end{titlepage}

\chapter*{Eidesstattliche Erklärung}

Nikolas Zimmermann erklärt hiermit, dass diese Dissertation und die darin dargelegten Inhalte die eigenen sind
und selbstständig, als Ergebnis der eigenen originären Forschung, generiert wurden. \\
\\
Hiermit erkläre ich an Eides statt
\begin{enumerate}
\item Diese Arbeit wurde vollständig oder größtenteils in der Phase als Doktorand dieser
Fakultät und Universität angefertigt;
\item Sofern irgendein Bestandteil dieser Dissertation zuvor für einen akademischen Abschluss
oder eine andere Qualifikation an dieser oder einer anderen Institution verwendet wurde,
wurde dies klar angezeigt;
\item Wenn immer andere eigene- oder Veröffentlichungen Dritter herangezogen wurden,
wurden diese klar benannt;
\item Wenn aus anderen eigenen- oder Veröffentlichungen Dritter zitiert wurde, wurde stets die
Quelle hierfür angegeben. Diese Dissertation ist vollständig meine eigene Arbeit, mit der
Ausnahme solcher Zitate;
\item Alle wesentlichen Quellen von Unterstützung wurden benannt;
\item Wenn immer ein Teil dieser Dissertation auf der Zusammenarbeit mit anderen basiert,
wurde von mir klar gekennzeichnet, was von anderen und was von mir selbst erarbeitet
wurde;
\item Teile dieser Arbeit wurden zuvor veröffentlicht und zwar in: Physical Review Letters~(\textbf{121} 051102, 2018)
\end{enumerate}

\bigskip
\bigskip
\bigskip
\bigskip
Aachen, 27. Februar 2020

\section*{Abstract}
{
  \medskip
  \begin{center}
    \begin{onehalfspace}
      \large\bfseries\thesistitle
    \end{onehalfspace}
  \end{center}

  \vspace*{-2ex}
  \noindent
  This thesis presents an analysis of the cosmic-ray electron and positron flux using the AMS-02 detector
  on the International Space Station as a function of time and energy. The time-averaged flux is integrated over
  6.5 years of AMS-02 science data and provides the electron and positron flux with unprecedented accuracy, covering the energy
  range from \SI{0.5}{\GeV} to \SI{1}{\TeV}. In total 28.39 million events were identified as electrons and 1.95 million as positrons.
  For each of the 88 Bartels rotation periods (27 days), within the 6.5 years, an individual electron and positron flux is derived
  spanning the energy range from \SIrange{1}{50}{\GeV}.

  \medskip
  \noindent
  The challenge of the analysis is to extract the small electron and positron signal in the overwhelming proton background present
  in cosmic rays. A detailed description of the analysis techniques is presented, including a thorough derivation of the systematic uncertainties.

  \medskip
  \noindent
  The main motivation for measuring the cosmic-ray electron and positron flux in a time-averaged way is to explore the energy
  dependence up to high energies in detail and search for structures in the spectrum. The traditional understanding is that electrons
  are primary cosmic rays, whereas positrons are believed to be secondaries, produced by collisions of primary protons with the interstellar medium.
  A clear deviation from the traditional understanding was discovered: the positron flux cannot be described by a single power
  law, nor by the sum of two power laws. The secondary production term plus an additional source term, with a
  finite cut-off energy, is necessary to describe the positron data. Above the cut-off energy, the positron flux is rapidly
  decreasing. The cut-off is established with a significance of $4\sigma$, providing strong evidence that a new source
  of cosmic-ray positrons was discovered, which is responsible for the rise of the positron flux, and its decrease at high
  energies when the source term contribution is vanishing. The origin of the source term remains unclear: both astrophysical
  sources, such as pulsars, and dark-matter annihilation are candidates to describe the positron flux data.

  \medskip
  \noindent
  The majority of the electrons is believed to come from one of the several astrophysical sources, each making a power law
  contribution to the electron flux. The electron flux was found to be well described by the sum of two power laws over
  the whole energy range, supporting the observation that more than one astrophysical source is responsible for the measured
  electron flux.

  \medskip
  \noindent
  For the first time, the charge-sign dependent modulation during solar maximum has been investigated by electrons and
  positrons alone, using the time-dependent fluxes derived in this thesis. Short-term effects such as Forbush decreases
  and solar flares were identified simultaneously in the electron and positron flux that cancel in the positron/electron
  ratio. Long-term effects are revealed in the positron/electron ratio: A smooth transition from one value to another, after
  the polarity reversal of the solar magnetic field in July 2013. The transition magnitude is decreasing as a function of
  energy, which was predicated by solar modulation models that incorporate drift effects. This novel dataset allows one to build
  sophisticated models of solar modulation that can predict the time-dependence of both the electron and positron flux in future.
  This knowledge will allow a precise modelling of the interstellar electron flux and positron flux from low energies in the \SI{}{\GeV} regime up to the
  \SI{}{\TeV} regime.
}

\clearpage

\section*{Zusammenfassung}
{
  \medskip
  \begin{center}
    \begin{onehalfspace}
      \large\bfseries\thesistitleGerman
    \end{onehalfspace}
  \end{center}

  \vspace*{-2ex}
  \noindent
  In der vorliegenden Dissertation wird die Analyse der kosmischen Elektron- und Positronflüsse, gemessen mit dem AMS-02 Detektor auf der
  Internationalen Raumstation, in zwei Varianten vorgestellt: zeitgemittelt und zeitabhängig. Die zeitgemittelten Flüsse
  decken 6.5 Jahre AMS-02 Daten ab und erlauben es die Elektron- und Positronflüsse mit unerreichter Präzision zu messen,
  im Energieintervall von \SI{0.5}{\GeV} bis \SI{1}{\TeV}. Insgesamt 28.39 Millionen Ereignisse wurden als Elektronen identifiziert
  und 1.95 Millionen als Positronen. Für jede der 88 Sonnenrotationsperioden (\enquote{Bartels rotation}) innerhalb der
  6.5 Jahre wird zusätzlich jeweils ein Elektron- und ein Positronfluss bestimmt im Energieintervall von \SIrange{1}{50}{\GeV}.

  \medskip
  \noindent
  Die Herausforderung der Analyse bestand darin, das kleine Elektronen- und Positronensignal aus dem großen Protonenuntergrund zu
  extrahieren, der die kosmische Strahlung dominiert. Eine detaillierte Beschreibung der Analysetechniken, sowie eine gründliche Diskussion
  aller relevanten systematischen Unsicherheiten wird präsentiert.

  \medskip
  \noindent
  Die Messung der zeitgemittelten kosmischen Elektron- und Positronflüsse erlaubt es die Spektren bis zu den höchsten Energien im
  Detail zu untersuchen und nach unbekannten Strukturen in der Energieabhängigkeit zu suchen. Nach traditionellem Verständnis sind
  Elektronen primäre kosmische Teilchen und Positronen sekundäre, die erst bei der Kollision von primären Protonen mit der Materie
  im interstellaren Raum entstehen. Eine klare Abweichung vom traditionellen Verständnis wurde beobachtet:
  Der Positronfluss lässt sich nicht als Potenzgesetz oder als Summe von zwei Potenzgesetzen beschreiben. Die Summe eines Potenzgesetzes
  mit einem zusätzlichem Quellterm, der zu hohen Energien exponentiell abfällt, beschreibt den gemessenen Positronfluss. Oberhalb der Energiegrenze
  verliert der Quellterm seine Bedeutung und der Positronfluss fällt stark ab. Die Existenz einer solchen Energiegrenze wurde
  mit einer Signifikanz von $4\sigma$ bestimmt, was ein klarer Hinweis auf eine neue Quelle von kosmischen Positronen ist.
  Der Ursprung des Quellterms ist unbekannt: Es könnte sich um eine astrophysikalisches Quelle, wie einen Pulsar, handeln oder
  die Signatur eines annihilierenden dunkle Materie Teilchens.

  \medskip
  \noindent
  Der Großteil der Elektronen sollte von einer der bekannten astrophysikalischen Quellen kommen, wobei das Spektrum jeder
  Quelle mit einem Potenzgesetz zum gemessenen Elektronfluss beiträgt. Der Elektronfluss lässt sich als Summe von nur
  zwei Potenzgesetzen über den gesamten Energiebereich beschreiben, welches die Beobachtung stützt, dass mehr als eine
  astrophysikalsiche Quelle zum gemessenen Elektronfluss beiträgt.

  \medskip
  \noindent
  Zum ersten Mal konnte die Ladungsabhängigkeit der solaren Modulation allein mit Elektronen und Positronen untersucht werden.
  Effekte auf kurzen Zeitskalen, wie der \enquote{Forbush-Effekt}
  oder Sonneneruptionen lassen sich sowohl in Elektronen als auch Positronen zeitgleich identifizieren. Diese kurzzeitigen Effekte heben
  sich auf, wenn man das Verhältnis Positronen zu Elektronen betrachtet und ein klarer Langzeiteffekt bleibt übrig: der Übergang von einem
  Plateau zu einem anderen, nach der Polaritätsänderung des Sonnenmagnetfeldes im Juli 2013. Die Amplitude des Übergangs nimmt als Funktion
  der Energie ab und deckt sich daher mit Vorhersagen von Modellen der solaren Modulation, die Drifteffekte berücksichtigen. Die zeitabhängigen
  Flüsse erlauben es detaillierte Modelle der solaren Modulation zu entwickeln, die es ermöglichen werden, die zukünftige Zeitabhängigkeit
  der Elektron- und Positronflüsse vorherzusagen. Damit können dann fundierte Modelle der interstellaren Elektron- und Positronflüsse erstellt werden, die
  den gesamten Energiebereich von wenigen \SI{}{\GeV} bis zu \SI{}{\TeV} Energien beschreiben.
}

\pagestyle{scrplain}
\tableofcontents

\mainmatter
\pagestyle{scrheadings}

%
\chapter{Introduction}
\label{sec:intro}

\chapquote{"... In order to make further progress, particularly in the field of cosmic rays, it will be necessary to apply all our resources and apparatus simultaneously and side-by-side; an effort which has not yet been made, or at least, only to a limited extent..."}{\textup{Victor F. Hess}}{Nobel Lecture - 1936}

The search for the fundamental building blocks of nature, their construction principles
and the understanding of their interactions occupied generations of
particle physicists. Nowadays large accelerators, such as the \gls{LHC} at CERN, are used
to collide particles: the collision products are analyzed using sophisticated detectors.
The decades of effort lead to the construction of a fundamental model,
the \textit{Standard Model}, describing the fundamental forces in the Universe including their
interactions on the smallest length scales. The \textit{Standard Model} was tested countless times
and is the most accurate description of the microscopic structure of nature today.

However the ultimate laboratory to study particles at the highest possible energies is the Universe itself.
In 1912 Victor F. Hess discovered the cosmic rays, when he found that an electroscope discharged
more rapidly as he ascended in a balloon. He attributed this to a source of radiation entering
the atmosphere from above, and in 1936 was awarded the Nobel prize for his discovery. The famous quote
from his Nobel Lecture - see above - paved the way for numerous experiments in the field of cosmic rays,
which all share the same goal: unveil the nature of cosmic rays, trace down their origin to gain understanding
of the fundamental physics behind cosmic rays, such as the production and propagation mechanisms.

It was clear to physicists since the discovery of cosmic rays that the best place to study cosmic rays
is not the Earth itself, since the atmosphere disturbs the measurement. However it took more than half a century
until a unique, space-borne experiment was proposed in 1994: the \gls{AMS}~\cite{Ahlen1994}, to be
installed on the \gls{ISS}. Previously only balloon-borne experiments were used to probe cosmic rays above
the atmosphere, which are unable to provide true long term measurements of the cosmic rays.

The precursor experiment AMS-01 flew on the shuttle Discovery on the STS-91 mission, from June, \nth{2}
to June \nth{12} 1998 demonstrating that it is possible to operate a particle physics detector in space,
which was believed to be extremely hard to realize, due to the harsh environment: enormous mechanical stress
during the launch on a rocket, ever changing temperature conditions, power consumption, etc.
The precursor experiment was successful, and already generated several interesting scientific results, see e.g.
Refs.~\cite{Alcaraz2000,Aguilar2007}.

After more than a decade of research and construction AMS-02 was launched into space with the Space Shuttle
STS-134 mission and installed onboard the ISS on May, \nth{19} 2011, to perform a long duration
mission of fundamental physics research in space. The primary scientific goal is to measure the composition
and energy spectra of charged cosmic rays, to look for primordial anti-matter and to explore the nature
of dark matter.

In this thesis, the precision measurement of the time dependence induced by solar activity on electron
and positron fluxes at energies between \SIrange{1}{50}{\GeV} is presented and a measurement of the time-averaged
electron and positron fluxes between \SIrange{0.5}{1000}{\GeV} based on 6.5 years of AMS-02 science data.
Electrons ($e^{-}$) and positrons ($e^{+}$) are a rare component of the cosmic rays since they present only
\SIapprox{1}{\percent} and \SIapprox{0.1}{\percent} of the total cosmic-ray flux, which is dominated
by protons (\SIapprox{90}{\percent}) and helium (\SIapprox{8}{\percent}). It is therefore challenging
to extract the $e^{\pm}$ signal from the overwhelming proton and helium background.

The time-dependent fluxes allow for comprehensive studies of the energy and charge-sign dependence on the
time scale of months, related to solar activity~\cite{Potgieter1993a,Potgieter1993b,Gopalswamy2003}, and long-term effects
related to the periodic 22 year cycle of the heliospheric magnetic field~\cite{Ferreira2003}.

Time-dependent structures in the energy spectra of all cosmic rays emerge from solar modulation
when charged particles enter the heliosphere~\cite{Potgieter2013} - whereas interstellar charged
cosmic rays are thought to be stable on the time scale of decades~\cite{Strong1998,Maurin2001}.

Solar modulation involves convective, diffusive, particle drift, and adiabatic energy loss processes.
It is well established that only the drift motion induces a dependence of solar modulation on the cosmic-ray
particle charge sign~\cite{Potgieter2014}. Measuring the electron and positron fluxes simultaneously with high statistics
offers a unique way to study charge-sign dependent solar modulation effects.

Furthermore the understanding of low-energy variations of the fluxes is important when modelling the high-energy
part of the fluxes, since the effect of solar modulation alters the energy dependence of cosmic-ray fluxes
up to \SIapprox{20}{\GeV}, which will be shown in this thesis. Therefore to derive cosmic-ray flux models describing
the full energy range \SIrange{0.5}{1000}{\GeV}, the low-energy variations need to be taken into account, properly.

\bigskip
In \cref{sec:cosmic-ray} an overview of the cosmic-ray physics is given, followed by an introduction
to the AMS-02 instrument in \cref{sec:detector}. A detailed presentation of the analysis techniques,
the arising challenges and the systematic uncertainties associated with the measurement is given in
\cref{sec:analysis}. A presentation and discussion of the results in \cref{sec:results} and a short
summary in \cref{sec:summary} will conclude this dissertation.

\chapter{Cosmic rays}
\label{sec:cosmic-ray}

The term \textit{cosmic rays} refers to high energy charged particles and $\gamma$-rays with
extraterrestrial origin. Charged cosmic rays mainly consist of protons (\SIapprox{90}{\percent}), helium (\SIapprox{8}{\percent}),
electrons (\SIapprox{1}{\percent}) and heavy nuclei (\SIapprox{1}{\percent}) and a small fraction
of anti-matter, such as positrons (see \cref{sec:cosmic-ray-composition}).

This chapter focuses on charged particles, especially electrons and positrons
produced from outside the solar system, which is the topic of this thesis.
Charged cosmic rays are messengers for the Universe, encoding information about their origin
and travel through the Universe, before reaching Earth. The fluxes of cosmic rays were
extensively studied during the past century, but the origin as well as the production
mechanism is still not completely understood. The leading explanation of the production of
cosmic rays is the \textit{supernova} scenario, which will be described in \cref{sec:cosmic-ray-production-and-acceleration}.

The understanding of the propagation is important, since cosmic rays travel a long distance
and continuously interact with the \gls{ISM} (\cref{sec:cosmic-ray-propagation-ism}), the
Sun's magnetic field (\cref{sec:cosmic-ray-propagation-heliosphere}) and the magnetic field
of the Earth (\cref{sec:cosmic-ray-propagation-magnetosphere}) before they eventually
reach Earth.

Apart from understanding their origin, the study on cosmic-ray production and propagation
helps to understand the composition of the Universe, especially at early times, right after
the \textit{Big Bang}~\cite{Mathews2017}. Many observations suggest that the observable matter
only consists of \SIapprox{5}{\percent} of the mass-energy density of Universe~\cite{Akrami2018}.
Part of the missing mass-energy density can be attributed to \textit{Dark Matter}. Indirect dark
matter searches, probing their existence by searching for the product of their decay and
annihilation, will be briefly explained in \cref{sec:cosmic-ray-dark-matter}.

\section{Composition and energy dependence}
\label{sec:cosmic-ray-composition}

\begin{figure}[H]
  \centering
  \includegraphics[width=0.8\linewidth]{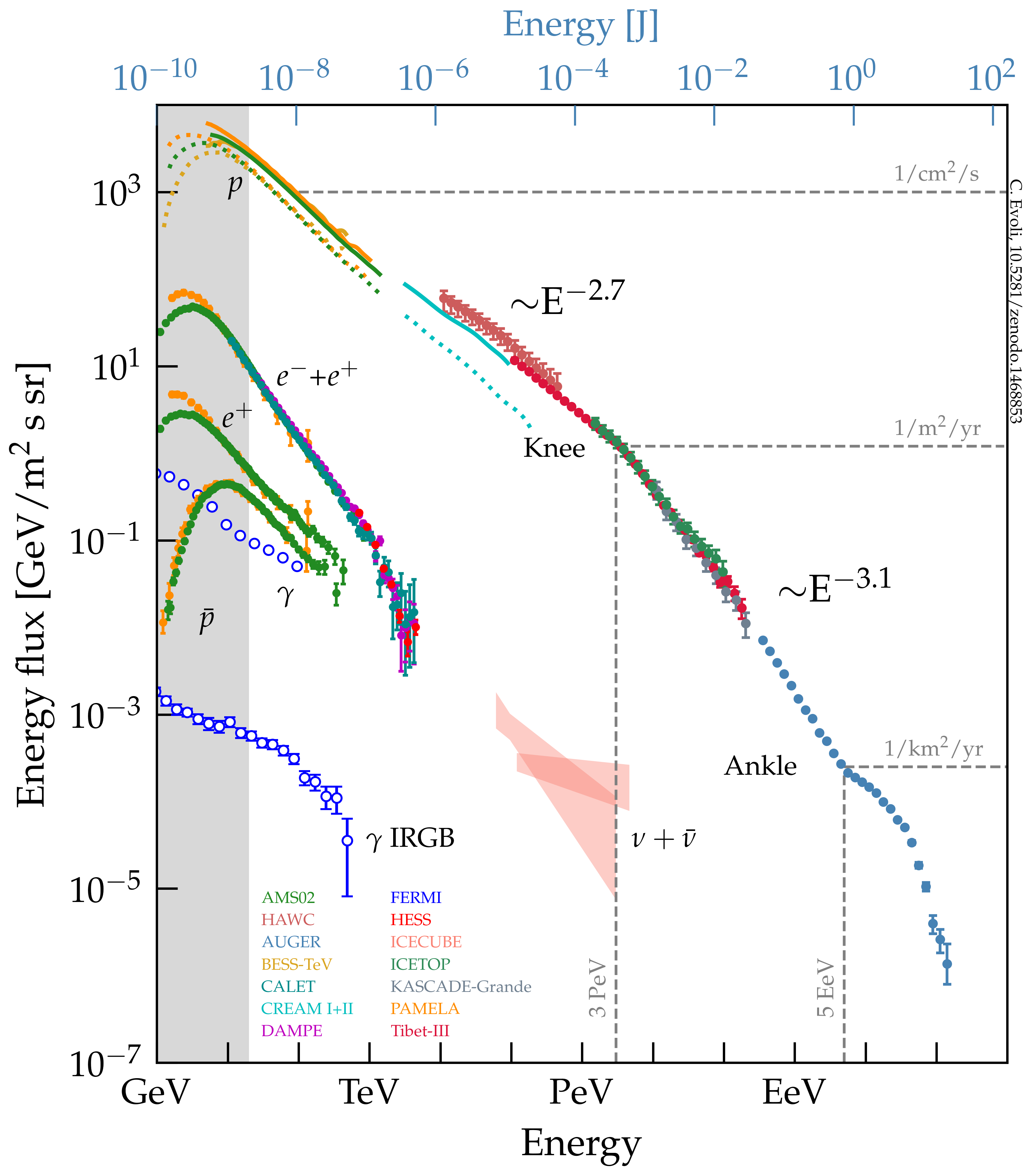}
  \caption{The cosmic-ray energy spectrum. Credit: Ref.~\cite{Evoli2018}}
  \label{fig:cosmic-rays-all-particle-spectrum}
\end{figure}

\Cref{fig:cosmic-rays-all-particle-spectrum} shows an overview of the cosmic-ray measurements
of the past two decades. Cosmic rays are categorized in two main categories: \textbf{Primaries} and \textbf{Secondaries}.
Primary cosmic rays are particles produced and accelerated at their sources, whereas secondary
cosmic rays are produced by primaries interacting with the \gls{ISM} during propagation.
A detailed introduction on cosmic rays can be found in Ref.~\cite{Longair2011}.

The energy dependence for most of the cosmic rays follows a power law:

\begin{equation}
  \label{eq:power-law}
  \frac{\diff\text{N}}{\diff\text{E}} \propto E^{-\gamma},
\end{equation}

where $\gamma$ is the \textit{spectral index}. The overall cosmic-ray flux is well
described with a spectral index of $\gamma \approx 2.7$ for energies between \SIapprox{1}{\GeV} and
\SIapprox{3}{\PeV} (\textit{knee}), as shown in \cref{fig:cosmic-rays-all-particle-spectrum}. Above
the \textit{knee} the spectral index softens to $\gamma \approx 3.1$, until the \textit{ankle} is reached
at around \SIapprox{1}{\EeV}.

The transitions of the spectral index is believed to originate in differences in the acceleration and
propagation processes, which will be discussed in the following sections.

\section{Production and acceleration}
\label{sec:cosmic-ray-production-and-acceleration}

Since the first detection of cosmic rays early in the \nth{20} century there have been many efforts
to find out the origin of these particles and their production mechanism. An intrinsic problem is that
charged cosmic rays are deflected in the interstellar and intergalactic magnetic field, which makes it
impossible to precisely track down their origin. However cosmic rays at energies above \SIScientific{1e5}{\TeV}
are most likely to have an extragalactic origin -- see recent anisotropy studies from the Auger Collaboration in Ref.~\cite{Aab2017}.

Particles relevant to the measurement with space-born experiments, such as AMS-02, are mainly of galactic
origin. The most compelling scenario for the production of Galactic cosmic rays are \textit{supernovae}.

A supernova~\cite{Bethe2003} is a powerful and luminous stellar explosion, with an optical luminosity comparable to that of an
entire Galaxy, lasting for weeks or longer.

According to Ref.~\cite{Raffelt2002} there are two entirely different classes of supernovae. One physical class
are the type Ia \gls{SN} explosions. These are thought to occur, when a carbon-oxygen white dwarf accretes
matter from a companion star until it reaches the Chandrasekhar limit~\cite{Chandrasekhar1932} ($M \gtrapprox 1.45 \cdot M_{\odot}$)
and begins to collapse. This triggers a nuclear explosion, powered by the fusion of carbon and oxygen to heavier nuclei.
The explosion disrupts the original star (\enquote{progenitor}) entirely. This process is called \textit{thermonuclear supernova}.

The other class of supernova explosions mark the evolutionary end of massive stars with a mass of more than $8 \cdot M_{\odot}$.
Such stars exhibit an onion structure with several burning shells and an iron core. The iron cores is supported by
electron degeneracy pressure, strongly cooled via neutrinos, and surrounded by burning shells of \ce{Si}, \ce{O}, \ce{C}, etc.
The complex \ce{Si} shell burning continues to grow iron cores up to the Chandrasekhar limit. Once an iron core attains this
critical mass, an unstable gravitational collapse starts~\cite{Raffelt2002,Bethe2003}, leading to a massive explosion.
The progenitor either collapses to a neutron star or a black hole, or it is completely destroyed. This processed is called \textit{hydrodynamic supernova}.

Thermonuclear supernovae occur typically once per 300 years, whereas hydrodynamic supernovae
happen once per 30 to 50 years in our Galaxy. Energy is released by supernovae into space with an
average rate of \SIapproxScientific{1e42}{\erg\per\second}. Only a few percent of the energy would be necessary
to accelerate primary particles up to \SIapproxScientific{1e20}{\eV} to describe the observed flux, outlined in \cref{fig:cosmic-rays-all-particle-spectrum}.

According to Refs.~\cite{Berezhko2008,Blasi2011} particles are accelerated in the shock wave inside the \gls{SNRs}.
The acceleration mechanism that is usually assumed to work in SNRs is \gls{DSA}~\cite{Bell1978,Blandford1987},
converting \SIrange{10}{20}{\percent} of the kinetic energy of the supernova shell into cosmic rays.

The \gls{DSA} mechanism produces power law energy spectra, consistent with the observations (\cref{fig:cosmic-rays-all-particle-spectrum}) and
is therefore the leading candidate to describe the cosmic-ray spectra. Furthermore recent observation of several \gls{SNRs} provide strong
evidence that charged particles are accelerated in the shock of \gls{SNRs}~\cite{Uchiyama2007,CassamChenai2008,Abdo2010}.

\section{Propagation through the interstellar medium}
\label{sec:cosmic-ray-propagation-ism}

When charged cosmic rays travel through the Galaxy, their spectra and their composition
changes due to a variety of physical processes that take place. Hadronic interactions of
primary protons and nuclei with the \gls{ISM} create secondary charged particles and $\gamma$-rays
by $\pi^0$ production. Electrons/positrons lose energy by bremsstrahlung (interaction with the \gls{ISM}),
by synchrotron radiation (interaction with the Galactic magnetic field), as well as inverse Compton
scattering (scattering on the photons of the \gls{CMB}~\cite{Penzias1965}).

Due to their charge, charged cosmic rays are scattered on magneto-hydrodynamic waves and discontinuities, which implies
that their trajectories do not follow straight lines. On large scales, charged particles effectively perform a random walk,
and therefore, a diffusion model is the appropriate approximative description of the propagation process.

The diffusion model with the inclusion of convection provides the most adequate description of cosmic-ray transport in the Galaxy at
energies below \SIapproxScientific{1e17}{\eV}, as described in Ref.~\cite{Strong2007}.
A general way to describe cosmic-ray propagation for a particular species is given by:

\begin{equation}
  \label{eq:prop-eq}
  \begin{aligned}
    \frac{\partial \psi(\vec{r}, p, t)}{\partial t} &= q(\vec{r}, p, t) + \vec{\nabla}\cdot(D_{xx} \vec{\nabla} \psi - \vec{V}\psi) \\
                                                    &+ \frac{\partial}{\partial p} p^2 D_{pp} \frac{\partial}{\partial p} \frac{1}{p^2} \psi
                                                    - \frac{\partial}{\partial p} \left[ \dot{p} \psi - \frac{p}{3} (\vec{\nabla} \cdot \vec{V}) \psi\right]
                                                    - \frac{1}{\tau_{f}} \psi - \frac{1}{\tau_{r}} \psi,
  \end{aligned}
\end{equation}

where $\psi(\vec{r}, p, t)$ is the cosmic-ray density per unit of total particle momentum $p$ at position $\vec{r}$,
$\psi(p) \diff p = 4 \phi p^2 f(\vec{p}) \diff p$ in terms of phase-space density $f(\vec{p})$.
The terms on the right hand side of \cref{eq:prop-eq} are explained in the following:

\begin{enumerate}
  \item $q(\vec{r},p)$ is the \textbf{source term} including primary, spallation and decay contribution. The injection spectrum of nucleons is assumed to be a power law in momentum,
    $\diff q(p) / \diff p \propto p^{-\gamma}$.

  \item $D_{xx}$ is the \textbf{spatial diffusion coefficient}. Typical numeric values are $D_{xx} \approx \SI{3e28}{\centi\meter\squared\per\second} - \SI{5e28}{\centi\meter\squared\per\second}$
    at \SI{1}{\GeV}$/n$. $D_{xx}$ increases with rigidity as $R^{\alpha}$, where $\alpha$ is in the range \SIrange{0.3}{0.6}{}, depending on the empirical diffusion model.

  \item Galactic winds give rise to a \textbf{convective transport} of charged particles and are described by the convection velocity $\vec{V}$. $\vec{V}$ is assumed to increase linearly
    with the distance from the Galactic plane, implying a constant adiabatic energy loss.

  \item Besides the spatial diffusion, the scattering of charged particles on randomly moving magneto-hydrodynamic waves leads to stochastic reacceleration.
    The \textbf{diffuse reacceleration} is described as diffusion in momentum space and is determined by the coefficient $D_{pp}$. The relation between the
    diffusion coefficient $D_{pp}$ and the spatial diffusion coefficient $D_{xx}$ is approximated by

   \begin{equation*}
     D_{pp} = \frac{p^2 v_{A}^2}{9 D_{xx}},
   \end{equation*}

   where the Alfvén speed $v_{A}$~\cite{Alfven1942} describes the characteristic velocity of the weak disturbances propagating in a magnetic field. Typical values
   are $v_{A}$ \SIapprox{30}{\kilo\meter\per\second}.

  \item $\dot{p} = \diff p / \diff t$ is the \textbf{momentum gain or loss rate}, and the term involving $\vec{\nabla} \cdot \vec{V}$ describes \textbf{adiabatic momentum gain or loss}
    in the non-uniform flow of gas with a frozen-in magnetic field whose inhomogeneities scatter the cosmic rays.

  \item $\tau_{f}$ is the time scale for loss by \textbf{fragmentation}, and $\tau_{r}$ is the time scale for \textbf{radioactive decay}.
\end{enumerate}

The best approach to solve the cosmic-ray propagation equation is to use a state of the art numerical model, such as
Galprop~\cite{Moskalenko1997,Strong1998,Moskalenko2002,Ptuskin2006,Strong2007}, which aims to give simultaneous
predictions of all relevant observations, while incorporating as much as current information as possible, e.g.
on source distribution and the galactic structure.

To compare the predictions of the Galprop model with the fluxes measured at Earth, e.g. using AMS-02, one first has to consider
the interactions of the cosmic rays with the solar wind and the magnetic field of the Earth, which will be explained
in the following sections.

\section{Propagation through the heliosphere}
\label{sec:cosmic-ray-propagation-heliosphere}

\begin{figure}[H]
  \centering
  \includegraphics[width=0.7\linewidth]{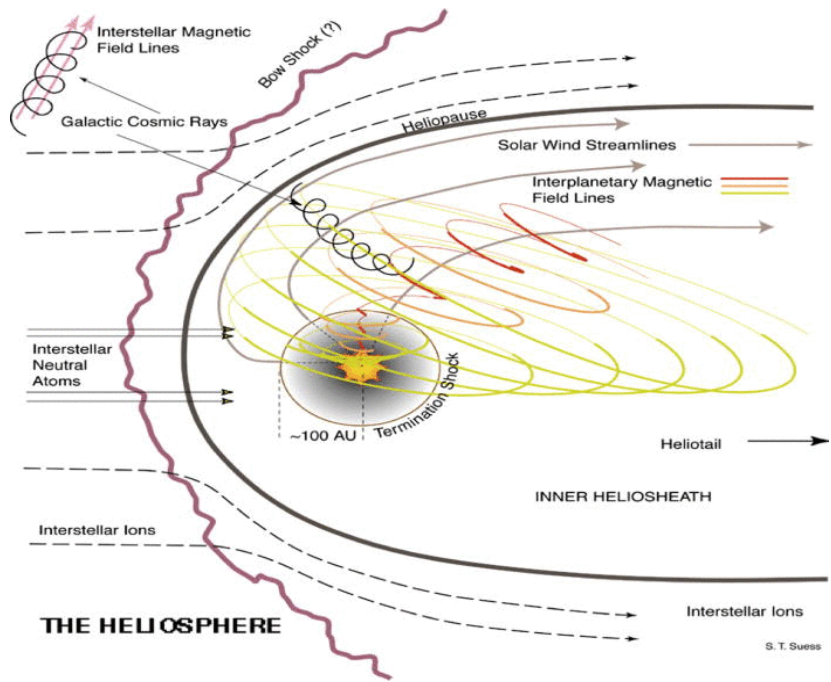}
  \caption{Sketch of the heliosphere. Credit: S. T. Suess.}
  \label{fig:heliosphere}
\end{figure}

When charged cosmic rays enter the solar system, they interact with the magnetic field originating from the Sun, as illustrated in \cref{fig:heliosphere}.
Furthermore the Sun emits a stream of charged particles from its upper atmosphere - the \textbf{solar wind}. It is a plasma
mostly consisting of electrons, protons and alpha particles with a kinetic energy between \SIrange{0.5}{10}{\keV}. The plasma is highly electrically conductive,
meaning the magnetic field lines and the plasma flows are effectively \enquote{frozen} together. The magnetic pressure greatly exceeds the plasma pressure
and thus the plasma is structured and confined by the magnetic field. The magnetic field is effectively carried out up to the termination shock where the plasma
slows down to subsonic speed: the coronal magnetic field is dragged out by the solar wind and forms the \textbf{\gls{HMF}}~\cite{Owens2013}.

The structure and dynamics of the \gls{HMF} are key to understanding and forecasting space weather, as the \gls{HMF} directly couples the Sun with planetary
magnetospheres, as well as directing the flow of charged particles through the heliosphere.

The magnetic field boundary separating oppositely directed magnetic field lines originating from the northern and southern poles of the Sun is
carried out by the solar wind to form the \textbf{\gls{HCS}}, a large scale boundary which extends throughout the Sun's equatorial plane in the heliosphere.
A small electrical current (\SIvarApproxScientific{I}{1e-10}{\ampere\per\meter\squared}) flows within the current sheet, which is approximately
\SIapprox{10000}{\kilo\meter} thick near the orbit of the Earth.

\begin{figure}[H]
  \centering
  \includegraphics[width=0.4\linewidth]{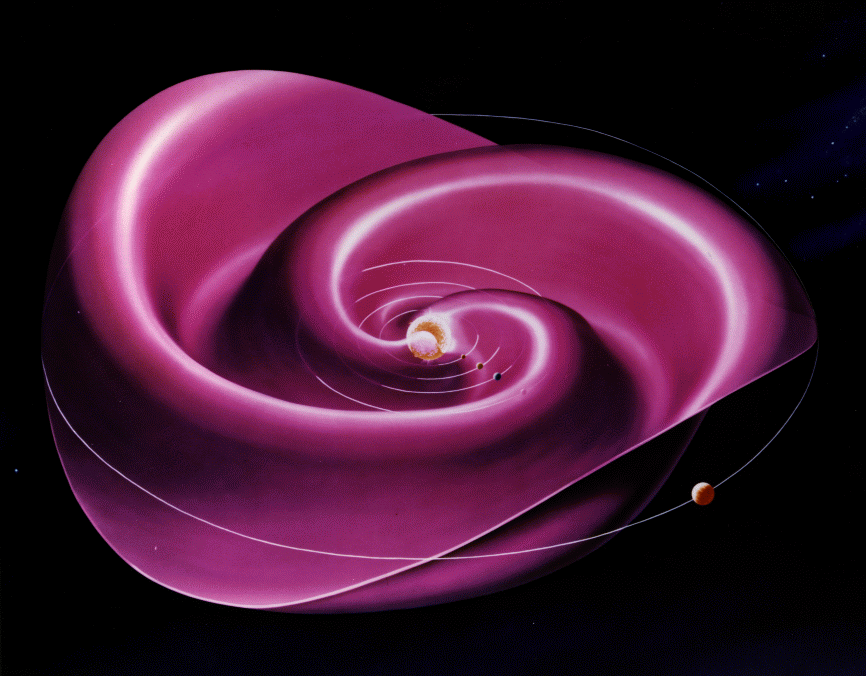}
  \caption{Heliospheric current sheet. Credit: NASA GSFC.}
  \label{fig:heliospheric-current-sheet}
\end{figure}

The shape results from the influence of the Sun's rotating magnetic field on the plasma in the interplanetary medium, as illustrated in \cref{fig:heliospheric-current-sheet}.
The current sheet evolves slowly, over timescales of months, as the Sun's magnetic field changes in response to the emergence and decay of solar active regions, disrupted on
short time scales by e.g. \gls{CME}~\cite{Gopalswamy2003}. An essential feature of the \gls{HCS} is the tilt of the Sun's magnetic dipole with respect to the rotation axis,
characterized by the \textbf{Tilt Angle $\theta_{t}$}. The tilt angle is strongly correlated with the Sun's magnetic activity: a small tilt angle (\SIapprox{5}{\degree})
corresponds to quiet periods and a large title angle to active periods.

The magnetic activity of the Sun follows a nearly periodic 11 year cycle~\cite{Ferreira2003}, where the magnetic field configuration oscillates between a toroidal and
a polodial configuration~\cite{Babcock1961}. The strength of the solar radiation, the amount of coronal mass ejections, the number of sunspots,
the number of solar flares, etc. are all modulated by the magnetic activity of the Sun, from active to quiet periods with a periodicity of 11 years.
Cosmic-ray particles entering the heliosphere are scattered on the \gls{HMF}, similar to the scattering on the galactic magnetic field, when cosmic rays travel through the \gls{ISM}.
To understand the predictions from galactic propagation models (e.g. Galprop, see \cref{sec:cosmic-ray-propagation-ism}), the solar modulation needs to be taken into
account.

A simple model was proposed in Ref.~\cite{Gleeson1968}, describing solar modulation as diffusion through the \gls{HMF}, including convection by the outward motion
of the solar wind and adiabatic declaration of the cosmic rays. In the \textit{force-field approximation}, that is used for the modelling of the measured fluxes in this
thesis (\cref{sec:results}), the effect of solar modulation can be described by a single parameter, the fisk potential $\phi$, that depends on the solar wind speed $V(\vec{r}, t)$ and the diffusion
coefficient $\kappa$ as given by:

\begin{equation}
  \label{eq:force-field-phi}
  \phi(\vec{r}, t) = \frac{(E + m) T}{3 E} \int\limits_{r_{E}}^{r_{b}} \frac{V(x, t)}{\kappa(x, E, t)} \diff x,
\end{equation}

where $E$ is the total energy, $T$ the kinetic energy and $m$ the mass of a cosmic-ray particle. The integral is evaluated from the location
of the Earth $r_{E}$ to the boundary of the heliosphere $r_{b}$. The interstellar cosmic-ray flux $J_{\text{IS}}$ is related to the locally
observed flux $J$ by:

\begin{equation}
  \label{eq:force-field-solar-mod}
  J(E) = \frac{E^2 - m^2}{(E + \abs{Z} \phi)^2 - m^2} \cdot J_{\text{IS}}(E + \abs{Z} \phi),
\end{equation}

where $Z$ is the charge of the cosmic-ray particle. The modulation parameter $\phi$ has the dimension of a rigidity and is of the order of
\SIapprox{500}{\mega\volt}. $\phi$ is a model dependent quantity and changes with time over a solar cycle. Furthermore it has to be derived
for each cosmic-ray species individually. The simple description allows one to easily treat the solar modulation when modelling the energy
dependence of cosmic-ray spectra, in an approximative way.

\section{Propagation through the magnetosphere}
\label{sec:cosmic-ray-propagation-magnetosphere}

\begin{figure}[H]
  \centering
  \includegraphics[width=0.4\linewidth]{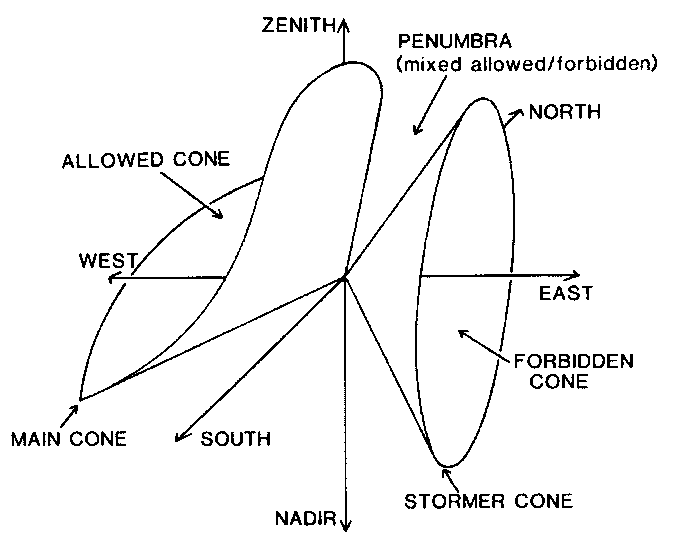}
  \caption{The geomagnetic cutoff as defined by Størmer~\cite{Stoermer1956}. Credit: Ref.~\cite{Smart2005}}
  \label{fig:cut-off-interpretation}
\end{figure}

The Earth's magnetosphere is the final barrier for cosmic-ray particles, before they can be measured at Earth or in low orbits,
such as the orbit of the \gls{ISS}. Low energetic particles cannot penetrate the magnetosphere and get deflected away from Earth.
This effect is known as \textbf{geomagnetic cut-off}. For a dipole field, Størmer derived an axial symmetric cone towards east direction
in which positively charged particles below a specific rigidity cannot enter (\textit{forbidden cone}), as illustrated in \cref{fig:cut-off-interpretation}.
The size of the \textit{forbidden cone} depends on the energy of the particle, a low energetic cosmic-ray particle corresponds to a large \textit{forbidden cone}.
For negatively charged particles the cones are mirrored: the \textit{forbidden cone} points towards the west direction.

The minimum rigidity needed to penetrate the magnetosphere is called the cut-off rigidity $R_{c}$. Following Ref.~\cite{Smart2005}, the
magnetic field can be modelled as dipole field in the vicinity of the Earth and the cut-off rigidity is given by:

\begin{equation}
  \label{eq:cut-off}
  R_{c} = \frac{M \cos^{4}{\lambda}}{r^2 \left(1 + \sqrt{1 - \sin{\epsilon} \sin{\delta} \cos^{3}{\lambda}}\right)^{2}},
\end{equation}

where $M$ is the dipole moment, $\lambda$ is the magnetic latitude, $\epsilon$ is zenith angle,
$\delta$ the azimuthal angle to the north and $r$ is the distance from the dipole center.

For particles entering radially to the dipole the vertical cut-off rigidity $R_{vc}$ can be approximated as: $R_{vc} = 14.9 \cos^{4}{\lambda} / r^2$,
where $r$ is in unit of earth radii.

Since the cut-off depends on the latitude $\lambda$ of the observer, the cut-off rigidity $R_{c}$ is lowest near the magnetic poles.
For particles arriving from any given direction, the cut-off depends on the azimuthal angle $\delta$ as well, leading to the
\textbf{east-west effect}~\cite{Ogawa1950}. A positively charged particle at the same zenith angle has a higher cut-off from the east
direction then for a negatively charged particle, and vice versa.

\begin{figure}[H]
  \centering
  \includegraphics[width=0.85\linewidth]{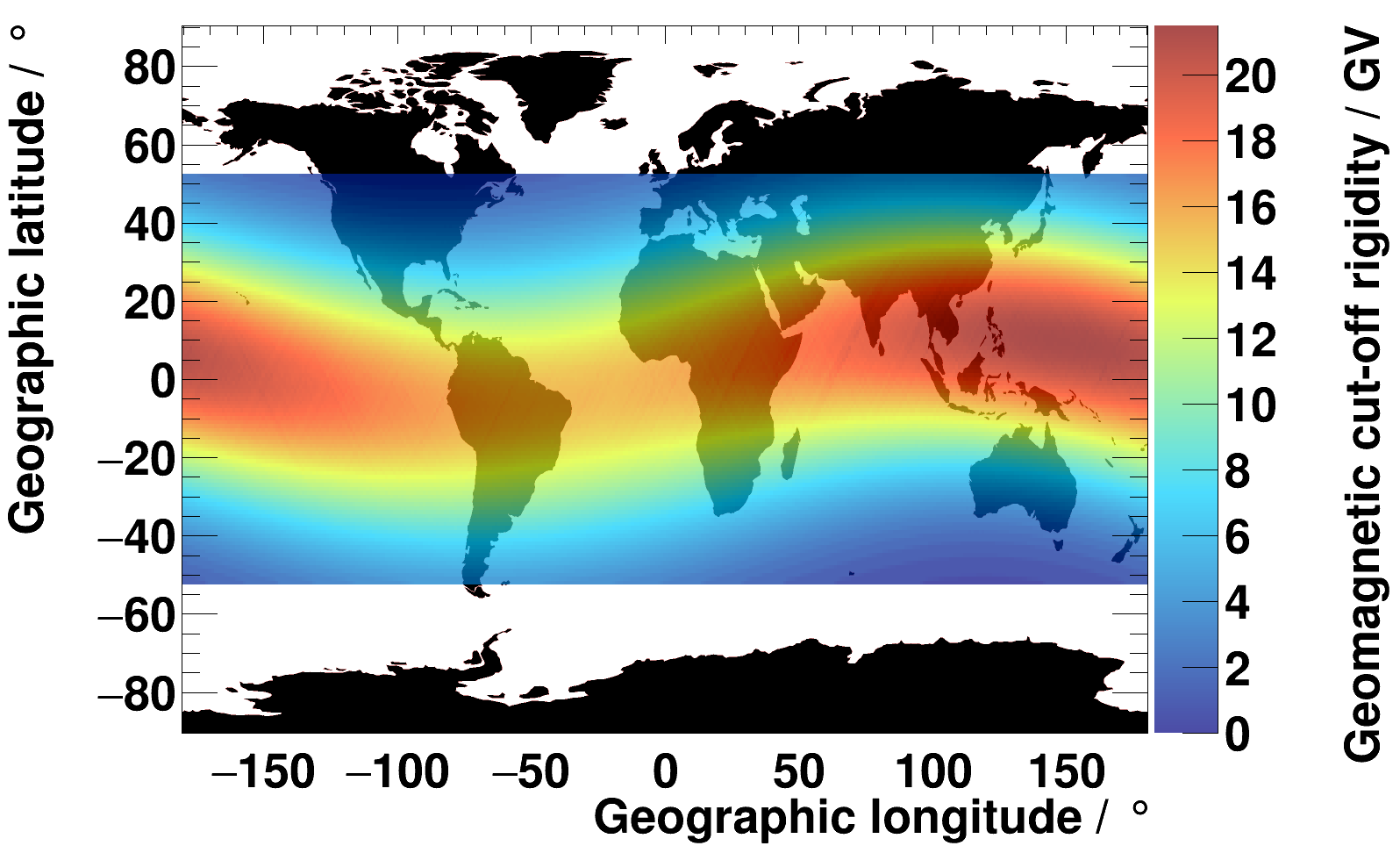}
  \caption{The plots shows the geomagnetic cut-off rigidity as function of the geomagnetic longitude/latitude.}
  \label{fig:cut-off-vs-iss-position}
\end{figure}

The geomagnetic cut-off effect must be taken into account when analyzing the fluxes, measured by AMS-02. AMS-02 is orbiting onboard the \gls{ISS}
between $\pm\,\SI{52}{\degree}$ latitude, with particle rates that vary from a few hundred \SI{}{\hertz} up to \SI{2000}{\hertz} close to the geomagnetic
poles. The geomagnetic cut-off rigidity $R_{c}$, computed using the Størmer formula \eqref{eq:cut-off} is
shown in \cref{fig:cut-off-vs-iss-position} as function of the geomagnetic longitude/latitude of the \gls{ISS}.

It is evident that low energetic particles can only be measured with AMS-02 near the pole regions, never in the equatorial region, where the cut-off rigidity is high.
Particles above \SIapprox{25}{\GeV} can be detected at any longitude/latitude. In order to measure primary cosmic rays, all particles must be excluded from analysis,
whose energy is smaller than the geomagnetic cut-off rigidity at a given geographic longitude/latitude.

\section{Electron and positron specific propagation aspects}
\label{sec:cosmic-ray-electron-positron-propagation}

In this section the electron and positron specific propagation aspects will be reviewed, since the electron and positron flux analysis
in the main topic of this thesis. The spallation and decay terms in the propagation equation \cref{eq:prop-eq} are irrelevant
for electrons or positrons, since there are no hadronic interactions.

As described in Ref.~\cite{Longair2011}, the main energy loss processes for electrons and positrons are
\textit{synchrotron radiation} and \textit{inverse Compton scattering}. The energy loss rate, in the ultra-relativistic limit $v \rightarrow c$, is given by:

\begin{equation}
  \begin{aligned}
    -&\left(\frac{\diff E}{\diff t}\right)_{\text{synch}} &= \frac{4}{3} \sigma_{T} c \gamma^2 U_{\text{mag}} &\approx \SI{6.6e4}{} \gamma^2 B^2, \\
    -&\left(\frac{\diff E}{\diff t}\right)_{\text{IC}}    &= \frac{4}{3} \sigma_{T} c \gamma^2 U_{\text{rad}} &,
  \end{aligned}
\end{equation}

for the \textit{synchrotron radiation} and the \textit{inverse Compton scattering} process, respectively.
$\sigma_{T}$ denotes the Thomson cross-section, $\gamma$ the Lorentz factor and $U$ the energy density
of the radiation fields. Typical values are $B \approx \SI{3e-10}{\tesla}$, $U_{\text{rad}} \approx \SI{6e5}{\eV\per\meter\cubed}$
and $U_{\text{rad}} \approx 3 \cdot U_{\text{mag}}$.

Inverse Compton losses and synchrotron losses are equally important for high energy electrons and positrons in the Galaxy.
Due to their small mass, electrons and positrons lose energy much faster compared to protons and other heavy nuclei.

According to numerical simulations (see Ref.~\cite{Panov2013}) a \SIapprox{1}{\TeV} electron or positron has a maximum life time
of $\tau \approx \SI{1e5}{\years}$ and a mean diffusion distance of only \SIapprox{1}{\kilo\parsec}. This suggests that the high
energy electrons and positrons measured at Earth, must come from nearby regions in the Galaxy.

\section{Electron and positron source candidates}
\label{sec:cosmic-ray-electron-positron-sources}

As described in \cref{sec:cosmic-ray-production-and-acceleration}, supernova remnants are widely believed to be responsible
for the origin of cosmic rays, as charged particles are accelerated in their strong shock wave. The main evidence for this
assumption comes from the observation of radio and $\gamma$-ray emission from \gls{SNRs}~\cite{Kobayashi2004}. The radio emission
stems from the synchrotron radiation of electrons in a strong magnetic field~\cite{KatzStone2000}, wheres $\gamma$-ray emission
has two possible sources: inverse Compton scattering~\cite{Aharonian2006} and the neutral pion decay~\cite{Ackermann2013}.

Only electrons are expected to be accelerated within \gls{SNRs}, not positrons. The flux of electrons injected into the \gls{ISM}
from \gls{SNRs} can be described as a power law: $\phi_{e^{-},\,\text{SNR}}(E) = A_{\text{SNR}} E^{-\gamma_{\text{SNR}}}$, where
$\gamma_{\text{SNR}} \approx 2$. The flux after propagation through the \gls{ISM} changes its energy dependence:
$\tilde{\phi}_{e^{-},\,\text{SNR}}(E) = A_{\text{SNR}} E^{-\gamma_{\text{SNR}} - \delta}$, where $\delta \approx 1$ due
to energy losses during propagation.

These assumptions reproduce the measured fluxes at Earth: the electron flux at energies between
\SIapprox{10}{\GeV} up to the \SI{}{\TeV} regime are known to follow a power law with a spectral index $\gamma \approx 3$~\cite{Aguilar2019a}.
Therefore electrons are considered as primary cosmic rays.

\bigskip
Positrons, on the other hand, are considered as secondary cosmic rays, since they are believed to be produced in collisions of
primary protons with the \gls{ISM}~\cite{Shen1968}. This leads to the creation of charged pion pairs, which further decay into
electron and positron pairs:

\begin{equation*}
  \begin{aligned}
    \pi^{-} \Rightarrow \mu^{-} + \bar{\nu_{\mu}}, \qquad & \mu^{-} \Rightarrow e^{-} + \bar{\nu_{e}} + \nu_{\mu} \\
    \pi^{+} \Rightarrow \mu^{+} + \nu_{\mu},       \qquad & \mu^{+} \Rightarrow e^{+} + \nu_{e} + \bar{\nu_{\mu}}
  \end{aligned}
\end{equation*}

These processes produce equal amounts of electrons and positrons following power laws:
\begin{equation}
  \label{eq:flux-secondary}
  \phi_{e^{+},\,\text{sec}}(E) = A_{\text{sec}} E^{-\gamma_{\text{sec}}}; \qquad\phi_{e^{-},\,\text{sec}}(E) = A_{\text{sec}} E^{-\gamma_{\text{sec}}}
\end{equation}

This scenario would lead to a positron flux and a positron fraction (ratio of positrons over positrons and electrons)
that would monotonically decrease with energy, in contrary to recent observations by PAMELA~\cite{Adriani2009} and AMS-02~\cite{Aguilar2013}.
Therefore an additional source of positrons must exist, besides the component associated with secondary production.

Astrophysical sources such as \textit{pulsars} or \textit{dark matter} annihilation are possible candidates for positron sources,
which will be explained in the following sections.

\subsection{Pulsars}
\label{sec:cosmic-ray-electron-positron-sources-pulsars}

A \textit{pulsar} is a rotating neutron star with a strong magnetic field that emits a beam of electromagnetic radiation~\cite{Hewish1968}.
This radiation can be observed only when the beam of emission is pointing towards Earth and is responsible for the pulsed appearance of the emission.
The electromagnetic radiation originates from the rotational energy of the neutron star, generating an electrical field from the movement of the
magnetic field. The protons and electrons from the surface of the star are accelerated and create an electromagnetic beam originating from
the poles of the magnetic field. The rotation periods of the pulsar are regular and the duration ranges from milliseconds to seconds, depending
on the age of the pulsar and its power. Due to the electromagnetic emission the rotation slows down over time.

A pulsar is an additional source of electrons and positrons, as the electrons extracted from the star surface propagate outwards the magnetic field
lines and emit high energy photons, which can convert into additional electron and positron pairs. These pairs can escape the magnetosphere and are injected
into the \gls{ISM}.

According to Ref.~\cite{Hooper2009}, the spectra from pulsars is harder than the secondary positron spectra and has a specific cut-off energy $E_{\text{cutoff}}$,
up to which the given pulsar can accelerate particles:

\begin{equation}
  \label{eq:flux-pulsar}
  \begin{aligned}
    \phi_{e^{+},\,\text{pul}}(E) &= A_{\text{pul}} E^{-\gamma_{\text{pul}}} E^{-E/E_{\text{cutoff}}} \\
    \phi_{e^{-},\,\text{pul}}(E) &= A_{\text{pul}} E^{-\gamma_{\text{pul}}} E^{-E/E_{\text{cutoff}}}
  \end{aligned}
\end{equation}

The sum of the positron flux from secondary production (\cref{eq:flux-secondary}) and the positron flux from pulsars (\cref{eq:flux-pulsar})
leads to a spectrum, which has an additional rise from \SI{}{\GeV} energies onwards, that gradually diminishes up to the \SI{}{\TeV} regime.
This observation makes \textit{pulsars} a promising candidate to explain the observed rise in the positron flux. According to Ref.~\cite{Verbiest2012}
there are a few candidate pulsars within the region of interest of \SIapprox{1}{\kilo\parsec}, as described in \cref{sec:cosmic-ray-electron-positron-propagation}.

However also \textit{dark matter} annihilation can lead to a similar rise in the positron flux, which will be shown in the following section.

\subsection{Dark matter}
\label{sec:cosmic-ray-dark-matter}

There is overwhelming evidence for the existence of a non-luminous form of matter, \textit{dark matter}, from the observation
of galactic rotation curves~\cite{Rubin1970}, the gravitational lensing effect~\cite{Einstein1936}, and the \gls{CMB}~\cite{Akrami2018}.
According to recent \gls{CMB} studies, dark matter makes up \SIapprox{85}{\percent} ($\Omega_{c}/(\Omega_{b} + \Omega_{c})$, according to $\Lambda$CDM model, see Ref.~\cite{Akrami2018})
of the total matter in the universe.

Almost nine decades have passed since the first suggestion of Zwicky that a non-luminous form of matter must exist, to explain the observations
of rotation curves of galaxies. The rotation curve of a Galaxy describes the rotation velocity $v(r)$ of objects at a distance $r$ from the galactic
centre. It it based on measurements of the Doppler shift of emissions or absorption lines. If all matter in a Galaxy
were located in the luminous disc, the rotation velocity would be $v(r) \propto r^{-1/2}$, in accordance with
Keplers third law. However the rotation curves were observed to be flat, $v = \text{const.}$, consistent with a mass
distribution of $\rho \propto r^{-2}$ that must extend beyond the visible disc. This fact gave rise to the idea
that a non-luminous form of matter, \textit{dark matter}, forms halos around the discs of galaxies.

The nature of dark matter still remains unexplained and constitutes one of the most exciting questions
in fundamental physics today. Most attempts to explain dark matter propose that it is made of one
or more new particles. The hunt for dark matter is an ongoing effort on both the experimental and
on the theoretical side. It is widely accepted that any candidate for dark matter must be only weakly
interacting, otherwise it would have been detected already after decades of direct~\cite{Sumner2011} and indirect~\cite{Gaskins2016} searches.
Furthermore it must be stable on cosmological timescales, or it would be long gone today - in contrary to astronomical observations.

Many dark matter candidates were proposed, such as \textit{sterile neutrinos}~\cite{Boyarsky2019}, \textit{gravitinos}~\cite{Steffen2006},
\textit{axions}~\cite{Duffy2009}, \textit{Kaluza-Klein dark matter}~\cite{Cheng2002} and \textit{\gls{WIMPs}}~\cite{Roszkowski2017}, to name the most common candidates.

According to Ref.~\cite{Bergstrom2013}, the annihilation of WIMP dark matter particles produce equal amounts of electrons and positrons.
For the annihilation/decay process the injected spectrum of cosmic-ray electrons or positrons per volume and time is given by:

\begin{equation}
  \label{eq:flux-dm}
  \begin{aligned}
    Q_\text{annihilation} &= \frac{1}{2} \langle \sigma v \rangle (\rho_{\chi} / m_{\chi})^2 \frac{\diff N}{\diff E}, \\
    Q_\text{decay}       &= \Gamma \rho_{\chi} / m_{\chi} \frac{\diff N}{\diff E},
  \end{aligned}
\end{equation}

where $\langle \sigma v \rangle$ is the velocity-averaged annihilation rate, $\Gamma$ the decay rate, $\rho_{\chi}$ the dark matter density,
$m_{\chi}$ the dark matter mass and $\diff N / \diff E$, the spectrum of electrons or positrons produced per annihilation/decay.

Electrons/positrons from dark matter annihilation or decay typically result from the decay of hadronic final states, like $\pi^{\pm}$, or the
leptonic decay of $\tau^{\pm}$ or $\mu^{\pm}$. In the first case, the high multiplicity of the processes will produce a continuous
and smooth spectra over a wide energy range. The direct production of electrons and positrons will result in a characteristic spectrum, with a sharp
edge-like feature at $E = m_{\chi}$ (for annihilation) or $E = \frac{1}{2} m_{\chi}$ (for decay).

As shown in Ref.~\cite{Bergstrom2013}, this can explain a rise in the positron fraction (and the positron flux), just like the pulsar hypothesis can.
Therefore high precision data up to the highest energies is vital to search for edge-like structures in the positron flux, which is one of the goals
for this thesis.

\chapter{Detector}
\label{sec:detector}

The AMS-02 detector is a general purpose magnetic spectrometer with a large acceptance~\cite{Ahlen1994,Kounine2011b}.
It was launched to space using the Space Shuttle \textit{Endeavor} during the \textit{STS-134} mission on May~\nth{16}~2011.
On May~\nth{19}~2011~it was installed on the \gls{ISS}.

It operates in a low earth orbit, with an orbital plane inclination of \SI{51.6}{\degree} with respect to the
Earth's equator. AMS-02 completes approximately 16 orbits of data taking every day, delivering a stable performance
since its installation and high-quality data~\cite{Ting2013}.

\begin{figure}[H]
  \centering
  \includegraphics[width=0.55\linewidth]{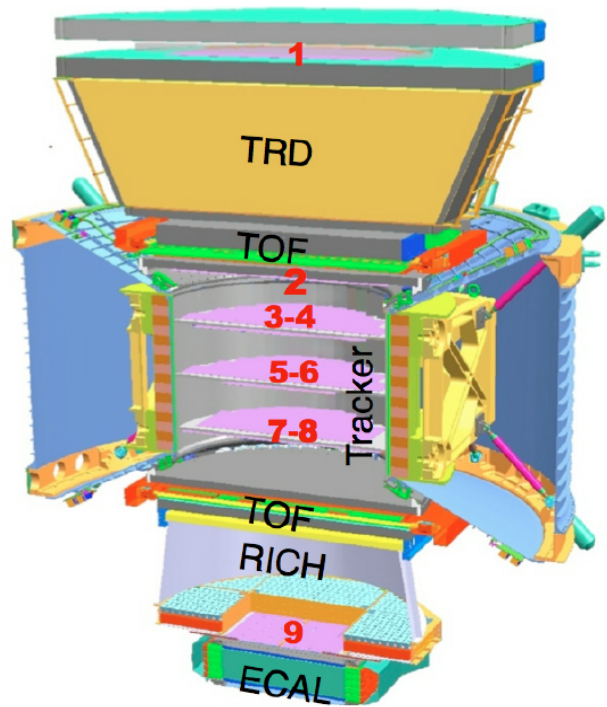}
  \caption{Schematic view of the AMS-02 detector. Credit: Ref.~\cite{AmsWebsite}}
  \label{fig:detector-schematic-view}
\end{figure}

The dimensions of the detector are approximately \SI{5 x 5 x 4}{\meter} and the total weight is \SI{7500}{\kilo\gram}. A schematic view of the instrument is shown in \cref{fig:detector-schematic-view}.

The key subdetectors to measure the electron and the positron flux are:

\begin{enumerate}
  \item\textbf{Silicon tracker} together with the \textbf{magnet} (\cref{sec:detector-magnet,sec:detector-tracker}), to separate negatively from positively charged particles
  \item\textbf{Transition Radiation Detector} (\cref{sec:detector-trd}), to distinguish electrons and positrons from protons
  \item\textbf{Electromagnetic calorimeter} (\cref{sec:detector-ecal}, to measure the electron or positron energy and to discriminate protons
    from electrons or positrons by their shower shape
\end{enumerate}

The strength of AMS-02 is the redundancy of the energy, momentum and charge measurements. Various subdetectors can be combined in the
data analysis, to select pure samples of cosmic-ray particles, e.g. positrons separated from the overwhelming proton background, which
is thousand times larger than the amount of positrons.

\section{Magnet}
\label{sec:detector-magnet}

\begin{figure}[H]
  \centering
  \includegraphics[width=0.55\linewidth]{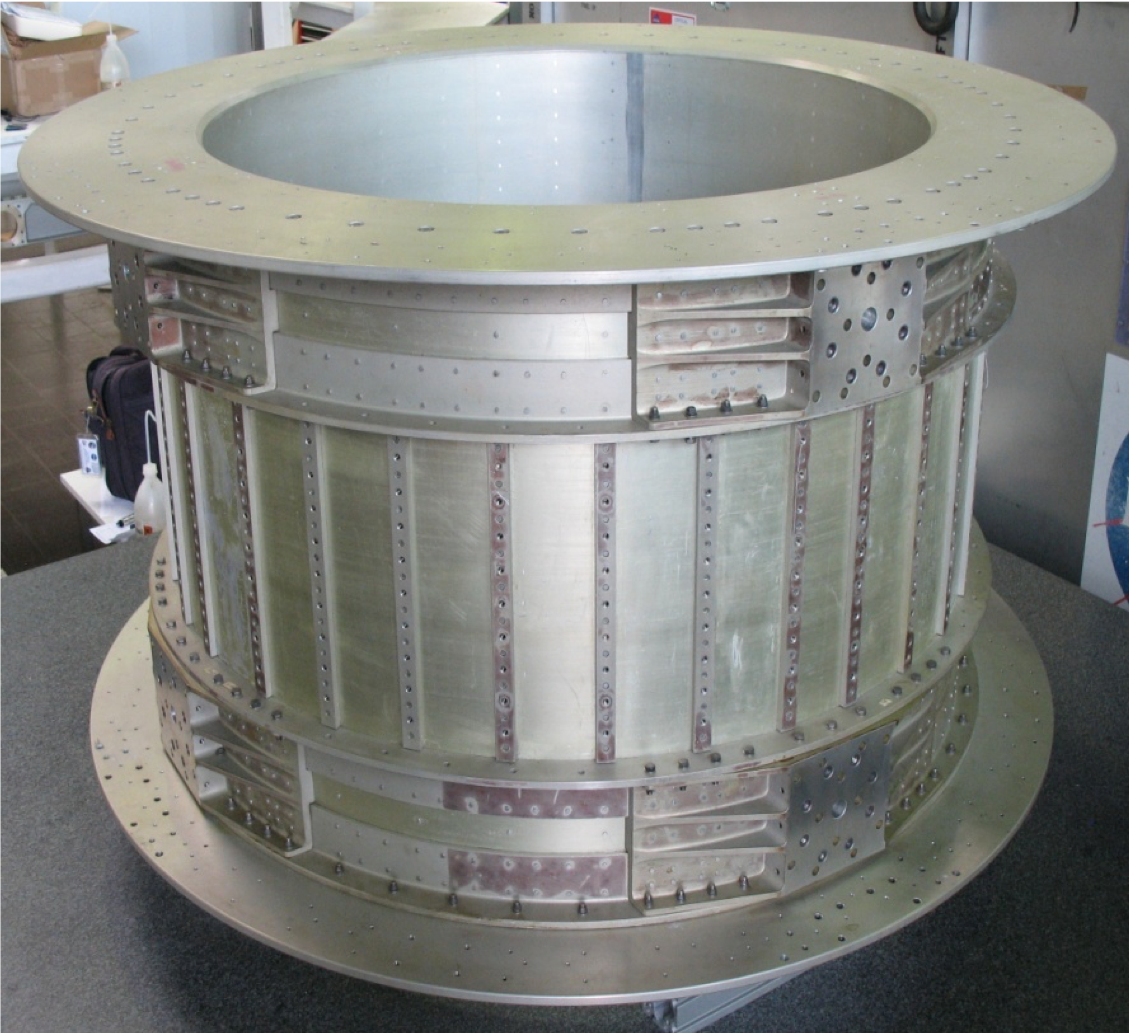}
  \caption{Picture of the permanent magnet used in AMS-02. Credit: Ref.~\cite{AmsWebsite}}
  \label{fig:detector-magnet}
\end{figure}

The magnet~\cite{Luebelsmeyer2011}, shown in \cref{fig:detector-magnet}, is made of 64 high-grade \ce{Nd}-\ce{Fe}-\ce{B} sectors
assembled in a cylindrical shell structure - \SI{0.8}{\meter} long - with an inner diameter of \SI{1.1}{\meter}. The magnet produces
a field of \SI{1.4}{\kilo\gauss} in the X direction at the center of the magnet and negligible dipole moment outside the magnet.
The dipole moment outside of the magnet needs to vanish to eliminate the effect of torque on the ISS.

\begin{figure}[H]
  \includegraphics[width=\linewidth]{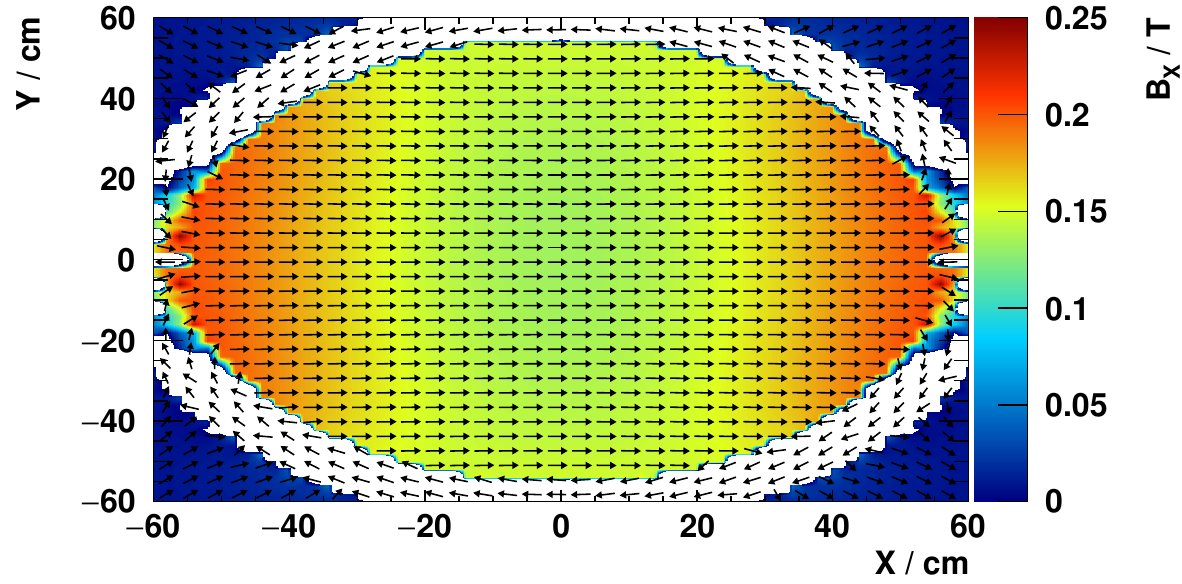}
  \caption{Visualization of the magnetic field in X-Y direction of the AMS-02 permanent magnet at \SIvarEquals{$Z$}{0}{\centi\meter} (center of AMS). Credit: B. Beischer.}
  \label{fig:detector-magnet-measurement}
\end{figure}

\Cref{fig:detector-magnet-measurement} shows the X-Y view of the magnetic field at \SIvarEquals{$Z$}{0}{\centi\meter} (center of AMS), as implemented
in the AMS-02 reconstruction software. The magnetic field of the permanent magnet was measured in 1997 and re-measured in 2010 in \SI{120000}{}
locations. The deviation of the field from both measurements is smaller than \SI{1}{\percent} in all locations~\cite{AmsWebsite}. The AMS reference coordinate
system is aligned with the center of the magnet, such that the x-axis is parallel to the magnetic field line in the center. Since the z-axis
points vertically, the Y-Z projection is the bending plane for incident particles.

\section{Tracker}
\label{sec:detector-tracker}

The AMS-02 tracker~\cite{Hass2004,Haino2013} is composed of 2284 double-sided silicon micro-strip detectors, with dimensions
of \SI{72 x 41 x 0.3}{\milli\meter}, covering an effective area of \SI{6.4}{\meter\squared}.

\begin{figure}[H]
  \centering
  \includegraphics[width=0.6\linewidth]{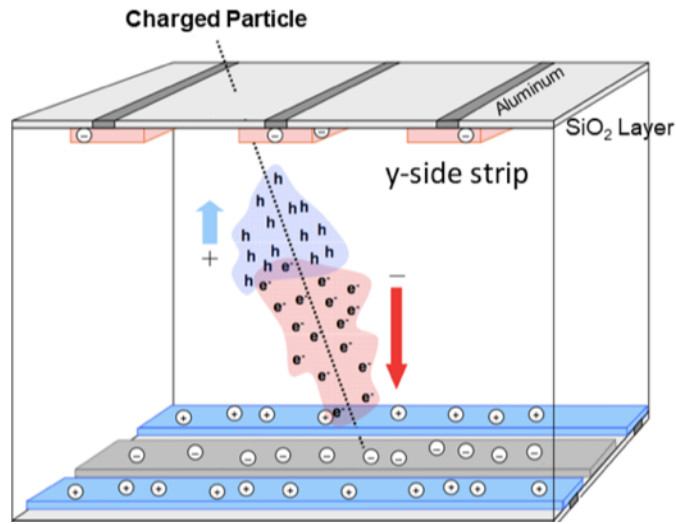}
  \caption{Schematic of the measurement principle of the \enquote{S side} silicon sensor. Credit: Ref.~\cite{AmsWebsite}}
  \label{fig:detector-tracker-principle}
\end{figure}

The high resistivity n-type sensors are inversely biased with an voltage of \SI{80}{\volt}. A traversing particle creates electron/hole
pairs in the sensors, as illustrated in \cref{fig:detector-tracker-principle}. The drifted electron/hole pairs are collected by the
$p^{+}$ strip with an implantation pitch of \SI{27.5}{\micro\meter} on one side (\textbf{S side} or \textbf{y-side}) and the $n^{+}$
strip with an implantation pitch of \SI{104}{\micro\meter} oriented orthogonally on the other side (\textbf{K side} or \textbf{x-side}).

7 to 15 silicon sensors are grouped together on a \textit{Ladder}, forming the building block of the AMS-02 tracker.
Each ladder has 1024 readout channels, 640 for the $p^{+}$ strips and 384 for the $n^{+}$ strips. All strips on the S side are readout
independently, whereas the K side strips are bonded together: the same strips on different sensors are linked together, to reduce the
overall amount of readout channels. This introduces an ambiguity on the K side, used to measured the non-bending plane coordinate, that
needs to be resolved during event reconstruction.

\begin{figure}[H]
  \centering
  \includegraphics[width=0.8\linewidth]{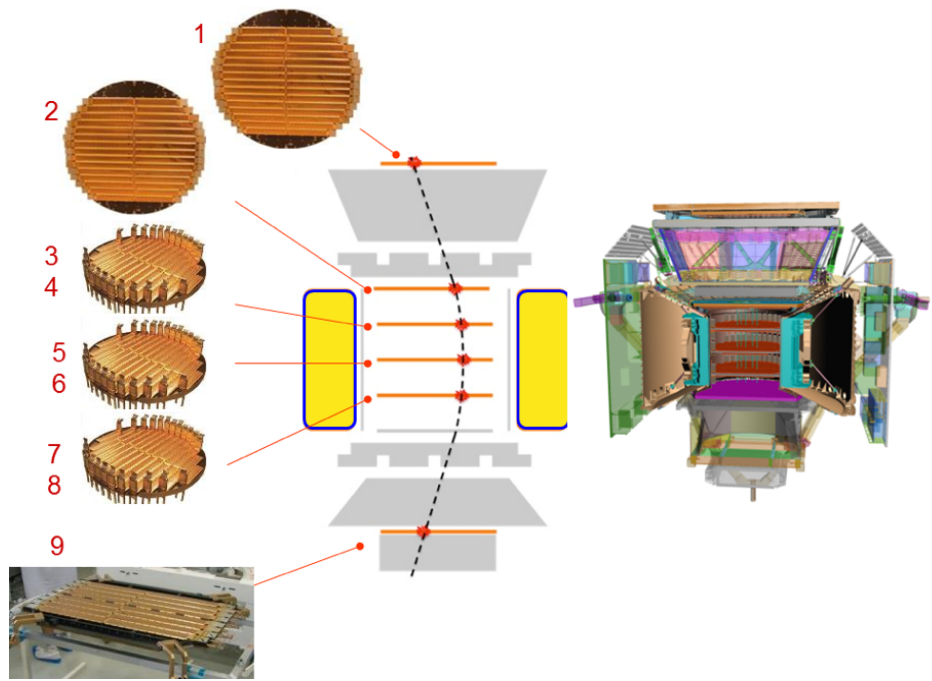}
  \caption{Schematic of the tracker showing the locations of the nine tracker planes. Credit: Ref.~\cite{AmsWebsite}}
  \label{fig:detector-tracker-layout}
\end{figure}

In total, there are 192 ladders arranged in nine layers, mounted on six mechanical supporting planes, placed at different vertical
positions in the detector, as shown in \cref{fig:detector-tracker-layout}. Six layers are arranged into three double-sided planes
(layer 3 to layer 8) placed inside the magnet, together with another single-sided plane (layer 2). The two remaining single-sided
planes (layer 1 and layer 9) are placed at the top and the bottom of the instrument, increasing the lever arm for the sagitta measurement.

The silicon sensors allows one to measure the particle position on the silicon surface, with an accuracy better than \SI{10}{\micro\meter}
in y-direction (S side) and \SI{30}{\micro\meter} in x-direction (K side). Moreover, the deposited ionization energy is proportional to the
square of the particle charge ($I~\propto~Z^2$) and can therefore be used to identify the type of the particle. Furthermore due to the magnetic
field, any charged particle traversing AMS-02 is bent and thus the charge sign can be reconstructed from a measurement of the curvature of the
trajectory.

\begin{figure}[H]
  \includegraphics[width=\linewidth]{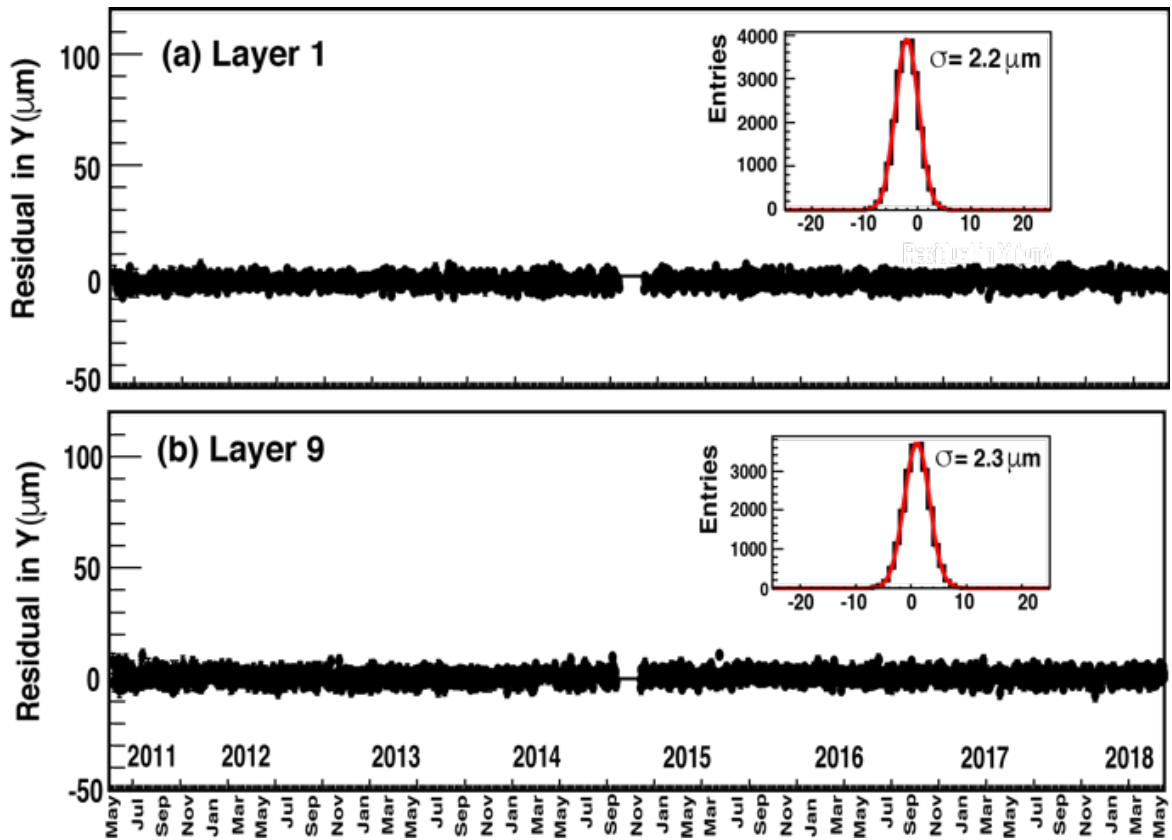}
  \caption{Schematic of the measurement principles of the silicon tracker. Credit: Ref.~\cite{AmsWebsite}}
  \label{fig:detector-tracker-alignment-stability}
\end{figure}

The inner tracker (formed by layer 2 to 8) is held stable by a carbon fiber structure with negligible coefficient of thermal expansion. The
stability of the inner tracker is monitored using the \gls{TAS}, which consists of 20 infrared laser beams that penetrate layers 2 through 8 and provide sub-micrometer
position measurements. Using cosmic rays over a two-minute window, the position of layer 1 is aligned with a
precision of \SI{5}{\micro\meter} with respect to the inner tracker and layer 9 with a precision of \SI{6}{\micro\meter}. The alignment
stability of layer 1 and layer 9 over seven years is shown in \cref{fig:detector-tracker-alignment-stability}. The stability\footnote{In
October and November 2014, the tracker was partially switched off due to detector studies.} is \SI{2.2}{\micro\meter} for layer 1 and
\SI{2.3}{\micro\meter} for layer 9.

Together with the magnet (\cref{sec:detector-magnet}), the tracker provides a maximum detectable rigidity of \SI{2}{\tera\volt} on average
for $Z = 1$ particles, over tracker planes 1-9, where rigidity is the momentum divided by the charge. Furthermore each layer of the tracker also
provides an independent measurement of $Z$. The charge resolution of the layers of the inner tracker together is $\Delta Z = 0.05$ for $Z = 1$ particles.

\clearpage
\section{\glsentryfull{TOF}}
\label{sec:detector-tof}

\begin{figure}[H]
  \includegraphics[width=\linewidth]{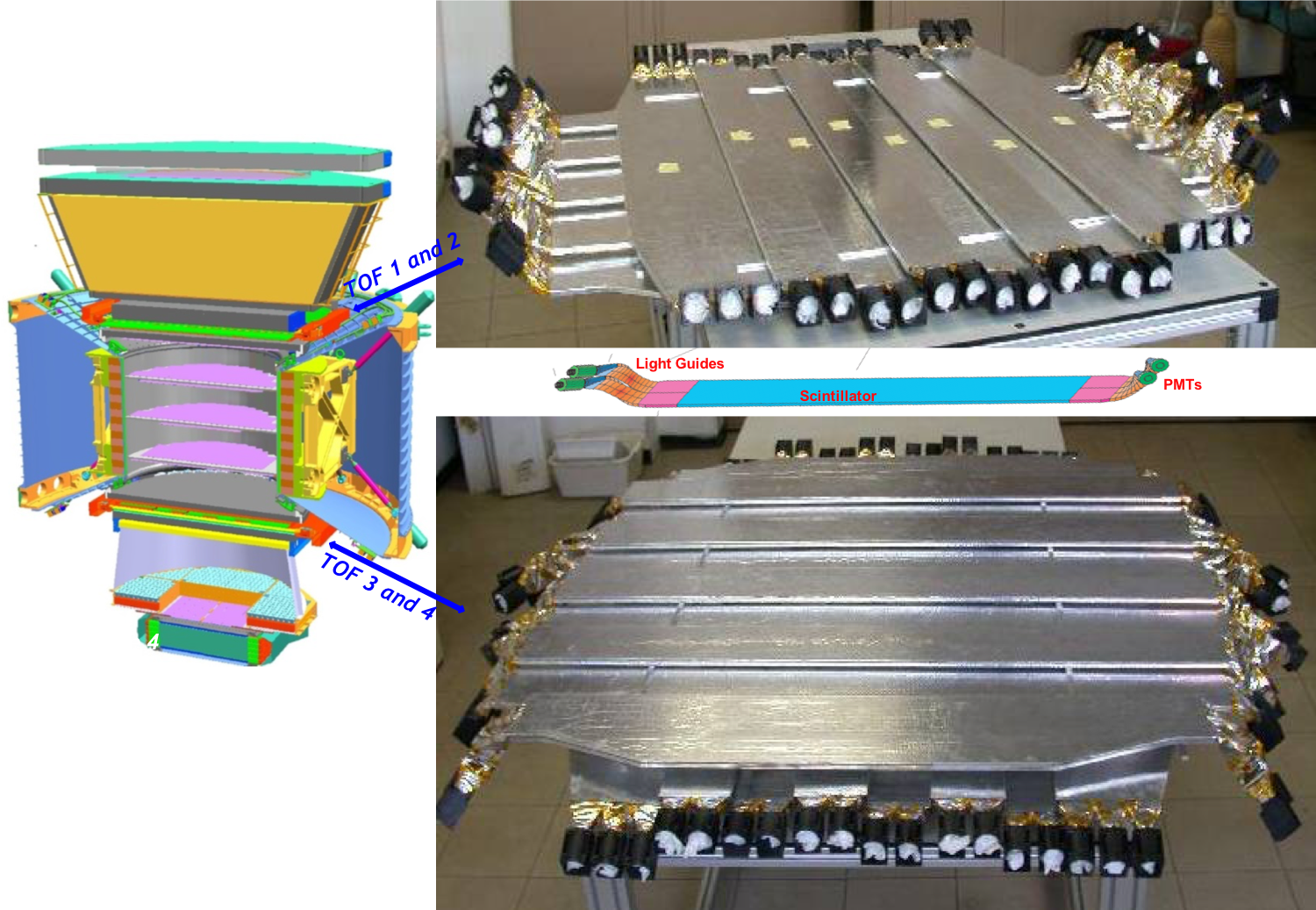}
  \caption{Picture of the Time-Of-Flight system. The upper and low TOF, each consists of two planes. Credit: Ref.~\cite{AmsWebsite}}
  \label{fig:detector-tof-layout}
\end{figure}

Two planes of \gls{TOF} counters, shown in \cref{fig:detector-tof-layout}, are located above the magnet, forming the \enquote{Upper TOF}
and two planes below the magnet forming the \enquote{Lower TOF}. Each plane contains eight or ten scintillating paddles. Each paddle is equipped with
two or three photomultiplier tubes on each end. To ensure efficient detection of traversing particles, the paddles in each plane overlaps with the
adjacent ones by \SI{5}{\milli\meter}.

The coincidence of signals from all four planes provides the main trigger (\cref{sec:detector-trigger})
for charged particles. By taking the time difference between signals from the upper and lower TOF planes a precise measurement of the particle velocity
can be obtained: the average time resolution is \SIapprox{160}{\pico\second} for $Z = 1$ particles.

\begin{figure}[H]
  \centering
  \includegraphics[width=0.75\linewidth]{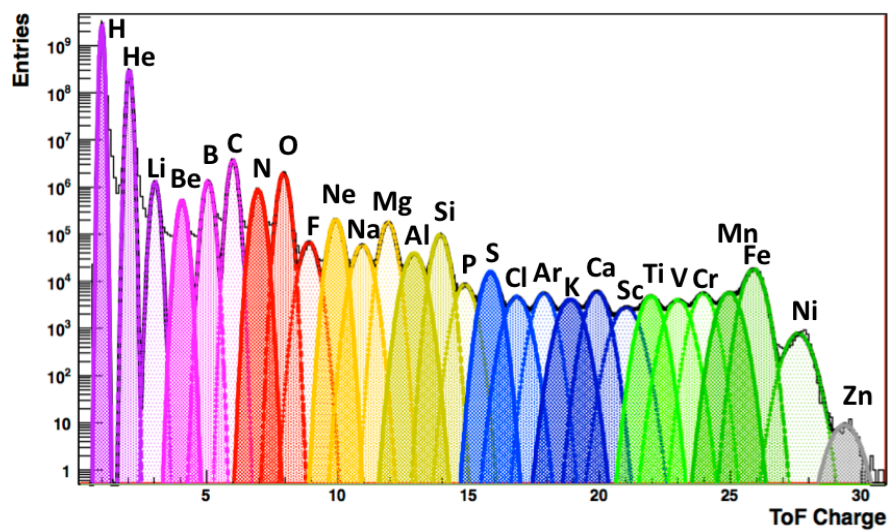}
  \caption{The charge distribution measured by the TOF, from $Z = 1$ (protons) up to $Z = 30$ (Zinc). Credit: Ref.~\cite{AmsWebsite}}
  \label{fig:detector-tof-charge}
\end{figure}

While the ionization energy deposition in the TOF counters by nuclei is proportional to $Z^2$, the scintillator light output slowly saturates and the
$Z$ dependence becomes almost linear as described by Birks law~\cite{Bindi2014}. The charge resolution was measured to be $\Delta Z = 0.05$
for $Z = 1$ particles. The TOF is able to differentiate charges up to Zinc, as shown in \cref{fig:detector-tof-charge}.

\section{\glsentryfull{TRD}}
\label{sec:detector-trd}

The \gls{TRD}~\cite{Kirn2013,Heil2013} is located at the top of the AMS-02 detector.
It is designed to discriminate between light and heavy particles, such as positrons and protons,
by exploiting the $\gamma$ dependence of the transition radiation effect.

Transition radiation~\cite{Ginzburg1945} is produced by highly relativistic charged particles, with a large Lorentz factor
$\gamma$, when crossing the boundary of two materials with different dielectric constants. The dipole
formed by the charged particle with its mirror charge will change as it passes the interface, and this
gives rise to emission of soft x-ray photons.

\begin{figure}[H]
  \centering
  \includegraphics[width=0.65\linewidth]{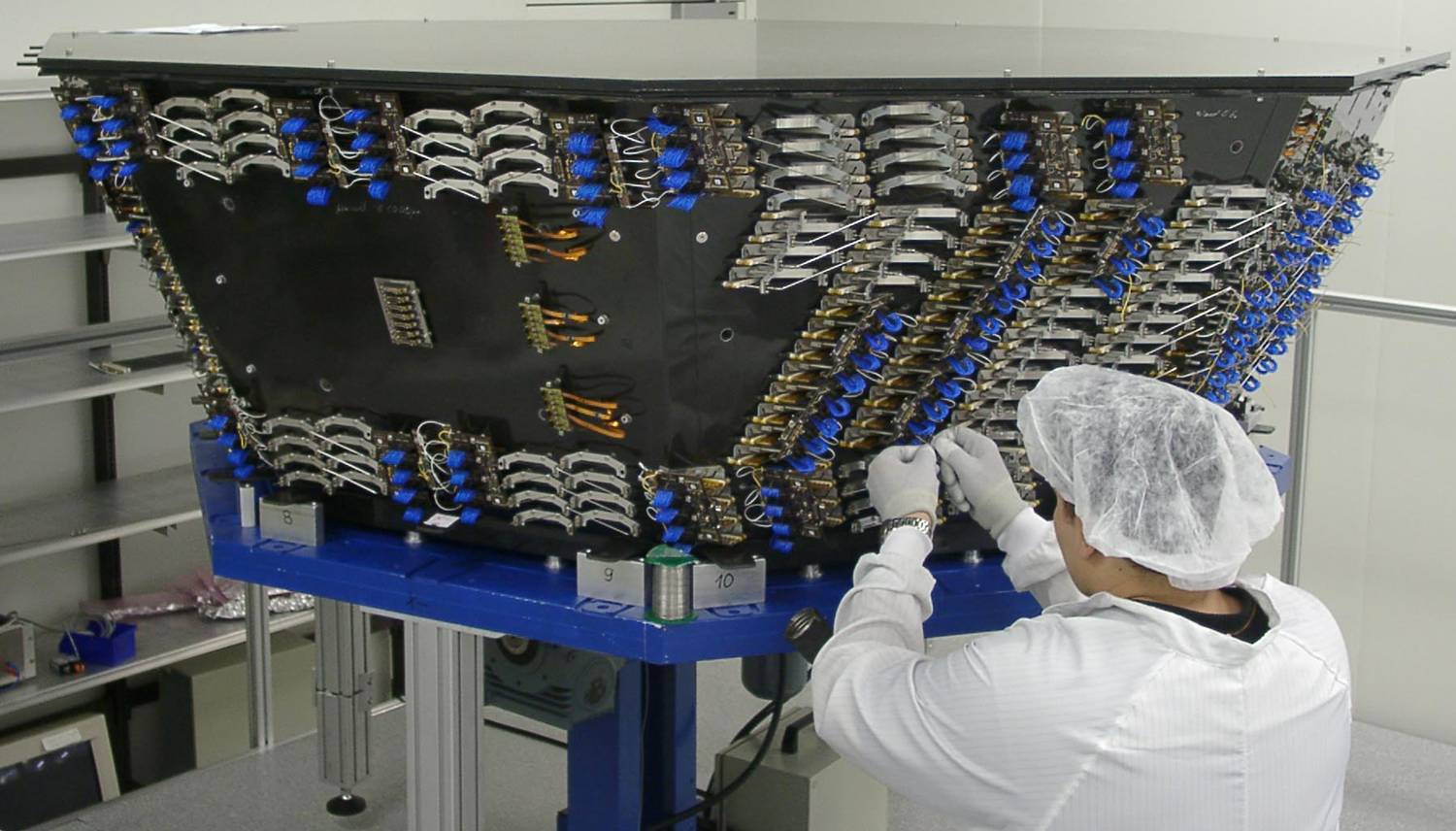}
  \caption{Picture of the TRD during assembly, showing the inverted octagonal pyramid shape. Credit: Ref.~\cite{AmsWebsite}}
  \label{fig:detector-trd-housing}
\end{figure}

The TRD has an inverted octagonal pyramid shape, as shown in \cref{fig:detector-trd-housing}.

It is mounted between the first layer of the tracker (\cref{sec:detector-tracker}) and the first layer of the
TOF (\cref{sec:detector-tof}). The TRD is \SI{80}{\centi\meter} high and spans \SI{2}{\meter} at its top plane
and \SI{1.5}{\meter} at its bottom plane. The detector is placed in a skeleton made of aluminium honeycomb walls. It is
populated with 5248 proportional tubes with a diameter of \SI{6}{\milli\meter}, made from \SI{72}{\milli\meter} thick
double-layered kapton-aluminium foils, which functions as the cathode of the proportional tubes. The tubes are grouped
into 328 modules, with lengths from \SI{0.8}{\meter} to \SI{2}{\meter}, depending on the position where the
module is mounted in the pyramid shape.

A \SI{30}{\milli\meter} thick gold-plated tungsten wire is mounted in the middle of each straw tube, acting as anode.
Each tube is filled with a gas mixture of approximately 90:10 ratio of xenon (\ce{Xe}) to carbon dioxide (\ce{CO2}).
When a traversing charged particle ionizes the gas in a tube, a voltage proportional to the number of ionized gas atoms
is produced on the output of the wire. This proportionality is called \textit{gas gain}.

Xenon is used to detect the ionization signal of crossing charged particles in addition to the low energy
photons of the transition radiation, while carbon dioxide acts as a quenching gas for charge multiplication, ensuring that
the gas returns to its initial state for the next measurement, after a particle traversal.

To compensate the gas gain change due to diffusion across tube walls, a daily high voltage adjustment is performed to
keep the gas gain at a predefined level, while the total pressure in the TRD decreases, due to the diffusion loss. To
keep the total pressure at around \SI{1}{\bar} gas is refilled using a gas supply system approximately once a month.
The TRD gas system is equipped with \SI{49}{\kilo\gram} xenon and \SI{5}{\kilo\gram} \ce{CO2}, allowing for a continuous
operation in space for 30 years.

\begin{figure}[H]
  \centering
  \includegraphics[width=0.40\linewidth]{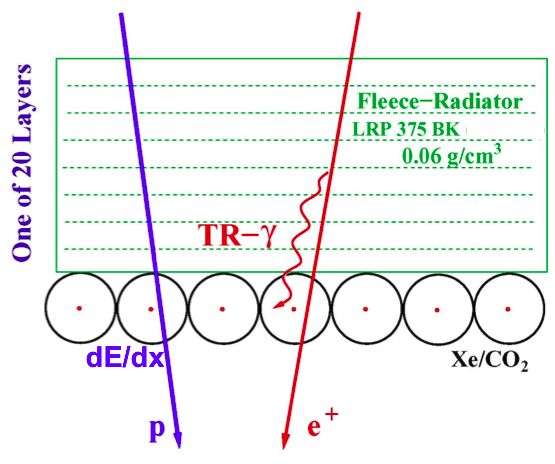}
  \caption{Working principle of the TRD. Credit: Ref.~\cite{AmsWebsite}}
  \label{fig:detector-trd-principle}
\end{figure}

The probability of emitting transition radiation photons increases with the number of boundaries crossed. For that reason the modules are arranged
into 20 vertical layers with \SI{22}{\milli\meter} of fleece radiator interleaved between each layer. The fleece radiator is composed of \SI{10}{\milli\meter}
polypropylene/polyethylene fibers with a density of \SI{0.06}{\gram\per\centi\meter\cubed}. The proportional tubes in the four highest and lowest layers
of the TAS are mounted parallel to the x-axis of the AMS-02 coordinate system (non-bending plane of the magnetic field), while the straw tubes in the middle
layers are parallel to the y-axis (bending plane of the magnetic field), to allow for a three-dimensional reconstruction of the particle trajectory.

The working principle of the TRD is illustrated in \cref{fig:detector-trd-principle}. A proton traversing the straw tube ionizes the gas. A positron
at the same energy will additionally emit transition radiation photons during the passage through the radiator, as the $\gamma$ factor is $~\approx~2000$ times
larger. The low energy photons will be absorbed by the xenon and enhance the signal measured on the tungsten wire.

\begin{figure}[H]
  \centering
  \includegraphics[width=0.50\linewidth]{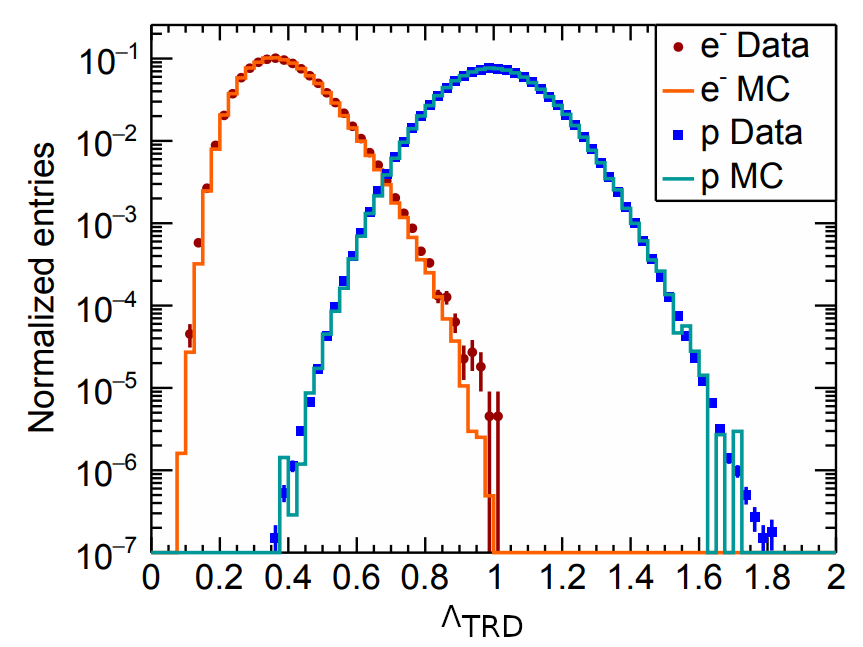}
  \caption{Example showing the TRD estimator, which combines the $\diff\text{E}/\diff\text{x}$ measurements from all 20 TRD layers into a single estimator, in the energy range \SIrange{10}{100}{\GeV}. Credit: Ref.~\cite{Raiha2017}}
  \label{fig:detector-trd-measurement}
\end{figure}

By combining the $\diff\text{E}/\diff\text{x}$ measurement from all 20 TRD layers a likelihood estimator can be constructed, that separates light from heavy particles,
such as electrons or positrons from protons, as illustrated in \cref{fig:detector-trd-measurement}. A detailed description of the TRD estimator $\Lambda_{\text{TRD}}$
is given in \cref{sec:analysis-event-reconstruction-trd-estimator}.

\begin{figure}[H]
  \begin{subfigure}{0.46\linewidth}
    \includegraphics[width=\linewidth]{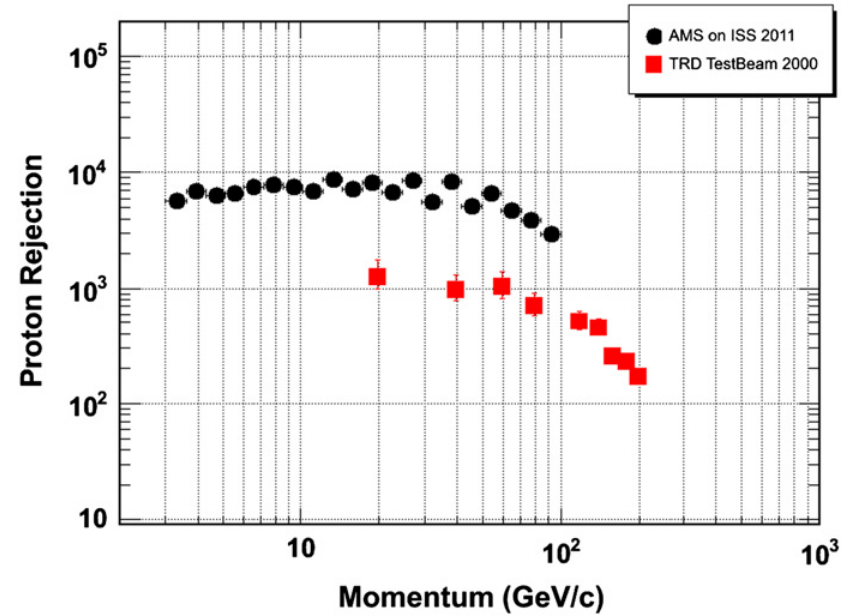}
    \caption{Credit: Ref.~\cite{Kirn2013}}
  \end{subfigure}
  \hfill
  \begin{subfigure}{0.54\linewidth}
    \includegraphics[width=\linewidth]{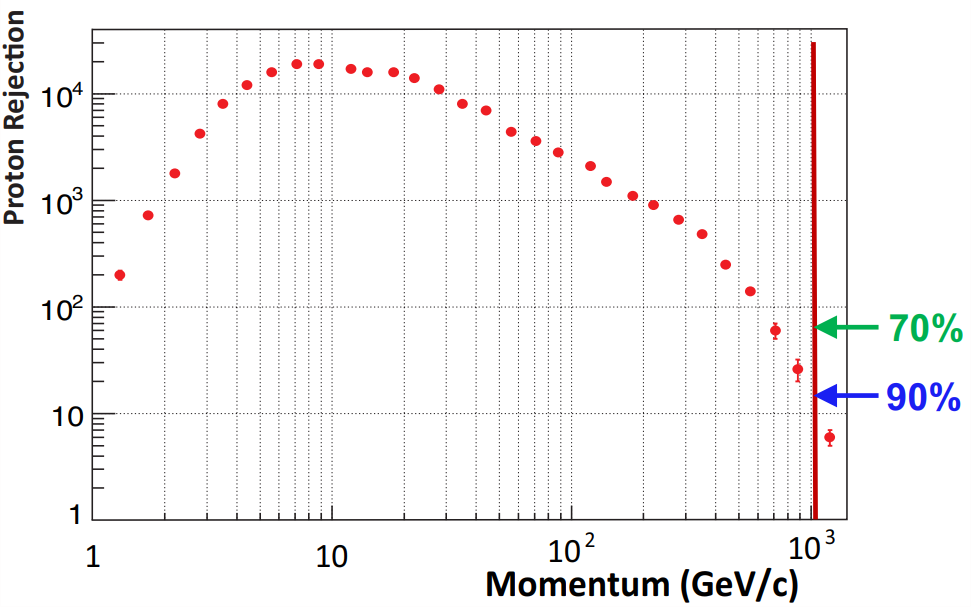}
    \caption{Credit: Ref.~\cite{Kounine2017b}}
  \end{subfigure}
  \caption{Proton rejection of the TRD as function of the $e^{\pm}$ momentum at \SI{90}{\percent} $e^{\pm}$ efficiency. The left plot shows the proton rejection determined during testbeam in 2000, in comparison with the proton rejection determined from ISS data in 2011. The right plot shows the most recent measurement of the proton rejection, which greatly improved with more ISS statistics and sophisticated reconstruction algorithms.}
  \label{fig:detector-trd-rejection}
\end{figure}

The primary purpose of the TRD is to identify positrons from the proton background by transition radiation. The abundance ratio between protons and positrons is
about 1000 at \SI{10}{\GeV} and it increases with energy. At an $e^{\pm}$ efficiency of \SI{90}{\percent}, the TRD proton rejection exceeds $10^4$ above \SI{4}{\GeV},
gradually decreasing to $10^3$ at \SIapprox{200}{\GeV}, until it reaches approximately 20 at \SIapprox{1}{\TeV}, as shown in \cref{fig:detector-trd-rejection}.

\section{\glsentryfull{ECAL}}
\label{sec:detector-ecal}

The AMS-02 ECAL consists of a multilayer sandwich of lead foils and \SIapprox{50000}{} scintillating fibers with an active area of \SI{648 x 648}{\milli\meter} and a
thickness of \SI{166.5}{\milli\meter}, corresponding to 17 radiation lengths $X_{0}$~\cite{RosierLees2012}. The calorimeter is composed of 9 superlayers, each
\SI{18.5}{\milli\meter} thick and made of 11 grooved, \SI{1}{\milli\meter} thick lead foils interleaved with 10 layers of \SI{1}{\milli\meter} diameter
scintillating fibers. In each superlayer, the fibers are oriented in only one direction. The 3D imaging capability of the detector is obtained by stacking alternating
superlayers with fibers parallel to the x-axis and y-axis (five and four superlayers, respectively).

\begin{figure}[H]
  \centering
  \includegraphics[width=0.9\linewidth]{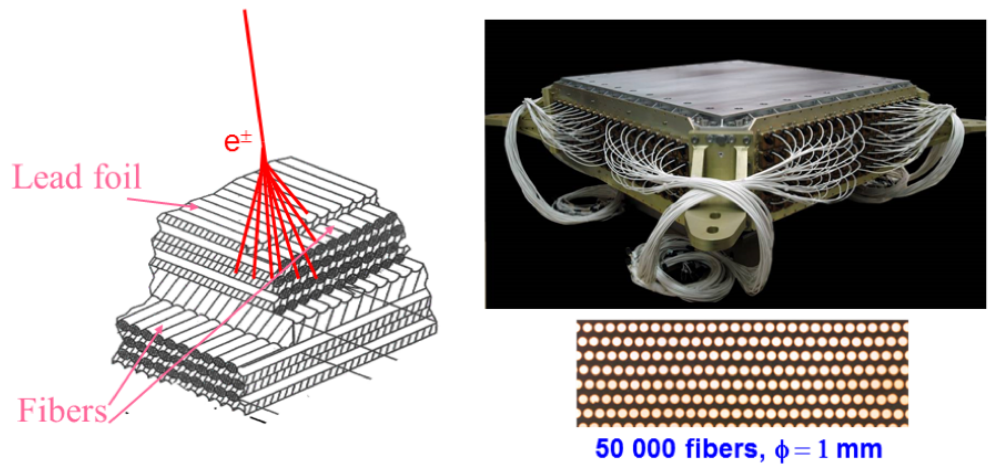}
  \caption{Schematic of the electromagnetic shower development in the ECAL. Credit: Ref.~\cite{AmsWebsite}}
  \label{fig:detector-ecal}
\end{figure}

The scintillating fibers are read out on one end by 324 photomultipliers (PMT). Each PMT has four anodes and is surrounded by a magnetic shield which contains
light guides, the frontend electronics and the PMT base. An ECAL cell is defined as an area of \SI{9 x 9}{\milli\meter}, covering approximately 35 fibers,
corresponding to one anode. In total there are 1296 cells segmented into 18 layers longitudinally, with 72 transverse cells in each layer. This design allows
a sampling of the shower in three dimensions with a fine granularity. The PMT / electronics are able to process the signals over a wide dynamic range, from
minimum ionizing particles (MIPs), which produce \SIapprox{10}{} photoelectrons per cell, up to the \SIapprox{60000}{} photoelectrons produced in one cell by the core of
an electromagnetic shower of a \SI{1}{\TeV} electron or positron~\cite{Kounine2017a}.

The primary purpose of the ECAL is to separate electrons and positrons from protons, as well as providing an accurate energy measurement for electron and positrons
up to the \SI{}{\TeV} regime. When an electron or positron traverses the first layer of the ECAL it starts to emit bremsstrahlung photons, which further convert into
additional electron and positron pairs through the interaction with the nuclei of the high-density material of the ECAL. The secondary
electrons and positrons also emit bremsstrahlung photons and an \textit{electromagnetic shower} cascade develops.
The two processes (bremsstrahlung and pair-production) continue until photons fall below the pair production threshold and
energy losses of electrons and positrons due to ionization start to dominate over the losses due to bremsstrahlung.

\begin{figure}[H]
  \centering
  \includegraphics[width=0.8\linewidth]{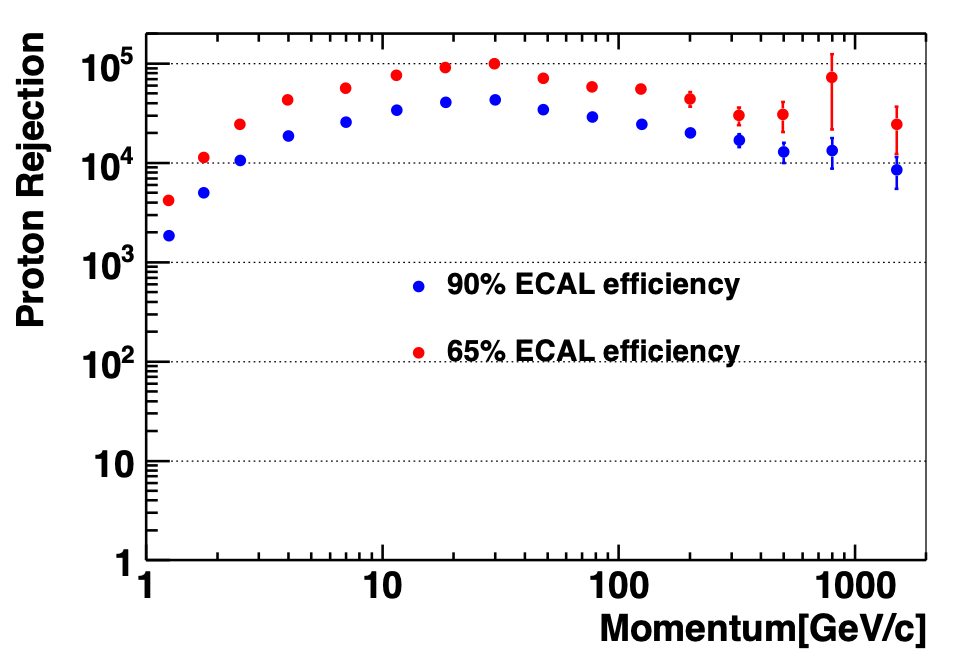}
  \caption{Proton rejection of the ECAL as function of the $e^{\pm}$ momentum. Credit: Ref.~\cite{AmsWebsite}}
  \label{fig:detector-ecal-rejection}
\end{figure}

The ECAL, combined with the Tracker ($E/\abs{R}$ cut), offers a proton rejection larger than $10^4$, at an $e^{\pm}$ efficiency of \SI{90}{\percent}, between
\SIrange{3}{1000}{\GeV}, as shown in \cref{fig:detector-ecal-rejection}. Approximately one order of magnitude of the rejection power stems from the combination of Tracker and
ECAL - the $E/\abs{R}$ cut. Another order of magnitude from the proton spectrum\footnote{When determining the proton rejection as function of ECAL energy, low energetic protons
are naturally suppressed, as protons do not deposit all of their energy in an ECAL. On average, the protons in the event sample, categorized as function of ECAL energy, have
a three times larger rigidity than measured ECAL energy. Thus the proton spectrum is sampled at higher rigidities, where the spectrum is rapidly falling (power law), leading to a smaller
proton background, compared to measuring the proton rejection as function of rigidity.} and the remaining 2-3 orders of magnitude from the ECAL itself, which is a common
magnitude of proton rejection for an ECAL.

In a dedicated testbeam the ECAL energy resolution~\cite{Gallucci2015} was measured to be

\begin{equation*}
  \frac{\sigma_{E}(E)}{E} = \frac{\SI{10.4(2)}{\percent}}{\sqrt{E}} \oplus \SI{1.4(1)}{\percent}.
\end{equation*}

\section{\glsentryfull{ACC}}
\label{sec:detector-acc}

\begin{figure}[H]
  \includegraphics[width=\linewidth]{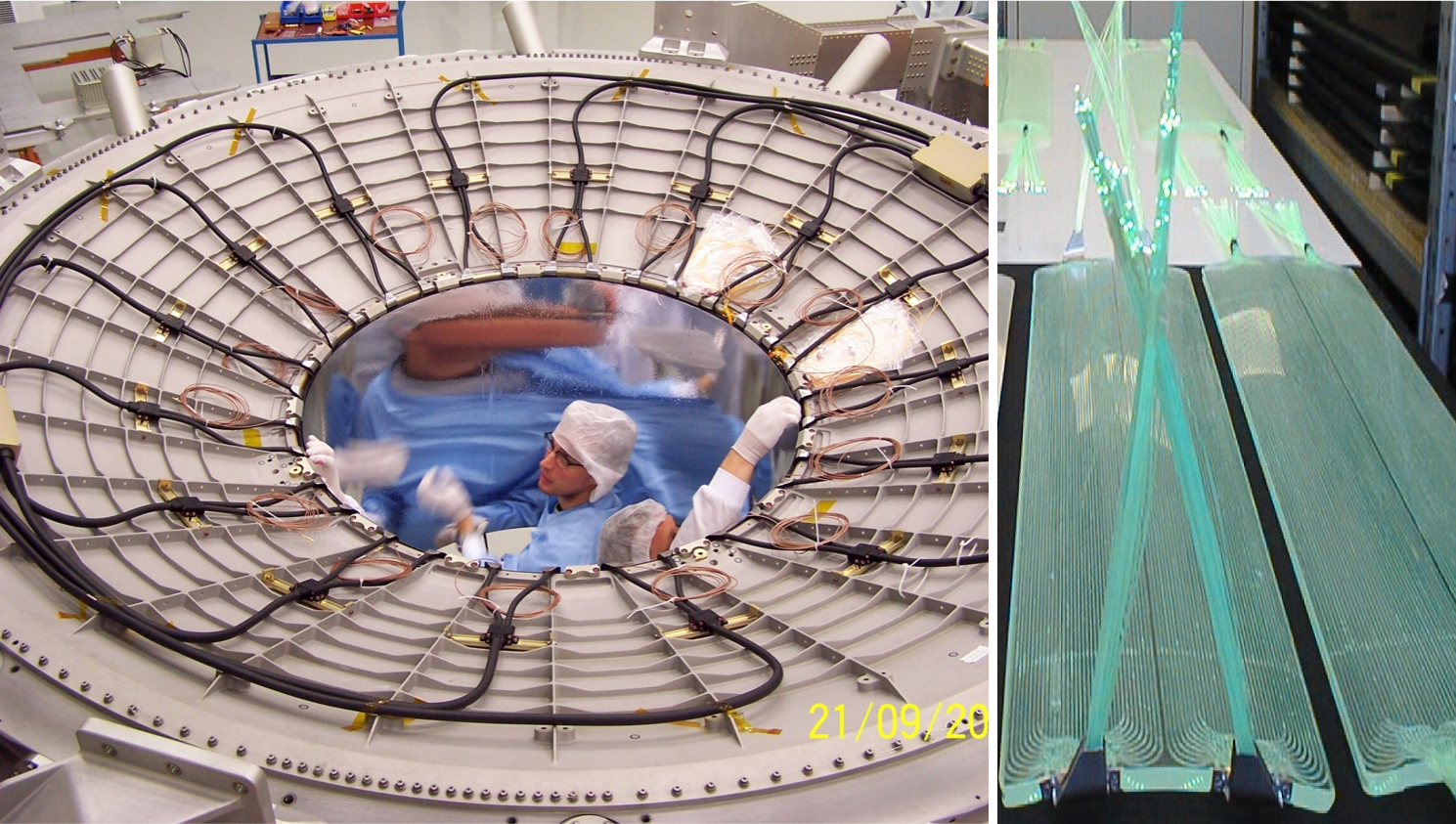}
  \caption{Picture of the ACC during construction. Credit: Ref.~\cite{AmsWebsite}}
  \label{fig:detector-acc}
\end{figure}

The ACC counters~\cite{Doetinchem2009}, shown in \cref{fig:detector-acc}, form a barrel surrounding the inner tracker, with overlapping coverage along
the vertical edges to ensure efficient detection of particles traversing in all directions. Their purpose is to detect events with unwanted particles
that enter or leave the inner tracker volume transversely.

\begin{figure}[H]
  \includegraphics[width=\linewidth]{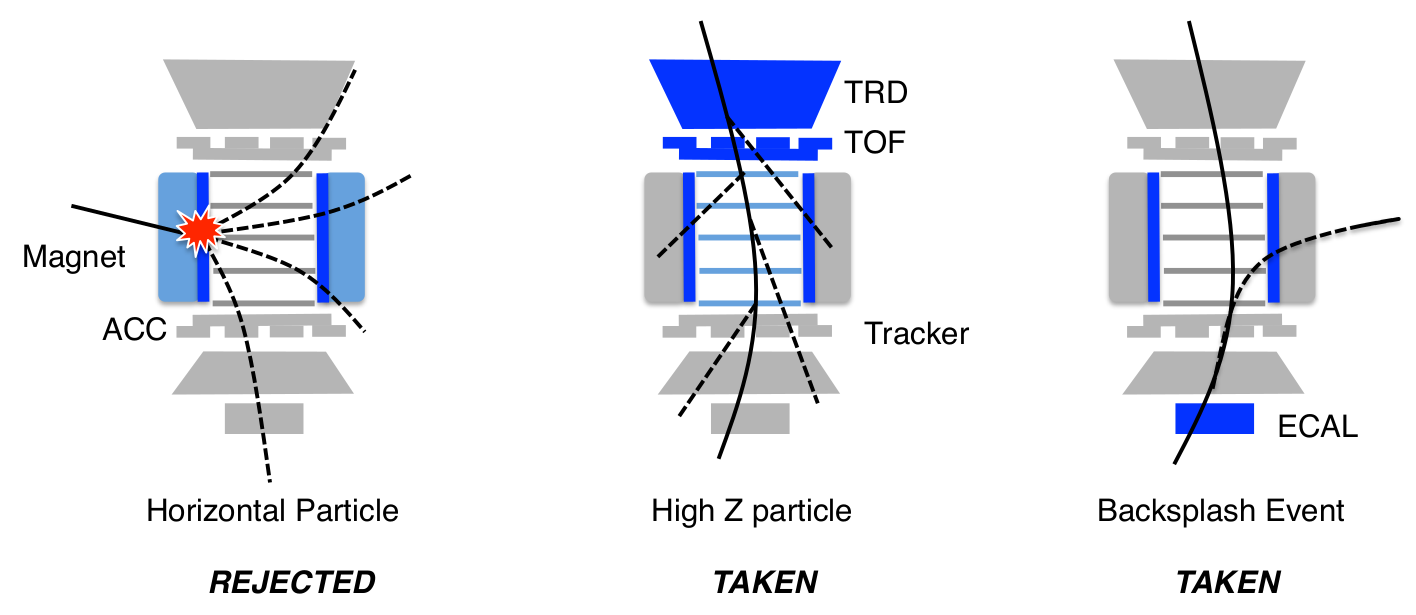}
  \caption{Illustration of the operation mode of the ACC. Credit: Ref.~\cite{AmsWebsite}}
  \label{fig:detector-acc-measurements}
\end{figure}

The ACC consists of sixteen curved scintillator panels of \SI{0.8}{\meter} length, instrumented with wavelength-shifting fibers to collect the scintillation
light. The ACC signal is an important part for the trigger logic, as it is used to veto particles coming from the side, as illustrated in
\cref{fig:detector-acc-measurements}.

The performance of the ACC was tested extensively using atmospheric muons: the upper limit on the ACC inefficiency are set to be
\SIvarOp{I}{<}{1.5e-5} at \SI{95}{\percent} confidence level.

\section{\glsentryfull{RICH}}
\label{sec:detector-rich}

\begin{figure}[H]
  \centering
  \includegraphics[width=0.7\linewidth]{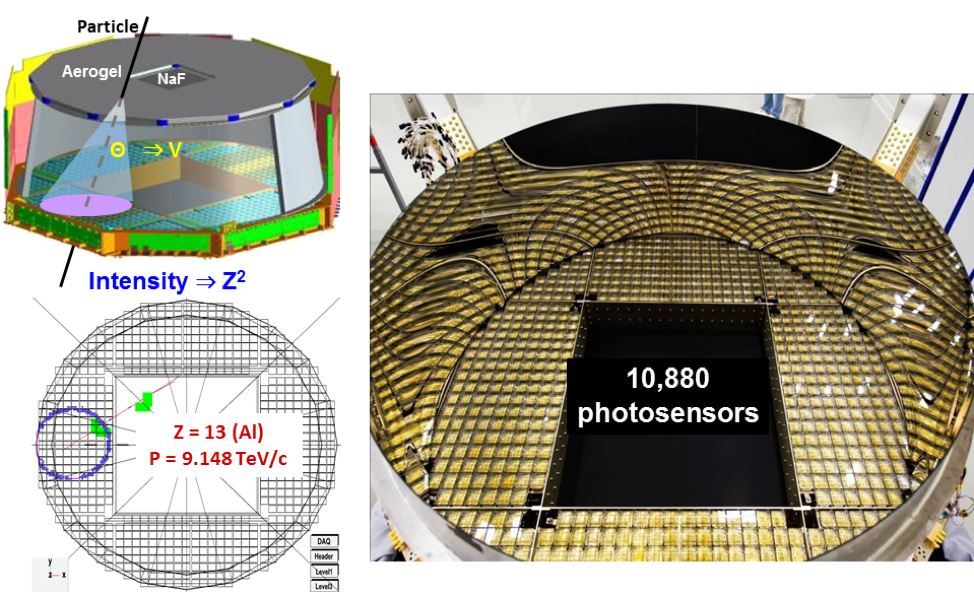}
  \caption{Picture of the RICH and its working principle. Credit: Ref.~\cite{AmsWebsite}}
  \label{fig:detector-rich}
\end{figure}

The AMS-02 RICH~\cite{Giovacchini2014}, shown in \cref{fig:detector-rich}, is mounted between the lower TOF planes and the ECAL. The RICH is
designed to measure the absolute charge of relativistic particles and their velocity with a precision of \SIapprox{0.1}{\percent} for protons.

The detector consists of a layer of radiator material, a conical reflector and a detection plane. The RICH has a truncated conical shape with a
\SI{60}{\centi\meter} upper radius, a \SI{67}{\centi\meter} lower radius and a height of \SI{47}{\centi\meter}. The upper plane contains the
radiator, which is made of two non-overlapping dielectric materials.

The dielectric radiators induce the emission of a cone of Cherenkov photons when traversed by charged particles with a velocity greater
than the velocity of light in the material. The central radiator is formed by 16 sodium fluoride, \ce{NaF}, tiles, each
\SI{8.5 x 8.5 x 0.5}{\centi\meter}, with a refractive index $n = 1.33$. These are surrounded by 92 tiles, each
\SI{11.5 x 11.5 x 2.5}{\centi\meter}, of silica aerogel with a refractive index $n = 1.05$. This allows the detection of particles with
velocities $\beta > 0.75$, for those which pass through the NaF radiator and $\beta > 0.953$ for those which pass through the aerogel radiator.

The Cherenkov photons are emitted only when velocity of incoming particle exceed the threshold velocity $1/n$. These photons are emitted under
an angle $\theta$ with respect to the particle direction, where $\theta$ is given by $\cos{\theta} = \beta \cdot n$. The number of generated
photons per track length is proportional to $Z^2$.

An expansion volume with a height of \SI{45}{\centi\meter} between the radiators and the photon detection plane allows the ring of Cherenkov photons
to expand. A high reflectivity mirror surrounds the expansion volume to reduce the lateral loss of photons. The Cherenkov photons are detected
by an array of \SI{10880}{} photosensors with a spatial granularity of \SI{8.5 x 8.5}{\milli\meter}. It is designed such that photons produced by the
NaF radiator in the middle have a larger cone size and therefore no photosensors are needed in the middle at the end of the expansion volume to detect
them. This is especially important as the ECAL is mounted in the middle block, at the end of the expansion volume, and material above the calorimeter
needs to be avoided to avoid pre-showering.

The $\beta$ resolution was measured to be $\Delta \beta / \beta \approx 0.001$ for $Z = 1$ particles in a dedicated testbeam study.

\section{Trigger}
\label{sec:detector-trigger}

The AMS-02 trigger logic~\cite{Lin2005} utilizes signals from TOF, ACC and ECAL to take a fast
decision whether an event should be recorded or not. The trigger logic takes \SIapprox{1}{\micro\second}
to make a decision, which represents a significant contribution to the dead time of the experiment.
The detector is in a busy state and cannot detect new particles during the dead time.

AMS-02 uses a complex decision tree architecture with three different stages: the \enquote{Fast trigger},
the \enquote{Level 1 trigger} and the \enquote{Level 3 trigger}.

Only if the conditions of the first trigger stage are fulfilled, the next is considered, to save processing time.
At the time of writing only the first two trigger stages are used - the Level 3 trigger stage is deactivated as
there is enough bandwidth available to transfer all events triggered by the Level 1 trigger to the ground.

The fast trigger conditions are designed for different particle types: FTC for charged particles,
FTZ for particles with high charge (or strangelets) and FTE for all particles that produce electromagnetic showers (photons / leptons).
If any of the three conditions are fulfilled, a fast trigger signal is generated.

If a fast trigger signal is generated, the dedicated JLV1 electronics board~\cite{Kounine2009} starts evaluating
15 different Level 1 trigger conditions, which can be grouped into five categories:

\begin{enumerate}[(a)]
  \item\textbf{Charged particles}\hfill\\
    Events with at least 3 out of 4 TOF panels with a signal over threshold are accepted.
  \item\textbf{Big Z particles}\hfill\\
    Events with large energy depositions in the TOF panels are accepted.
  \item\textbf{ACC}\hfill\\
    Provides a veto to reject particles traversing AMS horizontally.
  \item\textbf{ECAL-F / ECAL-A}\hfill\\
    Events with energy depositions in the ECAL layers are accepted.
  \item\textbf{External} (not used on the ISS)\hfill\\
    Accepts events if an external trigger condition is fulfilled, e.g. the Cherenkov counters in the beam test
    that were connected with the AMS-02 JLV1 trigger boards.
\end{enumerate}

From the 15 trigger signals, seven sub-triggers were carefully chosen for data taking onboard the ISS~\cite{Kounine2011a}:

\begin{enumerate}
  \item\textbf{Unbiased charged}\hfill\\
    Requires 3 out of 4 TOF panels in coincidence -- efficiency is close to unity.
    Only every \nth{100} event fulfilling this condition is recorded, to reduce the trigger rate and save downlink bandwidth.
  \item\textbf{Single charged}\hfill\\
    Requires 4 out of 4 TOF panels in coincidence and no ACC hits.
  \item\textbf{Normal ions}\hfill\\
    Requires 4 out of 4 TOF panels in coincidence with large energy deposition and less than five ACC hits.
  \item\textbf{Slow ions}\hfill\\
    Equal to \enquote{Normal ions}, except that the gate width is extended (see~\cite{Kounine2011a} for details) to account for the slow strangelet particle.
  \item\textbf{Electrons}\hfill\\
    Requires 3 out of 4 TOF panels in coincidence plus at least one ECAL shower reconstructed.
  \item\textbf{Photons}\hfill\\
    Requires energy depositions in the ECAL in both X/Y super-layers.
  \item\textbf{Unbiased EM}\hfill\\
    Requires only energy depositions in the ECAL.
    Only every \nth{1000} event fulfilling this condition is recorded, to reduce the trigger rate and save downlink bandwidth.
\end{enumerate}

The \textbf{Single charged}, \textbf{Normal ions}, \textbf{Slow ions}, \textbf{Electrons} and \textbf{Photons} trigger branches are called \textbf{physics triggers}.
The remaining trigger branches \textbf{Unbiased charge} and \textbf{Unbiased EM} are used to measure the physics trigger efficiency directly from ISS data.
The assumption - which was verified in dedicated studies - is that these unbiased triggers have an efficiency close to unity, such that no particle can traverse AMS-02
vertically without firing an unbiased trigger signal.

\chapter{Analysis}
\label{sec:analysis}

In this chapter the electron flux and the positron flux analysis is presented, as well as the dedicated
positron/electron ratio and positron fraction analysis.

At first the data analysis framework is introduced (\cref{sec:analysis-data-analysis-framework}), which is used to
analyze the AMS-02 data. It is shown which subdetectors are needed to separate the $e^{\pm}$ signal from the overwhelming proton
background in cosmic rays, and how to combine the subdetector measurements in an optimal way (\cref{sec:analysis-event-reconstruction}), including
a discussion of the \textit{charge-confusion}, the misreconstructing of a positively charged particle as negative, and vice-versa.
An overview of the data selection criteria is presented (\cref{sec:analysis-data-selection}) and the necessary techniques used to differentiate
between signal and background (\cref{sec:analysis-lepton-counts}). Afterwards the time-averaged fluxes and ratios are presented (\cref{sec:analysis-flux-time-averaged,sec:analysis-ratios-time-averaged})
including a discussion of all systematic uncertainties (\cref{sec:analysis-flux-time-averaged-sysunc,sec:analysis-ratios-time-averaged-sysunc}). The chapter is concluded
with the presentation of the time-dependent fluxes and ratios (\cref{sec:analysis-flux-time-dependent,sec:analysis-ratios-time-dependent}) and a discussion of their
uncertainties (\cref{sec:analysis-flux-time-dependent-sysunc,sec:analysis-ratios-time-dependent-sysunc}).

The fluxes of cosmic-ray electrons or cosmic-ray positrons for a time interval $i$ in the energy bin $E$ of width $\Delta E$
are given by

\begin{equation}
  \label{eq:isoflux}
  \Phi_{e^{\pm},\,i}(E) = \frac{N_{e^{\pm},\,i}(E)}{\Delta E \cdot T_{i}(E) \cdot A_{e^{\pm},\,i}(E) \cdot \epsilon_{e^{\pm},\,i}(E)},
\end{equation}

where $N_{e^{-},\,i}(E)$ and $N_{e^{+},\,i}(E)$ are the numbers of electrons and positrons, respectively. $A_{e^{\pm},\,i}(E)$ is the acceptance and $\epsilon_{e^{\pm},\,i}(E)$
the combined signal selection efficiency. Both quantities potentially differ for electrons and positrons. $T_{i}(E)$ refers to the energy-dependent measurement time in the given time interval $i$
and is independent of the particle species.

In the following sections detailed descriptions are given how to derive all ingredients necessary for the electron and positron fluxes and
which challenges arise.

\section{Data analysis framework}
\label{sec:analysis-data-analysis-framework}

The AMS-02 detector produces approximately \SI{3.7}{\mega\bit} of data for each event, containing information from the \SIapprox{300000}{} readout channels.
Events are recorded at rates up to \SI{2}{\kilo\hertz} resulting in a data rate of \SI[per-mode=symbol]{7}{\giga\bit\per\second}, which is reduced to an average
of \SI[per-mode=symbol]{10}{\mega\bit\per\second} without loss of physics information. The data is transmitted to the White Sands Ground Terminal in New Mexico
via the \gls{HRDL} link using the NASA \gls{TDRS} satellite system. Afterwards the data is forwarded to \gls{MSFC}, buffered and finally transmitted to the AMS-02 \gls{POCC} at CERN
via the Internet.

The incoming raw data is reconstructed using the AMS-02 \textit{gbatch} software package, which was developed by the collaboration during the past decade. The result no longer contains
raw data, but high-level reconstructed objects such as tracks, showers, particles, etc. The data files are written to disk as ROOT~\cite{Brun1997} trees and are called \textbf{AMS ROOT} files.
The 6.5 years dataset used in this thesis corresponds to \SIapprox{890}{\tera\byte} of data. Each AMS ROOT file covers approximately \SIapprox{22}{\minute}
of science data and has a size of \SIapprox{6}{\giga\byte} on average. In total \SI{143608}{} files have to be processed, to run once over the dataset. Due to limitations in the
\textit{gbatch} software framework, each file has to be processed using a single CPU core. The typical processing time of the full dataset is in the order of weeks when using a few thousand CPU
cores. Furthermore when e.g. running 1000 batch jobs at the same time on a supercomputer - each processing a single AMS ROOT file - the I/O
access pattern is inefficient. Many random accesses on the storage backends lead to I/O congestion and thus the available bandwidth is dramatically decreasing and the runtime of each job increases.

In order to speed up the turnaround time for analysis I developed a special data format together with Bastian Beischer and Henning Gast to overcome limitations of the
AMS ROOT files: the \textbf{ACQt files}. An ACQt file can be analyzed in parallel by multiple nodes in a supercomputer, using hundreds of CPU cores at the same time. The \textit{ACsoft}
software package produces ACQt files from AMS ROOT files, maintaining a 1:1 correspondence. After a full data production, a certain number of ACQt files are merged together
in so-called \textbf{Multi-ACQt files}. Each Multi-ACQt file covers a full day of AMS-02 data and has a typical size of \SIapprox{100}{\giga\byte}. Furthermore the individual ACQt files are much
smaller on disk than their AMS ROOT file counterparts, since the information is tightly encoded and optimized as much as possible\footnote{Each variable is transformed and rounded to fit into
the smallest possible bit representation: between \SIrange{1}{4}{} byte.}. This allows us to shrink the analysis turnaround time from weeks to days, which makes an
iterative analysis, like the one presented in this thesis, possible.

During the past seven years I developed the whole framework, in collaboration with Bastian Beischer, that automatizes the ACQt file production, ensures their consistency
and takes care of the bookkeeping. Each dataset is cataloged in a database: the number of events in each file, the production duration, the disk space consumption etc.
The whole AMS analysis group in Aachen uses this analysis framework and significant scientific results were produced using it, such as Refs.~\cite{Aguilar2014a,Aguilar2018}.

\section{Event reconstruction}
\label{sec:analysis-event-reconstruction}

In order to identify electrons or positrons in events recorded by AMS-02, the measurements of the individual subdetectors need to be
combined. This is a challenging task, as ambiguities in the event reconstruction need to be resolved. An incoming electron or positron
might emit bremsstrahlung photons, which convert into secondary electron and positron pairs. The secondary products as well as the
primary particle may leave hits in the silicon tracker and thus multiple tracker tracks might be reconstructed. Furthermore multiple ECAL
showers might be reconstructed, if more than one particle propagates towards the ECAL and the displacement between the individual particles
(e.g. the primary particle and a secondary electron or positron) is large enough - which frequently happens at low energies.

\subsection{TOF velocity and charge reconstruction}
\label{sec:analysis-event-reconstruction-tof}

In the AMS-02 TOF reconstruction all PMTs covering the TOF paddles are read out independently to collect the TOF PMT anode and dynode signal.
If a signal is present in any of the TOF paddles, a TOF cluster is formed, associated with a time measurement and a layer/bar index~\cite{Bindi2014}.

From the reconstructed TOF clusters a pattern recognition algorithm selects TOF clusters in each of the four TOF layers belonging
to the same trajectory, by utilizing external information e.g. from the tracker or the TRD, to resolve ambiguities. This might lead
to multiple combinations of TOF clusters, each potentially belonging to a different incident particle.

For each of the TOF cluster combinations the charge in the upper and lower TOF bars is computed, as well as the time of flight velocity $\beta$
of the incident particle. The velocity $\beta = \Delta x / (\SI{}{\clight} \cdot \Delta t)$ is measured using the particle time of
flight $\Delta t$ between the upper and lower planes and the path length $\Delta x$ - e.g. given by the tracker.
The time measurement is performed using the anode signals from the PMT (\cref{sec:detector-tof}).

The measurement of the charge from the PMT anode signals allows to determine the charge with high precision\footnote{The PMT anode signals
start to saturate around \SIvarApprox{Z}{4}{\elementarycharge}. The PMT dynode signals are used to extend the measurement range up to heavy nuclei, such as Iron.}.
The energy deposited in the TOF bars, $\diff\text{E}/\diff\text{x}$, by the passage of particles is proportional to $Z^2$. Furthermore it depends on
the velocity of the particle, such that: $\diff\text{E}/\diff\text{x} \propto f(\beta) \cdot Z^2$.

For the analysis the correct TOF cluster combination needs to be selected, matching the selected tracker track - to ensure the $\Delta x$ was
taken from the correct track, and thus the $\beta$ is correctly computed.

\subsection{Tracker track reconstruction}
\label{sec:analysis-event-reconstruction-tracker-track}

In the AMS-02 tracker track reconstruction~\cite{Hass2004,BazoAlba2013} all silicon ladders are read out independently in X and Y direction. For each ladder and each direction,
strips with the highest amplitudes - seed strips - are identified. The seed strips define the starting point for the cluster building procedure.
The energy depositions in the strips to the left and to the right of the seed strip are merged together with the seed strip energy deposition, as long
as their amplitude is higher than a predefined threshold. The silicon strips in the interval $[\text{seedStrip - nStripsLeft},~\text{seedStrip + nStripsRight}]$
then form a cluster. The first phase of the track reconstruction yields an independent list of X and Y clusters for each ladder which are then matched in the second
phase to form reconstructed hits.

A reconstructed tracker hit consists of a Y cluster and eventually a X cluster. In AMS-02 the X-side is read out using much less channels (described in
\cref{sec:detector-tracker}) than the Y-side, leaving an ambiguity of \SI{8}{\centi\meter} in the coordinate along the ladder, that needs
to be resolved during the track reconstruction. The multiplicity resolving depends on external information from other subdetectors, such as the TOF or the TRD.
If the multiplicity could be resolved the reconstructed hit additionally contains an associated X cluster.

A reconstructed tracker track is formed by applying a pattern recognition algorithm to select possible combinations of reconstructed hits, that may
form a track, spanning multiple tracker layers. This potentially leads to a reconstruction of multiple tracker tracks for an event.

\subsection{TRD track reconstruction}
\label{sec:analysis-event-reconstruction-trd-track}

The AMS-02 TRD track reconstruction~\cite{Gast2015} identifies segments (straight lines in X-Z / Y-Z projection) for all TRD layers in X-Z orientation and all TRD layers
in Y-Z orientation, independently. This procedure might identify multiple segments per projection, depending on the amount of hits in the TRD.

A reconstructed TRD track consists of a segment in X-Z orientation and another segment in Y-Z orientation. If an ECAL shower is present in the event,
the TRD segments will be extrapolated to the ECAL shower with the highest energy and the position is compared to the centre-of-gravity of the shower.
The segments with the lowest deviation from the shower centre-of-gravity are selected to form a reconstructed TRD track. If no ECAL shower is present
in the event, a TRD track is only reconstructed, if exactly one segment in X-Z direction and one segment in Y-Z direction is available.

\subsection{ECAL shower and energy reconstruction}
\label{sec:analysis-event-reconstruction-ecal-shower}

In the AMS-02 ECAL shower reconstruction~\cite{Gallucci2015,Kounine2017a} all cells are considered that recorded an energy deposition exceeding a predefined threshold, corresponding to a few \SI{}{\MeV}.
In each layer the cell with the largest energy deposition is identified, forming the start position of the cluster search. The energy depositions in the cells to the
left and to the right of the start cluster are merged together, as long as their amplitude is higher than a predefined threshold -- similar to the cluster finding
procedure used in the tracker track reconstruction. This yields a number of one-dimensional clusters for each layer. Since the layers are alternating in X-Z and Y-Z direction
two-dimensional clusters can be formed, by utilizing external information (e.g. a reconstructed tracker track). One or more two-dimensional clusters form an ECAL shower, depending
on the spatial difference between the reconstructed clusters. If the clusters are close, they are associated to the same ECAL shower. If the displacement is large enough,
multiple ECAL showers may be identified for an event.

The sum of all energy depositions in all cells of the clusters associated with the ECAL shower
is called \textit{deposited energy}. As described in Ref.~\cite{Gallucci2015} several corrections are applied on the cell level energy depositions: attenuation correction
(correcting the light attenuation along the fiber), equalization (ensure that the response for MIPs is identical in each cell) and anode efficiency (the true energy
deposition is underestimated towards the borders of the cell). The energy scale used for the analysis is the \textit{reconstructed energy} which corrects the
\textit{deposited energy} for rear leakage: the loss of energy in transversal direction, when the electromagnetic shower cascade is not fully contained anymore in the
calorimeter\footnote{Only \SIapprox{75}{\percent} of the energy released by a \SI{}{\TeV} $e^{\pm}$ is contained in the calorimeter, the rest leaks out.}. The effect
gains importance with increasing energy. The reconstructed energy scale is derived with the help of the Monte-Carlo simulation and was verified in dedicated testbeam
measurements.

\subsection{Combining subdetector measurements}
\label{sec:analysis-event-reconstruction-combining-subdetectors}

Custom particle reconstruction algorithms were developed to combine the subdetector information in an optimal way.
The particle reconstruction algorithm utilizes all necessary AMS-02 subdetectors to form a \textbf{primary particle}.
The \textbf{primary particle} groups together the individual subdetector measurements belonging to the same particle
that entered the instrument. For each event the particle reconstruction algorithm defines exactly one primary particle, by construction,
even if certain subdetector information are missing (e.g. no tracker track reconstructed, no TOF $\beta$ / charge measurement available,
no ECAL shower present, etc.) or multiple combinations could be used to define the primary particle.

As first step in the particle reconstruction, the list of all reconstructed tracker tracks is filtered, to reject those
without a corresponding TOF $\beta$ / charge measurement. Multiple TOF $\beta$ and charge measurements are available for each event,
as the charge is derived from the TOF $\diff\text{E}/\diff\text{x}$ measurements multiple variants are computed, depending on the $\diff\text{x}$,
as described in \cref{sec:analysis-event-reconstruction-tof}.

As second step in the particle reconstruction, the spatial difference between the centre-of-gravity of each ECAL shower in the event
and each tracker track (extrapolated to the Z position of the ECAL shower centre-of-gravity) is computed.
The pair (ECAL shower + tracker track with associated TOF $\beta$ measurement), for which the spatial
difference between the track extrapolation and the shower centre-of-gravity is smallest, is selected as \textbf{particle candidate} for further analysis.

Thus after running the first two steps of the particle reconstruction algorithm, a TOF $\beta$ and charge measurement, an ECAL shower and a tracker track is selected,
forming a particle candidate. If any of these quantities is missing, a fallback solution is implemented: selecting e.g. the TOF
$\beta$ measurement, with the highest $\abs{\beta}$, or the tracker track with the best goodness-of-fit parameter in bending plane,
or the ECAL shower with the largest energy deposition.

As third step in the particle reconstruction, each reconstructed TRD track in the event is extrapolated to the Z position of the ECAL
shower centre-of-gravity and the spatial difference between the TRD track extrapolation and the shower centre-of-gravity is computed.
The TRD track whose spatial difference is smallest is selected as TRD track for further analysis. The TRD track is reconstructed as a
straight line, as no magnetic field is present in the TRD, which might alter the trajectory. To extrapolate the TRD track to the ECAL,
the magnetic field, produced by the magnet surrounding the inner tracker, needs to be taken into account. The extrapolation is done
using two hypotheses\footnote{Both hypotheses need to be checked, otherwise the extrapolation is correlated with the charge-confusion
effect. A misreconstructed charge-sign would lead to a decreased TRD matching efficiency, which needs to be avoided.}: use $(+p)$, or $(-p)$
as momentum for the extrapolation, where $p$ is proportional to the rigidity of the tracker track belonging to the primary particle candidate.

After the association of a TRD track, the primary particle reconstruction is complete, but does not guarantee that all subdetector measurements
refer to the same incident particle. When applying the preselection cuts - \cref{sec:analysis-data-selection-selection-cuts} - during
analysis, events with inconsistent measurements are rejected.

\subsection{Construction of the TRD estimator $\Lambda_{\text{TRD}}$}
\label{sec:analysis-event-reconstruction-trd-estimator}

The TRD allows one to separate light from heavy particles, as described in \cref{sec:detector-trd}.

The energy depositions in a single straw tube for electrons or positrons are higher than for protons, due to the additional TR photons that
are absorbed in the xenon -- this allows one to build an estimator which can be used to distinguish electrons or positrons from protons.
A particle traversing the TRD may leave traces in all 20 layers and these measurements can be combined into a single estimator: the TRD estimator $\Lambda_{\text{TRD}}$.

According to the Neyman-Pearson-Lemma~\cite{Neyman1933} the best statistical test to decide between two hypothesis is the likelihood ratio, which
can be defined as:

\begin{equation}
  \label{eq:trd-llh}
  \Lambda_{\text{TRD}} = -\log \left(\mathcal{L}_{e^{-}}/(\mathcal{L}_{e^{-}}+\mathcal{L}_{p})\right).
\end{equation}

The determination of the PDFs $\mathcal{L}_{e^{-}}$ and $\mathcal{L}_{p}$ in \cref{eq:trd-llh} is the key ingredient\footnote{The set of
PDFs used in this analysis are called \enquote{Trd-P}~\cite{Gast2015} and were derived in the Aachen group in 2015.} to build the TRD estimator.

The PDFs depend on the gas composition (the partial pressure of the xenon - $\text{p}(\ce{Xe})$, the layer number $n$
(relevant for electrons as the amount of transition radiation depends on the number of layers crossed in the TRD), the
rigidity $R$ of the particle and the path length $\diff\text{x}$ traversed in the straw tube.

There are two independent ways to compute the path length $\diff\text{x}$. The first option is to determine the path length purely
from the TRD. In each projection X-Z and Y-Z track finding is performed, by combining the hits in the straw tubes into segments (\cref{sec:analysis-event-reconstruction-trd-track}).
The best matching segments in X-Z and Y-Z are then combined into a 3D trajectory from which the path length in each straw tube
can be calculated. The accuracy is limited due to the rather large diameter of the straw tubes of \SI{6}{\milli\meter}.

The second option is to rely on the extrapolation of a track reconstructed in the silicon tracker, which offers a much more
precise trajectory reconstruction, but fails at the lowest rigidity where the extrapolation of the tracker track through the TRD
is imprecise due to multiple scattering in the material above the silicon tracker. This distorts the trajectory and is relevant
for electrons below rigidities of \SIvarApprox{R}{8}{\giga\volt}.

Therefore the best option is to combine both approaches using an rigidity dependent weighted mean, and use the path length estimation
from the most precise source, depending on the reconstructed rigidity of the particle. This hybrid approach is used to form the
TRD estimator $\Lambda_{\text{TRD}}$, which offers a rejection power of greater than $10^2$ between \SIrange{0.5}{700}{\GeV}, as shown
in \cref{sec:detector-trd}.

\subsection{Construction of the ECAL estimator $\Lambda_{\text{ECAL}}$}
\label{sec:analysis-event-reconstruction-ecal-estimator}

The ECAL allows one to separate electrons or positrons from protons, as described in \cref{sec:detector-ecal}, by exploiting
the differences in the shower development in the electromagnetic calorimeter.

In contrary to electrons or positrons, protons traversing the ECAL do not develop electromagnetic showers. Since the nuclear radiation length $\lambda$ of the ECAL is
0.6, \SIapprox{50}{\percent} of the protons traverse the whole ECAL only losing energy via ionization. The typical
energy loss is in the order of a few hundred \SI{}{\MeV}. These protons are called \gls{MIP}.

Proton MIP events can be easily rejected by a minimum energy cut on the order of \SIapprox{0.5}{\GeV}. Protons that develop
\textit{hadronic showers} have a different characteristic with respect to \textit{electromagnetic showers}. The hadronic showering
process is dominated by inelastic hadronic interactions. High energetic protons may produce multiple particles (mostly charged pions and nucleons)
due to the inelastic interactions. Secondary particles are emitted originating from nuclear decay of excited nuclei.

Because of the $\pi^0$ creation that decay predominantly into $\gamma\gamma$, there is also an electromagnetic component
present in hadronic showers, usually starting deeper in the ECAL than a purely electromagnetic shower.

The differences in the shower development between electrons or positrons and protons were exploited to construct a multi-variate
estimator - $\Lambda_{\text{ECAL}}$ - utilizing 22 variables based on the longitudinal shower profile, the lateral shower profile
energy fractions (energy released in first layer divided by total energy deposition, etc.) and many more.
The ECAL estimator $\Lambda_{\text{ECAL}}$ combined with the Tracker ($E/\abs{R}$ cut) offers a rejection power greater than $10^4$ between \SIrange{3}{1000}{\GeV},
as shown in \cref{sec:detector-ecal}.

\subsection{Construction of the charge-confusion estimator $\Lambda_{\text{CC}}$}
\label{sec:analysis-event-reconstruction-construction-ccmva-estimator}

A crucial part of the electron and positron analysis is the understanding and treatment of the charge-confusion effect.
The charge sign of a particle might be misreconstructed due to interactions in the detector or the finite resolution of the silicon
tracker at the highest energies. In this analysis, special attention was given to construct a multi-variate estimator
which can be used to compare the charge-confusion prediction from the Monte-Carlo simulation with the ISS data: the CCMVA estimator $\Lambda_{\text{CC}}$.

It is important to note that the topology of the event is responsible for the amount of charge-confusion: events where only a single tracker
track was reconstructed, matching to a single TRD track and a single ECAL shower have the smallest probability to have a misreconstructed charge sign.
The probability of misidentification of the charge-sign increases with the track multiplicity: the more tracks are reconstructed, the more
probable it is for the track reconstruction to associate wrong hits to a track, eventually resulting in a flipped charge sign or a
wrong rigidity value.

In total 21 variables were identified that are sensitive to charge-confusion, 15 are relevant in both single-track and multi-tracks samples,
whereas six variables are only relevant in the multi-tracks sample. In the following all CCMVA input quantities are defined and the distributions
taken from the electron Monte-Carlo simulation are compared between the correct and wrong reconstructed charge sample. In \Cref{sec:appendix-ccmva-input-variables}
a comparison of all input variables between the electron Monte-Carlo simulation and the ISS negative rigidity data sample is shown, demonstrating
that all input variables are consistent between the Monte-Carlo simulation and ISS data.

The six most important input variables for the example energy bin \SIrange{17.98}{18.99}{\GeV} - identified by
TMVA~\cite{TMVA2007} during training - out of the 15 relevant for the single-track sample are presented - the remaining nine variables are described
in \cref{sec:appendix-ccmva-input-variables}.

\begin{enumerate}
  \item\textbf{Lower TOF charge}\hfill\\

    The lower TOF charge is computed by taking the average of the available charge measurements in the lower TOF clusters, as identified in the particle reconstruction, described in
    \cref{sec:analysis-event-reconstruction-combining-subdetectors}.

    \begin{figure}[H]
      \begin{subfigure}{0.50\linewidth}
        \includegraphics[width=\linewidth]{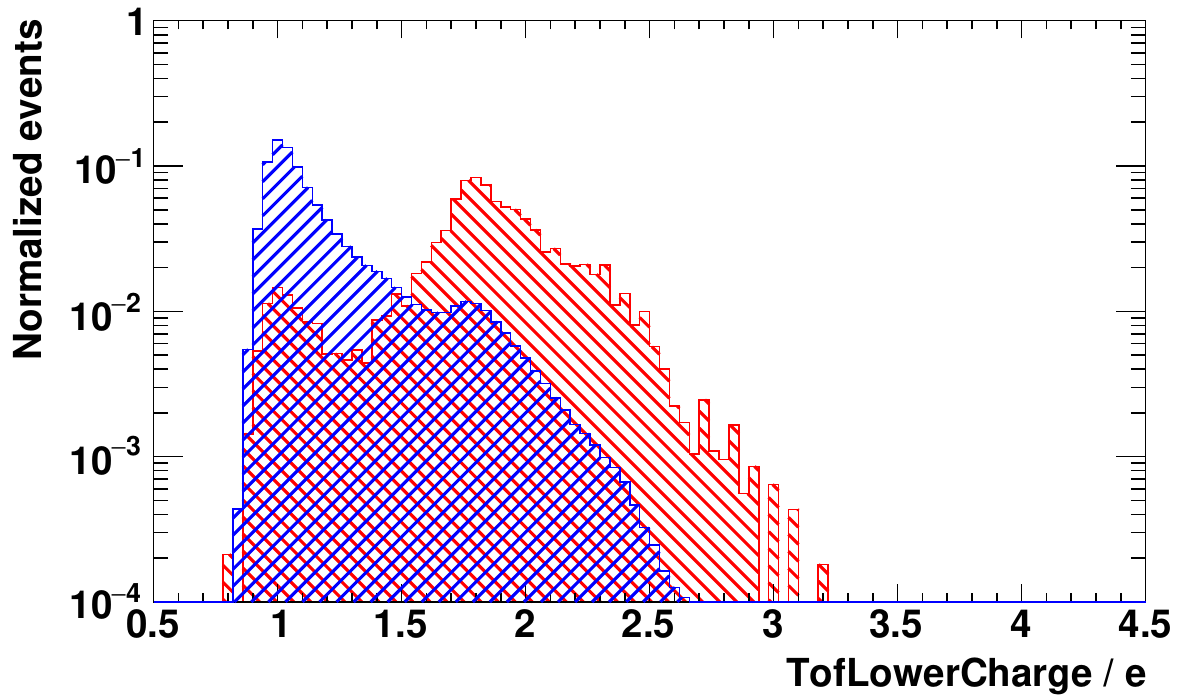}
        \caption{Single-track sample}
      \end{subfigure}
      \hfill
      \begin{subfigure}{0.50\linewidth}
        \includegraphics[width=\linewidth]{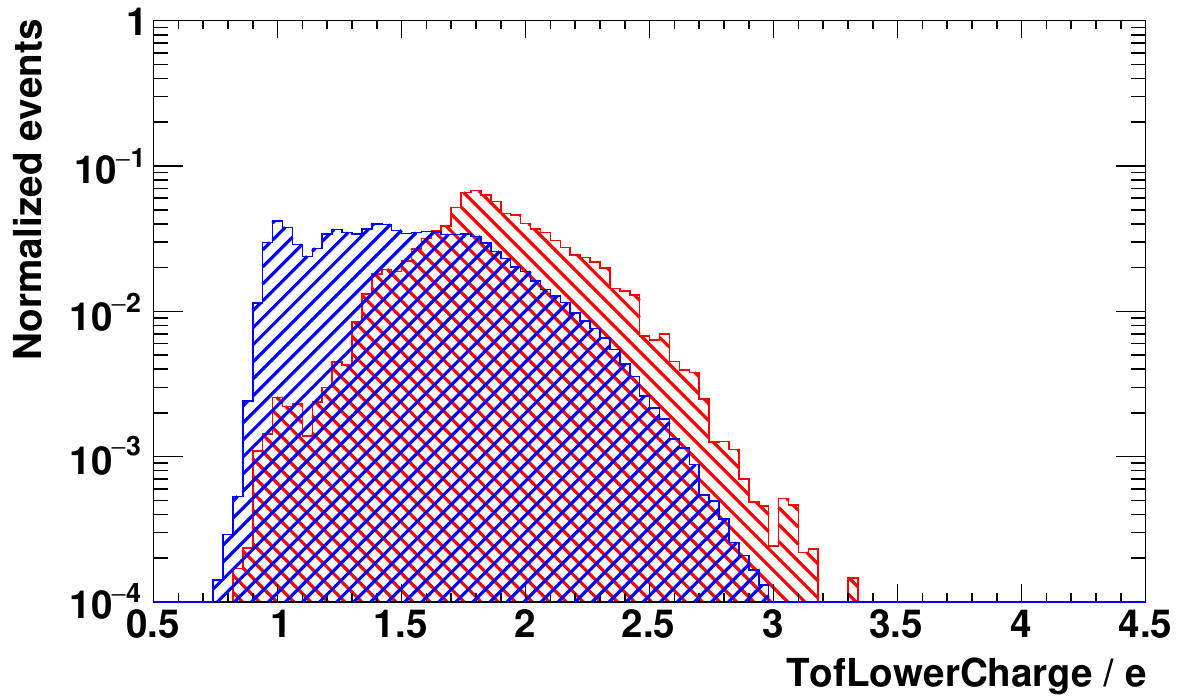}
        \caption{Multi-tracks sample}
      \end{subfigure}
      \caption{Example of TofLowerCharge distribution in the energy bin \SIrange{17.98}{18.99}{\GeV} in the electron Monte-Carlo simulation. The red histogram shows the charge-confused sample (R > 0) and the blue histogram the correct reconstructed sample (R < 0), after applying all preselection, selection and $e^{\pm}$ identification cuts.}
      \label{fig:ccmva-input-toflowercharge-correct-wrong-sign-mc-comparison}
    \end{figure}

    \Cref{fig:ccmva-input-toflowercharge-correct-wrong-sign-mc-comparison} shows the TofLowerCharge distribution in an example energy bin
    for both the correct and wrong rigidity sample in the electron Monte-Carlo simulation.

    The lower TOF charge is usually larger than the upper TOF charge as bremsstrahlung photons might convert in the inner tracker between the upper and lower TOF bars.
    The difference between the upper and lower TOF charges is an indication for the amount of interactions in the inner tracker and thus a useful quantity for the MVA estimator.

  \item\textbf{Energy/Rigidity matching}\hfill\\

    LogEnergyOverRigidity is defined as the logarithm of the energy deposited in the ECAL divided by the rigidity of the selected primary tracker track:

    \begin{equation*}
      \text{LogEnergyOverRigidity} = \log \left(E_{\text{ECAL}}/\abs{R_{\text{primary}}}\right)
    \end{equation*}

    When the primary particle interacted during the passage through AMS, by bremsstrahlung emission and conversions of the emitted photons, the secondary particles/photons usually
    travel collinear to the primary particle and are contained within the vicinity of the shower of the primary particle in the ECAL.

    \begin{figure}[H]
      \begin{subfigure}{0.50\linewidth}
        \includegraphics[width=\linewidth]{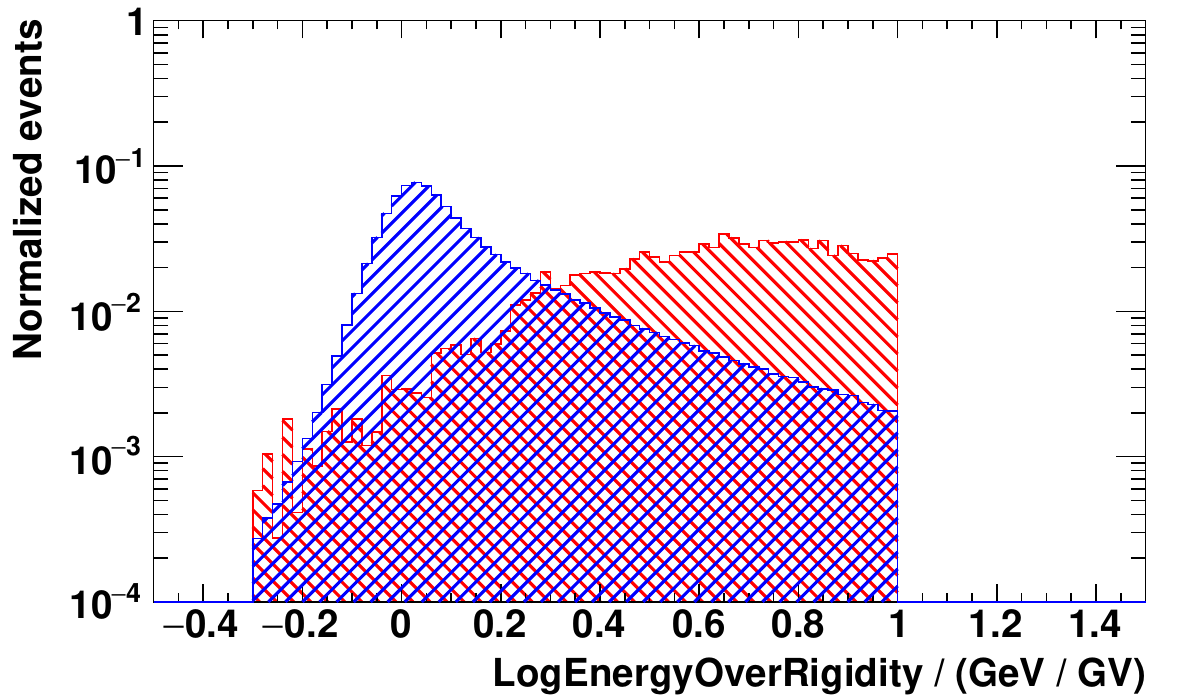}
        \caption{Single-track sample}
      \end{subfigure}
      \hfill
      \begin{subfigure}{0.50\linewidth}
        \includegraphics[width=\linewidth]{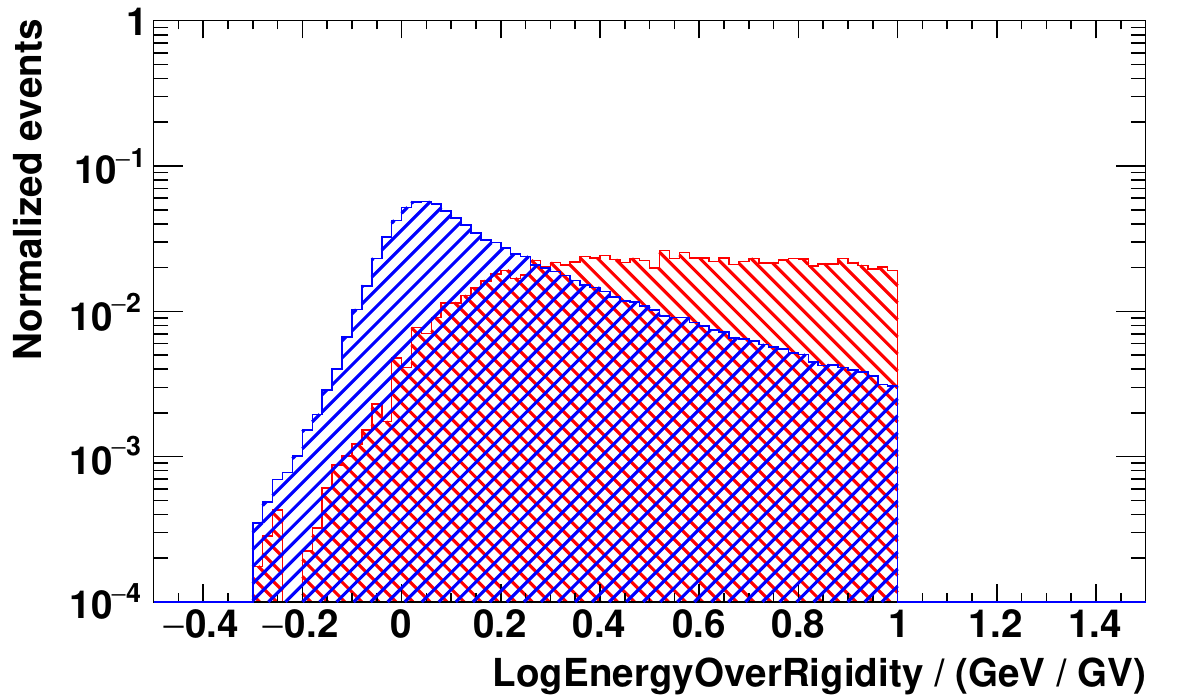}
        \caption{Multi-tracks sample}
      \end{subfigure}
      \caption{Example of LogEnergyOverRigidity distribution in the energy bin \SIrange{17.98}{18.99}{\GeV} in the electron Monte-Carlo simulation. Note that LogEnergyOverRigidity is bounded between $-0.3 < \log \left(E_{\text{ECAL}}/\abs{R_{\text{primary}}}\right) < 1$, due to the $E\/\abs{R}$ cut in the $e^{\pm}$ identification cuts.}
      \label{fig:ccmva-input-logenergyoverrigidity-correct-wrong-sign-mc-comparison}
    \end{figure}

    The rigidity on the other hand is more difficult to reconstruct, as in the case of multiple reconstructed tracks, a misassociation of individual hits is possible, leading to a distorted rigidity measurement.
    Correctly reconstructed events usually peak at 0, wrongly reconstructed events at larger or negative values, as shown in \cref{fig:ccmva-input-logenergyoverrigidity-correct-wrong-sign-mc-comparison}.

  \item\textbf{Influence of single tracker hits on rigidity reconstruction}\hfill\\

    In order to study the influence of single hits on the reconstructed rigidity, it is necessary to refit the track several times.
    In each iteration one of the layers is explicitly excluded from the track fit and the rigidity is determined. The difference in sagitta ($\propto$ inverse rigidity)
    with respect to the original track fit, which includes all available hits, is calculated for each layer. The maximum difference in sagitta
    enters the MVA estimator as input quantity:

    \begin{equation*}
      \text{TrkMaxDeltaSagittaFromRemovingHits} = \sign(R_{\text{primary}}) \\
                                                  \max_{l~\in~[1,9]} \abs{\frac{1}{R_{\text{primary}}} - \frac{1}{R_{\text{exclude~layer}~l}}}
    \end{equation*}

    The $\sign(R_{\text{primary}})$ factor ensures that the quantity is symmetric between correct reconstructed electrons and positrons.
    \Cref{fig:ccmva-input-trkmaxdeltasagittafromremovinghits-correct-wrong-sign-mc-comparison} shows the TrkMaxDeltaSagittaFromRemovingHits distribution in an example energy bin
    for both the correct and wrong rigidity sample in the electron Monte-Carlo simulation.

    \begin{figure}[H]
      \begin{subfigure}{0.50\linewidth}
        \includegraphics[width=\linewidth]{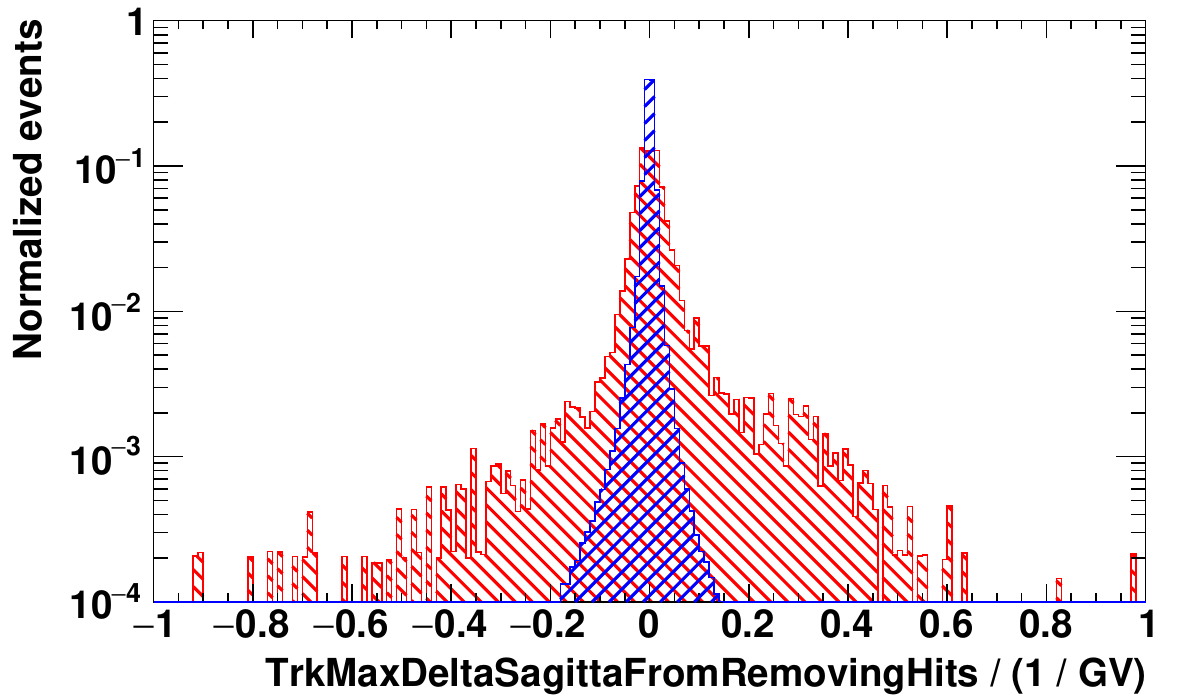}
        \caption{Single-track sample}
      \end{subfigure}
      \hfill
      \begin{subfigure}{0.50\linewidth}
        \includegraphics[width=\linewidth]{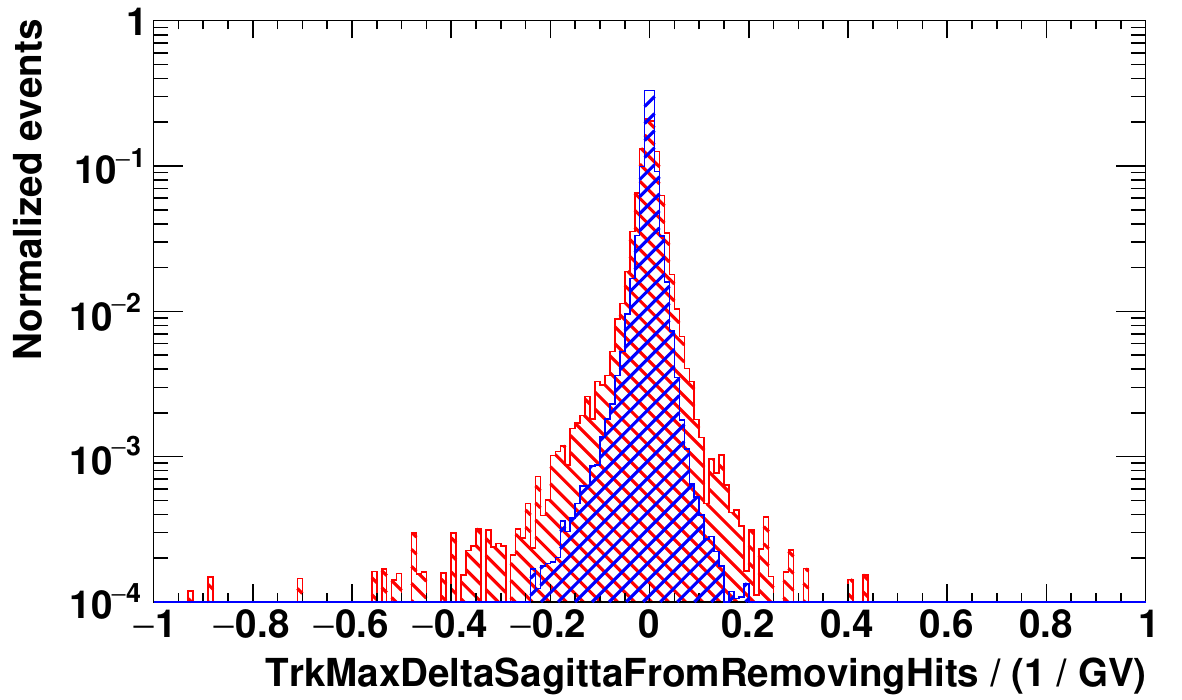}
        \caption{Multi-tracks sample}
      \end{subfigure}
      \caption{Example of TrkMaxDeltaSagittaFromRemovingHits distribution in the energy bin \SIrange{17.98}{18.99}{\GeV} in the electron Monte-Carlo simulation. The red histogram shows the charge-confused sample (R > 0) and the blue histogram the correct reconstructed sample (R < 0), after applying all preselection, selection and $e^{\pm}$ identification cuts.}
      \label{fig:ccmva-input-trkmaxdeltasagittafromremovinghits-correct-wrong-sign-mc-comparison}
    \end{figure}

  \item\textbf{Inner tracker charge}\hfill\\

    \begin{figure}[H]
      \begin{subfigure}{0.50\linewidth}
        \includegraphics[width=\linewidth]{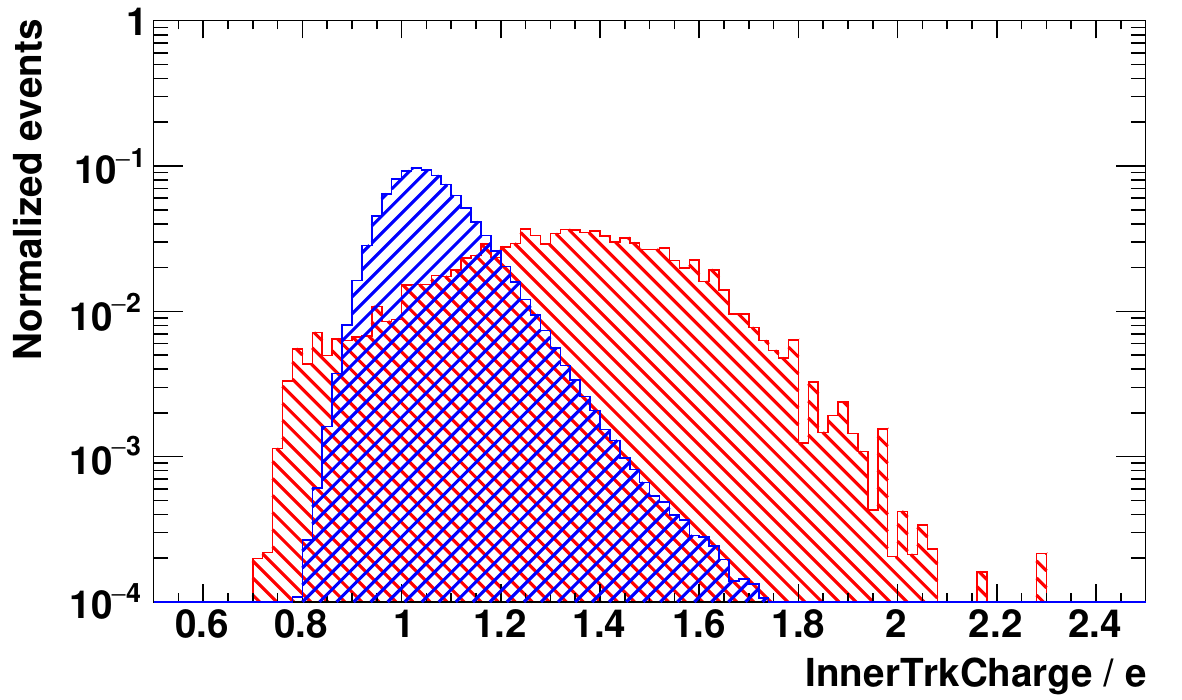}
        \caption{Single-track sample}
      \end{subfigure}
      \hfill
      \begin{subfigure}{0.50\linewidth}
        \includegraphics[width=\linewidth]{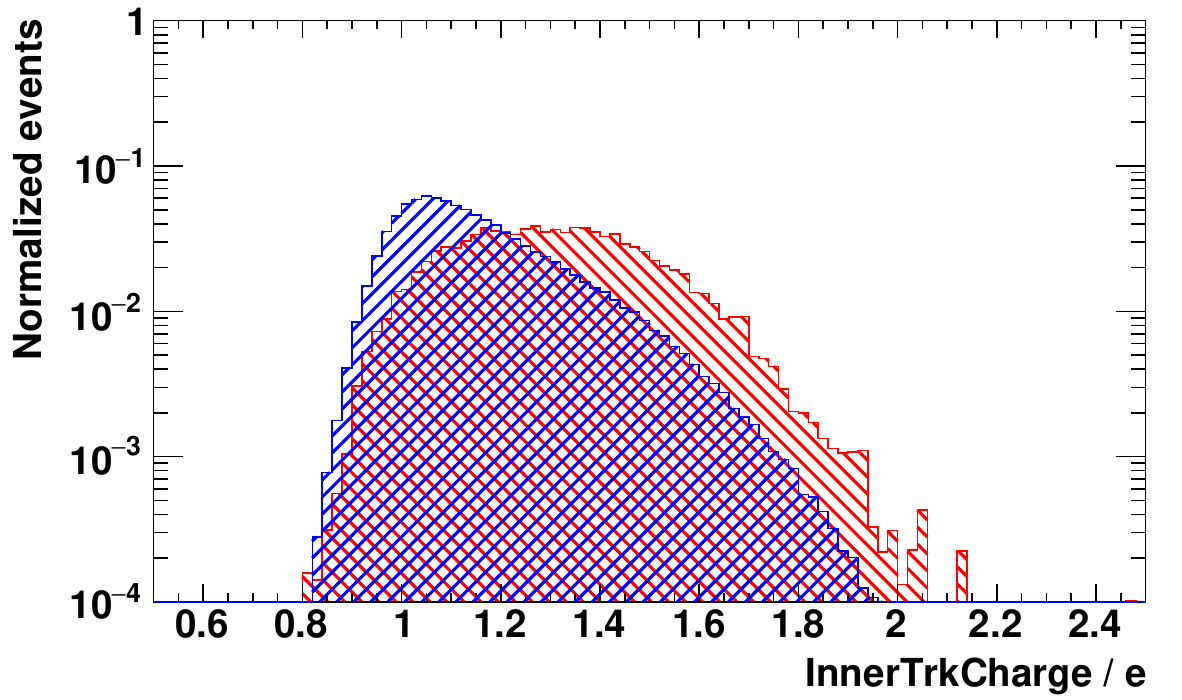}
        \caption{Multi-tracks sample}
      \end{subfigure}
      \caption{Example of InnerTrkCharge distribution in the energy bin \SIrange{17.98}{18.99}{\GeV} in the electron Monte-Carlo simulation. The red histogram shows the charge-confused sample (R > 0) and the blue histogram the correct reconstructed sample (R < 0), after applying all preselection, selection and $e^{\pm}$ identification cuts.}
      \label{fig:ccmva-input-innertrkcharge-correct-wrong-sign-mc-comparison}
    \end{figure}

    The inner tracker charge offers a handle to distinguish interactions within the inner tracker.
    It is computed based on the Y charges in the inner tracker layers using a truncated mean algorithm, ignoring the highest outliers.
    The wrong reconstructed sample tends to have a larger inner tracker charge, as shown in \cref{fig:ccmva-input-innertrkcharge-correct-wrong-sign-mc-comparison}.

  \item\textbf{Tracker track goodness-of-fit in X-Z orientation \&} \vspace{-1em} \item\textbf{Tracker track goodness-of-fit in Y-Z orientation}\hfill\\

    The logarithms of the track fit $\chi^2$ in X-Z and Y-Z projection enter the MVA as input quantities. It is expected that
    the wrong charge sign sample has a tendency to larger $\chi^2$ values as the correct charge sign sample, which offers
    separation power for the MVA.

    \Cref{fig:ccmva-input-logtrkchi2xy-correct-wrong-sign-mc-comparison} shows the LogTrkChi2X and LogTrkChi2Y distribution in an example energy bin
    for both the correct and wrong rigidity sample in the electron Monte-Carlo simulation.

    \begin{figure}[H]
      \begin{subfigure}{0.50\linewidth}
        \includegraphics[width=\linewidth]{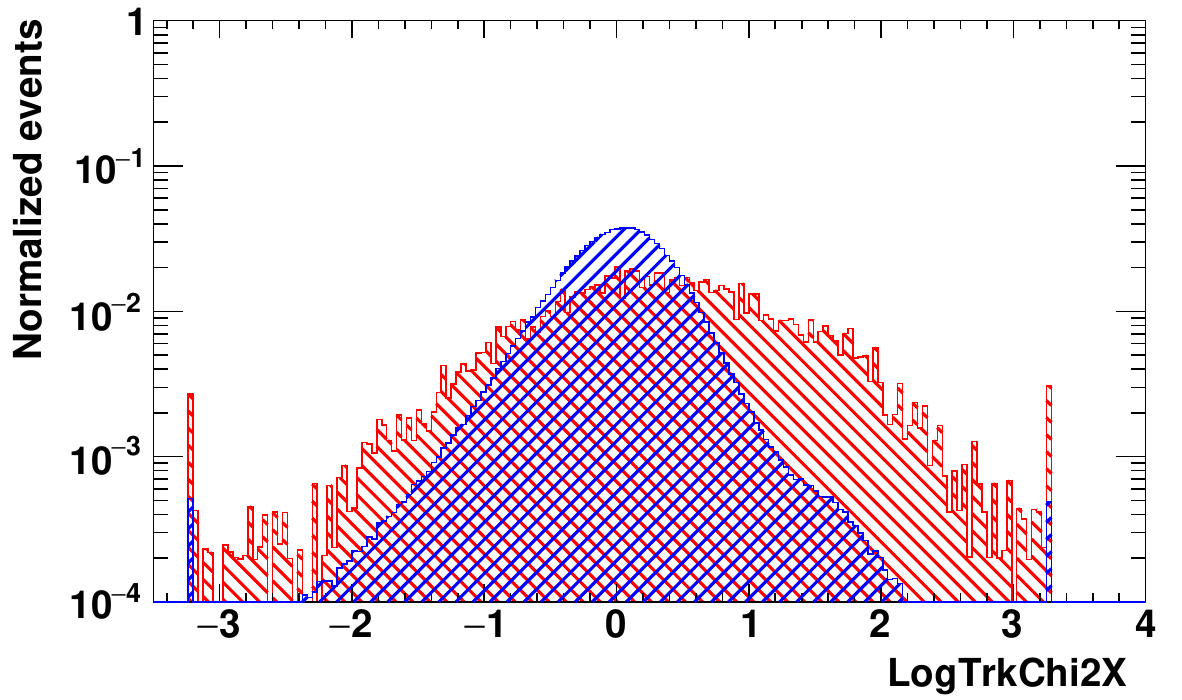}
        \caption{Single-track sample}
      \end{subfigure}
      \hfill
      \begin{subfigure}{0.50\linewidth}
        \includegraphics[width=\linewidth]{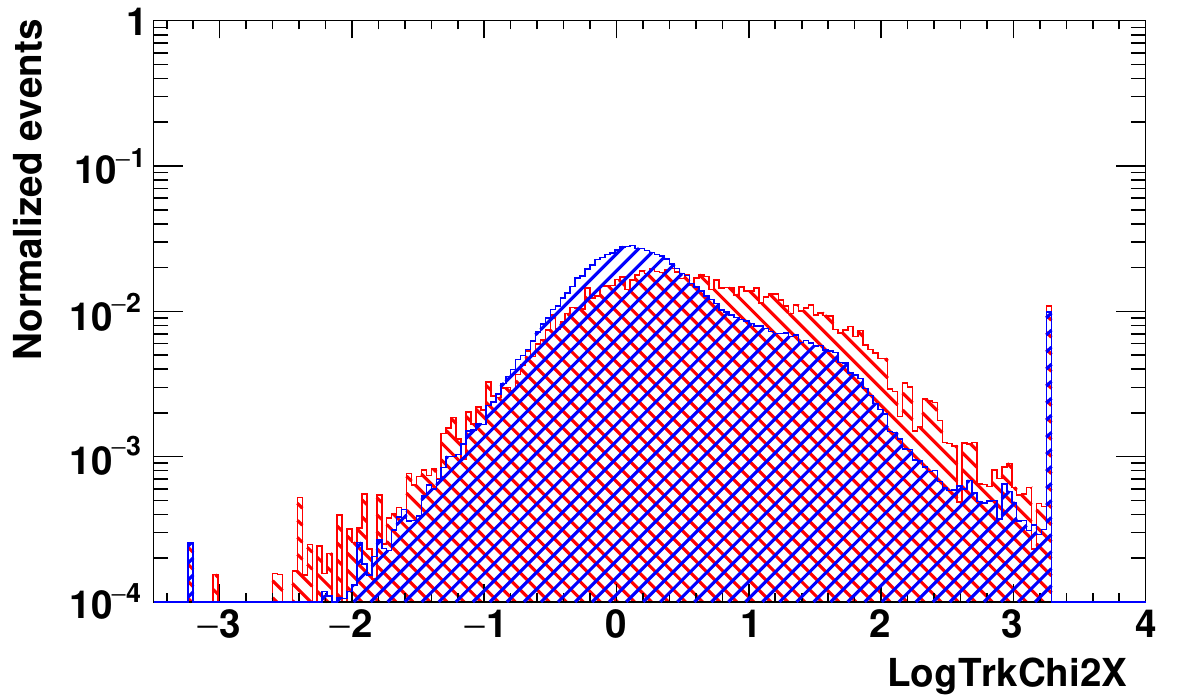}
        \caption{Multi-tracks sample}
      \end{subfigure}
      \begin{subfigure}{0.50\linewidth}
        \includegraphics[width=\linewidth]{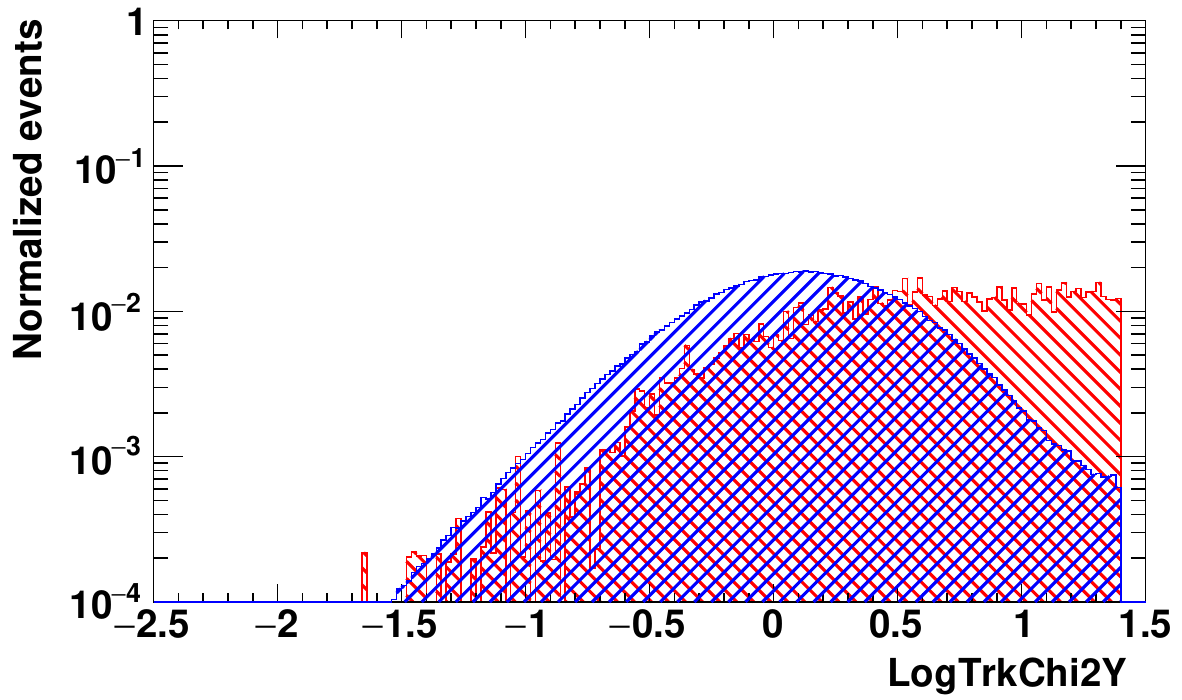}
        \caption{Single-track sample}
      \end{subfigure}
      \hfill
      \begin{subfigure}{0.50\linewidth}
        \includegraphics[width=\linewidth]{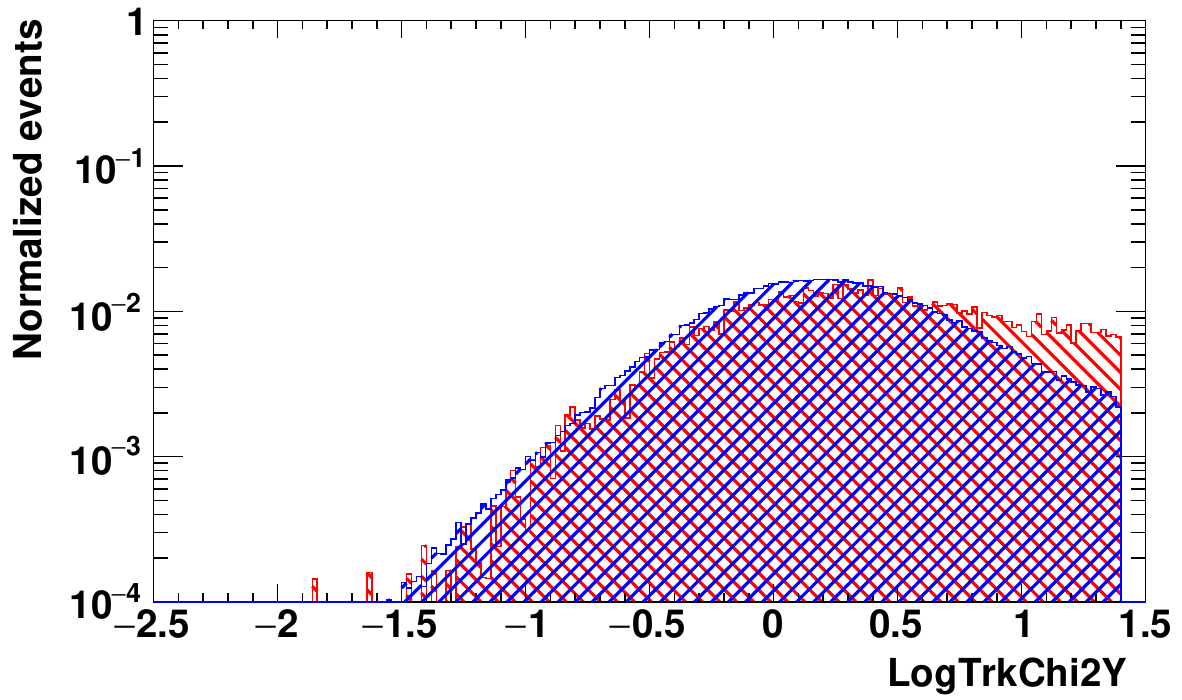}
        \caption{Multi-tracks sample}
      \end{subfigure}
      \caption{Example of LogTrkChi2X / LogTrkChi2Y distribution in the energy bin \SIrange{17.98}{18.99}{\GeV} in the electron Monte-Carlo simulation. The red histogram shows the charge-confused sample (R > 0) and the blue histogram the correct reconstructed sample (R < 0), after applying all preselection, selection and $e^{\pm}$ identification cuts.}
      \label{fig:ccmva-input-logtrkchi2xy-correct-wrong-sign-mc-comparison}
    \end{figure}
\end{enumerate}

The three most important input variables for the example energy bin \SIrange{17.98}{18.99}{\GeV} out of the six relevant
for the multi-tracks sample are presented in the following - the remaining three variables are described in \cref{sec:appendix-ccmva-input-variables}.

\begin{enumerate}
  \setcounter{enumi}{6}
  \item\textbf{Primary over sum of other rigidities}\hfill\\

    The logarithm of the ratio between the primary rigidity of the selected tracker track $R_{\text{primary}}$ and the reconstructed rigidities $R_{i}$ - associated with the secondary tracker tracks $i$ - is defined as LogPrimaryRigidityOverOtherRigiditiesRatio:

    \begin{equation*}
      \text{LogPrimaryRigidityOverOtherRigiditiesRatio} = \log \left(R_{\text{primary}}\bigg/\sum_{i = 2}^{N}{R_{i}}\right)
    \end{equation*}

    \Cref{fig:ccmva-input-logenergyoverrigiditiesratio-correct-wrong-sign-mc-comparison} shows the LogPrimaryRigidityOverOtherRigiditiesRatio distribution in an example energy bin
    for both the correct and wrong rigidity sample in the electron Monte-Carlo simulation.

    \begin{figure}[H]
      \centering
      \includegraphics[width=0.60\linewidth]{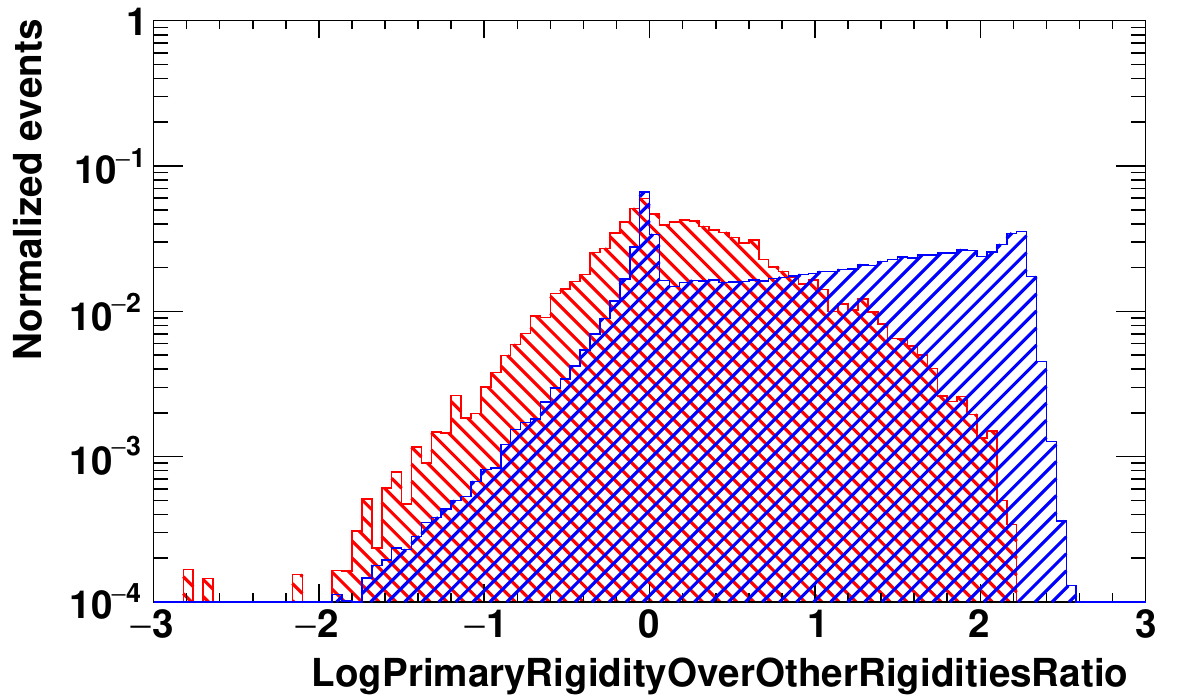}
      \caption{Example of LogPrimaryRigidityOverOtherRigiditiesRatio distribution in the energy bin \SIrange{17.98}{18.99}{\GeV} in the electron Monte-Carlo simulation. The red histogram shows the charge-confused sample (R > 0) and the blue histogram the correct reconstructed sample (R < 0), after applying all preselection, selection and $e^{\pm}$ identification cuts.}
      \label{fig:ccmva-input-logenergyoverrigiditiesratio-correct-wrong-sign-mc-comparison}
    \end{figure}

  \item\textbf{Tracker track bottom distance x-direction \&} \vspace{-0.75em} \item\textbf{Tracker track bottom distance y-direction}\hfill\\

    The primary tracker track is extrapolated to tracker layer 9 at \SIvarEquals{$z$}{113.55}{\centi\meter} and the result is denoted as $x_{\text{bottom, primary}}$~/~$y_{\text{bottom, primary}}$.
    The extrapolation is repeated for all other reconstructed tracker tracks in the event.

    TrkMinDXBottom is then defined as as the minimum distance in x-direction between the primary tracker track and any of the secondary tracks at the tracker layer 9 position.
    TrkMinDYBottom refers to the y-coordinate in the bending plane:

    \begin{equation*}
      \begin{aligned}
        \text{TrkMinDXTop} &= \min_{i~>~1} \abs{x_{i} - x_{\text{bottom, primary}}} \\
        \text{TrkMinDYTop} &= \min_{i~>~1} \abs{y_{i} - y_{\text{bottom, primary}}}
      \end{aligned}
    \end{equation*}

    \begin{figure}[H]
      \centering
      \begin{subfigure}{0.57\linewidth}
        \includegraphics[width=\linewidth]{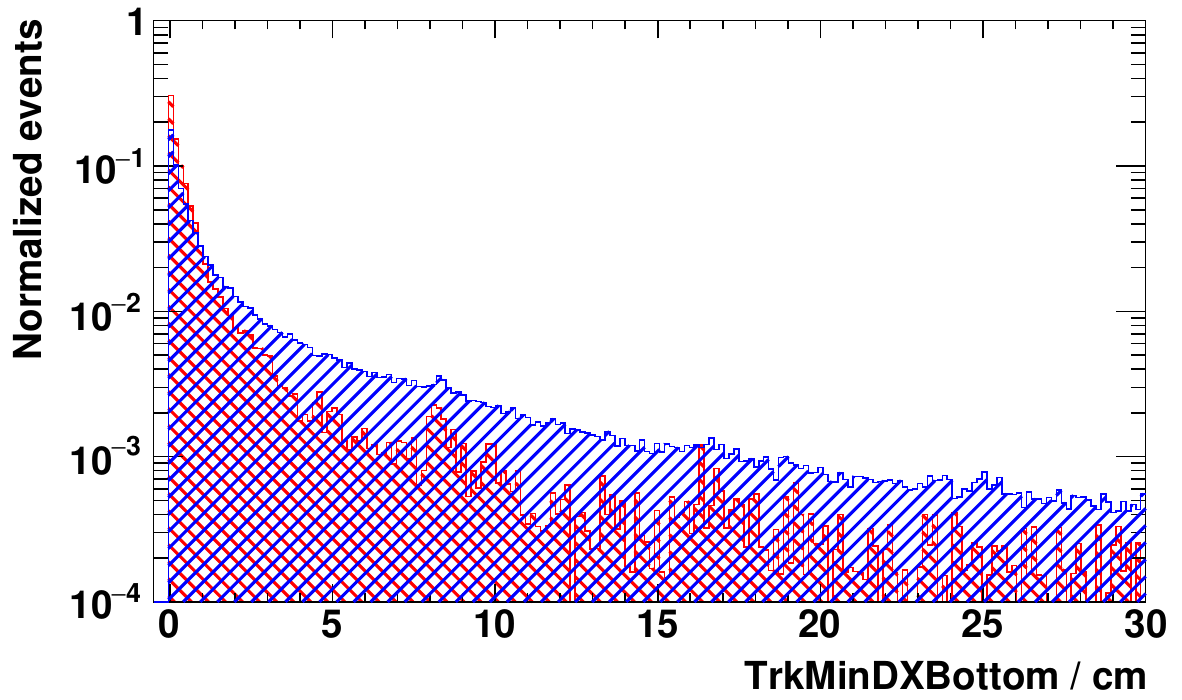}
        \caption{TrkMinDXBottom: Multi-tracks sample}
      \end{subfigure}
      \hfill
      \begin{subfigure}{0.57\linewidth}
        \includegraphics[width=\linewidth]{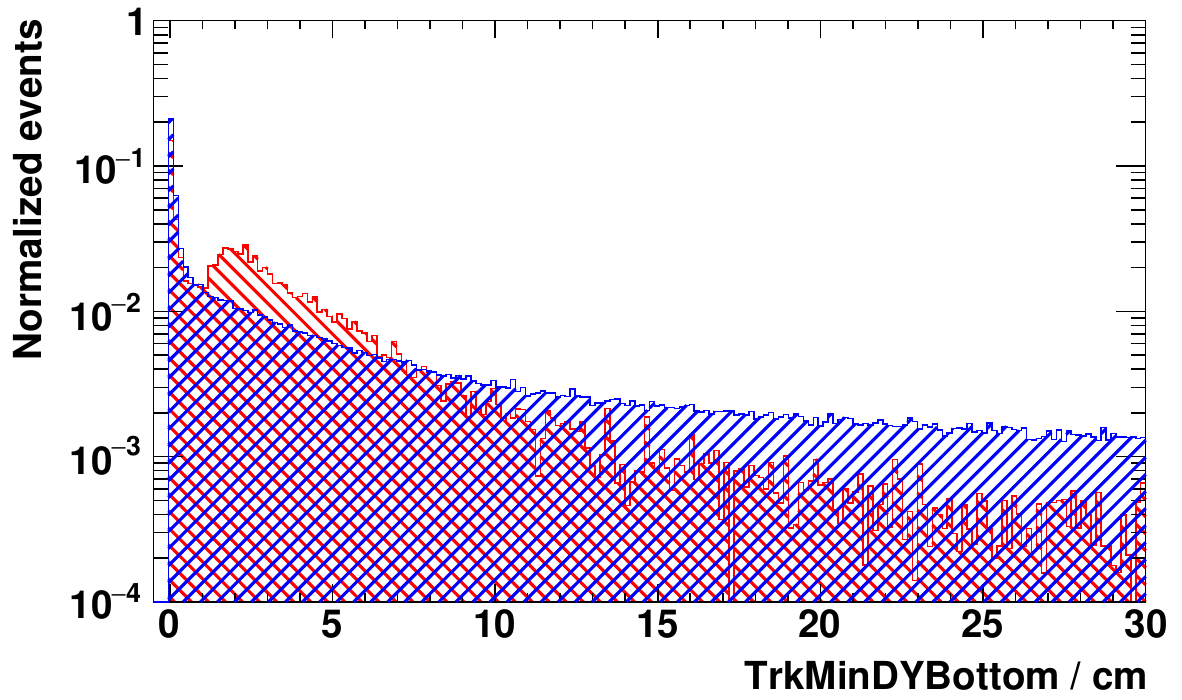}
        \caption{TrkMinDYBottom: Multi-tracks sample}
      \end{subfigure}
      \caption{Example of TrkMinDXBottom / TrkMinDYBottom distributions in the energy bin \SIrange{17.98}{18.99}{\GeV} in the electron Monte-Carlo simulation. The red histogram shows the charge-confused sample (R > 0) and the blue histogram the correct reconstructed sample (R < 0), after applying all preselection, selection and $e^{\pm}$ identification cuts.}
      \label{fig:ccmva-input-trkmindxdybottom-correct-wrong-sign-mc-comparison}
    \end{figure}

    \Cref{fig:ccmva-input-trkmindxdybottom-correct-wrong-sign-mc-comparison} shows the TrkMinDXBottom / TrkMinDYBottom distributions in an example energy bin
    for both the correct and wrong rigidity sample in the electron Monte-Carlo simulation.

    The shape of the TrkMinDYBottom distribution is clearly different for the correct and wrong rigidity sample and offers a unique way to discriminate the two samples in the MVA.
    It offers the highest separation power of all observables in the multi-tracks sample.
\end{enumerate}

\bigskip
In \cref{sec:appendix-ccmva-input-variables}, a comparison of all input variables between the electron and positron Monte-Carlo simulation can be found.
There are no charge-sign dependent differences in the input quantities -- by construction
the same MVA is applicable to electrons and positrons. An electron with negative charge sign appears identical as a positron with positive charge sign,
and vice-versa.

The CCMVA estimator $\Lambda_{\text{CC}}$ is trained individually for each energy bin in the analysis using the aforementioned input variables on
an electron Monte-Carlo simulation, using the TMVA~\cite{TMVA2007} package. Several classifier algorithms (Multi-Layer Perceptron,
Fisher-Likelihood, etc.) were tested and the \gls{BDT}~\cite{Roe2005} was selected for this analysis.
The output shape is gaussian like and suitable for a template fit. 300 individual decision trees were trained for each energy bin
using an adaptive gradient boosting technique.

All preselection, selection and $e^{\pm}$ identification cuts - \cref{sec:analysis-data-selection-preselection-cuts,sec:analysis-data-selection-selection-cuts,sec:analysis-data-selection-electron-positron-identification-cuts} are applied on the
electron Monte-Carlo simulation to prepare the training data sample.

If the reconstructed rigidity of the primary tracker track is negative it is classified as \textbf{\enquote{signal event}}
otherwise as \textbf{\enquote{background event}}. This is the only distinction\footnote{Other definitions were tested, such as
requiring the reconstructed rigidity to be compatible with the generated energy of the particle within the four times the
rigidity resolution. The rejection power of the so-obtained MVA did not improve or change, compared to the simple definition,
by only using the rigidity sign to discriminate signal / background events.} to differentiate between the correct and wrong
reconstructed rigidity sample. Furthermore the sample is split in two disjoint categories: the \textbf{\enquote{single-track sample}}
(TrkNumTracks = 1) and the \textbf{\enquote{multi-tracks sample}} (TrkNumTracks > 1). The training is applied individually for
each energy bin and each of the tracker pattern samples. Overtraining checks were performed, to ensure that the MVA is unbiased.

\begin{figure}[H]
  \begin{subfigure}{0.5\linewidth}
    \includegraphics[width=\linewidth]{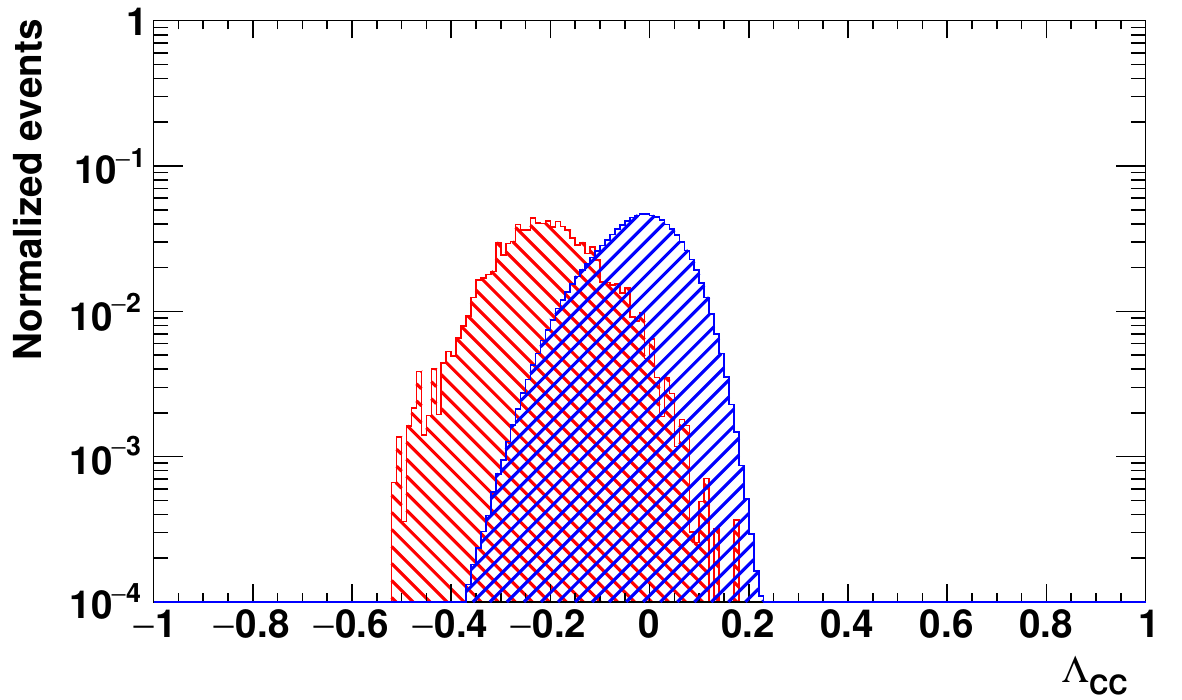}
    \caption{\SIrange{2}{2.31}{\GeV}}
  \end{subfigure}
  \hfill
  \begin{subfigure}{0.5\linewidth}
    \includegraphics[width=\linewidth]{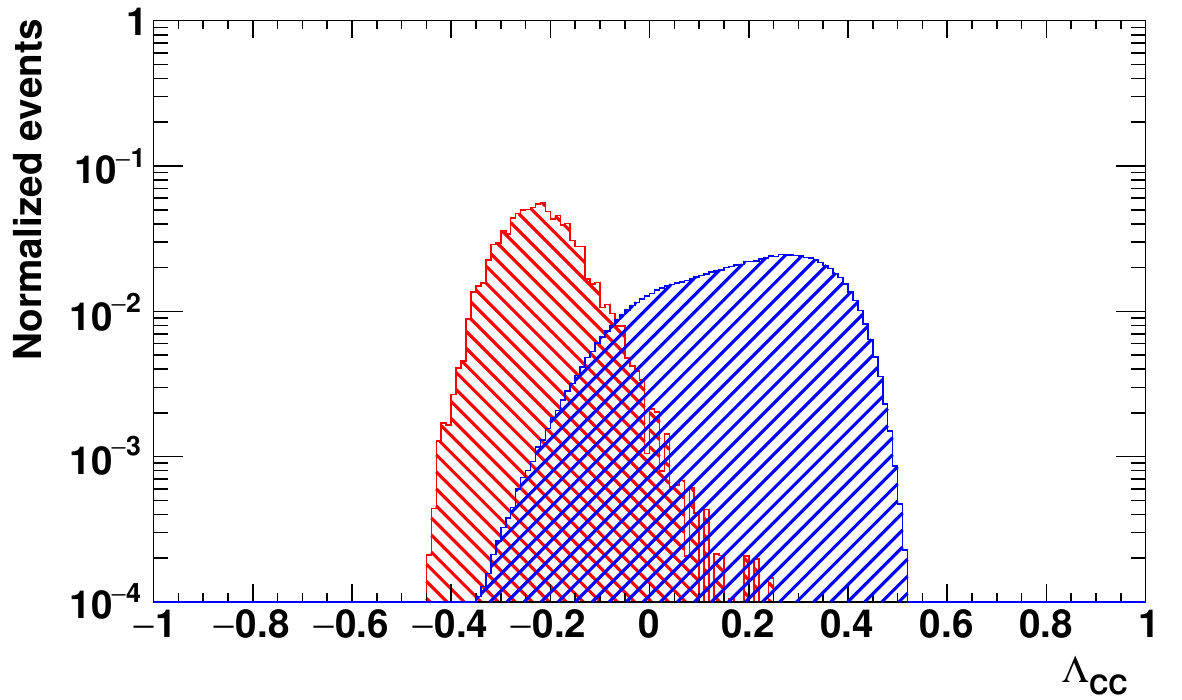}
    \caption{\SIrange{17.98}{18.99}{\GeV}}
  \end{subfigure}
  \hfill
  \begin{subfigure}{0.5\linewidth}
    \includegraphics[width=\linewidth]{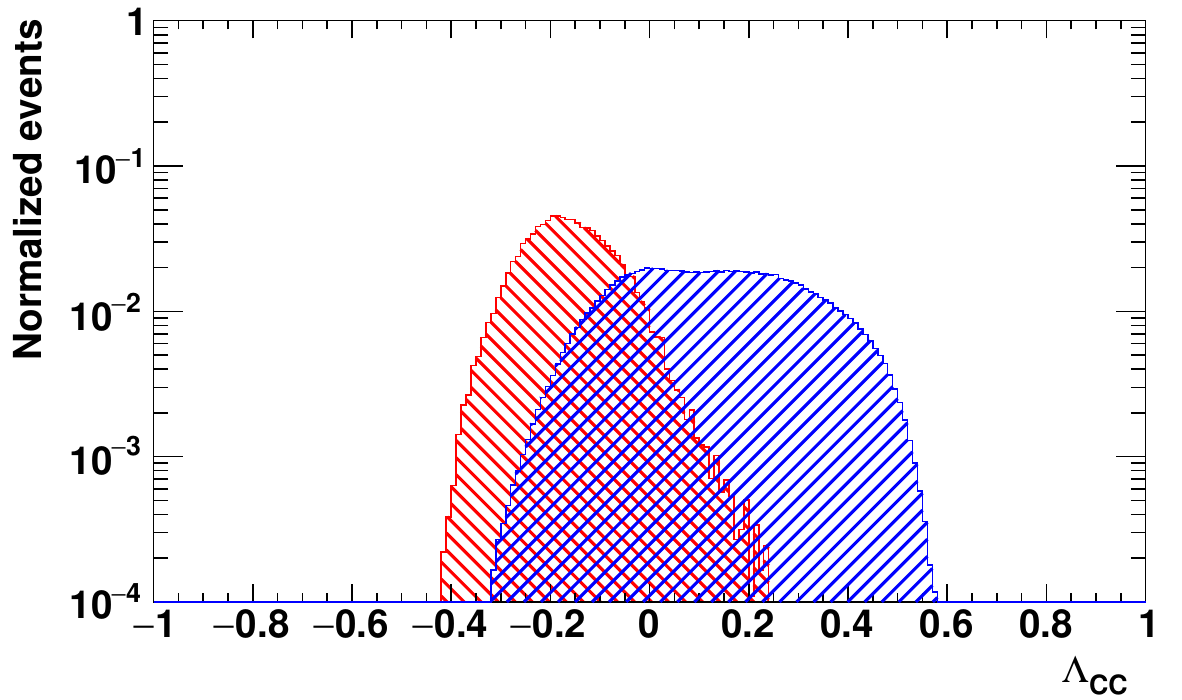}
    \caption{\SIrange{197.69}{237.16}{\GeV}}
  \end{subfigure}
  \hfill
  \begin{subfigure}{0.5\linewidth}
    \includegraphics[width=\linewidth]{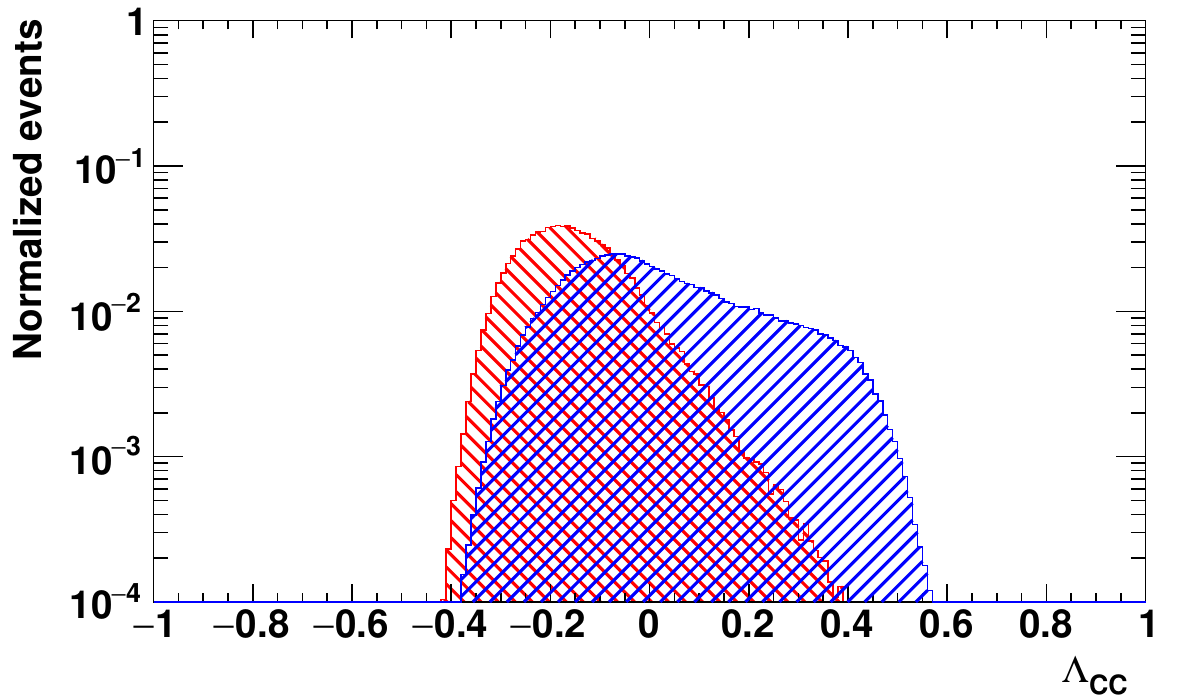}
    \caption{\SI{700}{\GeV}~--~\SI{1}{\TeV}}
  \end{subfigure}
  \caption{Overview of the CCMVA estimator in the single-track sample in four different energy intervals, from low energies to the highest energy in this analysis. The red histogram shows the charge-confused sample (R > 0) and the blue histogram the correct reconstructed sample (R < 0), after applying all preselection, selection and $e^{\pm}$ identification cuts.}
  \label{fig:ccmva-example-single-track-sample}
\end{figure}

\begin{figure}[H]
  \begin{subfigure}{0.5\linewidth}
    \includegraphics[width=\linewidth]{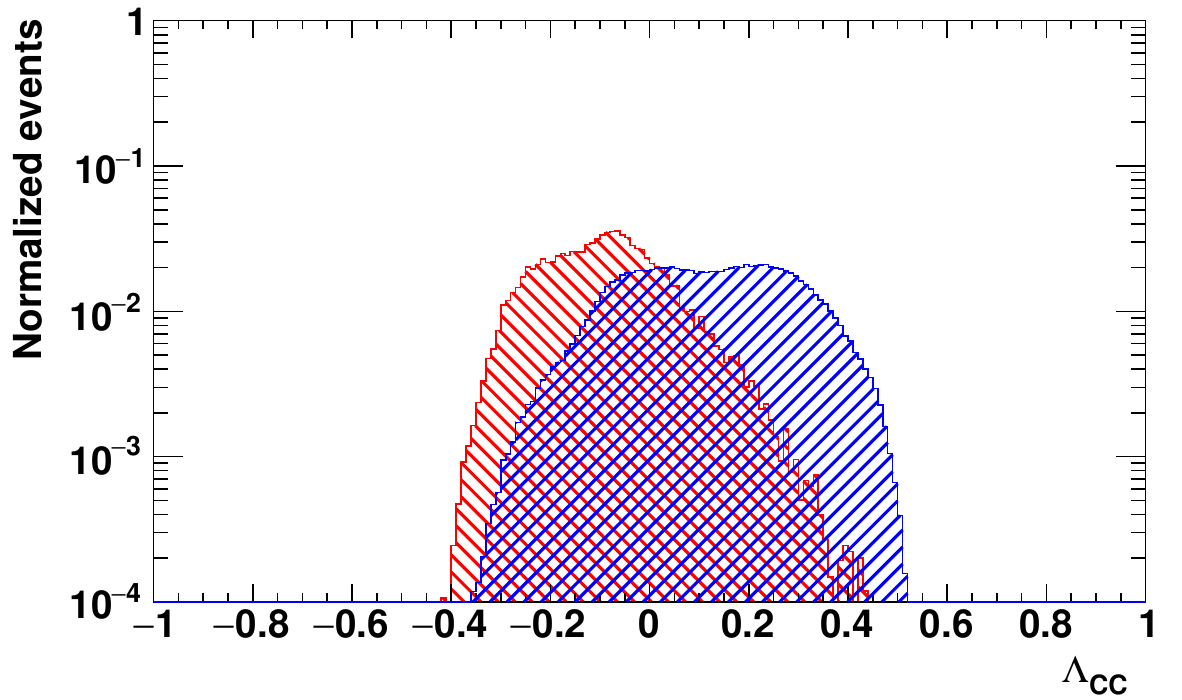}
    \caption{\SIrange{2}{2.31}{\GeV}}
  \end{subfigure}
  \hfill
  \begin{subfigure}{0.5\linewidth}
    \includegraphics[width=\linewidth]{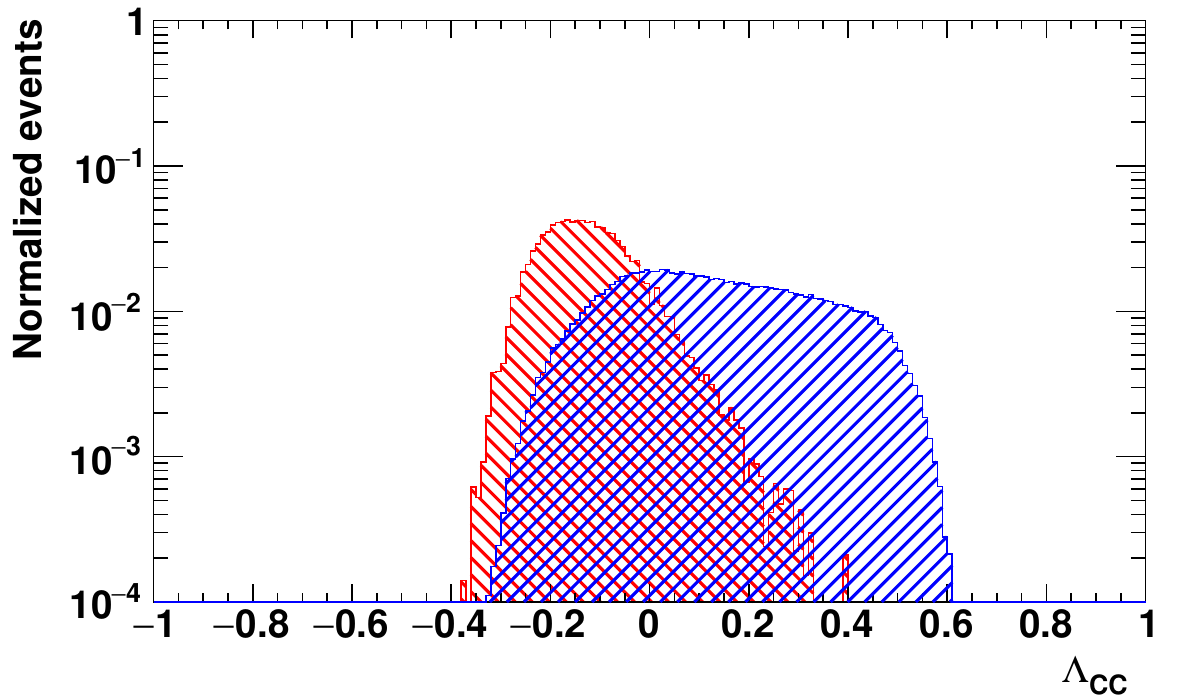}
    \caption{\SIrange{17.98}{18.99}{\GeV}}
  \end{subfigure}
  \hfill
  \begin{subfigure}{0.5\linewidth}
    \includegraphics[width=\linewidth]{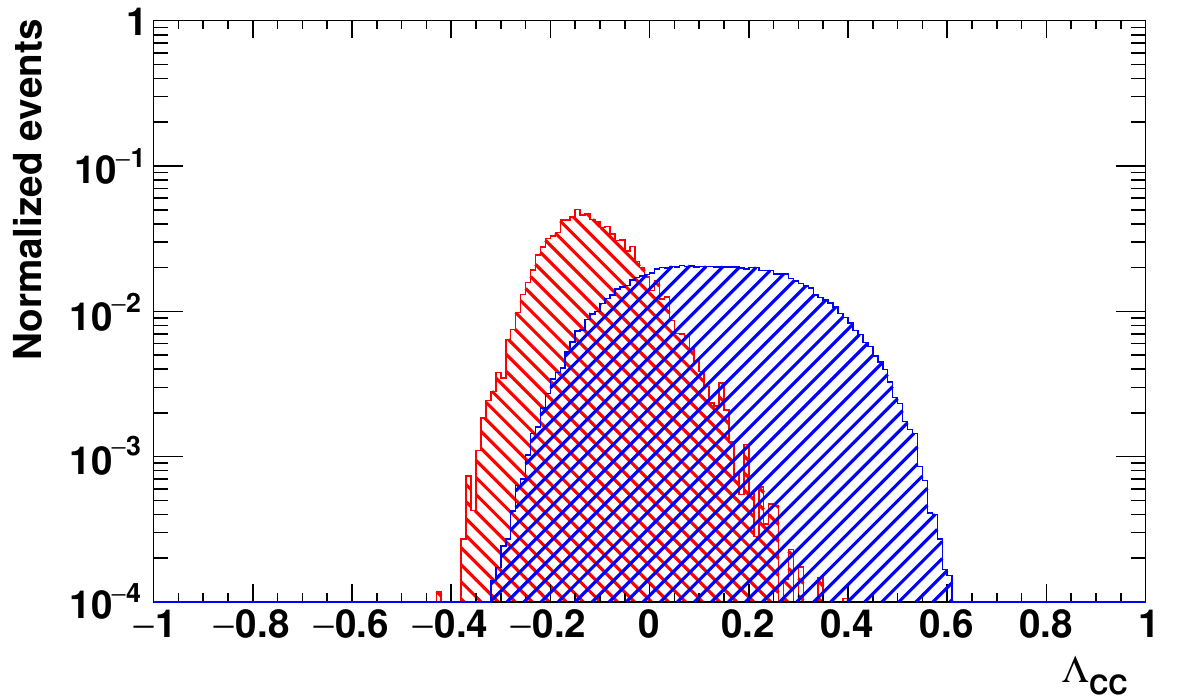}
    \caption{\SIrange{197.69}{237.16}{\GeV}}
  \end{subfigure}
  \hfill
  \begin{subfigure}{0.5\linewidth}
    \includegraphics[width=\linewidth]{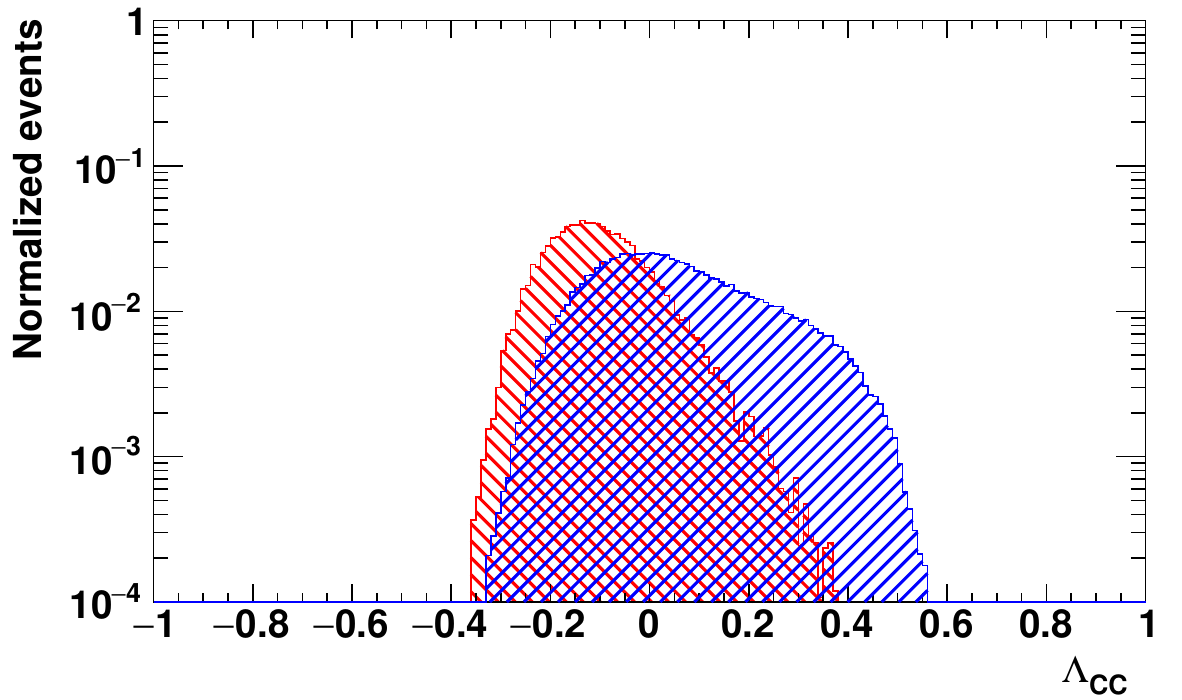}
    \caption{\SI{700}{\GeV}~--~\SI{1}{\TeV}}
  \end{subfigure}
  \caption{Overview of the CCMVA estimator in the multi-tracks sample in four different energy intervals, from low energies to the highest energy in this analysis.}
  \label{fig:ccmva-example-multi-tracks-sample}
\end{figure}

\Cref{fig:ccmva-example-single-track-sample} shows the CCMVA estimator in four example energy bins for the single-track sample. \Cref{fig:ccmva-example-multi-tracks-sample} shows
the same energy bins for the multi-tracks sample. It is evident that the constructed CCMVA estimator $\Lambda_{\text{CC}}$ can be used to discriminate the correct and wrong charge samples.

The shapes are suitable for template fits, which allows one to extract the magnitude of the charge-confusion effect directly on ISS data,
when applying the MVA estimator to an electron sample on ISS data. This will be shown later in \cref{sec:analysis-lepton-counts-2d-fit}.

To characterize the performance of the CCMVA estimator, \gls{ROC} curves are computed for all energy bins. A ROC curve is a plot of the background rejection
as function of the signal efficiency. On the electron Monte-Carlo simulation two histograms are filled for each energy bin, separately for the single-track
and the multi-tracks sample with the MVA estimator output for the positive rigidity sample (\enquote{background sample}) and the negative rigidity
sample (\enquote{signal sample}), respectively.

The signal efficiency $\epsilon_{\text{signal}}$ is scanned from \SIrange{0}{100}{\percent} to produce the ROC curve. A cut value $x_{\text{cut}}$,
is chosen corresponding to the given signal efficiency $\epsilon_{\text{signal}}$. The number of signal/background events for a given cut
value $x_{\text{cut}}$ can be computed by integrating the signal/background histograms:

\begin{equation}
  \label{eq:ccmva-roc-curve-definition}
  N_{\text{sig}} = \int\limits_{x_{\text{cut}}}^{\infty} H_{\text{sig}}(x) \diff x = \epsilon_{\text{signal}} \int_{-\infty}^{\infty} H_{\text{sig}}(x) \diff x; \qquad
  N_{\text{bkg}} = \int\limits_{x_{\text{cut}}}^{\infty} H_{\text{bkg}}(x) \diff x.
\end{equation}

The signal/background efficiency is defined as $N_{\text{sig}}$ / $N_{\text{bkg}}$ divided by all events in the signal/background histogram.

\begin{figure}[H]
  \centering
  \includegraphics[width=0.75\linewidth]{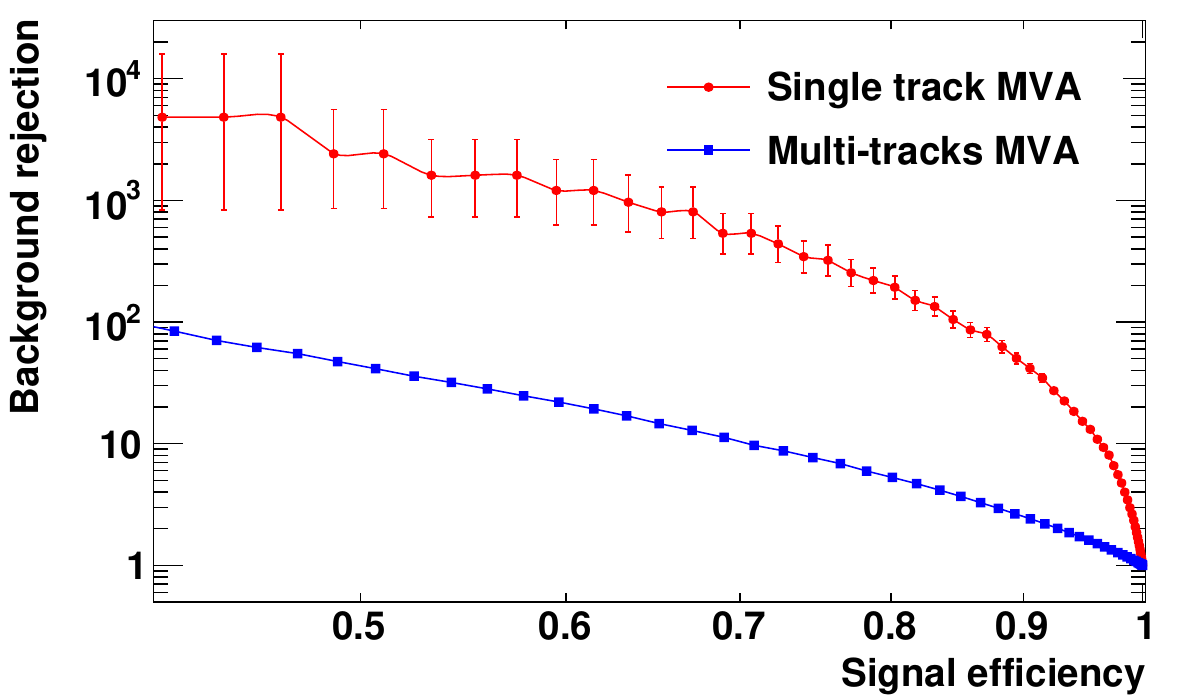}
  \caption{ROC curve of the CCMVA estimator in the energy bin \SIrange{17.98}{18.99}{\GeV}. The red curve represents the ROC curve for the single-track sample and the blue curve represents the multi-tracks sample.}
  \label{fig:ccmva-example-roc-curve}
\end{figure}

After repeating the procedure for all possible values of $\epsilon_{\text{signal}}$ the ROC curve can be computed, as shown in
\cref{fig:ccmva-example-roc-curve} for an example energy bin. The rejection power for the single-track MVA is intrinsically higher than
for the multi-tracks MVA estimator.

\begin{figure}[H]
  \centering
  \includegraphics[width=0.75\linewidth]{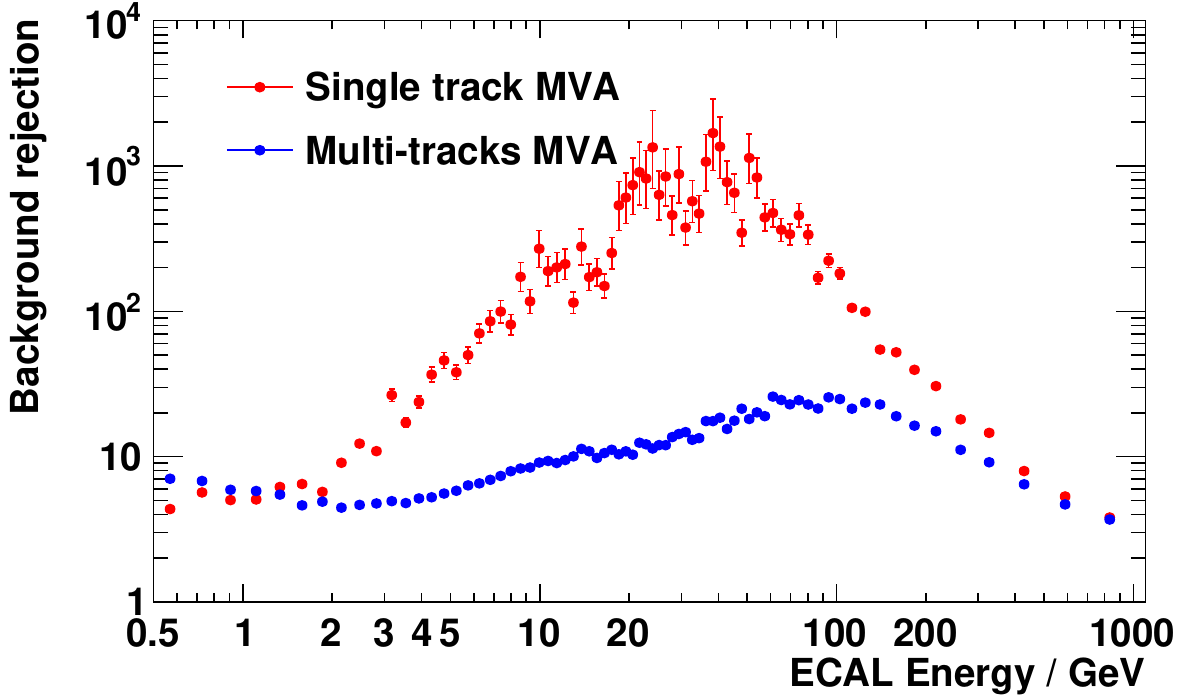}
  \caption{Rejection of the CCMVA estimator as function of energy. The background rejection exceeds $10^2$ in the energy range \SIrange{10}{100}{\GeV}, when choosing a fixed signal efficiency of \SI{70}{\percent}.}
  \label{fig:ccmva-performance}
\end{figure}

The ROC curves for all energy bins are evaluated at a fixed signal efficiency of \SI{70}{\percent} and the result is aggregated in a graph (\cref{fig:ccmva-performance}).
For both the single-track sample and the multi-tracks sample the CCMVA estimator offers a good background rejection power.

The CCMVA estimator performs best at intermediate energies between \SIrange{10}{100}{\GeV}. At high energies, when approaching \gls{MDR}, the intrinsic
charge-confusion due to the finite resolution of the silicon tracker is unavoidable. Thus the rejection power decreases towards high energies. At low energies
multiple scattering leads to an unavoidable amount of charge-confusion, also reducing the rejection power. The shape of the CCMVA rejection curve follows
the expectation from first principles.

\section{Data selection criteria}
\label{sec:analysis-data-selection}

For the electron and positron analysis, the overwhelming proton background in cosmic rays has to be suppressed as much as possible.
AMS-02 is equipped with several subdetectors to fulfill the task, most importantly the TRD~(\cref{sec:detector-trd}),
the ECAL~(\cref{sec:detector-ecal}) and the tracker~(\cref{sec:detector-tracker}).

As first step in the analysis flow, the \textbf{detector quality cuts} are applied, ensuring that only seconds are analyzed,
in which the detector is in nominal data-taking condition, as discussed in \cref{sec:analysis-data-selection-detector-quality-cuts}.
Afterwards the \textbf{preselection cuts} and \textbf{selection cuts} are applied, presented in \cref{sec:analysis-data-selection-preselection-cuts,sec:analysis-data-selection-selection-cuts}
This leaves a data sample where all events are guaranteed to be primary cosmic rays: events, whose
reconstructed energy is above the geomagnetic cut-off.
As last step the \textbf{$e^{\pm}$ identification cuts} - discussed in \cref{sec:analysis-data-selection-electron-positron-identification-cuts} -
are applied, which reduce the proton background and select a sample enhanced by $e^{\pm}$.

\subsection{Detector quality cuts}
\label{sec:analysis-data-selection-detector-quality-cuts}

All time intervals are excluded in which the detector is in a known, non-nominal condition, e.g. during TRD gas
refills or when certain detectors are not calibrated. The \textbf{detector quality cuts} are evaluated for each second of ISS data.
Only if all cuts pass for a given second, events recorded within that second are analyzed and contribute to the measuring time $T_{i}(E)$.
The list of all cuts and their description is given in \cref{sec:appendix-detector-quality-cuts}.

In total \SI{93.78}{\percent} of all seconds within the data taking period between \textbf{May~\nth{20},~2011} and \textbf{November~\nth{12},~2017}
pass the detector quality cuts.

\subsection{Preselection cuts}
\label{sec:analysis-data-selection-preselection-cuts}

As next step in the analysis flow, the \textbf{preselection cuts} are applied. These cuts ensure that a signal is present in all
subdetectors required for the electron and positron analysis and that all measurements referring to the primary particle, as defined in
\cref{sec:analysis-event-reconstruction-combining-subdetectors}, are consistent. The cuts are not $e^{\pm}$ specific
and must have a signal selection efficiency close to unity. An example of the signal\footnote{The trajectory of the particle is required
to pass through the TRD, the TOF, the inner tracker and the ECAL.} efficiency $\epsilon_{\text{sig}}$ determined from the electron Monte-Carlo simulation at a
fixed energy of \SI{20}{\GeV} is given for each cut, and an example of the background efficiency $\epsilon_{\text{bkg}}$ determined from the proton
Monte-Carlo simulation.

\begin{enumerate}
  \item\textbf{At least one ECAL shower, within fiducial volume} (\SIvarEquals{$\epsilon_{\text{sig}}$}{99.39}{\percent}; \SIvarEquals{$\epsilon_{\text{bkg}}$}{69.67}{\percent})\hfill\\
    As the ECAL provides the energy scale for the $e^{\pm}$ flux measurement, it is required that at least one
    shower is present in the calorimeter. The efficiency for this requirement is close to unity, if the particle
    traversing AMS-02 is leptonic, passes within the ECAL acceptance and has at least \SI{500}{\MeV} of energy.
    Furthermore it is required that the centre-of-gravity of the reconstructed shower is at least 1.3 Moli\`{e}re radii
    away from the borders of the ECAL, to ensure that only a small amount of energy may leak out of the fiducial
    volume of the ECAL.

  \item\label{enum:preselection-cut-at-least-one-useful-trd-track}\textbf{At least one useful TRD track} (\SIvarEquals{$\epsilon_{\text{sig}}$}{98.01}{\percent}; \SIvarEquals{$\epsilon_{\text{bkg}}$}{88.87}{\percent})\hfill\\
    In order to discriminate light from heavy particles such as protons and ions the TRD is a necessary
    subdetector for the $e^{\pm}$ flux analysis. Thus it is required that each event in the analysis, contains
    a reconstructed TRD track, whose direction is compatible with the reconstructed ECAL shower axis. This refines
    the selection criteria, described in the TRD track reconstruction (\cref{sec:analysis-event-reconstruction-trd-track}),
    to take into account not only the spatial difference between the ECAL shower centre-of-gravity and the extrapolated TRD track,
    but also the direction, with respect to the ECAL shower axis.

  \item\label{enum:preselection-cut-at-least-one-useful-tof-track}\textbf{At least one useful TOF cluster combination} (\SIvarEquals{$\epsilon_{\text{sig}}$}{99.36}{\percent}; \SIvarEquals{$\epsilon_{\text{bkg}}$}{97.86}{\percent})\hfill\\
    The TOF serves as trigger for all charged-particles within AMS-02. It is a crucial subdetector for the
    $e^{\pm}$ flux analysis, providing the direction of the trajectory of the incident particle (up- or down-going).
    For the $e^{\pm}$ flux measurement only particles coming from the top are considered and all others are rejected.

    Thus it is required that at least three out of the four TOF bars recorded a signal and the reconstructed
    incident direction is down-going. The time difference between the traversal of the upper and lower TOF bars
    has to be compatible with a relativistic electron or positron.

    Furthermore the previously selected TRD track is matched with all reconstructed TOF clusters (\cref{sec:analysis-event-reconstruction-tof}),
    to ensure that the correct TOF cluster combination was selected in the particle reconstruction (\cref{sec:analysis-event-reconstruction-combining-subdetectors}).

  \item\textbf{Highest energetic ECAL shower was selected} (\SIvarEquals{$\epsilon_{\text{sig}}$}{99.98}{\percent}; \SIvarEquals{$\epsilon_{\text{bkg}}$}{91.93}{\percent})\hfill\\
    Ensure that the particle reconstruction algorithm (\cref{sec:analysis-event-reconstruction-combining-subdetectors}) selected the ECAL shower containing
    the highest energy. If not, the primary particle definition is inconsistent and the event needs to be rejected.

  \item\textbf{Not in ISS solar array shadow}\hfill\\
    If a tracker track was reconstructed in the event, it must not point to the ISS solar arrays.
    When solar arrays are in the field of view of AMS, primary particles entering the solar arrays
    might produce secondary particles, which have to be rejected for the $e^{\pm}$ flux analysis.
\end{enumerate}

In total \SI{96.84}{\percent}\footnote{This is not the product of all individual signal efficiencies, as that would neglect correlations. Instead the number of passed events is determined after all preselection cuts passed.} of all signal events in the electron Monte-Carlo simulation survive the preselection cuts and only \SI{70.62}{\percent} of all background events in the proton Monte-Carlo simulation.

\subsection{Selection cuts}
\label{sec:analysis-data-selection-selection-cuts}

Only events which contain an ECAL shower, a compatible TRD track, matched with the TOF clusters
survive the preselection. The next step is to apply selection cuts, which specifically target a
leptonic $Z=1$ selection, reducing the background as much as possible, while keeping a high
signal efficiency.

\begin{enumerate}
  \item\textbf{TOF velocity measurement} (\SIvarEquals{$\epsilon_{\text{sig}}$}{99.85}{\percent}; \SIvarEquals{$\epsilon_{\text{bkg}}$}{99.66}{\percent})\hfill\\
    The measured velocity needs to be larger than \SI{0.8}{\clight}, to reject slow protons/ions.
    For the energies considered in this analysis, all leptonic particles have a velocity
    close to \SI{1}{\clight}, as shown in \cref{fig:analysis-data-selection-selection-cuts-tof-velocity}.
    The reconstructed TOF $\beta$ measurement does not exhibit an energy dependence, thus the cut value distribution is shown for the whole energy range of the analysis: \SIrange{0.5}{1000}{\GeV}.

    \begin{figure}[H]
      \centering
      \includegraphics[width=0.6\linewidth]{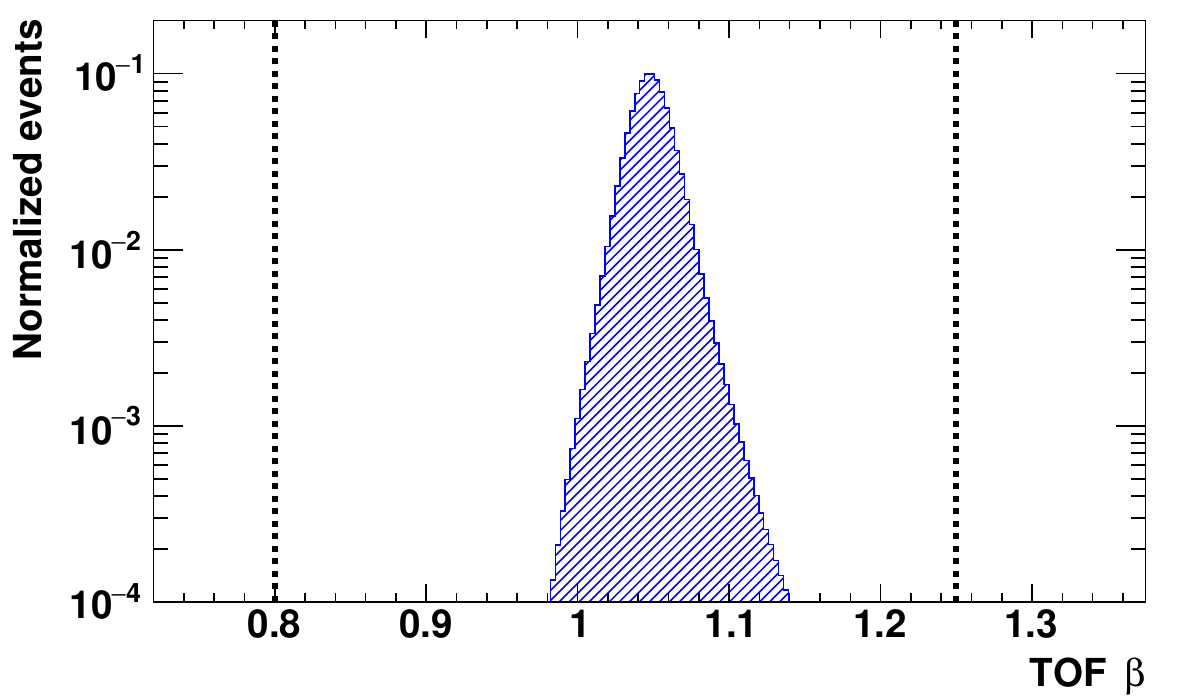}
      \caption{Plot of the TOF velocity $\beta$ in the energy range \SIrange{0.5}{1000}{\GeV} for the electron Monte-Carlo simulation. The applied lower and upper cut values are shown as black dashed lines. The chosen cut values are loose: all signal events are kept and only a few slow protons/ions at low energies are rejected.}
      \label{fig:analysis-data-selection-selection-cuts-tof-velocity}
    \end{figure}

  \item\label{enum:selection-cut-upper-tof-charge}\textbf{Upper TOF charge} (\SIvarEquals{$\epsilon_{\text{sig}}$}{93.73}{\percent}; \SIvarEquals{$\epsilon_{\text{bkg}}$}{97.42}{\percent})\hfill\\
    The charge reconstructed in the upper TOF bars must not exceed \SI{2}{\elementarycharge}. The loose
    charge requirement is connected to the nature of electrons and positrons which lose energy when traversing
    AMS-02 by emission of bremsstrahlung photons along the trajectory due to the magnetic field. These photons
    can convert into additional $e^{-}$~/~$e^{+}$ pairs in the detector material, for example
    in the upper TOF bars. When three particles - the primary electron or positron and an additional
    electron and positron pair - traverse a single TOF bar, the charge measurement will yield
    \SIvarApprox{Z}{1.7}{\elementarycharge}, as shown in \cref{fig:analysis-data-selection-selection-cuts-tof-upper-charge}.

    \begin{figure}[H]
      \centering
      \includegraphics[width=0.6\linewidth]{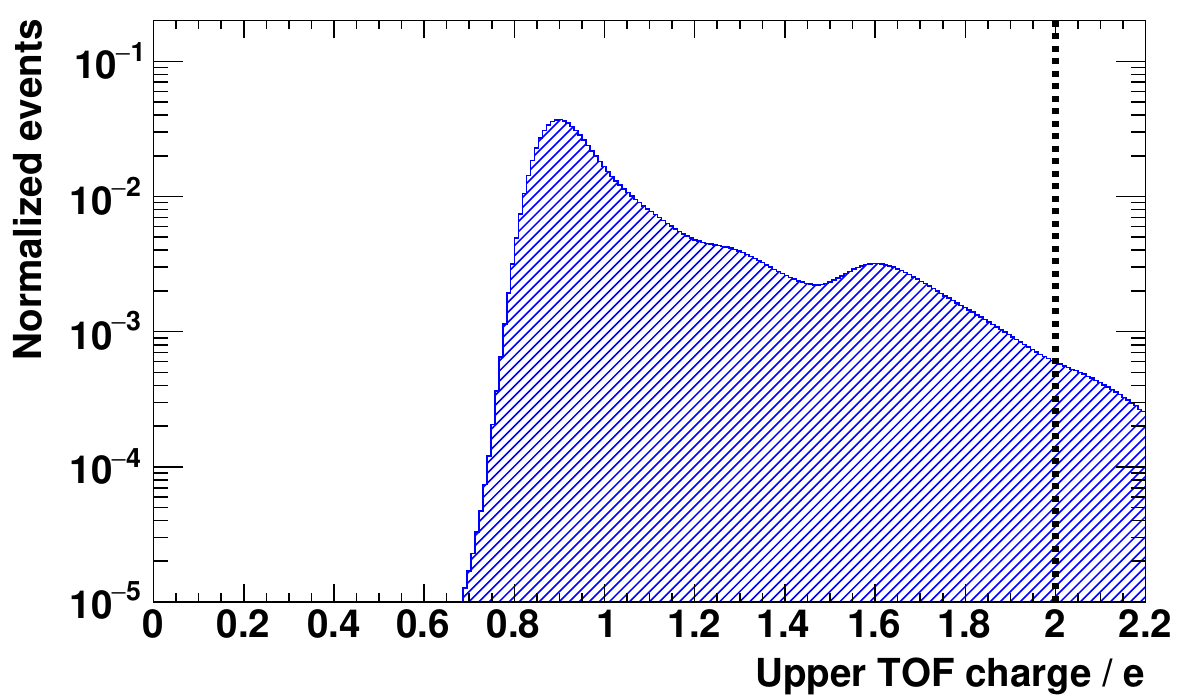}
      \caption{Plot of the charge measurement in the upper TOF in the energy range \SIrange{0.5}{1000}{\GeV} for the electron Monte-Carlo simulation. The applied upper cut value is shown as black dashed line.}
      \label{fig:analysis-data-selection-selection-cuts-tof-upper-charge}
    \end{figure}

    The reconstructed upper TOF charge measurement does not exhibit an energy dependence, thus the cut value distribution is shown for the whole energy range of the analysis: \SIrange{0.5}{1000}{\GeV}.

  \item\label{enum:selection-cut-trd-active-layers}\textbf{Enough active layers in TRD} (\SIvarEquals{$\epsilon_{\text{sig}}$}{89.82}{\percent}; \SIvarEquals{$\epsilon_{\text{bkg}}$}{87.58}{\percent})\hfill\\
    When a non-interacting particle traverses the TRD, a large number of TRD layers (on average > 18 out of 20)
    have recorded an energy deposition in the straw tubes. A cut requiring at least 16 active TRD layers is applied\footnote{The energy-dependence of the TRD active layers cut is flat because of the use of \textbf{hybrid hits}, as explained in \cref{sec:analysis-event-reconstruction-trd-estimator}. When using only the tracker to determine the path length, the efficiency for this cut would decrease to lower energies.}, as shown in \cref{fig:analysis-data-selection-selection-cuts-trd-active-layers}.
    \begin{figure}[H]
      \centering
      \includegraphics[width=0.7\linewidth]{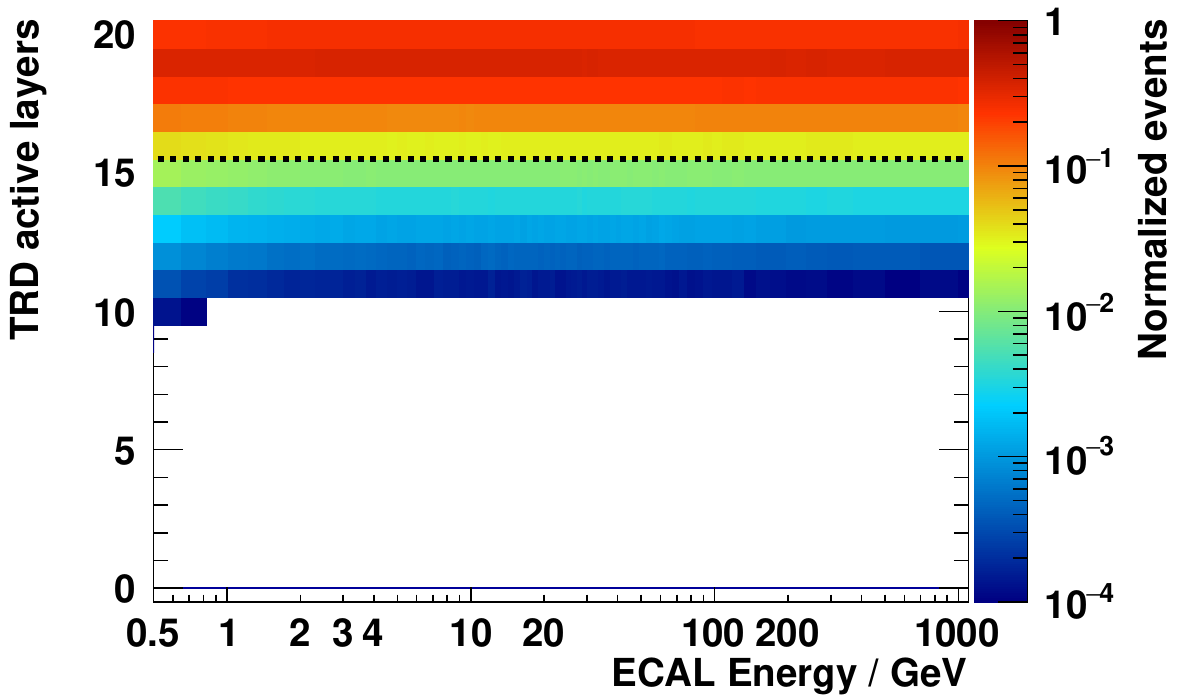}
      \caption{Plot of the active layers in the TRD as function of energy for the electron Monte-Carlo simulation. The applied lower cut value is shown as black dashed line.}
      \label{fig:analysis-data-selection-selection-cuts-trd-active-layers}
    \end{figure}

  \item\label{enum:selection-cut-trd-no-helium}\textbf{TRD helium rejection} (\SIvarEquals{$\epsilon_{\text{sig}}$}{96.64}{\percent}; \SIvarEquals{$\epsilon_{\text{bkg}}$}{70.96}{\percent})\hfill\\
    The $\diff\text{E}/\diff\text{x}$ measurements from all active straw tubes are combined into a single likelihood
    estimator $\mathcal{L}_{e^{-}/\ce{He}}$, which can be used to check the hypothesis whether the particle traversing
    the TRD is helium like or $e^{\pm}$ like.

    \begin{figure}[H]
      \begin{subfigure}{0.50\linewidth}
        \includegraphics[width=\linewidth]{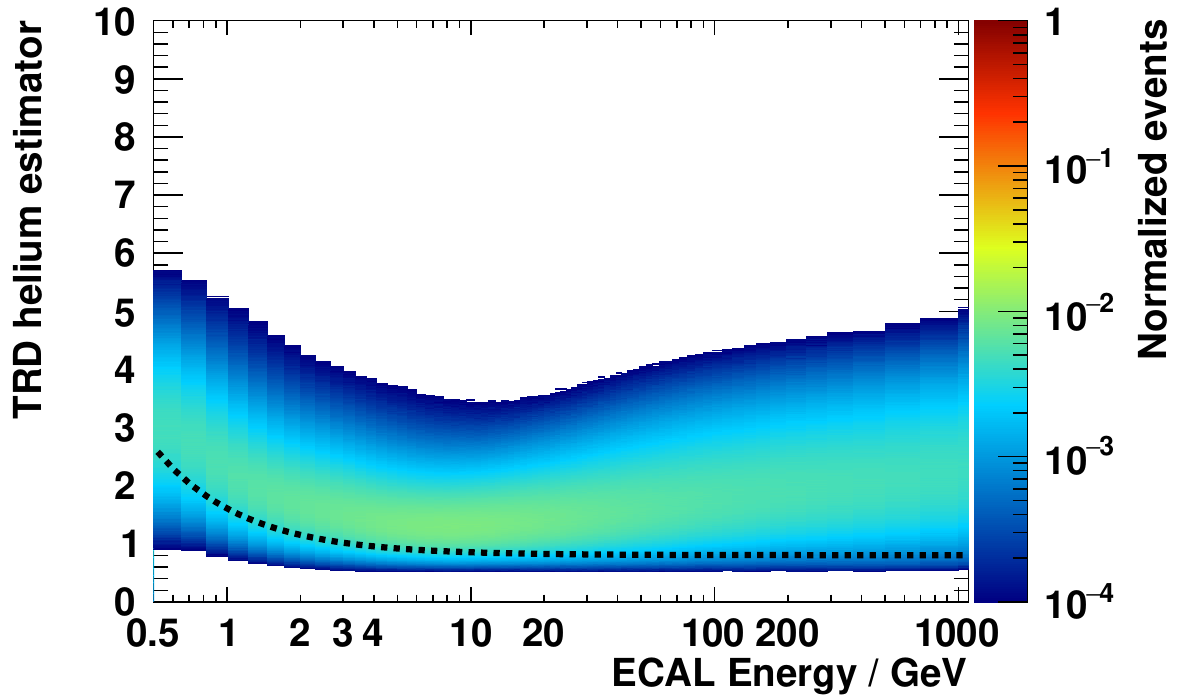}
        \caption{Electron Monte-Carlo}
        \label{fig:analysis-data-selection-selection-cuts-trd-no-helium-signal}
      \end{subfigure}
      \begin{subfigure}{0.50\linewidth}
        \includegraphics[width=\linewidth]{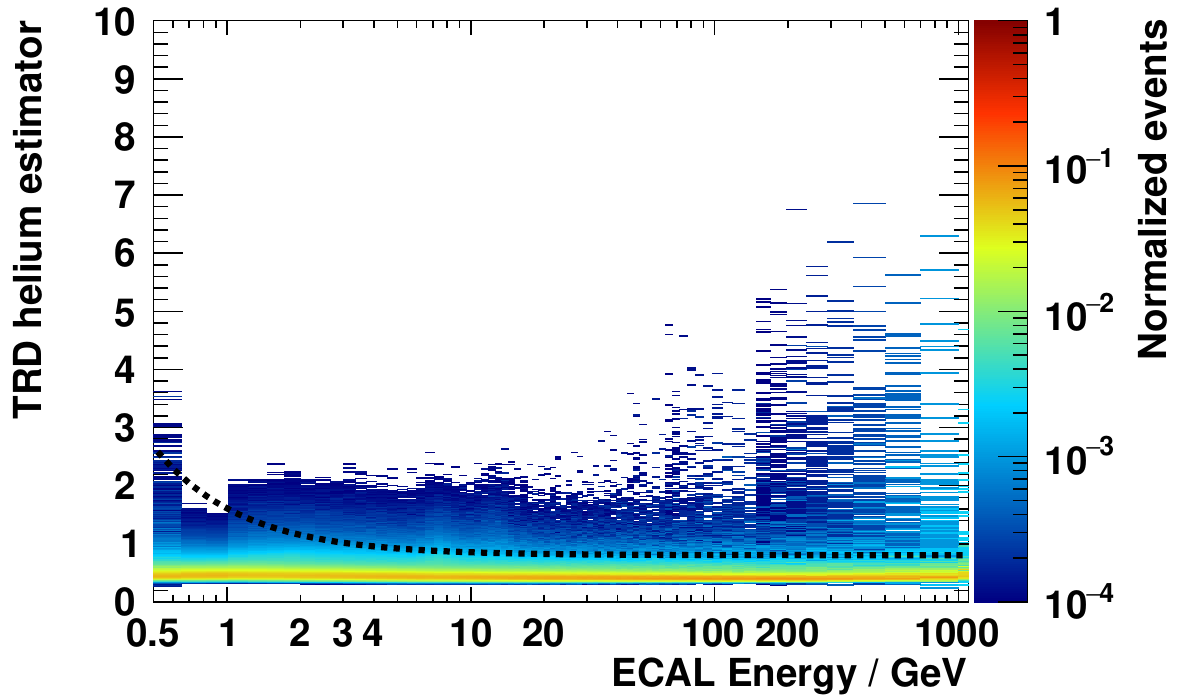}
        \caption{ISS Helium sample}
        \label{fig:analysis-data-selection-selection-cuts-trd-no-helium-background}
      \end{subfigure}
      \caption{Plot of the TRD likelihood estimator $\mathcal{L}_{e^{-}/\ce{He}}$ as function of energy for the electron Monte-Carlo simulation (left) and the ISS Helium sample (right). The applied lower cut value is shown as black dashed line.}
      \label{fig:analysis-data-selection-selection-cuts-trd-no-helium}
    \end{figure}

    \Cref{fig:analysis-data-selection-selection-cuts-trd-no-helium-background} shows that the energy dependent lower cut value removes most of the helium background events,
    while keeping the majority of all signal events, as shown in \cref{fig:analysis-data-selection-selection-cuts-trd-no-helium-signal}.

    The chosen cut value rejects a non-negligible amount of signal events below \SIapprox{10}{\GeV}. A tight cut is necessary to reach a sufficient helium rejection,
    which was determined in a dedicated study.

  \item\label{enum:selection-cut-trk-pattern}\textbf{Tracker hit pattern} (\SIvarEquals{$\epsilon_{\text{sig}}$}{86.80}{\percent}; \SIvarEquals{$\epsilon_{\text{bkg}}$}{85.70}{\percent})\hfill\\
    This analysis requires the tracker track to be well reconstructed in order to have a reliable charge-sign
    measurement, allowing to separate electrons from positrons. It is requested that at least one tracker track
    is reconstructed within the cone defined by the TRD track and the ECAL shower axis, with a certain number of
    tracker layers that have a signal. Besides the inner tracker (layer 3 or 4, layer 5 or 6, layer 7 or 8), there
    must be at least a hit in the layer 1, layer 9 or layer 2 in order to accept the event for further analysis.

    The maximum detectable rigidity for a so-called full-span event (layer 1 + 2 + 9 + inner tracker)
    is \SIapprox{1.2}{\TeV} for $e^{\pm}$. All other classes have a smaller MDR and a higher
    probability of misreconstructing the charge-sign. The allowed classes for this analysis, were chosen
    in order to maximize the signal efficiency, while keeping the misidentification probability under a tolerable
    level - which will be discussed in detail in later sections of this thesis (\cref{sec:analysis-lepton-counts-2d-fit}).

  \item\textbf{Inner tracker charge} (\SIvarEquals{$\epsilon_{\text{sig}}$}{99.99}{\percent}; \SIvarEquals{$\epsilon_{\text{bkg}}$}{99.99}{\percent})\hfill\\
    The charge measurements from all inner tracker layers are combined into a single charge measurement,
    by using a truncated mean, in order to remove outliers. A loose cut \SI{0.5}{\elementarycharge}~<~Z~<\SI{1.8}{\elementarycharge}
    is applied, to keep the three particle events (\SIvarApprox{Z}{1.7}{\elementarycharge}), as shown in \cref{fig:analysis-data-selection-selection-cuts-tracker-charge-signal},
    while removing the majority of the helium background events, as shown in \cref{fig:analysis-data-selection-selection-cuts-tracker-charge-background}.

    \begin{figure}[H]
      \begin{subfigure}{0.50\linewidth}
        \includegraphics[width=\linewidth]{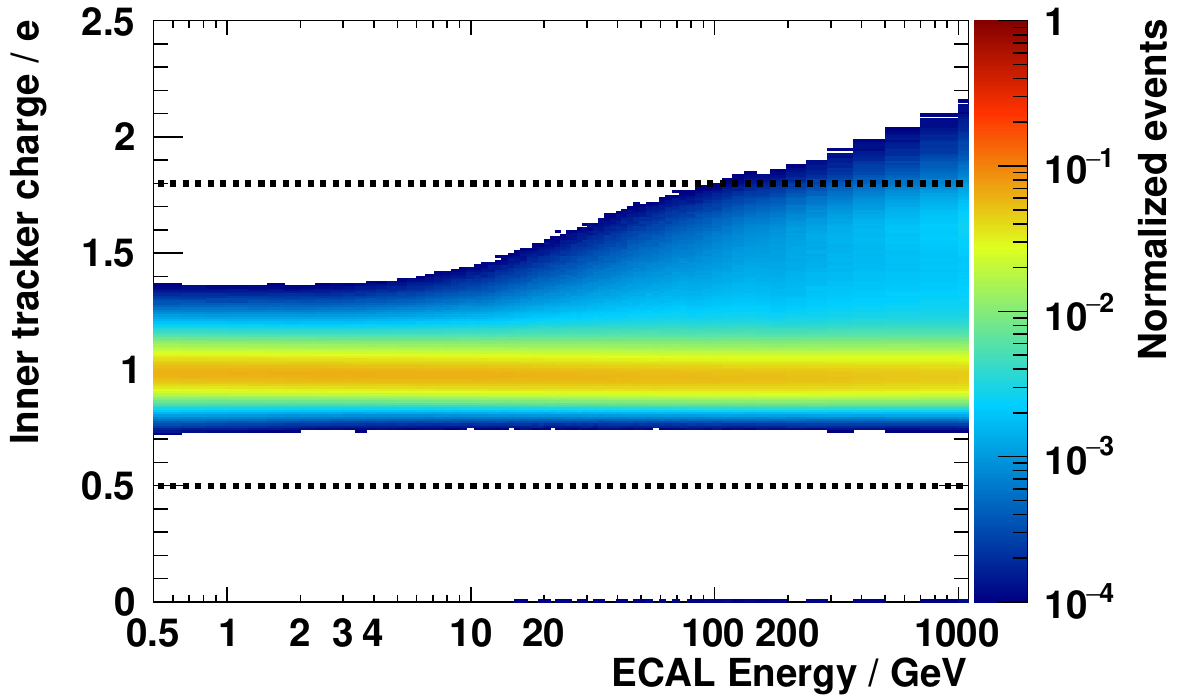}
        \caption{Electron Monte-Carlo}
        \label{fig:analysis-data-selection-selection-cuts-tracker-charge-signal}
      \end{subfigure}
      \begin{subfigure}{0.50\linewidth}
        \includegraphics[width=\linewidth]{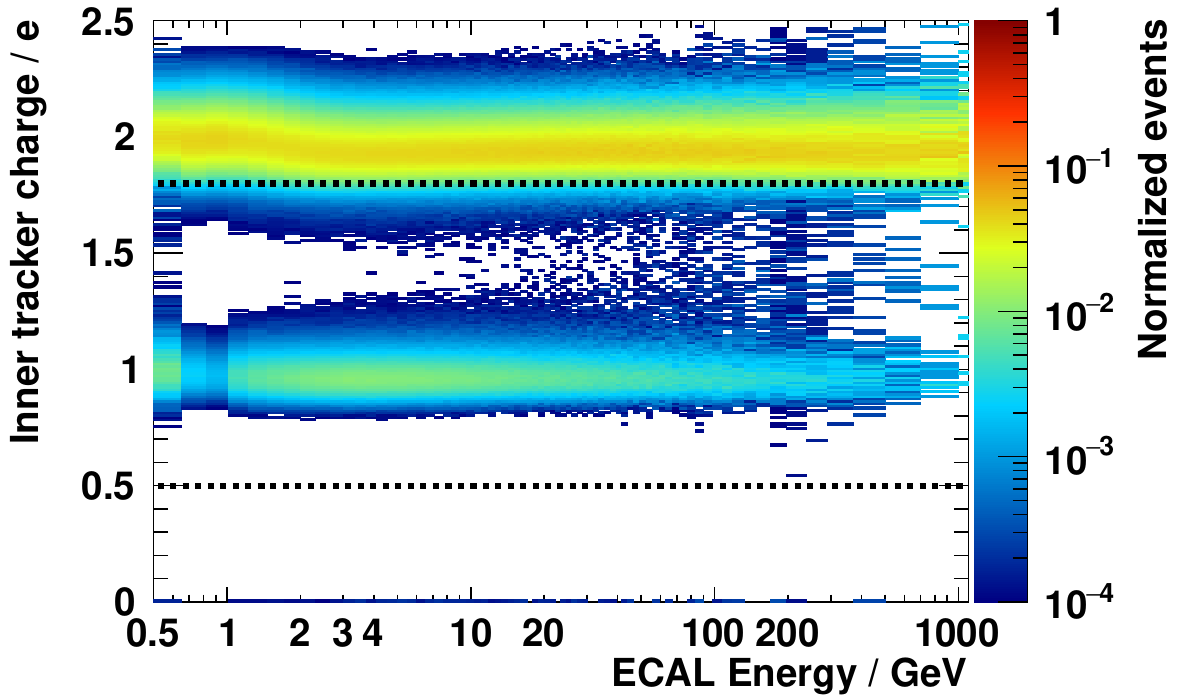}
        \caption{ISS R > 0 sample}
        \label{fig:analysis-data-selection-selection-cuts-tracker-charge-background}
      \end{subfigure}
      \caption{Plot of the charge measurement in the inner tracker as function of energy for the electron Monte-Carlo simulation (left) and the ISS R > 0 sample (right). The applied lower and upper cut values are shown as black dashed lines.}
      \label{fig:analysis-data-selection-selection-cuts-tracker-charge}
    \end{figure}

  \item\label{enum:selection-cut-trk-chi-square-y}\textbf{Tracker track goodness-of-fit in Y-projection} (\SIvarEquals{$\epsilon_{\text{sig}}$}{95.98}{\percent}; \SIvarEquals{$\epsilon_{\text{bkg}}$}{98.10}{\percent})\hfill\\
    A cut is applied on the tracker track goodness-of-fit ${\chi_{\text{y}}^{2}}$
    estimator. From the track fit procedure a $\chi^2$ can be computed, describing the goodness-of-fit in the bending
    plane. This quantity is connected with charge-confusion. The higher the $\chi^2$ value, the more
    likely a misreconstructed charge sign.

    \begin{figure}[H]
      \centering
      \includegraphics[width=0.7\linewidth]{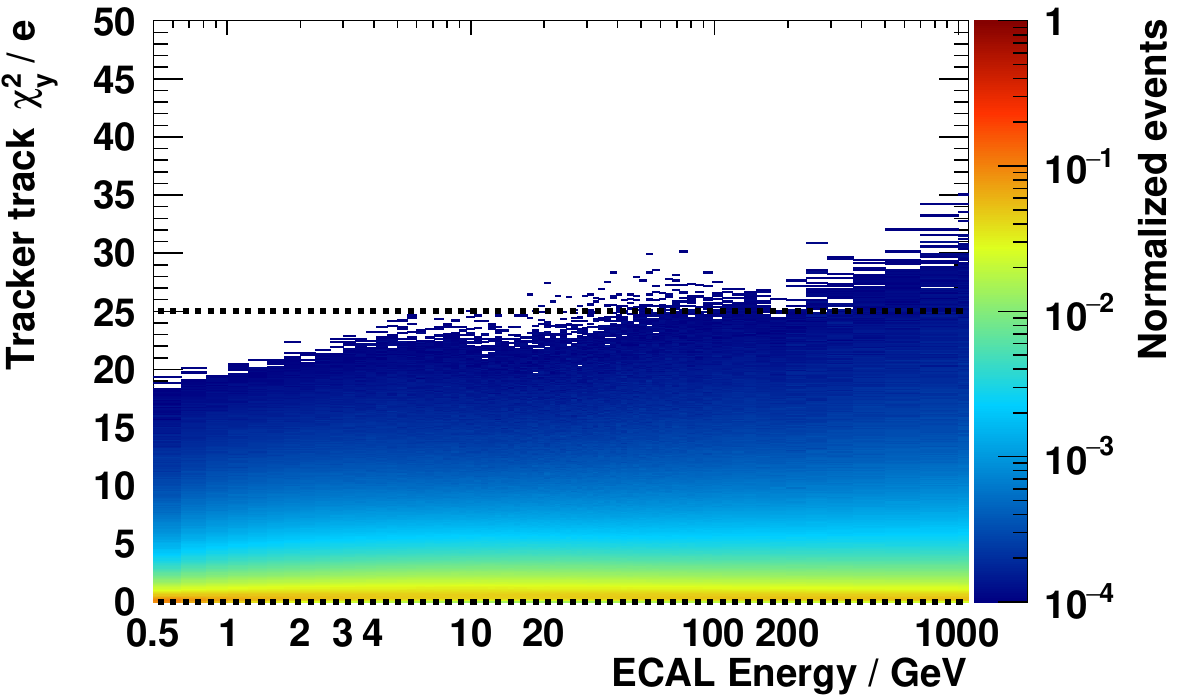}
      \caption{Plot of the tracker track goodness-of-fit ${\chi_{\text{y}}^{2}}$ as function of energy for the electron Monte-Carlo simulation. The applied lower and upper cut values are shown as black dashed lines.}
      \label{fig:analysis-data-selection-selection-cuts-tracker-chi-square-y}
    \end{figure}

    As compromise between high signal efficiency and a tolerable charge-confusion a cut value of ${\chi_{\text{y}}^{2}}$ < 20
    was chosen, as shown in \cref{fig:analysis-data-selection-selection-cuts-tracker-chi-square-y}.

  \item\textbf{Energy above geomagnetic cut-off}\hfill\\
    As last step in the selection, the energy reconstructed by the ECAL is required to be larger than 1.2 times
    the geomagnetic cut-off rigidity, computed in the current geomagnetic location AMS-02. The procedure is described
    in detail in \cref{sec:analysis-flux-time-averaged-measuring-time}.
\end{enumerate}

In total \SI{67.78}{\percent} of all signal events in the electron Monte-Carlo simulation survive the selection cuts
and only \SI{71.53}{\percent} of all background events in the proton Monte-Carlo simulation.

\subsection{Electron and positron identification cuts}
\label{sec:analysis-data-selection-electron-positron-identification-cuts}

All events passing the selection contain measurements in all subdetectors relevant for the $e^{\pm}$ flux analysis.
No cuts have been applied to reduce the proton background, except the presence of an ECAL shower, which remove
protons that pass the ECAL as MIPs (\cref{sec:analysis-event-reconstruction-ecal-estimator}). As next step
specific cuts can be applied that further suppress the proton background.

\begin{enumerate}
  \item\label{enum:electron-positron-identification-cut-energy-rigidity-matching}\textbf{Energy $\leftrightarrow$ rigidity matching} (\SIvarEquals{$\epsilon_{\text{sig}}$}{94.27}{\percent}; \SIvarEquals{$\epsilon_{\text{bkg}}$}{13.94}{\percent})\hfill\\
    The ratio of the measured energy in the calorimeter and the momentum measured by the tracker is required to fulfill: 0.5 < $E/\abs{R}$ < 10.
    Leptons up to the TeV regime release all their kinetic energy in the calorimeter.
    When the tracker rigidity is correctly reconstructed, $e^{\pm}$ peak at E/R values of $\approx$~1, or larger, when
    the rigidity is underestimated, due to bremsstrahlung loss or interactions in the detector.

    The lower cut value of 0.5 thus removes proton events, which deposited only parts of their energy in the calorimeter,
    as shown in \cref{fig:analysis-data-selection-identification-cuts-energy-over-rigidity-background}, while keeping
    as much signal events as possible (\cref{fig:analysis-data-selection-identification-cuts-energy-over-rigidity-signal}).

    \begin{figure}[H]
      \begin{subfigure}{0.50\linewidth}
        \includegraphics[width=\linewidth]{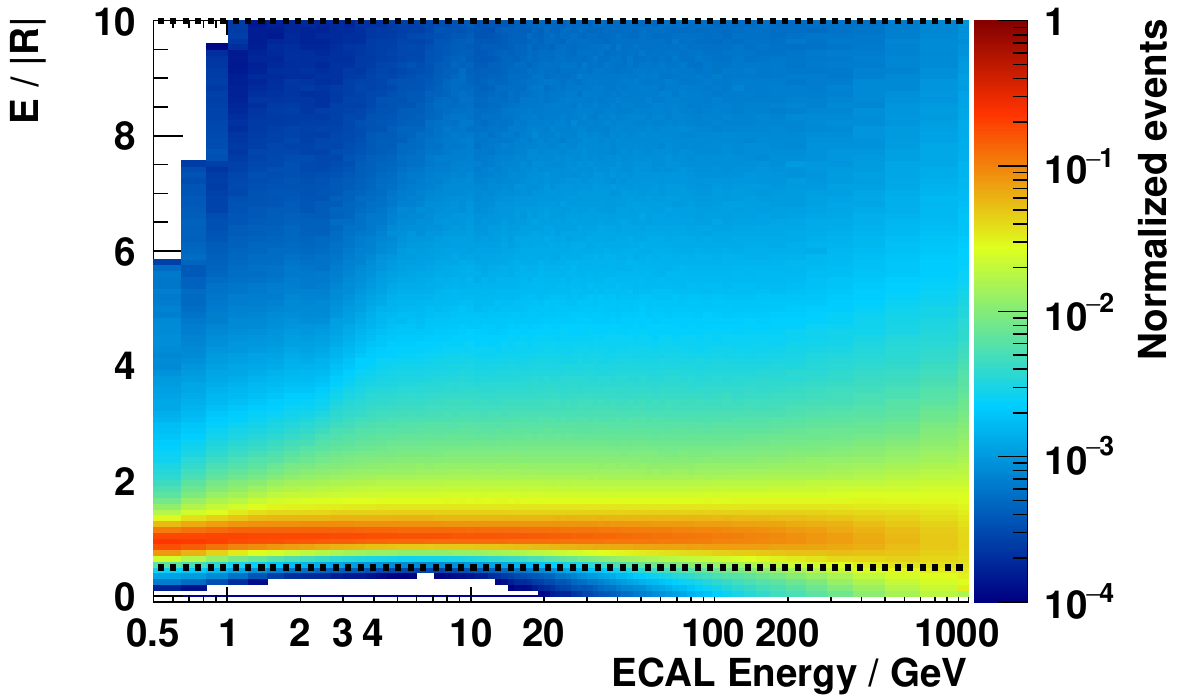}
        \caption{Electron Monte-Carlo}
        \label{fig:analysis-data-selection-identification-cuts-energy-over-rigidity-signal}
      \end{subfigure}
      \begin{subfigure}{0.50\linewidth}
        \includegraphics[width=\linewidth]{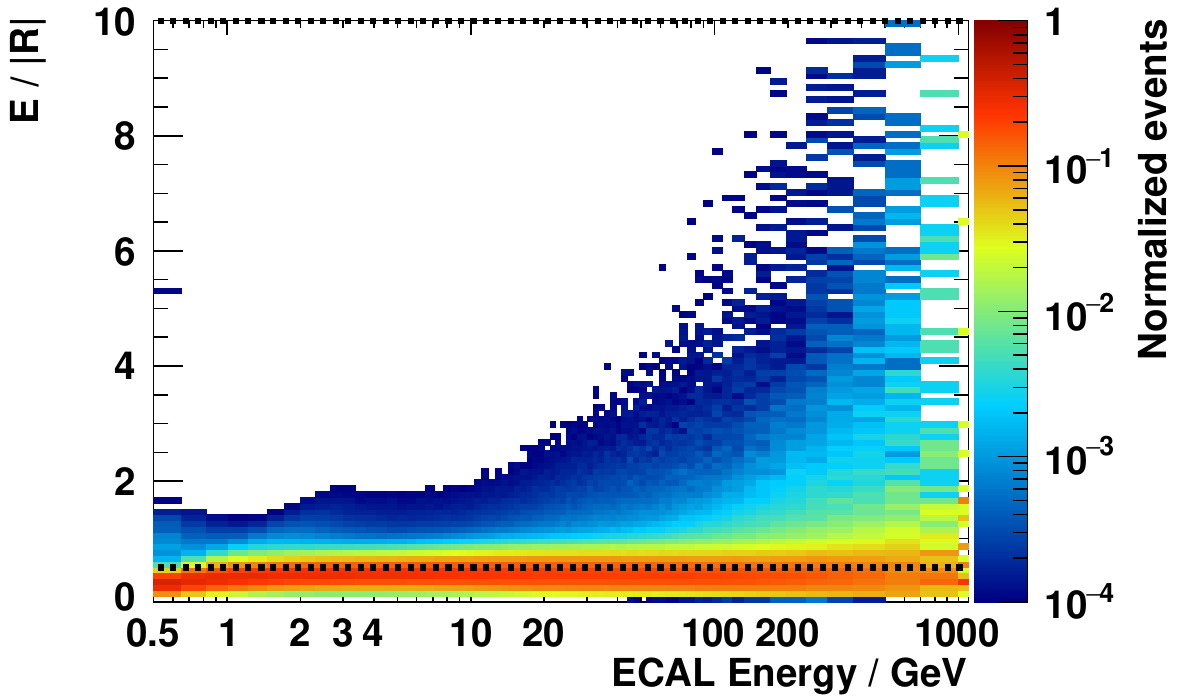}
        \caption{ISS R > 0 sample}
        \label{fig:analysis-data-selection-identification-cuts-energy-over-rigidity-background}
      \end{subfigure}
      \caption{Plot of the $E/\abs{R}$ distribution as function of energy for the electron Monte-Carlo simulation (left) and the ISS R > 0 sample (right). The applied lower and upper cut values are shown as black dashed lines.}
      \label{fig:analysis-data-selection-identification-cuts-energy-over-rigidity}
    \end{figure}

    Protons need to create $\pi^0$ in the ECAL which further decay into photons, in order to release a large amount of
    energy in the electromagnetic calorimeter. Most protons traverse the calorimeter losing only a small fraction of their
    energy via ionization losses.

    The upper cut value of 10 was chosen to maximize statistics, while keeping the charge-confusion under a tolerable
    level. The higher the E/R value for $e^{\pm}$, the more probable it becomes that the rigidity sign is misreconstructed.

  \item\label{enum:electron-positron-identification-cut-ecal-lateral-shower-shape}\textbf{ECAL lateral shower shape} (\SIvarEquals{$\epsilon_{\text{sig/bkg}}$}{100.00}{\percent})\hfill\\
    The lateral shower shape is a powerful tool to discriminate between hadronic and leptonic showers. Hadronic showers
    are wider and less concentrated than electromagnetic showers in the calorimeter and exhibit a more irregular shape.

    \begin{figure}[H]
      \begin{subfigure}{0.50\linewidth}
        \includegraphics[width=\linewidth]{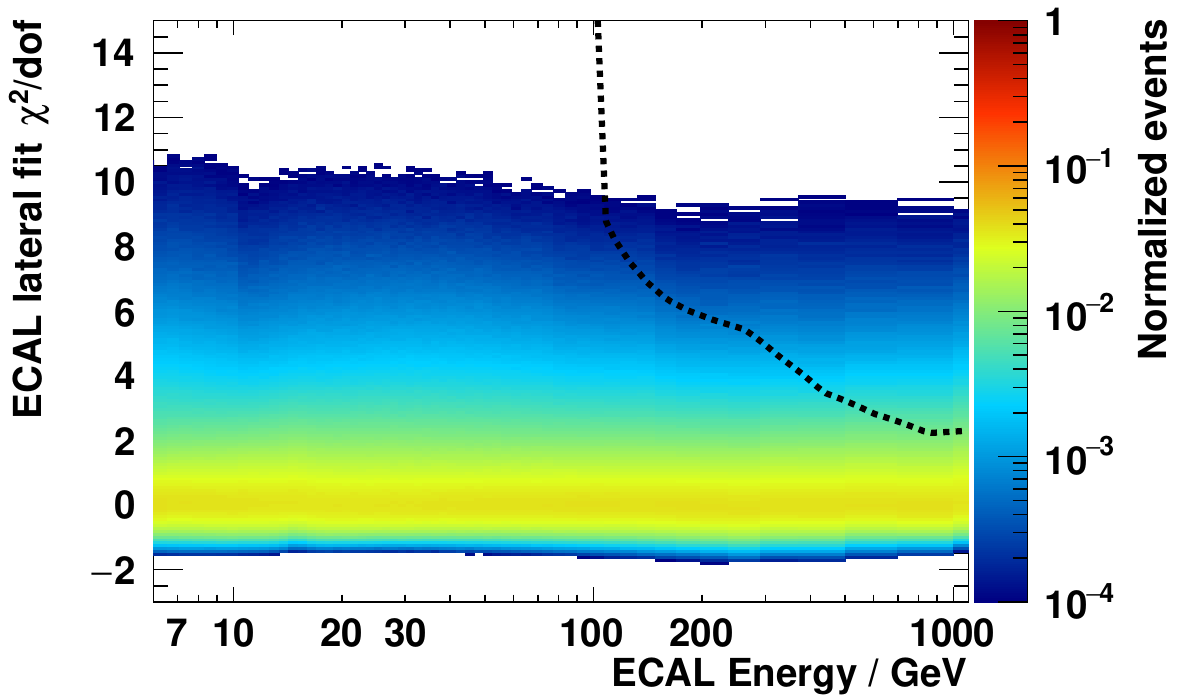}
        \caption{Electron Monte-Carlo}
        \label{fig:analysis-data-selection-identification-cuts-ecal-lateral-normalized-signal}
      \end{subfigure}
      \begin{subfigure}{0.50\linewidth}
        \includegraphics[width=\linewidth]{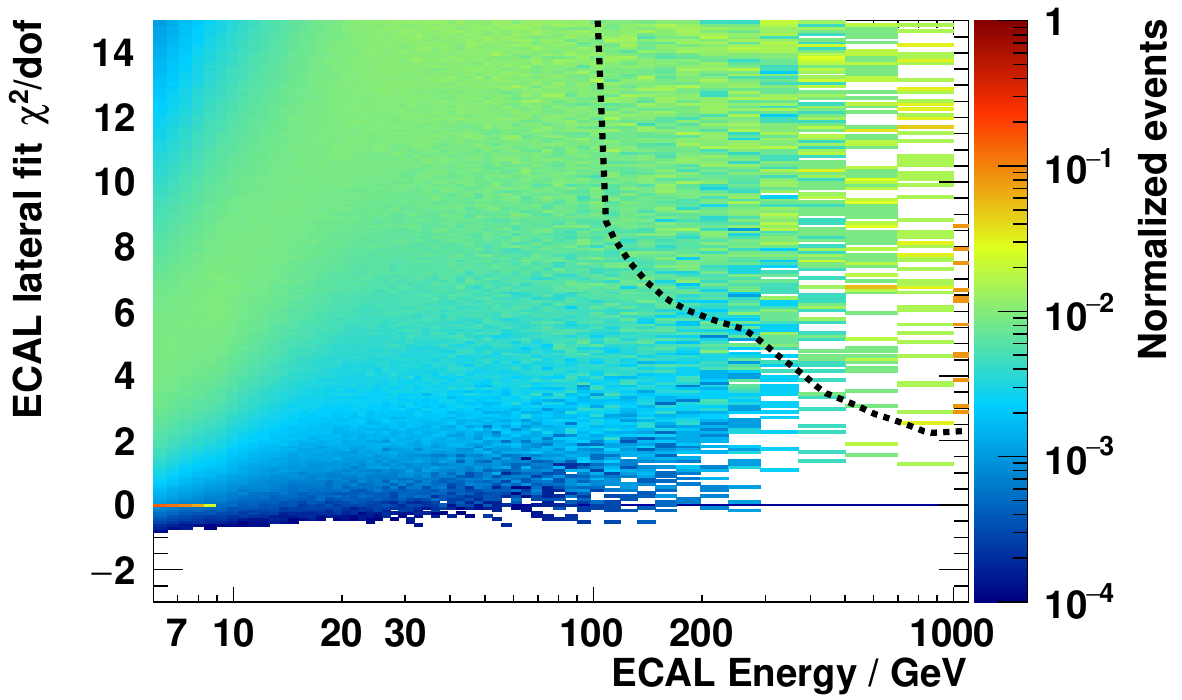}
        \caption{ISS R > 0 sample}
        \label{fig:analysis-data-selection-identification-cuts-ecal-lateral-normalized-background}
      \end{subfigure}
      \caption{Plot of the ECAL lateral goodness-of-fit distribution as function of energy for the electron Monte-Carlo simulation (left) and the ISS R > 0 sample (right). The applied upper cut value is shown as black dashed line.}
      \label{fig:analysis-data-selection-identification-cuts-ecal-lateral-normalized}
    \end{figure}

    A lateral shower fit is performed on the ECAL shower profile and an energy dependent cut is applied to reject proton events
    above \SIapprox{100}{\GeV}, as shown in \cref{fig:analysis-data-selection-identification-cuts-ecal-lateral-normalized-signal}.
    The shape of the cut value was optimized to keep all signal events below \SI{100}{\GeV} and to reject as many charge-confused protons
    as possible from the high-energy R < 0 sample. The R < 0 sample is used later on to define an optimal cut on the ECAL estimator to
    precisely control the proton and charge-confused proton background in the R > 0 and R < 0 sample, respectively. A soft cut on the lateral
    shower shape already removes many protons at high energy, as shown in \cref{fig:analysis-data-selection-identification-cuts-ecal-lateral-normalized-background}.

  \item\label{enum:electron-positron-identification-cut-tracker-track-ecal-cog-delta-x}\textbf{Tracker $\leftrightarrow$ ECAL matching in X-projection} (\SIvarEquals{$\epsilon_{\text{sig}}$}{97.46}{\percent}; \SIvarEquals{$\epsilon_{\text{bkg}}$}{85.00}{\percent}) \& \vspace{-1em}
  \item\textbf{Tracker $\leftrightarrow$ ECAL matching in Y-projection} (\SIvarEquals{$\epsilon_{\text{sig}}$}{99.99}{\percent}; \SIvarEquals{$\epsilon_{\text{bkg}}$}{90.64}{\percent})\hfill\\
    The tracker track is extrapolated to the Z position of the centre of gravity of the ECAL shower. Energy dependent
    cuts are applied on the $\Delta\text{X}/\Delta{\text{Y}}$ to ensure a good match between both detectors. While this is strictly
    not related to electron or positron identification, it is applied here to be able to define sensible cut values for a sample
    consisting of mostly $e^{\pm}$.

    The bands at $\abs{\Delta\text{X}} = \SI{8}{\centi\meter}$ are due to ambiguities in the tracker track reconstruction, as the X position is not
    precisely known due to the readout of the K side ladders (\cref{sec:detector-tracker}). For the $\Delta\text{Y}$ sample only an upper cut
    is applied, as the tail towards large negative $\Delta\text{Y}$ values is unavoidable, due to bremsstrahlung emission, which distorts the
    tracker track extrapolation. Note that for the positive rigidity sample the same cut is applied, but on $-\Delta\text{Y}$ instead of $\Delta\text{Y}$.

    \begin{figure}[H]
      \begin{subfigure}{0.50\linewidth}
        \includegraphics[width=\linewidth]{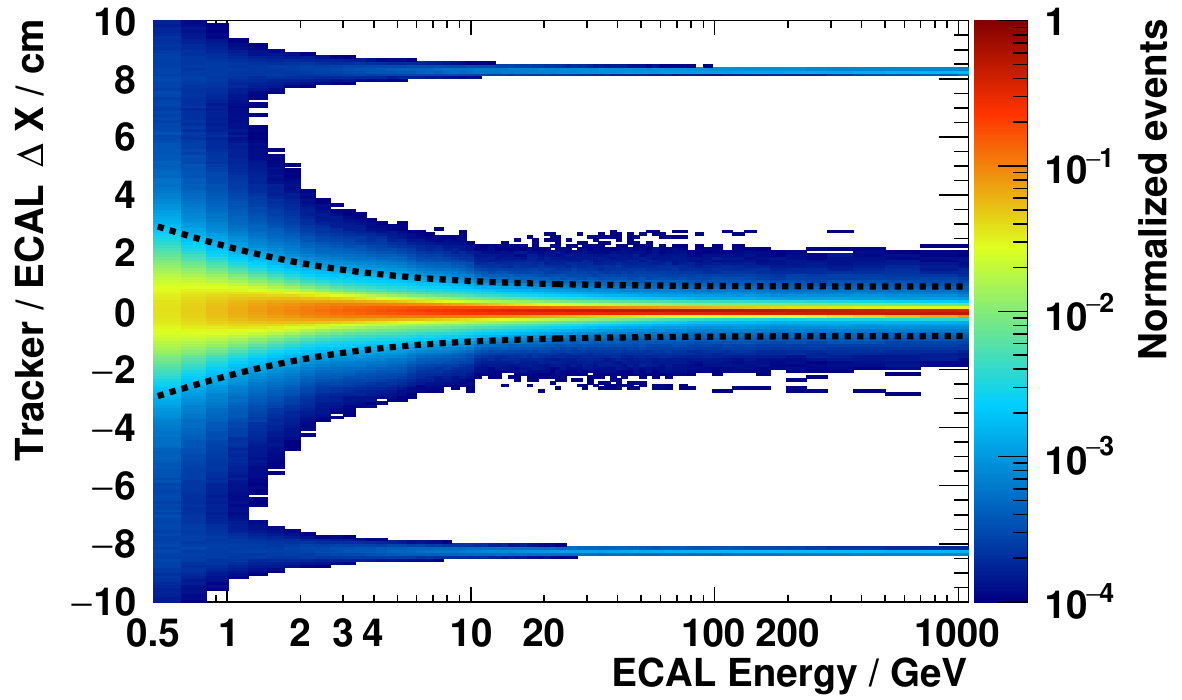}
        \caption{Tracker $\leftrightarrow$ ECAL - $\Delta\text{X}$}
        \label{fig:analysis-data-selection-identification-cuts-ecal-trk-delta-x}
      \end{subfigure}
      \begin{subfigure}{0.50\linewidth}
        \includegraphics[width=\linewidth]{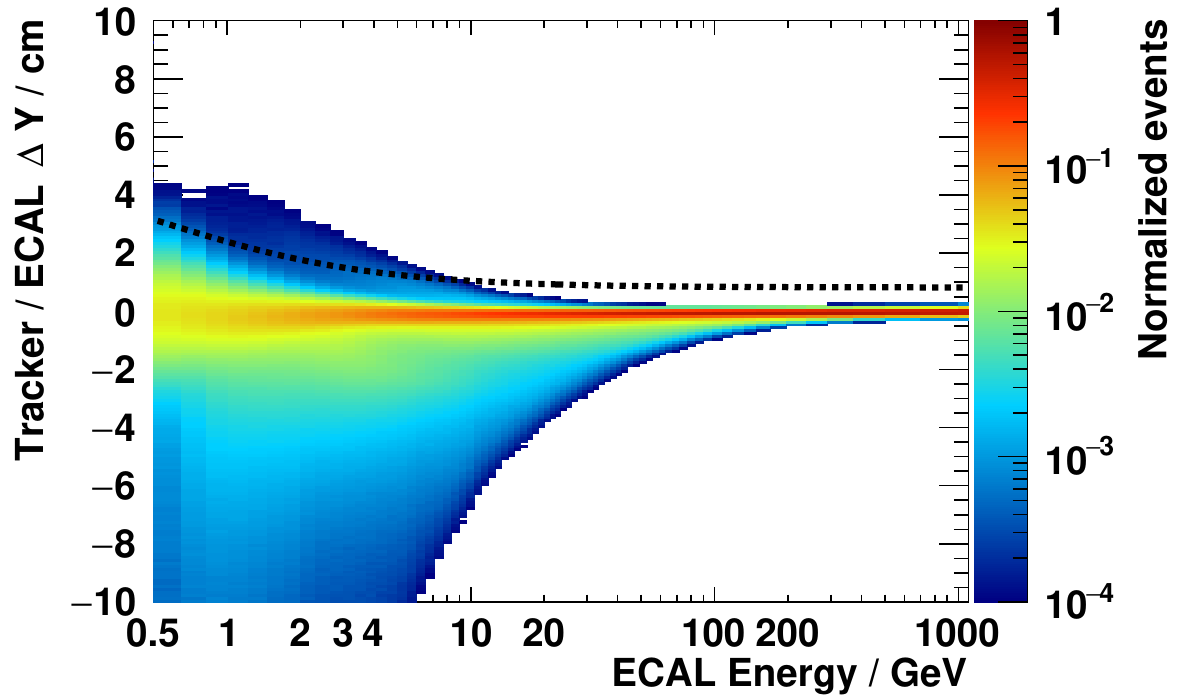}
        \caption{Tracker $\leftrightarrow$ ECAL - $\Delta\text{Y}$}
        \label{fig:analysis-data-selection-identification-cuts-ecal-trk-delta-y}
      \end{subfigure}
      \caption{Plot of the $\Delta\text{X}$ (left) and $\Delta\text{Y}$ (right) between the tracker track extrapolation and the ECAL shower centre of gravity as function of energy for the electron Monte-Carlo simulation. The applied lower and upper cut values are shown as black dashed lines.}
    \end{figure}
\end{enumerate}

In total \SI{91.92}{\percent} of all signal events in the electron Monte-Carlo simulation survive the $e^{\pm}$ identification cuts
and only \SI{10.29}{\percent} of all background events in the proton Monte-Carlo simulation.

After applying all aforementioned cuts a sample is left, which is enhanced by electrons and positrons and where many protons
are removed. The last cut that is applied is the requirement of a \textbf{\enquote{physics trigger}}. The purpose of this cut and
its implications will be discussed in \cref{sec:analysis-flux-time-averaged-trigger}.

\bigskip
The selection cuts already provide a factor $\approx 30$ proton rejection - only every \nth{30} proton passes the selection cuts.
The proton background will be further reduced due to cuts on the ECAL shower shape, which will be described later in \cref{sec:analysis-flux-time-averaged-ecal-estimator},
since the ECAL shower shape cut is not part of the $e^{\pm}$ identification cuts, but treated separately.

\bigskip
In the following sections it will be shown how to extract the electron and positron counts
as a function of kinetic energy and how the remaining proton background is treated.

\clearpage
\section{Electron and positron counts}
\label{sec:analysis-lepton-counts}

\begin{figure}[H]
  \includegraphics[width=\linewidth]{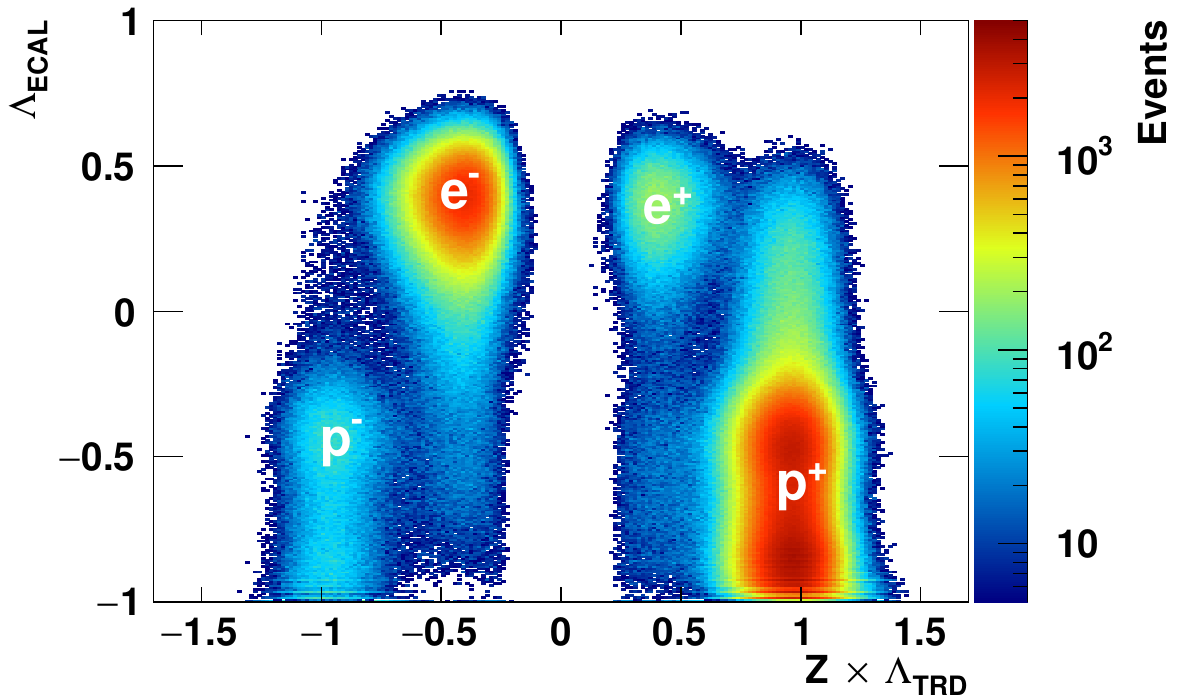}
  \caption{Overview of the ISS data sample after applying all selection cuts in the energy range \SIrange{20}{50}{\GeV}. The ECAL estimator $\Lambda_{\text{ECAL}}$ is drawn on the y-axis and the TRD estimator $\Lambda_{\text{TRD}}$ multiplied by the charge sign $Z$ of the primary tracker track on the x-axis. Clearly these two estimators allow to separate the different species: electrons, positrons, protons and charge-confused protons or antiprotons.}
  \label{fig:ecal-vs-trd-estimator}
\end{figure}

The main ingredient of a flux measurement is the number of particles in a given energy and time interval.
After applying all cuts described in the last chapter an electron and positron enhanced sample is left, as illustrated in
\cref{fig:ecal-vs-trd-estimator} for the energy range \SIrange{20}{50}{\GeV}.

The data sample presented in \cref{fig:ecal-vs-trd-estimator} contains many background events that can be removed
by imposing a cut on the ECAL estimator $\Lambda_{\text{ECAL}}$. A suitable cut value is chosen for each energy interval,
as discussed in detail in \cref{sec:analysis-flux-time-averaged-ecal-estimator}.

The leftover data sample contains electrons, positrons and a reduced amount of protons and antiprotons. Depending on the energy,
a non-negligible amount of charge-confused electrons, charge-confused positrons and charge-confused protons is also present in the data sample, which cannot be disentangled
using neither the ECAL estimator, nor the TRD estimator. To separate charge-confused electrons and charge-confused positrons from nominal electrons and positrons,
a multi-variate estimator - the CCMVA estimator $\Lambda_{\text{CC}}$ was constructed, which was described in detail in
\cref{sec:analysis-event-reconstruction-construction-ccmva-estimator}.

\begin{figure}[H]
  \includegraphics[width=\linewidth]{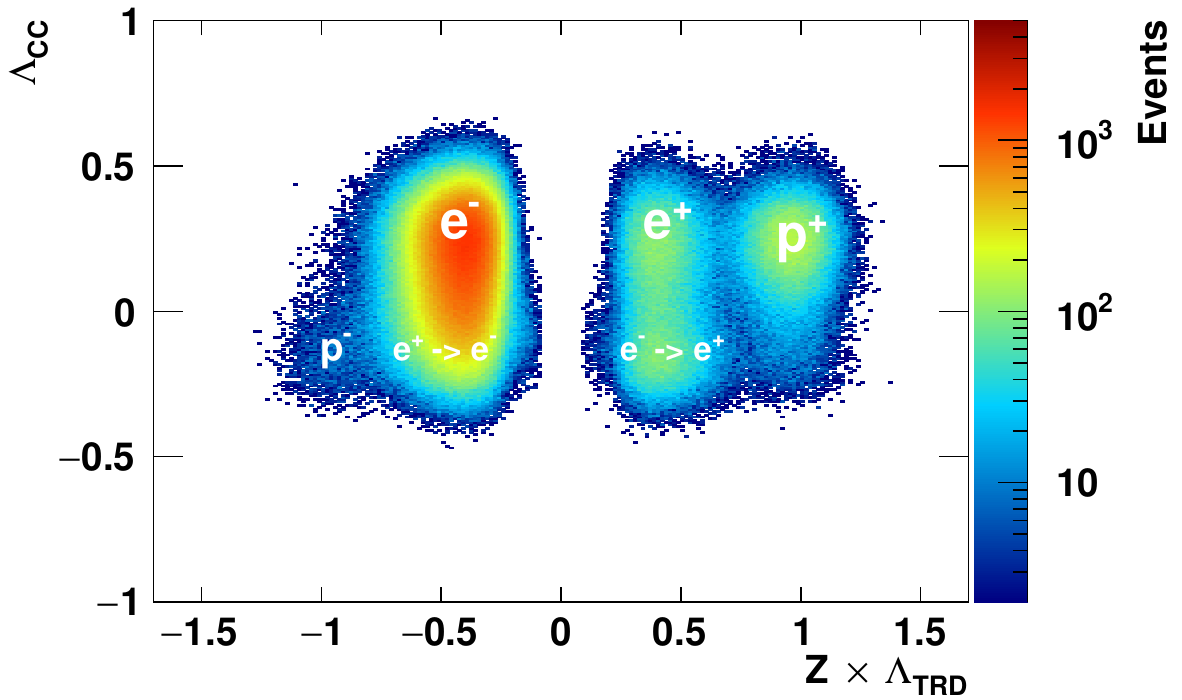}
  \caption{Overview of the ISS data sample after applying all selection cuts and an additional cut on the ECAL estimator $\Lambda_{\text{ECAL}}$ to reduce the proton background in the energy range \SIrange{20}{50}{\GeV}. The CCMVA estimator $\Lambda_{\text{CC}}$ is drawn on the y-axis and the TRD estimator $\Lambda_{\text{TRD}}$ multiplied by the charge sign $Z$ of the primary tracker track on the x-axis. Clearly these two estimators allow to separate the different species: electrons, positrons, protons and their charge-confused counterparts.}
  \label{fig:ccmva-vs-trd-estimator}
\end{figure}

\Cref{fig:ccmva-vs-trd-estimator} shows the ($Z \times \Lambda_{\text{TRD}}$ - $\Lambda_{\text{CC}}$) plane for the same data sample as
\cref{fig:ecal-vs-trd-estimator}, after imposing a cut on the ECAL estimator $\Lambda_{\text{ECAL}}$. In this representation six
components can be identified: electrons ($e^{-}$), charge-confused electrons ($e^{-} \rightarrow e^{+}$), positrons ($e^{+}$),
charge-confused positrons ($e^{+} \rightarrow e^{-}$), protons ($p^{+}$), antiprotons and charge-confused protons ($p^{-}$).
Antiprotons and charge-confused protons are not differentiated, because the CCMVA estimator was not trained or designed to resolve
the difference between antiprotons and charge-confused protons. For the electron and positron analysis there is no need to resolve them,
since both are background events. Furthermore the amount of charge-confused protons exceeds the amount of antiprotons, whose abundance is
suppressed by a factor of $\approx 10^4$ in cosmic rays, compared to protons. Thus the amount of charge-confused protons is expected
to be larger than the amount of antiprotons in the whole energy range in this analysis.

The true number of electrons is equal to the number of electrons that were correctly reconstructed with a negative charge sign plus the electrons
misreconstructed with positive charge sign (charge-confused electrons). Analogous, the true number of positrons is equal to the number of positrons
that were correctly reconstructed with a positive charge sign plus the positrons misreconstructed with a negative charge sign (charge-confused positrons).

To extract the true number of electrons or positrons a two-dimensional template fit is performed in the ($Z \times \Lambda_{\text{TRD}}$ - $\Lambda_{\text{CC}}$) plane
for a given energy and time interval, which will be described in detail in \cref{sec:analysis-lepton-counts-2d-fit}. To perform the template fit, reference distributions
for all components (electrons, positrons, protons, and their charge-confused counterparts) need to be obtained for both the TRD estimator $\Lambda_{\text{TRD}}$
and the CCMVA estimator $\Lambda_{\text{CC}}$.

\bigskip
In the following sections it will be presented how the TRD and CCMVA templates are obtained and how the two-dimensional template fit works in detail.

\subsection{Binning}
\label{sec:analysis-lepton-counts-binning}

The electron and positron flux, as well as the positron fraction and the positron/electron ratio will be determined in 74 energy bins, from
\SI{0.5}{\GeV} up to \SI{1}{\TeV} using the same binning as used in previous AMS-02 electron and positron flux publications~\cite{Aguilar2014a,Aguilar2019a,Aguilar2019b}.

The binning is chosen according to the energy resolution $\sigma_{\text{ECAL}}$ (\cref{sec:detector-ecal}): the bin width exceeds $3\cdot\sigma_{\text{ECAL}}$, to minimize
migration effects. Above \SIapprox{100}{\GeV} the criteria is relaxed and the bin width enlarged in order to have sufficient statistics in each high-energy analysis bin.

\subsection{TRD templates}
\label{sec:analysis-lepton-counts-trd-templates}

As shown in \cref{fig:ecal-vs-trd-estimator} the ECAL estimator, the TRD estimator and the charge sign of the reconstructed
tracker track can be used to separate electrons, positrons, protons, charge-confused protons. To obtain reference distributions
for the TRD estimator, pure data samples need to be selected using only the ECAL estimator and the charge sign of the reconstructed
tracker track. The purity of these data samples decrease with increasing energy -- the higher the energy, the more difficult it becomes
to obtain a pure sample with decent statistics.

For example, the negative rigidity sample contains mostly electrons, but also a non-negligible amount of charge-confused protons.
In order to extract a pure reference distribution for electrons one needs to take the charge-confused proton background into account
and properly subtract it.

Three ISS data samples are prepared, in order to extract the reference distributions for the TRD estimator.
All detector quality, preselection and selection cuts - \cref{sec:analysis-data-selection-detector-quality-cuts,sec:analysis-data-selection-preselection-cuts,sec:analysis-data-selection-selection-cuts} - are applied when selecting the ISS data samples.
Furthermore the same 0.5 < $E/\abs{R}$ < 1 cut as for the signal event selection is applied. The following enumeration lists the sample specific cuts used to select the data samples, from which the TRD templates are extracted:

\begin{enumerate}
  \item\textbf{Electron sample}\hfill
    \begin{itemize}
      \item ISS data: negative rigidity sample (R < 0)
      \item Electron like ECAL shower ($\Lambda_{\text{ECAL}}$ > 0)
      \item Energy dependent cut on the lateral shape
            (defined in \cref{sec:analysis-data-selection-electron-positron-identification-cuts} - \cref{enum:electron-positron-identification-cut-ecal-lateral-shower-shape})
    \end{itemize}

  \item\textbf{Charge-confused proton sample}\hfill
    \begin{itemize}
      \item ISS data: negative rigidity sample (R < 0)
      \item Hadron-like ECAL shower ($\Lambda_{\text{ECAL}}$ < 0)
    \end{itemize}

    \item\textbf{Proton sample}\hfill
    \begin{itemize}
      \item ISS data: positive rigidity sample (R > 0)
      \item Hadron-like ECAL shower ($\Lambda_{\text{ECAL}}$ < 0)
    \end{itemize}
\end{enumerate}

It is important to note that the shape of the TRD estimator depends on the topology of the event. Events where only a single tracker
track is reconstructed (\enquote{single-track sample}) and events with multiple reconstructed tracker tracks (\enquote{multi-tracks sample})
differ in the shape\footnote{In multi-track events, the path length in the TRD tubes ($\diff\text{x}$) is defined less precisely than in the single-track sample, which alters the TRD estimator shape. }
of the TRD estimator. Therefore all template extraction procedures are executed separately for both orthogonal event samples.

\bigskip
For both track samples, all reference distributions can be parameterized by a sum of analytical functions: a gaussian function $G(x; \mu, \sigma)$
and the Novosibirsk function $N(x; \mu, \sigma, \tau)$:

\begin{equation}
  \label{eq:trd-gaussian}
  G(x; \mu, \sigma) = \frac{1}{\sqrt{2 \pi \sigma^2}}\exp{\left(-\frac{(x - \mu)^2}{2\sigma^2}\right)},
\end{equation}

\begin{equation}
  \label{eq:trd-novosibirsk}
  \begin{aligned}
    N(x; \mu, \sigma, \tau)       &= \exp{\left(-\frac{1}{2} \left({\frac{\left(\ln{(\lambda(x; \mu, \sigma, \tau))}\right)^2}{\tau^2}} + \tau^2\right)\right)}, \text{where} \\
    \lambda(x; \mu, \sigma, \tau) &= 1 + \tau (x - \mu) \frac{\sinh{\left(\tau \sqrt{\ln{4}}\right)}}{\sigma \tau \sqrt{\ln{4}}}.
 \end{aligned}
\end{equation}

The following enumeration shows the parameterizations that are used to describe the TRD templates analytically for each energy bin in the analysis:

\vspace{-1mm}
\begin{enumerate}
  \item\textbf{Electron template}\hfill
    \begin{equation}
      \label{eq:trd-model-electrons}
      \begin{aligned}
        f_{\text{elec}}(x = \Lambda_{\text{TRD}}) &= \elecFractionNovoSymbol       & \cdot\  & N(x; \elecPeakNovoSymbol, \elecWidthNovoSymbol, \elecTailNovoSymbol) \\
                                                  &+ (1 - \elecFractionNovoSymbol) & \cdot\  & G(x; \elecPeakGausSymbol, \elecWidthGausSymbol)
      \end{aligned}
    \end{equation}

  \item\textbf{Charge-confused proton template}\hfill
    \begin{equation}
      \label{eq:trd-model-ccprotons}
      f_{\text{ccprot}}(x = \Lambda_{\text{TRD}}) = N(x; \ccProtPeakNovoSymbol, \ccProtWidthNovoSymbol, \ccProtTailNovoSymbol)
    \end{equation}

    \item\textbf{Proton template}
    \begin{equation}
      \label{eq:trd-model-protons}
      \begin{aligned}
        f_{\text{prot}}(x = \Lambda_{\text{TRD}}) &= \protFractionNovoSymbol       & \cdot\  & N(x; \protPeakNovoSymbol, \protWidthNovoSymbol, \protTailNovoSymbol) \\\
                                                  &+ (1 - \protFractionNovoSymbol) & \cdot\  & G(x; \protPeakGausSymbol, \protWidthGausSymbol), \text{where} \\
        \protWidthGausSymbol                      &= \protWidthNovoSymbol + \protWidthGausDeltaSymbol.
      \end{aligned}
    \end{equation}
\end{enumerate}
\vspace{-1mm}

\begin{figure}[H]
  \begin{subfigure}{0.50\linewidth}
    \includegraphics[width=\linewidth]{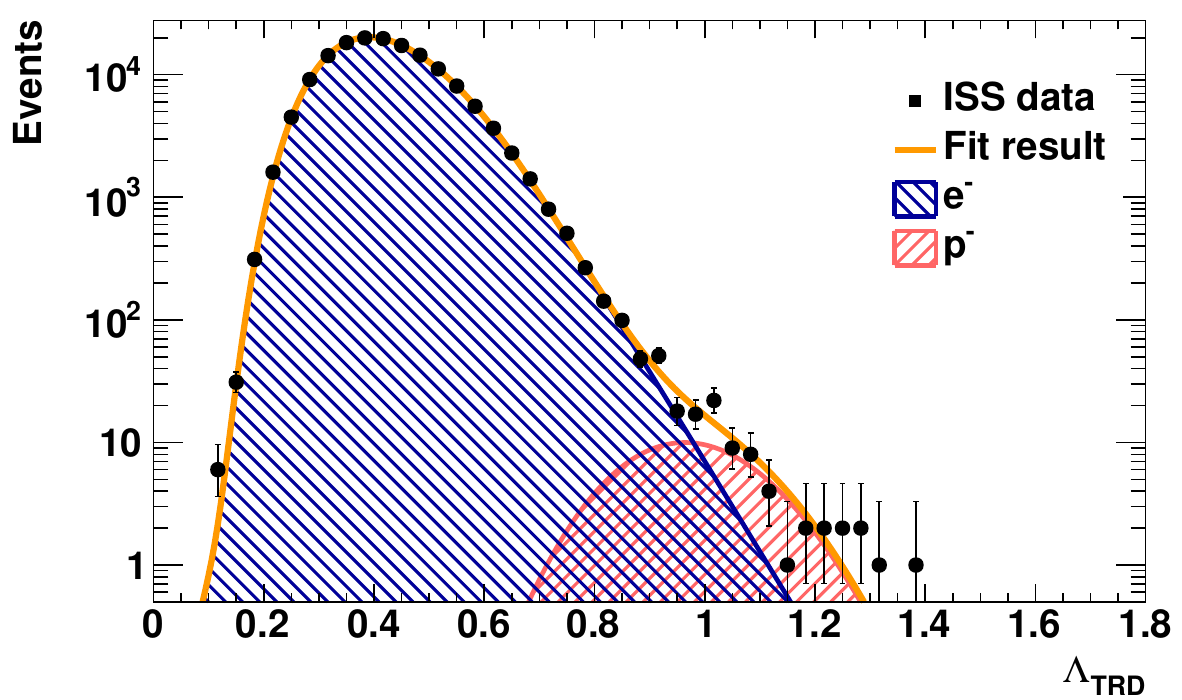}
    \caption{Electron sample - single-track}
  \end{subfigure}
  \begin{subfigure}{0.50\linewidth}
    \includegraphics[width=\linewidth]{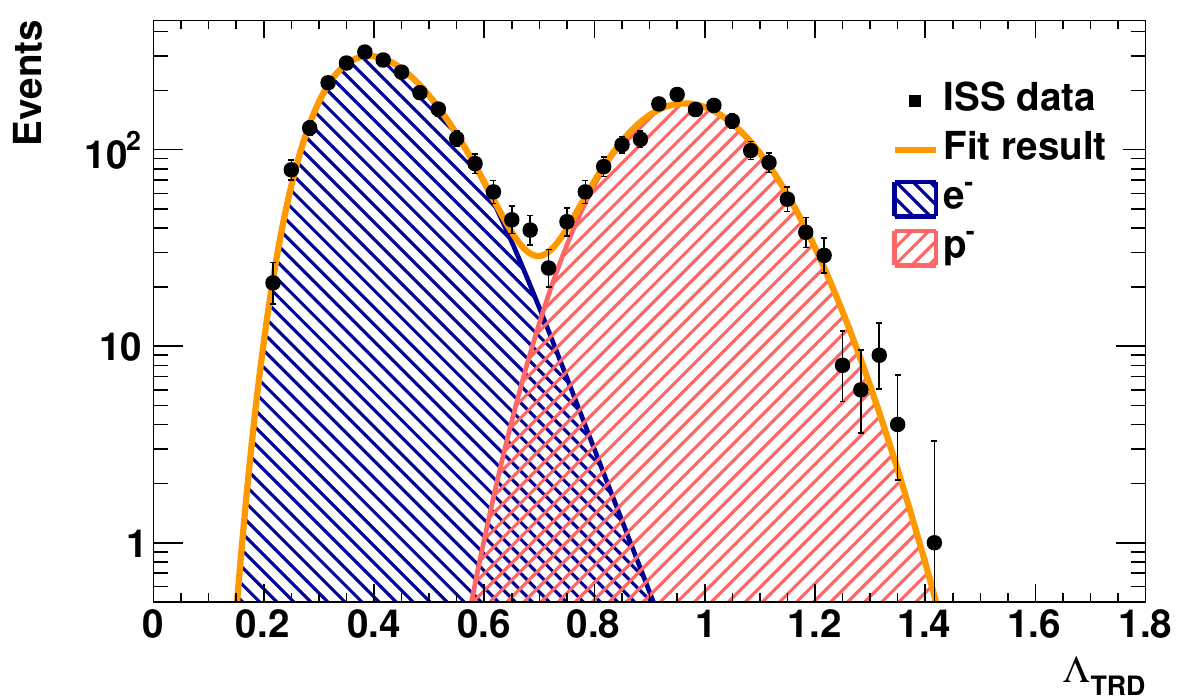}
    \caption{Charge-confused proton sample - single-track}
  \end{subfigure}
  \caption{Simultaneous fit of the electron template \cref{eq:trd-model-electrons} and the charge-confused proton template \cref{eq:trd-model-ccprotons} in the energy bin \SIrange{17.98}{18.99}{\GeV} for the single-track sample. The \elecColorText~area corresponds to the electron template, the \ccProtColorText~area to the charge-confused proton template and the \fitResultColorText~line to the sum of all templates.}
  \label{fig:trd-neg-template-single-track-bin-34}
\end{figure}

For each energy bin a maximum likelihood fit is performed, simultaneously on the \enquote{Electron sample} and the \enquote{Charge-confused proton sample},
to determine the parameters that describe the electron template - \cref{eq:trd-model-electrons} - and the charge-confused proton template - \cref{eq:trd-model-ccprotons}.

\begin{figure}[H]
  \begin{subfigure}{0.50\linewidth}
    \includegraphics[width=\linewidth]{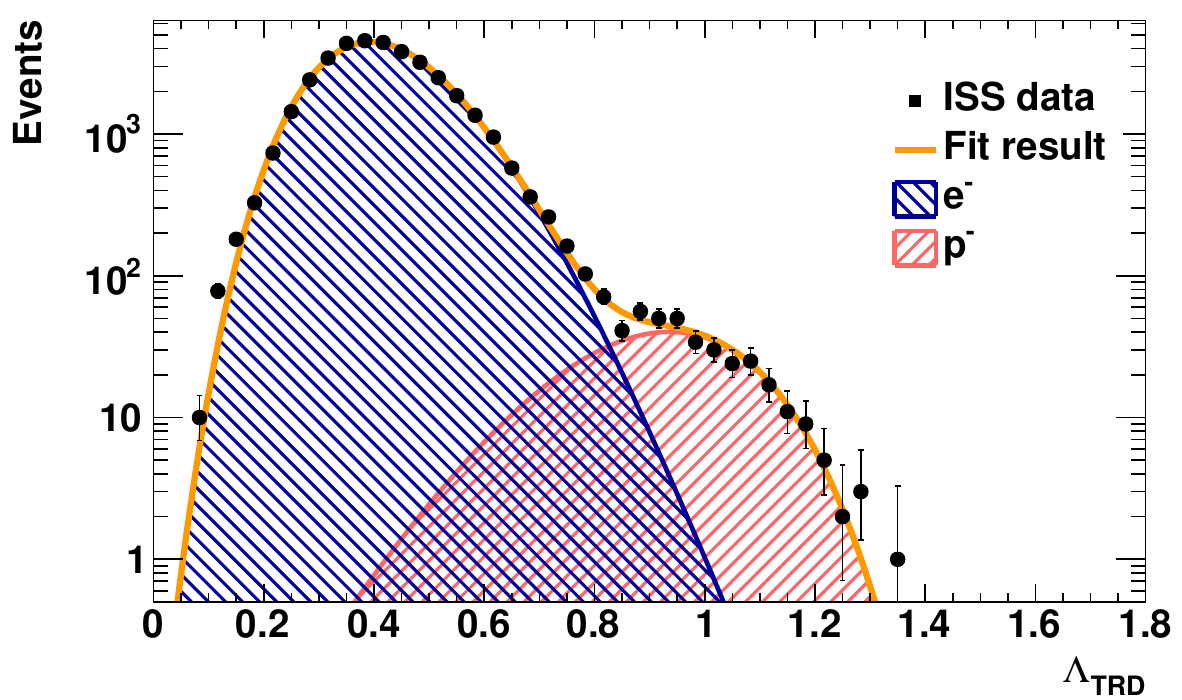}
    \caption{Electron sample - multi-tracks}
  \end{subfigure}
  \begin{subfigure}{0.50\linewidth}
    \includegraphics[width=\linewidth]{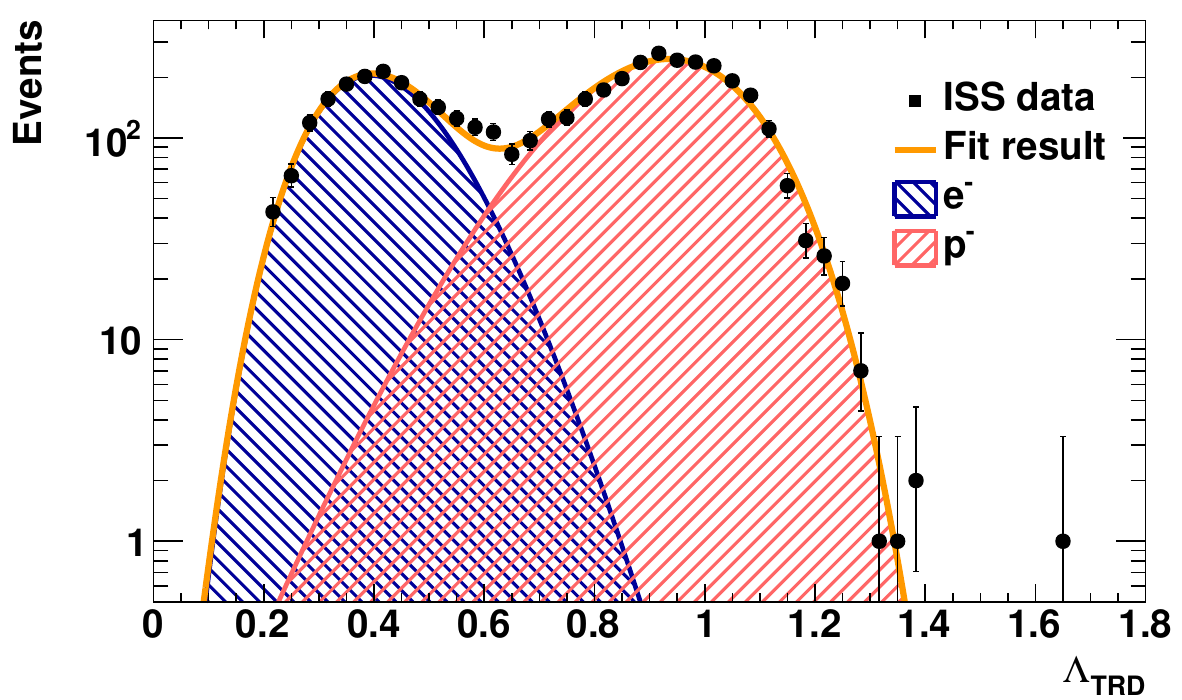}
    \caption{Charge-confused proton sample - multi-tracks}
  \end{subfigure}
  \caption{Simultaneous fit of the electron template \cref{eq:trd-model-electrons} and the charge-confused proton template \cref{eq:trd-model-ccprotons} in the energy bin \SIrange{17.98}{18.99}{\GeV} for the multi-tracks sample. The \elecColorText~area corresponds to the electron template, the \ccProtColorText~area to the charge-confused proton template and the \fitResultColorText~line to the sum of all templates.}
  \label{fig:trd-neg-template-multi-tracks-bin-34}
\end{figure}

\Cref{fig:trd-neg-template-single-track-bin-34,fig:trd-neg-template-multi-tracks-bin-34} show an example of the fit procedure for the single-track sample
and the multi-tracks sample, respectively, for an example energy bin. It is evident that the charge-confused proton component is enlarged in the
multi-tracks sample and thus it is important to extract the TRD templates separately for the single-track and multi-tracks sample.

\begin{figure}[H]
  \begin{subfigure}{0.50\linewidth}
    \includegraphics[width=\linewidth]{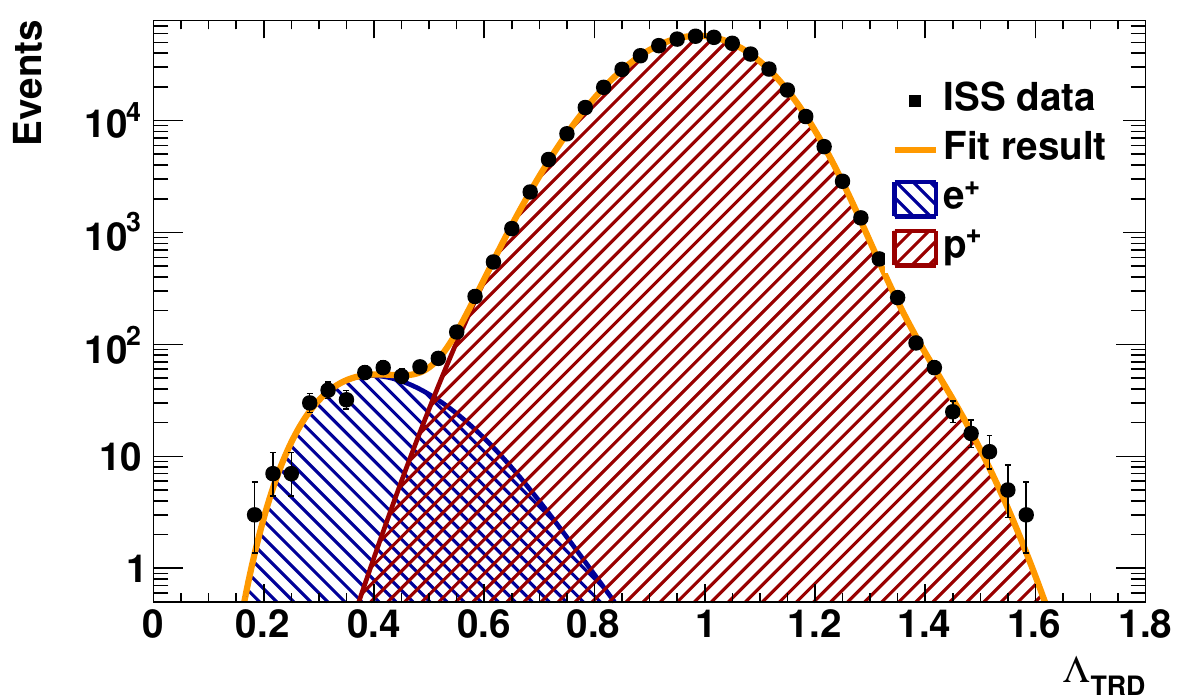}
    \caption{Single-track sample}
    \label{fig:trd-pos-template-single-track-bin-34}
  \end{subfigure}
  \begin{subfigure}{0.50\linewidth}
    \includegraphics[width=\linewidth]{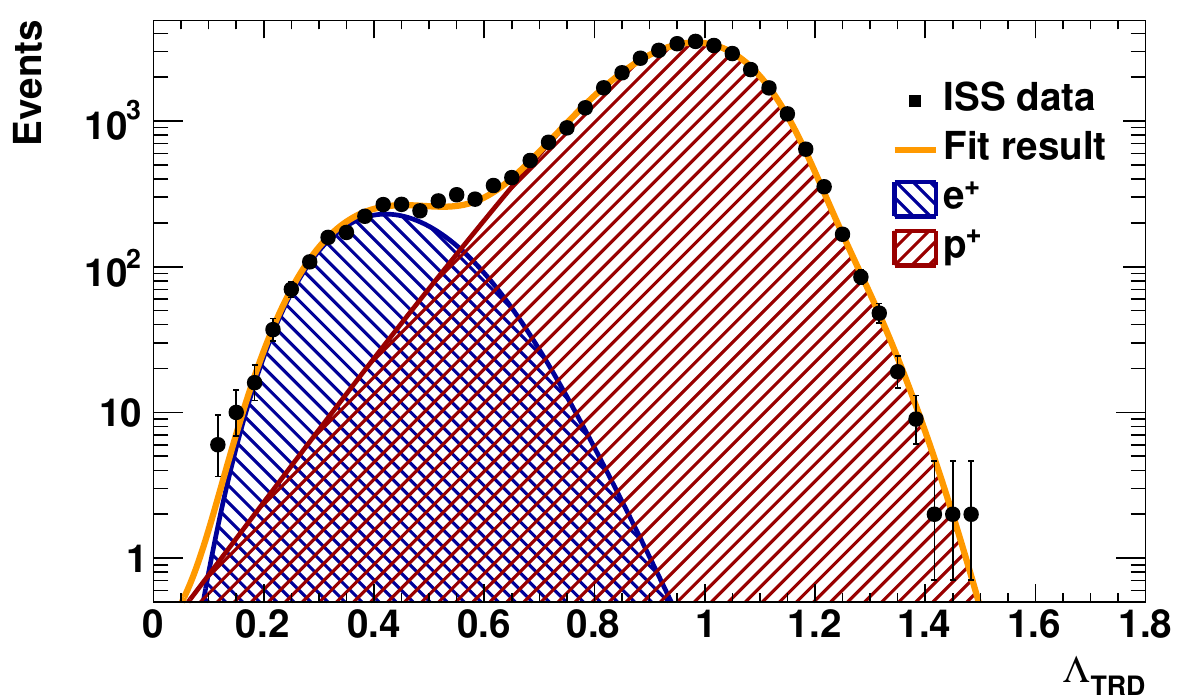}
    \caption{Multi-tracks sample}
    \label{fig:trd-pos-template-multi-tracks-bin-34}
  \end{subfigure}
  \caption{Fit of the proton template \cref{eq:trd-model-protons} in the energy bin \SIrange{17.98}{18.99}{\GeV} for the single-track sample and the multi-tracks sample. The \elecColorText~area corresponds to the electron template, the \ccProtColorText~area to the charge-confused proton template and the \fitResultColorText~line to the sum of all templates.}
  \label{fig:trd-pos-template-bin-34}
\end{figure}

Once the electron template is known, the positron template necessary to describe the positrons in the positive rigidity sample, is known as well, since
they are indistinguishable by the TRD and share the same template. Therefore the proton template - \cref{eq:trd-model-protons}, can be extracted, using a
maximum likelihood fit on the \enquote{Proton sample}, as shown in \cref{fig:trd-pos-template-single-track-bin-34} for the single-track sample and in
\cref{fig:trd-pos-template-multi-tracks-bin-34} for the multi-tracks sample for an example energy bin.

\medskip
The template extraction procedure is repeated independently for all energy bins in the analysis, yielding a list of parameters for each energy bin
and each template component. The energy dependence of each of the parameters can be parameterized by a smooth function: a spline~\cite{Birkhoff1960}.

\begin{figure}[H]
  \begin{subfigure}{0.48\linewidth}
    \includegraphics[width=\linewidth]{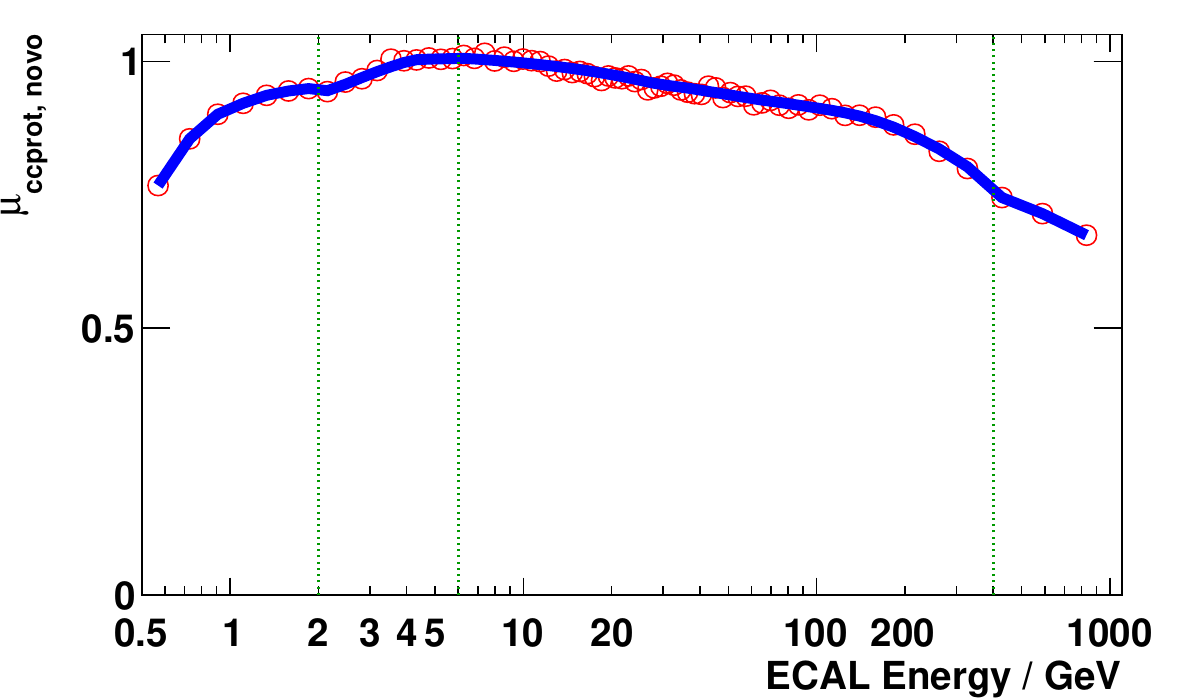}
    \caption{Template parameter~\ccProtPeakNovo}
  \end{subfigure}
  \hfill
  \begin{subfigure}{0.48\linewidth}
    \includegraphics[width=\linewidth]{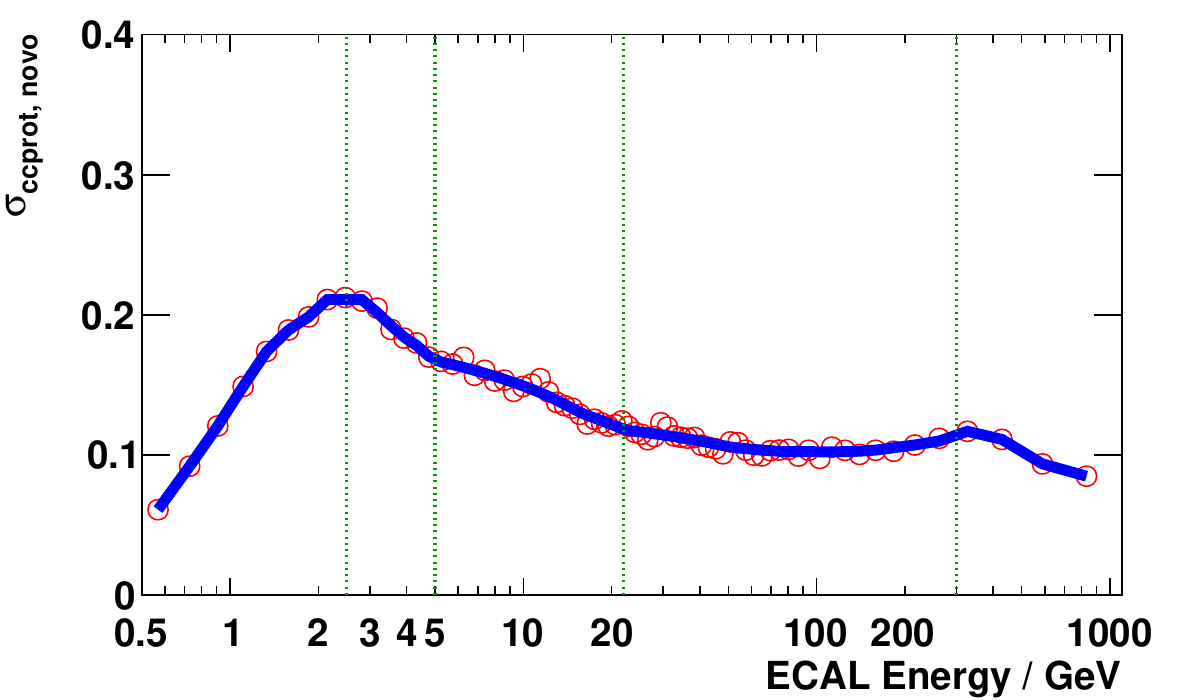}
    \caption{Template parameter~\ccProtWidthNovo}
  \end{subfigure}
  \hfill
  \begin{subfigure}{0.48\linewidth}
    \includegraphics[width=\linewidth]{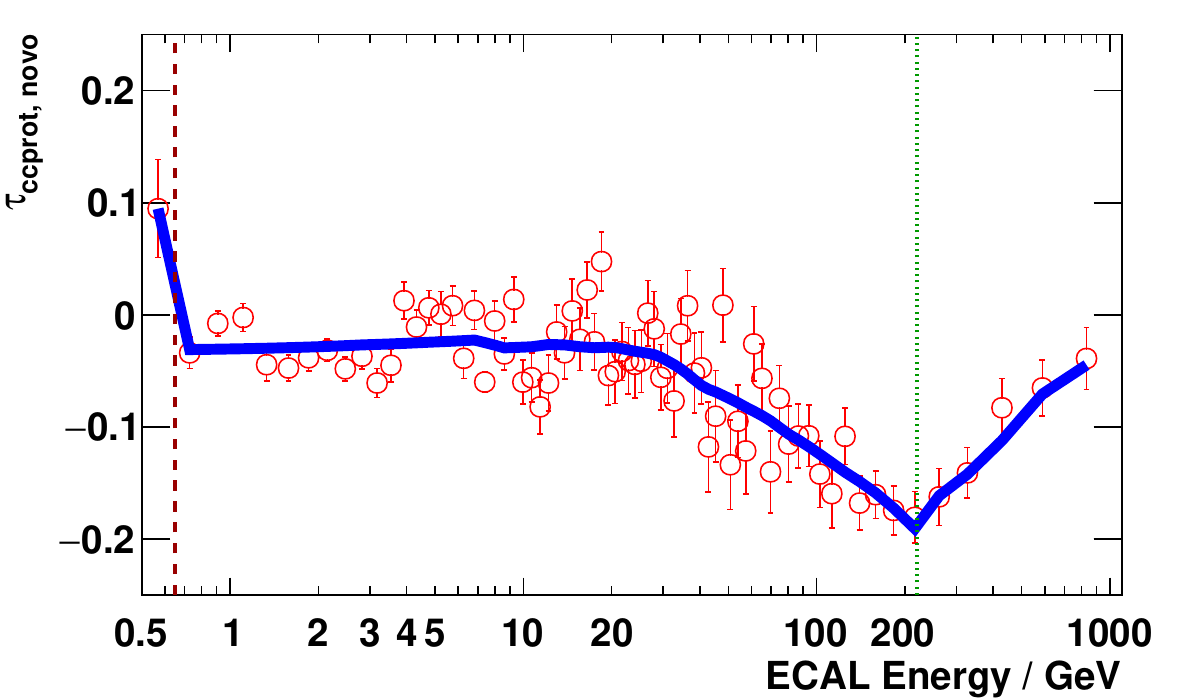}
    \caption{Template parameter~\ccProtTailNovo}
  \end{subfigure}
  \hfill
  \begin{subfigure}{0.48\linewidth}
    \includegraphics[width=\linewidth]{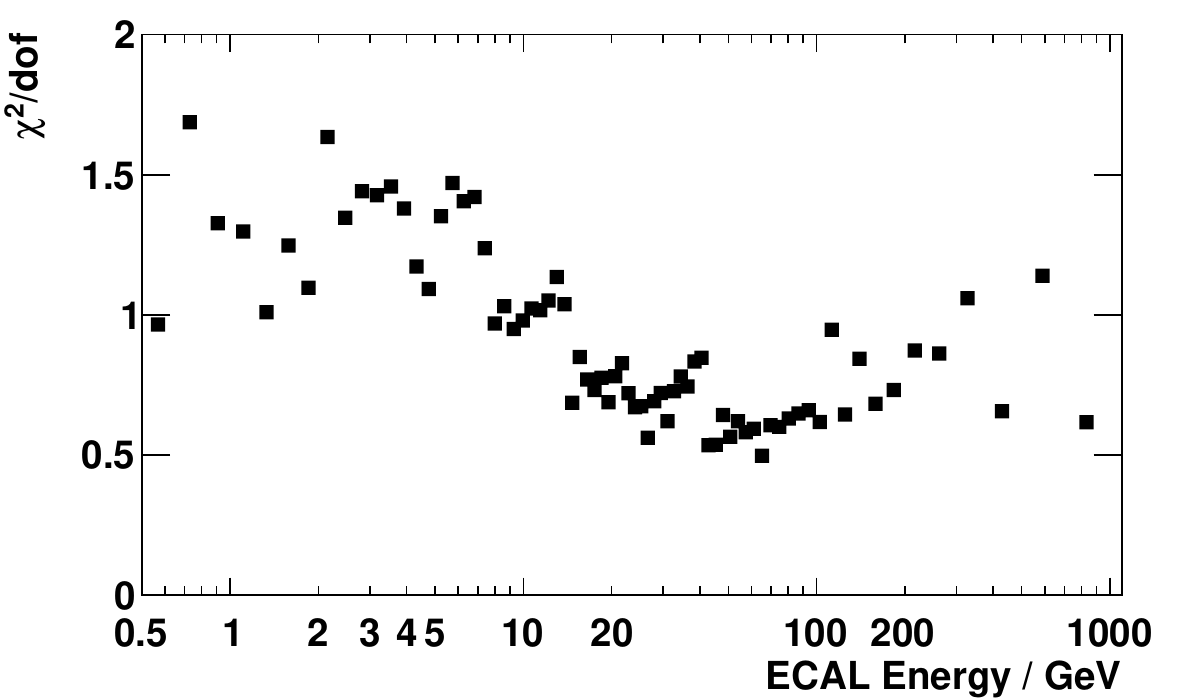}
    \caption{Goodness-of-fit}
    \label{fig:trd-template-parameters-ccprotons-chisquare-single-track-vs-energy}
  \end{subfigure}
  \caption{Analytical TRD template parameters describing the evolution of the charge-confused proton template as function of energy for the single-track sample. In each energy bin the analytical function -~\cref{eq:trd-model-ccprotons}~- is fit to the template data sample, yielding the red points. A smoothing procedure yields the blue curves, which are used as template parameters for the analysis. The vertical dashed green lines mark areas where different isolated smoothing procedures are applied. The goodness-of-fit is indicated by the black points in the lower right plot.}
  \label{fig:trd-template-parameters-ccprotons-single-track-vs-energy}
\end{figure}

\Cref{fig:trd-template-parameters-ccprotons-single-track-vs-energy} shows the charge-confused proton template parameters as function of energy for the single-track sample
and \cref{fig:trd-template-parameters-ccprotons-multi-tracks-vs-energy} for the multi-tracks sample.

Only three parameters are necessary to describe the template analytically for all energies: the peak position~\ccProtPeakNovo, the width~\ccProtWidthNovo~and the tail parameter~\ccProtTailNovo.

The $\chi^2/\text{dof}$ of the fit is shown in \cref{fig:trd-template-parameters-ccprotons-chisquare-single-track-vs-energy} for the single-track sample. The goodness-of-fit of the single-track sample exhibits a $\chi^2/\text{dof} > 1$ towards low energies, which can be cured by introducing another degree of freedom: an additional Novosibirsk component below \SI{10}{\GeV}.

While this improves the $\chi^2$, it has no impact on the template fits that will be executed using these templates -- for that reason the additional Novosibirsk component was omitted, to stabilize the minimization procedure.

The results for the multi-tracks sample are presented in \cref{sec:appendix-trd-templates}. The conclusions are identical as for the single-track sample.

\begin{figure}[H]
  \begin{subfigure}{0.48\linewidth}
    \includegraphics[width=\linewidth]{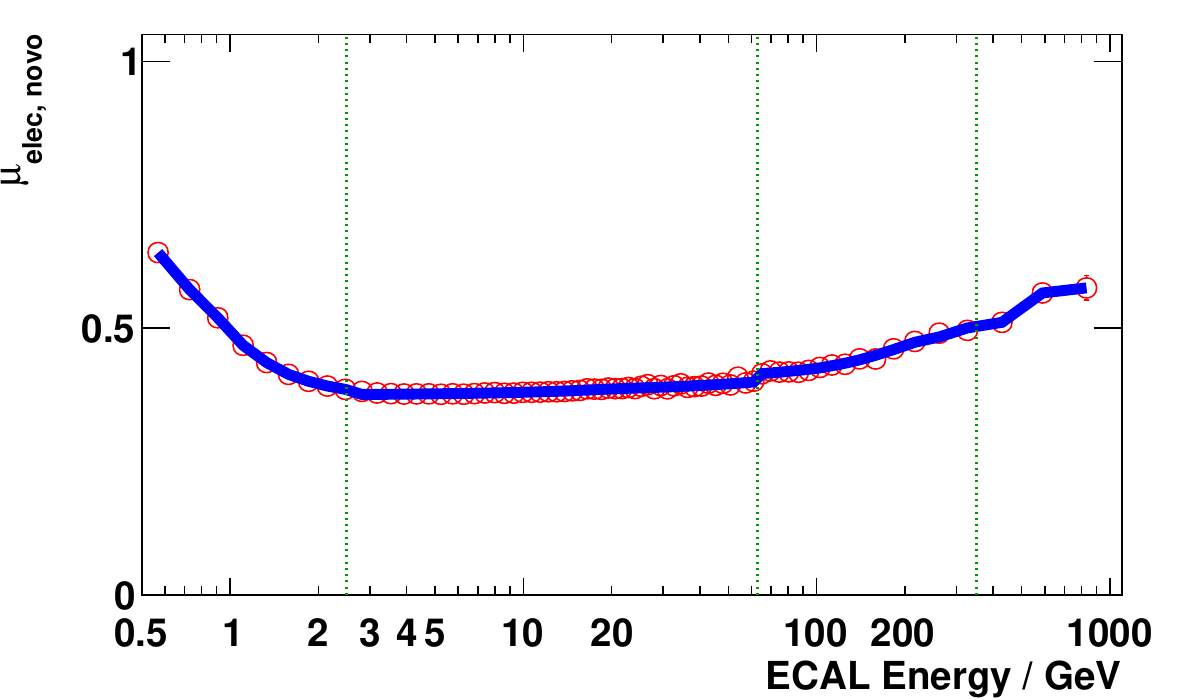}
    \caption{Template parameter~\elecPeakNovo}
  \end{subfigure}
  \hfill
  \begin{subfigure}{0.48\linewidth}
    \includegraphics[width=\linewidth]{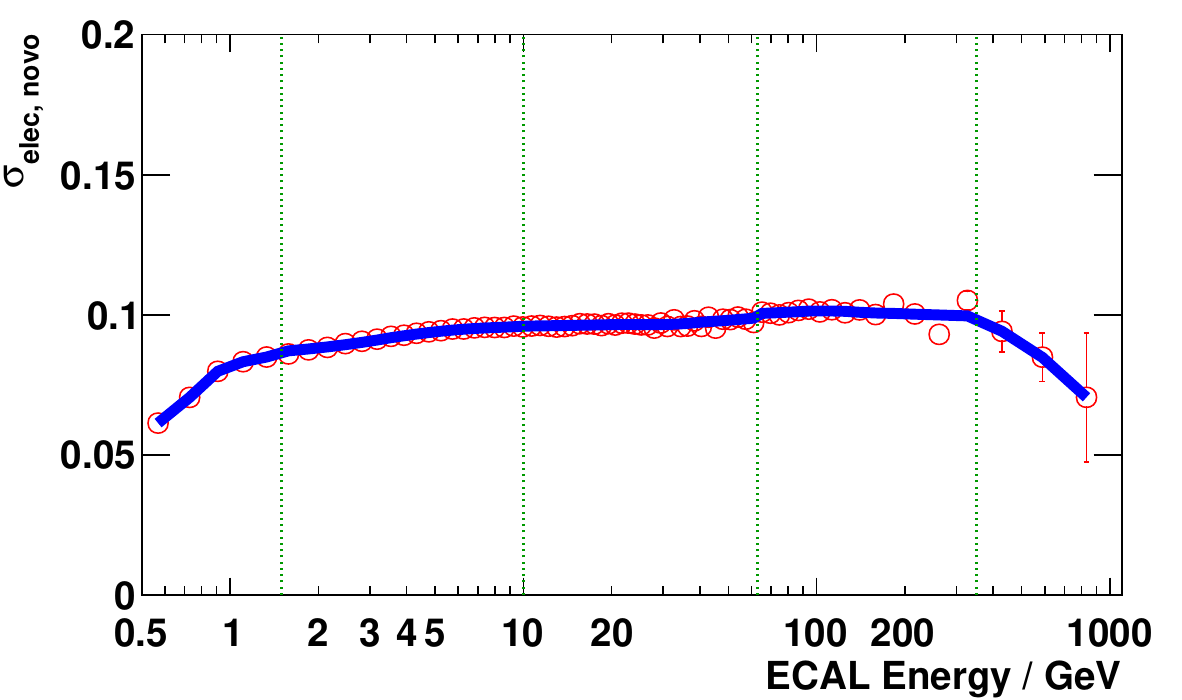}
    \caption{Template parameter~\elecWidthNovo}
  \end{subfigure}
  \hfill
  \begin{subfigure}{0.48\linewidth}
    \includegraphics[width=\linewidth]{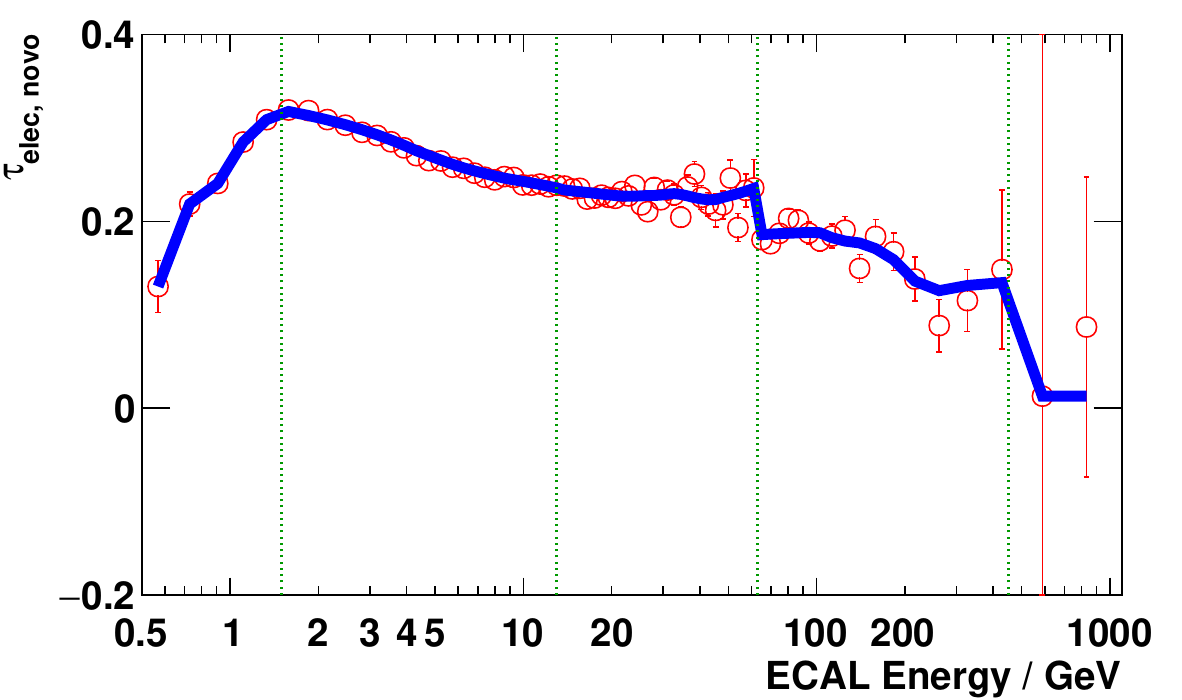}
    \caption{Template parameter~\elecTailNovo}
  \end{subfigure}
  \hfill
  \begin{subfigure}{0.48\linewidth}
    \includegraphics[width=\linewidth]{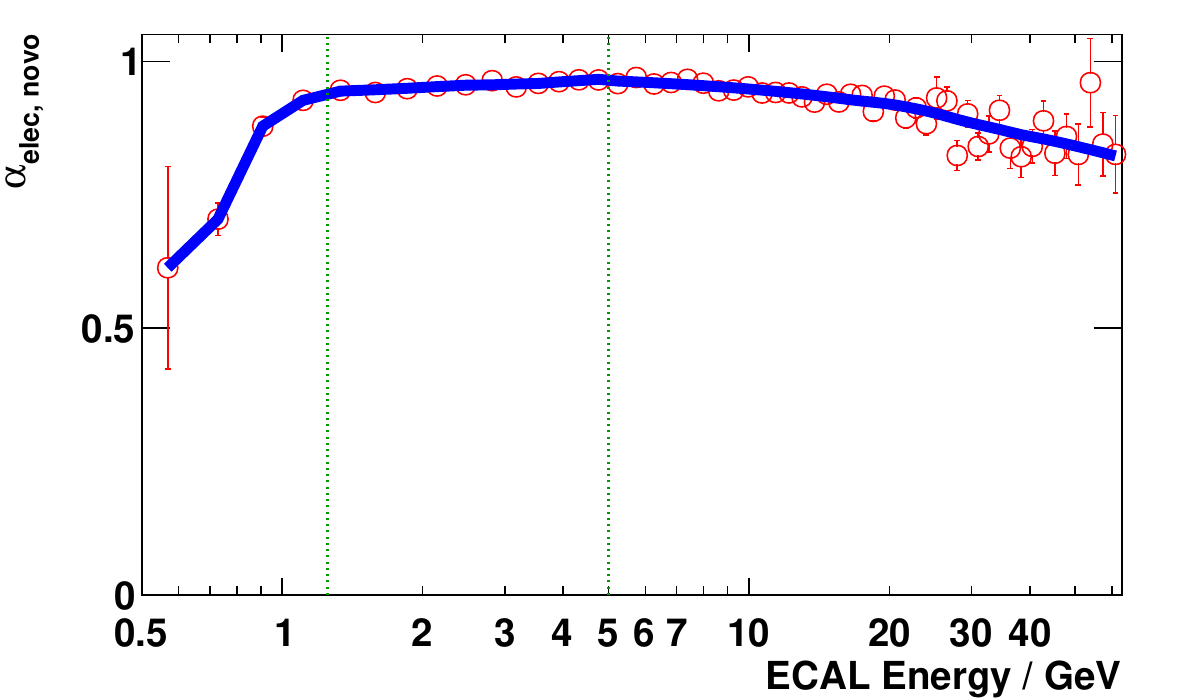}
    \caption{Template parameter~\elecFractionNovo}
    \label{fig:trd-template-parameters-electrons-elecfractionnovo-single-track-vs-energy}
  \end{subfigure}
  \hfill
  \begin{subfigure}{0.48\linewidth}
    \includegraphics[width=\linewidth]{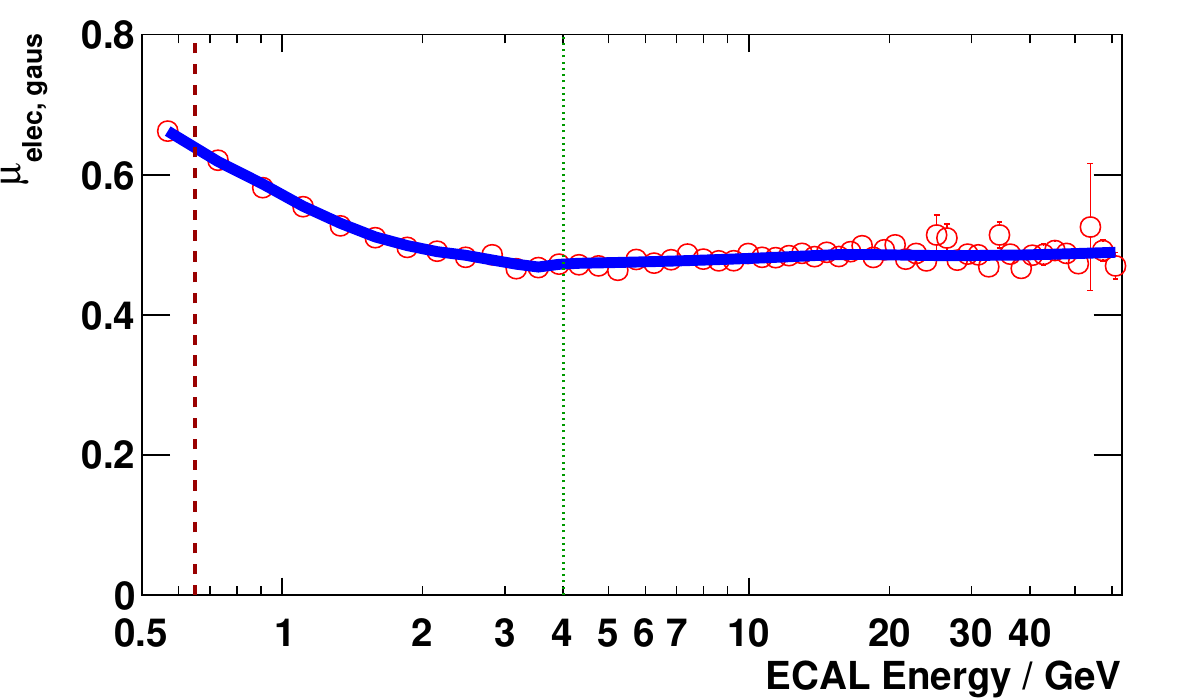}
    \caption{Template parameter~\elecPeakGaus}
    \label{fig:trd-template-parameters-electrons-elecpeakgaus-single-track-vs-energy}
  \end{subfigure}
  \hfill
  \begin{subfigure}{0.48\linewidth}
    \includegraphics[width=\linewidth]{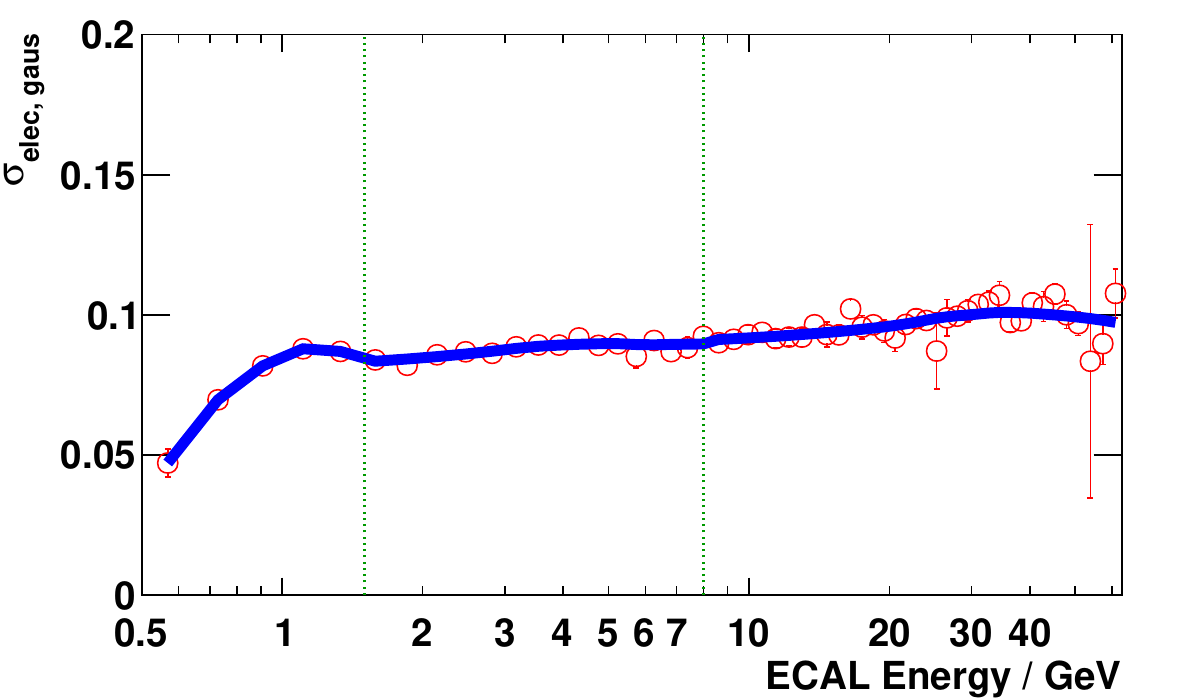}
    \caption{Template parameter \elecWidthGaus}
    \label{fig:trd-template-parameters-electrons-elecwidthgaus-single-track-vs-energy}
  \end{subfigure}
  \caption{Analytical TRD template parameters describing the evolution of the electron template as function of energy for the single-track sample. In each energy bin the analytical function -~\cref{eq:trd-model-electrons}~- is fit to the template data sample, yielding the red points. A smoothing procedure yields the blue curves, which are used as template parameters for the analysis. The vertical dashed green lines mark areas where different isolated smoothing procedures are applied. The vertical dashed red lines mark energies where the smoothing procedure starts -- points before this border are preserved.}
  \label{fig:trd-template-parameters-electrons-single-track-vs-energy}
\end{figure}

\clearpage
A similar smoothing procedure is applied for the electron template, as shown in \cref{fig:trd-template-parameters-electrons-single-track-vs-energy} for the single-track sample and in \cref{fig:trd-template-parameters-electrons-multi-tracks-vs-energy} for the multi-tracks sample. In total six parameters are relevant to describe the template analytically for all energies: three parameters corresponding to the
Novosibirsk function in the electron model (peak position~\elecPeakNovo, width~\elecWidthNovo~and tail parameter~\elecTailNovo), two for
the gaussian part (peak position~\elecPeakGaus, width~\elecWidthGaus) and a ratio describing the relative contribution of the
Novosibirsk needed to fit the template (fraction~\elecFractionNovo).

The three template parameters~\elecFractionNovo, \elecPeakGaus~and~\elecWidthGaus~presented in \cref{fig:trd-template-parameters-electrons-elecfractionnovo-single-track-vs-energy,fig:trd-template-parameters-electrons-elecpeakgaus-single-track-vs-energy,fig:trd-template-parameters-electrons-elecwidthgaus-single-track-vs-energy} are only relevant below \SI{65}{\GeV}.
Above this energy threshold the electron template can be described with a single Novosibirsk, without a gaussian component.

\begin{figure}[H]
  \centering
  \includegraphics[width=0.7\linewidth]{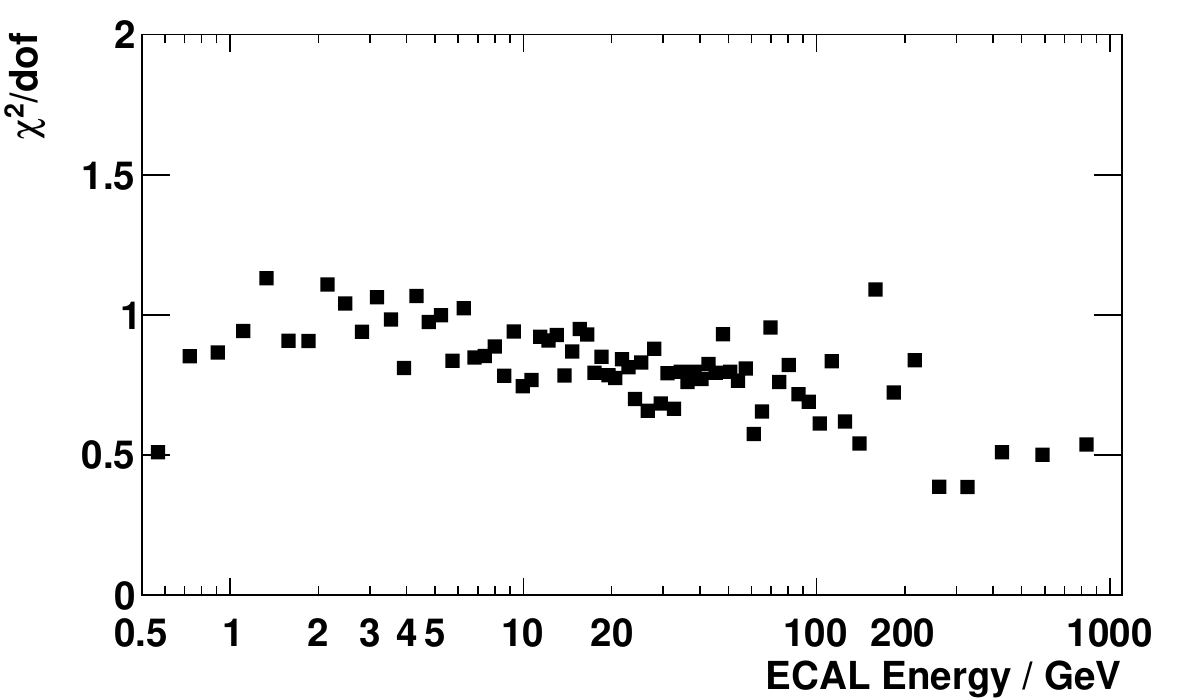}
  \caption{Goodness of fit of the model \cref{eq:trd-model-electrons} with respect to the template data in the single-track sample. The black points show the $\chi^2/\text{dof}$ as function of energy. The result is consistent with 1 for all energies for the single-track sample.}
  \label{fig:trd-template-parameters-electrons-chisquare-single-track-vs-energy}
\end{figure}

The $\chi^2/\text{dof}$ of the fit for the single-track sample, as shown in \cref{fig:trd-template-parameters-electrons-chisquare-single-track-vs-energy}, is consistent with 1 over the whole energy range.

The results for the multi-tracks sample are presented in \cref{sec:appendix-trd-templates}. The conclusions are identical as for the single-track sample.

\begin{figure}[H]
  \begin{subfigure}{0.48\linewidth}
    \includegraphics[width=\linewidth]{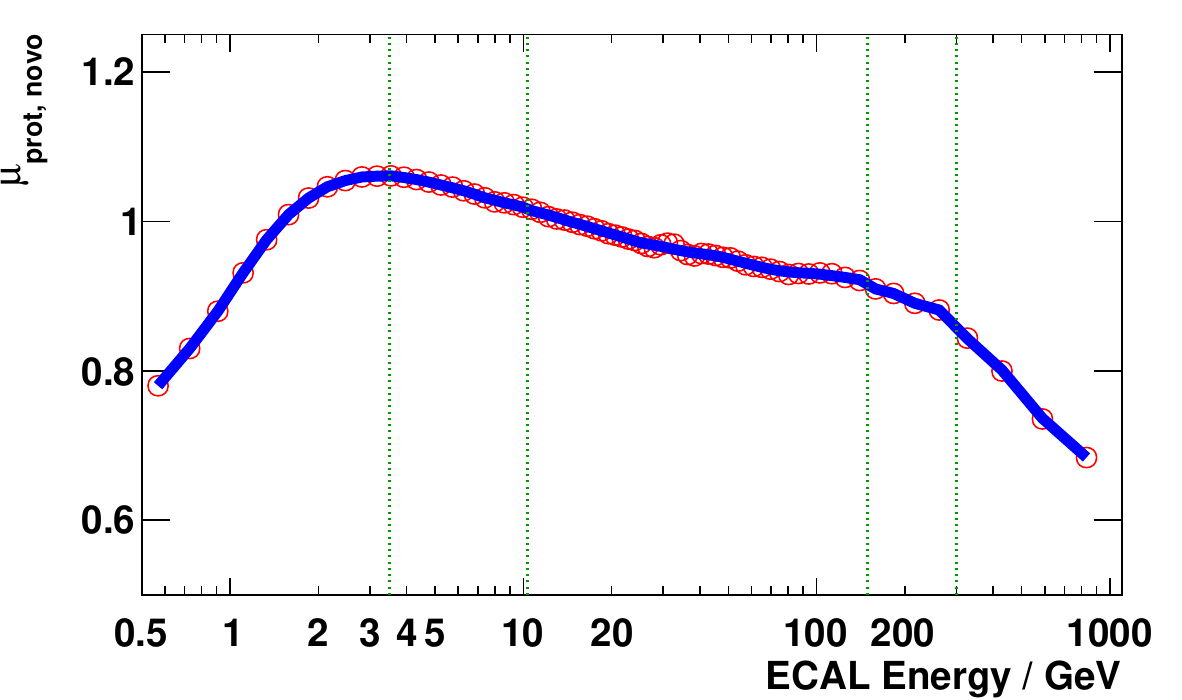}
    \caption{Template parameter~\protPeakNovo}
  \end{subfigure}
  \hfill
  \begin{subfigure}{0.48\linewidth}
    \includegraphics[width=\linewidth]{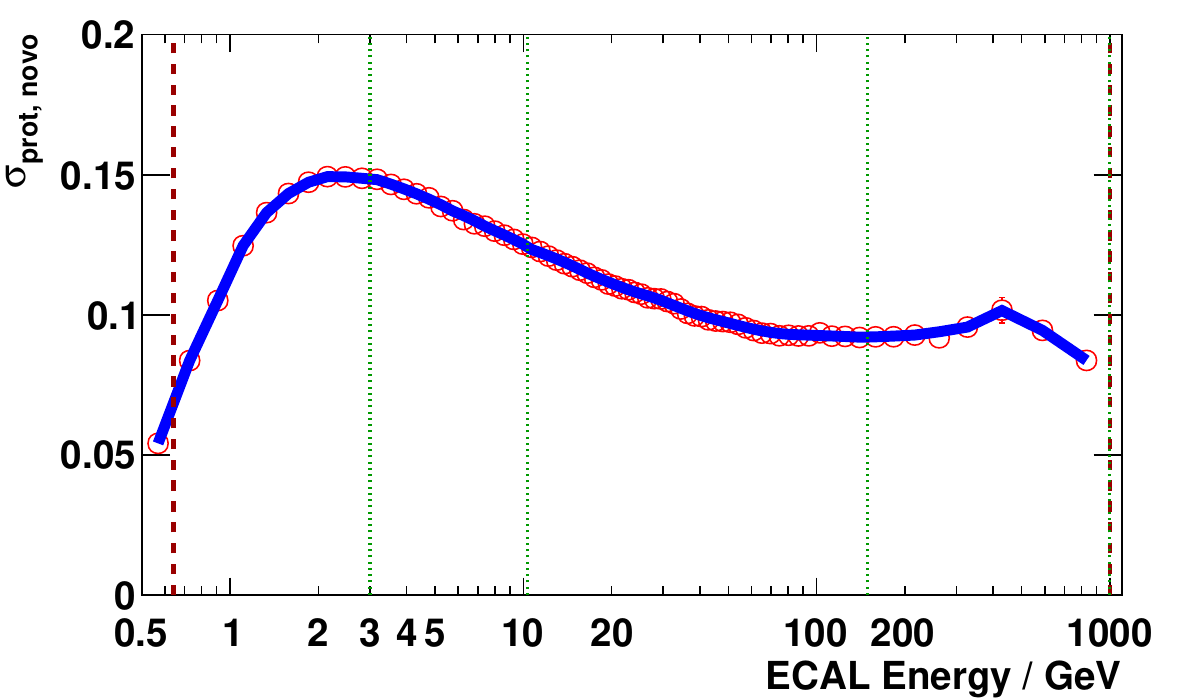}
    \caption{Template parameter~\protWidthNovo}
  \end{subfigure}
  \hfill
  \begin{subfigure}{0.48\linewidth}
    \includegraphics[width=\linewidth]{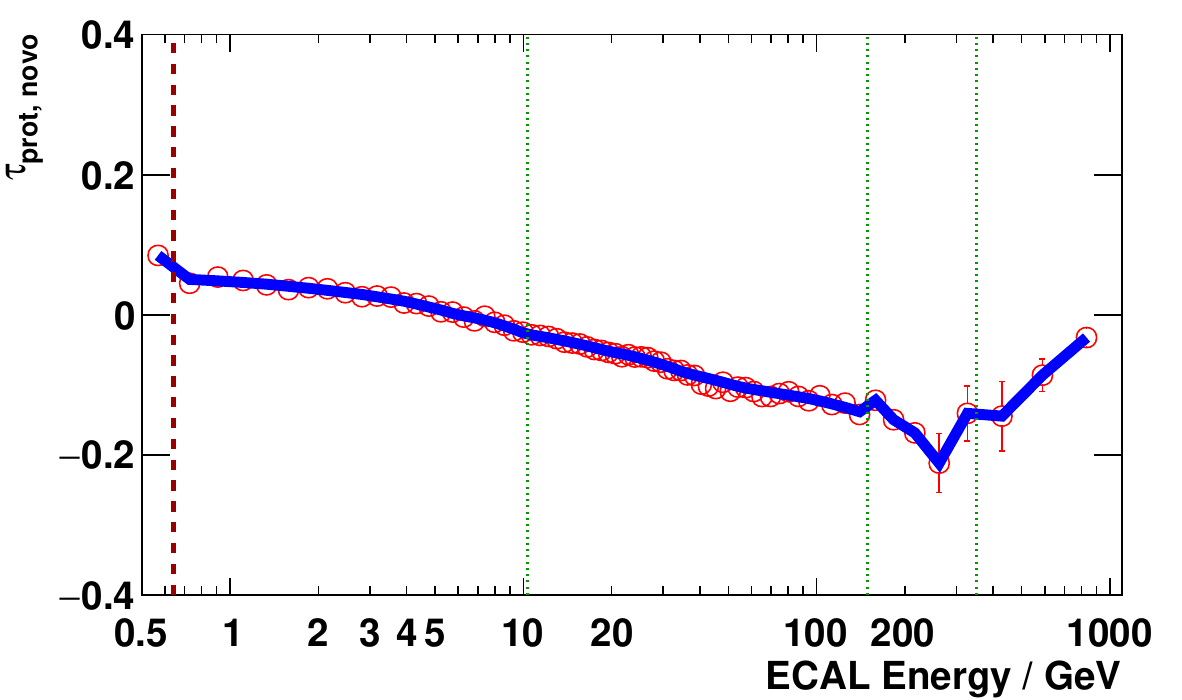}
    \caption{Template parameter~\protTailNovo}
  \end{subfigure}
  \hfill
  \begin{subfigure}{0.48\linewidth}
    \includegraphics[width=\linewidth]{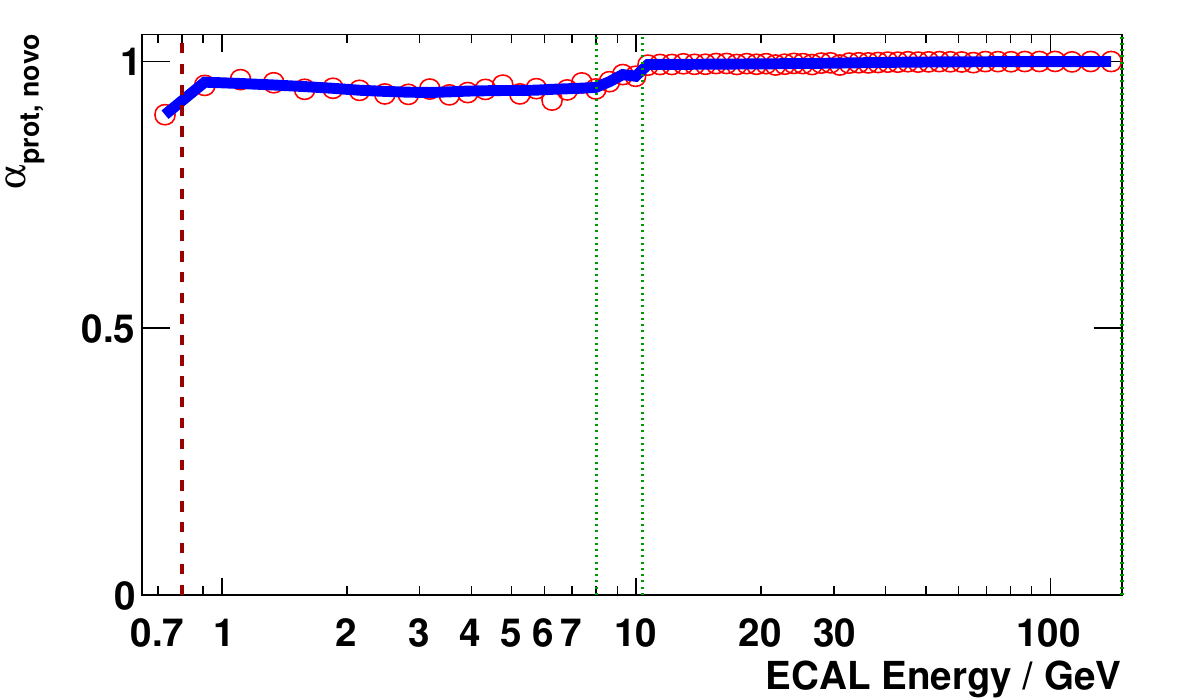}
    \caption{Template parameter~\protFractionNovo}
    \label{fig:trd-template-parameters-protons-protfractionnovo-single-track-vs-energy}
  \end{subfigure}
  \hfill
  \begin{subfigure}{0.48\linewidth}
    \includegraphics[width=\linewidth]{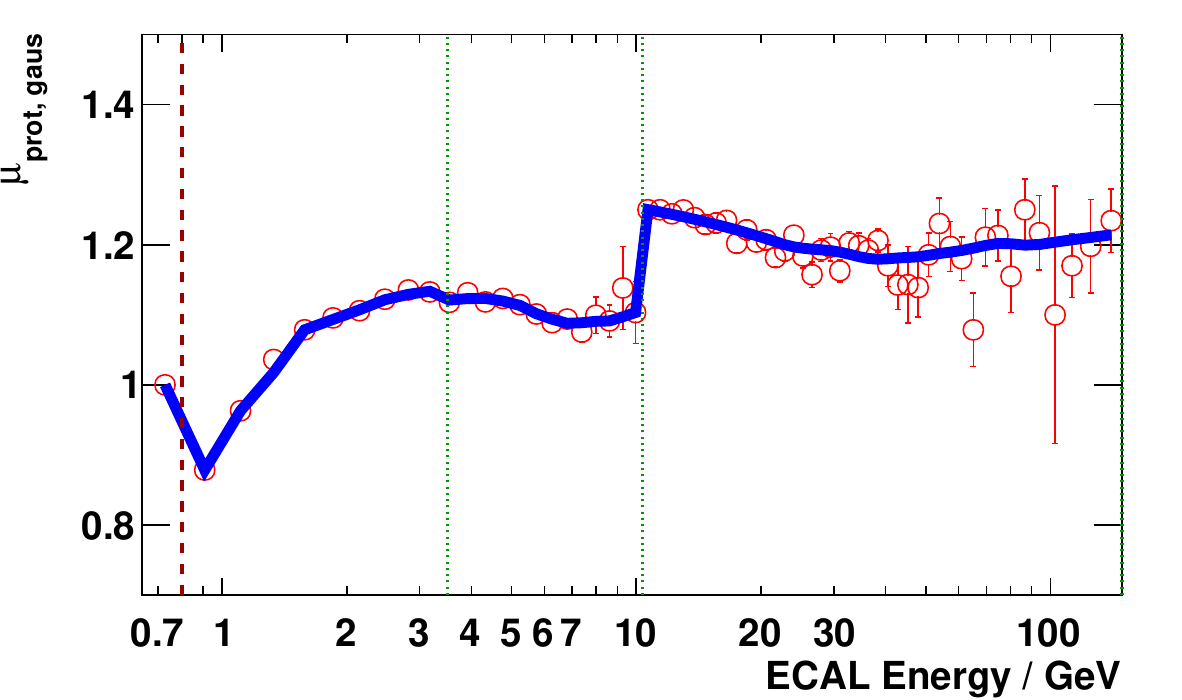}
    \caption{Template parameter~\protPeakGaus}
    \label{fig:trd-template-parameters-protons-protpeakgaus-single-track-vs-energy}
  \end{subfigure}
  \hfill
  \begin{subfigure}{0.48\linewidth}
    \includegraphics[width=\linewidth]{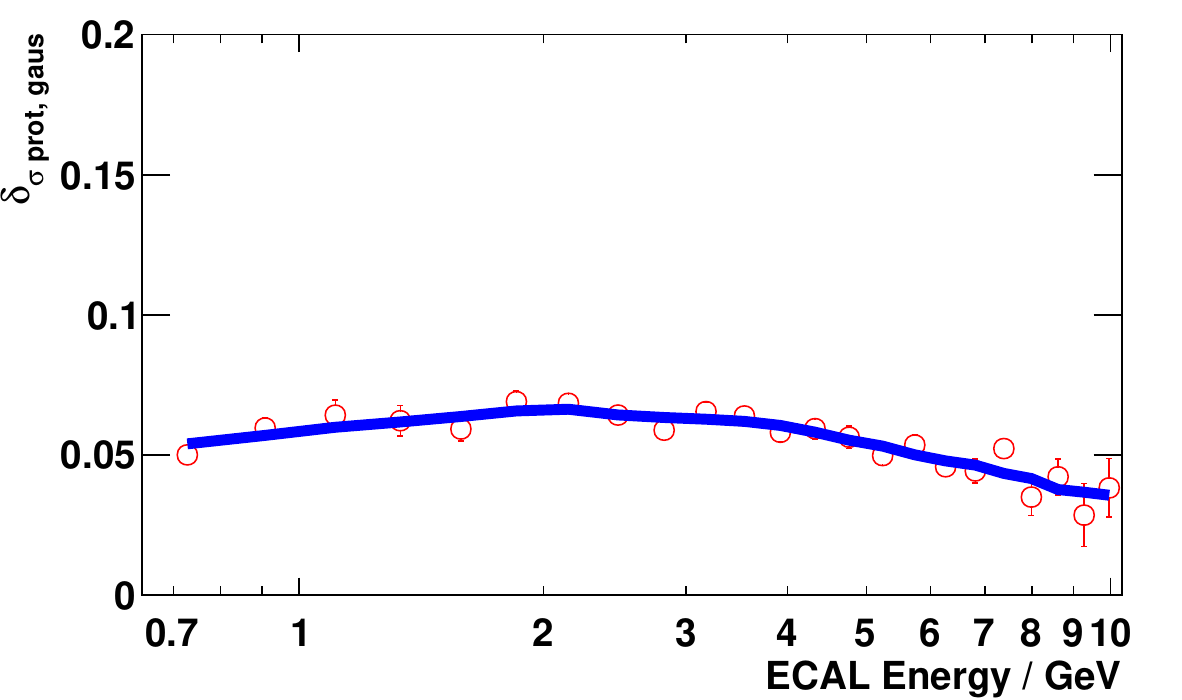}
    \caption{Template parameter~\protWidthGausDelta}
    \label{fig:trd-template-parameters-protons-protwidthgausdelta-single-track-vs-energy}
  \end{subfigure}
  \caption{Analytical TRD template parameters describing the evolution of the proton template as function of energy for the single-track sample. In each energy bin the analytical function -~\cref{eq:trd-model-protons}~- is fit to the template data sample, yielding the red points. A smoothing procedure yields the blue curves, which are used as template parameters for the analysis. The vertical dashed green lines mark areas where different isolated smoothing procedures are applied. The vertical dashed red lines mark energies where the smoothing procedure starts/ends -- points before/after this border are preserved.}
  \label{fig:trd-template-parameters-protons-single-track-vs-energy}
\end{figure}

\clearpage
An analogous smoothing procedure is applied for the proton template, as shown in \cref{fig:trd-template-parameters-protons-single-track-vs-energy} for the single-track sample.
In total six parameters are relevant to describe the template analytically for all energies: three parameters corresponding to the
Novosibirsk function in the proton model (peak position~\protPeakNovo, width~\protWidthNovo~and tail parameter~\protTailNovo), two for
the gaussian part (peak position~\protPeakGaus, width delta~\protWidthGausDelta) and a ratio describing the relative contribution of the
Novosibirsk needed to fit the template (fraction~\protFractionNovo), corresponding to \cref{eq:trd-model-protons}.

The two template parameters~\protFractionNovo~and~\protPeakGaus~presented in \cref{fig:trd-template-parameters-protons-protfractionnovo-single-track-vs-energy,fig:trd-template-parameters-protons-protpeakgaus-single-track-vs-energy} are only relevant below \SI{110}{\GeV}. Above this energy threshold the proton template can be described with a single Novosibirsk, without a gaussian component.

The template parameter~\protWidthGausDelta~shown in \cref{fig:trd-template-parameters-protons-protwidthgausdelta-single-track-vs-energy} mainly improves the goodness-of-fit of the fit procedure at low energies and remains constant zero above \SI{10}{\GeV} for the single-track sample. For the multi-tracks sample the parameter is absent because its introduction lead to numerical instabilities in the fit procedure.

\begin{figure}[H]
  \centering
  \includegraphics[width=0.7\linewidth]{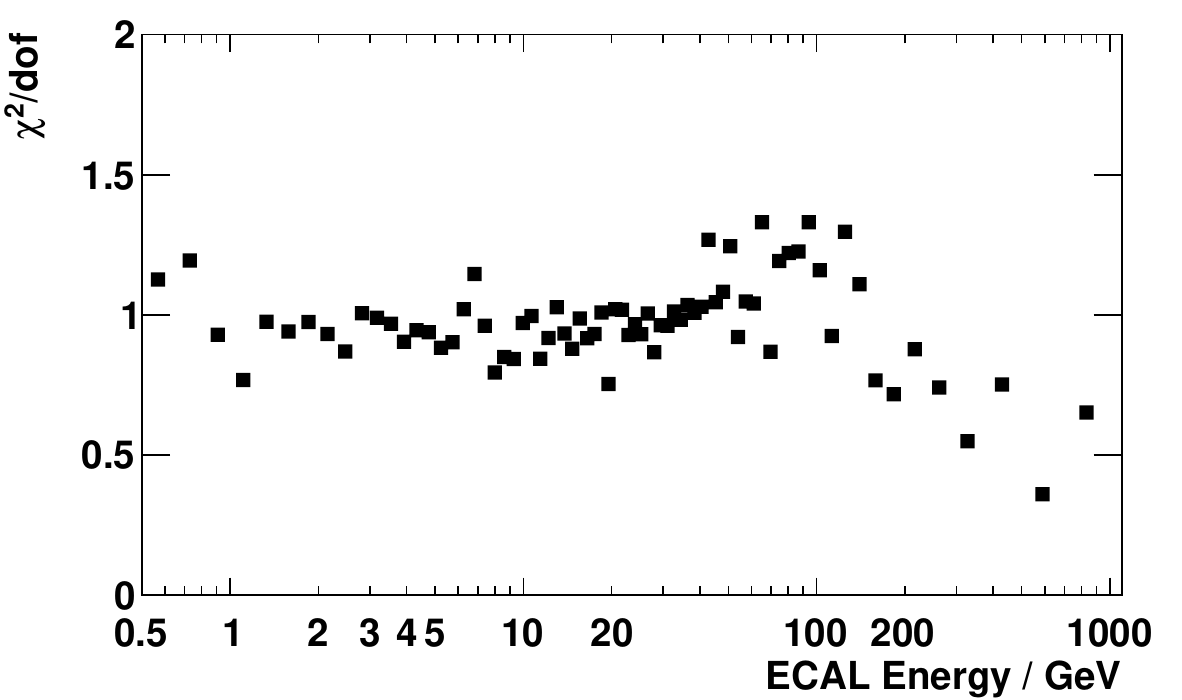}
  \caption{Goodness of fit of the model \cref{eq:trd-model-protons} with respect to the template data in the single-track sample. The black points show the $\chi^2/\text{dof}$ as function of energy. The result is consistent with 1 for all energies for the single-track sample.}
  \label{fig:trd-template-parameters-protons-chisquare-single-track-vs-energy}
\end{figure}

The $\chi^2/\text{dof}$ of the fit, \cref{fig:trd-template-parameters-protons-chisquare-single-track-vs-energy}, is consistent with 1 over the whole energy range.

The results for the multi-tracks sample are presented in \cref{sec:appendix-trd-templates}. The conclusions are identical as for the single-track sample.

\bigskip
The set of parameters presented in \cref{fig:trd-template-parameters-ccprotons-single-track-vs-energy} for the charge-confused protons,
\cref{fig:trd-template-parameters-electrons-single-track-vs-energy} for the electrons and \cref{fig:trd-template-parameters-protons-single-track-vs-energy} for the protons
for the single-track sample and for the multi-tracks sample (\cref{fig:trd-template-parameters-ccprotons-multi-tracks-vs-energy,fig:trd-template-parameters-electrons-multi-tracks-vs-energy,fig:trd-template-parameters-protons-multi-tracks-vs-energy}) form a fully analytical description of the TRD templates as function of energy.
This is a major building block in the electron and positron analysis.

\subsection{Charge-confusion templates}
\label{sec:analysis-lepton-counts-ccmva-templates}

The CCMVA estimator $\Lambda_{\text{CC}}$, described in \cref{sec:analysis-event-reconstruction-construction-ccmva-estimator}, will be used together with the
TRD estimator $\Lambda_{\text{TRD}}$ to perform two-dimensional template fits on ISS data to extract the number of electrons and positrons in
different energy intervals on ISS data, taking the charge-confusion effect into account.

To perform the two-dimensional template fit, templates for protons, charge-confused protons, electrons, positrons, charge-confused
electrons and charge-confused positrons need to be known in two variants: for the single-track sample and for the multi-tracks sample. The distinction between the two
different track topologies is equally important as for the TRD estimator, as the amount of charge-confusion in both samples is vastly different.
Another important difference with respect to the TRD estimator is that the CCMVA estimator shape cannot be derived from first principles, thus no
physically motivated analytical description of the shape is available. In order to construct templates either histograms containing the output of
the CCMVA distribution will be used, if the statistics are sufficient, or analytical approximations, by utilizing methods such
as \gls{KDE}~\cite{Rosenblatt1956,Parzen1962}.

The electron, positron, proton and charge-confused proton templates can be extracted from ISS data. The charge-confused electron and positron template can only be derived
from the Monte-Carlo simulation, as there is no way to select a pure sample of charge-confused electrons or charge-confused positrons on ISS data. Charge-confused protons
are selectable on ISS data, as the amount of antiprotons that could be misreconstructed as charge-confused protons is negligible. The charge-confused electron sample
on ISS data would be dominated by correctly reconstructed positrons plus an additional contribution from charge-confused electrons. This ambiguity forces one to use
the Monte-Carlo simulation to extract the charge-confused electron and charge-confused positron templates. The positron and charge-confused positron templates are identical to their
electron counterparts as both particles behave in the same way with respect to the CCMVA estimator, which was verified using dedicated electron and positron
Monte-Carlo simulations.

The cuts to prepare the template samples are similar as for the TRD estimator, presented in \cref{sec:analysis-lepton-counts-trd-templates}.
All detector quality, preselection and selection cuts - \cref{sec:analysis-data-selection-detector-quality-cuts,sec:analysis-data-selection-preselection-cuts,sec:analysis-data-selection-selection-cuts} - are applied when selecting
the data samples. Furthermore the same 0.5 < $E/\abs{R}$ < 1 cut as for the signal event selection is applied.
The following enumeration lists the sample specific cuts:

\begin{enumerate}
  \item\textbf{Proton sample}\hfill
    \begin{itemize}
      \item ISS data, positive rigidity sample (R > 0)
      \item Hadron-like ECAL shower ($\Lambda_{\text{ECAL}}$ < 0)
      \item TRD electron/proton likelihood ratio $\Lambda_{\text{TRD}}$ > 0.8
    \end{itemize}

  \item\textbf{Charge-confused proton sample}\hfill
    \begin{itemize}
      \item ISS data, negative rigidity sample (R < 0)
      \item Hadron-like ECAL shower ($\Lambda_{\text{ECAL}}$ < 0)
      \item TRD electron/proton likelihood ratio $\Lambda_{\text{TRD}}$ > 0.8
    \end{itemize}

  \item\textbf{Electron sample}\hfill
    \begin{itemize}
      \item ISS data, negative rigidity sample (R < 0)
      \item Energy dependent cut on the lateral shape
            (defined in \cref{sec:analysis-data-selection-electron-positron-identification-cuts} - \cref{enum:electron-positron-identification-cut-ecal-lateral-shower-shape})
      \item Electron like ECAL shower ($\Lambda_{\text{ECAL}}$ > 0)
      \item TRD electron/proton likelihood ratio $\Lambda_{\text{TRD}}$ < 0.75
    \end{itemize}

  \item\textbf{Charge-confused electron sample}\hfill
    \begin{itemize}
      \item Electron Monte-Carlo simulation, positive rigidity sample (R > 0)
      \item Energy dependent cut on the lateral shape
            (defined in \cref{sec:analysis-data-selection-electron-positron-identification-cuts} - \cref{enum:electron-positron-identification-cut-ecal-lateral-shower-shape})
      \item Electron like ECAL shower ($\Lambda_{\text{ECAL}}$ > 0)
      \item TRD electron/proton likelihood ratio $\Lambda_{\text{TRD}}$ < 0.75
    \end{itemize}
\end{enumerate}

\begin{figure}[H]
  \begin{subfigure}{0.49\linewidth}
    \includegraphics[width=\linewidth]{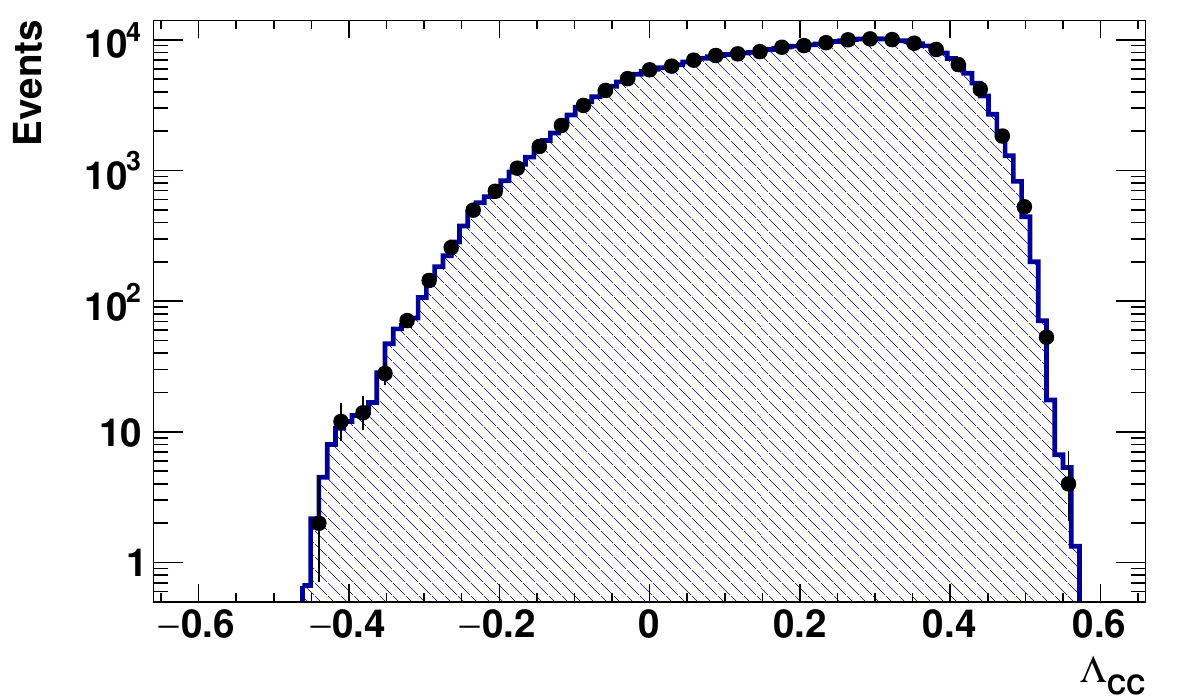}
    \caption{Electrons (single-track sample)}
  \end{subfigure}
  \hfill
  \begin{subfigure}{0.49\linewidth}
    \includegraphics[width=\linewidth]{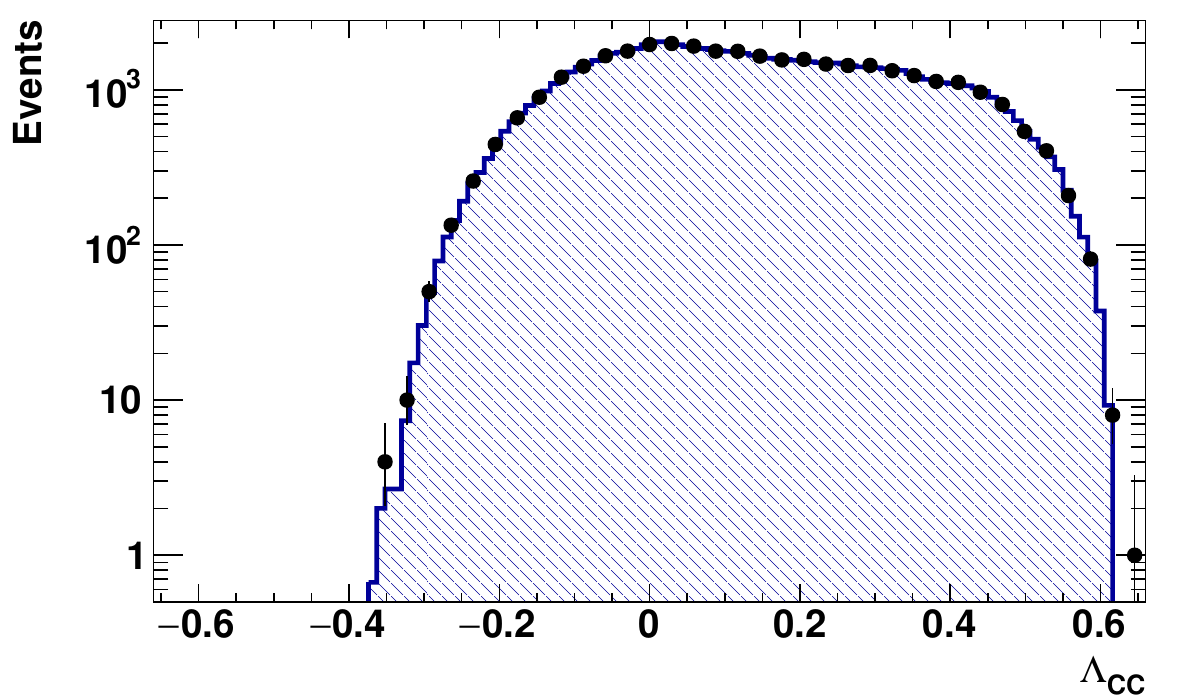}
    \caption{Electrons (multi-tracks sample)}
  \end{subfigure}
  \hfill
  \begin{subfigure}{0.49\linewidth}
    \includegraphics[width=\linewidth]{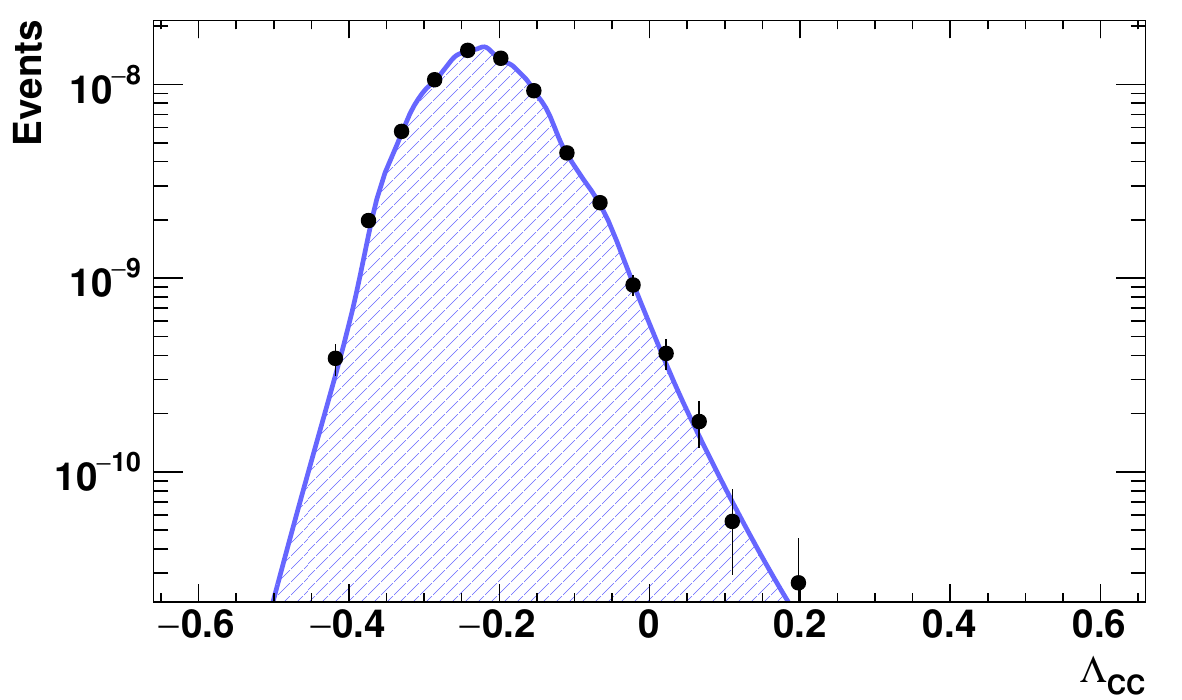}
    \caption{Charge-confused electrons (single-track sample)}
  \end{subfigure}
  \hfill
  \begin{subfigure}{0.49\linewidth}
    \includegraphics[width=\linewidth]{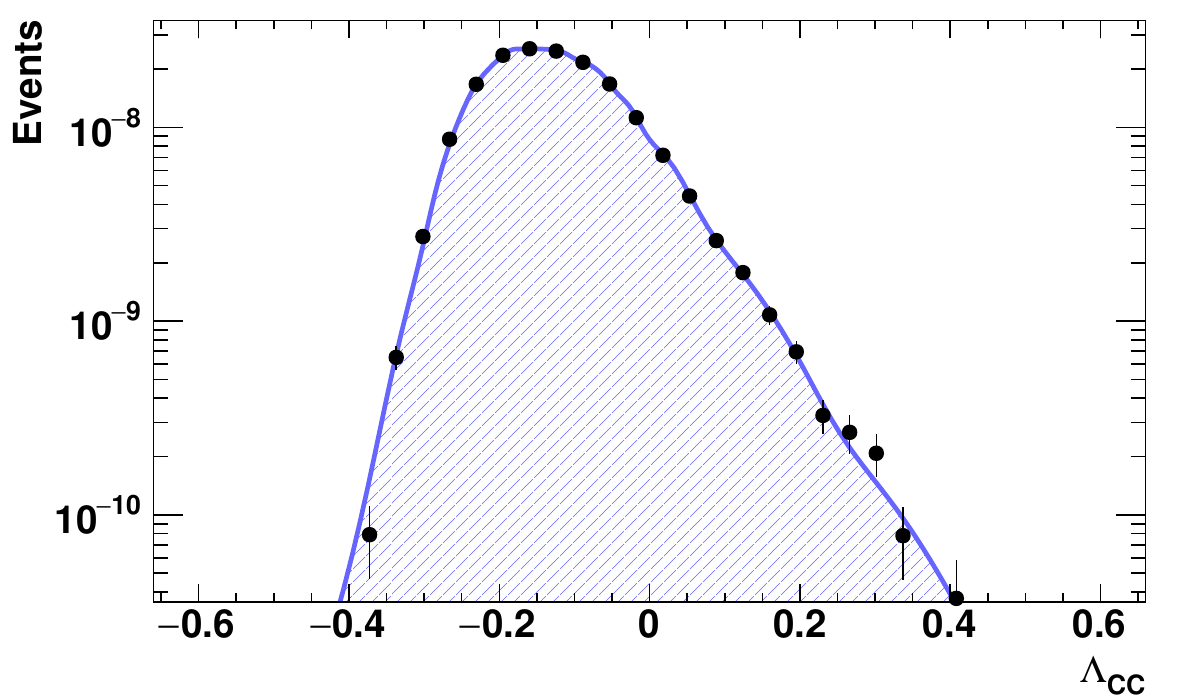}
    \caption{Charge-confused electrons (multi-tracks sample)}
  \end{subfigure}
  \begin{subfigure}{0.49\linewidth}
    \includegraphics[width=\linewidth]{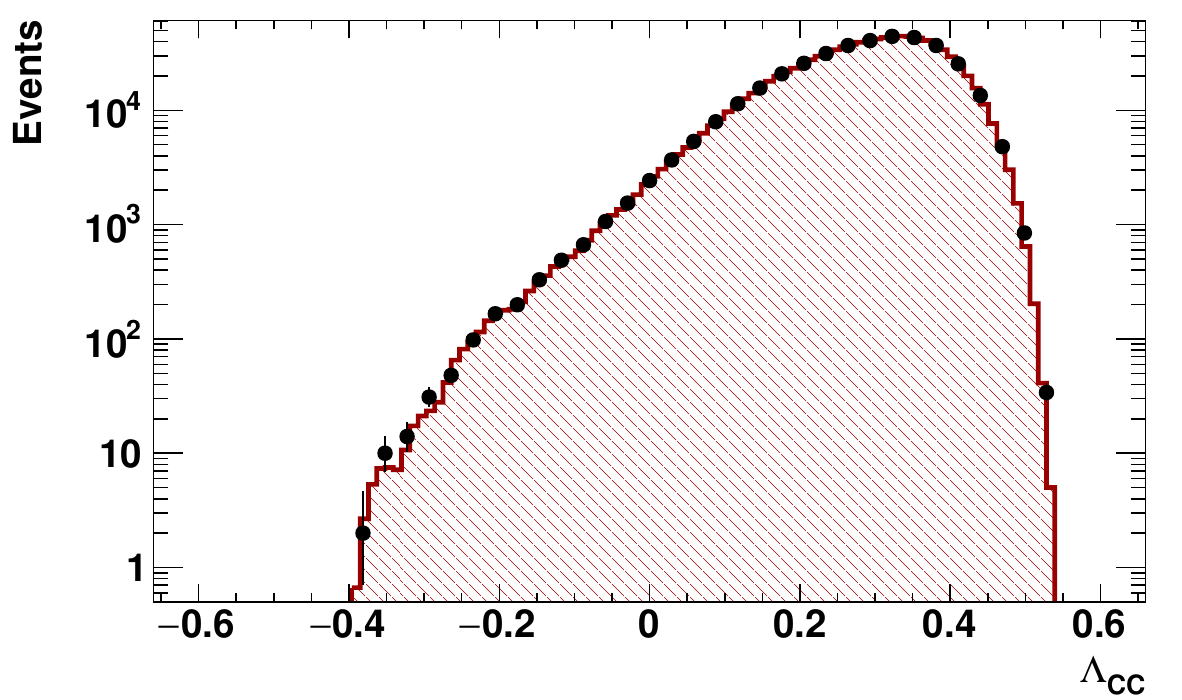}
    \caption{Protons (single-track sample)}
  \end{subfigure}
  \hfill
  \begin{subfigure}{0.49\linewidth}
    \includegraphics[width=\linewidth]{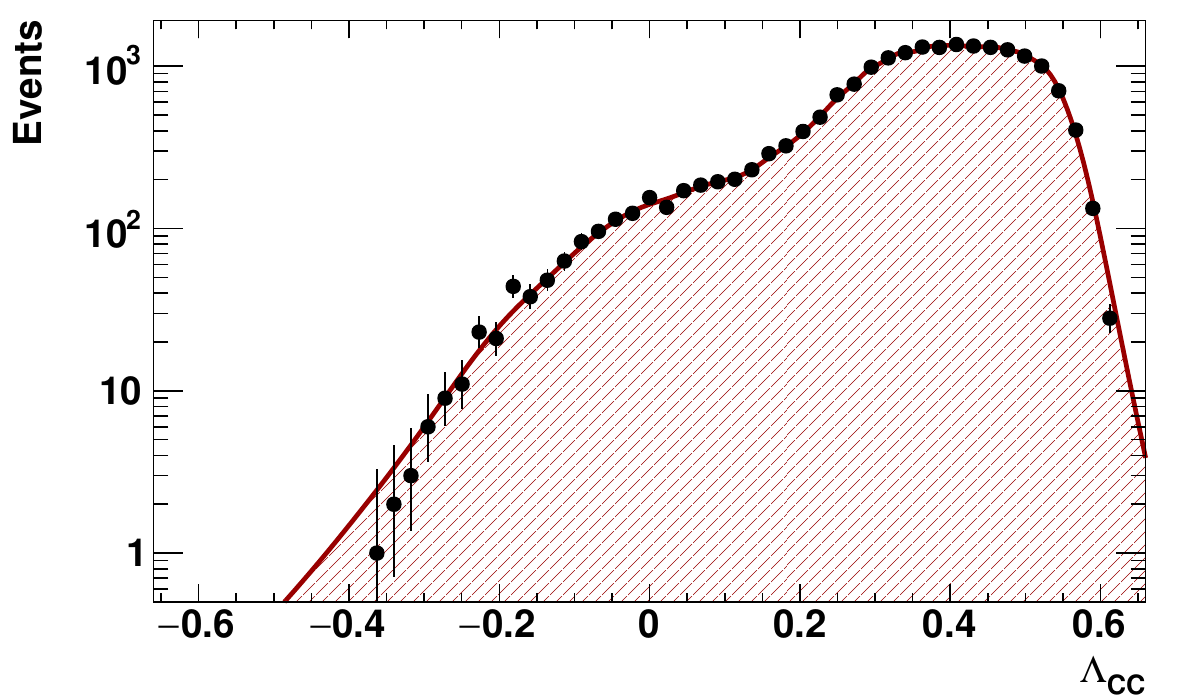}
    \caption{Protons (multi-tracks sample)}
  \end{subfigure}
  \hfill
  \begin{subfigure}{0.49\linewidth}
    \includegraphics[width=\linewidth]{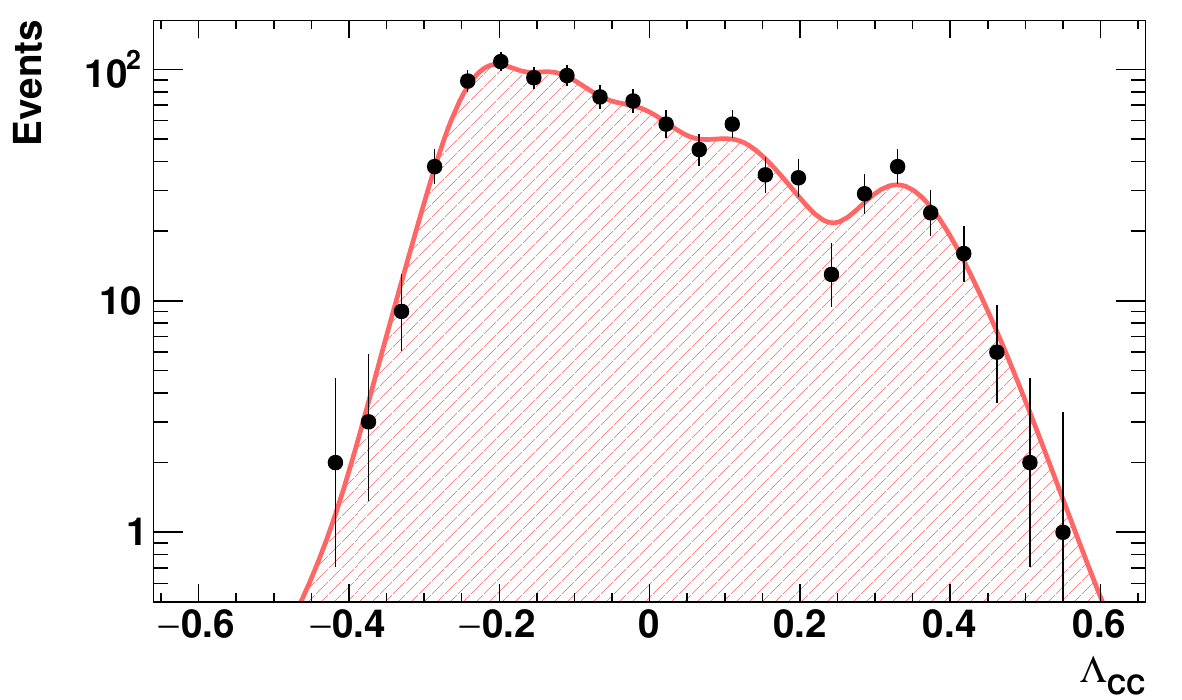}
    \caption{Charge-confused protons (single-track sample)}
  \end{subfigure}
  \hfill
  \begin{subfigure}{0.49\linewidth}
    \includegraphics[width=\linewidth]{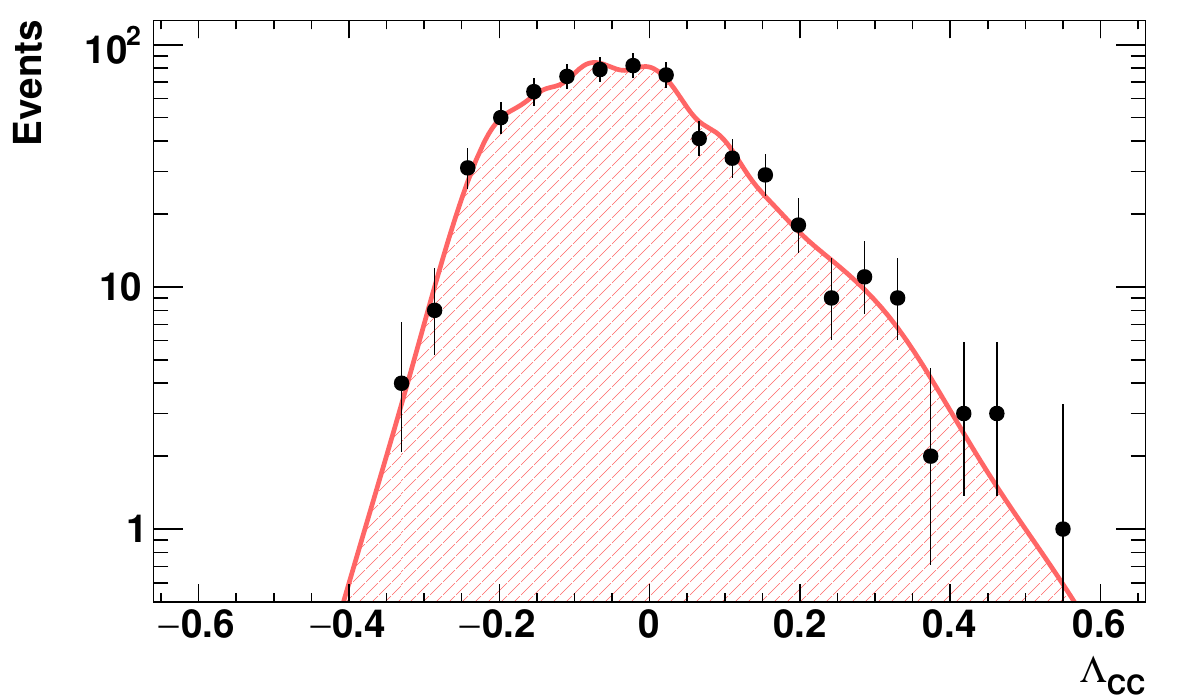}
    \caption{Charge-confused protons (multi-tracks sample)}
  \end{subfigure}
  \caption{Overview of the CCMVA templates in the energy bin \SIrange{17.98}{18.99}{\GeV}: electrons (\elecColorText~area), charge-confused electrons (\ccElecColorText~area), protons (\protColorText~area) and charge-confused protons (\ccProtColorText~area). The black points show the CCMVA estimator output, either obtained from electron Monte-Carlo simulation or ISS data - depending on the data sample.}
  \label{fig:ccmva-template-parameters}
\end{figure}

\Cref{fig:ccmva-template-parameters} shows the electron template, the charge-confused electron template, the proton template and the
charge-confused proton template in an example energy bin. Depending on the statistics in the examined data sample, either a histogram is
used as template PDF, or if the statistics are too low, a \gls{KDE} technique is applied to extract an analytical approximation as template PDF.
This is necessary to avoid the propagation of statistical fluctuations from the template data samples to the template PDF.

The procedure to extract the charge-confusion templates is repeated for all energy bins in the analysis.

The positron and charge-confused positron templates are exactly identical to the electron and charge-confused electron templates by construction.
This was verified on ISS data up to the highest energies for the electron template and using positron Monte-Carlo simulation for the charge-confused electron and charge-confused positron template.

\subsection{Construction of two-dimensional TRD / CCMVA templates}
\label{sec:analysis-lepton-counts-construction-2d-trd-ccmva-estimator}

The previously constructed one-dimensional TRD and CCMVA templates can be combined into two-dimensional templates according to \cref{eq:2d-pdf-construction},
as they are assumed to be uncorrelated - which was verified in dedicated Monte-Carlo simulations. The correlation is negligible, as
the TRD estimator $\Lambda_{\text{TRD}}$ is computed using the ECAL energy as input, not the tracker rigidity.

\begin{equation}
  \label{eq:2d-pdf-construction}
  \mathcal{P}_{\text{2D}}(x, y) = \mathcal{P}_{\text{TRD}}(x) \cdot \mathcal{P}_{\text{CC}}(y)
\end{equation}

\begin{figure}[H]
  \begin{subfigure}{0.50\linewidth}
    \includegraphics[width=\linewidth]{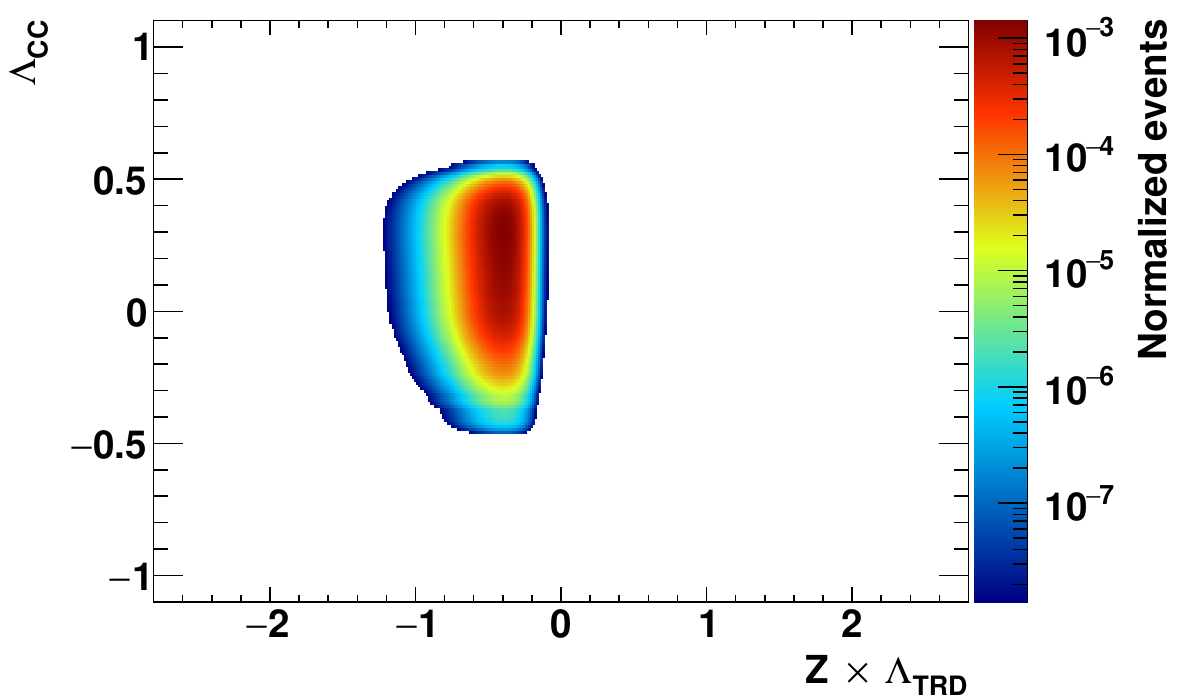}
    \caption{Electrons}
    \label{fig:2d-templates-single-track-electrons}
  \end{subfigure}
  \hfill
  \begin{subfigure}{0.50\linewidth}
    \includegraphics[width=\linewidth]{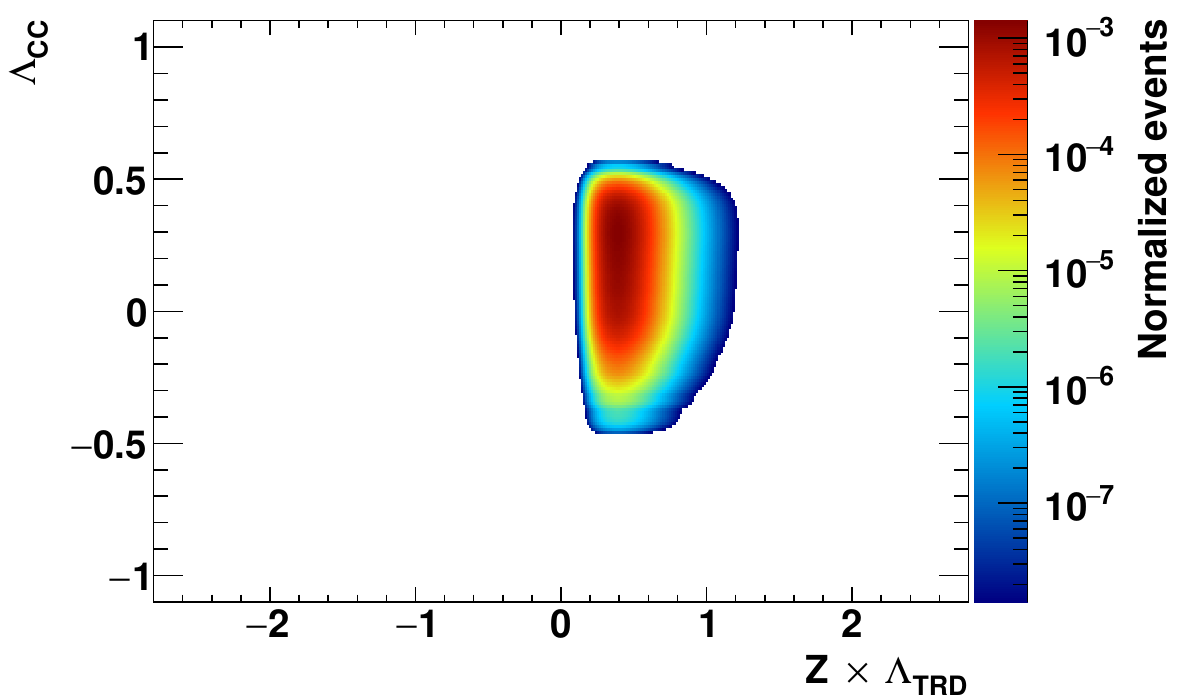}
    \caption{Positrons}
    \label{fig:2d-templates-single-track-positrons}
  \end{subfigure}
  \caption{Two-dimensional TRD / CCMVA templates in the energy bin \SIrange{17.98}{18.99}{\GeV} for the electrons and positrons in the single-track sample. The color encodes the amount of normalized entries in the histogram, a red color corresponds to a large number of events, a blue color to a small number of events.}
  \label{fig:2d-templates-single-track-electrons-positrons}
\end{figure}

\begin{figure}[H]
 \begin{subfigure}{0.50\linewidth}
    \includegraphics[width=\linewidth]{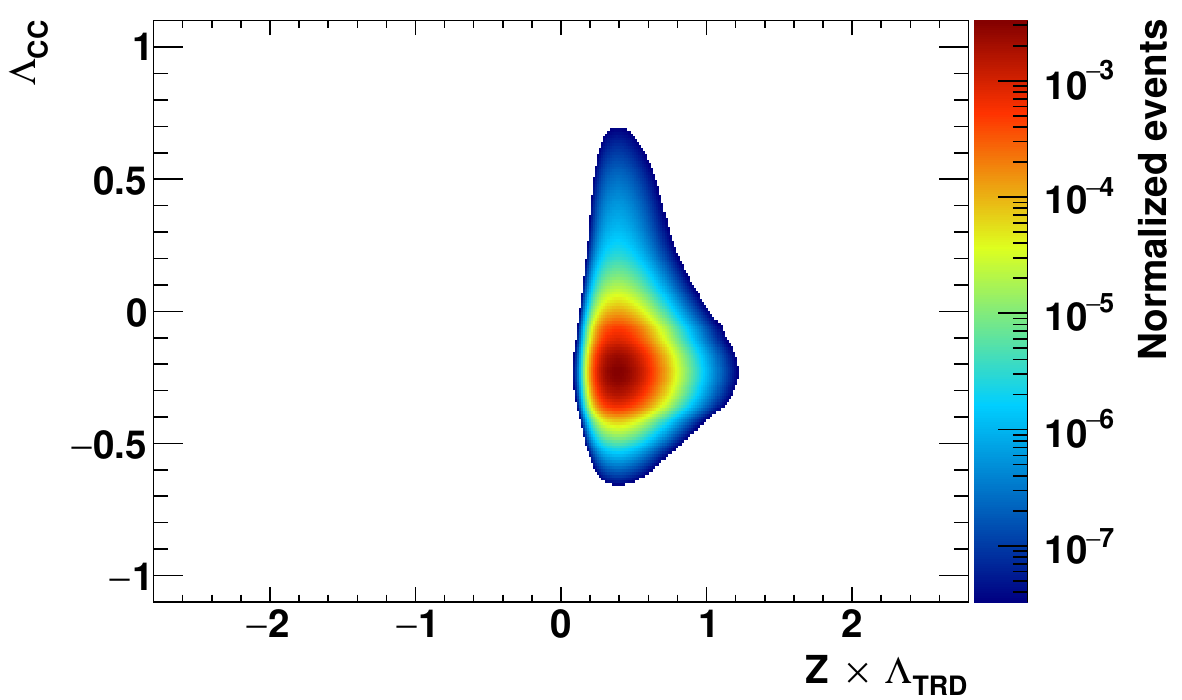}
    \caption{Charge-confused electrons}
    \label{fig:2d-templates-single-track-cc-electrons}
  \end{subfigure}
  \hfill
  \begin{subfigure}{0.50\linewidth}
    \includegraphics[width=\linewidth]{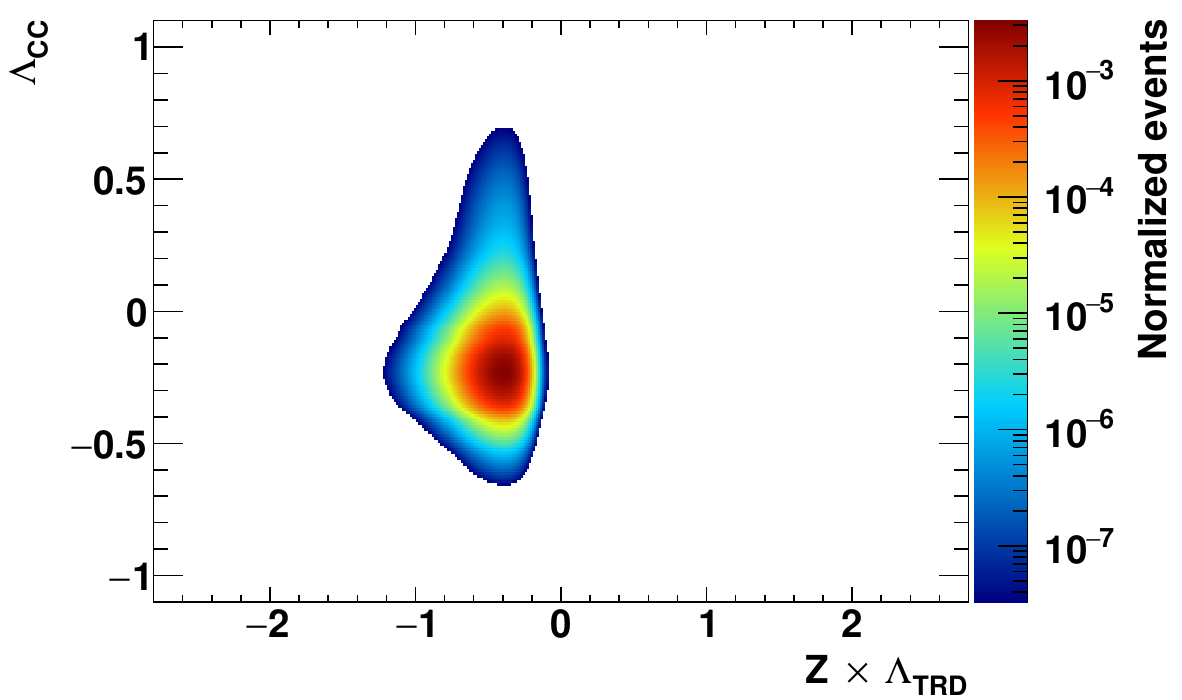}
    \caption{Charge-confused positrons}
    \label{fig:2d-templates-single-track-cc-positrons}
  \end{subfigure}
  \caption{Two-dimensional TRD / CCMVA templates in the energy bin \SIrange{17.98}{18.99}{\GeV} for the charge-confused electrons and charge-confused positrons in the single-track sample. The color encodes the amount of normalized entries in the histogram, a red color corresponds to a large number of events, a blue color to a small number of events.}
  \label{fig:2d-templates-single-track-cc-electrons-cc-positrons}
\end{figure}

\begin{figure}[H]
  \begin{subfigure}{0.50\linewidth}
    \includegraphics[width=\linewidth]{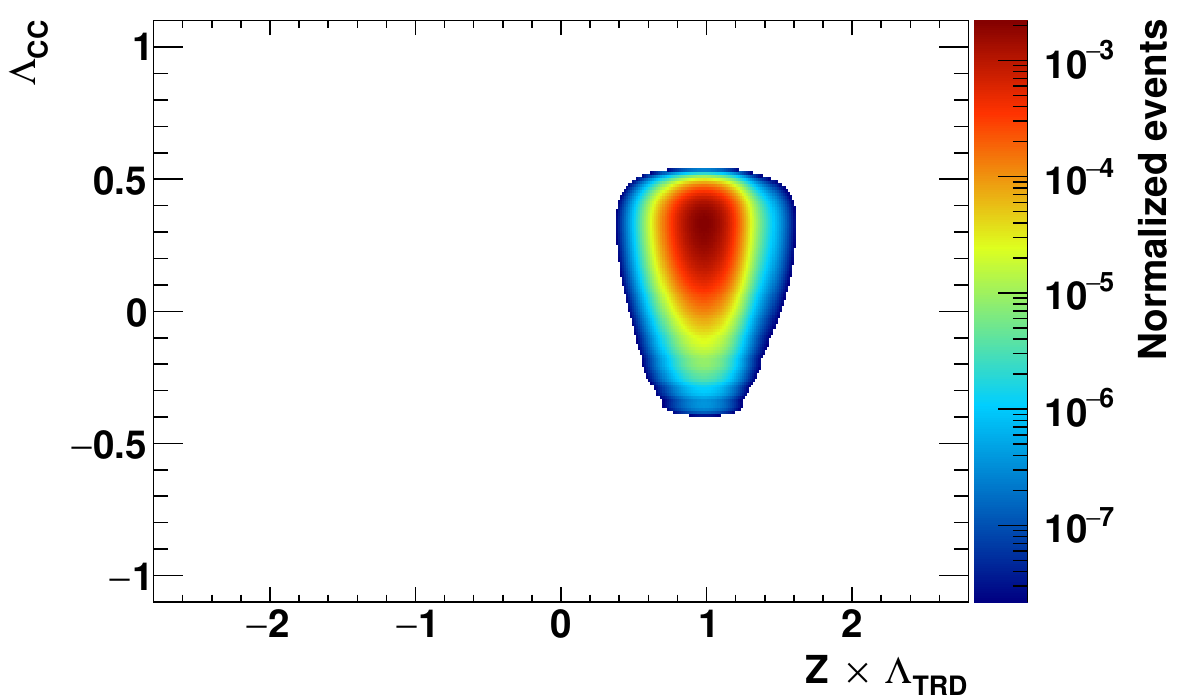}
    \caption{Protons}
  \end{subfigure}
  \hfill
  \begin{subfigure}{0.50\linewidth}
    \includegraphics[width=\linewidth]{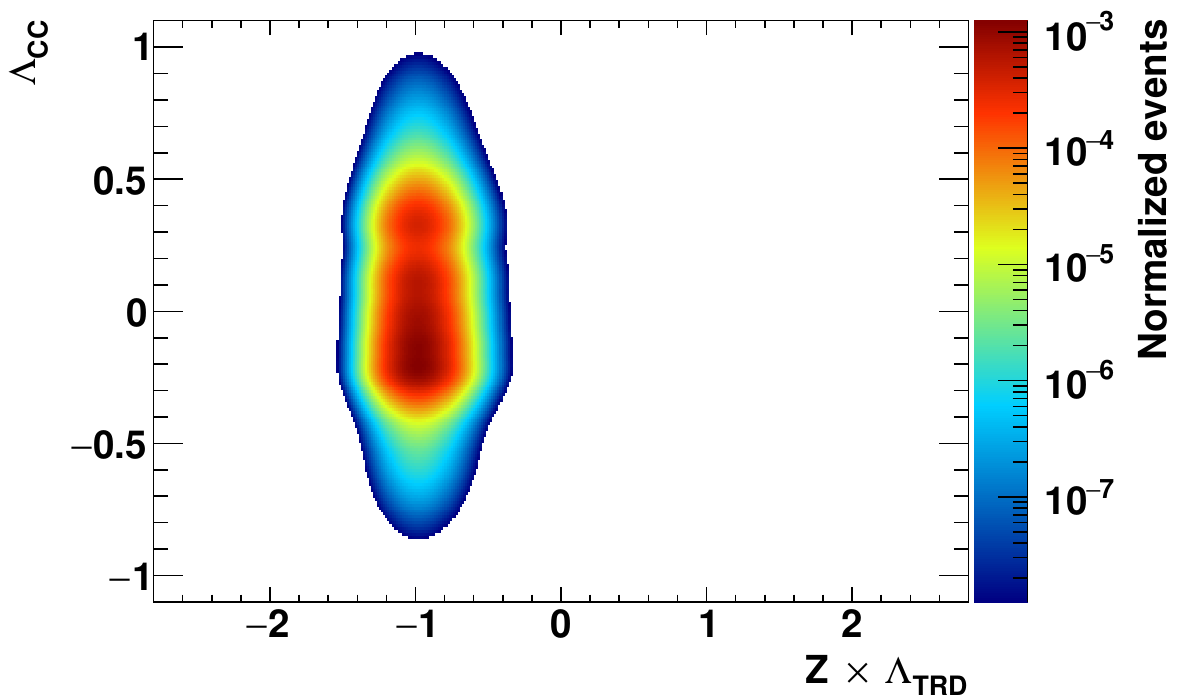}
    \caption{Charge-confused protons}
  \end{subfigure}
  \hfill
  \caption{Two-dimensional TRD / CCMVA templates in the energy bin \SIrange{17.98}{18.99}{\GeV} for the protons and charge-confused protons in the single-track sample. The color encodes the amount of normalized entries in the histogram, a red color corresponds to a large number of events, a blue color to a small number of events.}
  \label{fig:2d-templates-single-track-protons-cc-protons}
\end{figure}

\Cref{fig:2d-templates-single-track-electrons-positrons,fig:2d-templates-single-track-cc-electrons-cc-positrons,fig:2d-templates-single-track-protons-cc-protons}
shows the two-dimensional templates in an example energy bin for the single-track sample. The same overview is presented in \cref{fig:2d-templates-multi-tracks}
for the multi-tracks sample. The TRD component is only slightly different compared to the single-track sample, whereas the CCMVA distribution is vastly different,
as the response of the CCMVA is different for the multi-tracks sample.

The two-dimensional positron template (\cref{fig:2d-templates-single-track-positrons}) is constructed by mirroring the two-dimensional
electron template (\cref{fig:2d-templates-single-track-electrons}) around the y-axis.

Analogous the charge-confused positron template (\cref{fig:2d-templates-single-track-cc-positrons}) is constructed by mirroring the
charge-confused electron template (\cref{fig:2d-templates-single-track-cc-electrons}).

The construction of the two-dimensional templates is repeated for each energy interval in the analysis. These templates will be used
in a fit procedure to extract the number of electrons and positrons in each energy interval, which will be shown in the next section.

\clearpage
\subsection{Two-dimensional template fit}
\label{sec:analysis-lepton-counts-2d-fit}

\begin{figure}[H]
  \begin{subfigure}{0.50\linewidth}
    \includegraphics[width=\linewidth]{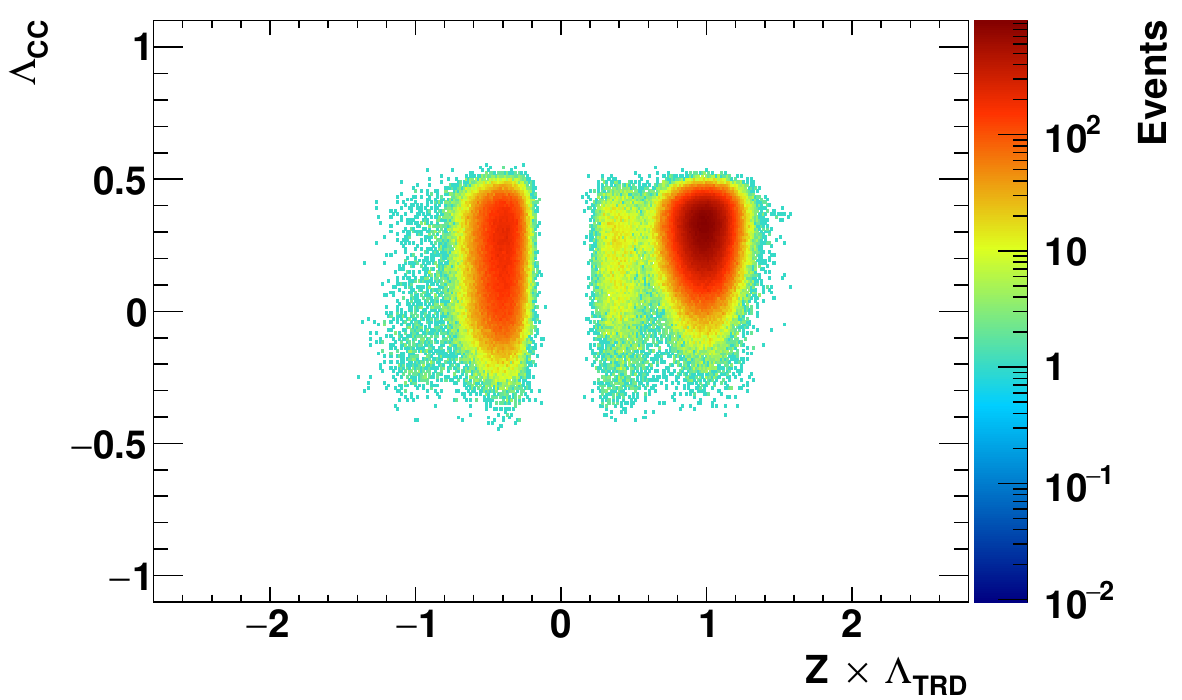}
    \caption{Before cut on ECAL estimator}
    \label{fig:2d-iss-data-single-track-before-ecal-cut}
  \end{subfigure}
  \hfill
  \begin{subfigure}{0.50\linewidth}
    \includegraphics[width=\linewidth]{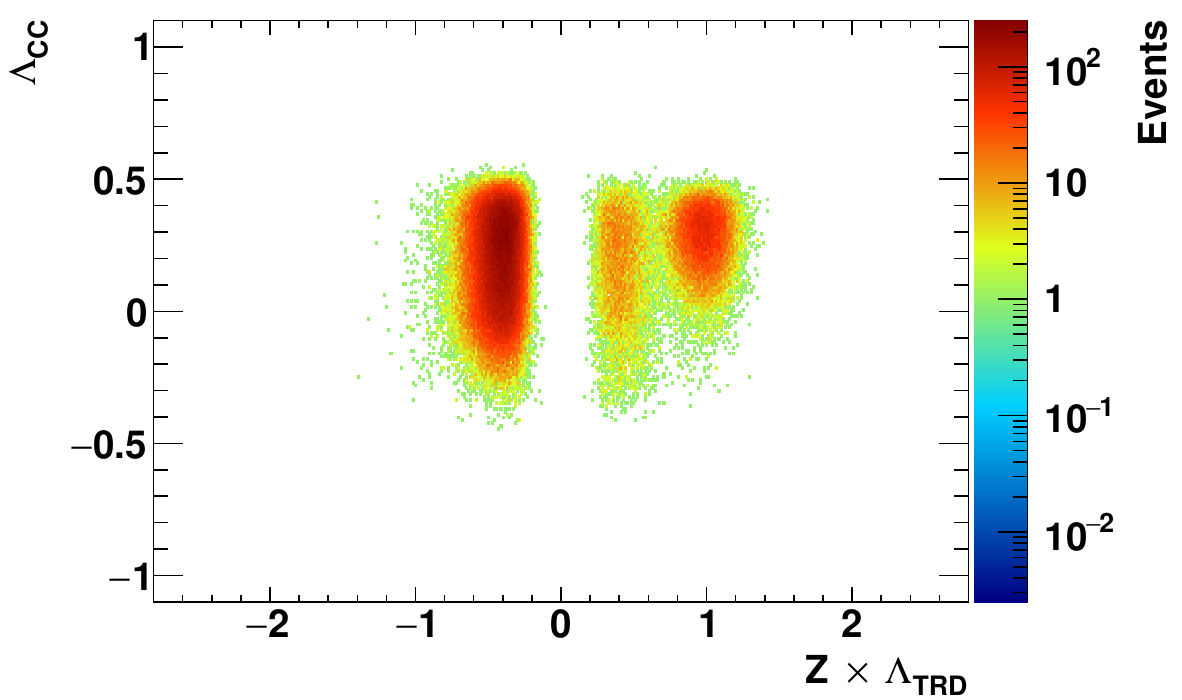}
    \caption{After cut on ECAL estimator}
    \label{fig:2d-iss-data-single-track-after-ecal-cut}
  \end{subfigure}
  \caption{ISS data distribution in the energy bin \SIrange{17.98}{18.99}{\GeV} for the single-track sample, as example, before and after applying a cut on the ECAL estimator to reduce the proton background.}
  \label{fig:2d-iss-data-single-track-before-after-ecal-cut}
\end{figure}

\Cref{fig:2d-iss-data-single-track-before-ecal-cut} shows a typical ISS data distribution in an example energy bin, \SIrange{17.98}{18.99}{\GeV},
after applying all data quality, preselection, selection and $e^{\pm}$ identification cuts (\cref{sec:analysis-data-selection}).
\Cref{fig:2d-iss-data-single-track-after-ecal-cut} shows the same data sample, after applying a cut on the ECAL estimator $\Lambda_{\text{ECAL}}$ to
reduce the proton background. The choice of the ECAL estimator cut value will be discussed separately in \cref{sec:analysis-flux-time-averaged-ecal-estimator}.

To extract the number of electrons and positrons an extended binned maximum likelihood fit~\cite{Cowan1998} is performed on the data sample in each energy interval.
The binned two-dimensional distribution of the data sample is denoted as $\mathcal{D}(x, y)$, where $x$ refers to the charge sign multiplied with the TRD estimator - $Z \times \Lambda_{\text{TRD}}$ -
and $y$ to the charge-confusion estimator $\Lambda_{\text{CC}}$. The fit is performed, by minimizing the negative log-likelihood function, \cref{eq:trd-2d-llh-binned}, denoted as $\mathcal{L}(x, y)$.

The goal is to extract five parameters for each energy interval: the number of electrons
$\textcolor{elecColor}{N_{e^{-}}}$, the number of positrons $\textcolor{posiColor}{N_{e^{+}}}$, the number of protons $\textcolor{protColor}{N_{p^{+}}}$,
the number of charge-confused protons $\textcolor{ccProtColor}{N_{p^{-}}}$ and the amount of charge-confusion $\textcolor{ccColor}{f_{\text{cc}}}$ in the leptonic sample.

In the fit procedure all particle counts are represented as fractions of the total number of events, e.g. $\textcolor{elecColor}{N_{e^{-}}} = \textcolor{elecColor}{f_{\text{elec}}} \cdot N_{\text{events}}$, where $N_{\text{events}} = \sum_x \sum_y D(x, y)$. The negative log-likelihood function - \cref{eq:trd-2d-llh-binned} - is minimized using the Minuit~\cite{James1975} minimizer
and the best-fit parameters are extracted. The last additive term in the likelihood functions differs from the standard maximum likelihood function: the normalisation of the
probability distribution function is allowed to vary. The number of events $N_{\text{events}}$ follows a Poissonian distribution and the last additive term in the likelihood function takes this into account.

\begin{equation}
  \label{eq:trd-2d-llh-binned}
  \begin{aligned}
    \mathcal{L}(x, y) &= -\left(\sum_{x} \sum_{y} \mathcal{D}(x, y) \cdot \log (\mathcal{P}(x, y))\right) \\
                      &+ \underbrace{(\textcolor{protColor}{f_{\text{prot}}} + \textcolor{ccProtColor}{f_{\text{ccprot}}} + \textcolor{elecColor}{f_{\text{elec}}} + \textcolor{posiColor}{f_{\text{posi}}}) \cdot N_{\text{events}}}_\text{extended likelihood term}
  \end{aligned}
\end{equation}

The normalized likelihood value $\mathcal{P}(x, y)$ is defined in \cref{eq:ccmva-trd-2d-pdf} as sum of the products
of the particle fraction $f_{k}$ with the template $\mathcal{P}_{k}(x, y)$, for each template component $k$.

The definition of the normalized likelihood value is the crucial part of the two-dimensional fit procedure. By the introduction of the amount of charge-confusion
$\textcolor{ccColor}{f_{\text{cc}}}$, four out of the six PDFs $\textcolor{elecColor}{\mathcal{P}_{\text{elec}}}(x, y)$, $\textcolor{ccElecColor}{\mathcal{P}_{\text{ccelec}}}(x, y)$,
$\textcolor{posiColor}{\mathcal{P}_{\text{posi}}}(x, y)$, $\textcolor{ccPosiColor}{\mathcal{P}_{\text{ccposi}}}(x, y)$ are no longer independent in the fit procedure
but connected via $\textcolor{ccColor}{f_{\text{cc}}}$, reducing the free parameters of the fit.
\vspace{-2mm}
\begin{equation}
  \label{eq:ccmva-trd-2d-pdf}
  \begin{aligned}
    \mathcal{P}(x, y) =\ &                                                \textcolor{protColor}  {f_{\text{prot}}}   & \cdot\ && \textcolor{protColor}  {\mathcal{P}_{\text{prot}}}(x, y)   &&\ + &&
                         &                                                \textcolor{ccProtColor}{f_{\text{ccprot}}} & \cdot\ && \textcolor{ccProtColor}{\mathcal{P}_{\text{ccprot}}}(x, y) &&\ + && \\
                         & (1 - \textcolor{ccColor}{f_{\text{cc}}}) \cdot \textcolor{elecColor}  {f_{\text{elec}}}   & \cdot\ && \textcolor{elecColor}  {\mathcal{P}_{\text{elec}}}(x, y)   &&\ + &&
                         & \textcolor{ccColor}{f_{\text{cc}}}       \cdot \textcolor{elecColor}  {f_{\text{elec}}}   & \cdot\ && \textcolor{ccElecColor}{\mathcal{P}_{\text{ccelec}}}(x, y) &&\ + && \\
                         & (1 - \textcolor{ccColor}{f_{\text{cc}}}) \cdot \textcolor{posiColor}  {f_{\text{posi}}}   & \cdot\ && \textcolor{posiColor}  {\mathcal{P}_{\text{posi}}}(x, y)   &&\ + &&
                         & \textcolor{ccColor}{f_{\text{cc}}}       \cdot \textcolor{posiColor}  {f_{\text{posi}}}   & \cdot\ && \textcolor{ccPosiColor}{\mathcal{P}_{\text{ccposi}}}(x, y)
  \end{aligned}
\end{equation}

\begin{figure}[H]
  \centering
  \begin{subfigure}{\linewidth}
    \centering
    \includegraphics[width=0.5\linewidth]{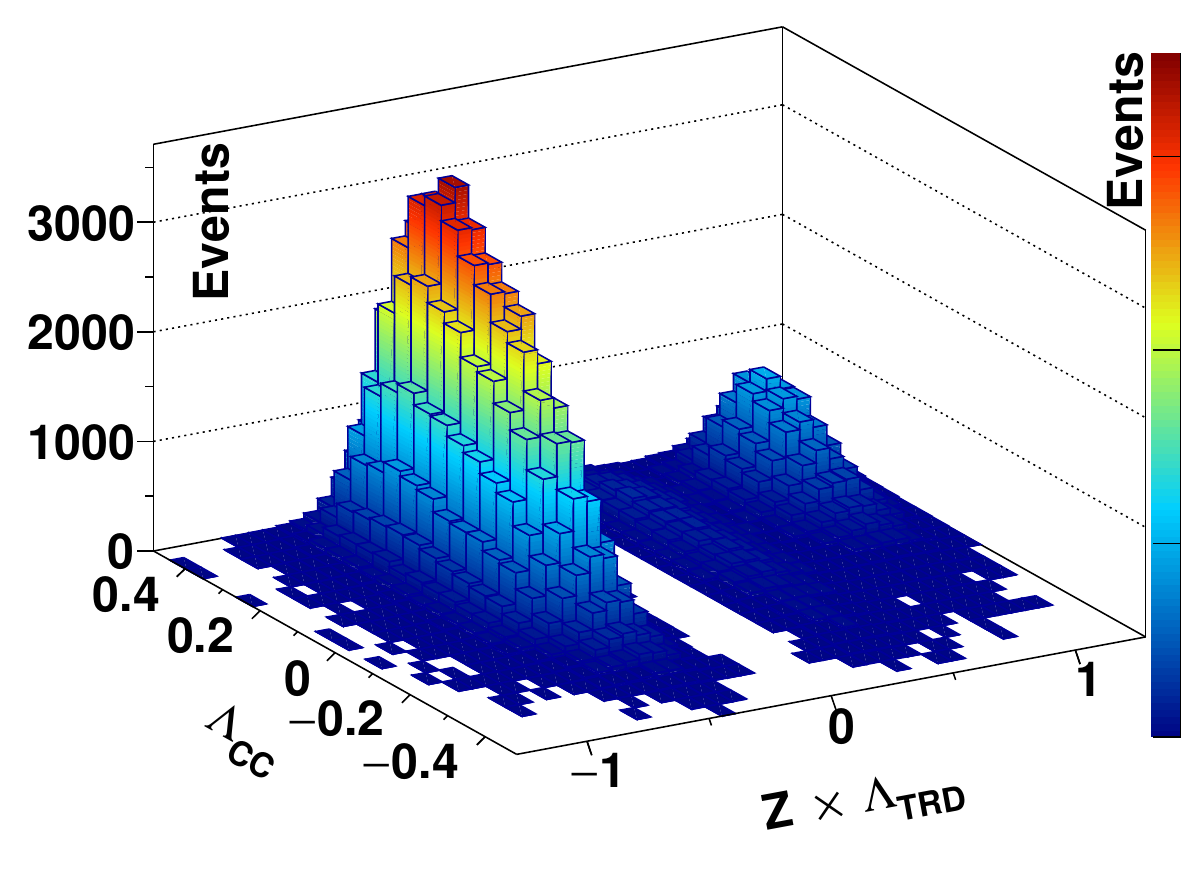}
    \caption{($Z \times \Lambda_{\text{TRD}}$ - $\Lambda_{\text{CC}}$) data distribution}
    \label{fig:2d-iss-data-fit-results-data}
  \end{subfigure}
  \hfill
  \begin{subfigure}{0.48\linewidth}
    \includegraphics[width=\linewidth]{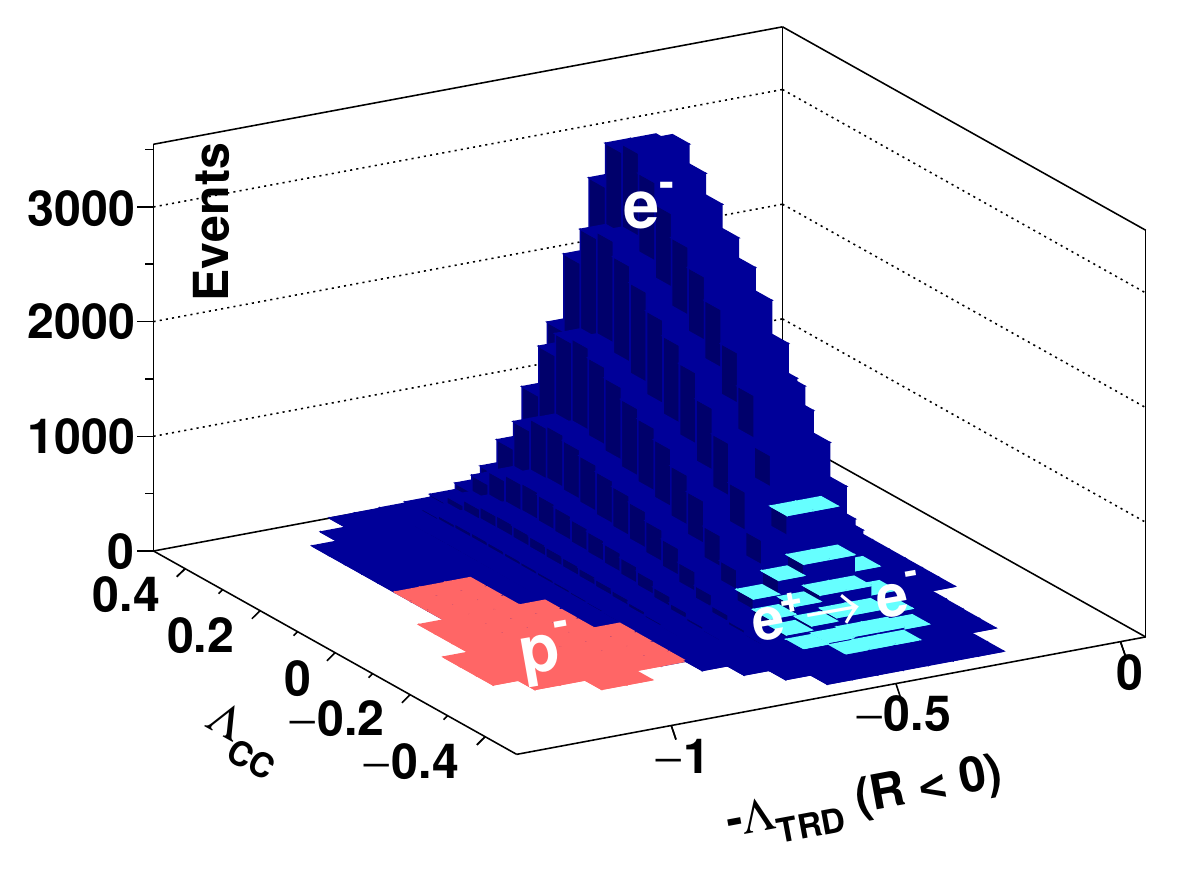}
    \caption{Fit results for R < 0 sample}
    \label{fig:2d-iss-data-fit-results-neg-fit}
  \end{subfigure}
  \begin{subfigure}{0.48\linewidth}
    \includegraphics[width=\linewidth]{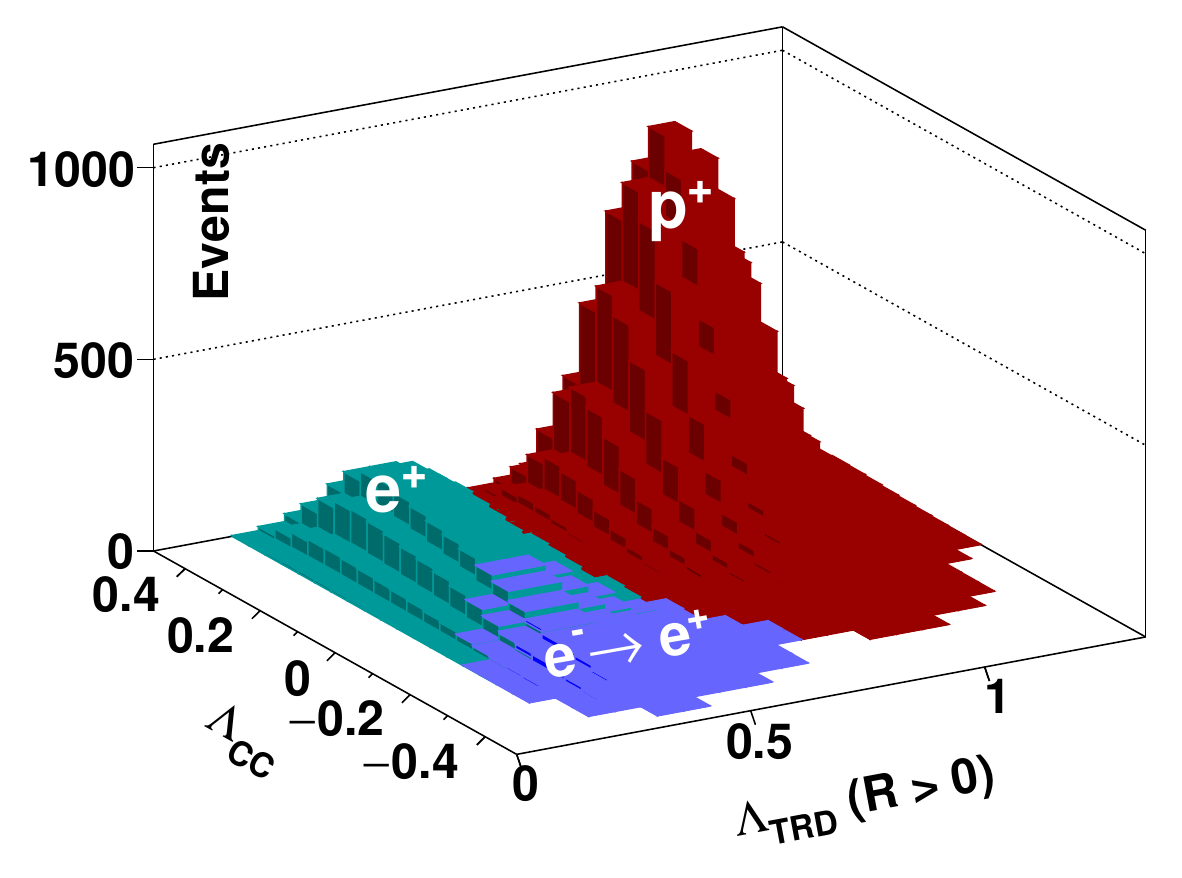}
    \caption{Fit results for R > 0 sample}
    \label{fig:2d-iss-data-fit-results-pos-fit}
  \end{subfigure}
  \caption{ISS data distribution in the energy bin \SIrange{17.98}{18.99}{\GeV} for the single-track sample as three-dimensional plot (top) along with the results of the fit procedure, split in two parts: one plot showing the fit results in the negative rigidity sample (lower left) and another plot showing the fit results in the positive rigidity sample (lower right).}
  \label{fig:2d-iss-data-fit-results}
\end{figure}

The ISS data distribution for the single-track sample (\cref{fig:2d-iss-data-single-track-after-ecal-cut}), is fit using the single-track templates shown in
\cref{fig:2d-templates-single-track-electrons-positrons,fig:2d-templates-single-track-cc-electrons-cc-positrons,fig:2d-templates-single-track-protons-cc-protons},
for an example energy bin \SIrange{17.98}{18.99}{\GeV}. The ISS data distribution is presented in a three-dimensional plot in \cref{fig:2d-iss-data-fit-results} along with the
fit results, split in the negative rigidity sample (\cref{fig:2d-iss-data-fit-results-neg-fit}) and the positive rigidity sample (\cref{fig:2d-iss-data-fit-results-pos-fit}).

\begin{figure}[H]
  \begin{subfigure}{\linewidth}
    \includegraphics[width=\linewidth,height=5cm]{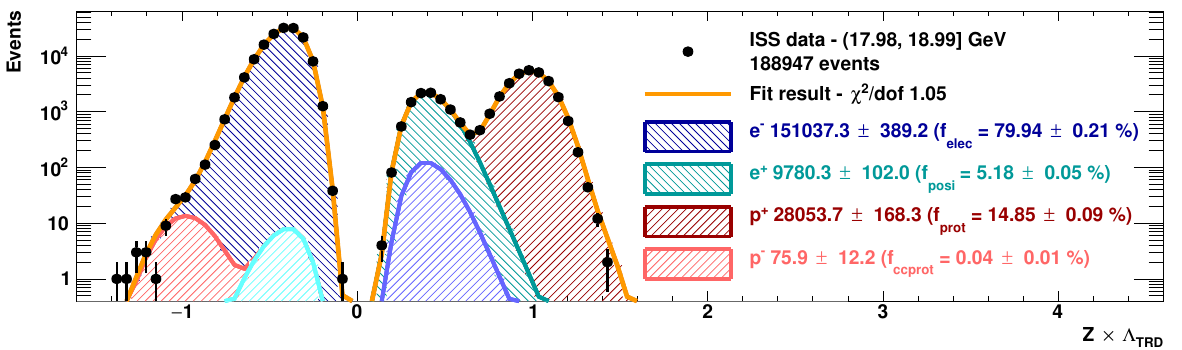}
    \caption{$Z \times \Lambda_{\text{TRD}}$ projection}
    \label{fig:2d-all-tracks-analysis-2d-single-track-sample-cc-fitted-from-iss-trd-proj}
  \end{subfigure}
  \hfill
  \begin{subfigure}{0.50\linewidth}
    \includegraphics[width=\linewidth,height=5cm]{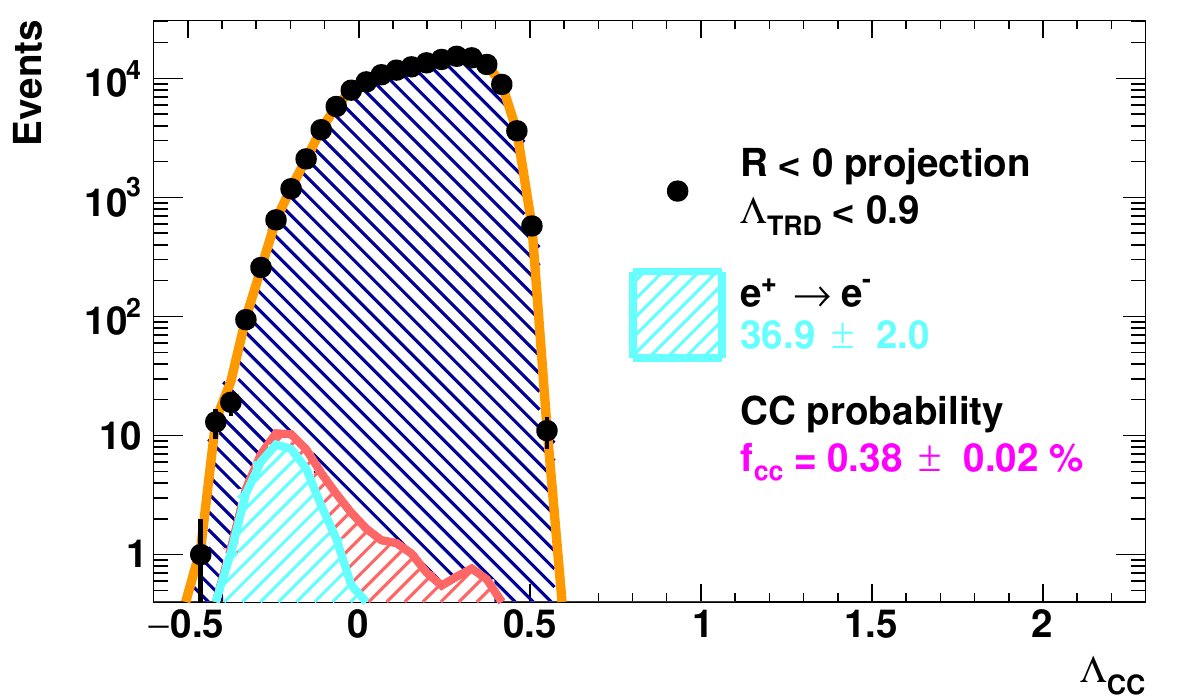}
    \caption{$\Lambda_{\text{CC}}$ projection - R < 0}
    \label{fig:2d-all-tracks-analysis-2d-single-track-sample-cc-fitted-from-iss-ccmva-neg-rig-proj}
  \end{subfigure}
  \hfill
  \begin{subfigure}{0.50\linewidth}
    \includegraphics[width=\linewidth,height=5cm]{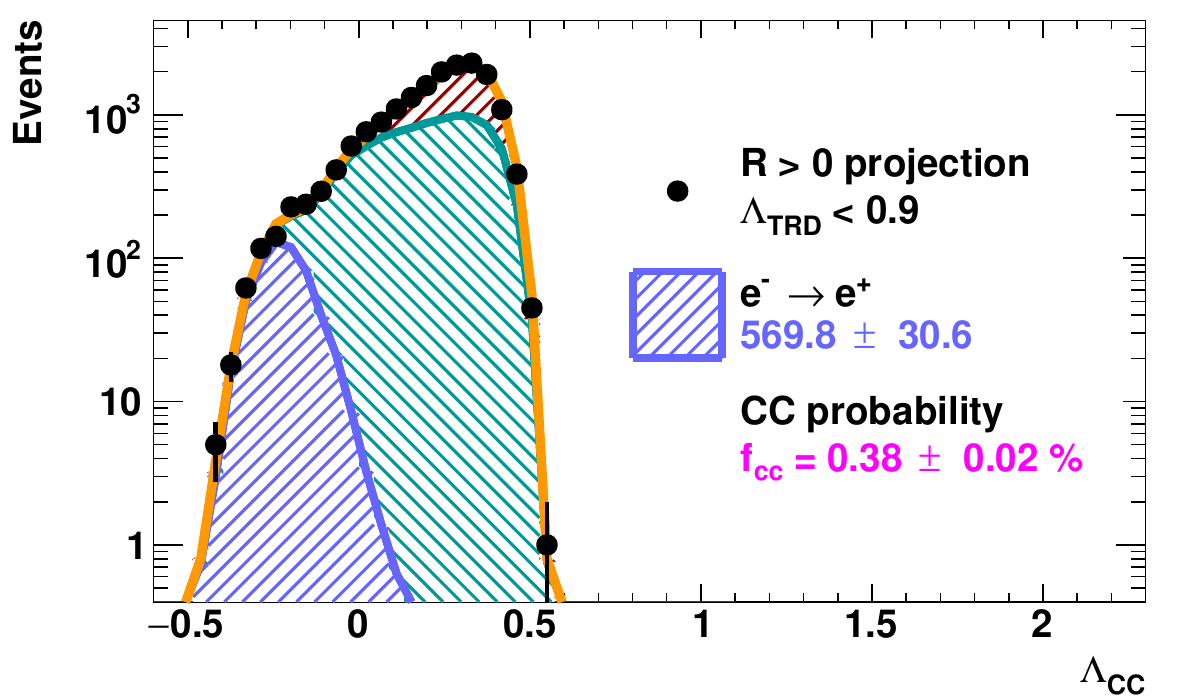}
    \caption{$\Lambda_{\text{CC}}$ projection - R > 0}
    \label{fig:2d-all-tracks-analysis-2d-single-track-sample-cc-fitted-from-iss-ccmva-pos-rig-proj}
  \end{subfigure}
  \caption{Results of the two-dimensional TRD / CCMVA template fit in the energy bin \SIrange{17.98}{18.99}{\GeV} for the single-track sample drawn as stacked histograms. The upper panel shows the projection of the two-dimensional fit to the x-axis. The lower left panel shows the projection to the y-axis, containing only events with R < 0. In this plot \textbf{\textcolor{elecColor}{electrons}}, \textbf{\textcolor{ccPosiColor}{charge-confused positrons}} and \textbf{\textcolor{ccProtColor}{charge-confused protons}} are visible. The lower right panel shows the projection to the y-axis, containing only events with R > 0. In this plot \textbf{\textcolor{posiColor}{positrons}}, \textbf{\textcolor{ccElecColor}{charge-confused electrons}} and \textbf{\textcolor{protColor}{protons}} are visible.}
  \label{fig:2d-all-tracks-analysis-2d-single-track-sample-cc-fitted-from-iss}
\end{figure}

\Cref{fig:2d-all-tracks-analysis-2d-single-track-sample-cc-fitted-from-iss} shows a projection of the three-dimensional fit result representation
to the x-axis (\enquote{$Z \times \Lambda_{\text{TRD}}$}) and the y-axis (\enquote{$\Lambda_{\text{CC}}$}), separately. The $Z \times \Lambda_{\text{TRD}}$ projection
(\cref{fig:2d-all-tracks-analysis-2d-single-track-sample-cc-fitted-from-iss-trd-proj}) is suitable to identify the different particle species: \textcolor{elecColor}{electrons},
\textcolor{posiColor}{positrons}, \textcolor{protColor}{protons}, and \textcolor{ccProtColor}{charge-confused protons}, whereas the $\Lambda_{\text{CC}}$ projection
is helpful to visualize the amount of charge-confusion in the electron or positron component. Therefore the $\Lambda_{\text{CC}}$ projection was obtained twice, once for
the negative rigidity sample (to identify \textcolor{elecColor}{electrons} and \textcolor{ccPosiColor}{charge-confused positrons} - see
\cref{fig:2d-all-tracks-analysis-2d-single-track-sample-cc-fitted-from-iss-ccmva-neg-rig-proj}) and once for the positive rigidity sample (to identify the
\textcolor{posiColor}{positrons}, and \textcolor{ccElecColor}{charge-confused electrons} - see
\cref{fig:2d-all-tracks-analysis-2d-single-track-sample-cc-fitted-from-iss-ccmva-pos-rig-proj}). Furthermore an additional cut on the TRD estimator was applied
when obtaining the $\Lambda_{\text{CC}}$ projection, to reduce the proton background, which is not helpful to disentangle electrons or positrons from their
charge-confused counterparts.

It is immediately apparent that the amount of charge-confusion in the leptonic sample is mostly determined by the amount of charge-confused
electrons in the positive rigidity sample, as shown in \cref{fig:2d-all-tracks-analysis-2d-single-track-sample-cc-fitted-from-iss-ccmva-pos-rig-proj}.
When studying the \enquote{$Z \times \Lambda_{\text{TRD}}$} projection it is obvious that the charge-confusion effect is significant and needs to be
taken into account for this analysis. Without a proper treatment of the charge-confusion effect, all positron-like events in the region around
x~$\approx$~0.4 would be identified as positrons, but in fact a significant fraction are electrons that are misreconstructed with positive
rigidity: charge-confused electrons. The same argument holds for the region around x~$\approx$~-0.4, however less pronounced. Most events are
electrons but a non-negligible amount are positrons that are charge-confused.

It is important to understand that the charge-confusion does not affect the energy measurement in the ECAL, but only the reconstructed rigidity -
otherwise the analysis strategy would be completely different, since the charge-confused events from a true energy $E$ would not be contained in the
data distribution for the same bin, but elsewhere. Therefore the CCMVA estimator plays the key role in this analysis allowing to separate the
charge-confused from the correctly reconstructed $e^{\pm}$.

The $\chi^2/\text{dof} = 1.05$ is computed using the full two-dimensional model and shows that a good description of the data is available. Empty
bins are skipped and do not contribute to the $\chi^2$ computation.

\begin{figure}[H]
  \centering
  \includegraphics[width=0.7\linewidth]{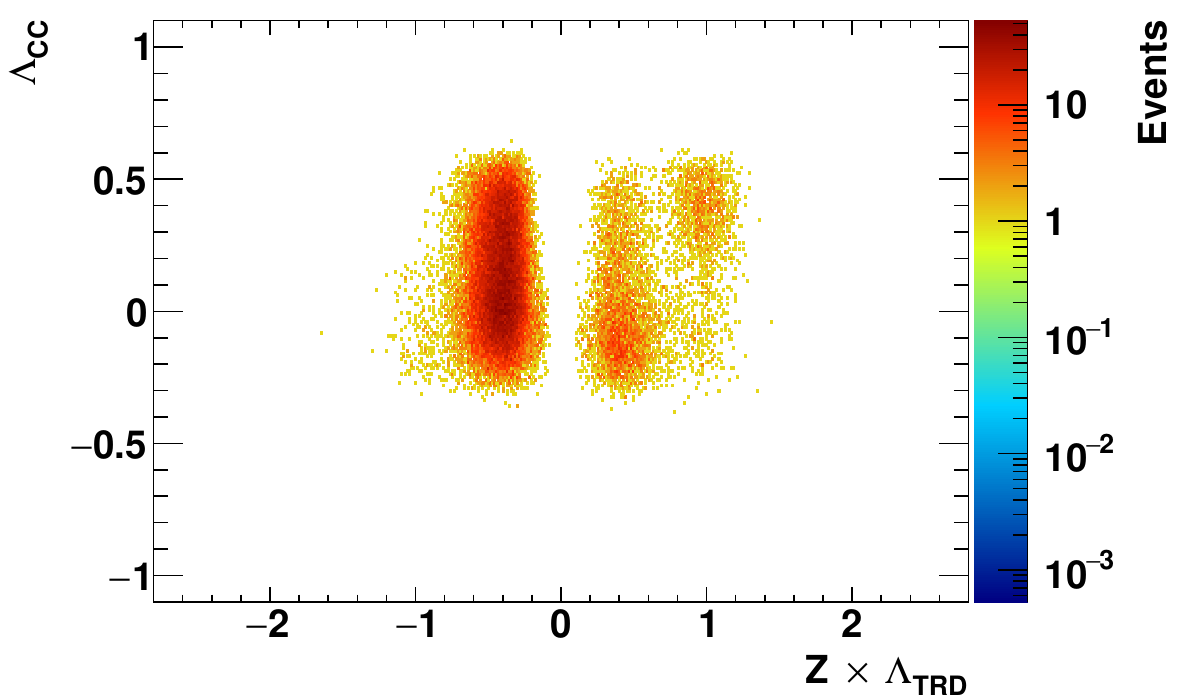}
  \caption{ISS data distribution in the energy bin \SIrange{17.98}{18.99}{\GeV} for the multi-tracks sample, after applying a cut on the ECAL estimator to reduce the proton background.}
  \label{fig:2d-iss-data-multi-tracks}
\end{figure}

\begin{figure}[H]
  \begin{subfigure}{\linewidth}
    \includegraphics[width=\linewidth,height=5cm]{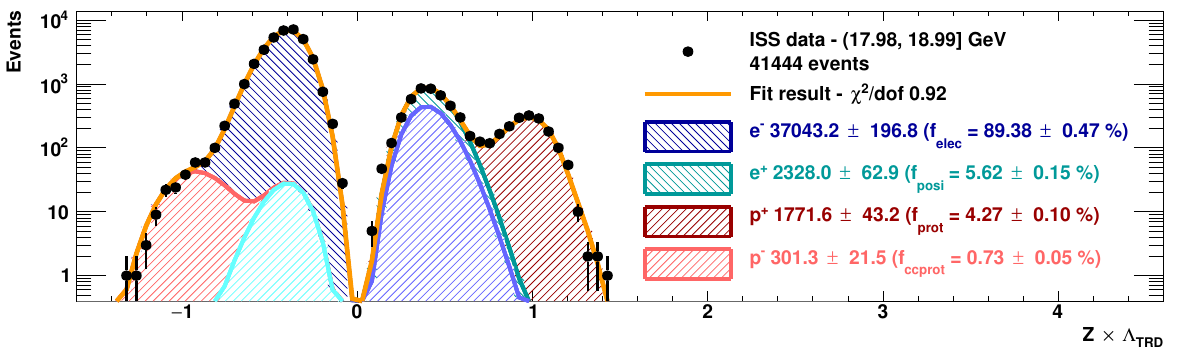}
    \caption{$Z \times \Lambda_{\text{TRD}}$ projection}
    \label{fig:2d-all-tracks-analysis-2d-multi-tracks-sample-cc-fitted-from-iss-trd-proj}
  \end{subfigure}
  \hfill
  \begin{subfigure}{0.50\linewidth}
    \includegraphics[width=\linewidth,height=5cm]{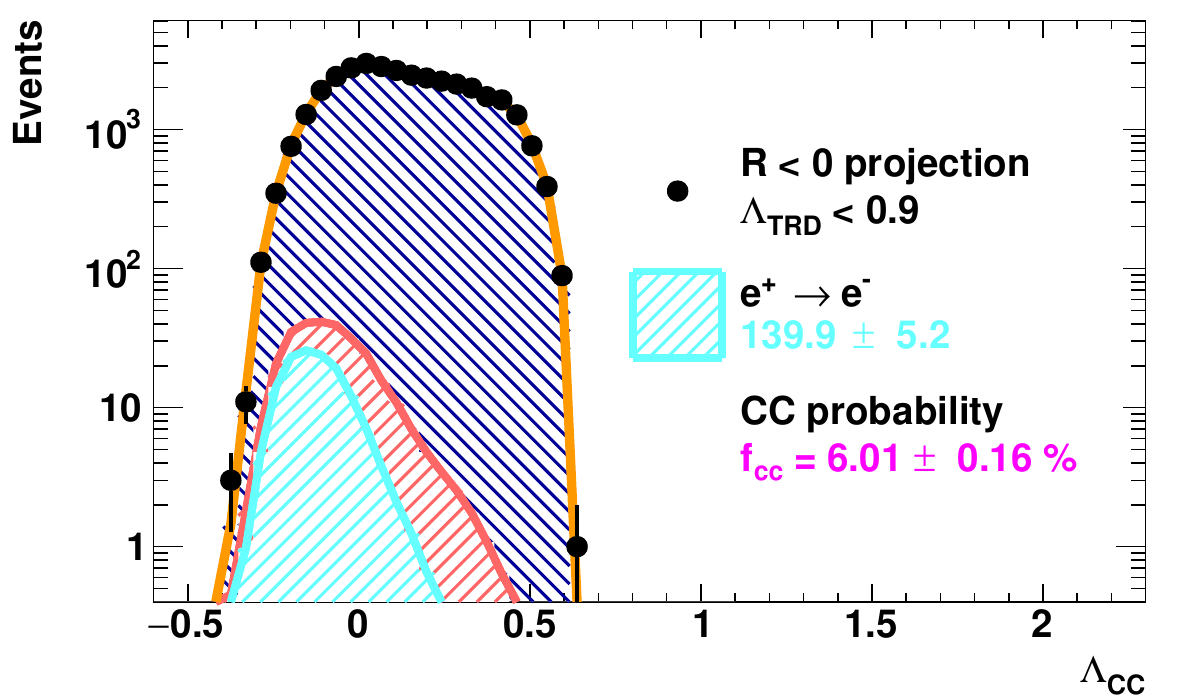}
    \caption{$\Lambda_{\text{CC}}$ projection - R < 0}
    \label{fig:2d-all-tracks-analysis-2d-multi-tracks-sample-cc-fitted-from-iss-ccmva-neg-rig-proj}
  \end{subfigure}
  \hfill
  \begin{subfigure}{0.50\linewidth}
    \includegraphics[width=\linewidth,height=5cm]{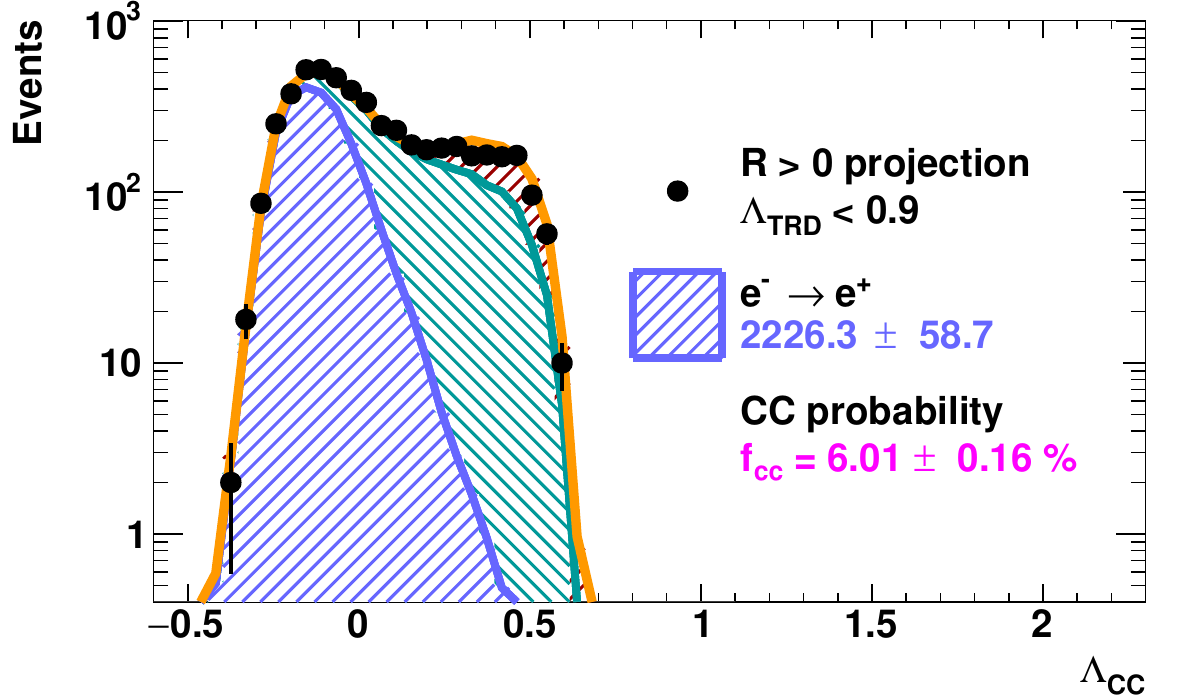}
    \caption{$\Lambda_{\text{CC}}$ projection - R > 0}
    \label{fig:2d-all-tracks-analysis-2d-multi-tracks-sample-cc-fitted-from-iss-ccmva-pos-rig-proj}
  \end{subfigure}
  \caption{Results of the two-dimensional TRD / CCMVA template fit in the energy bin \SIrange{17.98}{18.99}{\GeV} for the multi-tracks sample drawn as stacked histograms. The upper panel shows the projection of the two-dimensional fit to the x-axis. The lower left panel shows the projection to the y-axis, containing only events with R < 0. In this plot \textbf{\textcolor{elecColor}{electrons}}, \textbf{\textcolor{ccPosiColor}{charge-confused positrons}} and \textbf{\textcolor{ccProtColor}{charge-confused protons}} are visible. The lower right panel shows the projection to the y-axis, containing only events with R > 0. In this plot \textbf{\textcolor{posiColor}{positrons}}, \textbf{\textcolor{ccElecColor}{charge-confused electrons}} and \textbf{\textcolor{protColor}{protons}} are visible.}
  \label{fig:2d-all-tracks-analysis-2d-multi-tracks-sample-cc-fitted-from-iss}
\end{figure}

As next step the multi-tracks sample, \cref{fig:2d-iss-data-multi-tracks} is fit using the multi-tracks templates shown in \cref{fig:2d-templates-multi-tracks}.
The fit results are presented in \cref{fig:2d-all-tracks-analysis-2d-multi-tracks-sample-cc-fitted-from-iss} for an example energy bin.

As in the single-track sample case the $\chi^2/\text{dof} = 1.02$ proofs that a good description of the multi-tracks data sample is available.

It is important to note that the number of charge-confused electrons - 2226.3 - is comparable to the number of correctly reconstructed positrons - 2188.1 - in this energy bin. This is a general feature of the multi-tracks sample: the charge-confusion is an order of magnitude larger than in the single-track sample (\SI{6.01}{\percent} in the multi-tracks sample vs. \SI{0.38}{\percent} in the single-track sample).

For the flux analysis, the gain in statistics by adding the multi-tracks sample, outweighs the increase of systematic uncertainties. For the determination of the positron fraction, or the positron over electron ratio, this is not the case, as important uncertainties of the flux analysis cancel in the ratios. This topic will be revisited and discussed in detail in \cref{sec:analysis-flux-time-averaged-sysunc}.

The results of the single-track sample and the multi-tracks sample can be added to compute the results for the all-tracks sample.
For historical\footnote{It is sufficient to combine the results of the single-track and the multi-tracks fit to extract the number of electrons and positrons in the all-tracks data sample.
To extract the correlation coefficient $\rho$, denoting the correlation between the fitted number of electrons and positrons, it is more convenient to execute a separate template fit for the
all-tracks data sample and directly extract the parameter $\rho$. $\rho$ is needed when computing e.g. the uncertainty of the positron fraction or the positron/electron ratio from the fit
results, by error propagation.} reasons a separate template fit on the all-tracks sample is performed and the results of that fit will be used to derive the electron flux and the positron flux.

In order to describe the data in the all-tracks sample, templates for the all-tracks sample have to be constructed from the single-track templates and the multi-tracks templates. For each template
component $k$ the fraction of single-track events $\alpha_{k}$ in the all-tracks sample needs to be determined, by counting. Using this information all-tracks templates $\mathcal{P}_{k,\ \text{all-tracks}}(x, y)$, can be derived:

\begin{equation}
  \label{eq:trd-2d-all-tracks-template}
  \mathcal{P}_{k,\ \text{all-tracks}}(x, y) = \alpha_{k} \cdot \mathcal{P}_{k,\ \text{single-track}}(x, y) + (1 - \alpha_{k}) \cdot \mathcal{P}_{k,\ \text{multi-tracks}}(x, y).
\end{equation}

After performing the single-track sample fit and the multi-tracks sample fit the number of events $n_{k}$ in each template component $k$ is known.
Thus the fraction of single-track events $\alpha_{k}$ in each template component $k$ can be derived: $\alpha_{k} = n_{k,\ \text{single-track}} / (n_{k,\ \text{single-track}} + n_{k,\ \text{multi-tracks}})$.
\cref{tab:2d-fit-single-track-fractions-per-component} shows the so-obtained fractions of single-track events per template component. Using these $\alpha_k$ values, templates for the all-tracks sample can be constructed from the single-track templates and the multi-tracks templates, following the prescription in \cref{eq:trd-2d-all-tracks-template}.

\medskip
\begin{minipage}{\linewidth}
  \centering
  \begin{tabular}{lc}
    \toprule
                                                                        Component & Fraction of single-track events $\alpha_k$ \\
    \midrule
    \rowcolor{black!20}        \textcolor{elecColor}{Electrons}                   & 0.81 \\
                               \textcolor{ccElecColor}{Charge-confused electrons} & 0.20 \\
    \rowcolor{black!20}        \textcolor{posiColor}{Positrons}                   & 0.81 \\
                               \textcolor{ccPosiColor}{Charge-confused positrons} & 0.21 \\
    \rowcolor{black!20}        \textcolor{protColor}{Protons}                     & 0.94 \\
                               \textcolor{ccProtColor}{Charge-confused protons}   & 0.20 \\
  \end{tabular}
  \captionof{table}{Single-track event fractions for all template components in the two-dimensional fit of the all-tracks sample in the energy bin \SIrange{17.98}{18.99}{\GeV}. For comparison, the overall fraction of single-track events in this energy bin is 0.82.}
  \label{tab:2d-fit-single-track-fractions-per-component}
\end{minipage}

\begin{figure}[H]
  \begin{subfigure}{\linewidth}
    \includegraphics[width=\linewidth,height=5cm]{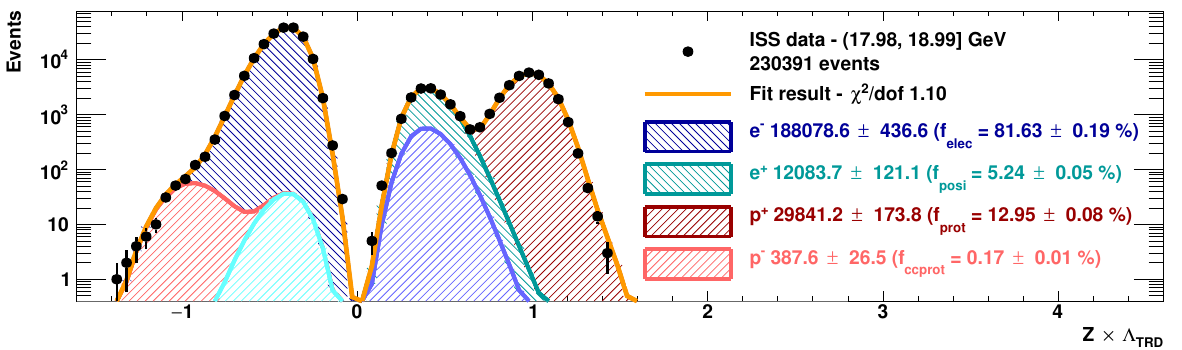}
    \caption{$Z \times \Lambda_{\text{TRD}}$ projection}
    \label{fig:2d-all-tracks-analysis-2d-all-tracks-sample-cc-fitted-from-iss-trd-proj}
  \end{subfigure}
  \hfill
  \begin{subfigure}{0.50\linewidth}
    \includegraphics[width=\linewidth,height=5cm]{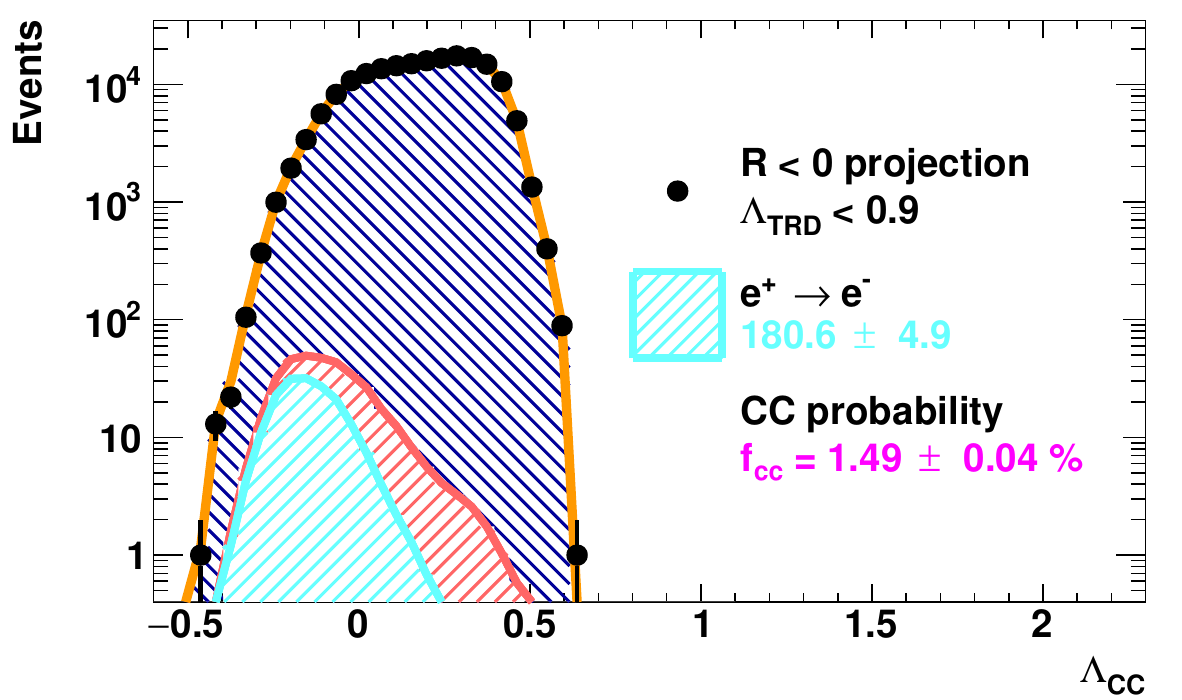}
    \caption{$\Lambda_{\text{CC}}$ projection - R < 0}
    \label{fig:2d-all-tracks-analysis-2d-all-tracks-sample-cc-fitted-from-iss-ccmva-neg-rig-proj}
  \end{subfigure}
  \hfill
  \begin{subfigure}{0.50\linewidth}
    \includegraphics[width=\linewidth,height=5cm]{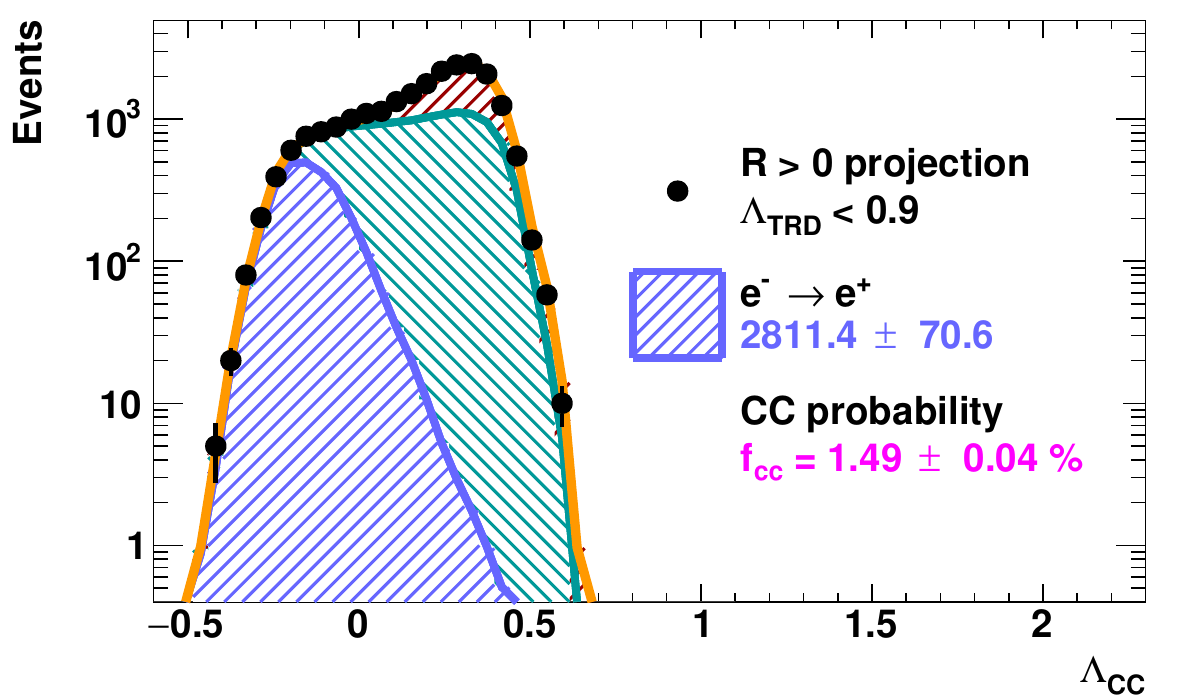}
    \caption{$\Lambda_{\text{CC}}$ projection - R > 0}
    \label{fig:2d-all-tracks-analysis-2d-all-tracks-sample-cc-fitted-from-iss-ccmva-pos-rig-proj}
  \end{subfigure}
  \caption{Results of the two-dimensional TRD / CCMVA template fit in the energy bin \SIrange{17.98}{18.99}{\GeV} for the all-tracks sample drawn as stacked histograms.}
  \label{fig:2d-all-tracks-analysis-2d-all-tracks-sample-cc-fitted-from-iss}
\end{figure}

It is important to note that the global fraction of single-track events in the all-tracks sample cannot be used to derive the templates for the all-tracks sample, as the amount of charge-confusion for the single-track and multi-tracks sample is different. Instead the single-track fraction in each component must be determined separately.

The fit result of the all-tracks sample is presented in \cref{fig:2d-all-tracks-analysis-2d-all-tracks-sample-cc-fitted-from-iss}.
The $\chi^2/\text{dof} = 1.10$ shows that the model is in very good agreement with the data and with the single-track / multi-tracks fit results - as expected.
The overall amount of charge-confusion in the all-tracks sample is approximately four times larger than in the single-track sample (\SI{1.49}{\percent} in the all-tracks sample vs. \SI{0.38}{\percent} in the single-track sample).

\begin{figure}[H]
  \begin{subfigure}{0.50\linewidth}
    \includegraphics[width=\linewidth]{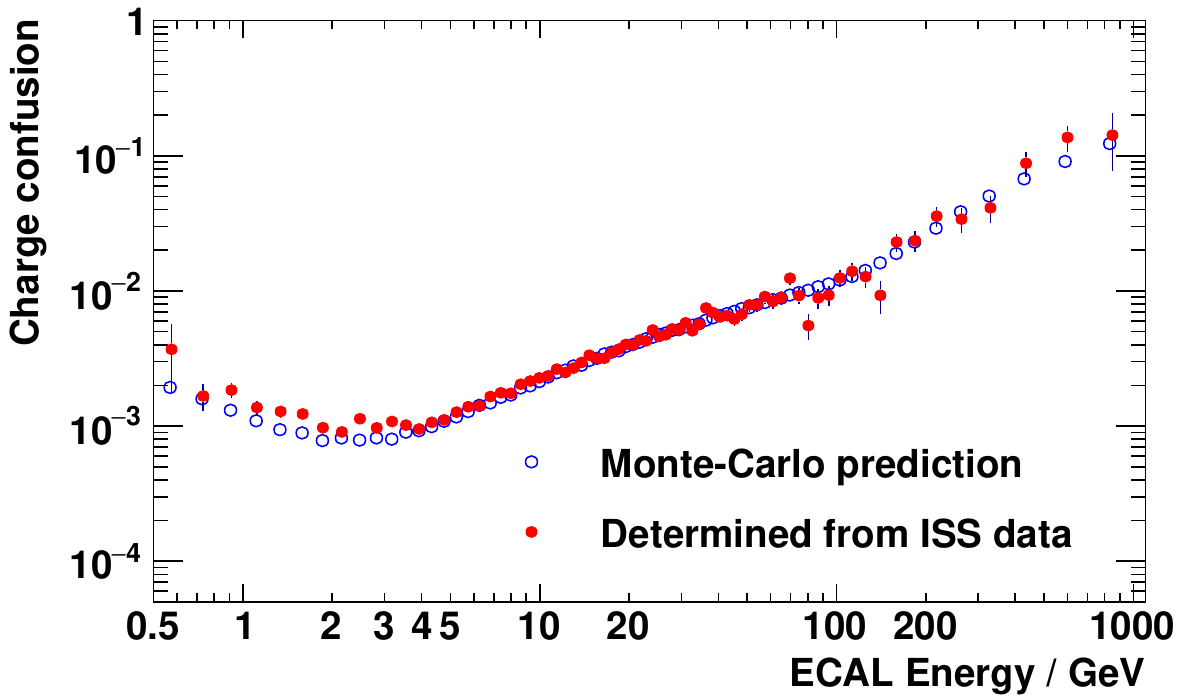}
    \caption{Single-track sample}
  \end{subfigure}
  \hfill
  \begin{subfigure}{0.50\linewidth}
    \includegraphics[width=\linewidth]{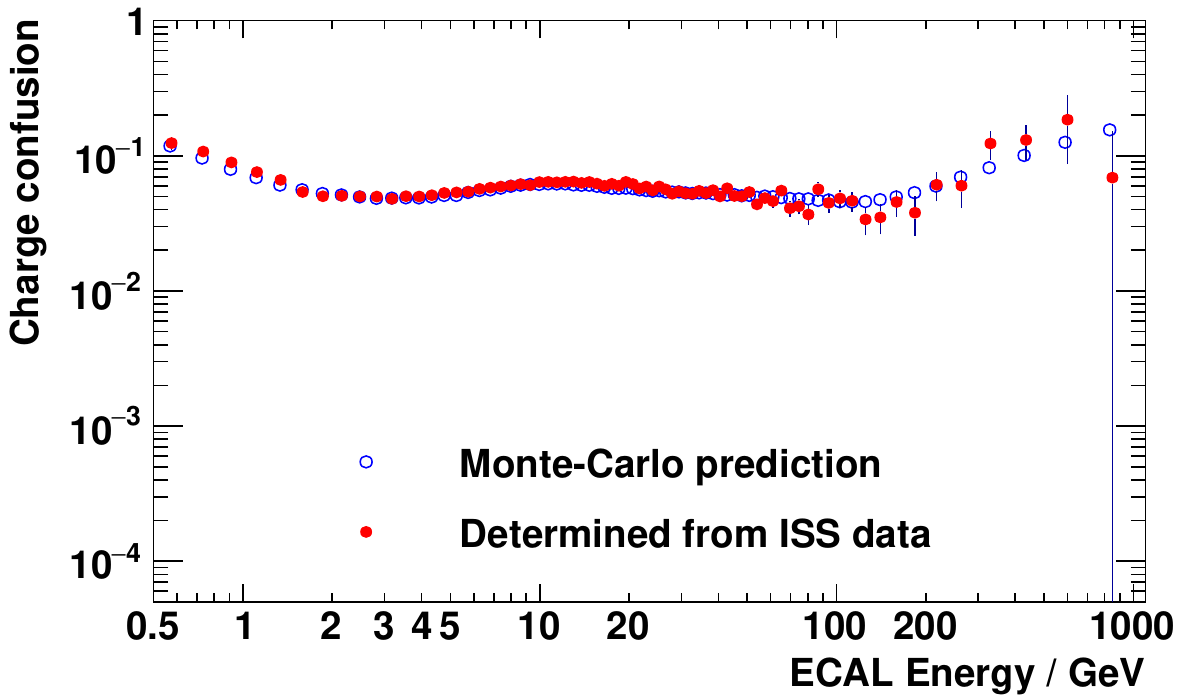}
    \caption{Multi-tracks sample}
  \end{subfigure}
  \hfill
  \begin{subfigure}{\linewidth}
    \centering
    \includegraphics[width=0.55\linewidth]{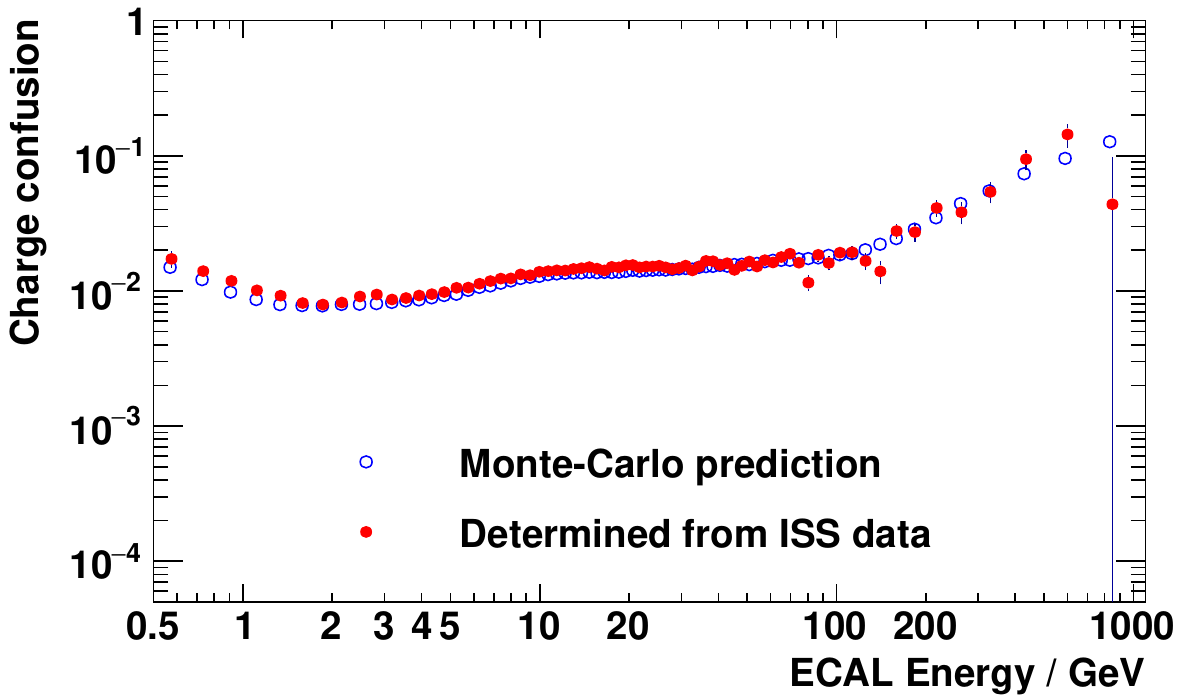}
    \caption{All-tracks sample}
  \end{subfigure}
  \caption{Comparison of the amount of charge-confusion determined from ISS data with the prediction from the Monte-Carlo simulation for the single-track sample, the multi-tracks sample and the all-tracks sample.}
  \label{fig:2d-all-tracks-analysis-2d-cc-comparison}
\end{figure}

The whole fit procedure is repeated for all energy bins in the analysis, for the single-track sample, the multi-tracks sample and the all-tracks sample.
Afterwards the amount of charge-confusion for the different samples can be compared with the Monte-Carlo prediction, as presented in
\cref{fig:2d-all-tracks-analysis-2d-cc-comparison}. All samples show excellent agreement with the prediction from the Monte-Carlo simulation.

Most importantly the shapes of the charge-confusion curves match the expectation: at low energies the charge-confusion rises, due to multiple scattering.
At high energies - around \SIapprox{150}{\GeV} - an additional component contributing to the charge-confusion is emerging: the finite resolution of the
silicon tracker, when approaching \gls{MDR}. Furthermore the charge-confusion curves are all smooth, without distinct structures. The transitions at low
and high energies are smooth.

\begin{figure}[H]
  \begin{subfigure}{\linewidth}
    \includegraphics[width=\linewidth,height=5cm]{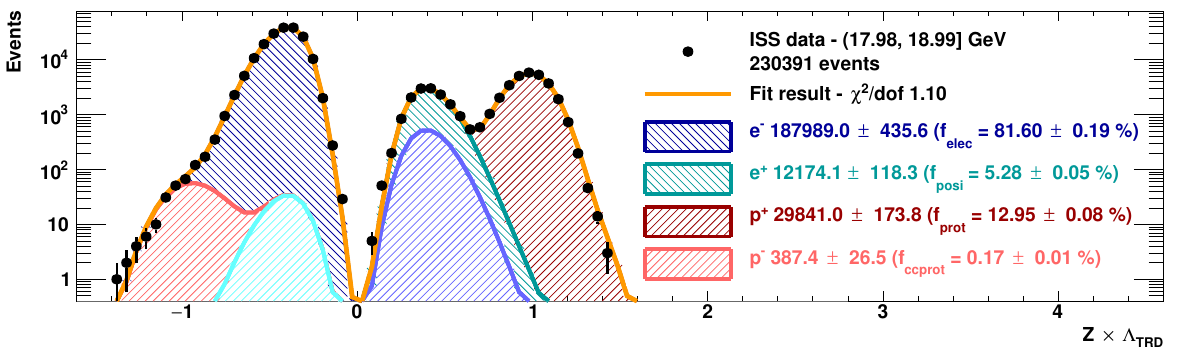}
    \caption{$Z \times \Lambda_{\text{TRD}}$ projection}
    \label{fig:2d-all-tracks-analysis-2d-all-tracks-sample-cc-fixed-from-mc-trd-proj}
  \end{subfigure}
  \hfill
  \begin{subfigure}{0.50\linewidth}
    \includegraphics[width=\linewidth,height=5cm]{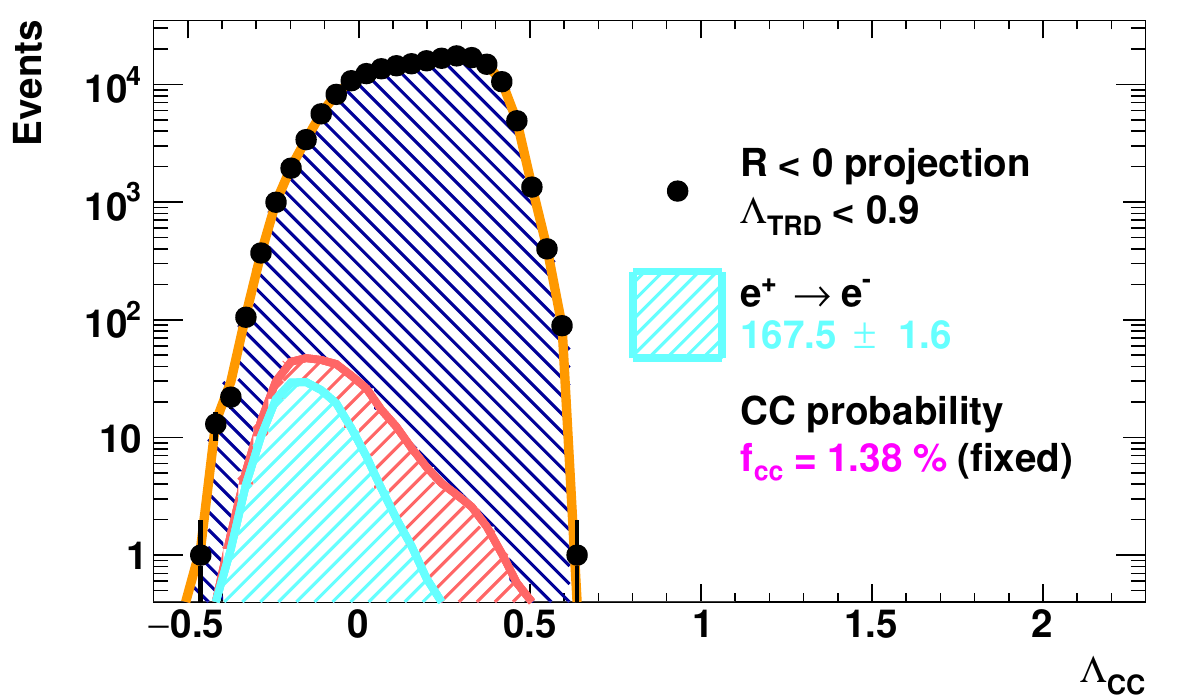}
    \caption{$\Lambda_{\text{CC}}$ projection - R < 0}
    \label{fig:2d-all-tracks-analysis-2d-all-tracks-sample-cc-fixed-from-mc-ccmva-neg-rig-proj}
  \end{subfigure}
  \hfill
  \begin{subfigure}{0.50\linewidth}
    \includegraphics[width=\linewidth,height=5cm]{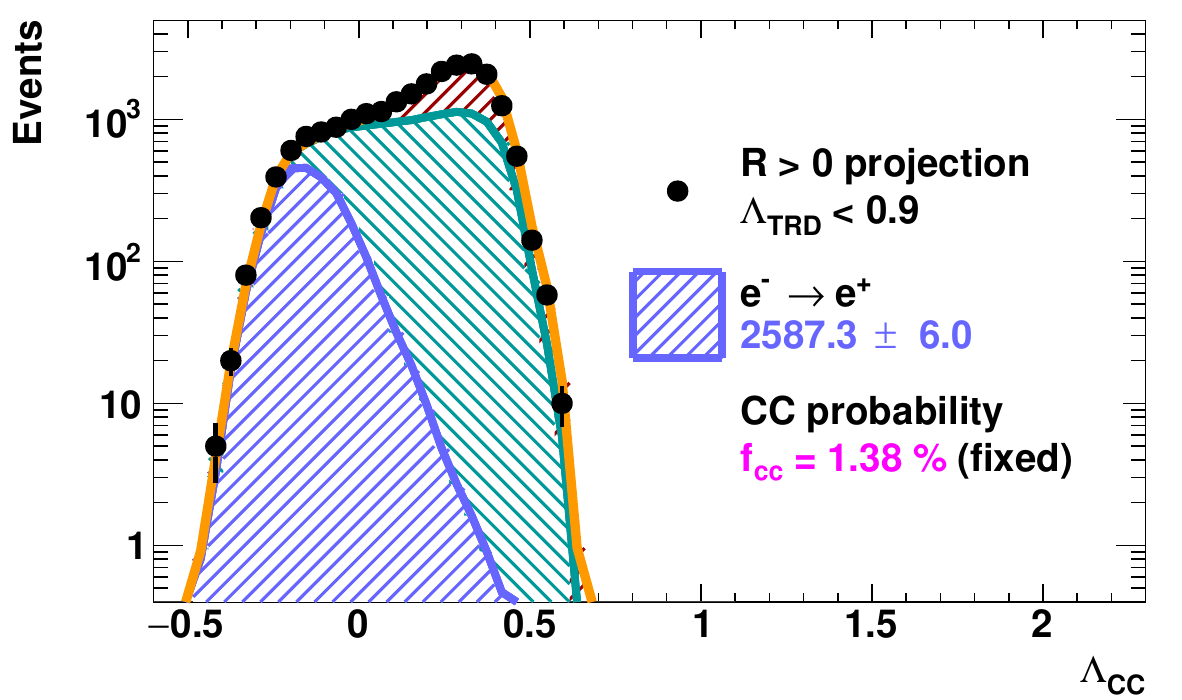}
    \caption{$\Lambda_{\text{CC}}$ projection - R > 0}
    \label{fig:2d-all-tracks-analysis-2d-all-tracks-sample-cc-fixed-from-mc-ccmva-pos-rig-proj}
  \end{subfigure}
  \caption{Results of the two-dimensional TRD / CCMVA template fit in the energy bin \SIrange{17.98}{18.99}{\GeV} for the all-tracks sample drawn as stacked histograms with the amount of charge-confusion fixed by the prediction from the Monte-Carlo simulation.}
  \label{fig:2d-all-tracks-analysis-2d-all-tracks-sample-cc-fixed-from-mc}
\end{figure}

The additional degree of freedom in the fit procedure to determine the charge-confusion leads to an increase of the uncertainties on the number of electrons and positrons, which
should be minimized for the flux analysis. Since the charge-confusion fitted from ISS data is in agreement with the Monte-Carlo simulation it opens the possibility to fix the amount of
charge-confused from the Monte-Carlo prediction and repeat the fit procedure. The fit results are presented in \cref{fig:2d-all-tracks-analysis-2d-all-tracks-sample-cc-fixed-from-mc}
for the all-tracks sample.

The number of positrons is equal to \textbf{12083.7 $\pm$ 121.1} when the amount of charge-confusion is a free fit parameter. The smallest uncertainty, if the separation
between all components would be perfect, is $\sqrt{N}$, corresponding to an expected Poissonian uncertainty in the limit of sufficiently high statistics. The reported
uncertainty of \textbf{121.1} is \SIapprox{10.2}{\percent} larger than the ideal uncertainty of $\sqrt{12083.7} \approx 109.9$.

\medskip
However, if the amount of charge-confusion is fixed to the prediction from the Monte-Carlo simulation the number of positrons is equal to \textbf{12174.1 $\pm$ 118.3}, which
is only \SIapprox{7.6}{\percent} larger than the smallest uncertainty and shows the gain of fixing the amount of charge-confusion from the Monte-Carlo prediction. The absolute
difference in the number of positrons is smaller than one sigma, not only in this energy bin but for the whole energy range of the analysis.

Therefore the charge-confusion value $\textcolor{ccColor}{f_{\text{cc}}}$ will be fixed to the prediction from the Monte-Carlo simulation, when extracting
the number of electrons and positrons for the flux and ratio analysis

Therefore the event counts from the all-tracks analysis will be used, with the charge-confused fixed to the Monte-Carlo prediction, to derive the electron and positron flux.
The systematic uncertainty due to the fixing of the charge-confusion will be discussed in detail in \cref{sec:analysis-flux-time-averaged-sysunc-charge-confusion-estimator}.

\begin{figure}[H]
  \centering
  \begin{subfigure}{0.80\linewidth}
    \includegraphics[width=\linewidth]{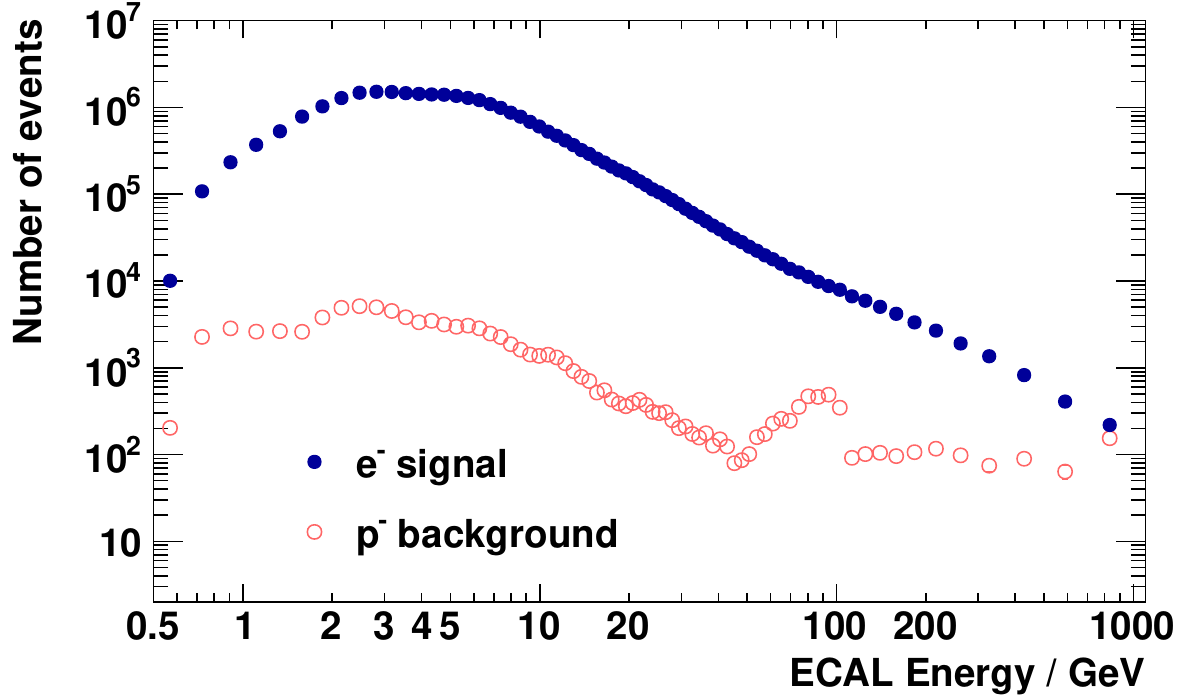}
    \caption{\textcolor{elecColor}{Electrons} and \textcolor{ccProtColor}{charge-confused protons}}
  \end{subfigure}
  \hfill
  \begin{subfigure}{0.80\linewidth}
    \includegraphics[width=\linewidth]{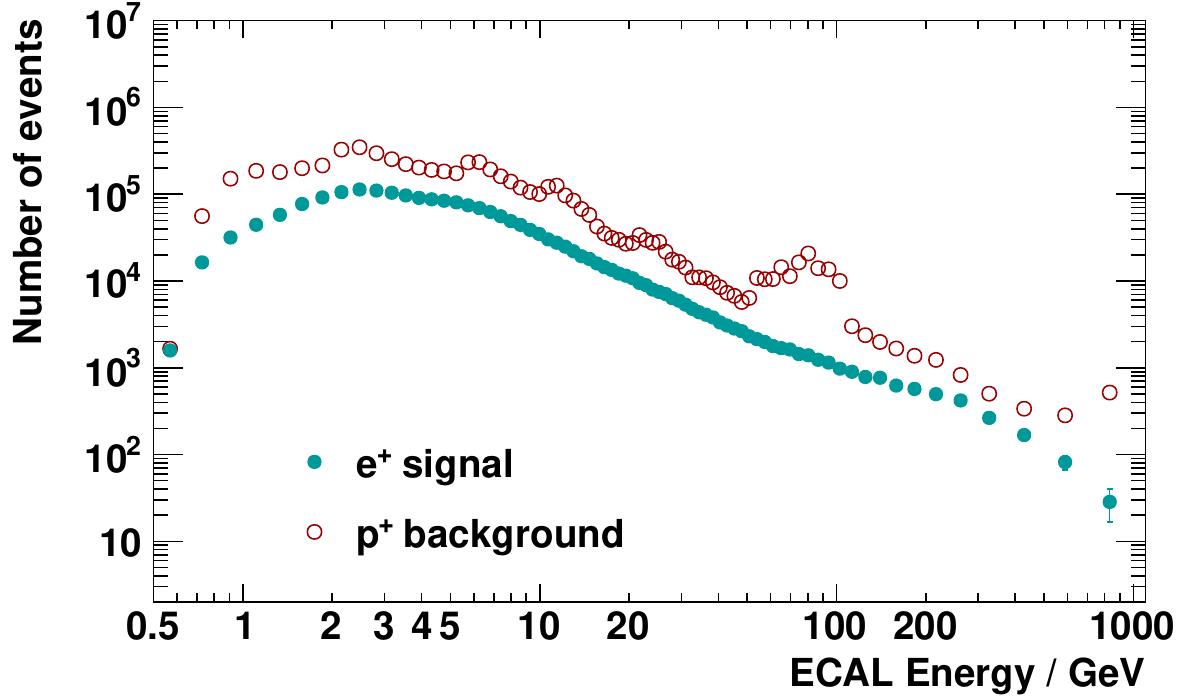}
    \caption{\textcolor{posiColor}{Positrons} and \textcolor{protColor}{protons}}
  \end{subfigure}
  \caption{Number of signal and background events for the all-tracks sample, extracted from the two-dimensional fit procedure.}
  \label{fig:2d-all-tracks-analysis-signal-and-background-events}
\end{figure}

The final event counts are presented in \cref{fig:2d-all-tracks-analysis-signal-and-background-events}. The electron and positron signal events follow smooth
curves, even though no corrections for the selection efficiency nor the geomagnetic cut-off is applied at this stage of the analysis.

The number of background events for the negative rigidity sample is always below the number of signal events in each energy bin. The negative rigidity sample
is naturally enhanced by electrons and the number of charge-confused protons is much smaller up to the highest energies.

The situation is different for the positive rigidity sample, where the number of background events is always larger than the number of signal events. It is possible
to keep the number of background events smaller than the signal events, but this costs signal efficiency and causes an increase of statistical uncertainties. The
presented event counts are the result of an iterative optimization procedure. The goal of the optimization procedure is to minimize the statistical uncertainties
of the positrons, by increasing the signal efficiency as much as possible. The consequence is a larger background, which is acceptable if the separation of the
signal / background components in the template fit is sufficient, which is the case for this analysis. The increase of background events between \SIrange{50}{100}{\GeV}
is a consequence of only applying a loose cut on the ECAL estimator to reject background events. Above \SI{100}{\GeV} the cut on the ECAL estimator has to be gradually hardened
to keep the separation power large enough that a minimal statistical uncertainty for the positrons is reached. The details of choosing the cut on the ECAL estimator will be
discussed in detail in \cref{sec:analysis-flux-time-averaged-ecal-estimator}.

\clearpage
\section{Time-averaged flux calculation}
\label{sec:analysis-flux-time-averaged}

In this section time-averaged fluxes will be derived, which means that there is only a single time interval $i$ to consider. The time interval
starts at the beginning of the data taking period - \textbf{May~\nth{20},~2011}, including all data up to \textbf{November~\nth{12},~2017} -
corresponding to the last day of ISS data that is analyzed in this work.

In total 74 energy bins will be analyzed, from \SI{0.5}{\GeV} up to \SI{1}{\TeV}, as described in \cref{sec:analysis-lepton-counts-binning}.
The 74 flux data points will be combined into a single energy-dependent flux. $\Phi_{e^{\pm}}(E)$ describes the electron or positron flux in
the energy bin $E$ of width $\Delta E$:

\begin{equation}
  \Phi_{e^{\pm}}(E) = \frac{N_{e^{\pm}}(E)}{\Delta E \cdot T(E) \cdot A_{e^{\pm}}(E) \cdot \epsilon_{e^{\pm}}(E)}.
  \label{eq:isoflux-time-averaged}
\end{equation}

$N_{e^{\pm}}(E)$ are the numbers of electrons or positrons, respectively, as determined in \cref{sec:analysis-lepton-counts-2d-fit} using the
two-dimensional fit procedure. $A_{e^{\pm}}(E)$ is the acceptance, $\epsilon_{e^{\pm}}(E)$
the signal selection efficiency and $T(E)$ the measurement time.

The efficiency $\epsilon_{e^{\pm}}(E) = \epsilon(E)$ was found to be independent of the particle species\footnote{This was
tested in dedicated studies up to \SI{200}{\GeV} where the positron ISS statistics is sufficient to perform the cross-check.} and
is defined as the product of the trigger efficiency and the ECAL estimator efficiency:

\begin{equation}
  \label{eq:efficiency}
  \epsilon(E) = \epsilon_{\text{trigger}}(E) \cdot \epsilon_{\text{ecal}}(E)
\end{equation}

Both efficiencies will be derived directly from ISS data, without involving any Monte-Carlo simulations.
The acceptance $A_{e^{\pm}}(E)$ is defined as the product of the acceptance derived from the electron or positron Monte-Carlo simulation
$A_{e^{\pm}}^{\text{MC}}(E)$ and a correction factor $(1 + \delta_{e^{\pm}}(E))$, which is used to correct for minor differences
between the prediction of the acceptance from the Monte-Carlo simulation and the ISS data:

\begin{equation}
  \label{eq:acceptance-initial}
  A_{e^{\pm}}(E) = A_{e^{\pm}}^{\text{MC}}(E) \cdot (1 + \delta_{e^{\pm}}(E)).
\end{equation}

The derivation of the correction factor $(1 + \delta_{e^{\pm}}(E))$ is explained in detail in \cref{sec:analysis-flux-time-averaged-acceptance}.
The acceptance $A_{e^{\pm}}^{\text{MC}}(E)$ is slightly different for electrons and positrons at low energies, due to differences in the
cross-sections, which will be explained in \cref{sec:analysis-flux-time-averaged-acceptance-asymmetry}.

The correction factor $(1 + \delta_{e^{\pm}}(E)) = (1 + \delta(E))$ is identical for electrons and positrons at least up to
\SI{100}{\GeV}, where the positron ISS statistics is sufficient to perform the cross-check precisely. Since no deviation is exhibited
and no differences between electrons and positrons is expected at higher energies from first principles the correction factor
is assumed to be identical for $e^{\pm}$ for all energies.

Thus the resulting acceptance $A_{e^{\pm}}(E)$ is given by:

\begin{equation}
  \label{eq:acceptance}
  A_{e^{\pm}}(E) = A_{e^{\pm}}^{\text{MC}}(E) \cdot (1 + \delta(E)).
\end{equation}

In the following sections it will be shown how $T(E)$, $A_{e^{\pm}}(E)$, $\epsilon(E)$ are derived in order to compute electron and positron fluxes
and why an unfolding procedure is necessary.

\clearpage
\subsection{ECAL estimator efficiency}
\label{sec:analysis-flux-time-averaged-ecal-estimator}

The ECAL estimator is the main handle to control the proton background in the ISS data samples.
A hard cut on the ECAL estimator significantly reduces the proton background in the positive rigidity sample, but at the same time
positrons are lost. A compromise needs to be found to keep the electron and positron signal efficiency as high as possible while rejecting
as many proton background events as necessary to keep the template fits stable.

Therefore, as already mentioned in \cref{sec:analysis-lepton-counts-2d-fit}, an iterative procedure was developed to optimize the cut on the ECAL estimator.
The procedure is illustrated in \cref{fig:ecal-estimator-optimization-procedure}. It requires an initial guess of the ECAL estimator
efficiency $\epsilon_{\text{ecal}}(E) = \epsilon_{\text{desired}}(E)$ as starting point. During the procedure the desired efficiency gets
refined until the proton rejection is sufficient in each energy bin.

\tikzstyle{decision} = [diamond, draw, fill=blue!20, text width=5em, text badly centered, node distance=3cm, inner sep=0pt]
\tikzstyle{block} = [rectangle, draw, fill=blue!20, text width=15em, text centered, rounded corners, minimum height=3em]
\tikzstyle{line} = [draw, -latex']

\begin{figure}[H]
  \centering
  \begin{tikzpicture}[scale=0.77, every node/.style={transform shape}, thick, node distance = 2.0cm, auto]
    \node [block] (init) {Choose ECAL estimator efficiency for each energy bin};
    \node [block, below of=init] (determine-cut) {Determine cut on ECAL estimator to achieve desired efficiency};
    \node [block, below of=determine-cut] (repeat-fits) {Repeat two-dimensional template fits and evaluate results};
    \node [block, right of=repeat-fits, node distance=7cm] (update) {Refine desired ECAL estimator efficiency};
    \node [decision, below of=repeat-fits] (decide) {Is proton rejection sufficient?};
    \node [block, below of=decide, node distance=3cm] (stop) {Ready};

    \path [line] (init) -- (determine-cut);
    \path [line] (determine-cut) -- (repeat-fits);
    \path [line] (repeat-fits) -- (decide);
    \path [line] (decide) -| node [near start] {no} (update);
    \path [line] (update) |- (determine-cut);
    \path [line] (decide) -- node {yes}(stop);
  \end{tikzpicture}
  \caption{Illustration of the iterative procedure used to determine the ECAL estimator cut value / efficiency.}
  \label{fig:ecal-estimator-optimization-procedure}
\end{figure}
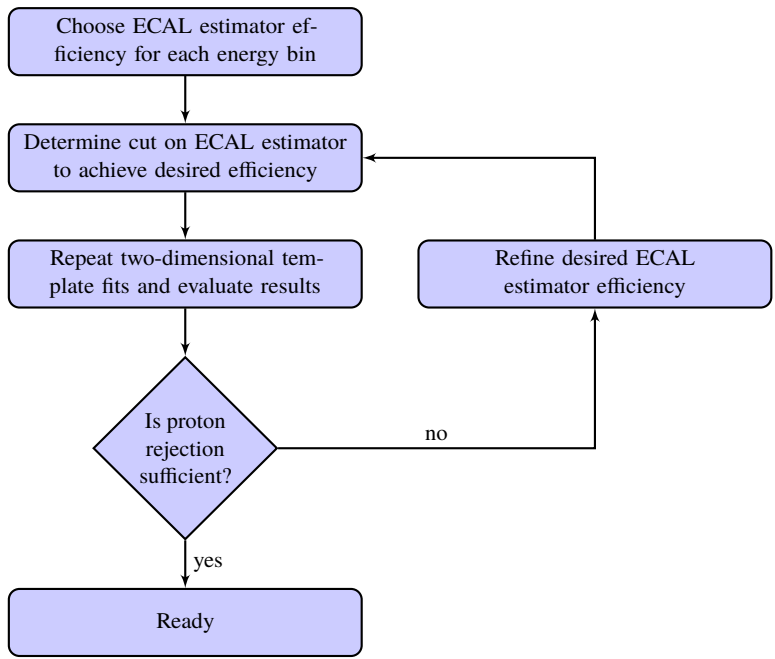

The initial guess for all energies is $\epsilon_{\text{desired}}(E) = 0.99$. If 1.0 was chosen no cut on the ECAL estimator would be applied.
Thus a value smaller than 1.0 has to be used, corresponding to a soft cut on the ECAL estimator, which already rejects the majority of the proton background
events in the positive rigidity sample.
After selecting an initial guess, multiple one-dimensional template fits in the TRD estimator are executed on the negative rigidity sample. At first a fit
is executed without imposing a cut on the ECAL estimator, to extract the number of electrons with \SI{100}{\percent} signal efficiency: $N_{e^{-}}^{0}$.
As the negative rigidity sample does not contain protons but only charge-confused protons, which are far less frequent than protons, it is possible to
extract the number of electrons with \SI{100}{\percent} signal efficiency in all energy bins without imposing a cut on the ECAL estimator.
For the positive rigidity sample the amount of protons is orders of magnitudes higher than the amount of positrons, if no ECAL estimator cut is imposed.
Thus it is not possible to extract a reliable estimate of the true number of positrons at \SI{100}{\percent} signal efficiency. This is the reason
why the procedure to determine the ECAL estimator cut value is performed only on the negative rigidity sample.

After determining the number of electrons with \SI{100}{\percent} signal efficiency - $N_{e^{-}}^{0}$ - the template fit will be repeated multiple times, gradually
hardening the cut on the ECAL estimator. For each iteration $m$ the number of electrons is extracted as $N_{e^{-}}^{m}$. The ratio of $N_{e^{-}}^{m} / N_{e^{-}}^{0}$
yields the ECAL estimator efficiency $\epsilon_{\text{ecal}}^{m}(E)$ for a given ECAL estimator cut value $\eta_{\text{ecal}}^{m}$.

\begin{figure}[H]
  \begin{subfigure}{0.5\linewidth}
    \includegraphics[width=\linewidth]{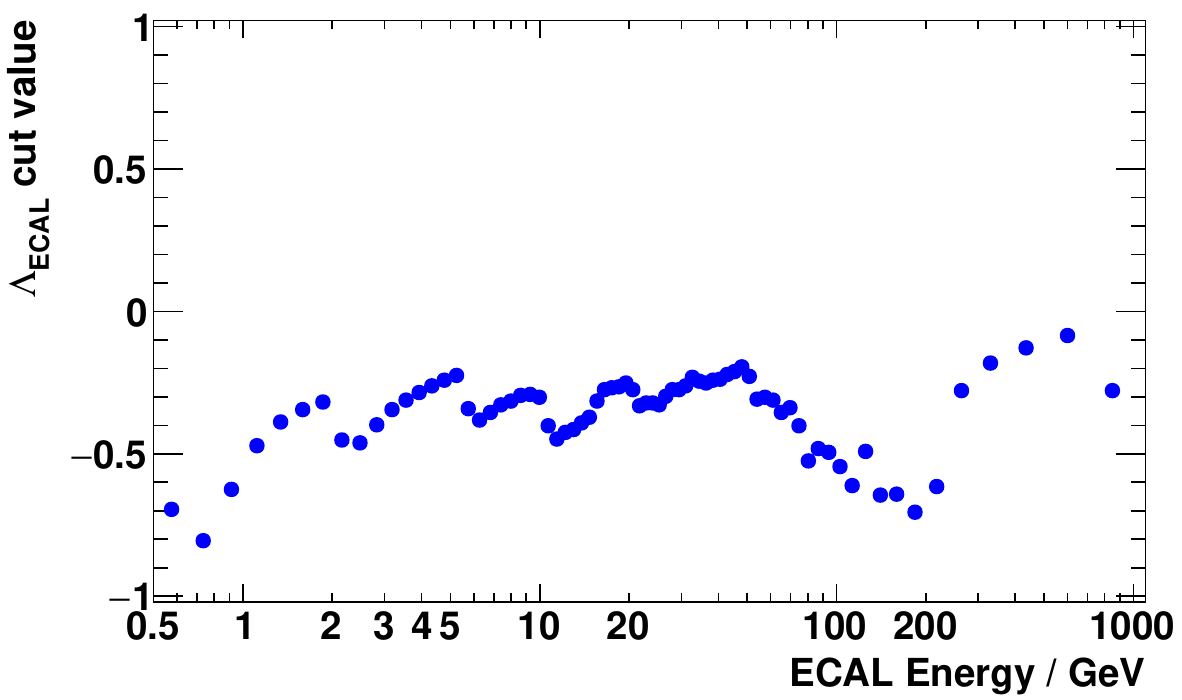}
    \caption{ECAL estimator cut value}
    \label{fig:ecal-estimator-ecal-bdt-cut-value}
  \end{subfigure}
  \hfill
  \begin{subfigure}{0.5\linewidth}
    \includegraphics[width=\linewidth]{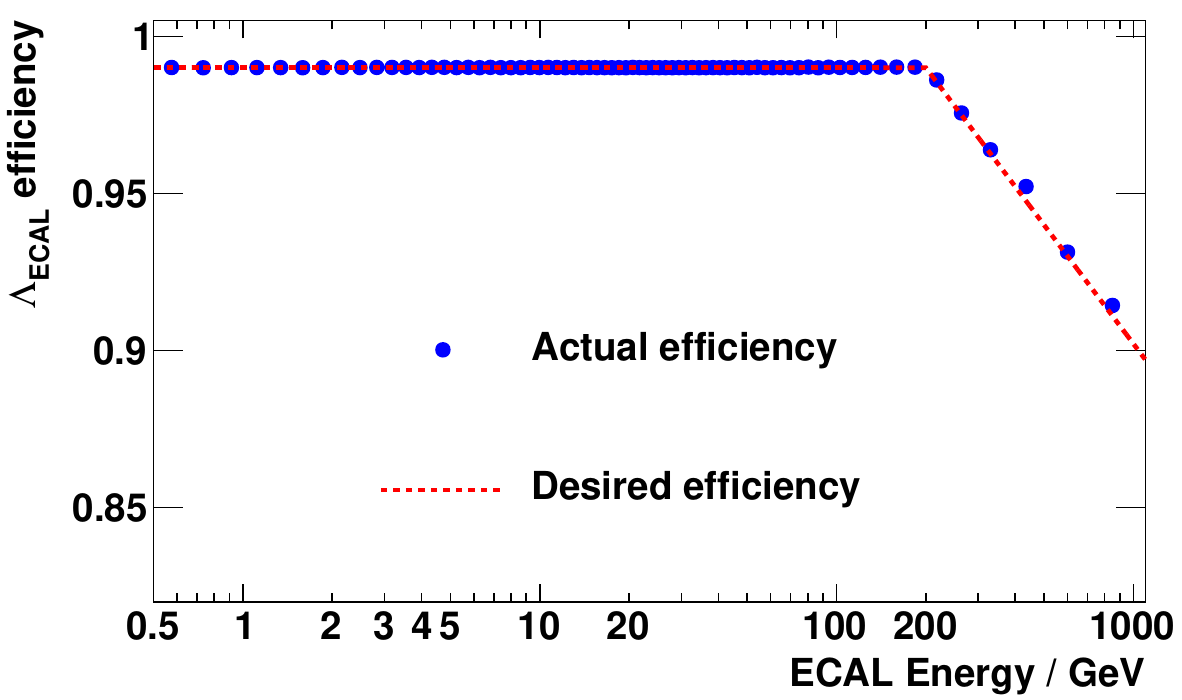}
    \caption{ECAL estimator cut efficiency}
    \label{fig:ecal-estimator-ecal-bdt-cut-efficiency}
  \end{subfigure}
  \caption{Chosen ECAL estimator efficiency for the analysis of the all-tracks sample. The left plot shows the ECAL estimator cut value that is used in the analysis. The visible structures originate in different training intervals used to train the ECAL estimator. The right plot shows the ECAL estimator efficiency $\epsilon_{\text{ecal}}(E)$ as function of energy. The dashed red line in the right plot represents the desired ECAL estimator efficiency and the blue symbols the actual efficiency in each energy bin, determined by a one-dimensional TRD template fit on the negative rigidity sample on ISS data.}
  \label{fig:ecal-estimator-ecal-bdt-cut}
\end{figure}

This allows to choose the cut value which corresponds to the desired efficiency $\epsilon_{\text{desired}}(E)$.
\Cref{fig:ecal-estimator-ecal-bdt-cut} presents the result of the optimization procedure: the chosen ECAL estimator cut value $\eta_{\text{ecal}}$
and its associated efficiency $\epsilon_{\text{ecal}}(E)$.

\Cref{fig:ecal-estimator-ecal-bdt-cut-efficiency} shows that the efficiency in the highest energy bin in the analysis is \SIapprox{70}{\percent}. The event counts
extracted from the two-dimensional fit procedure must be increased by approximately \SIapprox{30}{\percent} to account for the loss of statistics due to
the imposed cut on the ECAL estimator. Likewise all other energy bins need a correction, which decreases with decreasing energy.

It is important to note that the procedure to extract this efficiency from ISS data is necessary as the Monte-Carlo simulation cannot reproduce
the ECAL estimator efficiency in the current generation of the simulation software (B1092), as the lateral shower shape is not correctly
described, above \SIapprox{200}{\GeV}. Therefore the ISS data needs to be used to extract the efficiency of the ECAL estimator.

The presented optimization procedure is very useful to determine the cut value $\eta_{\text{ecal}}$ for the ECAL estimator. However there is no uncertainty
associated with the efficiency $N_{e^{-}} / N_{e^{-}}^{0}$ that stem from the two independent template fits. This explains the absence of errors bars on the
blue symbols in \cref{fig:ecal-estimator-ecal-bdt-cut-efficiency}. The number of events after applying the ECAL estimator cut is a subset of the number
of events without imposing a cut on the ECAL estimator.

Therefore there is no straight-forward solution to extract the efficiency with an associated uncertainty using this technique.

\clearpage
A statistically sound method to extract the ECAL estimator efficiency with uncertainty is to carry out \textit{one} template fit simultaneously
on two disjoint samples: those events that pass the ECAL estimator cut (\enquote{passed sample}) and those events that do not pass the cut (\enquote{failed sample}).

\begin{equation}
  \label{eq:ecal-bdt-efficiency-extraction}
  \begin{aligned}
    N_{e^{-}\text{, passed}} &= N_{e^{-}} \cdot      \epsilon_{\text{sig}}; & N_{e^{-}\text{, failed}} &= N_{e^{-}} \cdot (1 - \epsilon_{\text{sig}}), \\
    N_{p^{-}\text{, passed}} &= N_{p^{-}} \cdot      \epsilon_{\text{bkg}}; &N_{p^{-}\text{, failed}} &= N_{p^{-}} \cdot (1 - \epsilon_{\text{bkg}}).
  \end{aligned}
\end{equation}

Four fit parameters are used to describe both data samples: two event counts $N_{e^{-}}$, $N_{p^{-}}$ and two efficiencies $\epsilon_{\text{sig}}$,
$\epsilon_{\text{bkg}}$ - coupling the disjoint data samples (\cref{eq:ecal-bdt-efficiency-extraction}). The cut value of the ECAL estimator -
$\eta_{\text{ecal}}$ - for which the signal efficiency and uncertainty is determined by the simultaneous fit is given
by the aforementioned optimization procedure. The goal of the simultaneous fit method is to confirm the signal efficiency
as determined by the optimization procedure and to additionally quote an uncertainty on the efficiency value.

\begin{figure}[H]
  \begin{subfigure}{0.5\linewidth}
    \includegraphics[width=\linewidth]{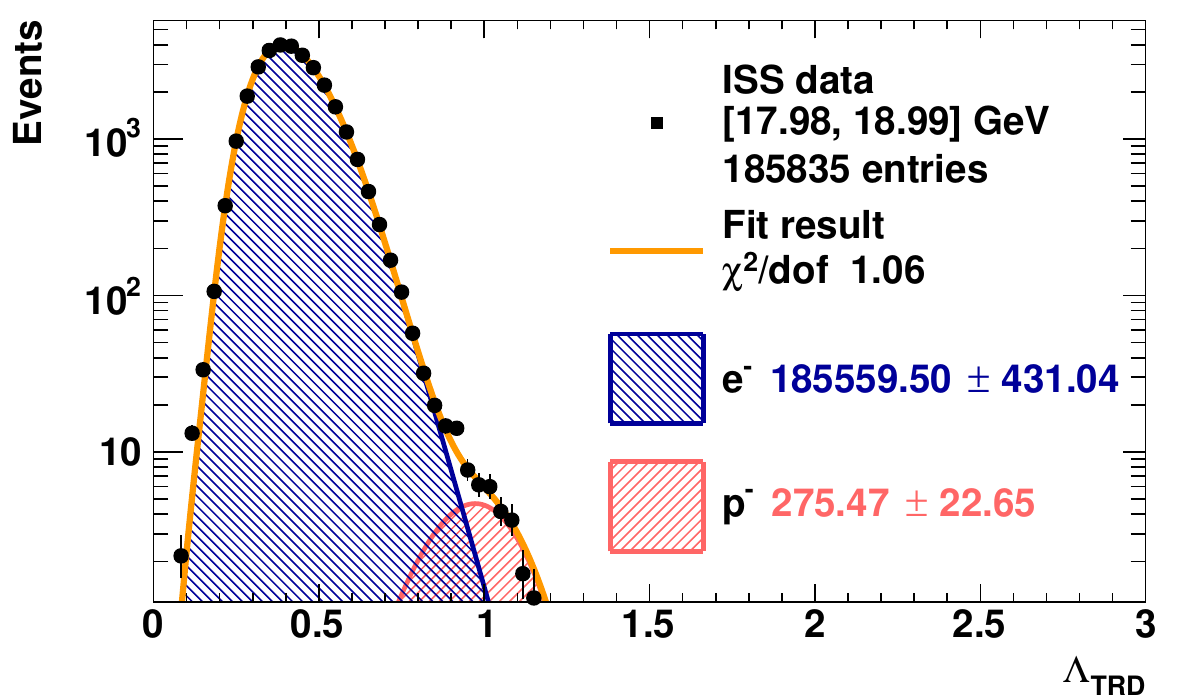}
    \caption{Events that pass the ECAL estimator cut}
  \end{subfigure}
  \hfill
  \begin{subfigure}{0.5\linewidth}
    \includegraphics[width=\linewidth]{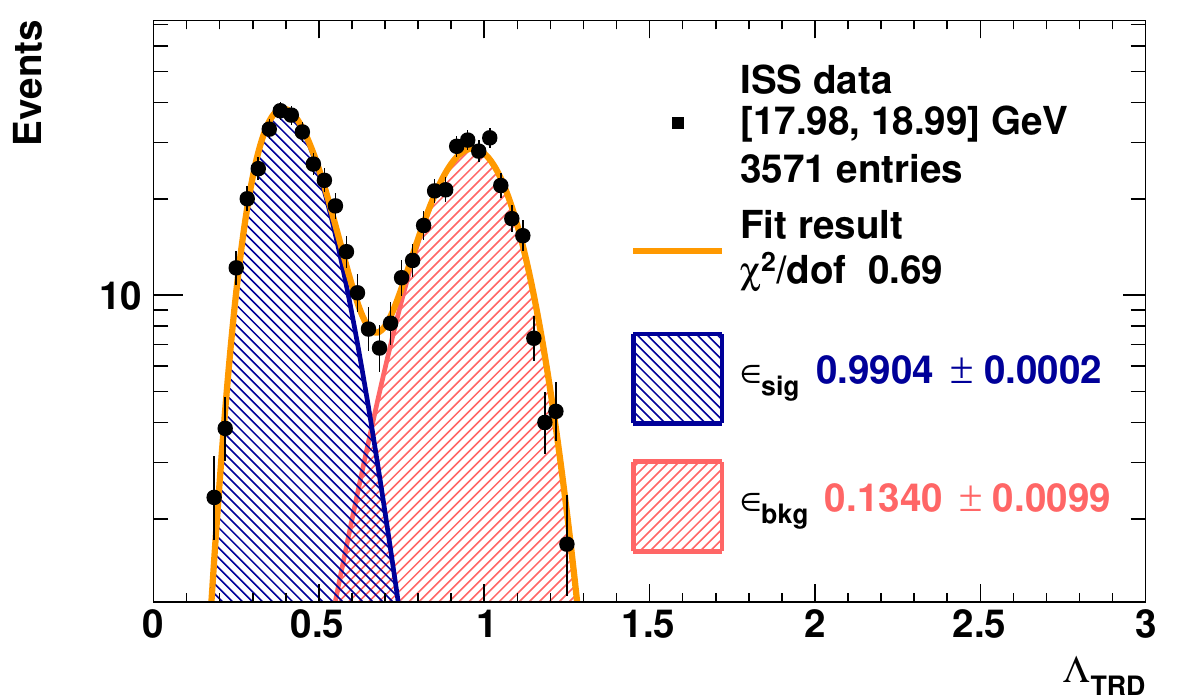}
    \caption{Events that do not pass the ECAL estimator cut}
  \end{subfigure}
  \caption{Results of the simultaneous fit to extract the ECAL estimator efficiency in the energy bin \SIrange{17.98}{18.99}{\GeV} for the all-tracks sample. The \elecColorText~area corresponds to the electron template, the \ccProtColorText~area to the charge-confused proton template and the \fitResultColorText~line to the sum of all templates.}
  \label{fig:ecal-estimator-ecal-bdt-cut-passed-failed}
\end{figure}

\Cref{fig:ecal-estimator-ecal-bdt-cut-passed-failed} shows the fit results in an example energy bin.
Both samples are described accurately, confirmed by good $\chi^2/\text{dof}$ values in both cases. The simultaneous fit procedure directly yields the
number of electrons / charge-confused protons without imposing a cut on the ECAL estimator as well as the efficiency on the signal / background simple, if the cut is applied.

To verify the absence of a bias in the simultaneous fit method, a bootstrap~\cite{Efron1979} method is used:
The idea of bootstrapping is to repeat the fit procedure $M$ times on sub-samples of the original data sample - which contains $N_{\text{all}}$ events.
The number of events $N$ in each sub-sample is generated by a random number generator following a Poisson distribution, such that
$\text{Prob}(N) = 1/N! \cdot \exp(-N_{\text{all}}) \cdot N_{\text{all}}^N$.
$N$ events are randomly selected out of the $[1, N_{\text{all}}]$ in the original data sample. These $N$ events form the sub-sample, specific for
the $m^{\text{th}}$ iteration of the bootstrap procedure.

The fit procedure is now executed $M \approx \mathcal{O}(100)$ times. In each iteration the signal efficiency is recorded. The RMS
of the signal efficiency distribution is an estimate of the true uncertainty of the signal efficiency. If the original uncertainty estimation, performed on the
full data sample, is unbiased, it should be approximately equal to the RMS of the signal efficiency distribution. Furthermore the
mean value of the signal efficiency distribution should be approximately equal to the results of a single fit, obtained on the full data sample, within the quoted uncertainties.

\begin{figure}[H]
  \centering
  \includegraphics[width=0.7\linewidth]{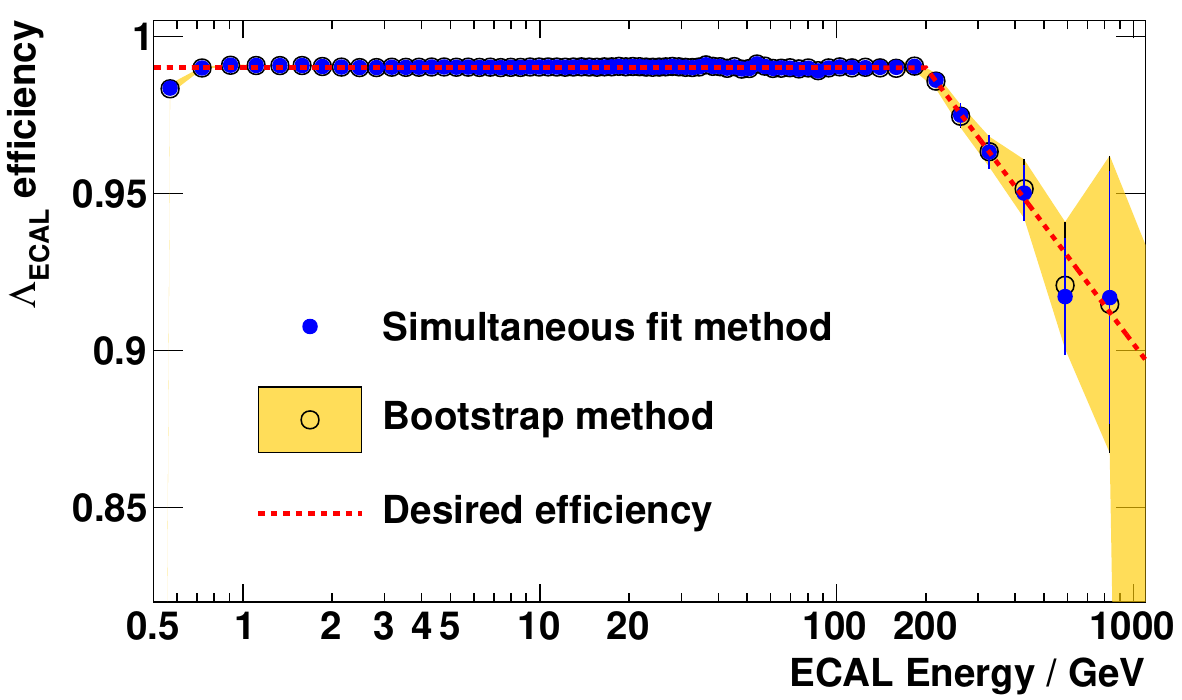}
  \caption{Final result showing the ECAL estimator efficiency $\epsilon_{\text{ecal}}(E)$ as function of energy derived using two different methods. The dashed red line represents the desired ECAL estimator efficiency and the blue symbols the actual efficiency in each energy bin, determined by a simultaneous fit on the passed/failed data sample. The black open symbols belong to the efficiency determined by the bootstrap method and the orange band represents the associated uncertainty.}
  \label{fig:ecal-estimator-ecal-bdt-cut-final-test}
\end{figure}

\Cref{fig:ecal-estimator-ecal-bdt-cut-final-test} shows a comparison between the desired efficiency, the efficiency from the simultaneous fit
method and the efficiency from the bootstrap method. All results are in excellent agreement. The efficiency and uncertainty presented here is
unbiased and a correct measure of the ECAL estimator efficiency on ISS data.

\subsection{Trigger efficiency}
\label{sec:analysis-flux-time-averaged-trigger}

\begin{figure}[H]
  \centering
  \includegraphics[width=0.7\linewidth]{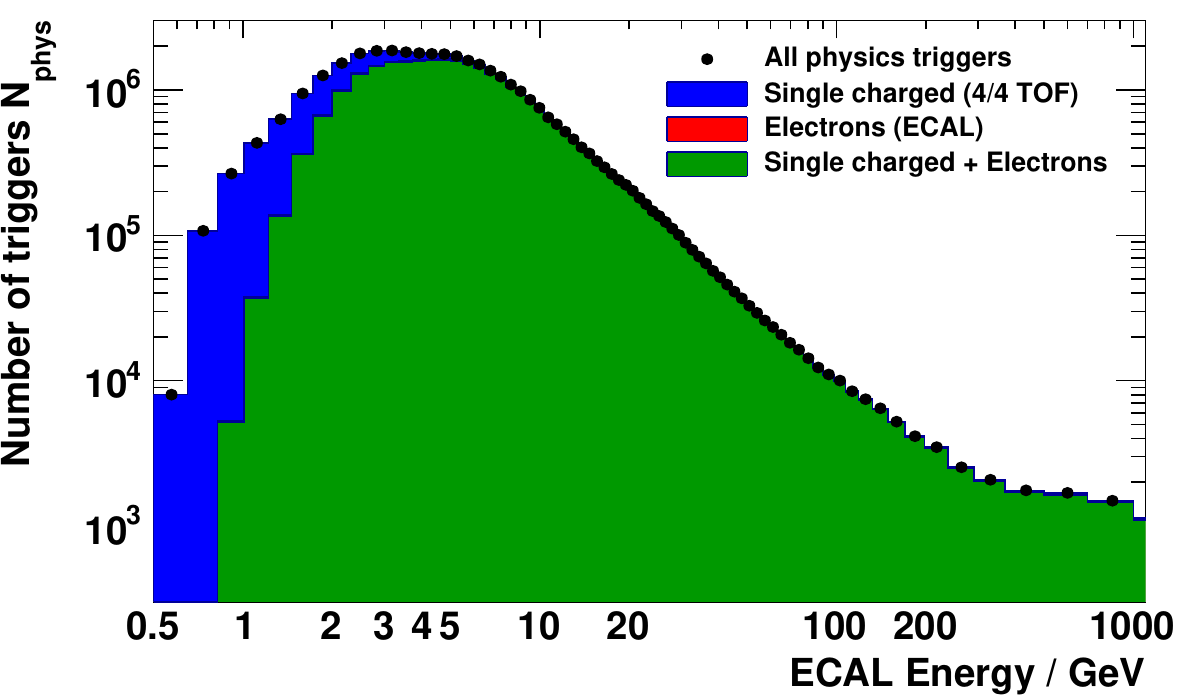}
  \caption{Decomposition of the physics triggers as stacked histograms. Two main trigger branches form the physics trigger for the $e^{\pm}$ analysis: the \textbf{Single charged} trigger and the \textbf{Electrons} trigger.}
  \label{fig:physics-trigger-decomposition}
\end{figure}

In \cref{sec:detector-trigger} the AMS-02 logic was described in detail. In the electron and positron analysis the data sample is
mainly triggered by the \textbf{Single charged} and the \textbf{Electrons} trigger, as shown in \cref{fig:physics-trigger-decomposition}.

The efficiency for the \textbf{Electrons} trigger rapidly decreases below \SI{2}{\GeV}, thus $e^{\pm}$ at these energies are only triggered
by the \textbf{Single charged} trigger. For all energies above \SI{2}{\GeV} the \textbf{Electrons} trigger is not visible on a logarithmic
scale since most events are triggered using both trigger branches: the combined \textbf{Single charged + Electrons} trigger (dark green).

The trigger efficiency $\epsilon_{\text{trigger}}(E)$ is approximated by the ratio between all events triggered by any of the physics triggers
and all available triggers, including the unbiased triggers.

By intention an unbiased event should only be triggered either via the \textbf{Unbiased charge} trigger or via the \textbf{Unbiased EM} trigger, otherwise the prescaling factor to
use is undefined. The analysis shows that in the final data sample for the electron and positron analysis there are $N_{\text{phys}}$ events with physics trigger,
$N_{\text{tof}}$ events with an \textbf{Unbiased charge} trigger (prescaled by $f_{\text{tof}} = 100$) and $N_{\text{ecal}}$ events with an \textbf{Unbiased EM}
trigger (prescaled by $f_{\text{ecal}} = 1000$). Due to a bug in the trigger hardware there is a non-negligible amount of events that have both trigger bits set: $N_{\text{tof+ecal}}$. The prescaling factor
$f_{\text{tof+ecal}}$ for these events can be calculated via probability theory:

\begin{equation}
  \label{eq:trigger-tof-ecal-prescaling}
  \begin{aligned}
    p(\text{tof}~\cup~\text{ecal}) &= p(\text{tof}) + p(\text{ecal}) - p(\text{tof}~\cap~\text{ecal}) = \frac{1}{f_{\text{tof}}} + \frac{1}{f_{\text{ecal}}} - \frac{1}{f_{\text{tof}} \cdot f_{\text{ecal}}}, \\
    f_{\text{tof+ecal}}            &= 1 / p(\text{tof}~\cup~\text{ecal}) \approx 90.99
  \end{aligned}
\end{equation}

\begin{figure}[H]
  \begin{subfigure}{0.33\linewidth}
    \includegraphics[width=\linewidth]{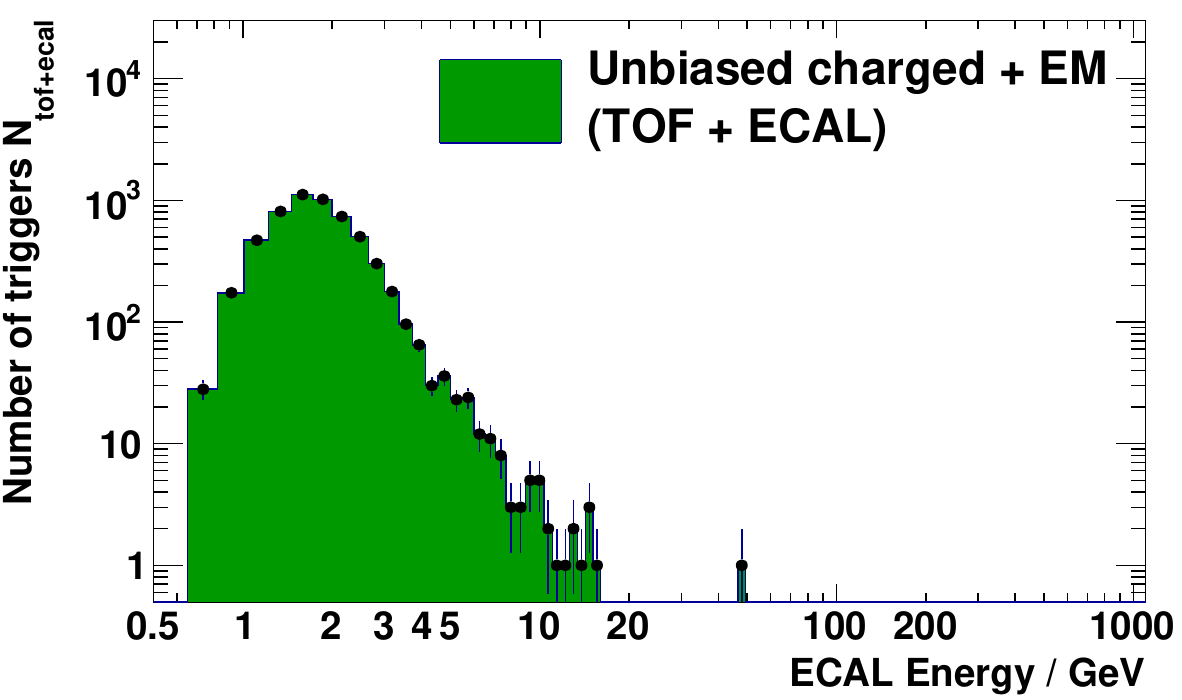}
    \caption{\textbf{Unbiased charged + EM} triggers}
  \end{subfigure}
  \begin{subfigure}{0.33\linewidth}
    \includegraphics[width=\linewidth]{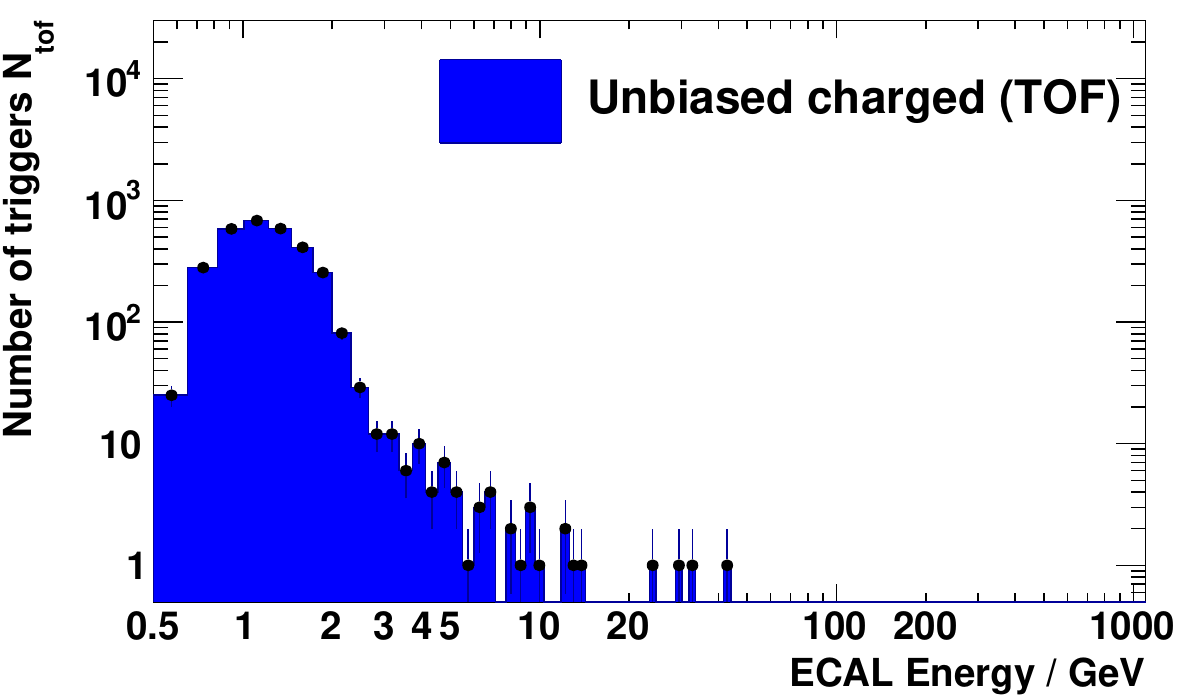}
    \caption{\textbf{Unbiased charged} triggers}
  \end{subfigure}
  \begin{subfigure}{0.33\linewidth}
    \includegraphics[width=\linewidth]{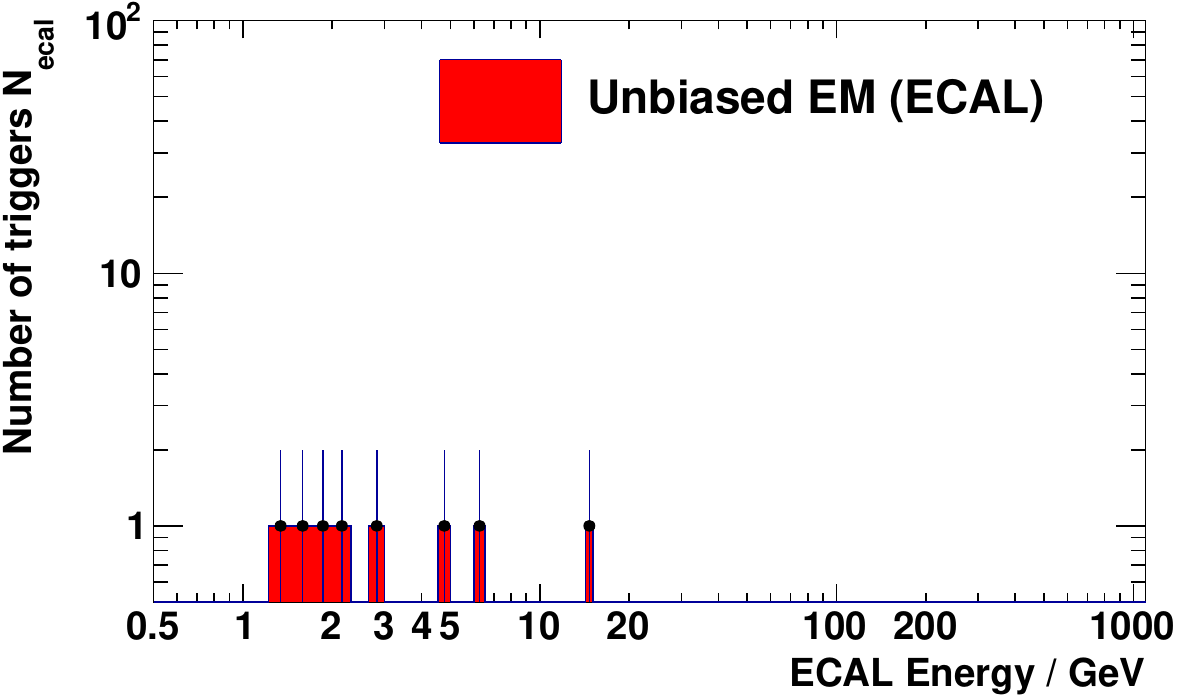}
    \caption{\textbf{Unbiased EM} triggers}
  \end{subfigure}
  \caption{Distribution of the unbiased triggers as function of energy. The most abundant unbiased triggers are the \textbf{Unbiased charged + Unbiased EM} triggers, followed by the \textbf{Unbiased charged} triggers. The least abundant triggers are the \textbf{Unbiased EM} triggers.}
\label{fig:unbiased-trigger-decomposition}
\end{figure}

\Cref{fig:unbiased-trigger-decomposition} shows the distribution of the unbiased triggers as function of energy.

After knowing the prescaling factors and the number of unbiased and physics triggers as function of energy, the trigger efficiency can be approximated from ISS data:

\begin{equation}
  \label{eq:trigger-efficiency}
  \epsilon_{\text{trigger}}(E) = \frac{N_{\text{phys}}}{N_{\text{phys}} + f_{\text{tof}} \cdot N_{\text{tof}} + f_{\text{ecal}} \cdot N_{\text{ecal}} + f_{\text{tof+ecal}} \cdot N_{\text{tof+ecal}}}.
\end{equation}

\begin{figure}[H]
  \centering
  \includegraphics[width=0.7\linewidth]{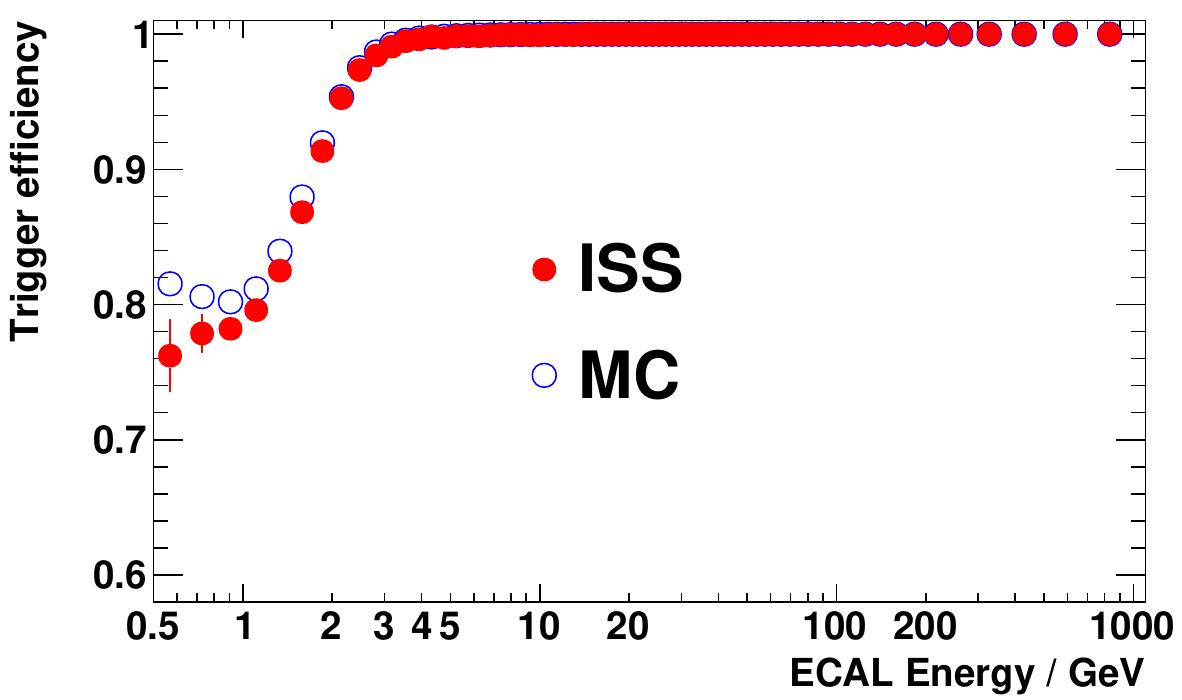}
  \caption{Comparison of the trigger efficiency $\epsilon_{\text{trigger}}(E)$ as function of energy between ISS data and Monte-Carlo simulation. The red symbols represent the trigger efficiency determined from ISS data according to \cref{eq:trigger-efficiency}. The blue symbols show the trigger efficiency determined by the Monte-Carlo simulation, by counting the number of unbiased and physics triggers.}
  \label{fig:trigger-iss-mc-comparison}
\end{figure}

The results are shown in \cref{fig:trigger-iss-mc-comparison} including a comparison with the Monte-Carlo simulation. There are discrepancies below
\SI{2}{\GeV}, where the Monte-Carlo simulation predicts a higher efficiency compared to ISS data, due to an imperfect simulation of the \textbf{Unbiased EM} trigger.
Overall the results are in good agreement.

\subsection{Measuring time}
\label{sec:analysis-flux-time-averaged-measuring-time}

To determine the measuring time the amount of seconds between the first event of the data taking period $T_{0}$
and the last event of the data taking period $T_{1}$ is counted. This defines the \enquote{exposure time},
where AMS-02 was operational and could record events. Afterwards each second will be tested if it is usable
for physics analysis, by applying the detector quality cuts (\cref{sec:analysis-data-selection-detector-quality-cuts}).
The amount of seconds that pass the detector quality cuts define the time period for which the \enquote{measuring time} $T(E$)
will be computed.

\begin{figure}[H]
  \begin{subfigure}{0.50\linewidth}
    \includegraphics[width=\linewidth]{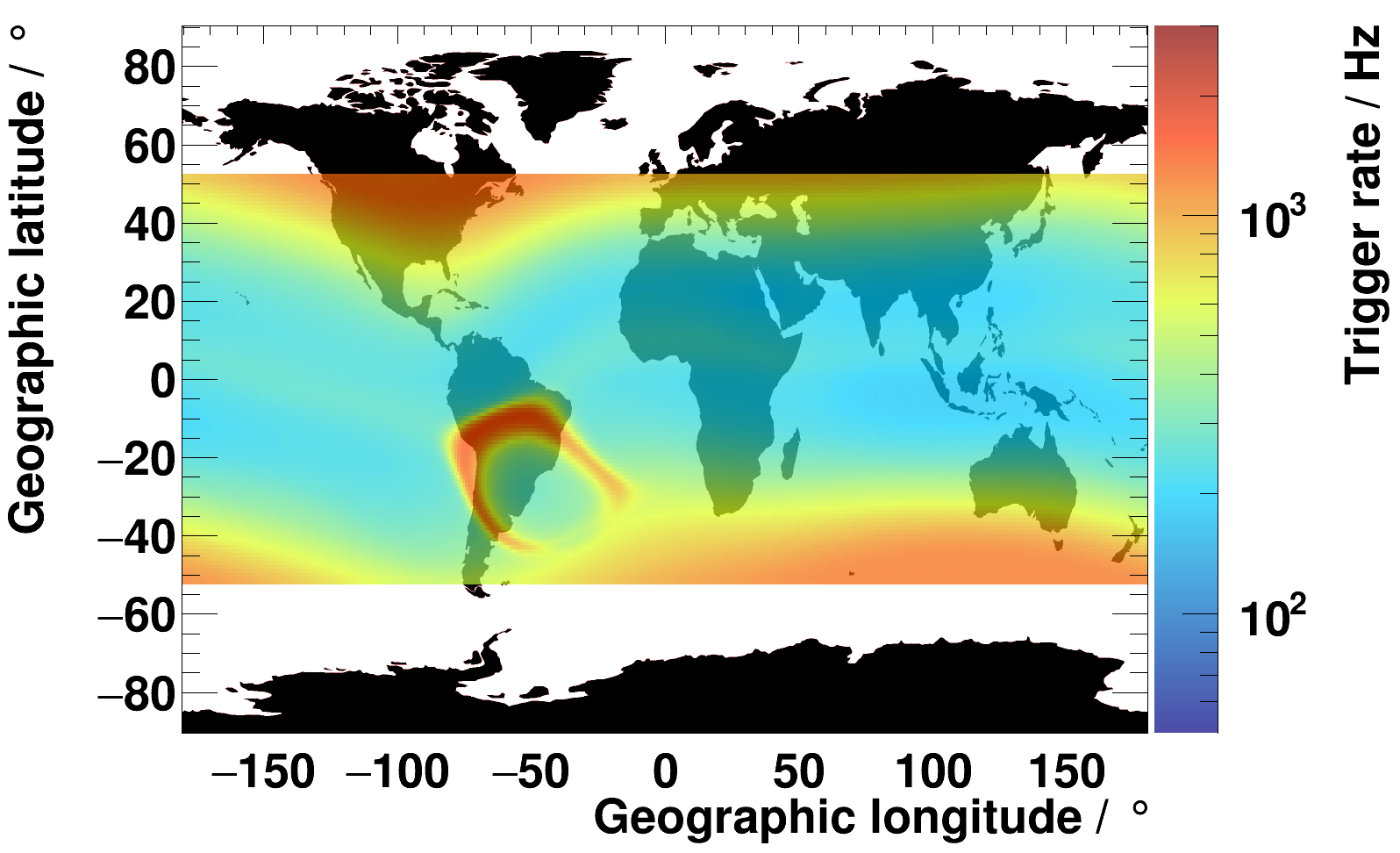}
    \caption{Trigger rate vs. ISS position}
  \end{subfigure}
  \hfill
  \begin{subfigure}{0.50\linewidth}
    \includegraphics[width=\linewidth]{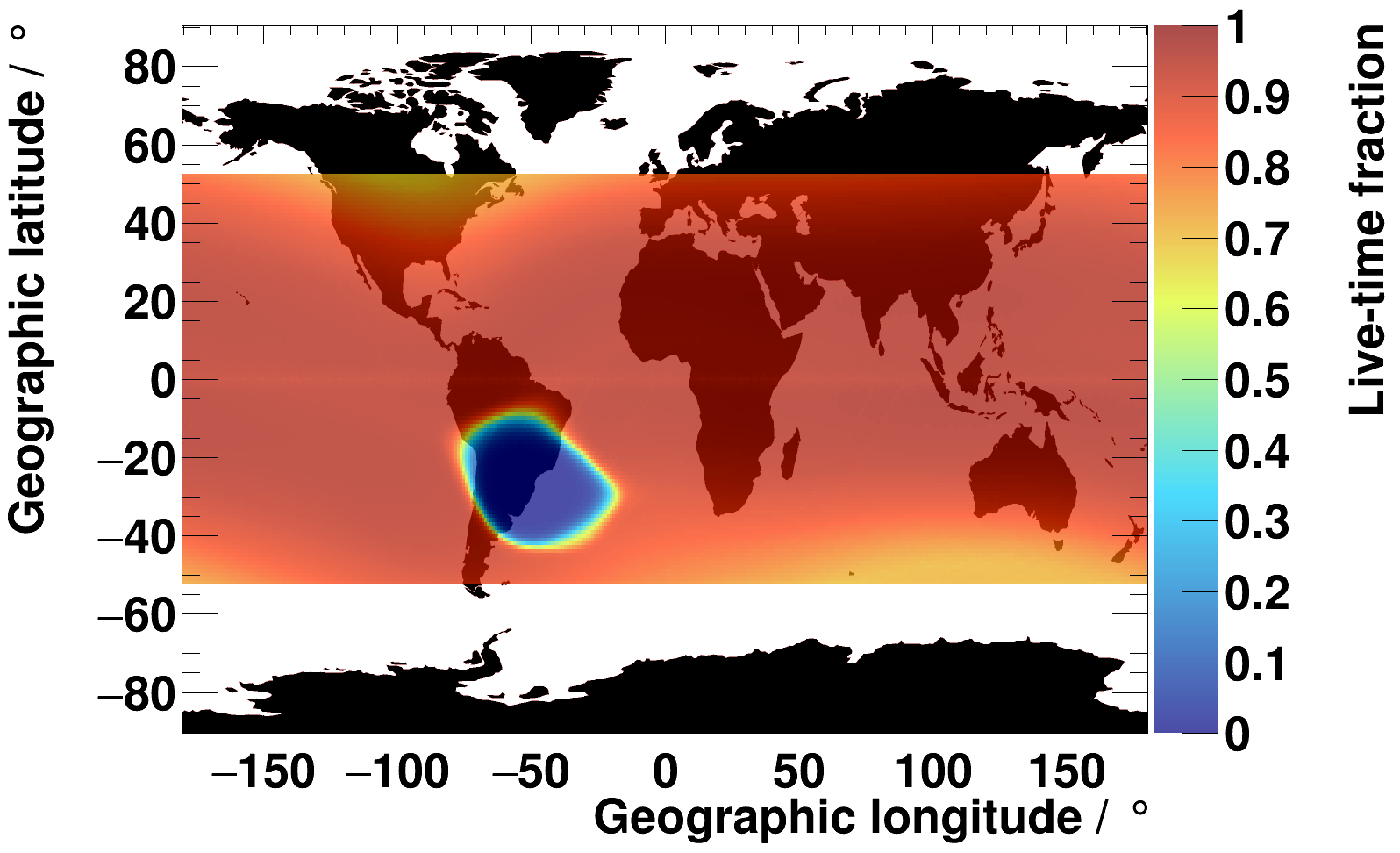}
    \caption{Live-time fraction vs. ISS position}
  \end{subfigure}
  \caption{The left plot shows the trigger rate of the experiment vs. ISS position and the right plot shows the live-time fraction of the experiment vs. ISS position. The live-time fraction is exceptionally low at the \gls{SAA}~\cite{Kurnosova1962} due to an intense flux of low-energy particles, filling the detector.}
  \label{fig:measuring-time-trigger-rate-and-live-time-vs-iss-position}
\end{figure}

The live-time fraction $f_{\text{live-time}}(t)$, defined in \cref{sec:detector-trigger}, needs to be taken into
account for each of the seconds. \Cref{fig:measuring-time-trigger-rate-and-live-time-vs-iss-position} shows the trigger rate and live-time as
function of the ISS position, projected to geographic longitude/latitude.

In certain geographic locations - like the \gls{SAA}~\cite{Kurnosova1962} - the live-time fraction can become exceptionally low and the trigger
rate increases. The live-time fraction is correlated with the geomagnetic cut-off rigidity $R_{c}$: if $R_{c}$ is low (e.g. near the poles) many more low energetic
particles can reach AMS-02 and thus the live-time fraction decreases, due to an increase in the trigger rate. The detector quality cuts reject periods where the
live-time fraction is below \SI{50}{\percent}, which effectively masks out the \gls{SAA} for analysis, where the flux of low-energy particles is intense.

The energy-independent measuring time is equal to the sum of the live-time fraction of all seconds in the measuring time period:

\begin{equation}
  \label{eq:measuring-time-without-cutoff}
  T = \sum_{t = T_{0}}^{T_{1}} f_{\text{live-time}}(t).
\end{equation}

\begin{figure}[H]
  \centering
  \includegraphics[width=0.75\linewidth]{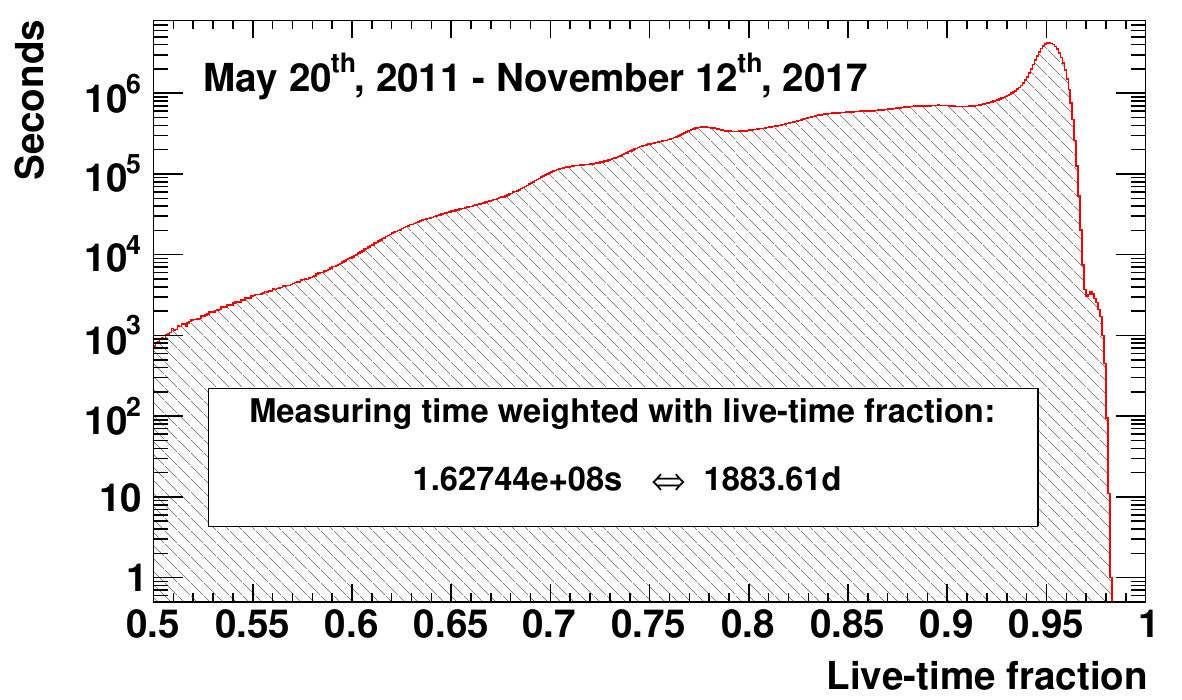}
  \caption{The plot shows the distribution of the live-time fraction of the experiment for all seconds that form the measuring time period. The distribution peaks at \SI{95}{\percent} with a long tail towards smaller live-time fractions. Live-time fractions below \SI{50}{\percent} are excluded for analysis by the detector quality cuts.}
  \label{fig:measuring-time-live-time-distribution}
\end{figure}

\Cref{fig:measuring-time-live-time-distribution} shows the live-time fraction distribution of all seconds that form the measuring time.
For the period \textbf{May~\nth{20},~2011 - November~\nth{12},~2017} the measuring time - weighted by the live-time fraction - is \textbf{1883.61 days}.

The energy dependence of the measuring time $T(E)$ arises from the geomagnetic cut-off, as presented in \cref{sec:cosmic-ray-propagation-magnetosphere}.
It was shown that the geomagnetic cut-off rigidity $R_{c}$ changes as function of the geomagnetic longitude/latitude.

Particles below a certain rigidity - $R_{c}$ - cannot penetrate the magnetosphere which surrounds the earth and thus cannot be measured with AMS-02.
The geomagnetic cut-off rigidity $R_{c}$ changes as function of longitude and latitude. At the pole regions the cut-off rigidity is minimal, due
to the direction of the magnetic fields lines, which are directed towards earth. This means that low-rigidity particles
can be detected effectively only at the pole regions, not in equatorial regions. This leads to an energy dependence of the measuring time $T(E)$.
Since the maximum value of the geomagnetic cut-off rigidity $R_{c}$ is \SIapprox{25}{\giga\volt}, the energy dependence of the measuring time vanishes
above this energy.

For the $e^{\pm}$ analysis, only events are allowed where the measured ECAL energy $E_{\text{measured}}$ exceeds the maximum geomagnetic cut-off rigidity
$R_{c}$ times a safety factor $f_\text{safety}$ in a \SI{25}{\degree} field-of-view. The requirement imposed for the analysis is:
$E_{j\text{, low}} > f_\text{safety} \cdot R_{c}$, where $E_{j\text{, low}}$ denotes the minimum bin energy of bin $j$, in which
the energy $E_{\text{measured}}$ was classified.
In this work the cut-off rigidity is calculated using the Størmer formula\footnote{Several methods were tested to calculate the geomagnetic cut-off
rigidity, such as the IGRF~\cite{Thebault2015} cut-off evaluated using a backtracing technique, as described in Ref.~\cite{Fiandrini2015}. All results are compatible and the Størmer approach was chosen, which is easy to compute.} (\cref{sec:cosmic-ray-propagation-magnetosphere}).

For each energy bin $j$ of the analysis the energy-dependent measuring time $T_{j}$ can be calculated:

\begin{equation}
  \label{eq:measuring-time-with-cutoff}
  \begin{aligned}
    T_{j} = \sum_{t = T_{0}}^{T_{1}}
    \begin{cases}
      f_{\text{live-time}}(t) & E_{j\text{, low}} > f_\text{safety} \cdot R_{c} \\
      0                       & \text{otherwise}
    \end{cases}
  \end{aligned}
\end{equation}

\begin{figure}[H]
  \centering
  \includegraphics[width=0.9\linewidth]{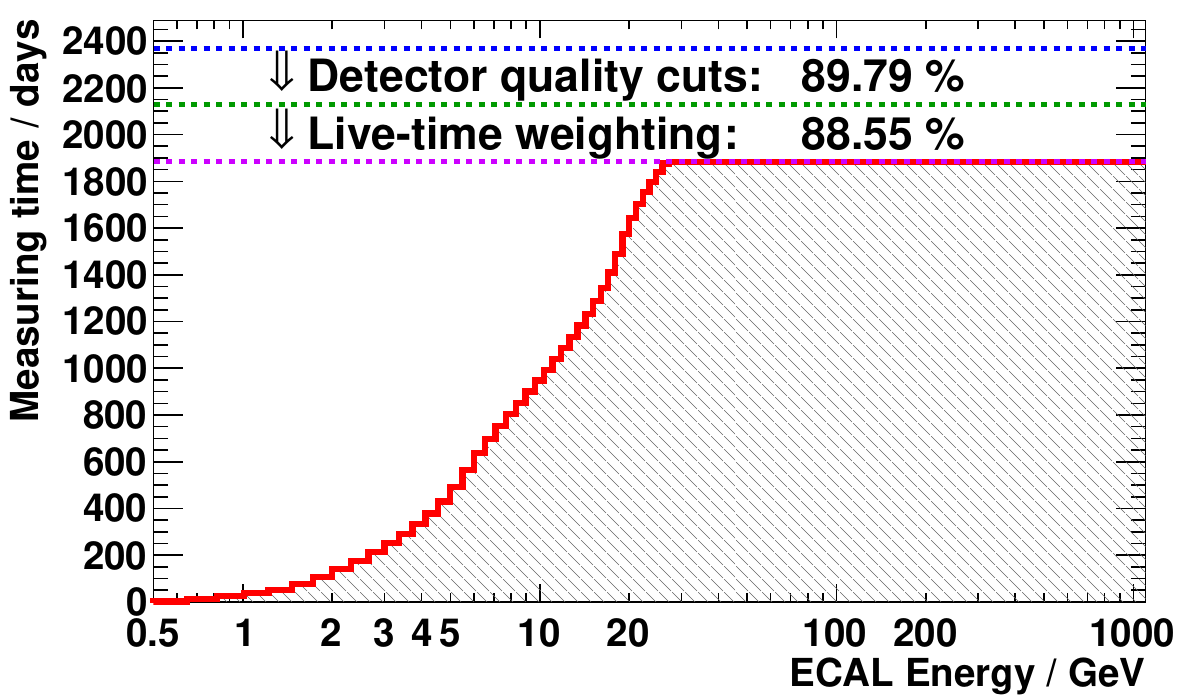}
  \caption{Visualization of the measurement time $T(E)$ as function of energy. The dashed blue line marks the nominal exposure time for the period \textbf{May~\nth{20},~2011 - November~\nth{11},~2017}. The dashed green line denotes the measuring time after applying the detector quality cuts. The dashed violet line marks the energy-independent measuring time, after accounting for the live-time fraction. Finally the \textbf{red line} represents the measuring time $T(E)$ after taking the geomagnetic cut-off into account.}
  \label{fig:measuring-time-decomposition}
\end{figure}

The measuring time $T_{j}$ for each energy bin $j$ can be combined into a single histogram $T(E)$, which is shown in
\cref{fig:measuring-time-decomposition}. The detector quality cuts, the live-time fraction of the experiment and most importantly
the geomagnetic cut-off, which is the reason for the energy dependence, are all taken into account.

The exposure time for the period \textbf{May~\nth{20},~2011 - November~\nth{11},~2017} equals to \textbf{\textcolor{blue}{2368.94 days}}. After applying the
detector quality cuts the remaining measuring time equals to \textbf{\textcolor{darkGreen}{2127.17 days}}. When taking the live-time
into account a total of \textbf{\textcolor{violet}{1883.61 days}} remain. This number remains constant above \SIapprox{25}{\GeV}.

\clearpage
\subsection{Acceptance}
\label{sec:analysis-flux-time-averaged-acceptance}

The acceptance is one of the most important ingredients in the $e^{\pm}$ flux analysis: a large acceptance corresponds to a large number of events
that can be collected. Thus the effective acceptance - including all selection efficiencies - should be as high as possible. The choice of subdetectors
that participate in the data analysis impose a limit on the maximum acceptance that can be achieved: the geometrical acceptance $A_{\text{geom}}$.
The geometrical acceptance can be calculated either using a standalone simulation that accurately models the geometry of all subdetectors, or by
using the AMS-02 Monte-Carlo simulation, based on \textsc{Geant4}~\cite{Agostinelli2003,Allison2006,Allison2016}.

The full AMS-02 model is placed inside a hypothetical cube with an edge length of \SI{3.9}{\meter}. Above the top surface $S_{\text{generator}}$ an isotropic
flux is generated of so-called \textbf{\enquote{charged geantinos}}. Charged geantinos are artificial particles that can be used as geometrical probe. They
are only transported through the AMS-02 model, but do not interact. The direction and rigidity is chosen randomly above the top surface to simulate an isotropic flux.

According to Sullivan~\cite{Sullivan1971} the only ingredients needed to compute the geometrical acceptance are the number of events that triggered the
experiment $N_{\text{triggered}}$ and the number of generated events $N_{\text{generated}}$ above the top surface $S_{\text{generator}} = 3.9 \cdot \SI{3.9}{\meter\squared}$.
Using these numbers the geometrical acceptance can be computed according to:

\begin{equation*}
  A_{\text{geom}} = \pi \cdot S_{\text{generator}} \cdot \frac{N_{\text{triggered}}}{N_{\text{generated}}}.
\end{equation*}

No selection cuts are applied in order to extract $N_{\text{triggered}}$ from the Monte-Carlo simulation for this study, but it is required that the geantino
can geometrically reach the TRD, the TOF, the tracker and the ECAL. The resulting geometrical acceptance is $A_{\text{geom}}$~\SIapprox{735}{\centi\meter\squared\steradian}.
Thus the acceptance $A_{e^{\pm}}^{\text{MC}}(E)$ is strictly smaller than or equal to $A_{\text{geom}}$, since $A_{e^{\pm}}^{\text{MC}}(E)$ includes the efficiency of several selection cuts
that further reduce the acceptance.

$A_{e^{\pm}}^{\text{MC}}(E)$ is derived in the same way as the geometrical acceptance, but using electrons or positrons instead of charged geantinos.
The main difference is that interactions are allowed and all selection cuts are applied, except the \textbf{\enquote{physics trigger}} requirement and the
final ECAL estimator cut (derived using the iterative procedure described in \cref{sec:analysis-flux-time-averaged-ecal-estimator}). The physics trigger
requirement is not part of the acceptance as its efficiency $\epsilon_{\text{trigger}}(E)$ is determined directly from ISS data and thus should not enter the
acceptance, to avoid double-counting. The ECAL estimator cut is not applied in the Monte-Carlo simulation, since its sole purpose is to reduce proton background,
which is not present in the electron Monte-Carlo simulation. Furthermore its efficiency $\epsilon_{\text{ecal}}$(E) is determined directly from the ISS data and
thus its efficiency is not part of the acceptance.

\begin{figure}[H]
  \centering
  \includegraphics[width=0.9\linewidth]{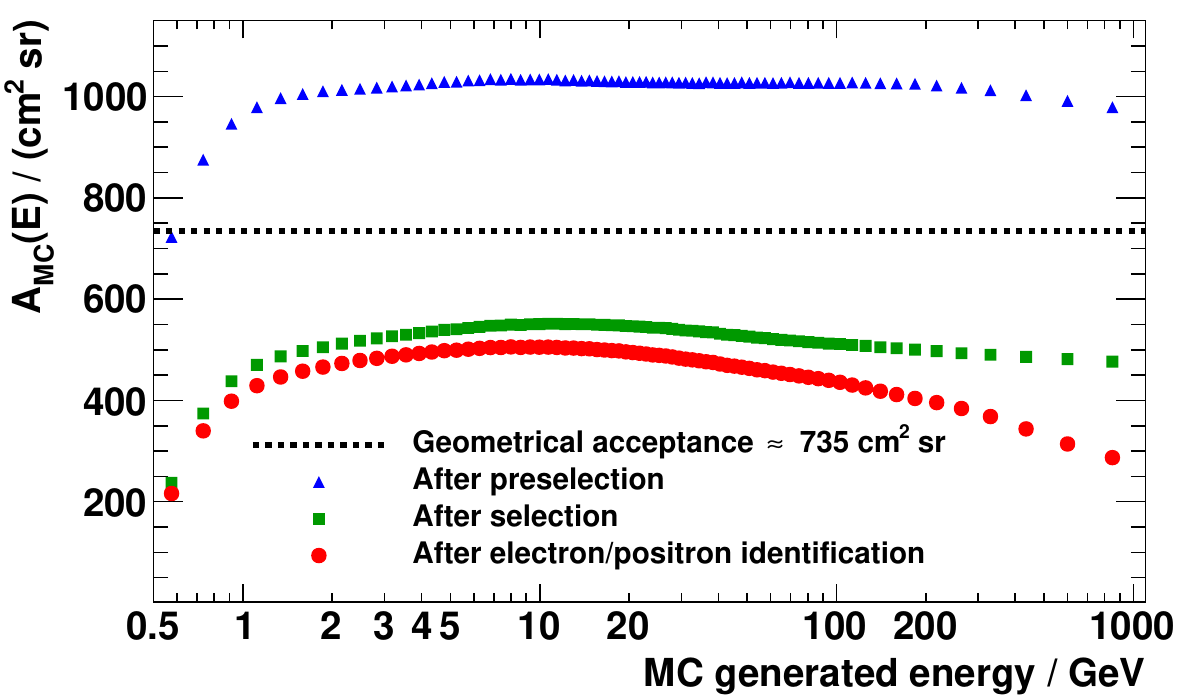}
  \caption{Determination of the acceptance $A_{e^{-}}^{\text{MC}}(E)$ from the electron Monte-Carlo simulation as function of the generated energy of the simulated particles. The black dashed line shows the geometrical acceptance $A_{\text{geom}}$. The other symbols show the acceptance after applying the preselection, selection and $e^{\pm}$ identification cuts. The acceptance after applying only the preselection cuts is higher than the geometrical acceptance, since in the preselection tracker requirements are not imposed yet, unlike in the definition of the geometrical acceptance. After applying the selection cuts, the acceptance curve is smaller than the geometrical acceptance.}
  \label{fig:acceptance-without-corrections}
\end{figure}

\Cref{fig:acceptance-without-corrections} shows the acceptance $A_{e^{-}}^{\text{MC}}(E)$ from the electron Monte-Carlo simulation as function of the generated
energy of the simulated particles. The positron acceptance $A_{e^{+}}^{\text{MC}}(E)$ is slightly higher than the electron acceptance at low energies, due to differences
in the scattering cross-sections, which will be explained in \cref{sec:analysis-flux-time-averaged-acceptance-asymmetry}.

To ensure the validity of the acceptance $A_{e^{-}}^{\text{MC}}(E)$ from the electron Monte-Carlo simulation every cut efficiency needs to be verified using ISS data.
The goal is to derive a correction factor $(1 + \delta(E))$, which absorbs any differences between the ISS data and the Monte-Carlo simulation, to ensure
that the acceptance describes the ISS data properly.

In order to determine the correction $(1 + \delta^{c}(E))$, a negative rigidity sample (\textbf{\enquote{tag sample}}) is selected for every cut $c$ using information from the
detectors unrelated to that cut. The efficiency of each cut $c$ is compared between ISS data and electron Monte-Carlo simulation, by taking the ratio ISS efficiency over MC efficiency.
If the ratio differs from unity, it is parameterized using a simple model, e.g. a constant or a straight line. The best matching model is fit to the ratio. Afterwards it will be tested
if the deviation from unity is significant, according to the uncertainty of the fit parameters.

Since all correction factors $(1 + \delta^{c}(E))$ are uncorrelated by a careful construction of the tag samples, the final correction $(1 + \delta(E))$ is the product of all
individual significant correction factors $(1 + \delta^{c}(E))$:

\begin{equation}
  \label{eq:acceptance-correction-prod}
  1 + \delta(E) = \prod_{c} (1 + \delta^{c}(E)).
\end{equation}

All tag cuts used to prepare the tag samples, for each cut $c$, are listed in \cref{sec:appendix-tag-and-probe-cuts}.
Seven cuts exhibit a significant ISS/Monte-Carlo deviation and in all cases two simple models were sufficient to parameterize the ISS over Monte-Carlo cut efficiency ratio:

\begin{enumerate}[(a)]
  \item\textbf{Straight line in logarithmic scale}\hfill\\
    Parameters: $p_{0} = \text{Offset}$, $p_{1} = \text{Slope}$

    The correction is significant, if $\abs{1 - p_{0}} \geq 2 \cdot \sigma_{p_{0}}$ or $\abs{1 - p_{0}} \geq 0.005$.

    \begin{equation}
      \label{eq:straight-line-in-log-scale}
      m_{1}(E; p_{0}, p_{1}) = p_{0} + p_{1} \cdot \logten{E}
    \end{equation}

  \item\textbf{Two straight lines with a break in logarithmic scale}\hfill\\
    Parameters: $p_{0} = \text{Offset of \nth{1} line}$, $p_{1} = \text{Slope of \nth{1} line}$, $p_{2} = \text{Slope of \nth{2} line}$, $p_{3} = \text{Break}$

    The correction is always considered significant

    \begin{equation}
      \label{eq:straight-line-with-break-in-log-scale}
      m_{2}(E; p_{0}, p_{1}, p_{2}, p_{3}) =
        \begin{cases}
          p_{0} + p_{1} \cdot \logten{E}                                                  & E \leq p_{3} \\
          \left(m_{2}(p_{3}) - p_{2} \cdot \logten{p_{3}}\right) + p_{2} \cdot \logten{E} & E > p_{3}
        \end{cases}
    \end{equation}
\end{enumerate}

The seven cuts with a significant ISS/Monte-Carlo correction will be presented in the following:

\bigskip
\noindent
{\large1. Preselection cut: \textbf{\enquote{At least one useful TRD track}} \&} \\
{\large2. Preselection cut: \textbf{\enquote{At least one useful TOF cluster combination}}} \\

\begin{figure}[H]
  \begin{subfigure}{0.50\linewidth}
    \includegraphics[width=\linewidth]{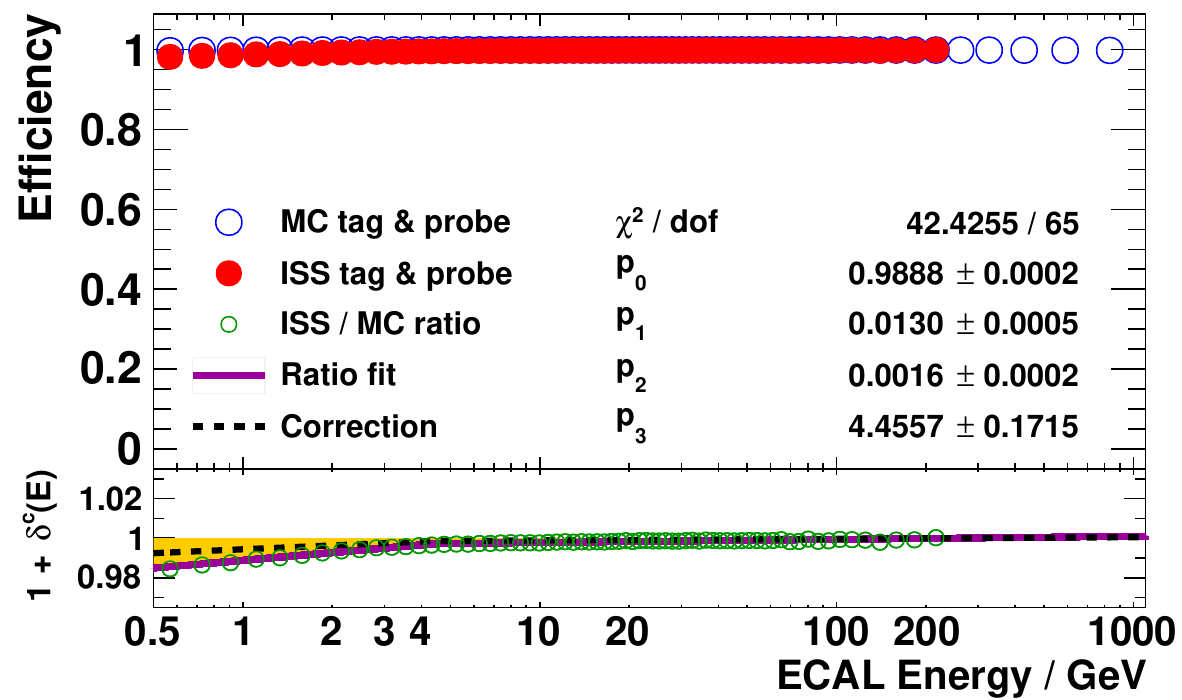}
    \caption{\textbf{\enquote{At least one useful TRD track}}}
    \label{fig:acceptance-preselection-tag-and-probe-trd-has-useful-track}
  \end{subfigure}
  \hfill
  \begin{subfigure}{0.50\linewidth}
    \includegraphics[width=\linewidth]{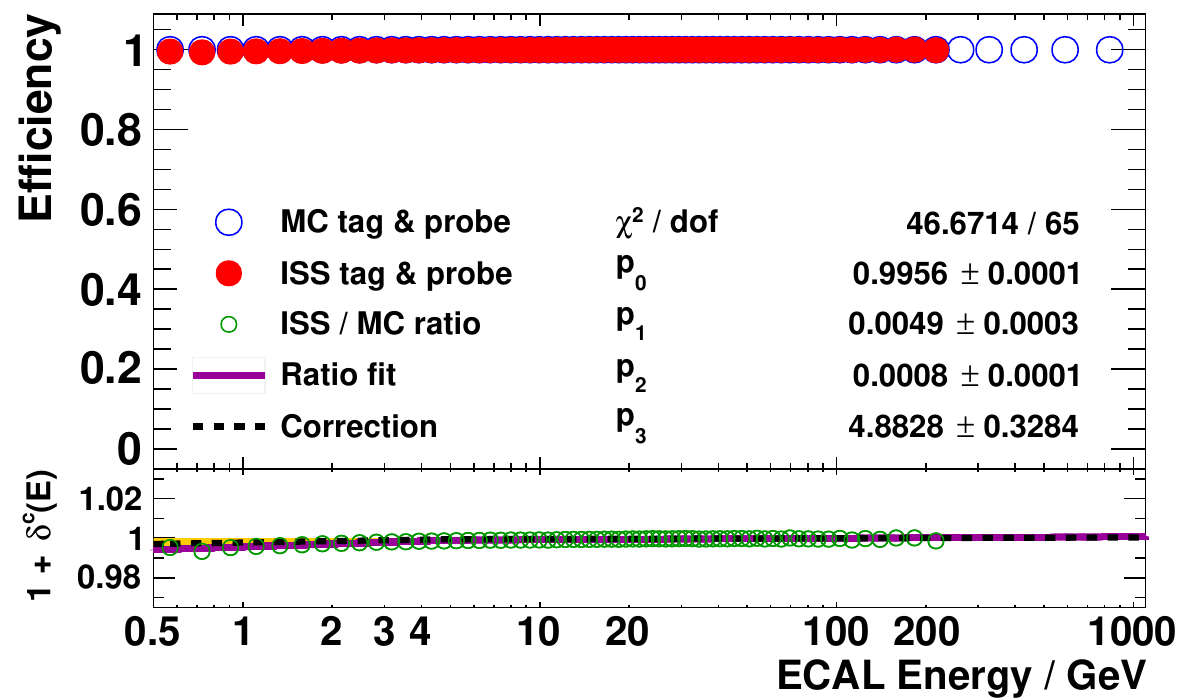}
    \caption{\textbf{\enquote{At least one useful TOF cluster combination}}}
    \label{fig:acceptance-preselection-tag-and-probe-tof-has-useful-track}
  \end{subfigure}
  \caption{Determination of the Data/Monte-Carlo correction factors $(1 + \delta^{c}(E))$ for the preselection cuts: \textbf{\enquote{At least one useful TRD track}} (left) and \textbf{\enquote{At least one useful TOF cluster combination}} (right) (\cref{sec:analysis-data-selection-preselection-cuts} - \cref{enum:preselection-cut-at-least-one-useful-trd-track,enum:preselection-cut-at-least-one-useful-tof-track}). The open blue symbols in the upper plots show the cut efficiency determined on the electron Monte-Carlo simulation on a dedicated sample: the tag sample. The filled red symbols in the upper plots show the cut efficiency determined from ISS data using the same tag cuts. The lower plots show the ratio ISS efficiency over MC efficiency as open green symbols. \Cref{eq:straight-line-with-break-in-log-scale} is fit to the ratios, represented as magenta lines. The dashed black lines show that half of the deviation from unity is taken as correction to the Monte-Carlo acceptance $A_{e^{-}}^{\text{MC}}(E)$. The associated systematic uncertainty is visualized by the orange bands.}
  \label{fig:acceptance-preselection-tag-and-probe-trd-tof-has-useful-track}
\end{figure}

\Cref{fig:acceptance-preselection-tag-and-probe-trd-tof-has-useful-track} shows the summary of the tag \& probe method for both cuts.
The tag sample selection cuts are given in \cref{sec:appendix-tag-and-probe-samples-at-least-one-useful-trd-track,sec:appendix-tag-and-probe-samples-at-least-one-useful-tof-track}.
The cut efficiency is compared between ISS data and Monte-Carlo simulation up to \SI{250}{\GeV} - above that energy a cut-based identification of electrons is not pure enough anymore.

At \SI{0.5}{\GeV} a deviation of \SIapprox{1.5}{\percent} between the ISS and Monte-Carlo efficiency is exhibited for the TRD track requirement, which gradually decreases until \SI{5}{\GeV}.
The TOF track requirement shows a similar shape, albeit with a smaller deviation of below \SIapprox{1}{\percent} at \SI{0.5}{\GeV}.

The Monte-Carlo efficiency is higher than the observed efficiency on ISS data for both cuts.
Since it is not known whether the description of the efficiency in the Monte-Carlo simulation is correct or whether the ISS data efficiency is correct, a conservative
approach is chosen: use half of the deviation as correction to the Monte-Carlo acceptance. The uncertainty on the correction due to the fit parameter uncertainty
is added in quadrature with the deviation from unity and treated as systematic uncertainty.

\clearpage
\noindent
{\large3. Selection cut: \textbf{\enquote{Upper TOF charge}} \&} \\
{\large4. Selection cut: \textbf{\enquote{Enough active layers in TRD}}} \\

\begin{figure}[H]
  \begin{subfigure}{0.50\linewidth}
    \includegraphics[width=\linewidth]{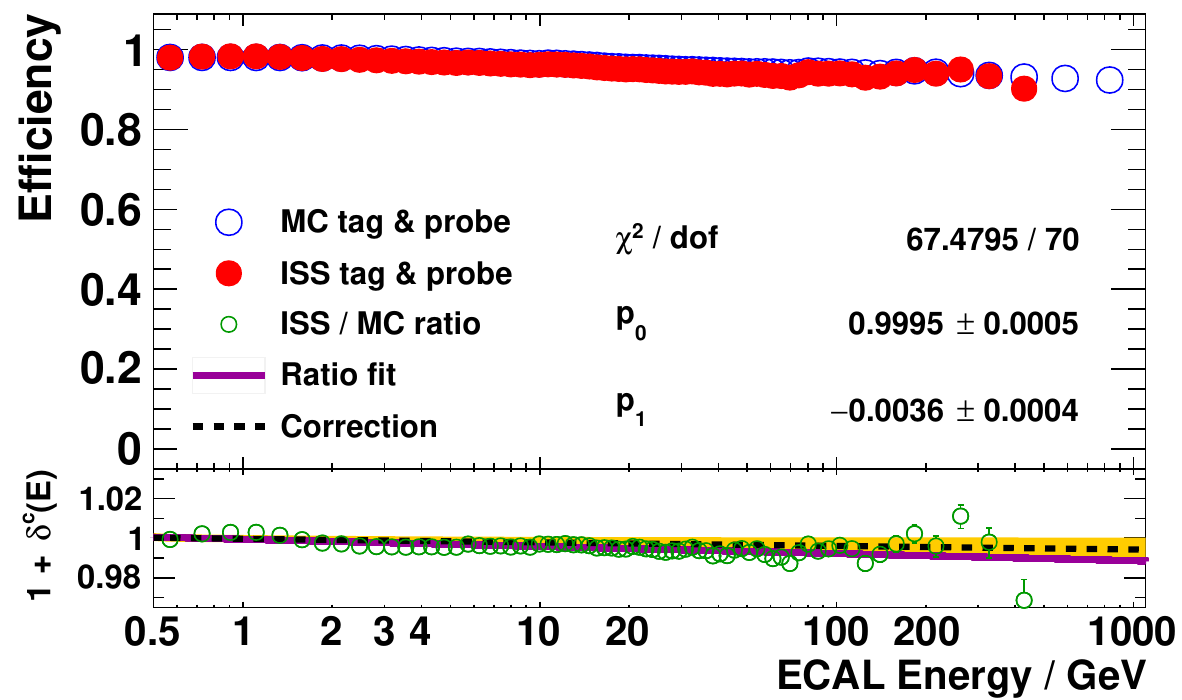}
    \caption{\textbf{\enquote{Upper TOF charge}}}
    \label{fig:acceptance-selection-tag-and-probe-upper-tof-charge}
  \end{subfigure}
  \hfill
  \begin{subfigure}{0.50\linewidth}
    \includegraphics[width=\linewidth]{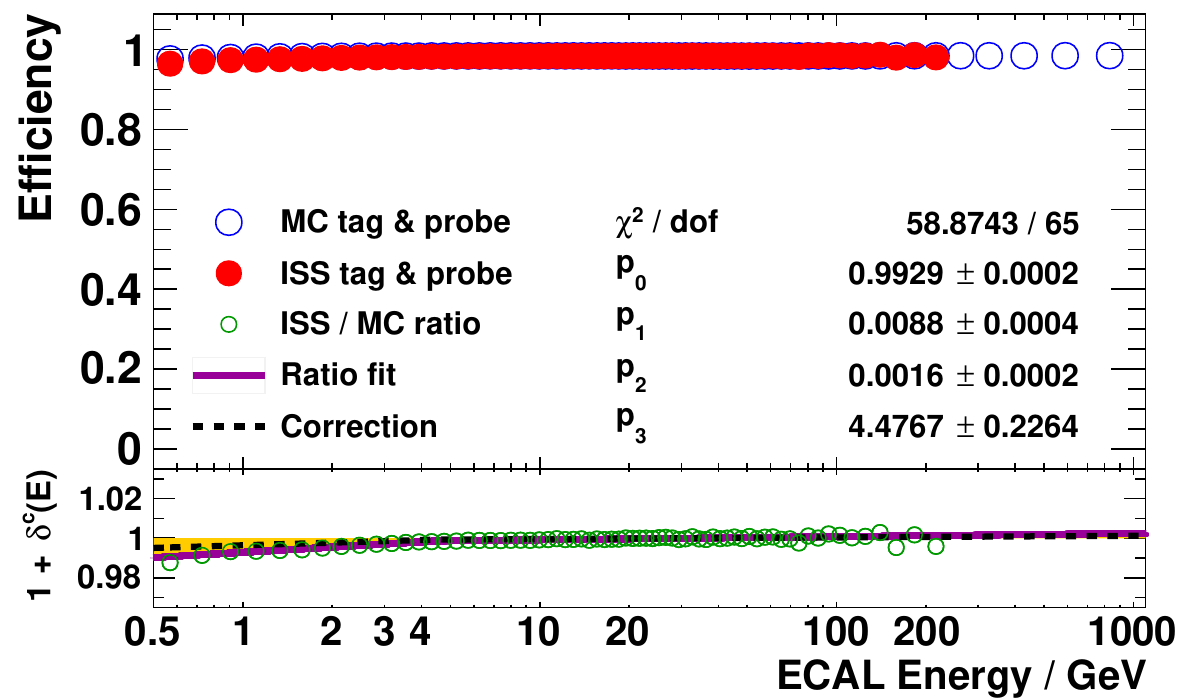}
    \caption{\textbf{\enquote{Enough active layers in TRD}}}
    \label{fig:acceptance-selection-tag-and-probe-trd-active-layers}
  \end{subfigure}
  \caption{Determination of the Data/Monte-Carlo correction factors $(1 + \delta^{c}(E))$ for the selection cuts: \textbf{\enquote{Upper TOF charge}} (left) and \textbf{\enquote{Enough active layers in TRD}} (right) (\cref{sec:analysis-data-selection-selection-cuts} - \cref{enum:selection-cut-upper-tof-charge,enum:selection-cut-trd-active-layers}). The open blue symbols in the upper plots show the cut efficiency determined on the electron Monte-Carlo simulation on a dedicated sample: the tag sample. The filled red symbols in the upper plots show the cut efficiency determined from ISS data using the same tag cuts. The lower plots show the ratio ISS efficiency over MC efficiency as open green symbols. \Cref{eq:straight-line-in-log-scale} is fit to the left ratio, \cref{eq:straight-line-with-break-in-log-scale} is fit to the right ratio, represented as magenta lines. The dashed black lines show that half of the deviation from unity is taken as correction to the Monte-Carlo acceptance $A_{e^{-}}^{\text{MC}}(E)$. The associated systematic uncertainty is visualized by the orange bands.}
  \label{fig:acceptance-selection-tag-and-probe-upper-tof-charge-trd-active-layers}
\end{figure}

\Cref{fig:acceptance-selection-tag-and-probe-upper-tof-charge-trd-active-layers} shows the summary of the tag \& probe method for both cuts.
The tag sample selection cuts are given in \cref{sec:appendix-tag-and-probe-samples-upper-tof-charge,sec:appendix-tag-and-probe-samples-trd-active-layers}.
The cut efficiency is compared between ISS data and Monte-Carlo simulation up to \SI{250}{\GeV} - above that energy a cut-based identification of electrons is not pure enough anymore.

Both cuts exhibit only a small deviation from the predicted efficiency by the Monte-Carlo simulation. The upper TOF charge measurement ISS/MC ratio is best parameterized with a straight line in logarithmic scale.
The TRD active layers requirement shows a deviation of \SIapprox{1}{\percent} at \SI{0.5}{\GeV}, which vanishes above \SI{5}{\GeV} and thus the associated ISS/MC ratio is best described with two connected straight lines including a break.

Since it is not known whether the description of the efficiency in the Monte-Carlo simulation is correct or whether the ISS data sample is pure, a conservative
approach is chosen: use half of the deviation as correction to the Monte-Carlo acceptance. The uncertainty on the correction due to the fit parameter uncertainty
is added in quadrature with the deviation from unity and treated as systematic uncertainty.

\bigskip
\noindent
{\large5. Selection cut: \textbf{\enquote{Tracker track goodness-of-fit in Y-projection}}} \\

\begin{figure}[H]
  \centering
  \includegraphics[width=\linewidth]{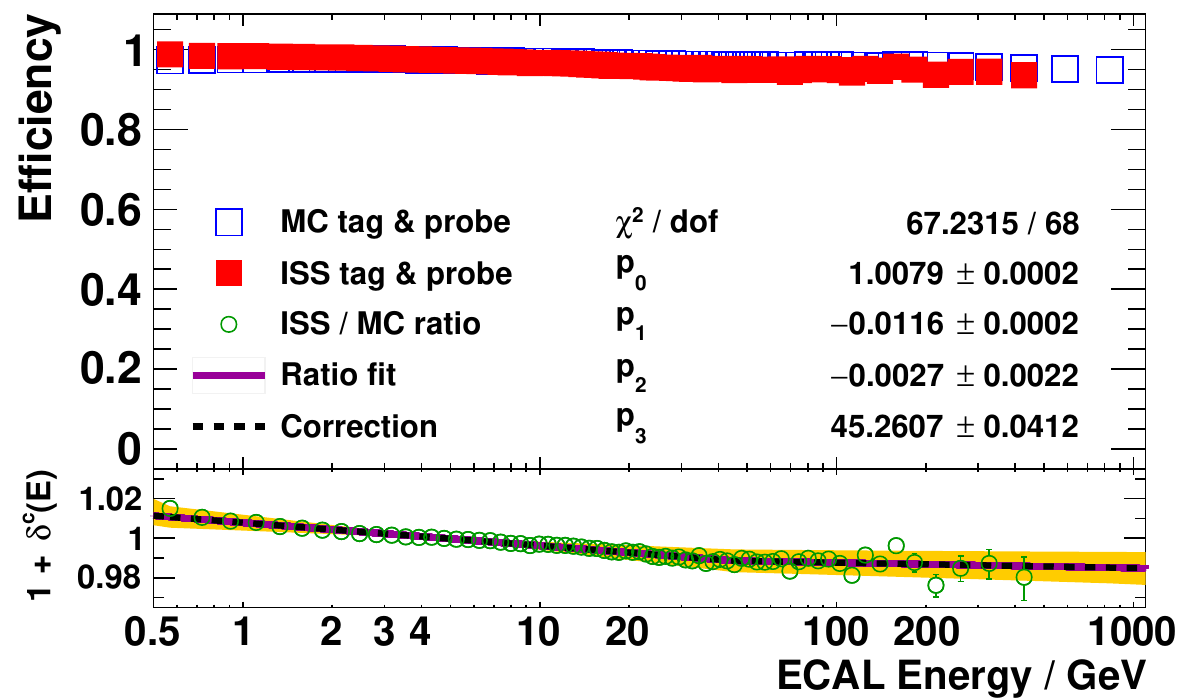}
  \caption{Determination of the Data/Monte-Carlo correction factor $(1 + \delta^{c}(E))$ for the selection cut: \textbf{\enquote{Tracker track goodness-of-fit in Y-projection}} (\cref{sec:analysis-data-selection-selection-cuts} - \cref{enum:selection-cut-trk-chi-square-y}). The open blue symbols in the upper plots show the cut efficiency determined on the electron Monte-Carlo simulation on a dedicated sample: the tag sample. The filled red symbols in the upper plots show the cut efficiency determined from ISS data using the same tag cuts. The lower plots show the ratio ISS efficiency over MC efficiency as open green symbols. \Cref{eq:straight-line-with-break-in-log-scale} is fit to the ratio, represented as magenta line. The dashed black line shows that half of the deviation from unity is taken as correction to the Monte-Carlo acceptance $A_{e^{-}}^{\text{MC}}(E)$. The associated systematic uncertainty is visualized by the orange band.}
  \label{fig:acceptance-selection-tag-and-probe-tracker-chi-square-y}
\end{figure}

\Cref{fig:acceptance-selection-tag-and-probe-tracker-chi-square-y} shows the summary of the tag \& probe method for this cut.
The tag sample selection cuts are given in \cref{sec:appendix-tag-and-probe-samples-tracker-chi-square-y}.
The remaining three cuts to be tested (including this one) will yield the largest Data/Monte-Carlo correction factors $(1 + \delta^{c}(E))$. It is crucial to ensure
that they are valid up to the highest energies and the associated systematic uncertainties are as small as possible.

At this stage of the analysis all subdetector information to select $e^{-}$ enhanced samples are available - most noticeable: two out of three tracker quality
cuts were applied (charge measurement and tracker hit pattern requirements).
This gives the opportunity to avoid cut based tag \& probe methods, but instead perform template fits in the TRD estimator to determine the efficiencies.
The benefit is that one no longer has to ensure a minimum contamination of charge-confused protons in the negative rigidity sample,
since the template fit is able to discriminate between electrons and charge-confused protons.

Squares as markers instead of circles for the efficiencies in the tag \& probe method summary plot - \cref{fig:acceptance-selection-tag-and-probe-tracker-chi-square-y} -
indicate that the method to determine them has changed from a cut-based approach to a template-fit approach. Furthermore the energy range that is tested could be doubled from \SI{250}{\GeV} to \SI{500}{\GeV}.

Due to the introduction of the template-fit method to extract the efficiencies it is no longer necessary to question the purity of the ISS tag sample.
The full deviation of the ISS efficiency with respect to the Monte-Carlo efficiency is used as correction factor for the Monte-Carlo acceptance. The systematic
uncertainty no longer includes the deviation from unity, but only represents the uncertainty of the correction due to the uncertainty of the model fit parameters.

\clearpage
\noindent
{\large6. $e^{\pm}$ identification cut: \textbf{\enquote{Energy $\leftrightarrow$ rigidity matching}} \&} \\
{\large7. $e^{\pm}$ identification cut: \textbf{\enquote{Tracker $\leftrightarrow$ ECAL matching in X-projection}}} \\

\begin{figure}[H]
  \begin{subfigure}{0.50\linewidth}
    \includegraphics[width=\linewidth]{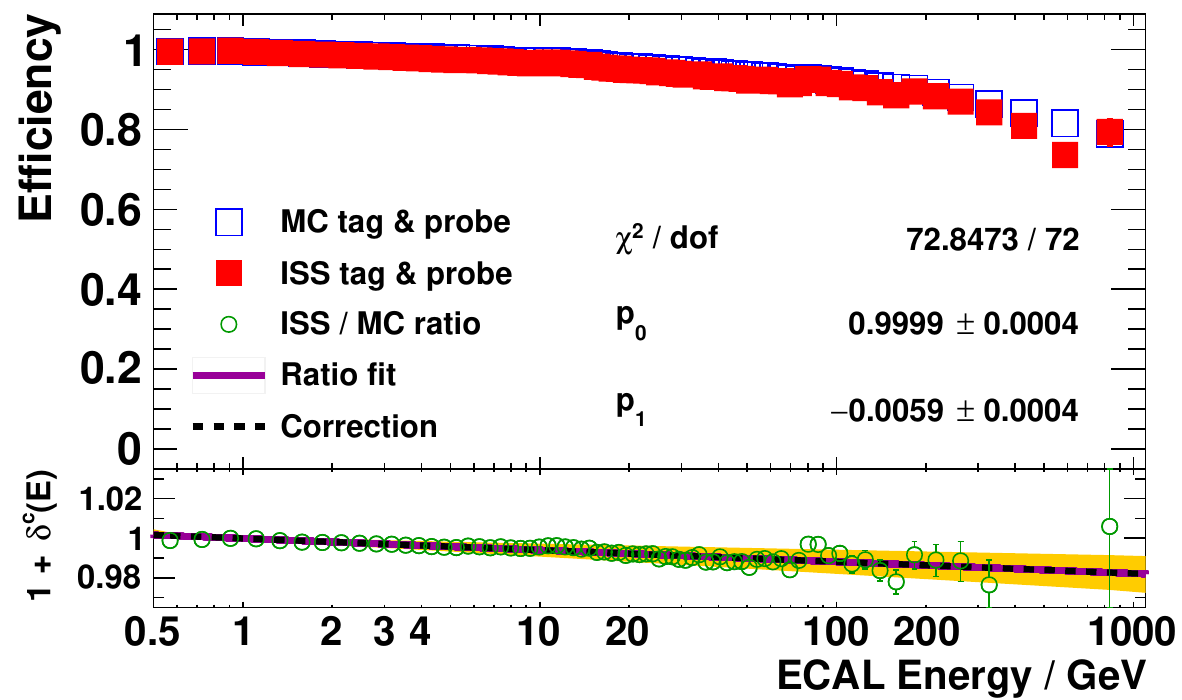}
    \caption{\textbf{\enquote{Energy $\leftrightarrow$ rigidity matching}}}
    \label{fig:acceptance-identification-energy-over-rigidity}
  \end{subfigure}
  \hfill
  \begin{subfigure}{0.50\linewidth}
    \includegraphics[width=\linewidth]{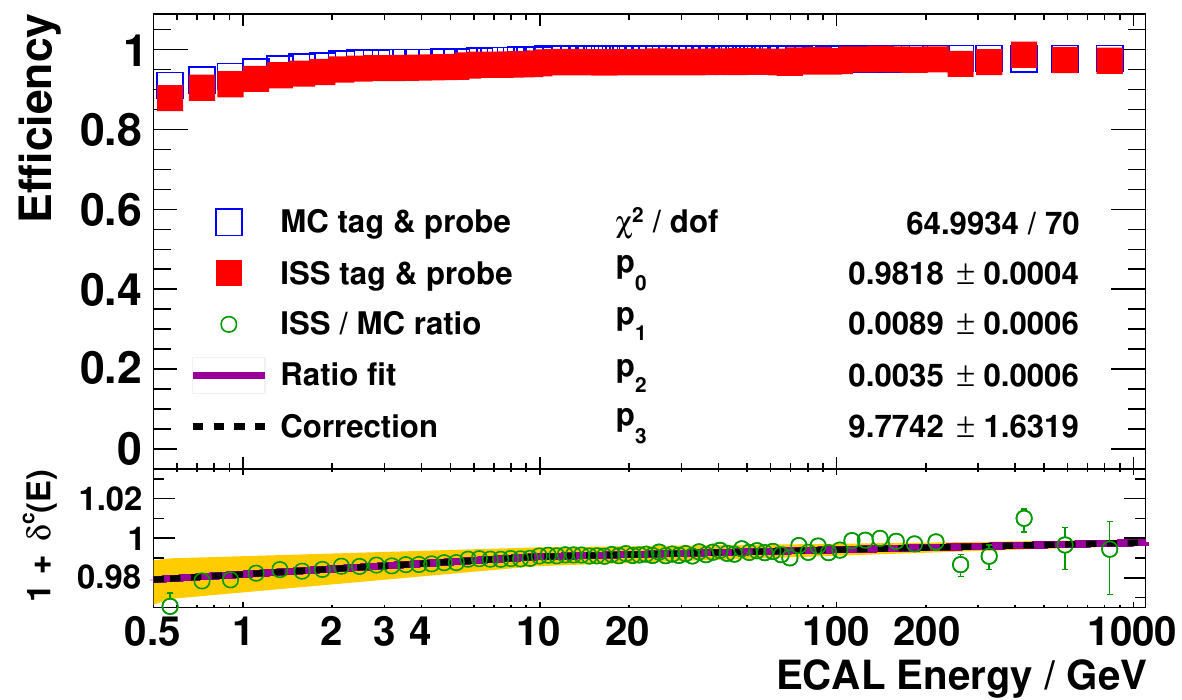}
    \caption{\textbf{\enquote{Tracker $\leftrightarrow$ ECAL matching in X-projection}}}
    \label{fig:acceptance-identification-tag-and-probe-tracker-track-ecal-cog-delta-x}
  \end{subfigure}
  \caption{Determination of the Data/Monte-Carlo correction factors $(1 + \delta^{c}(E))$ for the $e^{\pm}$ identification cuts: \textbf{\enquote{Energy $\leftrightarrow$ rigidity matching}} (left) and \textbf{\enquote{Tracker $\leftrightarrow$ ECAL matching in X-projection}} (right) (\cref{sec:analysis-data-selection-electron-positron-identification-cuts} - \cref{enum:electron-positron-identification-cut-energy-rigidity-matching,enum:electron-positron-identification-cut-tracker-track-ecal-cog-delta-x}). The open blue symbols in the upper plots show the cut efficiency determined on the electron Monte-Carlo simulation on a dedicated sample: the tag sample. The filled red symbols in the upper plots show the cut efficiency determined from ISS data using the same tag cuts. The lower plots show the ratio ISS efficiency over MC efficiency as open green symbols. \Cref{eq:straight-line-in-log-scale} is fit to the left ratio, \cref{eq:straight-line-with-break-in-log-scale} is fit to the right ratio, represented as magenta lines. The dashed black lines show that half of the deviation from unity is taken as correction to the Monte-Carlo acceptance $A_{e^{-}}^{\text{MC}}(E)$. The associated systematic uncertainty is visualized by the orange bands.}
  \label{fig:acceptance-identification-tag-and-probe-energy-rigidity-matching-tracker-track-ecal-cog-delta-x}
\end{figure}

\Cref{fig:acceptance-identification-tag-and-probe-energy-rigidity-matching-tracker-track-ecal-cog-delta-x} shows the summary of the tag \& probe method for both cuts.
The tag sample selection cuts are given in \cref{sec:appendix-tag-and-probe-samples-energy-rigidity-matching,sec:appendix-tag-and-probe-samples-tracker-track-ecal-cog-delta-x}.

Both ISS/Monte-Carlo efficiency ratios are well parameterized by the chosen models.
The ISS efficiency was extracted using the template-fit method and the full deviation of the ISS efficiency with respect to the
Monte-Carlo efficiency is used as correction factor for the Monte-Carlo acceptance.

\begin{figure}[H]
  \centering
  \includegraphics[width=0.90\linewidth]{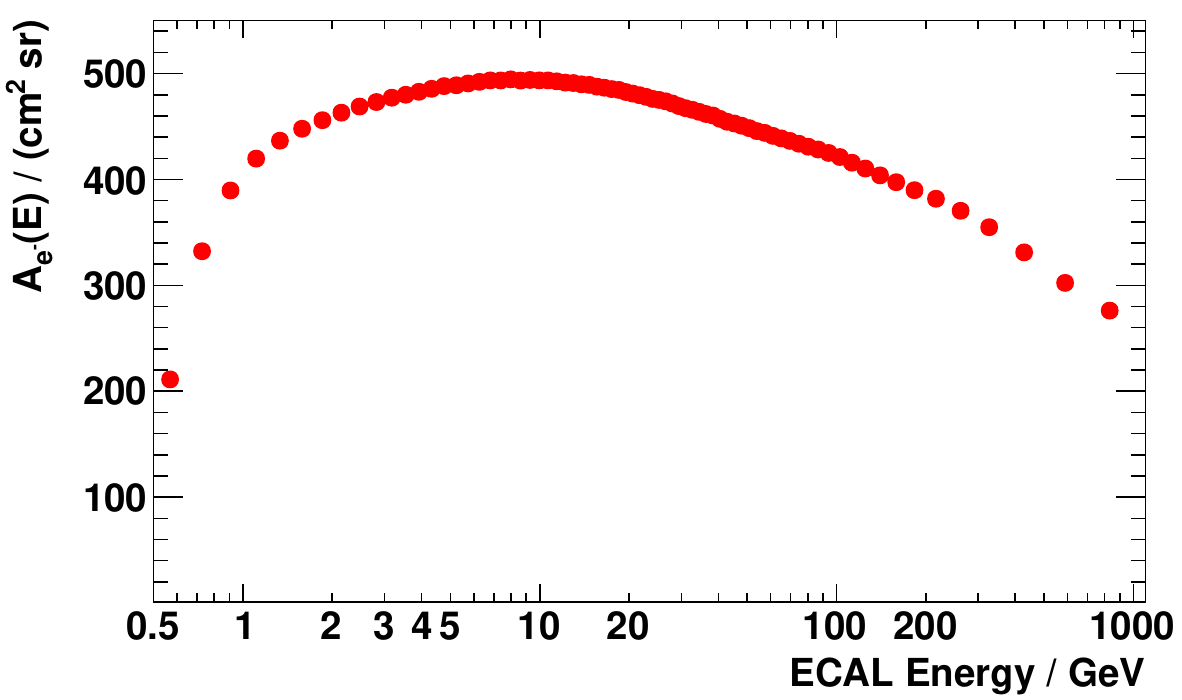}
  \caption{Final result of the acceptance $A_{e^{-}}(E)$ as function of energy, including all Data/Monte-Carlo corrections $(1 + \delta(E))$. The red symbols show that the acceptance is a smooth function with a well-behaving energy dependence.}
  \label{fig:acceptance-final}
\end{figure}

\begin{figure}[H]
  \centering
  \includegraphics[width=0.90\linewidth]{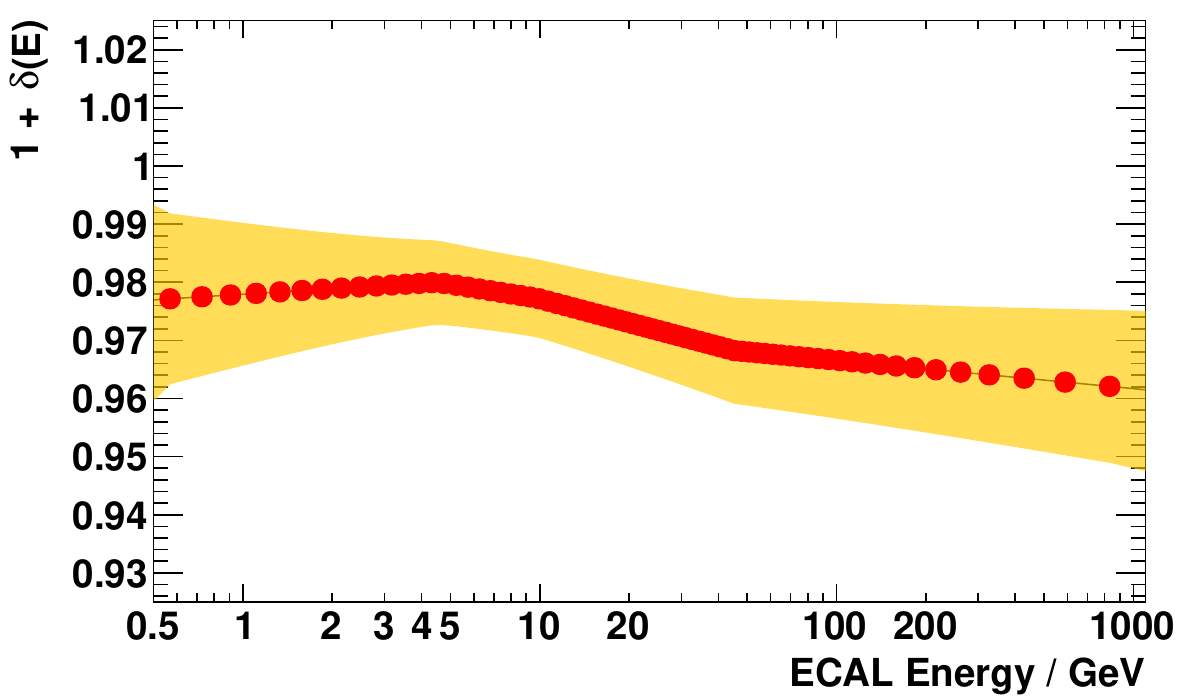}
  \caption{The magnitude of the combined Data/Monte-Carlo correction factor $(1 + \delta(E))$ is shown as red symbols. The correction applied to the Monte-Carlo acceptance varies from \SI{2}{\percent} at low energies up to \SI{4}{\percent} at high energies and is known for all energies better than \SI{1.5}{\percent}, indicated by the orange band.}
  \label{fig:acceptance-final-correction}
\end{figure}

Since all correction factors $(1 + \delta^{c}(E))$ were determined, the final correction to the acceptance $A(E)$ can be computed, according to \cref{eq:acceptance-correction-prod}.
The uncertainty on the combined Data/Monte-Carlo correction factor $(1 + \delta(E))$ is obtained by adding all the uncertainty of all individual
correction factors $(1 + \delta^{c}(E))$ in quadrature. The resulting acceptance $A(E)$ is shown in \cref{fig:acceptance-final} and the correction
factor in \cref{fig:acceptance-final-correction}.

Note that the x-axis title no longer states \enquote{MC generated energy}, as in \cref{fig:acceptance-without-corrections}, but \enquote{ECAL Energy}.
The reason for this is that the Data/Monte-Carlo correction factor $(1 + \delta(E))$ was determined as function of the ECAL energy, since it involves ISS
data where a generated energy is not available. The ECAL energy must be used as best estimator of the true particle energy on ISS data.
If the correction factor $(1 + \delta(E))$, \cref{fig:acceptance-final-correction}, would be large and/or a steeply falling or rising curve as function
of energy, this could possibly introduce a bias. To avoid a possible bias the number of passed events and the number of total events as function of ECAL
energy need to be unfolded to the \enquote{Top-Of-Instrument energy}, before taking the ratio passed over total events to determine the efficiency for each cut.
This was tested for the \enquote{Energy $\leftrightarrow$ rigidity matching} cut: the effect is small and thus can be neglected.

\subsection{Acceptance asymmetry}
\label{sec:analysis-flux-time-averaged-acceptance-asymmetry}

The acceptance derived from the electron Monte-Carlo simulation is suitable for electrons but slightly differs for positrons.
The scattering cross-section of positrons differs from electrons for scattering by a nuclear coulomb field.
The penetration depth in various materials, e.g. \ce{C} or \ce{Si}, as well as the multiple scattering is different~\cite{Rohrlich1954}.
Thus the acceptance determination was repeated using a positron Monte-Carlo simulation, up to \SI{200}{\GeV}.

\begin{figure}[H]
  \centering
  \includegraphics[width=\linewidth]{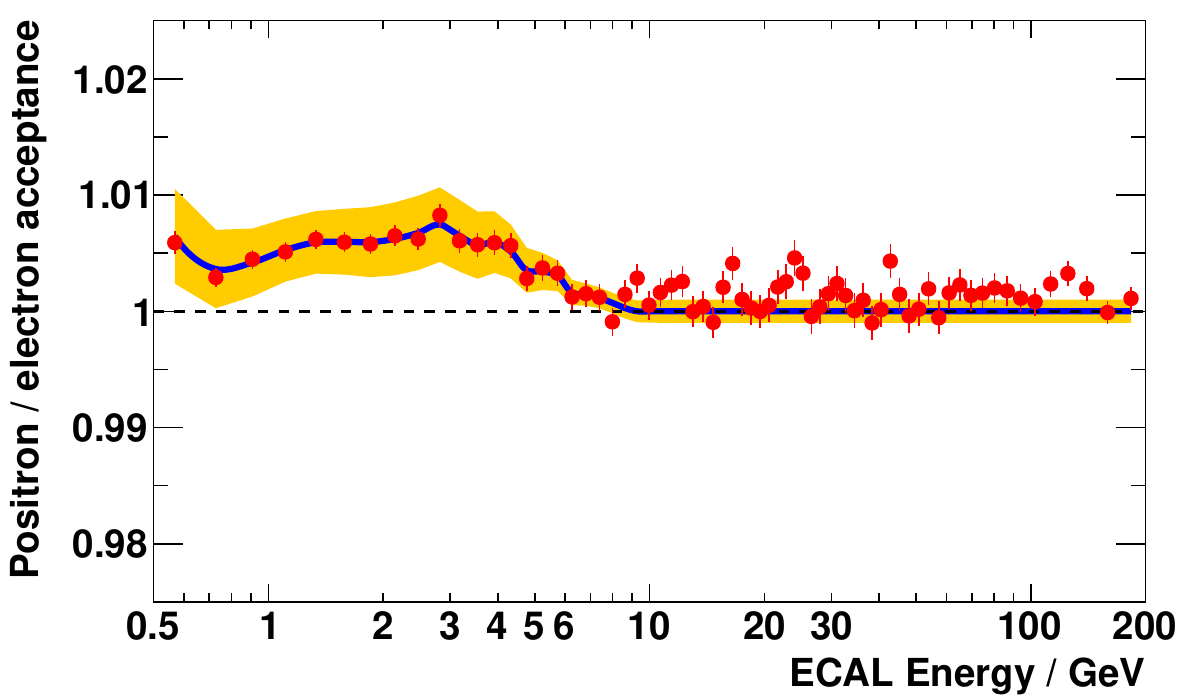}
  \caption{Visualization of the positron over electron acceptance ratio $A_{e^{+}}(E) / A_{e^{-}}(E)$, shown as red symbols. The difference is around \SI{1}{\percent} at \SI{3}{\GeV} where the difference between the electron and the positron acceptance is largest. The blue curve multiplied to the electron acceptance yields the positron acceptance.}
  \label{fig:acceptance-asymmetry-ratio}
\end{figure}

\Cref{fig:acceptance-asymmetry-ratio} shows the positron over electron acceptance ratio $A_{e^{+}}(E) / A_{e^{-}}(E)$. The acceptances differ at most by
\SI{0.6}{\percent} over all energies. At energies below \SIapprox{8}{\GeV} the acceptances are no longer indistinguishable, but show a small difference in the energy dependence.

To derive the electron flux the acceptance $A_{e^{-}}(E)$ is used and $A_{e^{+}}(E)$ to derive the positron flux. For the positron over electron ratio
and the positron fraction analysis, the acceptance asymmetry ratio $A_{e^{+}}(E) / A_{e^{-}}(E)$ is used as correction factor.

The uncertainty indicated by the orange band in \cref{fig:acceptance-asymmetry-ratio} is propagated to the uncertainty of the positron acceptance $A_{e^{+}}(E)$ - the electron
acceptance $A_{e^{-}}(E)$ is not affected. When computing flux ratios, such as the positron/electron ratio or the positron fraction the systematic uncertainty of the acceptance
cancels. In those cases the systematic uncertainty of the acceptance asymmetry will be included as extra systematic uncertainty on the positron/electron ratio and
the positron fraction. This topic will be revisited once the positron/electron ratio and the positron fraction - \cref{sec:analysis-ratios-time-averaged} - is derived.

\subsection{Unfolding}
\label{sec:analysis-flux-time-averaged-unfolding}

In \cref{sec:analysis-lepton-counts-2d-fit} a two-dimensional template fit was used to extract the electron and positron event counts as function of the measured
ECAL energy. For the flux measurement the event counts need to be obtained as function of the \gls{TOI} energy, to correct for the remaining difference
between the measured ECAL energy and the true energy.

The applied rear/side leakage correction may overestimate the energy of the incident particle and thus
an energy higher than the true energy might be reconstructed. At low energies, emitted bremsstrahlung photons of the primary particle often convert into additional
electron and positron pairs, where the momentum distribution is asymmetric: one particle has a higher kinetic energy, than the other. The secondary with
the lower energy might leave the detector, due to the bending of the magnetic field. If that happens, the reconstructed energy underestimates the true energy of
the primary particle. These kind of effects need to be corrected using the unfolding procedure\footnote{The unfolding procedure is based on work from Henning Gast
and was finalized in close collaboration with him.}.

\begin{figure}[H]
  \centering
  \includegraphics[width=0.85\linewidth]{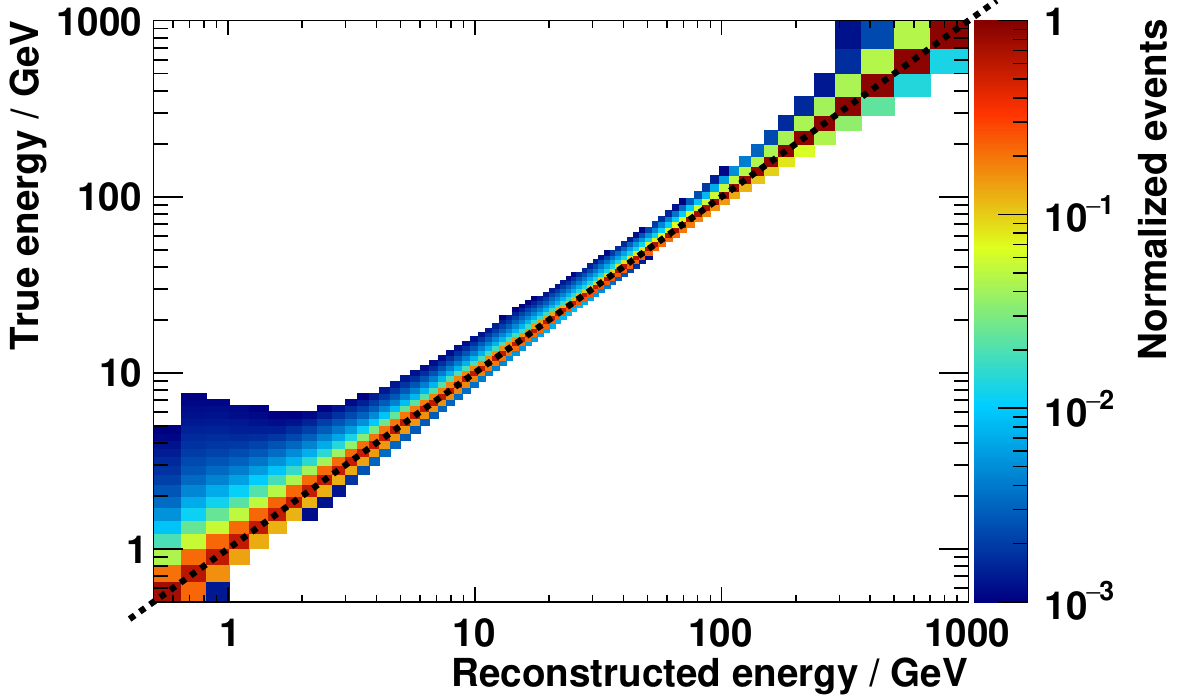}
  \caption{Migration matrix $\mathbf{M}$ obtained from the electron Monte-Carlo simulation, using the same binning as the flux analysis.}
  \label{fig:unfolding-migration-matrix}
\end{figure}

\Cref{fig:unfolding-migration-matrix} shows the normalized migration matrix $\textbf{M}$, which shows the measured ECAL energy (\enquote{Reconstructed energy})
vs.~the Monte-Carlo generated energy (\enquote{True energy}). The plot shows that for each slice in true energy an interval of possible energies can be reconstructed.
The most probable reconstructed energy matches the true energy (diagonal elements of the matrix). It is evident that at low true energies, the reconstructed energy
often underestimates the true energy. The probability to underestimate the true energy is generally higher than to overestimate the energy, over the whole energy range.

Mathematically the unfolding process can be written as matrix multiplication:

\begin{equation*}
  \mathbf{\hat{n}} = \mathbf{U} \cdot \mathbf{n},
\end{equation*}

where $\mathbf{n}$ denotes the observed distribution in data and $\mathbf{\hat{n}}$ the unfolded, true distribution.

A matrix inversion is enough to solve the equation, but is numerically unstable. In the literature several stable unfolding methods were proposed, such as
the Bayesian unfolding (described in \cref{sec:appendix-unfolding-iterative-bayesian-method}).

The goal for the unfolding is to reconstruct the true energy distribution of the events $\mathbf{\hat{n}}$ from the measured energy distribution $\mathbf{n}$.
In total 76 energy bins are available both for the true and measured energy distributions. The actual analysis only uses 74 energy bins (\cref{sec:analysis-lepton-counts-binning}),
but the event counts are determined in two more bins: one extra low energy bin (\SIrange{0.25}{0.5}{\GeV}) and one extra high energy bin (\SIrange{1}{1.5}{\TeV}), to account for
migration from/to these \enquote{overflow bins}\footnote{In a toy Monte-Carlo simulation it was shown that without using the extra overflow bins the unfolded event counts deviate
from the true input flux by more than \SI{50}{\percent}. Thus it is important to perform the unfolding process including the overflow bins, to reduce the bias in the first quoted
analysis bins.}

The probability of measuring a particle with energy $n_{j}$ given the particle had the true energy $\hat{n}_{i}$ is already provided by the
elements of the migration matrix $\mathbf{M}_{ij}$ (\cref{fig:unfolding-migration-matrix}). The input for the unfolding procedure in \textit{RooUnfold} is
thus the migration matrix $\mathbf{M}$ as two-dimensional histogram and the distribution of the measured events $\mathbf{n}$ as one-dimensional histogram.

The distribution of the measured events $\mathbf{n}$ is affected by the geomagnetic cut-off, since only particles above the cut-off rigidity are analyzed:
$E_{j\text{, low}} > f_\text{safety} \cdot R_{c}$ (\cref{sec:analysis-flux-time-averaged-measuring-time}).

\begin{figure}[H]
  \begin{subfigure}{0.50\linewidth}
    \includegraphics[width=\linewidth]{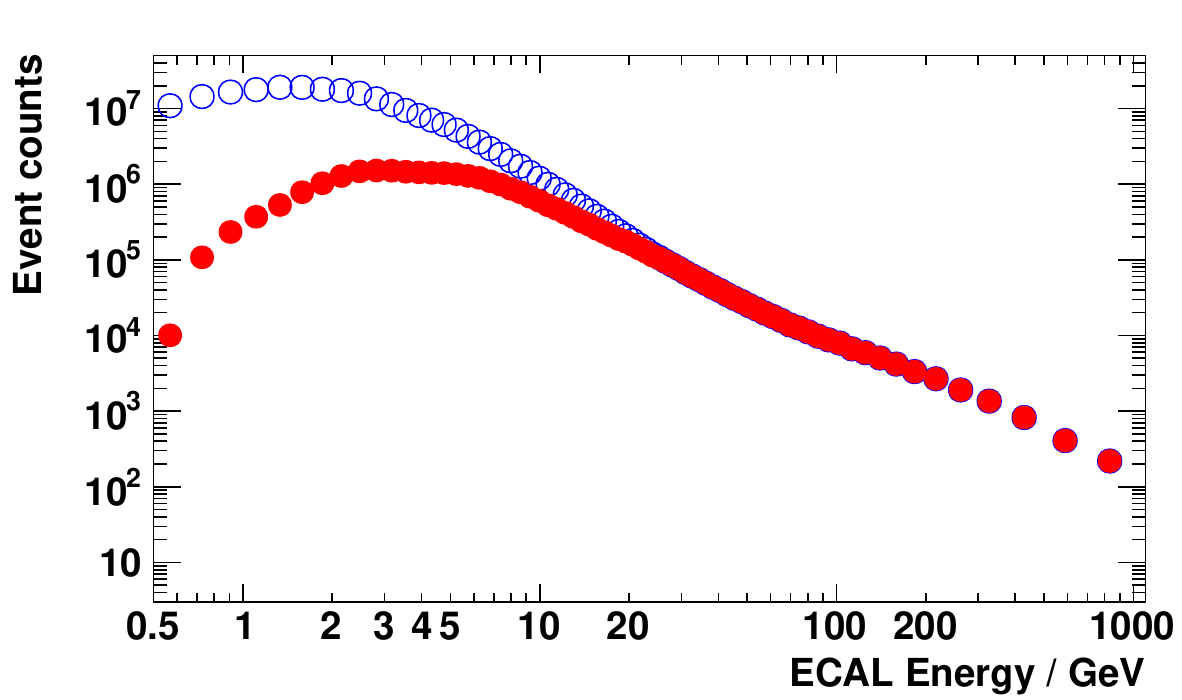}
    \caption{Electron event counts}
  \end{subfigure}
  \hfill
  \begin{subfigure}{0.50\linewidth}
    \includegraphics[width=\linewidth]{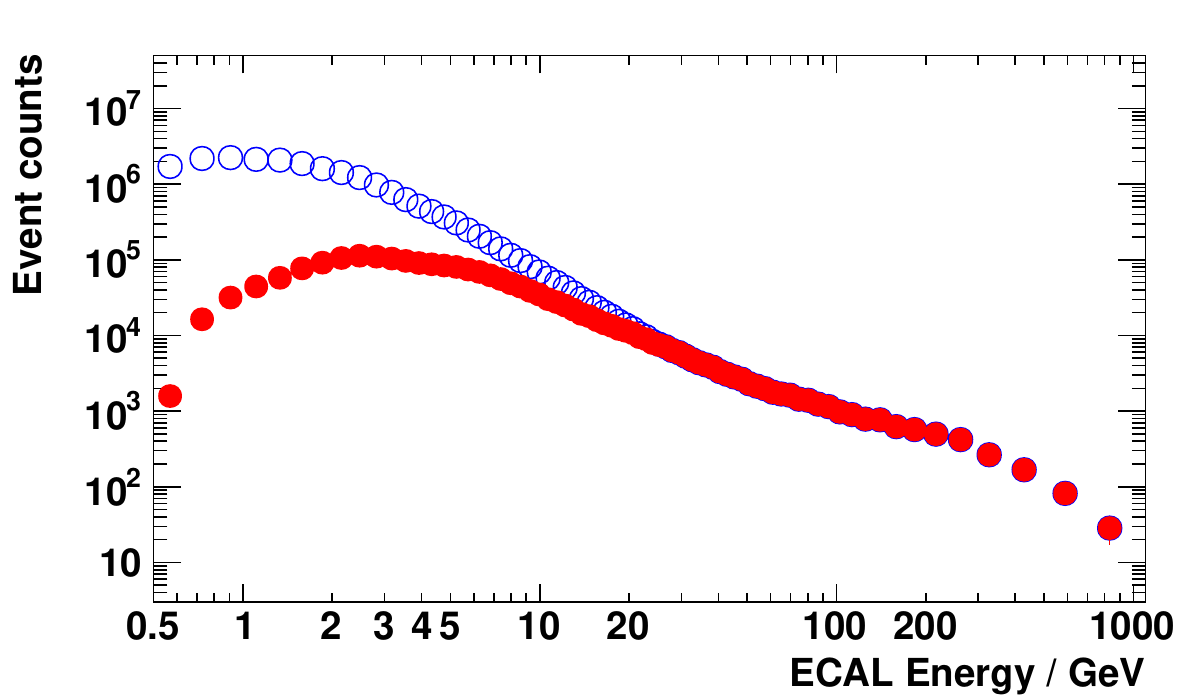}
    \caption{Positron event counts}
  \end{subfigure}
  \caption{Comparison of the event counts $\mathbf{n}$ determined from the two-dimensional template fit (red symbols) with the artificial event counts $\mathbf{n}_{\text{no cut-off}}$ that AMS-02 would measure if there was no geomagnetic cut-off effect (blue symbols). The left plot shows the electron event counts, the right plot the positron event counts.}
  \label{fig:unfolding-event-counts-without-cutoff}
\end{figure}

\Cref{fig:unfolding-event-counts-without-cutoff} shows the event counts for electrons and positrons $\mathbf{n}$ extracted from the two-dimensional template fit
(\cref{sec:analysis-lepton-counts-2d-fit}) as red symbols. To visualize the effect of the geomagnetic cut-off the event counts $\mathbf{n}$ are divided
by the cut-off weight: $w_{\text{cut-off}}(E) = T(E)\ /\ T(> \SI{25}{\GeV})$. The event counts $\mathbf{n}_{\text{no cut-off}} = \mathbf{n} / w_{\text{cut-off}}(E)$,
so-obtained, are shown as blue symbols.

\begin{figure}[H]
  \centering
  \includegraphics[width=0.90\linewidth]{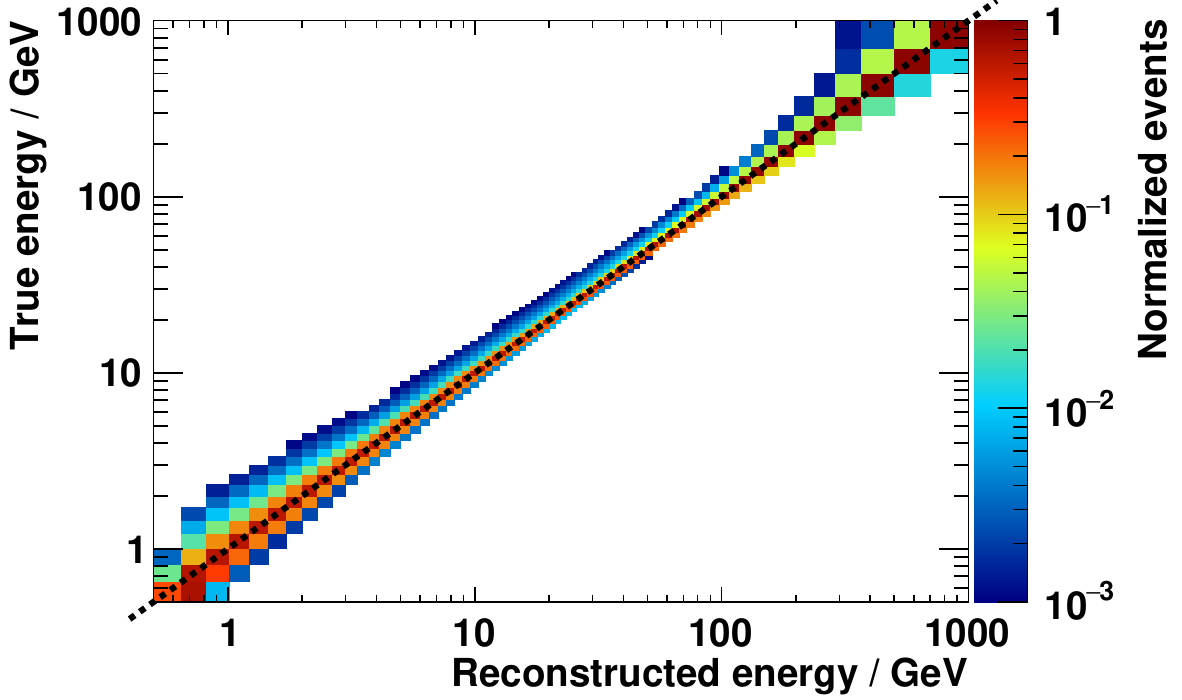}
  \caption{Plot of the migration matrix, additionally weighted using the geomagnetic cut-off weight $w_{\text{cut-off}}(E)$ determined from ISS data.}
  \label{fig:unfolding-migration-matrix-with-cutoff-weight}
\end{figure}

The migration matrix is obtained from the Monte-Carlo simulation and does not contain any information about
the loss of events as function of the \textbf{\textit{measured}} energy due to the geomagnetic cut-off. To take the geomagnetic cut-off into account the migration
matrix needs to be modified. For each bin in x-direction every slice in y-direction needs to be weighted using $w_{\text{cut-off}}(E)$. The resulting migration matrix
$\mathbf{\tilde{M}}$ is shown in \cref{fig:unfolding-migration-matrix-with-cutoff-weight}.

After performing the iterative Bayesian unfolding procedure with four iterations\footnote{Dedicated studies revealed that $N_{\text{iter}} = 4$ is an optimal choice of the parameter.}
using the measured event counts $\mathbf{n}$ and the migration matrix $\mathbf{\tilde{M}}$ the true event count distribution $\mathbf{\hat{n}}$ is available and can be compared to the
measured $\mathbf{n}$ event count distribution to quantify the magnitude of the unfolding.

\begin{figure}[H]
  \begin{subfigure}{0.50\linewidth}
    \includegraphics[width=\linewidth]{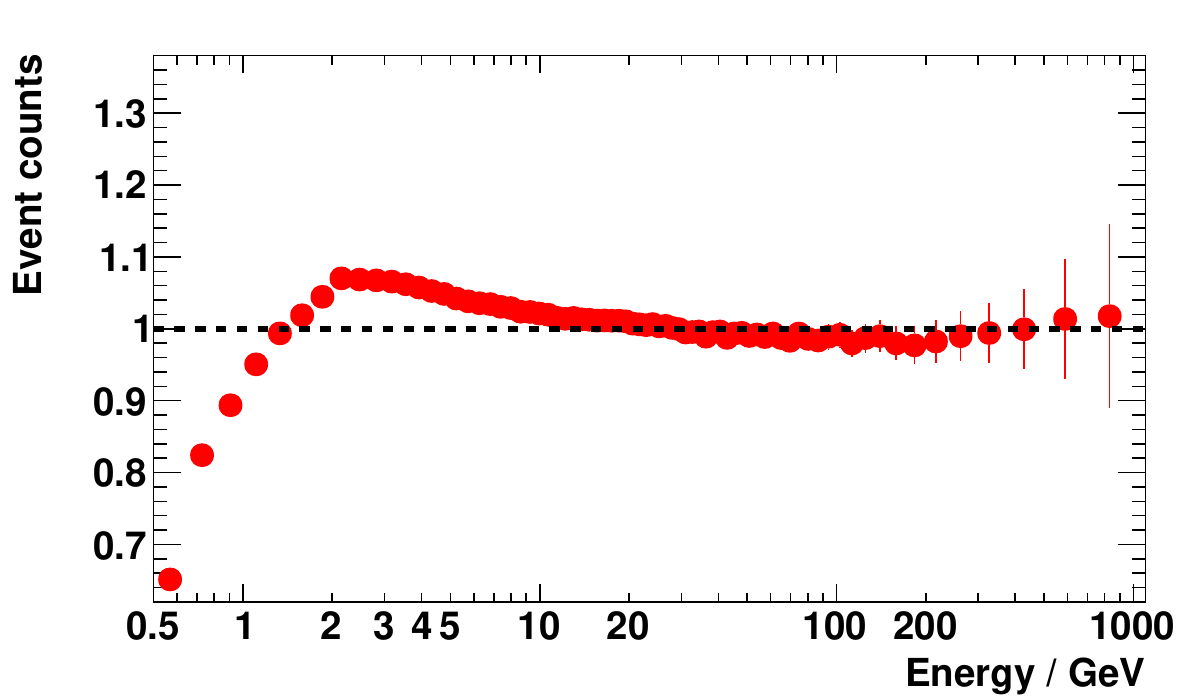}
    \caption{Unfolded / measured electron event counts}
  \end{subfigure}
  \hfill
  \begin{subfigure}{0.50\linewidth}
    \includegraphics[width=\linewidth]{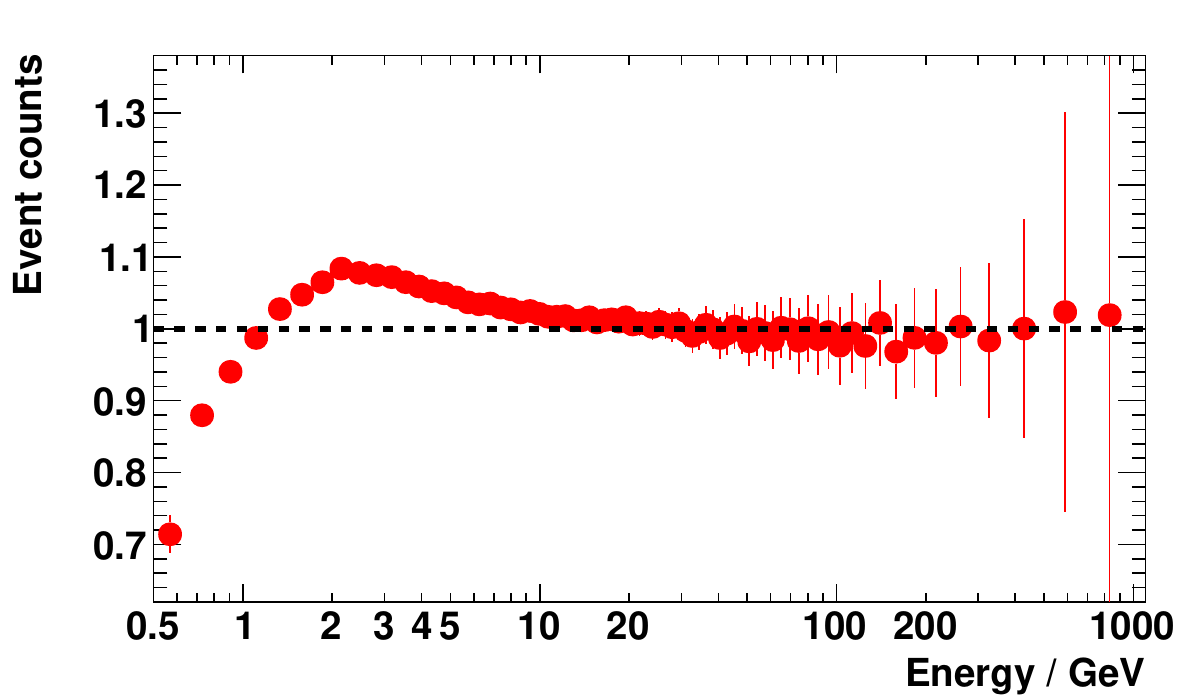}
    \caption{Unfolded / measured positron event counts}
  \end{subfigure}
  \caption{Plot of the ratio of unfolded event counts over measured event counts to quantity the magnitude of unfolding. Above \SI{100}{\GeV} the effect of the unfolding becomes negligible. At low energies the effect is large with corrections up to \SI{60}{\percent} in the first energy bin: \SIrange{0.5}{0.65}{\GeV}. With increasing energy the magnitude of the unfolding decreases, but it is still a sizeable effect over the whole energy range.}
  \label{fig:unfolding-count-comparison-ratio}
\end{figure}

\Cref{fig:unfolding-count-comparison-ratio} shows the ratios of the unfolded event counts $\mathbf{\hat{n}}$ divided by the measured event counts $\mathbf{n}$
as function of energy for electrons and positrons, respectively.

Below \SI{10}{\GeV} the effect of the unfolding is most pronounced and it gradually decreases towards high energies. The reason is a combination of two effects:
the shape of the event counts as function of energy (flat steeply rising/falling) and the bin width affect the amount of migration. The large bin width at high energies,
which exceeds the ECAL energy resolution (\cref{sec:detector-ecal}) leads to a minimum amount of migration. The smaller the bin width, the larger is
the unfolding effect.

Finally an important detail has to be mentioned: the migration matrix $\mathbf{\tilde{M}}$ is derived in two variants, one for unfolding the electron event counts
and another variant for unfolding the positron event counts. The difference is the Monte-Carlo event weight that is used when filling the matrix. The Monte-Carlo
$e^{\pm}$ datasets used in this work are simulated with a flat energy spectrum in $\log(E)$, which does not represent the energy distribution observed in ISS data.
The Monte-Carlo simulation needs to be re-weighted to ensure that the energy distribution follows the one observed in ISS data. To achieve that goal the electron
matrix $\mathbf{\tilde{M}}_{\text{elec}}$ is weighted using the final electron flux and the positron matrix $\mathbf{\tilde{M}}_{\text{posi}}$
using the final positron flux. Obviously this is an iterative procedure since the final flux is not known a-priori. As starting point the electron and positron flux
from previous publications was used and then refined using the fluxes derived in this work.

The unfolding procedure was repeated using another method: the \enquote{folded acceptance method} (bin-by-bin unfolding) which gives consistent results
with the Bayesian unfolding, but wrongly estimates the statistical uncertainties of the event counts after unfolding. Further checks are presented in
\cref{sec:analysis-flux-time-averaged-sysunc-unfolding}.

\clearpage
\section{Time-averaged systematic uncertainties for flux analysis}
\label{sec:analysis-flux-time-averaged-sysunc}

In this section all systematic uncertainties that contribute to the time-averaged fluxes are summarized.
The most important systematic uncertainty for the flux analysis is the acceptance uncertainty, described in \cref{sec:analysis-flux-time-averaged-sysunc-acceptance},
followed by the unfolding uncertainty, described in \cref{sec:analysis-flux-time-averaged-sysunc-unfolding}. Above \SI{500}{\GeV} the knowledge of the
TRD estimator and the associated uncertainty is the dominant systematic uncertainty, which is presented in the next section.

\subsection{TRD estimator}
\label{sec:analysis-flux-time-averaged-sysunc-trd-estimator}

In this section the contribution to the overall systematic uncertainty on the fluxes, induced by the choice of the TRD templates, will be quantified.
The idea is to randomize the TRD template parameters e.g. $M = 100$ times and repeat the two-dimensional fit procedure with the \enquote{smeared}
templates. At each iteration $m$ the number of electrons and the number of protons is recorded. When the fit procedure is repeated, a distribution
of these numbers emerges. The RMS of the electron and proton distributions is a measure of the systematic uncertainty induced by the limited
knowledge of the shape of TRD estimator.

All TRD templates are described analytically as a function of energy with a set of parameters for each energy bin, as described in
\cref{sec:analysis-lepton-counts-trd-templates}. These parameters have associated uncertainties extracted from the fit procedures.
Furthermore the fit procedure not only yields the best-fit parameters with their uncertainties but also the correlations between them,
encoded in the covariance matrix $\mathbf{V}$.

The covariance matrix in energy bin $j$ of the fit result for negative rigidities (the fit result of the sample where the electron and
charge-confused proton template was determined) is denoted as $\mathbf{V}_{j\text{, neg}}$. The values of the $N_{\text{neg}}$ electron/charge-confused proton
template parameters are denoted as $\mathbf{p}_{\text{neg}}$. The covariance matrix of the fit result for positive rigidities is
denoted as $\mathbf{V}_{j\text{, pos}}$ and the values of the $N_{\text{pos}}$ proton template parameters as $\mathbf{p}_{\text{pos}}$, respectively.

Smeared TRD template parameters for the electron template and charge-confused proton template, for the $m^{\text{th}}$ iteration in energy bin $j$ are obtained by:

\begin{equation*}
  \mathbf{\tilde{p}}_{\text{neg}} = \mathbf{p}_{\text{neg}} + (\mathbf{v}_{m\text{, neg}} \cdot \mathbf{L}_{j\text{, neg}}^{\intercal}).
\end{equation*}

$\mathbf{v}_{m\text{, neg}}$ is a vector of $N_{\text{neg}}$ random variables, following a normal distribution:

\begin{equation*}
  \mathbf{v} = \begin{pmatrix} v_{0} \\ \vdots \\ v_{N} \end{pmatrix} = \begin{pmatrix} \mathcal{N}(\mu = 0,\,\sigma^{2} = 1) \\ \vdots \\ \mathcal{N}(\mu = 0,\,\sigma^{2} = 1) \end{pmatrix}.
\end{equation*}

In the same way new TRD proton template parameters can be generated:

\begin{equation}
  \mathbf{\tilde{p}}_{\text{pos}} = \mathbf{p}_{\text{pos}} + (\mathbf{v}_{m\text{, pos}} \cdot \mathbf{L}_{j\text{, pos}}^{\intercal}).
\end{equation}

The matrices $\mathbf{L}_{j\text{, neg}}$ and $\mathbf{L}_{j\text{, pos}}$ can be derived from the covariance matrices $\mathbf{V}_{j\text{, neg}}$
and $\mathbf{V}_{j\text{, pos}}$, respectively. The covariance matrix $\mathbf{V}$ can be factored uniquely into a product - \enquote{Cholesky decomposition}\ ~\cite{Golub1996} - such that::

\begin{equation*}
  \mathbf{L}^{\intercal} \cdot \mathbf{L} = \mathbf{V}.
\end{equation*}

For each energy bin $j$ the two-dimensional fit procedure is repeated $M = 100$ times and at each iteration $m$ all three TRD templates -
electron template, charge-confused proton template and proton template - are varied. From these \enquote{smeared} one-dimensional TRD templates
two-dimensional templates are constructed, by taking the product of the \enquote{smeared} one-dimensional TRD templates with the regular one-dimensional
CCMVA templates. The two-dimensional positron template is built by mirroring the electron template around the y-axis, identical to the
construction of the regular two-dimensional positron template from the regular electron template. By varying only the TRD part of the two-dimensional
templates, the limited knowledge of the TRD templates can be quantified in isolation from the CCMVA estimator. Thus a systematic uncertainty
can be extracted, quantifying the knowledge of the shapes of the TRD estimator. Note that the amount of charge-confusion is fixed from
the Monte-Carlo simulation (\cref{sec:analysis-lepton-counts-2d-fit}), when repeating the two-dimensional template fits.

\begin{figure}[H]
  \begin{subfigure}{0.50\linewidth}
    \includegraphics[width=\linewidth]{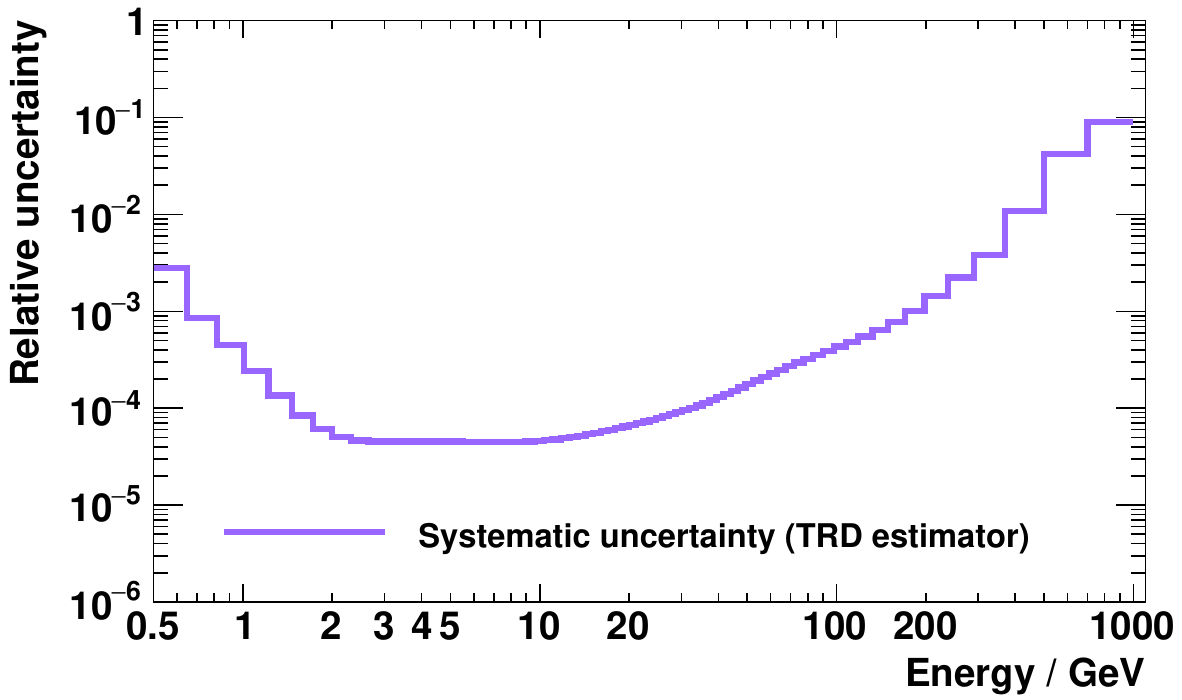}
    \caption{For electron flux}
  \end{subfigure}
  \hfill
  \begin{subfigure}{0.50\linewidth}
    \includegraphics[width=\linewidth]{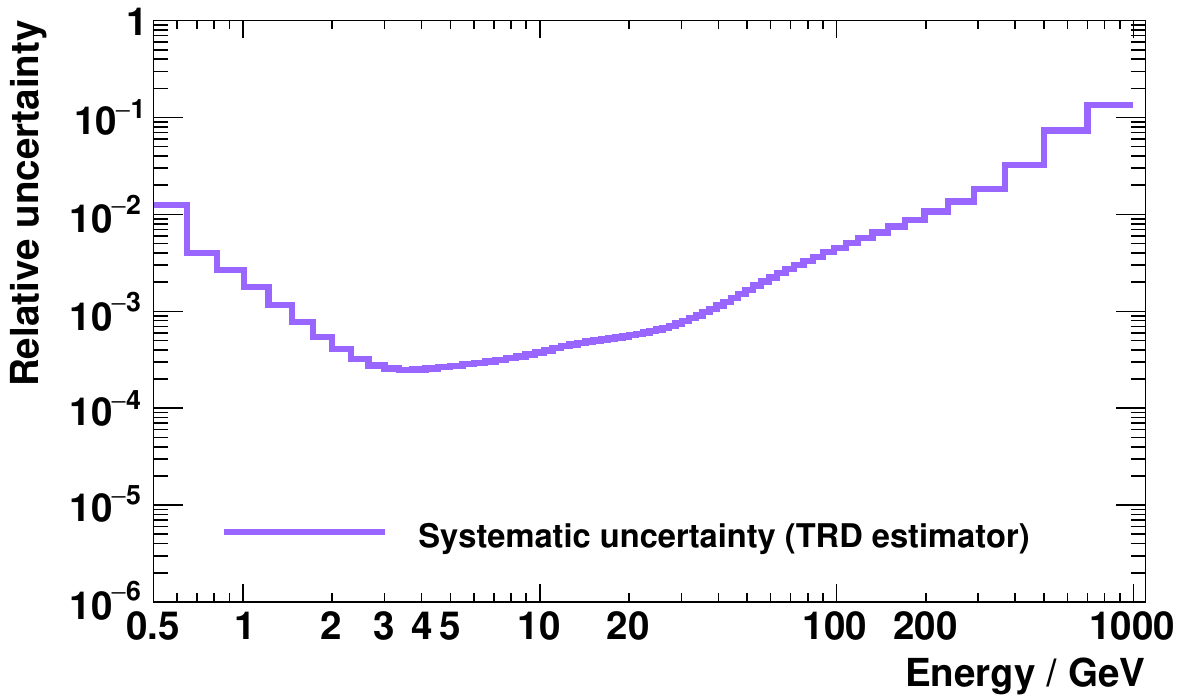}
    \caption{For positron flux}
  \end{subfigure}
  \caption{Visualization of the relative systematic uncertainty associated with the electron flux (left plot) and associated with the positron flux (right plot) due to the knowledge of the TRD estimator.}
  \label{fig:time-averaged-fluxes-syst-uncertainty-trd-estimator}
\end{figure}

\Cref{fig:time-averaged-fluxes-syst-uncertainty-trd-estimator} shows the resulting relative systematic uncertainty $\sigma_{\text{trd}}(E) / \Phi_{e^{\pm}}(E)$ for
the electron flux and the positron flux. The uncertainty associated with the positron flux is larger than the one associated with the electron flux.
The variation of the charge-confused proton template is almost negligible, since the amount of charge-confused protons is small,
compared to the number of electrons. The consequence is that when determining the number of electrons at each iteration $k$,
the RMS is dominated by the knowledge of the electron template. On the other hand, when determining the number of positrons, the
positron template and the proton template are equally important, as the number of protons and positrons are comparable in each
energy bin. This difference is the main reason why the uncertainty from the TRD estimator for the positron flux is larger than for the electron flux.

The shape of the relative systematic uncertainty follows the statistical uncertainty of the data sample, from which the templates
are extracted. It is evident that this is the limiting factor on the knowledge of the TRD templates at high energies and thus
reflected in this uncertainty, which goes up to \SIapprox{10}{\percent} in the highest energy bin.

\subsection{CCMVA estimator}
\label{sec:analysis-flux-time-averaged-sysunc-charge-confusion-estimator}

The two-dimensional template fit presented in \cref{sec:analysis-lepton-counts-2d-fit} is executed twice for each of the single-track / multi-tracks / all-tracks samples.
In the first iteration the charge-confused is a free fit parameter. In the second iteration the charge-confusion is fixed to the Monte-Carlo prediction, as good
agreement was found between ISS data and the Monte-Carlo simulation (\cref{fig:2d-all-tracks-analysis-2d-cc-comparison}).

\begin{figure}[H]
  \begin{subfigure}{0.5\linewidth}
    \includegraphics[width=\linewidth]{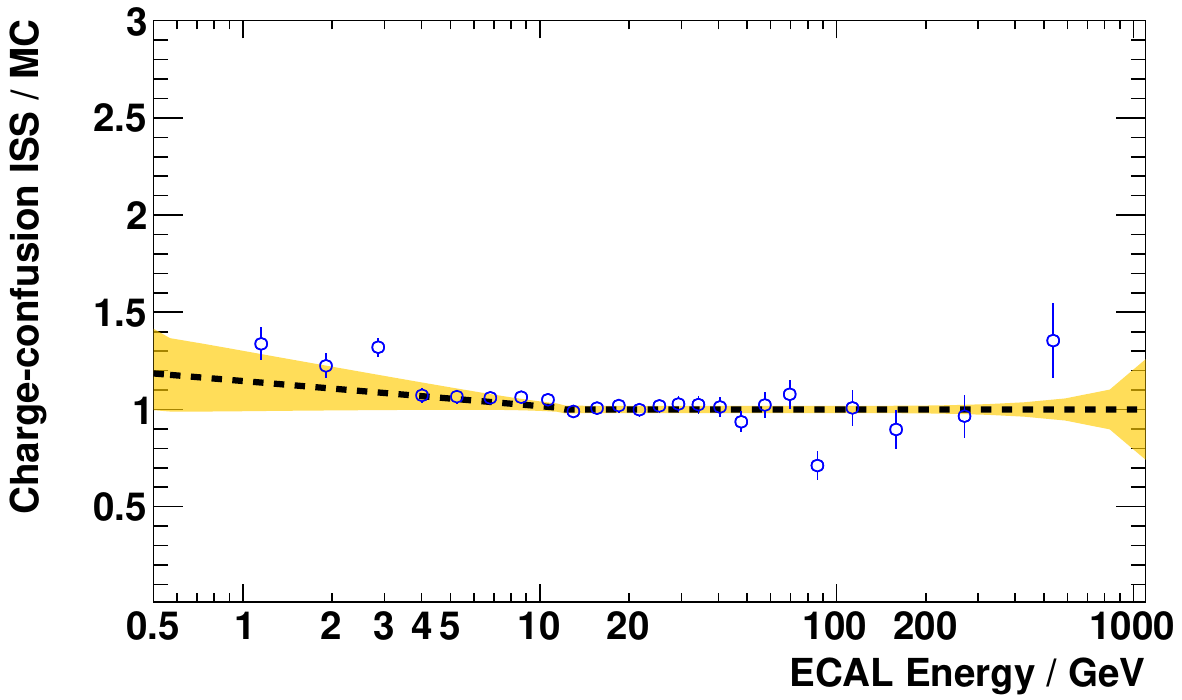}
    \caption{Single-track sample}
  \end{subfigure}
  \hfill
  \begin{subfigure}{0.5\linewidth}
    \includegraphics[width=\linewidth]{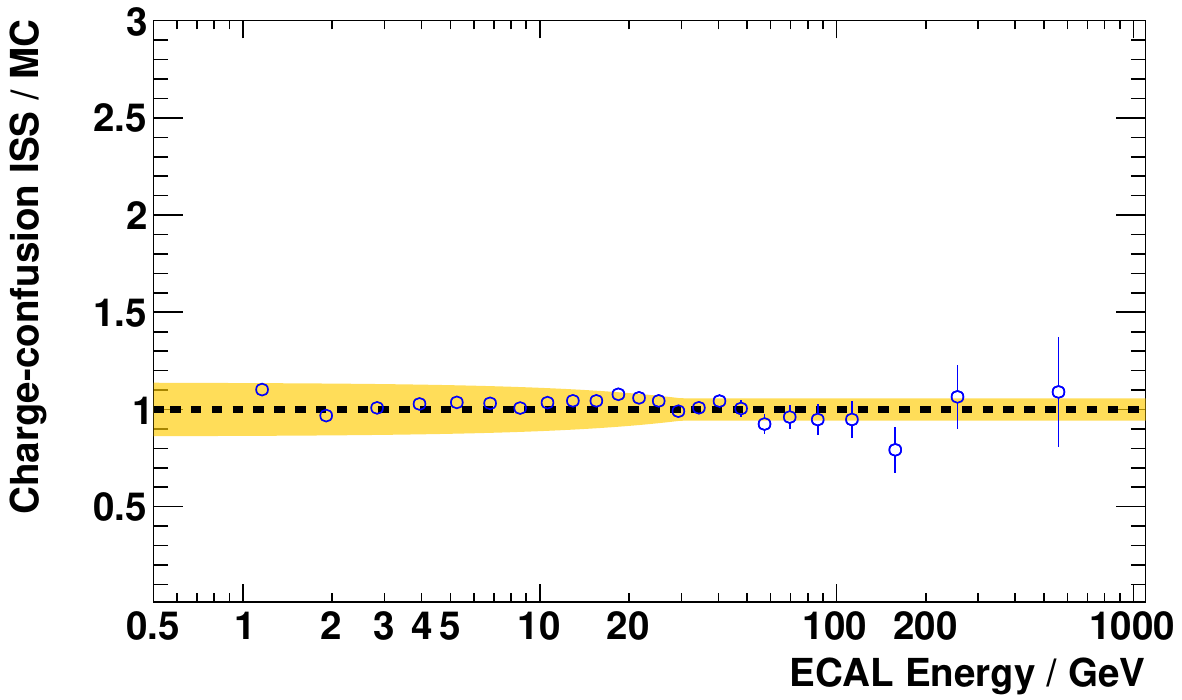}
    \caption{Multi-tracks sample}
  \end{subfigure}
  \hfill
  \begin{subfigure}{0.85\linewidth}
    \includegraphics[width=\linewidth]{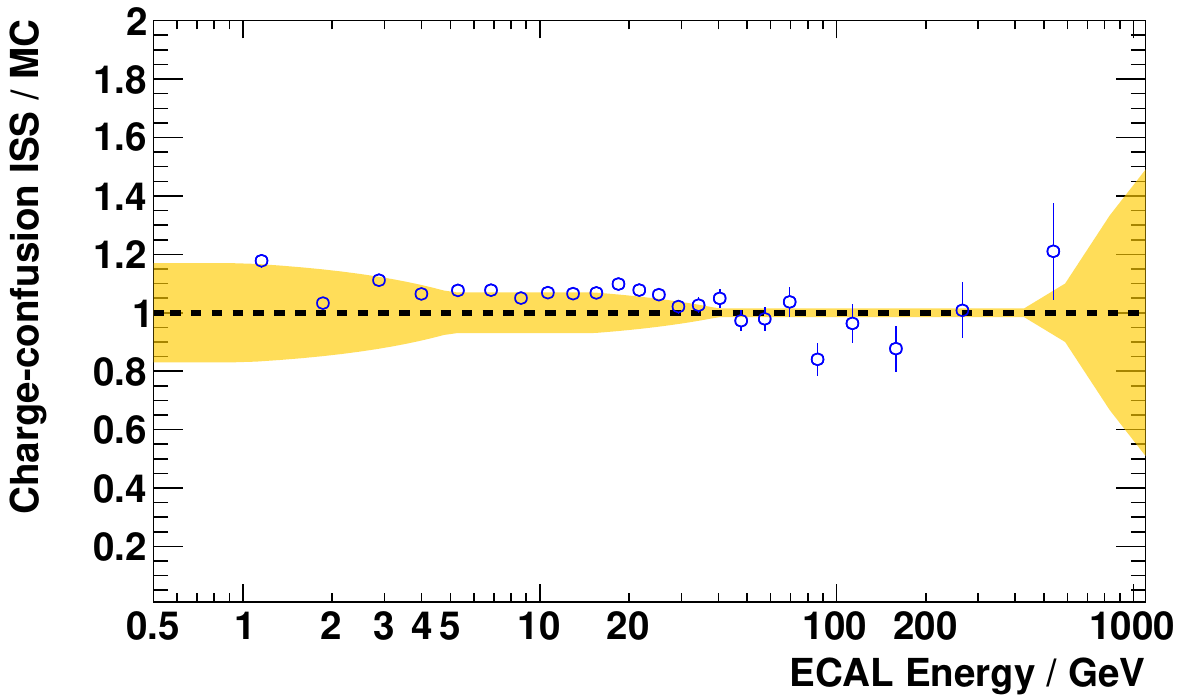}
    \caption{All-tracks sample}
  \end{subfigure}
  \caption{Comparison of the ratio of charge-confusion determined from ISS data divided by the prediction from the Monte-Carlo simulation (blue symbols) for the single-track sample, the multi-tracks sample and the all-tracks sample. The orange band represents the chosen systematic uncertainty to cover the differences between the ISS data and the Monte-Carlo simulation.}
  \label{fig:2d-all-tracks-analysis-2d-cc-comparison-ratio}
\end{figure}

From the difference in the magnitude of charge-confusion a systematic uncertainty can be derived. \Cref{fig:2d-all-tracks-analysis-2d-cc-comparison-ratio}
shows the ratio: ISS charge-confusion over Monte-Carlo prediction for the three different track samples. The uncertainty was chosen by hand to ensure that
at least \SIapprox{67}{\percent} of all data points are covered by the uncertainty band. Since the charge-confusion curves (\cref{fig:2d-all-tracks-analysis-2d-cc-comparison})
are smooth as function of energy and show no prominent structures, statistical fluctuations were minimized, by averaging over three analysis energy bins.

For the flux analysis the all-tracks samples is used and thus the systematic uncertainty from the lower plot in \cref{fig:2d-all-tracks-analysis-2d-cc-comparison-ratio}
needs to be examined. The uncertainty is \SI{17}{\percent} at \SI{0.5}{\GeV} and decreases to \SI{7}{\percent} at \SI{5}{\GeV}. Between \SIrange{15}{40}{\GeV} it gradually decreases
to \SI{1.5}{\percent} and stays constant until \SI{500}{\GeV}, where it increases again up to to \SI{35}{\percent} at \SI{1}{\TeV}.

\begin{figure}[H]
  \begin{subfigure}{0.50\linewidth}
    \includegraphics[width=\linewidth]{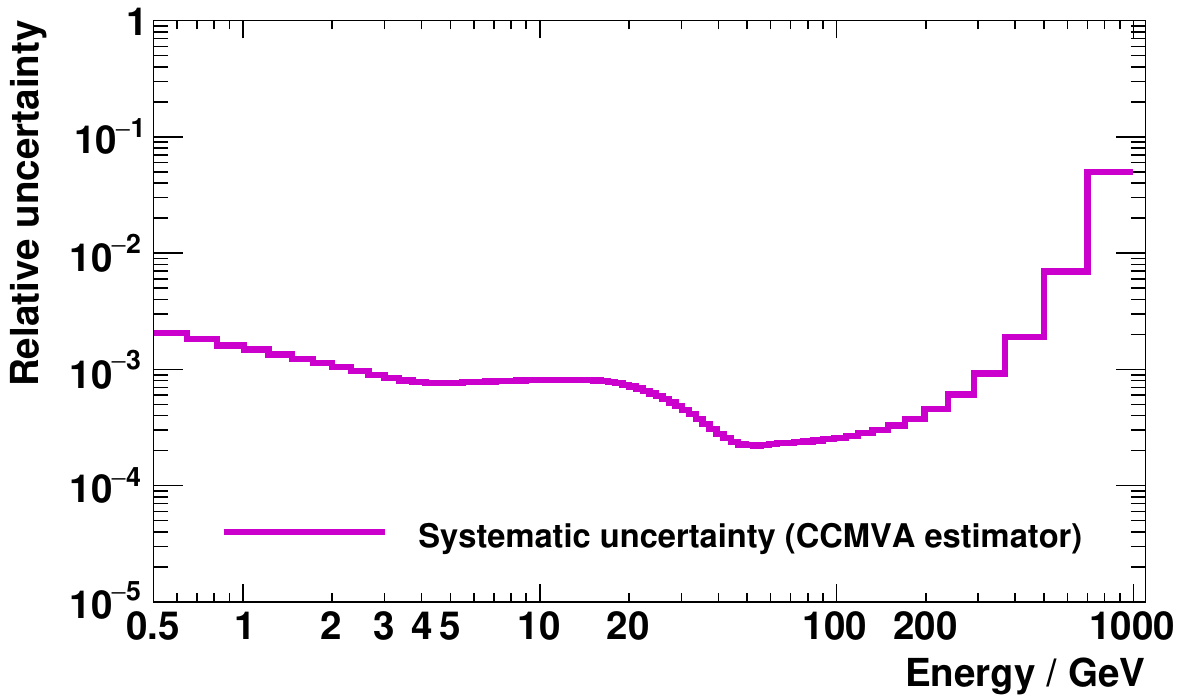}
    \caption{For electron flux}
  \end{subfigure}
  \hfill
  \begin{subfigure}{0.50\linewidth}
    \includegraphics[width=\linewidth]{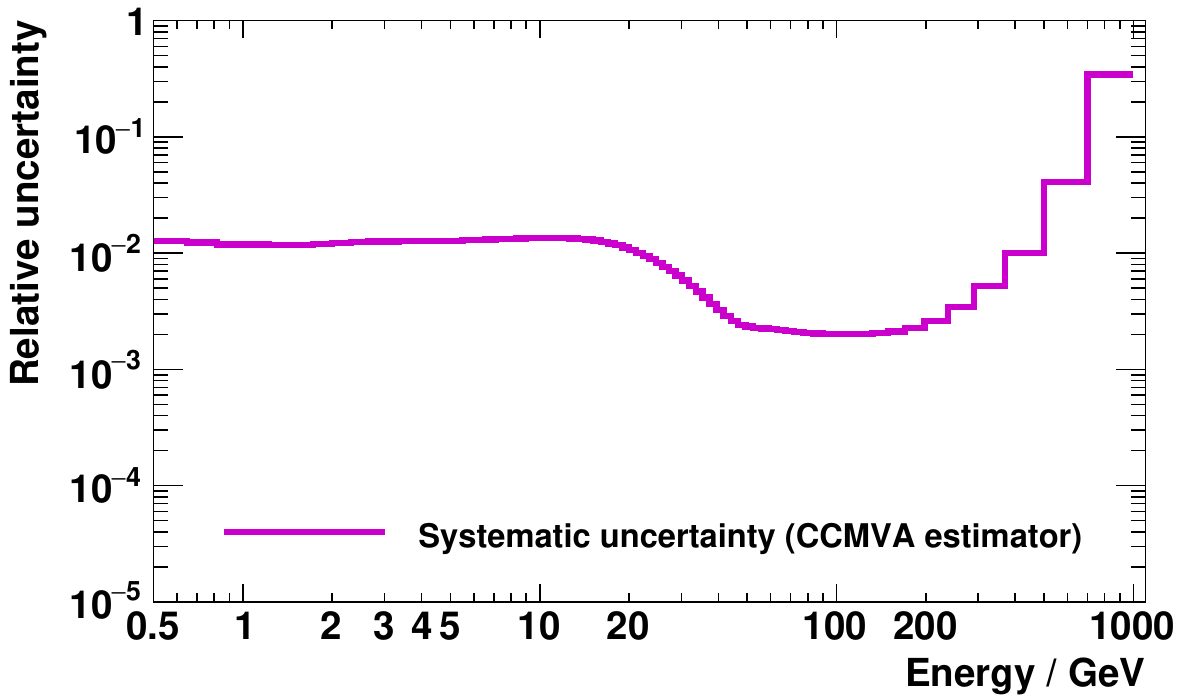}
    \caption{For positron flux}
  \end{subfigure}
  \caption{Visualization of the relative systematic uncertainty associated with the electron flux (left plot) and associated with the positron flux (right plot) due to the knowledge of the charge-confusion.}
  \label{fig:time-averaged-fluxes-syst-uncertainty-charge-confusion}
\end{figure}

Since the systematic uncertainty of the charge-confusion determination - $\sigma_{\text{cc}}(E)$ - is known, the effect on the fluxes can be quantified, by error propagation.
\Cref{fig:time-averaged-fluxes-syst-uncertainty-charge-confusion} shows the resulting relative systematic uncertainty $\sigma_{\text{cc, }e^{\pm}}(E) / \Phi_{e^{\pm}}(E)$ as function of energy.

The systematic uncertainty due to the knowledge of the charge-confusion affects the positron flux stronger than the electron flux.
For the electron flux the systematic uncertainty gradually increases from the permille level up to \SIapprox{5}{\percent} at \SI{1}{\TeV}.
For the positron flux the situation is different: the charge-confusion systematic uncertainty starts at \SIapprox{1}{\percent}, slowly decreasing
to \SI{0.25}{\percent} at \SIapprox{150}{\GeV} and then gradually increases up to to \SI{35}{\percent} at \SI{1}{\TeV}.

This behaviour matches the expectation since there are many more charge-confused electrons that can be misreconstructed as positrons, than
charge-confused positrons that can be be misreconstructed as electrons. Due to the larger abundance of electrons compared to positrons in cosmic rays
the electrons are less affected by the charge-confusion, which is reflected in the systematic uncertainty.

\subsection{ECAL estimator efficiency}
\label{sec:analysis-flux-time-averaged-sysunc-ecal-estimator-efficiency}

The derivation of the ECAL estimator efficiency including its uncertainty was already discussed in \cref{sec:analysis-flux-time-averaged-ecal-estimator}.
The uncertainty determined using the \enquote{Simultaneous fit method} (\cref{fig:ecal-estimator-ecal-bdt-cut-final-test}) is used as
systematic uncertainty of the ECAL estimator efficiency.

\begin{figure}[H]
  \centering
  \includegraphics[width=0.5\linewidth]{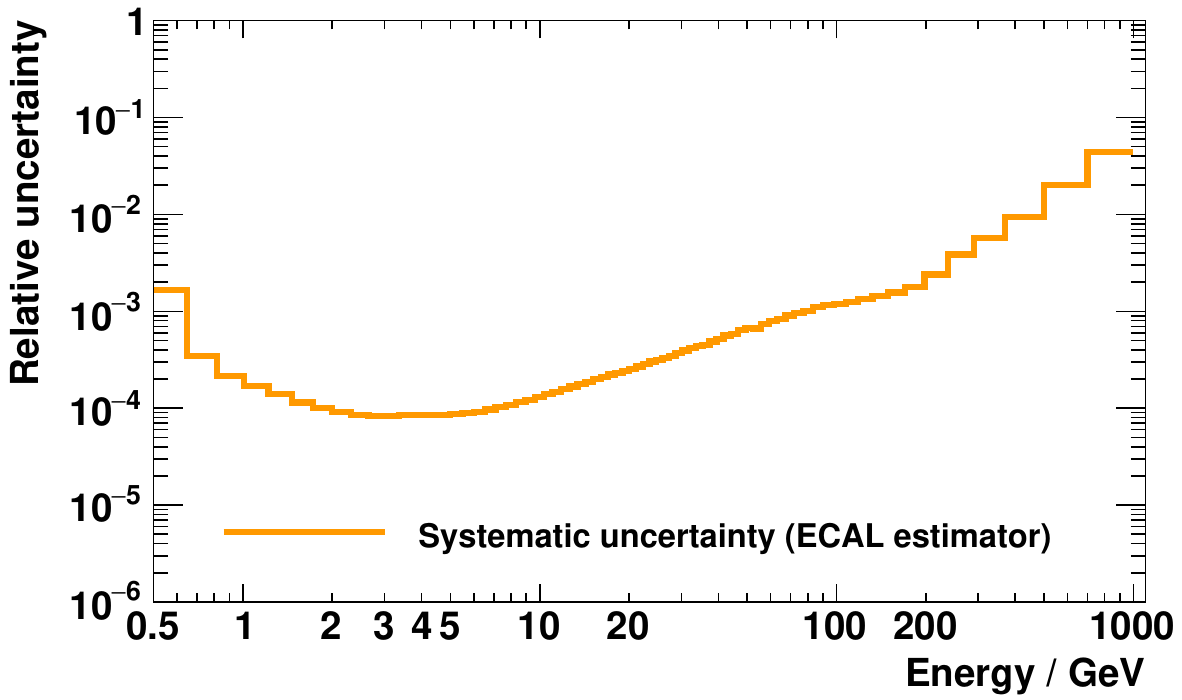}
  \caption{Visualization of the relative systematic uncertainty associated with the electron flux and the positron flux due to the knowledge of the ECAL estimator efficiency.}
  \label{fig:time-averaged-fluxes-syst-uncertainty-ecal-estimator}
\end{figure}

\Cref{fig:time-averaged-fluxes-syst-uncertainty-ecal-estimator} shows the resulting relative systematic uncertainty $\sigma_{\text{ecal}}(E) / \Phi_{e^{\pm}}(E)$ as function of energy.
The uncertainty associated with the electron flux is identical to the uncertainty associated with the positron flux.

\subsection{Trigger efficiency}
\label{sec:analysis-flux-time-averaged-sysunc-trigger-efficiency}

The uncertainty of the trigger efficiency stems from the number of unbiased triggers, which rapidly decreases with increasing energy.
The physics trigger histogram as function of energy (all physics triggers sample) is divided by the total trigger histogram as function of energy
(all physics and unbiased triggers after prescaling is applied) to deduce the trigger efficiency. The associated uncertainty is computed according
to Clopper-Pearson~\cite{ClopperPearson1934}.

Above \SIapprox{15}{\GeV} only six unbiased events remain in the ISS data sample. The lower limit on the trigger efficiency above \SI{15}{\GeV} is 0.9999 at \SI{95}{\percent} confidence level.
The systematic uncertainty on the trigger efficiency above \SI{15}{\GeV} is extrapolated from the data below \SI{15}{\GeV}, by hand. This does not impose any bias for the flux analysis,
since the contribution to the overall systematic uncertainty is below the permille level at energies above \SI{15}{\GeV}.

\begin{figure}[H]
  \centering
  \includegraphics[width=0.7\linewidth]{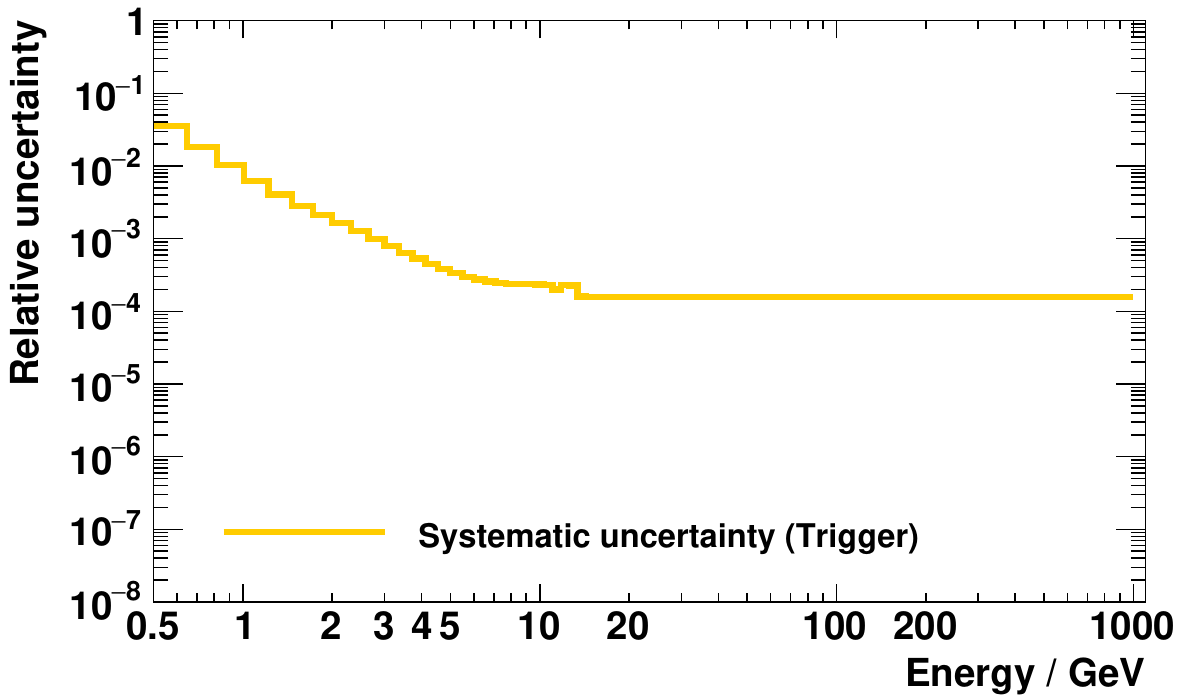}
  \caption{Visualization of the relative systematic uncertainty associated with the electron flux and the positron flux due to the knowledge of the trigger efficiency.}
  \label{fig:time-averaged-fluxes-syst-uncertainty-trigger}
\end{figure}

\Cref{fig:time-averaged-fluxes-syst-uncertainty-trigger} shows the resulting relative systematic uncertainty $\sigma_{\text{trigger}}(E) / \Phi_{e^{\pm}}(E)$ as function of energy.
The uncertainty associated with the electron flux is identical to the uncertainty associated with the positron flux.

\subsection{Measuring time}
\label{sec:analysis-flux-time-averaged-sysunc-measuring-time}

The measuring time $T(E)$ was derived in \cref{sec:analysis-flux-time-averaged-measuring-time}. The energy-independent part of the measuring time
is free of any systematic uncertainty since all ingredients are known accurately: the seconds between the start and the end of data taking and the live-time of the experiment
in each second. A possible source of a systematic uncertainty arises in the energy-dependence: the geomagnetic cut-off calculation.

Different geomagnetic cut-off modes were tested: Størmer~\cite{Stoermer1956} cut-off in \SI{25}{\degree}, \SI{30}{\degree}, \SI{35}{\degree} and \SI{40}{\degree} field of view.
With an increased field of view the geomagnetic cut-off in each second is enlarged. Therefore the measured event counts at low energies decrease since less particles at a given
energy fulfill the requirement of having a measured energy above the geomagnetic cut-off, which is the necessary requirement to classify them as primary cosmic rays. With the
chosen safety factor $f_\text{safety} = 1.2$ the fluxes in each field of view were identical, except for an increased statistical error, as expected from the decreased event counts.

Furthermore the Størmer cut-off was exchanged for the more sophisticated IGRF cut-off while scanning the safety factor from $0.9 - 1.4$. With the chosen
nominal safety factor $f_\text{safety} = 1.2$ all fluxes were identical to those obtained using the Størmer cut-off, except for the first energy
bin \SIrange{0.5}{0.7}{\GeV} which has zero events when using the IGRF cut-off. Since AMS already published electron and positron fluxes starting at
\SI{0.5}{\GeV} and since the remaining flux data points are consistent between IGRF / Størmer, the Størmer cut-off was chosen for this work.

The aforementioned cross-checks all yield consistent results, therefore no additional systematic uncertainty for the measuring time $T(E)$ is needed.

\subsection{Acceptance}
\label{sec:analysis-flux-time-averaged-sysunc-acceptance}

\begin{figure}[H]
  \centering
  \includegraphics[width=0.85\linewidth]{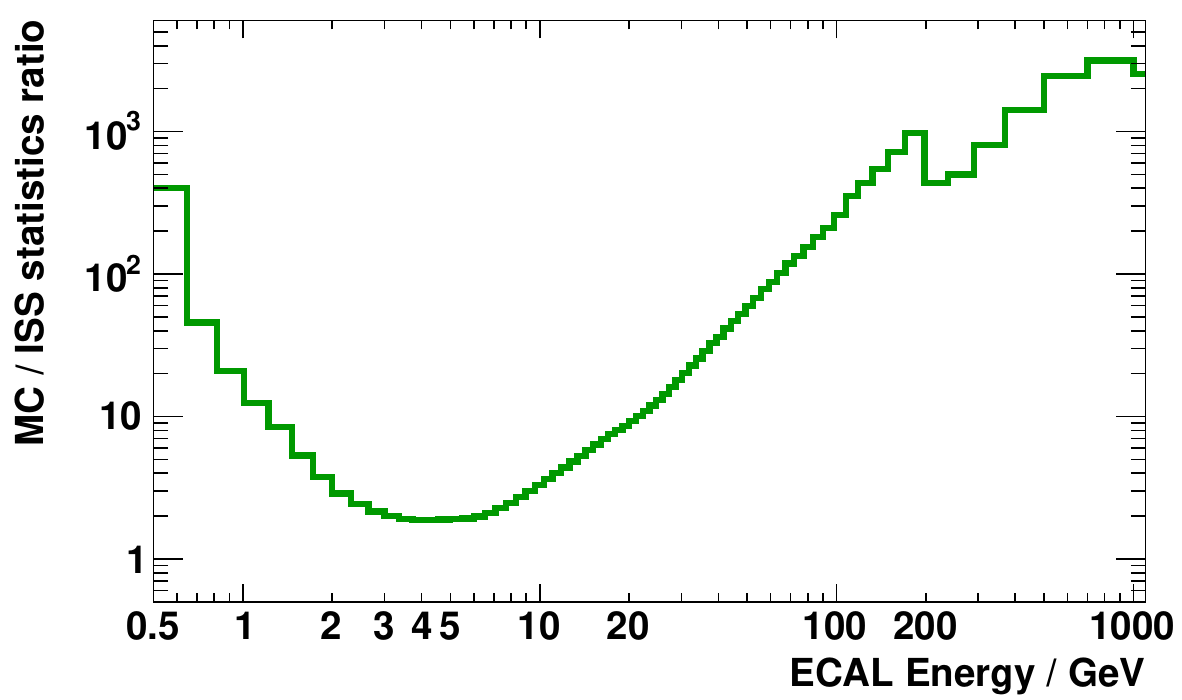}
  \caption{Comparison of the available Monte-Carlo statistics to the ISS statistics after applying all data quality, preselection, selection and identification cuts. The green line shows the ratio of negative rigidity Monte-Carlo event counts divided by ISS event counts.}
  \label{fig:mc-iss-statistics}
\end{figure}

The acceptance was derived in \cref{sec:analysis-flux-time-averaged-acceptance}. Two possible sources of systematic uncertainties arise: the finite
Monte-Carlo statistics used to derived the acceptance from the Monte-Carlo simulation and the Data/Monte-Carlo correction factor $(1 + \delta(E))$. The
Monte-Carlo statistics exceeds the ISS statistics over all energies, as shown in \cref{fig:mc-iss-statistics}. Especially at high energies the Monte-Carlo
contains orders of magnitudes more events than the ISS data. The uncertainty from the Monte-Carlo statistics is negligible over all energies.

Thus the only relevant systematic uncertainty for the acceptance stems from the uncertainty on the correction factor $(1 + \delta(E))$.
The uncertainty on the correction factor was described in detail in \cref{sec:analysis-flux-time-averaged-acceptance}.
The magnitude of the combined Data/Monte-Carlo correction factor $(1 + \delta(E))$ was presented in \cref{fig:acceptance-final-correction}.
The correction applied to the Monte-Carlo acceptance varies from \SI{2}{\percent} at low energies up to \SI{4}{\percent} at high energies and is
known for all energies better than \SI{1.5}{\percent}, indicated by the orange band in the plot. This directly translates
into the systematic uncertainty used for the $e^{\pm}$ fluxes.

\begin{figure}[H]
  \centering
  \includegraphics[width=0.95\linewidth]{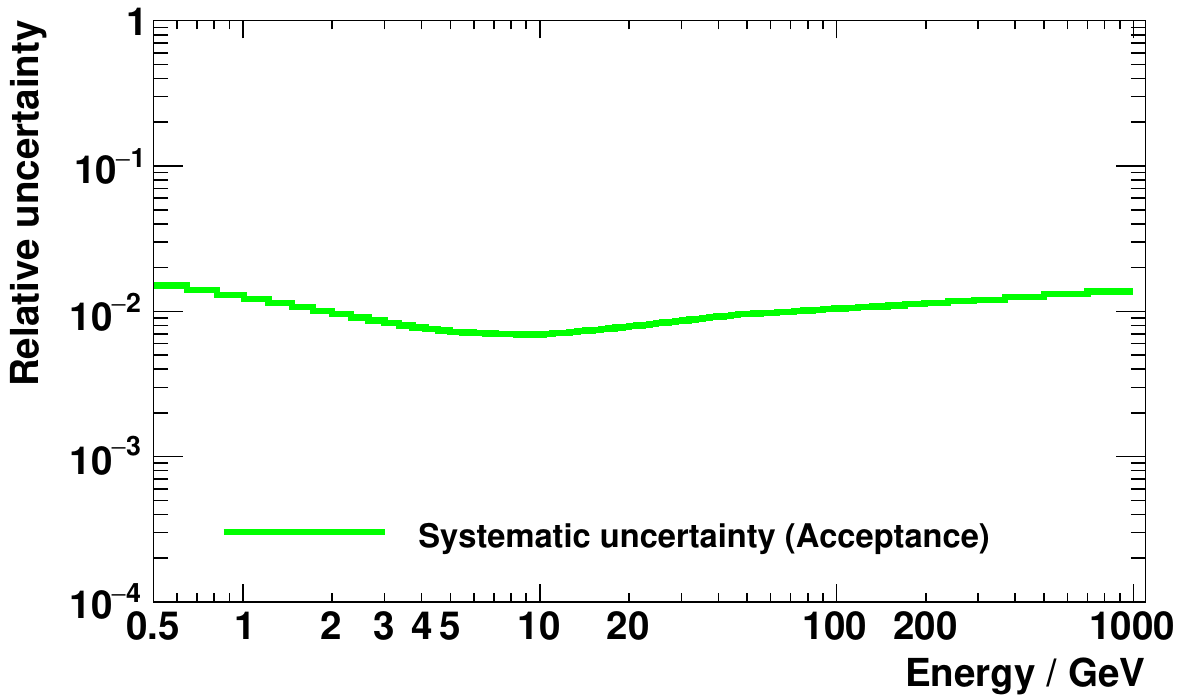}
  \caption{Visualization of the relative systematic uncertainty associated with the electron flux and the positron flux due to the knowledge of the acceptance.}
  \label{fig:time-averaged-fluxes-syst-uncertainty-acceptance}
\end{figure}

\Cref{fig:time-averaged-fluxes-syst-uncertainty-acceptance} shows the resulting relative systematic uncertainty $\sigma_{\text{acc}}(E) / \Phi_{e^{\pm}}(E)$ as function of energy.
The uncertainties associated with the electron flux and the positron flux is identical.

\subsection{Unfolding}
\label{sec:analysis-flux-time-averaged-sysunc-unfolding}

The unfolding procedure was described in detail in \cref{sec:analysis-flux-time-averaged-unfolding}. Two possible sources of systematic uncertainties arise:
the degree to which the migration matrix is known and the stability of the unfolding procedure.

\subsubsection{Knowledge of the migration matrix}
\label{sec:analysis-flux-time-averaged-sysunc-unfolding-knowledge-migration-matrix}

To obtain the migration matrix one has to rely on the Monte-Carlo simulation. As AMS-02 was extensively tested in a beam test, the migration matrix can be
cross-checked at different energies and compared with the Monte-Carlo simulation, for both electrons and positrons. This allows to asses the uncertainty
of the unfolding method, due to the knowledge of the migration matrix.

To ease the comparison between the Monte-Carlo simulation and the test beam data, the migration matrix will be parameterized using an analytical model.
The migration matrix, shown in \cref{fig:unfolding-migration-matrix}, is projected to the x-axis (reconstructed energy axis) for each bin in y-direction
(true energy axis). In each of the projections an analytical model of the reconstructed energy migration is derived, to create a full model of the
reconstructed energy as function of the true energy.

A sum of a \textit{Landau} distribution and a \textit{Crystal Ball} function forms the analytical model:

\begin{equation}
  \label{eq:unfolding-analytical-migration-model}
  f_{\text{model}}(E) = \xi \cdot p_{\text{landau}}(E) + \nu \cdot f_{\text{cb}}(E),
\end{equation}

where $\xi$ and $\nu$ denote the relative contributions of both functions to the model. The analytical model depends on seven parameters, two of them describing
the shape of the \textit{Landau} distribution, four associated with the \textit{Crystal Ball} function and one additional parameter $n_{\text{lf}}$, which controls how much the
landau distribution contributes to the sum ($\xi \propto n_{\text{lf}}$).

The \textit{Landau} distribution~\cite{Landau1944} is characterized by the \gls{MPV} $\mu$ and the width $\sigma$.
The \textit{Crystal Ball} function~\cite{Skwarnicki1986} consists of a gaussian core and a power law tail below a certain threshold. It depends
on four parameters: $f_{\text{cb}}(E; \alpha, n, \bar{x}, \sigma)$.

\begin{figure}[H]
  \includegraphics[width=\linewidth]{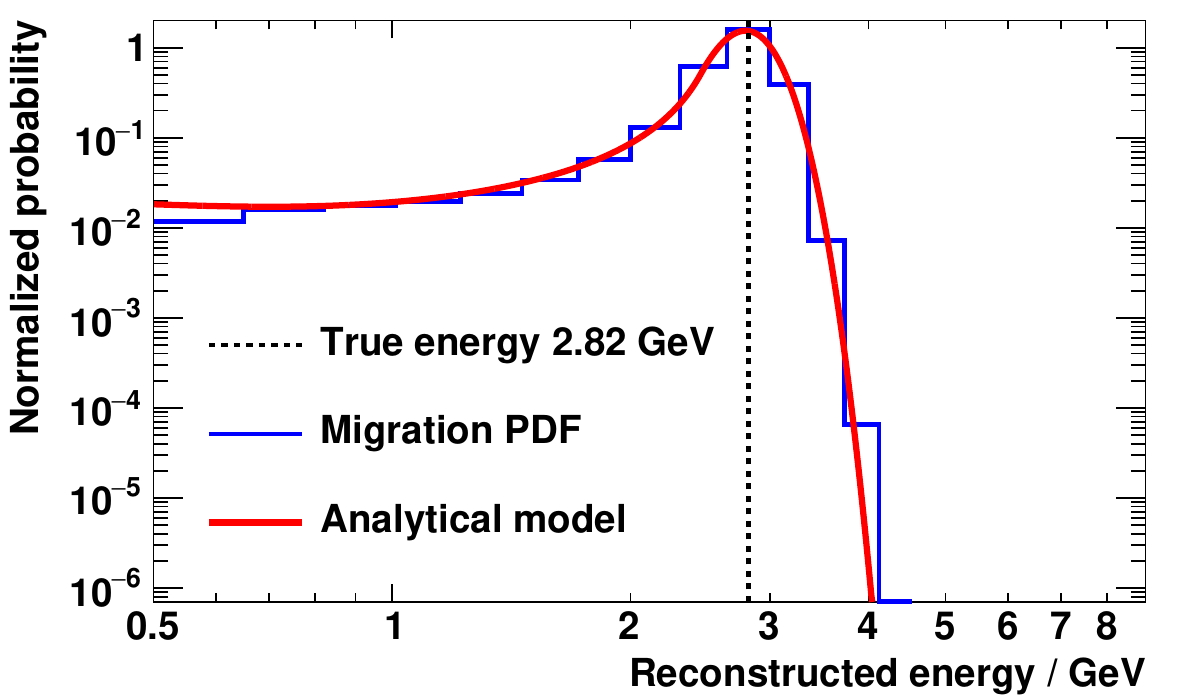}
  \caption{Projection of the migration matrix, \cref{fig:unfolding-migration-matrix}, to the x-axis (reconstructed energy) for a given true energy. The projection is shown as blue histogram and the red line shows the analytical model - \cref{eq:unfolding-analytical-migration-model} - fit to the blue histogram. The black dashed line shows the true energy (\SIvarEquals{y}{2.82}{\GeV}) for comparison.}
  \label{fig:unfolding-test-migration-parameterization-in-bin-10}
\end{figure}

\Cref{fig:unfolding-test-migration-parameterization-in-bin-10} shows the analytical model of the migration matrix in an example energy bin.
The analytical model is a PDF -- the integral over the curve yields 1.

The analytical model - \cref{eq:unfolding-analytical-migration-model} - is fitted to all bins in y-direction (true energy) of the migration matrix.

\begin{figure}[H]
  \centering
  \includegraphics[width=0.9\linewidth]{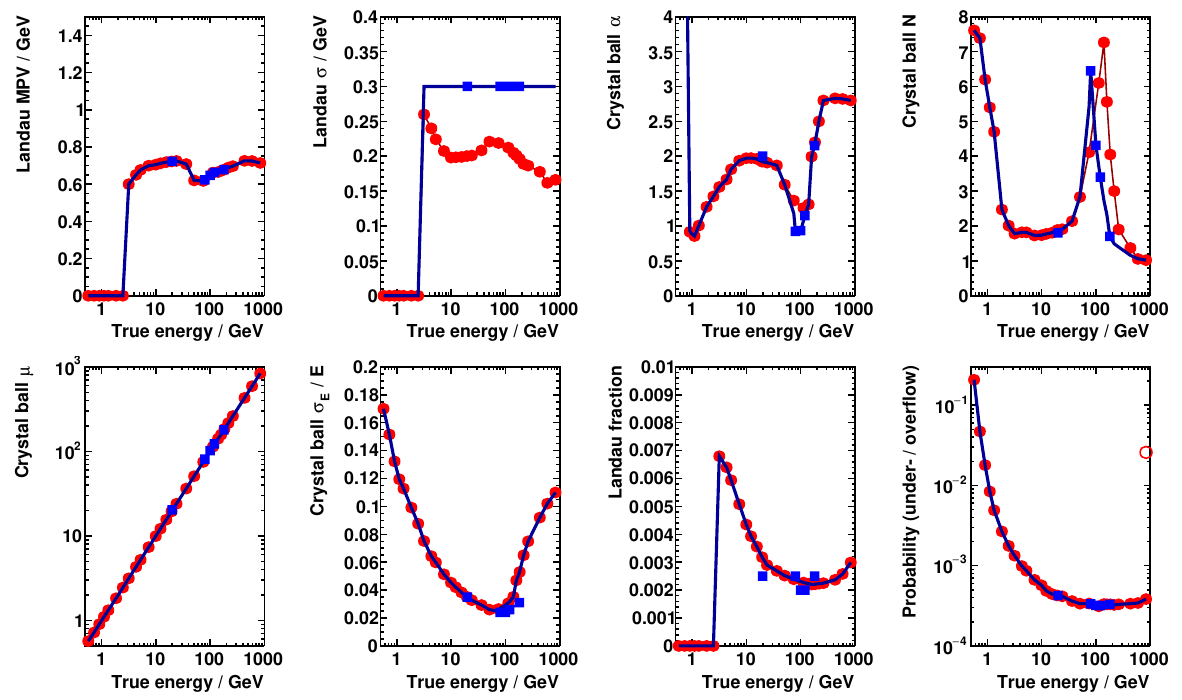}
  \caption{Results of the fit of the analytical model - \cref{eq:unfolding-analytical-migration-model} - to all bins in y-direction (true energy) of the migration matrix obtained from the Monte-Carlo simulation (red symbols), compared to the analytical model fit to the migration matrices obtained from the mono-energetic testbeam data sets (blue symbols). The testbeam data sets that were analyzed are $e^{+}$ at 20 and \SI{80}{\GeV}, $e^{-}$ at 100, 120 and \SI{180}{\GeV}, yielding five different migration matrices. The analytical model is fit to all five migration matrices and the parameters are extracted, shown as blue symbols. The testbeam results can be compared to the prediction from the Monte-Carlo simulation: Three parameters show a significant deviation: the Landau $\sigma$, the Crystal Ball $\alpha$ and $N$ parameters. The dark blue line marks the extrapolation of the parameters obtained from the discrete test beam data points towards the whole energy range.}
  \label{fig:unfolding-parameterization-bt-parameters}
\end{figure}

\Cref{fig:unfolding-parameterization-bt-parameters} shows a comparison of the analytical model parameters for the migration matrix obtained from the Monte-Carlo simulation with the parameters obtained for the migration matrices from the testbeam data. All except three parameters show a good agreement with the Monte-Carlo simulation. The Landau $\sigma$ differs by more than \SI{50}{\percent} from the Monte-Carlo prediction and the energy dependence appears flat. The origin of this difference has to be clarified in a dedicated study. The Crystal Ball $\alpha$ parameter shows a deviation from the Monte-Carlo simulation at \SI{100}{\GeV}, whereas the other data points agree with the Monte-Carlo simulation. The Crystal Ball $N$ parameter shows a similar magnitude and energy dependence but is slightly shifted towards smaller energies, with respect to the parameters obtained from the Monte-Carlo simulation. Note that the $\alpha$ and $N$ parameters are highly correlated, leading to a degeneracy in the fit procedure. The dark blue line represents the extrapolation of the test beam parameters to the whole energy range.

After finishing the fit procedure two distinctive set of parameters - describing the migration matrix analytically - are available: the parameters obtained from the Monte-Carlo simulation and the parameters obtained with the help of the testbeam data. The next step is to generate toy fluxes according to a predefined model of the electron or positron flux and fold them once with the migration matrix from the Monte-Carlo simulation and once with the analytical migration matrix from the test beam. The ratio of the two fluxes shows the impact of the choice of the parameterization on the resulting fluxes and thus is a measure of the systematic uncertainty induced by the migration matrix.

\begin{figure}[H]
  \centering
  \includegraphics[width=0.7\linewidth]{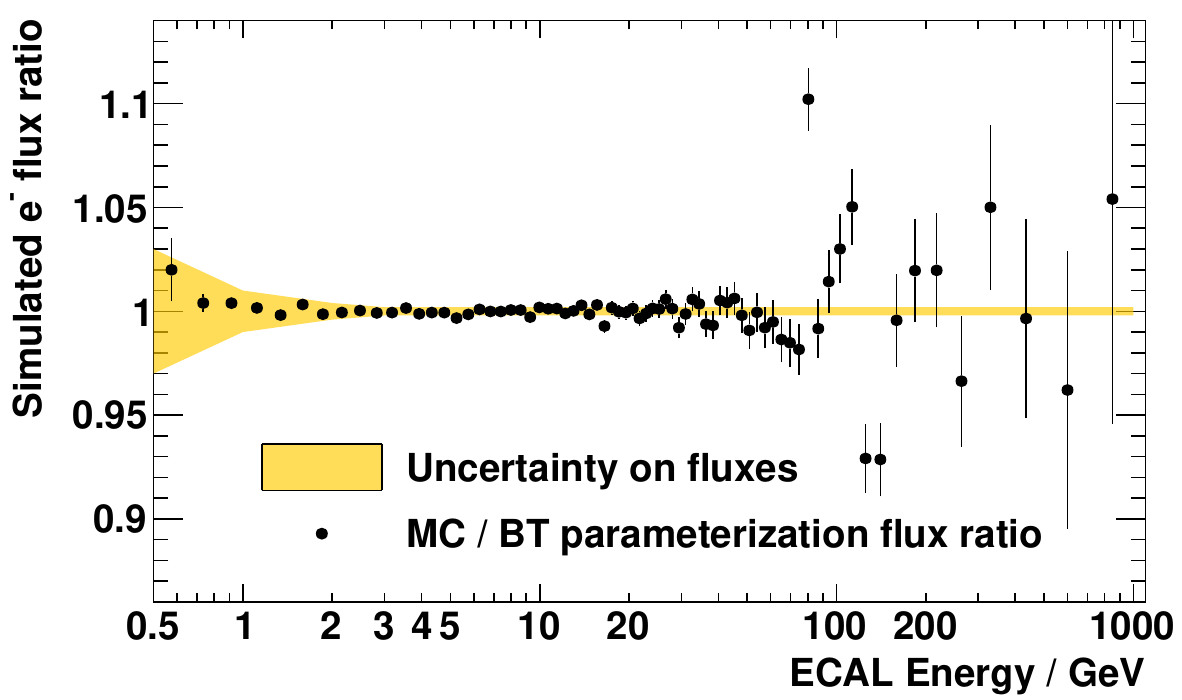}
  \caption{Ratio of the toy $e^{-}$ flux forward folded with the Monte-Carlo parameterization over the toy $e^{-}$ flux forward-folded with the testbeam parameterization - shown as black symbols. The orange band represents the systematic uncertainty on the fluxes due to the knowledge of the migration matrix.}
  \label{fig:unfolding-bt-parameters-syst-error-electron-average}
\end{figure}

\begin{figure}[H]
  \centering
  \includegraphics[width=0.7\linewidth]{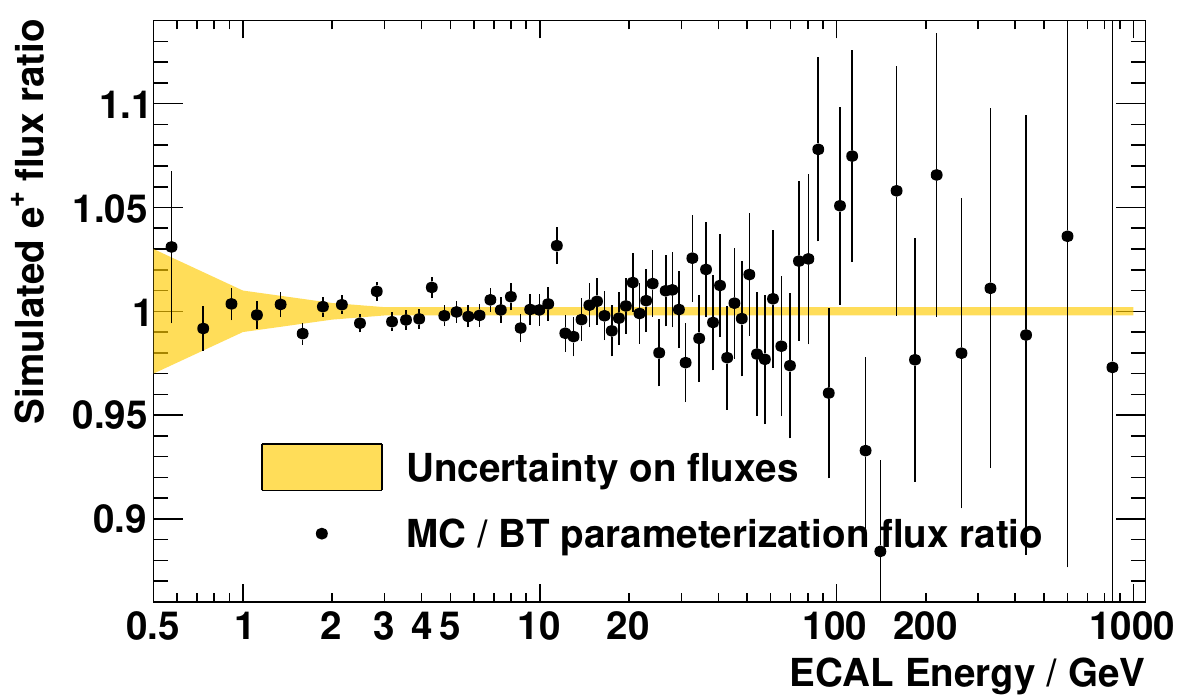}
  \caption{Ratio of the toy $e^{+}$ flux forward folded with the Monte-Carlo parameterization over the toy $e^{+}$ flux forward-folded with the testbeam parameterization - shown as black symbols. The orange band represents the systematic uncertainty on the fluxes due to the knowledge of the migration matrix.}
  \label{fig:unfolding-bt-parameters-syst-error-positron-average}
\end{figure}

\Cref{fig:unfolding-bt-parameters-syst-error-electron-average} shows the ratio of the toy $e^{-}$ flux forward folded with the migration matrix obtained from the Monte-Carlo parameterization over the toy $e^{-}$ flux forward-folded with the migration matrix obtained from the testbeam parameterization. \Cref{fig:unfolding-bt-parameters-syst-error-positron-average} shows the same ratio for the $e^{+}$ toy fluxes. From the data points that are incompatible with unity a systematic uncertainty can be derived, due to the limited knowledge of the migration matrix. This is shown as orange band and is equal for $e^{\pm}$.

All data points above \SIapprox{5}{\GeV} are consistent with unity, except for a few outliers around \SIapprox{100}{\GeV}. In this energy range the simple analytical model - \cref{eq:unfolding-analytical-migration-model} - does not fully describe the true migration PDF. Using a better parameterization these outliers can be avoided. Therefore this does not represent a real effect, but only reveals a deficiency in the choice of the parameterization.


\subsubsection{Stability of unfolding procedure}
\label{sec:analysis-flux-time-averaged-sysunc-unfolding-stability-unfolding-procedure}

\Cref{fig:unfolding-procedure-stability} shows an overview of the procedure that was developed to assess
the stability of the unfolding. The idea is to generate a toy flux many times, according to a predefined
model of the true $e^{\pm}$ flux. This toy flux is then forward folded using the migration matrix and unfolded afterwards
using the Bayesian unfolding method. This yields $O$ unfolded fluxes which can be compared with the single true
input flux model. Any deviation from the true flux has to be attributed to a systematic uncertainty in the unfolding procedure.

\tikzstyle{decision} = [diamond, draw, fill=red!20, text width=7em, text badly centered, node distance=4cm, inner sep=0pt]
\tikzstyle{block} = [rectangle, draw, fill=blue!20, text width=14em, text centered, rounded corners, minimum height=3em]
\tikzstyle{line} = [draw, -latex']

\begin{figure}[H]
  \centering
  \begin{tikzpicture}[scale=0.73, every node/.style={transform shape}, thick, node distance = 2.0cm, auto]
    \node [block] (generate-matrix) {1. Generate migration matrix};
    \node [block, below of=generate-matrix] (simulate-measurement) {2. Simulate measurement};
    \node [block, below of=simulate-measurement] (unfold-flux) {3. Unfold measured toy flux};
    \node [block, right of=unfold-flux, node distance=7cm] (repeat) {Repeat procedure};
    \node [decision, below of=unfold-flux] (decide) {Reached $P=2500$ iterations?};
    \node [block, below of=decide, node distance=4cm] (stop) {Ready};

    \path [line] (generate-matrix) -- (simulate-measurement);
    \path [line] (simulate-measurement) -- (unfold-flux);
    \path [line] (unfold-flux) -- (decide);
    \path [line] (decide) -| node [near start] {no} (repeat);
    \path [line] (repeat) |- (generate-matrix);
    \path [line] (decide) -- node {yes}(stop);
  \end{tikzpicture}
  \caption{Illustration of the procedure to assess the stability of the unfolding procedure.}
  \label{fig:unfolding-procedure-stability}
\end{figure}
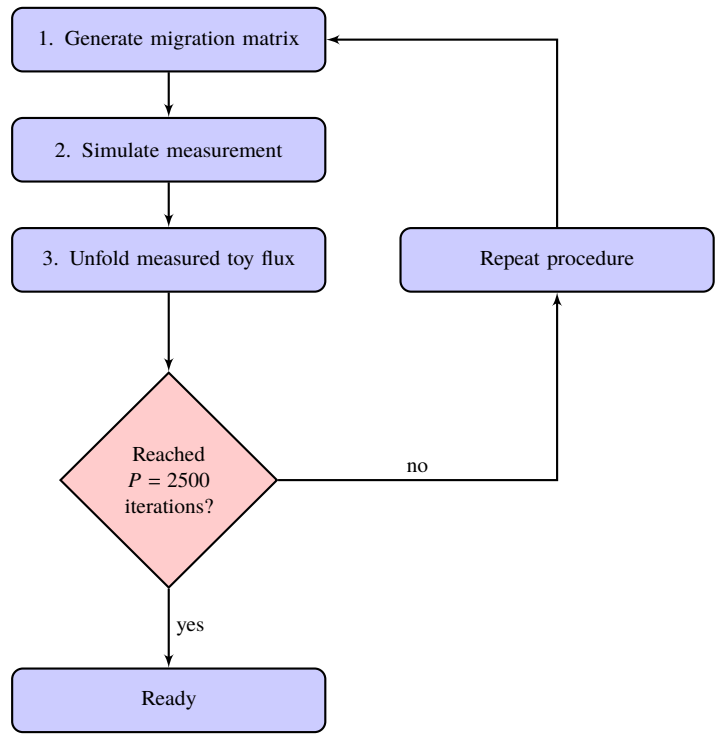

The procedure outlined in \cref{fig:unfolding-procedure-stability} is repeated $P=2500$ times.
At each iteration $p$ the following steps are executed:

\begin{enumerate}
  \item\textbf{Generate migration matrix}\hfill\\
    A new migration matrix $\textbf{M}$ is filled with $M\approx 10^9$ entries. For each entry $m$ a true energy $E_{\text{true}}^{m}$
    is drawn from a random number generator, following a power law of $E^{-1}$ between \SI{0.25}{\GeV} and \SI{1.5}{\TeV}.
    The reconstructed energy is derived by smearing the true energy according to the analytical model from
    \cref{eq:unfolding-analytical-migration-model}: $E_{\text{reconstructed}}^{m} = f_{\text{model}}(E_{\text{true}}^{m})$ and the migration
    matrix is filled.

    A predefined flux $\Phi_{\text{true}}(E_{\text{true}})$ is used as weight, to correctly model the true energy distribution of the cosmic-ray
    $e^{\pm}$ fluxes. The predefined $e^{\pm}$ fluxes are modelled as broken power laws with different breaks and spectral indices, closely matching the
    final $e^{\pm}$ fluxes derived in this work.

    This recipe leads to a unique toy migration matrix $\textbf{M}^{p}$ for each iteration $p$ of the procedure illustrated in
    \cref{fig:unfolding-procedure-stability}.

  \item\textbf{Simulate measurement}\hfill\\
    The toy flux $\Phi_{\text{true}}(E_{\text{true}})$ is multiplied with the acceptance, the trigger efficiency and scaled by the measuring
    time over cut-off (which is an energy-independent scalar above \SIapprox{25}{\GeV}) to compute the number of true event counts as function
    of the true energy: $N_{\text{true}}(E_{\text{true}})$, explicitly ignoring the geomagnetic cut-off. The geomagnetic cut-off must be
    factored out at this point, since it is only known as function of the reconstructed energy. In each energy bin the true event counts
    $N_{\text{true}}(E_{\text{true}})$ are then smeared according to a Poisson distribution to obtain $\hat{N}_{\text{true}}(E_{\text{true}})$,
    to simulate an actual measurement.

    As next step the number of measured event counts as function of the reconstructed energy $\hat{N}_{\text{rec}}(E_{\text{rec}})$
    can be computed. For each true energy bin $k$ the energy $E_{\text{true}}^{k}$ and the number of true event counts $\hat{N}_{\text{true}}^{k}$ are
    known.

    The reconstructed energy is now computed $\hat{N}_{\text{true}}^{k}$ times by smearing the true energy according to the analytical model from
    \cref{eq:unfolding-analytical-migration-model}: $E_{\text{rec}}^{k} = f_{\text{model}}(E_{\text{true}}^{k})$.

    This yields $\hat{N}_{\text{true}}^{k}$ reconstructed energies $E_{\text{rec}}^{k}$, corresponding to the $\hat{N}_{\text{true}}^{k}$
    number of true event counts before smearing. The so-obtained reconstructed energies are filled into a one-dimensional histogram storing the
    reconstructed event counts $N_{\text{rec}}(E_{\text{rec}})$, if the reconstructed energy is above the geomagnetic cut-off.
    The geomagnetic cut-off is simulated by drawing a random number between 0 and 1 and comparing if the random number is smaller
    than the cut-off probability $w_{\text{cut-off}}(E) = T(E)\ /\ T(> \SI{25}{\GeV})$ (defined in
    \cref{sec:analysis-flux-time-averaged-unfolding}). Only if that condition is fulfilled, the
    reconstructed energy is filled into the aforementioned histogram $N_{\text{rec}}(E_{\text{rec}})$.

    This procedure is repeated for all true energy bins $\forall k~\in~[1,76]$ in the analysis. Afterwards the simulation of a measurement is completed:
    both the true event counts $\hat{N}_{\text{true}}(E_{\text{true}})$ and the reconstructed event counts $N_{\text{rec}}(E_{\text{rec}})$,
    including the simulation of the geomagnetic cut-off are available.

    The reconstructed event counts are then divided by the bin widths, the trigger efficiency, the acceptance and the energy-dependent measuring time to
    obtain the reconstructed toy flux $\Phi_{\text{rec}}^{p}(E_{\text{rec}})$ for iteration $p$, as for a real measurement.

  \item\textbf{Unfold measured toy flux}\hfill\\
    The standard unfolding procedure, described in \cref{sec:analysis-flux-time-averaged-unfolding}, is executed using the migration matrix
    $\textbf{M}^{p}$ from step 1, and the reconstructed flux $\Phi_{\text{rec}}^{p}(E_{\text{rec}})$ from step 2. This yields
    the unfolded flux: $\tilde{\Phi}_{\text{unfolded}}^{p}(E_{\text{true}})$.

    The deviation of the unfolded flux $\tilde{\Phi}_{\text{unfolded}}^{p}(E_{\text{true}})$ from the true flux $\Phi_{\text{true}}(E_{\text{true}})$
    divided by the true flux is calculated and recorded for each iteration $p$.
\end{enumerate}

\begin{figure}[H]
  \centering
  \includegraphics[width=0.7\linewidth]{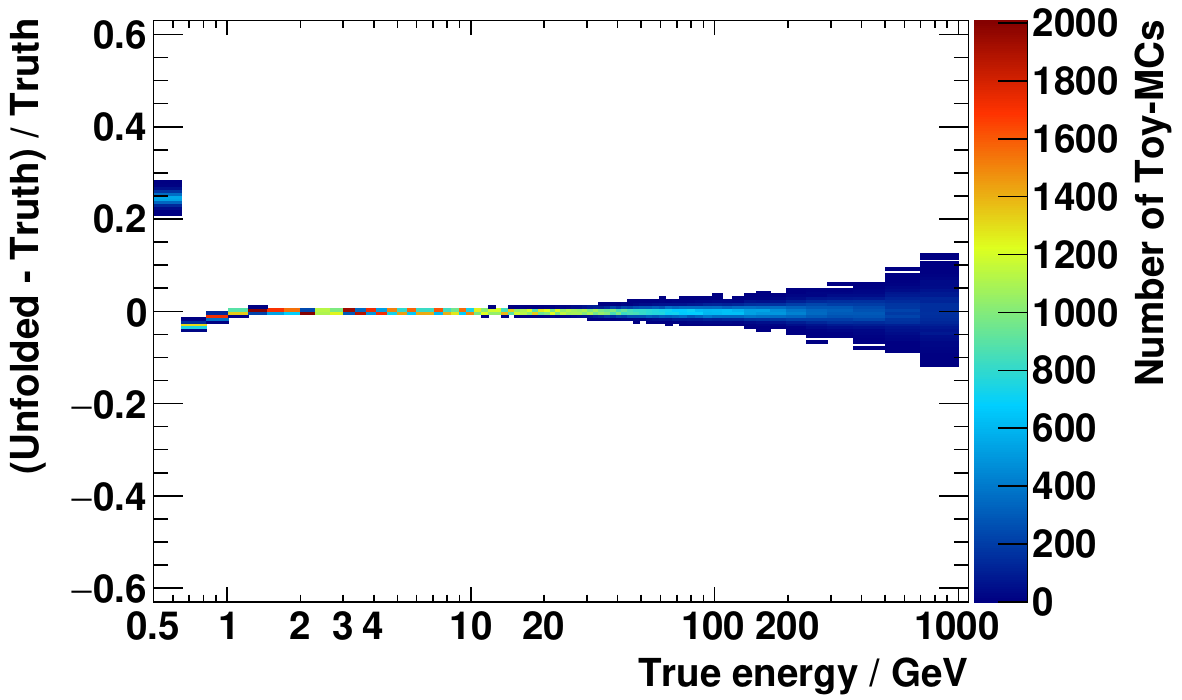}
  \caption{Distribution of the deviation of the unfolded $e^{-}$ flux $\tilde{\Phi}_{\text{unfolded}}$ from the true flux $\Phi_{\text{true}}$ divided by the true flux for $P$ Toy-MCs as function of the true energy.}
  \label{fig:unfolding-toymc-unfolded-truth-ratio-2d-electron-average}
\end{figure}

After repeating the aforementioned procedure $P$ times a distribution of the deviation of the unfolded flux $\tilde{\Phi}_{\text{unfolded}}$
from the true flux $\Phi_{\text{true}}$ divided by the true flux can be obtained (\cref{fig:unfolding-toymc-unfolded-truth-ratio-2d-electron-average}).
The ratio is compatible with unity above \SIapprox{1}{\GeV} indicating that the unfolding procedure is free of a bias. Below that energy threshold
a large deviation is exposed, which gets larger with decreasing energy.

Note that the plot starts at \SI{0.5}{\GeV} even though the unfolding procedure operates on one more bin on the left side - \SIrange{0.25}{0.5}{\GeV} - and one more bin
on the right side of the analysis binning: \SIrange{1}{1.5}{\TeV}. Without the extra bin on the left side the bias in the first analysis bin \SIrange{0.5}{0.65}{\GeV} is even larger
than the \SIapprox{25}{\percent} that is visible in the plot. This shows the limits of the unfolding procedure: not all energy bins that are unfolded are free of a bias.
Especially the border regions are problematic and need to be excluded for further analysis. This is the main reason why the unfolding procedure uses 76 energy bins,
out of which 74 are used to produce the fluxes. The problem is less severe at high energies, simply due to the spectral index of the fluxes. There are much less events
at high energies that could migrate to lower energy bins and thus the problem is less pronounced than at low energies.

\begin{figure}[H]
  \centering
  \includegraphics[width=0.8\linewidth]{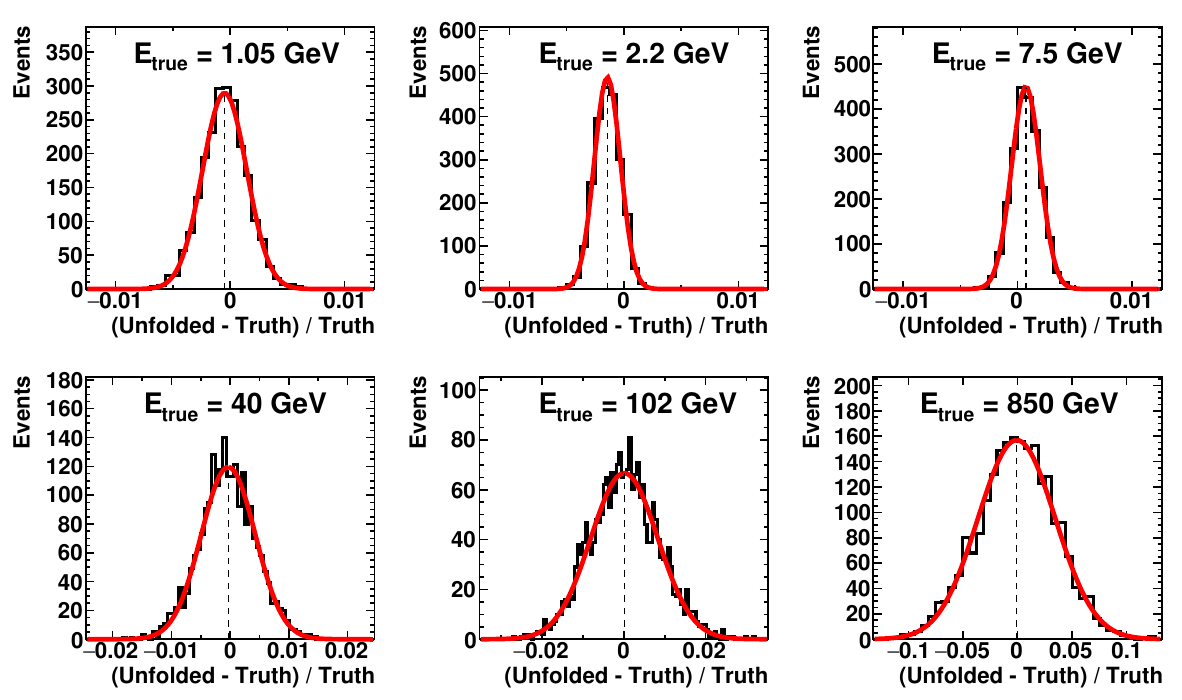}
  \caption{Example projections in several energy bins from \cref{fig:unfolding-toymc-unfolded-truth-ratio-2d-electron-average} to the y-axis for the $e^{-}$ fluxes. The black histogram shows the projection and the red line a gaussian fit to the histogram. The black dashed line indicates the mean of the gaussian, which is a measure of the bias.}
  \label{fig:unfolding-toymc-unfolded-truth-ratio-projections-electron-average}
\end{figure}

To quantify the bias, projections to the y-axis are generated for each true energy bin. \Cref{fig:unfolding-toymc-unfolded-truth-ratio-projections-electron-average}
shows several example projections, from which the bias and the width can be deduced.

\begin{figure}[H]
  \centering
  \includegraphics[width=0.9\linewidth]{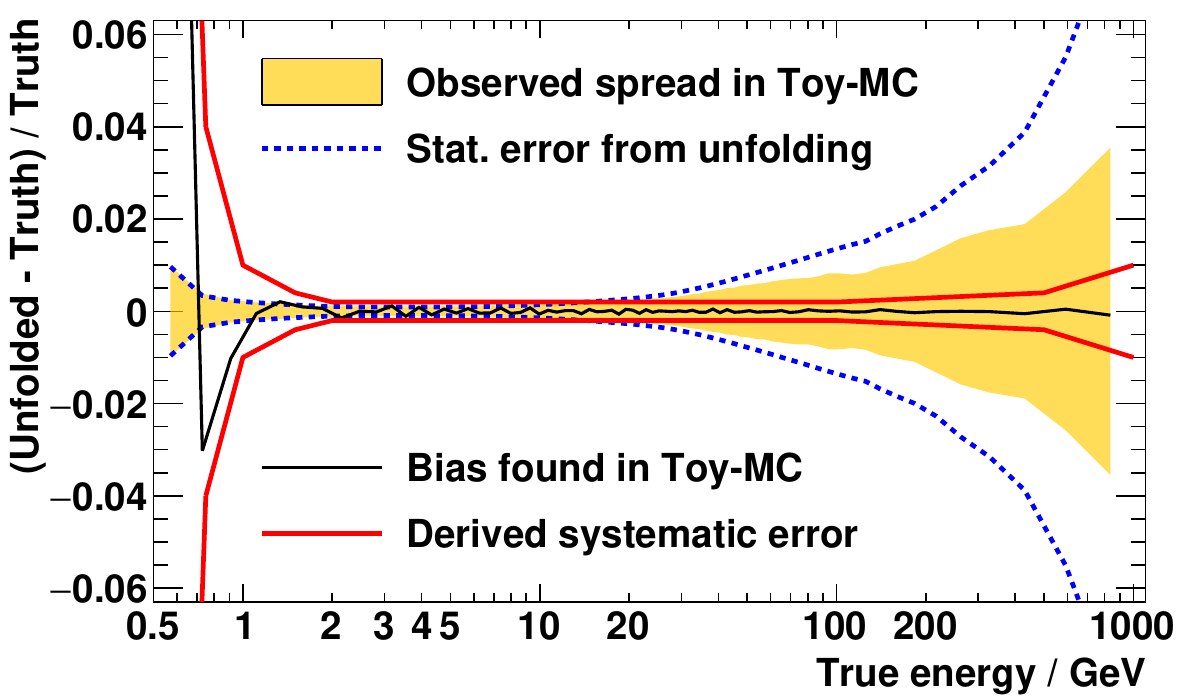}
  \caption{Visualization of the bias of the unfolding procedure for the $e^{-}$ flux, extracted from the gaussian fits of \cref{fig:unfolding-toymc-unfolded-truth-ratio-projections-electron-average}, as function of the true energy. The black line shows the bias of the unfolded flux with respect to the true flux. The red line shows the associated systematic error, which by construction encloses the black line. The blue line shows the statistical uncertainty of the unfolded flux and the orange band the observed spread from the toy Monte-Carlo simulation - corresponding to the width of the gaussians in \cref{fig:unfolding-toymc-unfolded-truth-ratio-projections-electron-average}.}
  \label{fig:unfolding-toymc-bias-syst-error-electron-average}
\end{figure}

After performing gaussian fits to the projections in all true energy bins the bias as function of the true energy can be quantified,
as shown in \cref{fig:unfolding-toymc-bias-syst-error-electron-average} for the $e^{-}$ flux.

\begin{figure}[H]
  \centering
  \includegraphics[width=0.7\linewidth]{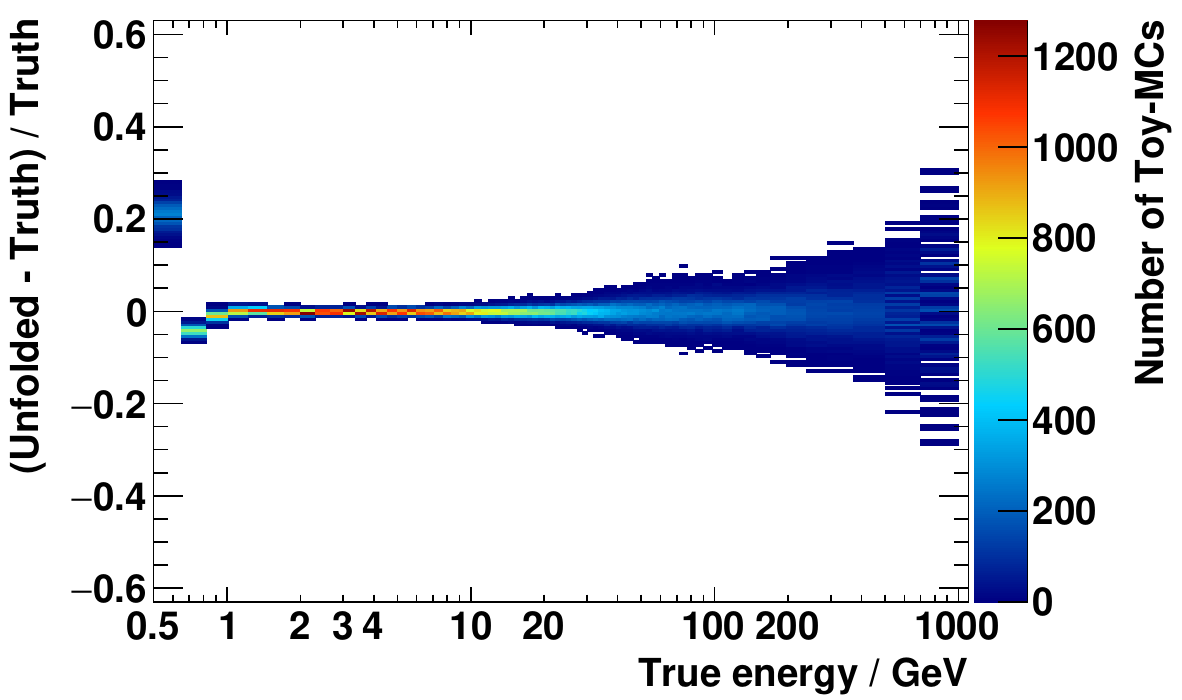}
  \caption{Distribution of the deviation of the unfolded $e^{+}$ flux $\tilde{\Phi}_{\text{unfolded}}$ from the true flux $\Phi_{\text{true}}$ divided by the true flux for $P$ Toy-MCs as function of the true energy.}
  \label{fig:unfolding-toymc-unfolded-truth-ratio-2d-positron-average}
\end{figure}

The same study can be performed for the $e^{+}$ flux: \cref{fig:unfolding-toymc-unfolded-truth-ratio-2d-positron-average}
shows the deviation of the unfolded flux $\tilde{\Phi}_{\text{unfolded}}$ from the true flux $\Phi_{\text{true}}$ divided by the true flux.

The ratio is compatible with unity above \SIapprox{1}{\GeV} indicating that the unfolding procedure is free of a bias. Below that energy threshold
a large deviation is exposed, which gets larger with decreasing energy. This behaviour is identical as for the electrons.

\begin{figure}[H]
  \centering
  \includegraphics[width=0.8\linewidth]{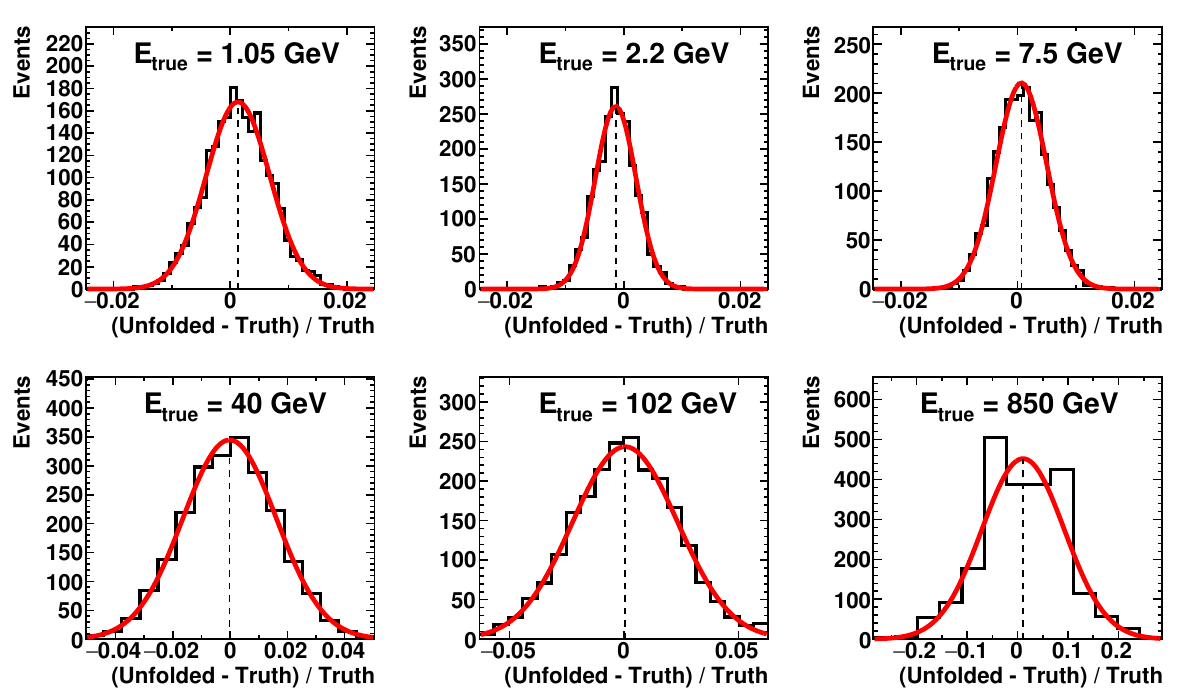}
  \caption{Example projections in several energy bins from \cref{fig:unfolding-toymc-unfolded-truth-ratio-2d-positron-average} to the y-axis for the $e^{+}$ fluxes.}
  \label{fig:unfolding-toymc-unfolded-truth-ratio-projections-positron-average}
\end{figure}

\begin{figure}[H]
  \centering
  \includegraphics[width=0.8\linewidth]{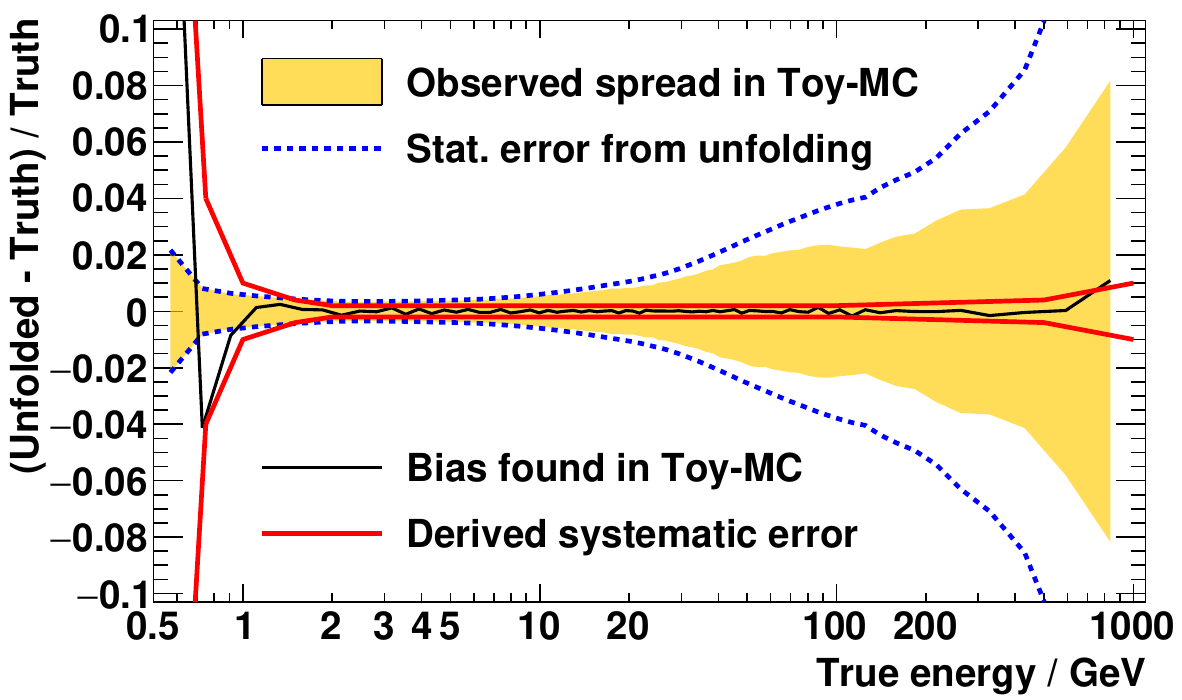}
  \caption{Visualization of the bias of the unfolding procedure for the $e^{+}$ flux, extracted from the gaussian fits of \cref{fig:unfolding-toymc-unfolded-truth-ratio-projections-positron-average}, as function of the true energy.}
  \label{fig:unfolding-toymc-bias-syst-error-positron-average}
\end{figure}

To quantify the bias, projections to the y-axis are performed for each true energy bin. \Cref{fig:unfolding-toymc-unfolded-truth-ratio-projections-positron-average}
shows several example projections, from which the bias can be deduced. After performing gaussian fits to the projections in all true energy bins the bias as function of the true energy can be quantified,
as shown in \cref{fig:unfolding-toymc-bias-syst-error-positron-average} for the $e^{+}$ flux.

Note that the observed spread from the toy Monte-Carlo simulation (orange band) should match the statistical uncertainty from the unfolding procedure (blue line).
Up to \SIapprox{10}{\GeV} both methods yield the same results, at higher energies the spread in the Toy-MC is smaller than the expectation, which is in indication
that the unfolding procedure slightly overestimates the true statistical uncertainty.

\subsubsection{Derivation of the final unfolding uncertainty}
\label{sec:analysis-flux-time-averaged-sysunc-combined-uncertainty}

\begin{figure}[H]
  \centering
  \includegraphics[width=0.9\linewidth]{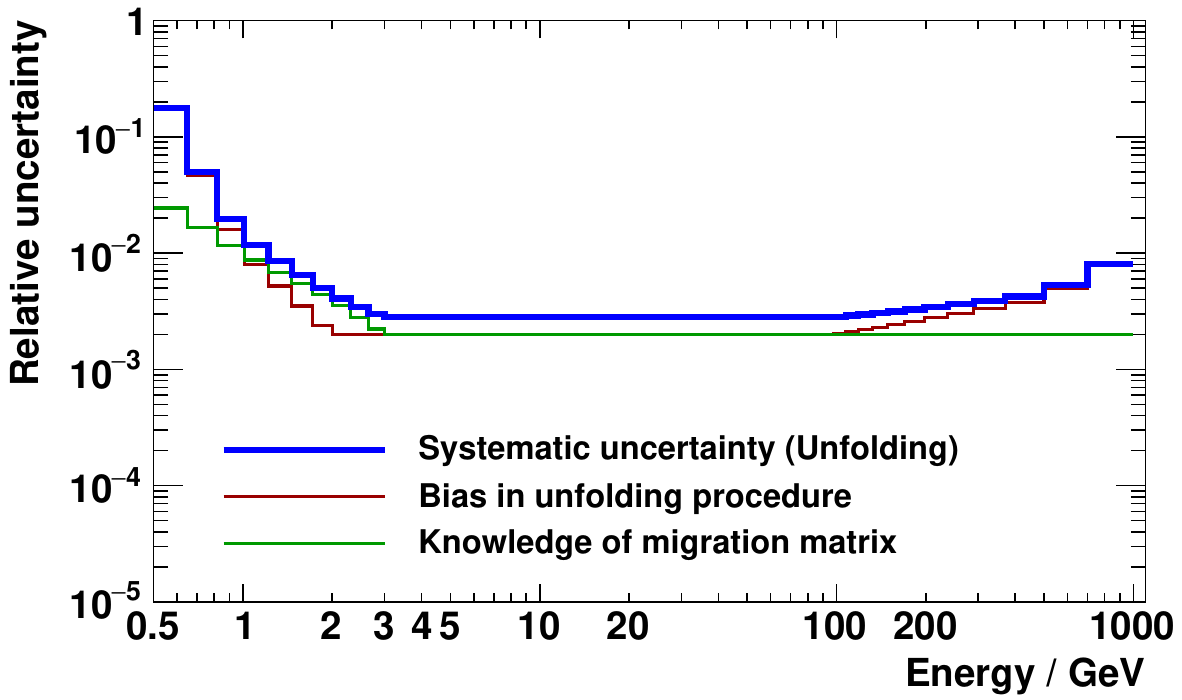}
  \caption{Visualization of the relative systematic uncertainty associated with the electron flux and the positron flux due to the unfolding procedure.}
  \label{fig:time-averaged-fluxes-syst-uncertainty-unfolding}
\end{figure}

Both components that contribute to the unfolding systematic uncertainty were determined: the limited knowledge of the migration matrix and the bias in the unfolding procedure.
\Cref{fig:time-averaged-fluxes-syst-uncertainty-unfolding} shows the resulting relative systematic uncertainty $\sigma_{\text{unf}}(E) / \Phi_{e^{\pm}}(E)$ as function of energy.
The uncertainty associated with the electron flux is identical to the uncertainty associated with the positron flux.

From \SIrange{3}{100}{\GeV} the uncertainty is constant at the permille level and thus negligible. It slightly increases towards high energies, due to the bias
in the unfolding procedure and reaches less than \SI{1}{\percent} at \SI{1}{\TeV}. At low energies the systematic uncertainty increases up to \SIapprox{20}{\percent}
at \SI{0.5}{\GeV}.

\subsection{Summary}
\label{sec:analysis-flux-time-averaged-sysunc-summary}

\begin{figure}[H]
  \includegraphics[width=\linewidth]{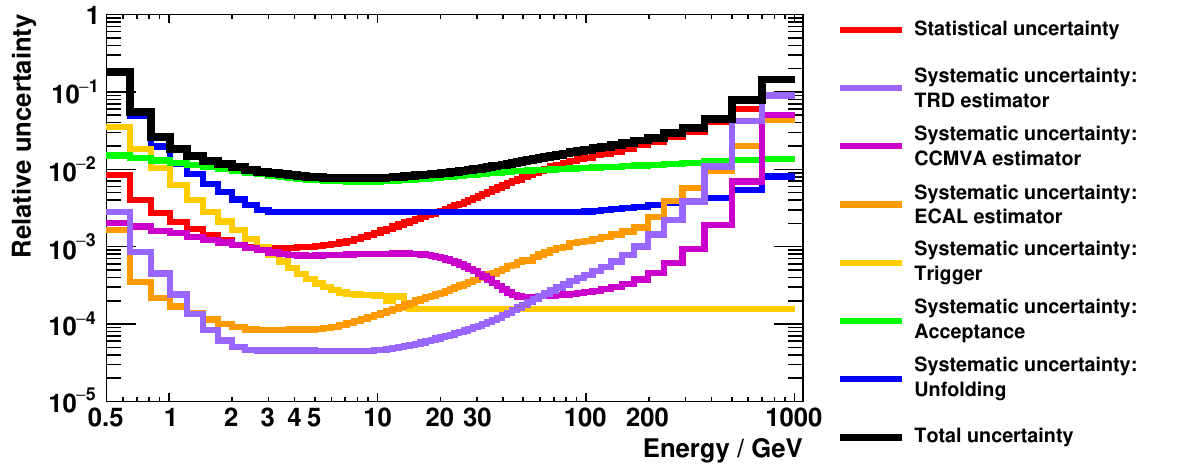}
  \caption{Composition of the relative systematic uncertainty associated with the electron flux.}
  \label{fig:time-averaged-fluxes-syst-uncertainty-composition-electrons}
\end{figure}

\Cref{fig:time-averaged-fluxes-syst-uncertainty-composition-electrons} shows the composition of the relative systematic uncertainty $\sigma(E) / \Phi_{e^{-}}(E)$ as function of energy
for the $e^{-}$ flux.

Above \SIapprox{70}{\GeV} the statistical uncertainty dominates the uncertainty of the $e^{-}$ flux measurement. Below this energy, the systematic uncertainty always
exceeds the statistical uncertainty. The acceptance uncertainty $\sigma_{\text{acc}}(E)$, described in \cref{sec:analysis-flux-time-averaged-sysunc-acceptance},
is the largest contribution to the overall systematic uncertainty between \SIrange{1}{600}{\GeV}. Between \SIrange{0.5}{1}{\GeV} the unfolding uncertainty $\sigma_{\text{unf}}(E)$,
described in \cref{sec:analysis-flux-time-averaged-sysunc-unfolding}, dominates the systematic uncertainty.

\begin{figure}[H]
  \includegraphics[width=\linewidth]{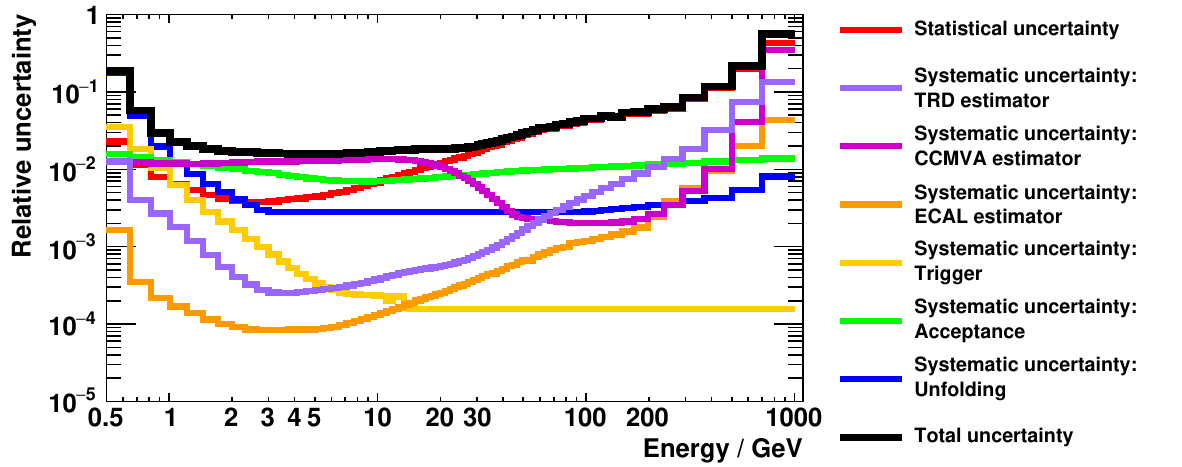}
  \caption{Composition of the relative systematic uncertainty associated with the positron flux.}
  \label{fig:time-averaged-fluxes-syst-uncertainty-composition-positrons}
\end{figure}

\Cref{fig:time-averaged-fluxes-syst-uncertainty-composition-positrons} shows the composition of the relative systematic uncertainty $\sigma(E) / \Phi_{e^{+}}(E)$ as function of energy
for the $e^{+}$ flux. The statistical uncertainty dominates the uncertainty of the $e^{+}$ flux measurement, above \SI{20}{\GeV}. This observation is identical as for the
$e^{-}$ flux, with the difference that statistical limitations are already exposed at lower energies, as there a significantly less positrons in cosmic rays than electrons.
Between \SIrange{0.5}{1}{\GeV} the unfolding uncertainty $\sigma_{\text{unf}}(E)$, described in \cref{sec:analysis-flux-time-averaged-sysunc-unfolding}, dominates the systematic
uncertainty - identical as for the electrons. The charge-confusion uncertainty $\sigma_{\text{cc}}(E)$, described in \cref{sec:analysis-flux-time-averaged-sysunc-charge-confusion-estimator},
and the acceptance uncertainty $\sigma_{\text{acc}}(E)$, described in \cref{sec:analysis-flux-time-averaged-sysunc-acceptance}, are the largest contributions to the overall systematic
uncertainty at energies between \SIrange{1}{200}{\GeV}. Above \SIapprox{200}{\GeV} the uncertainty $\sigma_{\text{trd}}(E)$, associated with the knowledge of the shape of the TRD estimator
(\cref{sec:analysis-flux-time-averaged-sysunc-trd-estimator}), dominates the systematic uncertainty.

\clearpage
\section{Time-averaged positron/electron ratio and positron fraction calculation}
\label{sec:analysis-ratios-time-averaged}

The time-averaged positron/electron ratio in the energy bin $E$ of width $\Delta E$ is given by

\begin{equation}
  \label{eq:positron-electron-ratio}
  R_{e}(E) = \frac{\Phi_{e^{+}}(E)}{\Phi_{e^{-}}(E)},
\end{equation}

\noindent
and the positron fraction by

\begin{equation}
  \label{eq:positron-fraction}
  p(E) = \frac{\Phi_{e^{+}}(E)}{\Phi_{e^{+}}(E) + \Phi_{e^{-}}(E)}.
\end{equation}

When inserting the $e^{\pm}$ fluxes $\Phi_{e^{\pm},\,i}(E)$ from \cref{eq:isoflux} into the equations for the positron/electron ratio \cref{eq:positron-electron-ratio}
and the positron fraction \cref{eq:positron-fraction} it is obvious that the bin width $\Delta E$ and the measuring time $T(E)$ as well as the efficiencies $\epsilon(E)$
cancel. The acceptance does not cancel strictly, since it slightly differs for positrons/electrons at low energies:

\begin{equation}
  \label{eq:positron-electron-ratios-expanded}
  \begin{aligned}
    R_{e}(E) &= \frac{N_{e^{+}}(E)}{N_{e^{-}}(E)} \cdot \mathbf{\frac{A_{e^{-}}(E)}{A_{e^{+}}(E)}}, \\
    p(E)     &= \frac{N_{e^{+}}(E)}{N_{e^{+}}(E) + N_{e^{-}}(E) \cdot \mathbf{\frac{A_{e^{+}}}{A_{e^{-}}}}}.
  \end{aligned}
\end{equation}

The positron over electron acceptance ratio

\begin{equation}
  \label{eq:positron-electron-acceptance-ratio-expanded}
  \frac{A_{e^{+}}(E)}{A_{e^{-}}(E)} = \frac{A^{\text{MC}}_{e^{+}}(E)}{A^{\text{MC}}_{e^{-}}(E)} \cdot \frac{1 + \delta_{e^{+}}(E)}{1 + \delta_{e^{-}}(E)} = \frac{A^{\text{MC}}_{e^{+}}(E)}{A^{\text{MC}}_{e^{-}}(E)}
\end{equation}

was determined from the electron and positron Monte-Carlo simulation, as described in \cref{sec:analysis-flux-time-averaged-acceptance-asymmetry}, and
is used to correct the positron/electron ratio and the positron fraction results. Since in the positron/electron ratio and the positron fraction the uncertainty
on the absolute acceptance is not present, the acceptance asymmetry uncertainty $\sigma_{\text{asymm}}(E)$ must be taken into account.

The number of electrons and positrons $N_{e^{\pm}}(E)$ can be extracted from the two-dimensional template fit procedure, with the charge-confusion value fixed to the
Monte-Carlo prediction, as described in \cref{sec:analysis-lepton-counts-2d-fit}. However there are important differences in the data samples that are analyzed between the
flux and the ratio analysis. A summary is given in \cref{tab:analysis-strategies} and will be explained in the following.

\bigskip
\noindent
\begin{minipage}{\linewidth}
  \centering
  \begin{tabular}{lcc}
    \toprule
                               Component & Single-track analysis                                        & All-tracks analysis                                          \\
    \midrule
    \rowcolor{black!20} Statistics       & nominal                                                      & increased by \SIapprox{20}{\percent}                         \\
                        Acceptance       & $(1 + \delta(E))$ \SIapprox{5}{\percent} larger              & $(1 + \delta(E))$ nominal                                    \\
    \rowcolor{black!20} Charge-confusion & $f_{\text{CC}} = \SI{0.38}{\percent}$ at \SIapprox{20}{\GeV} & $f_{\text{CC}} = \SI{1.49}{\percent}$ at \SIapprox{20}{\GeV} \\
                        Suitable for     & $R_{e}(E)$, $p(E)$ analysis                                  & $\Phi_{e^{+}}(E)$, $\Phi_{e^{-}}(E)$ analysis                \\
  \end{tabular}
  \captionof{table}{Comparison of the analysis strategies for the electron and positron flux analysis with respect to the positron/electron ratio and positron fraction analysis. The single-track analysis is chosen for the positron/electron ratio and positron fraction analysis and the all-tracks analysis for the electron and positron flux analysis.}
  \label{tab:analysis-strategies}
\end{minipage}
\bigskip

For the flux analysis the \enquote{all-tracks sample}, a union of the \enquote{single-track sample} and the \enquote{multi-tracks sample}, is used to determine the number of
electrons and the number of positrons. Using the all-tracks sample the smallest possible acceptance systematic uncertainty can be achieved, since imposing a single tracker track cut
leads to a large deviation between the ISS selection efficiency and the Monte-Carlo prediction. This enlarges the Data/Monte-Carlo correction factor $(1 + \delta(E))$ by
\SIapprox{5}{\percent} and increases the systematic uncertainty by almost \SI{3}{\percent}. As side-effect the statistics can be increased by almost \SIapprox{20}{\percent}.
The drawback of the all-tracks sample is the enlarged charge-confusion and its associated uncertainty. However the overall uncertainty of the flux analysis
based on the all-tracks sample is smaller compared to using the single-track sample. As consequence the all-tracks sample is used for the $\Phi_{e^{\pm}}(E)$ flux analysis.

For the positron/electron ratio $R_{e}(E)$ and the positron fraction $p(E)$ this argument does not hold, since the acceptance uncertainty cancels in the ratios.
The charge-confusion in the single-track sample is considerable smaller and the magnitude of the unfolding is smaller, since the single-track sample is less affected by
migration effects, due to a better energy reconstruction in these events. For these reasons the ratio analysis will be performed using the event counts $N_{e^{\pm}}(E)$
from the single-track sample.

\section{Time-averaged systematic uncertainties for positron/electron ratio and positron fraction}
\label{sec:analysis-ratios-time-averaged-sysunc}

In this section all systematic uncertainties that contribute to the time-averaged positron/electron ratio and positron fraction are summarized.

The most important systematic uncertainty for the ratio analysis is the unfolding uncertainty, described in \cref{sec:analysis-ratios-time-averaged-sysunc-unfolding},
dominating the overall systematic uncertainty up to \SI{50}{\GeV}. Between \SIrange{2}{4}{\GeV} the acceptance asymmetry uncertainty is the largest contribution, described in
\cref{sec:analysis-ratios-time-averaged-sysunc-acceptance-asymmetry}. Above \SI{50}{\GeV} the knowledge of the TRD estimator and the associated uncertainty is the
dominant systematic uncertainty, presented in \cref{sec:analysis-ratios-time-averaged-sysunc-trd-estimator}.

\subsection{Unfolding}
\label{sec:analysis-ratios-time-averaged-sysunc-unfolding}

The number of electrons or positrons as function of energy, are unfolded as described in \cref{sec:analysis-flux-time-averaged-unfolding} and
the uncertainty was redetermined using the single-track sample and is slightly smaller as for the flux analysis presented in \cref{sec:analysis-flux-time-averaged-sysunc-unfolding}.

\begin{figure}[H]
  \centering
  \includegraphics[width=0.7\linewidth]{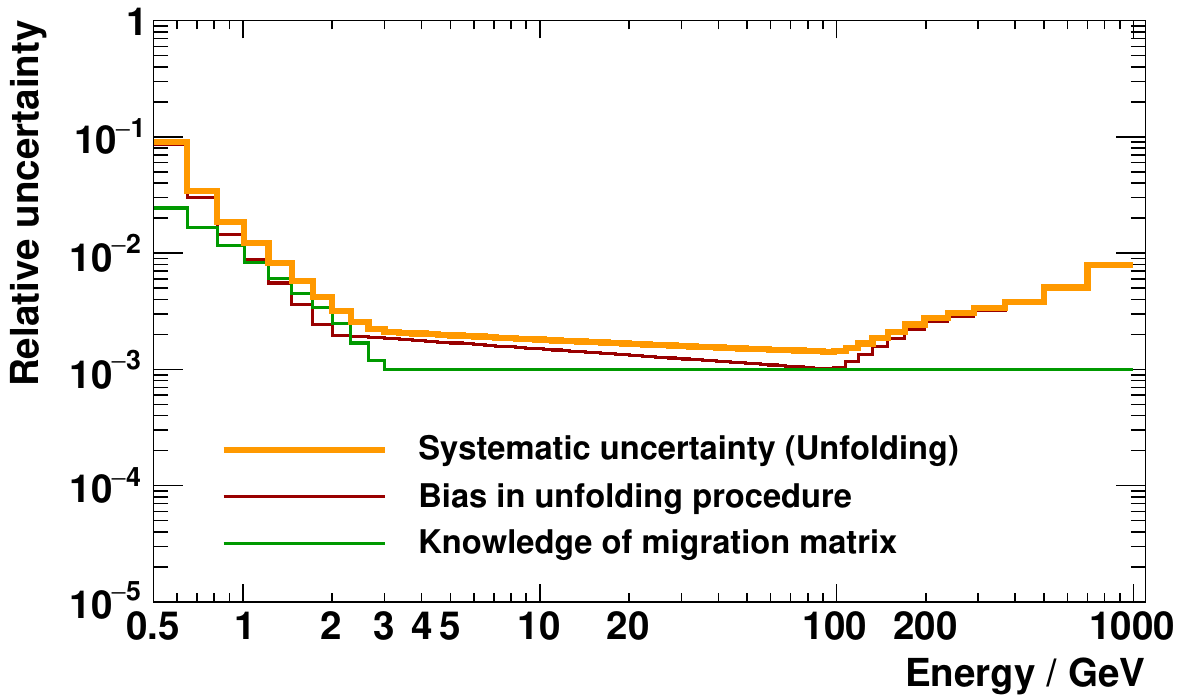}
  \caption{Visualization of the relative systematic uncertainty associated with the positron/electron ratio and the positron fraction due to unfolding.}
  \label{fig:time-averaged-ratios-syst-uncertainty-unfolding}
\end{figure}

\Cref{fig:time-averaged-ratios-syst-uncertainty-unfolding} shows the resulting relative systematic uncertainty $\sigma_{\text{unf}}(E) / R_{e}(E)$ as function of energy
for the positron/electron ratio. The uncertainty is identical for the positron fraction.

\subsection{Acceptance asymmetry}
\label{sec:analysis-ratios-time-averaged-sysunc-acceptance-asymmetry}

The uncertainty was derived for the $e^{\pm}$ fluxes in \cref{sec:analysis-flux-time-averaged-acceptance-asymmetry}.
For the flux analysis the small uncertainty due to the acceptance asymmetry was neglected, since the acceptance uncertainty $\sigma_{\text{acc}}(E)$ is considerable larger.
For the positron/electron ratio and the positron fraction the acceptance uncertainty cancels by definition, such that the acceptance asymmetry uncertainty $\sigma_{\text{asymm}}(E)$
becomes important.

\begin{figure}[H]
  \centering
  \includegraphics[width=0.7\linewidth]{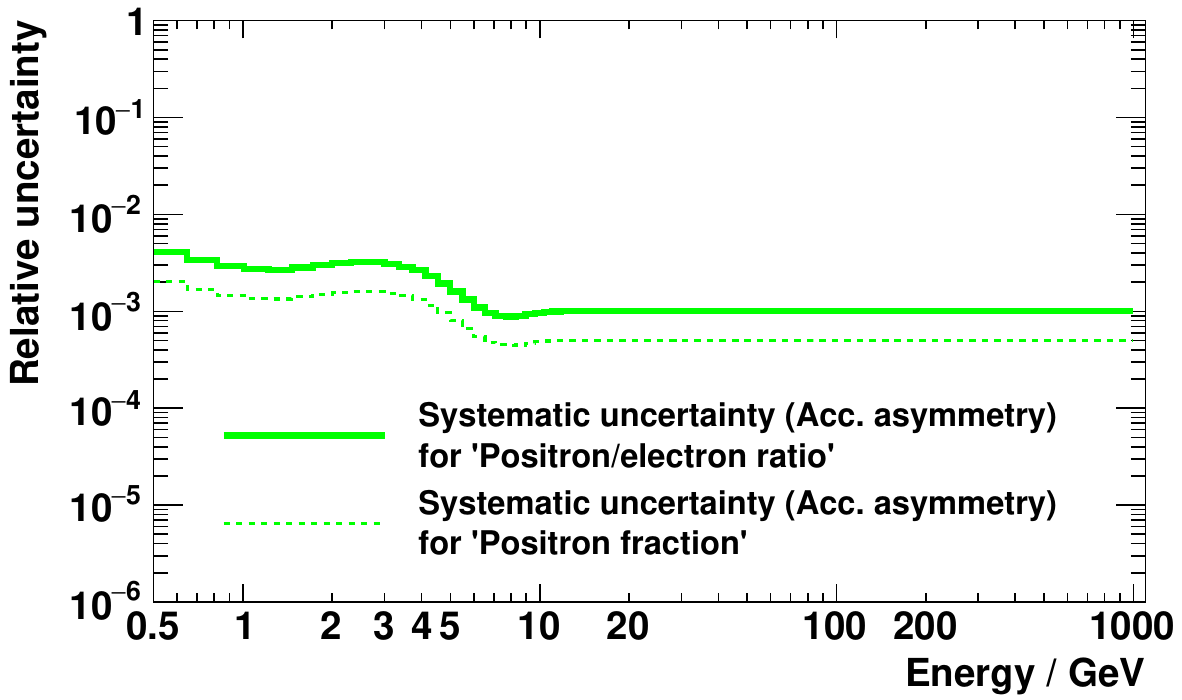}
  \caption{Visualization of the relative systematic uncertainty associated with the positron/electron ratio and the positron fraction due to the acceptance asymmetry between $e^{-}$ / $e^{+}$.}
  \label{fig:time-averaged-ratios-syst-uncertainty-acceptance-asymm}
\end{figure}

\Cref{fig:time-averaged-ratios-syst-uncertainty-acceptance-asymm} shows the resulting relative systematic uncertainty $\sigma_{\text{asymm}}(E) / R_{e}(E)$ as function of energy
for the positron/electron ratio and the relative systematic uncertainty $\sigma_{\text{asymm}}(E) / p(E)$ as function of energy for the positron fraction.

\subsection{CCMVA estimator}
\label{sec:analysis-ratios-time-averaged-sysunc-charge-confusion-estimator}

The charge-confusion uncertainty was derived for the $e^{\pm}$ fluxes in \cref{sec:analysis-flux-time-averaged-sysunc-charge-confusion-estimator}.
To study the influence of the amount of charge-confusion on the positron/electron ratio $R_{e}(E)$ and the positron fraction $p(E)$ the same procedure
as for the flux analysis (\cref{sec:analysis-flux-time-averaged-sysunc-charge-confusion-estimator}) will be repeated.

\begin{figure}[H]
  \includegraphics[width=\linewidth]{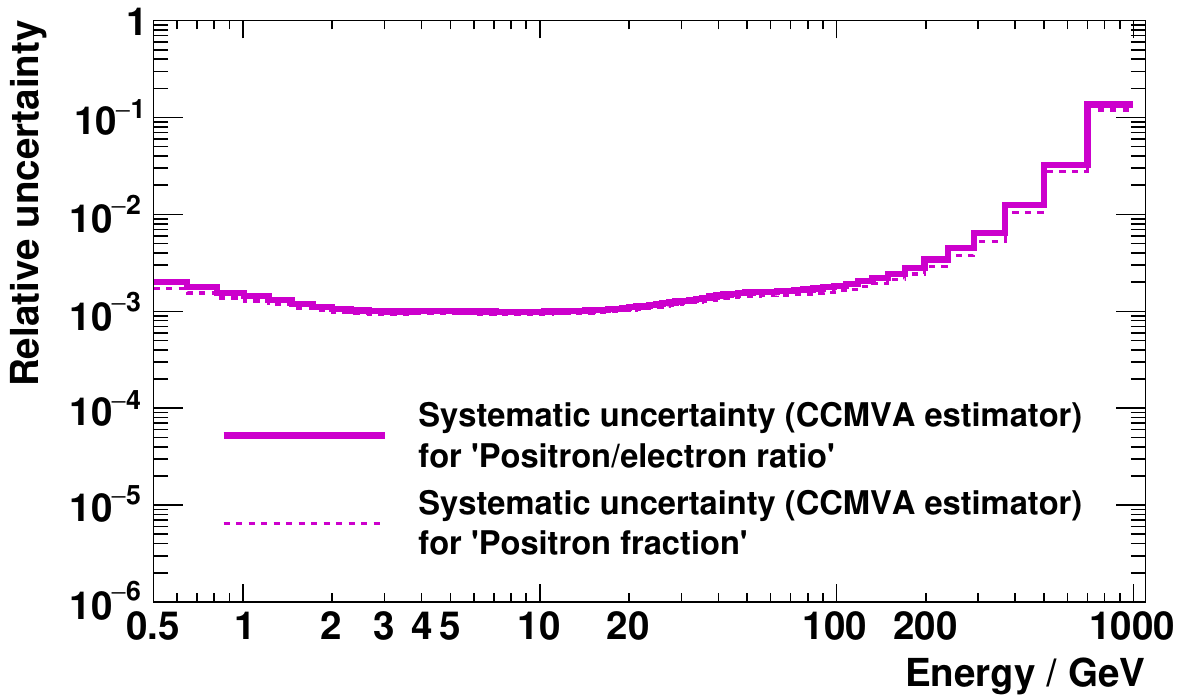}
  \caption{Visualization of the relative systematic uncertainty associated with the positron/electron ratio and the positron fraction due to the knowledge of the charge-confusion.}
  \label{fig:time-averaged-ratios-syst-uncertainty-charge-confusion}
\end{figure}

\Cref{fig:time-averaged-ratios-syst-uncertainty-charge-confusion} shows the resulting relative systematic uncertainty $\sigma_{\text{cc}}(E) / R_{e}(E)$ as function of energy
for the positron/electron ratio and the relative systematic uncertainty $\sigma_{\text{cc}}(E) / p(E)$ as function of energy for the positron fraction.

\subsection{TRD estimator}
\label{sec:analysis-ratios-time-averaged-sysunc-trd-estimator}

The TRD estimator uncertainty was derived for the $e^{\pm}$ fluxes in \cref{sec:analysis-flux-time-averaged-sysunc-trd-estimator}.
The same procedure is repeated for the positron/electron ratio and the positron fraction, with the only difference that the multi-tracks sample does not need to be considered: only the TRD
templates for the single-track sample are varied.

\begin{figure}[H]
  \centering
  \includegraphics[width=0.9\linewidth]{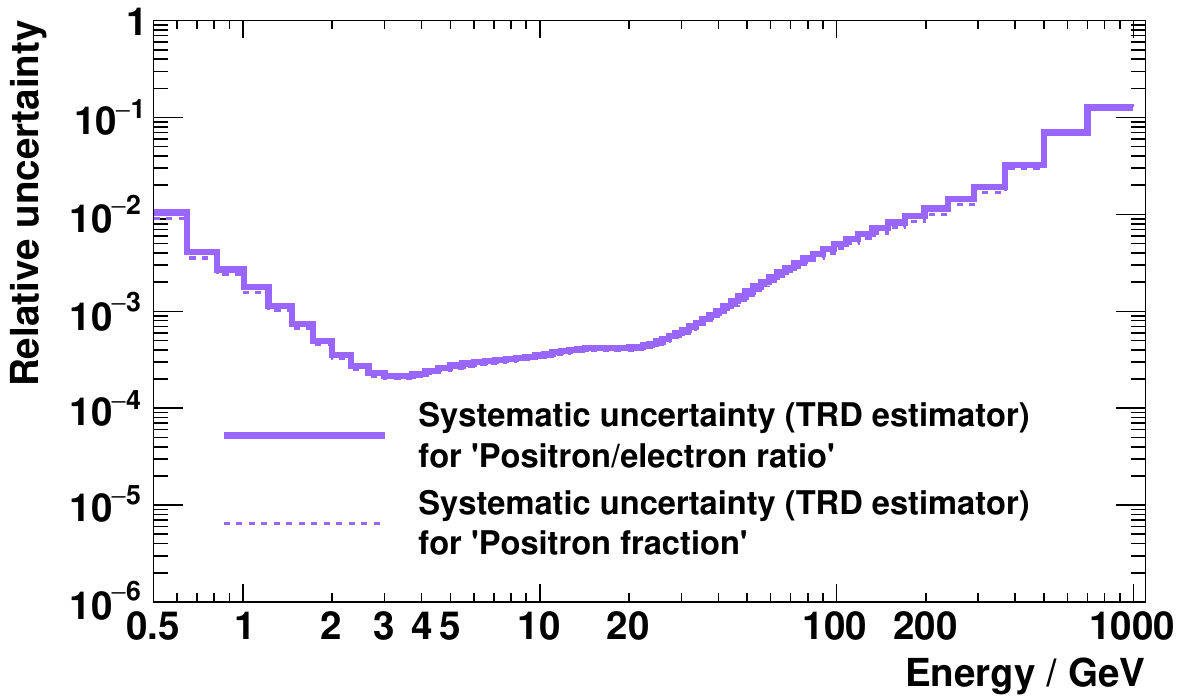}
  \caption{Visualization of the relative systematic uncertainty associated with the positron/electron ratio and the positron fraction due to the knowledge of the TRD estimator.}
  \label{fig:time-averaged-ratios-syst-uncertainty-trd-estimator}
\end{figure}

\Cref{fig:time-averaged-ratios-syst-uncertainty-trd-estimator} shows the resulting relative systematic uncertainty $\sigma_{\text{trd}}(E) / R_{e}(E)$ as function of energy
for the positron/electron ratio and the relative systematic uncertainty $\sigma_{\text{trd}}(E) / p(E)$ as function of energy for the positron fraction.

\subsection{Summary}
\label{sec:analysis-ratios-time-averaged-sysunc-summary}

\begin{figure}[H]
  \includegraphics[width=\linewidth]{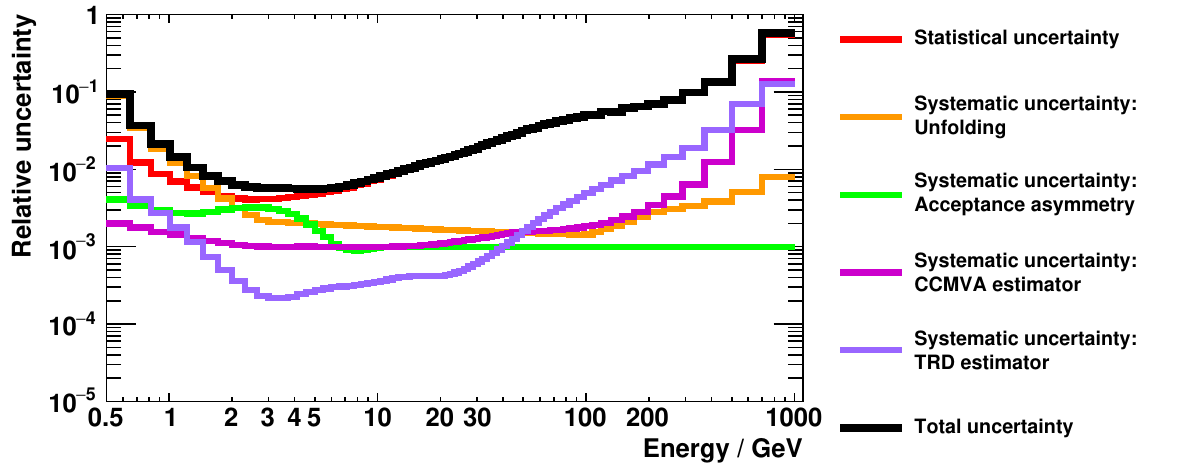}
  \caption{Composition of the relative systematic uncertainty associated with the positron/electron ratio.}
  \label{fig:time-averaged-ratios-syst-uncertainty-composition}
\end{figure}

\Cref{fig:time-averaged-ratios-syst-uncertainty-composition} shows the composition of the relative systematic uncertainty $\sigma(E) / R_{e}(E)$ as function of energy
for the positron/electron ratio. The composition of the relative systematic uncertainty $\sigma(E) / p(E)$ as function of energy for the positron fraction is almost identical
and thus was omitted.

For both the positron/electron ratio and the positron fraction the unfolding uncertainty dominates at low energies, below \SIapprox{2}{\GeV} where the migration
effect is strong. Above \SIapprox{2}{\GeV} the statistical uncertainty dominates the overall uncertainty.

\clearpage
\section{Time-dependent flux calculation}
\label{sec:analysis-flux-time-dependent}

The equation for the flux in time bin $i$ is given by

\begin{equation}
  \label{eq:isoflux-time-dependent}
  \Phi_{e^{\pm},\,i}(E) = \frac{N_{e^{\pm},\,i}(E)}{\Delta E \cdot T_{i}(E) \cdot A_{e^{\pm},\,i}(E) \cdot \epsilon_{i}(E)}.
\end{equation}

The efficiency $\epsilon_{i}(E)$ is defined in \cref{eq:efficiency-time-dependent} as the product of the trigger
efficiency and the ECAL estimator efficiency. Both efficiencies will be derived directly from
ISS data, without involving any Monte-Carlo simulations - just like for the time-averaged flux analysis.

\begin{equation}
  \label{eq:efficiency-time-dependent}
  \epsilon_{i}(E) = \epsilon_{\text{trigger},\,i}(E) \cdot \epsilon_{\text{ecal},\,i}(E)
\end{equation}

For the time-dependent analysis the whole time interval - \textbf{May~\nth{20},~2011} until \textbf{November~\nth{12},~2017} - is divided into
88 periods that last exactly 27 days - corresponding to one \enquote{Bartels rotation}~\cite{Bartels1934}. The Bartels rotation number is a
monotonically increasing number that marks the apparent rotations of the Sun as viewed from Earth.
The first Bartels rotation starts on \textbf{February,~\nth{8}~1832} - an arbitrary assignment proposed by Julius Bartels.

In total 49 energy intervals are analyzed spanning the range \SIrange{1.01}{49.33}{\GeV}. The same energy intervals are used
as for the time-averaged analysis, just in a reduced range, to account for the reduced statistics with respect to the full
time range.

To derive time-dependent fluxes the question arises, which components of the flux \cref{eq:isoflux-time-dependent} are
time-dependent and which ones can be taken from the time-averaged analysis. Clearly the measuring time $T_{i}(E)$ and the number of
electrons or positrons $N_{e^{\pm},\,i}(E)$ have to be computed for each Bartels rotation $i$ separately.

\subsection{ECAL estimator efficiency}
\label{sec:analysis-flux-time-dependent-ecal-estimator}

The ECAL estimator efficiency $\epsilon_{\text{ecal},\,i}(E)$ is determined individually in each Bartels rotation $i$, following the recipe given in the time-averaged flux analysis,
described in \cref{sec:analysis-flux-time-averaged-ecal-estimator}.

The results of the bootstrap method are compared between the time-averaged flux analysis and the time-dependent flux analysis for all energy bins
and Bartels rotations. For all energy bins the average of the time-dependent ECAL estimator efficiency is compatible with the time-averaged
ECAL estimator efficiency and thus $\epsilon_{\text{ecal},\,i}(E) = \epsilon_{\text{ecal}}(E)$.

\subsection{Trigger efficiency}
\label{sec:analysis-flux-time-dependent-trigger}

The trigger efficiency $\epsilon_{\text{trigger},\,i}(E)$ is determined individually in each Bartels rotation $i$, following the recipe given in the time-averaged flux analysis,
described in \cref{sec:analysis-flux-time-averaged-trigger}.

\begin{figure}[H]
  \includegraphics[width=\linewidth]{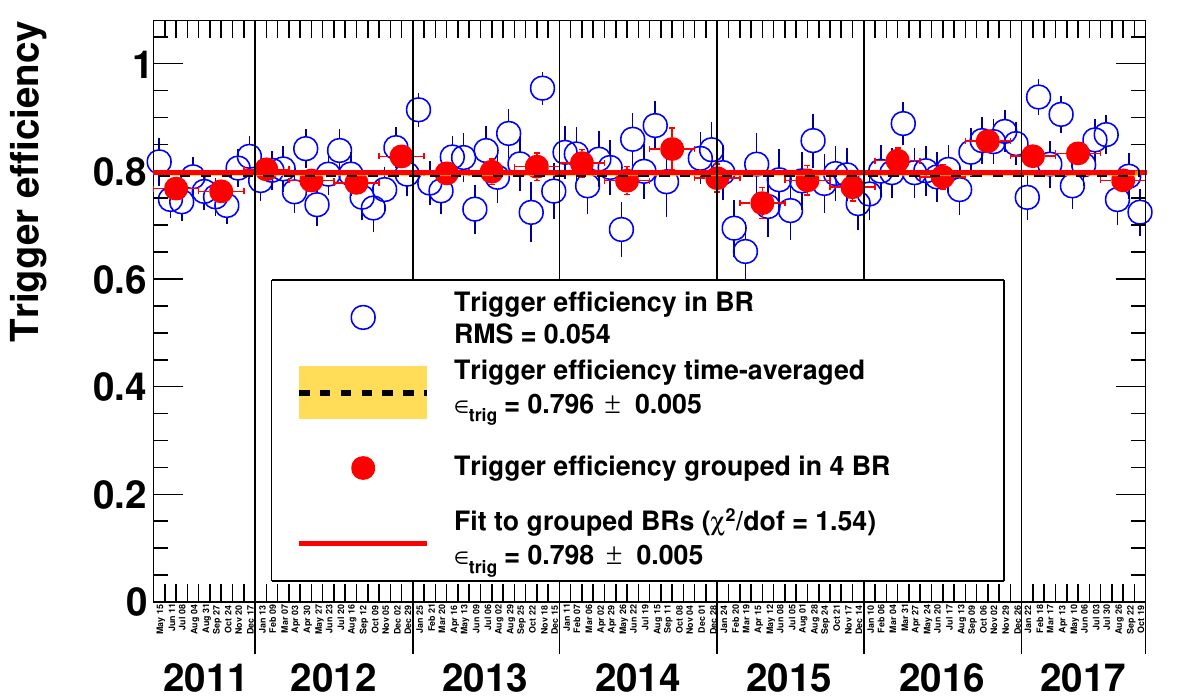}
  \caption{Time-dependent trigger efficiency for the energy bin \SIrange{1.01}{1.22}{\GeV}. The x-axis spans 88 Bartels rotations and each bin label denotes the date when the Bartels rotation started (in UTC). The blue symbols show the trigger efficiency in the specific energy bin as function of time. The red symbols group four Bartels rotation into one point in time to reduce the statistical fluctuations. The black dashed line shows the time-averaged efficiency for comparison. The red line shows a fit of a constant to the red symbols, yielding an acceptable $\chi^2/\text{dof} = 1.37$ value. Within the uncertainties the trigger efficiency is independent of time.}
  \label{fig:time-dependent-trigger}
\end{figure}

\Cref{fig:time-dependent-trigger} show the trigger efficiency as function of time for the first energy bin of the time-dependent analysis.
For all energy bins the average of the time-dependent trigger efficiency is compatible with the time-averaged
trigger efficiency and thus $\epsilon_{\text{trigger},\,i}(E) = \epsilon_{\text{trigger}}(E)$.

Since both the trigger efficiency and the ECAL estimator efficiency are independent of time the whole efficiency is identical to the time-averaged case (\cref{eq:efficiency} in \cref{sec:analysis-flux-time-averaged}):

\begin{equation*}
  \epsilon(E) = \epsilon_{\text{trigger}}(E) \cdot \epsilon_{\text{ecal}}(E).
\end{equation*}

\subsection{Acceptance}
\label{sec:analysis-flux-time-dependent-acceptance}

The time-dependent acceptance $A_{e^{\pm},\,i}(E)$ entering the flux \cref{eq:isoflux-time-dependent} is given by

\begin{equation}
  \label{eq:acceptance-time-dependent}
  A_{e^{\pm},\,i}(E) = A_{e^{\pm}}^{\text{MC}}(E) \cdot (1 + \delta_{i}(E)),
\end{equation}

where $\delta_{i}(E)$ denotes a time-dependent Data/Monte-Carlo correction factor for Bartels rotation $i$.

For each cut in the preselection category (defined in \cref{sec:analysis-data-selection-preselection-cuts}), the selection category (defined in
\cref{sec:analysis-data-selection-selection-cuts}) and the $e^{\pm}$ identification category (defined in
\cref{sec:analysis-data-selection-electron-positron-identification-cuts}) and each Bartels rotation $i$ the tag \& probe method - described in
\cref{sec:analysis-flux-time-averaged-acceptance} for the time-averaged flux analysis - is performed.

A negative rigidity sample (\textbf{\enquote{tag sample}}) is selected for every cut $c$ using information from detectors unrelated to
that cut. The efficiency of each cut $c$ is compared between ISS data in Bartels rotation $i$ and the electron Monte-Carlo simulation. The ratio
ISS efficiency divided by Monte-Carlo efficiency - as function of energy - is computed. If the ratio differs from unity, it is parameterized using a
simple model, e.g. a constant or a straight line. The best matching model is fit to the ratio. Afterwards it will be tested if the deviation from unity
is significant, according to the uncertainty of the fit parameters.

It turns out that the energy dependence of each cut $c$ in each Bartels rotation $i$ is identical to the time-averaged energy dependence. Therefore
it is not necessary to compare the ISS data in each Bartels rotation $i$ with the electron Monte-Carlo simulation and parameterize the ISS efficiency
over Monte-Carlo efficiency ratio in order to extract the correction factor $(1 + \delta_{i}^{c}(E))$. Instead the cut efficiency of each cut $c$ in each
Bartels rotation $i$ can be directly compared to the ISS time-averaged cut efficiency, by computing the ratio: efficiency in Bartels rotation $i$ divided
by time-averaged ISS cut efficiency.

\begin{figure}[H]
  \includegraphics[width=\linewidth]{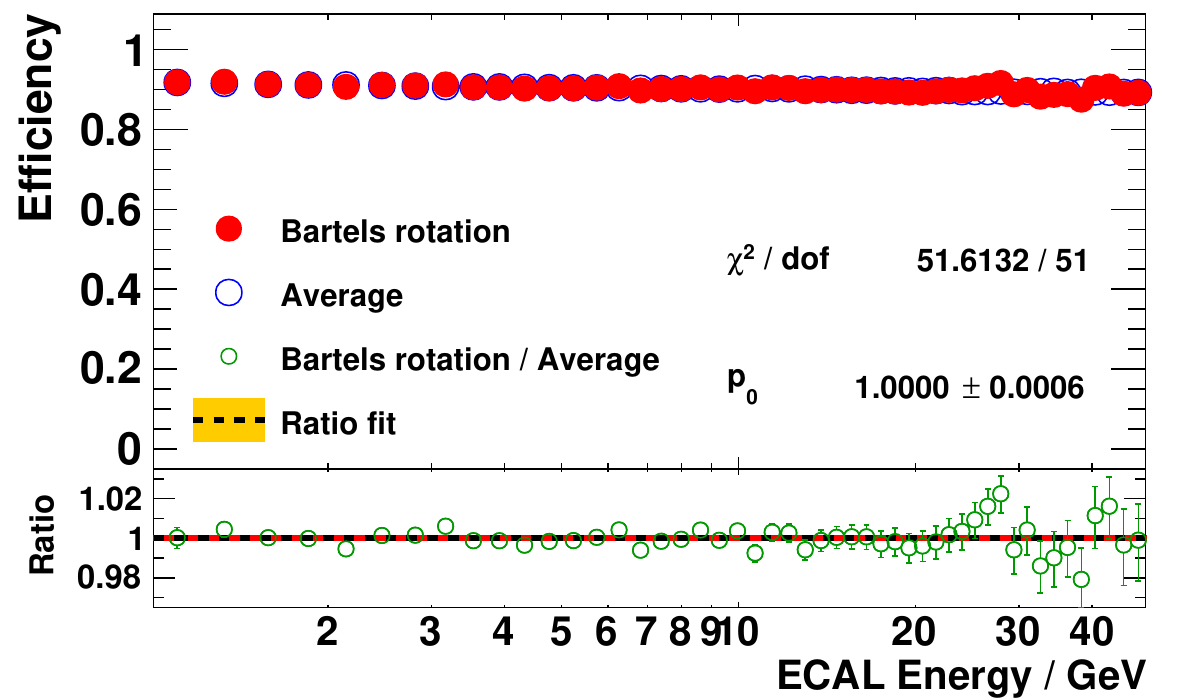}
  \caption{Determination of the ratio: time-averaged cut efficiency $(1 + \delta^{c}(E))$ over time-dependent cut efficiency $(1 + \delta_{i}^{c}(E))$ in Bartels rotation $i = 26$ (Apr~\nth{16},~2013 - May~\nth{13},~2013) for the selection cut: \textbf{\enquote{Tracker hit pattern}} (\cref{sec:analysis-data-selection-selection-cuts} - \cref{enum:selection-cut-trk-pattern}). The blue symbols in the upper plot show the cut efficiency determined from ISS data for the time-averaged analysis on a dedicated sample: the tag sample. The red symbols in the upper plot show the cut efficiency determined from ISS data in Bartels rotation $i = 26$ using the same tag cuts. The lower plot shows the ratio time-averaged efficiency over efficiency in Bartels rotation $i = 26$ as green symbols. A constant is fit to the ratio, shown as dashed black line including the orange uncertainty band.}
  \label{fig:time-dependent-acceptance-tracker-pattern-br-to-average-example}
\end{figure}

\Cref{fig:time-dependent-acceptance-tracker-pattern-br-to-average-example} shows the comparison between the time-averaged cut efficiency
and the time-dependent cut efficiency in a single Bartels rotation, as example, for the \enquote{Tracker hit pattern} cut. The structure in the ratio plot
around \SIapprox{25}{\GeV} is a statistical fluctuation and is not present in other Bartels Rotations.

\begin{figure}[H]
  \includegraphics[width=\linewidth]{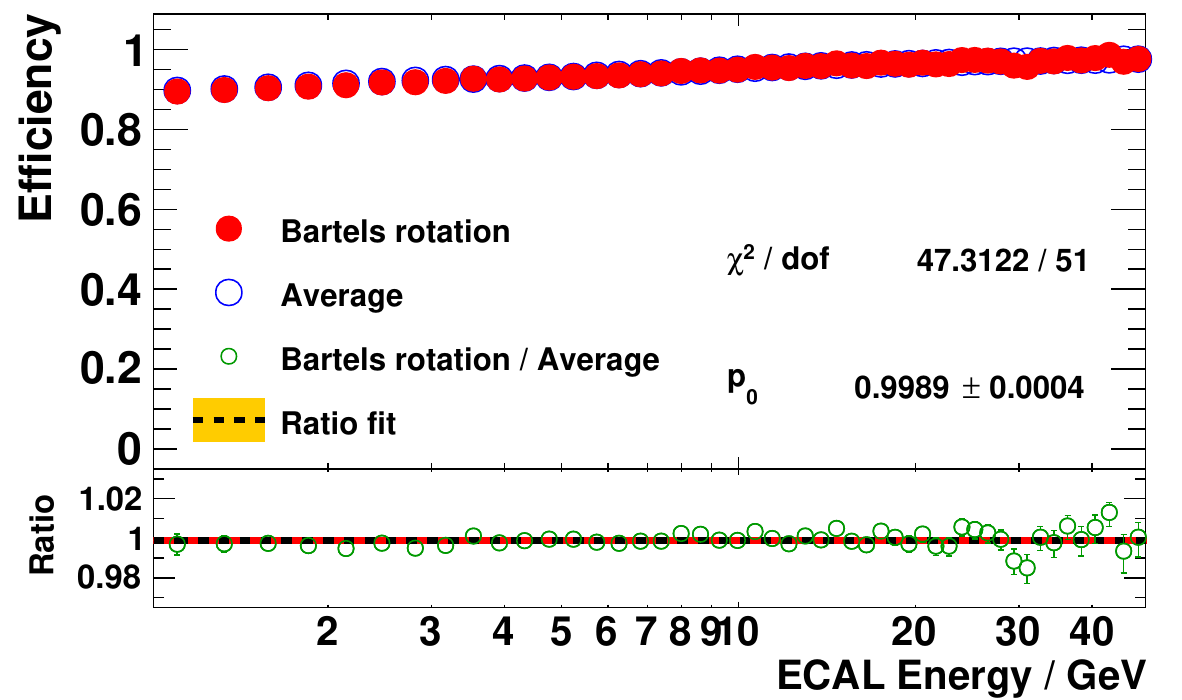}
  \caption{Determination of the ratio: time-averaged cut efficiency $(1 + \delta^{c}(E))$ over time-dependent cut efficiency $(1 + \delta_{i}^{c}(E))$ in Bartels rotation $i = 26$ (Apr~\nth{16},~2013 - May~\nth{13},~2013) for the selection cut: \textbf{\enquote{TRD helium rejection}} (\cref{sec:analysis-data-selection-selection-cuts} - \cref{enum:selection-cut-trd-no-helium}). The blue symbols in the upper plot show the cut efficiency determined from ISS data for the time-averaged analysis on a dedicated sample: the tag sample. The red symbols in the upper plot show the cut efficiency determined from ISS data in Bartels rotation $i = 26$ using the same tag cuts. The lower plot shows the ratio time-averaged efficiency over efficiency in Bartels rotation $i = 26$ as green symbols. A constant is fit to the ratio, shown as dashed black line including the orange uncertainty band.}
  \label{fig:time-dependent-acceptance-trd-no-helium-br-to-average-example}
\end{figure}

\Cref{fig:time-dependent-acceptance-trd-no-helium-br-to-average-example} shows the same comparison for the \enquote{TRD helium rejection}
cut in a single Bartels rotation. These are the only cuts that exhibit a significant time dependence, that needs to be corrected for.

Since the energy dependence of each cut $c$ in each Bartels rotation $i$ is identical to the time-averaged energy dependence, the correction factor
$(1 + \delta_{i}^{c}(E))$ can be factored into the time-averaged part $(1 + \delta^{c}(E))$ and an additional time-dependent correction factor
$(1 + \hat{\delta}(t))$:

\begin{equation*}
  1 + \delta_{i}^{c}(E) = (1 + \delta^{c}(E)) \cdot (1 + \hat{\delta}^{c}(t)).
\end{equation*}

The time-averaged part $(1 + \delta^{c}(E))$ stems from the time-averaged flux analysis and is derived by comparing the predicted cut efficiency
from the electron Monte-Carlo simulation to the time-averaged ISS data for each cut $c$. The time-dependent part $(1 + \hat{\delta}^{c}(t))$ is the
ratio of the time-dependent ISS cut efficiency in each Bartels rotation $i$ to the time-averaged ISS cut efficiency.

Since all correction factors $(1 + \delta_{i}^{c}(E))$ are uncorrelated by a careful construction of the tag samples, the correction
$(1 + \delta_{i}(E))$ is a product of all individual correction factors $(1 + \delta_{i}^{c}(E))$:

\begin{equation*}
  1 + \delta_{i}(E) = \prod_{c} (1 + \delta_{i}^{c}(E)) = \prod_{c} (1 + \delta^{c}(E)) \cdot (1 + \hat{\delta}^{c}(t)) = (1 + \delta(E)) \cdot (1 + \hat{\delta}(t)).
\end{equation*}

\begin{figure}[H]
  \includegraphics[width=\linewidth]{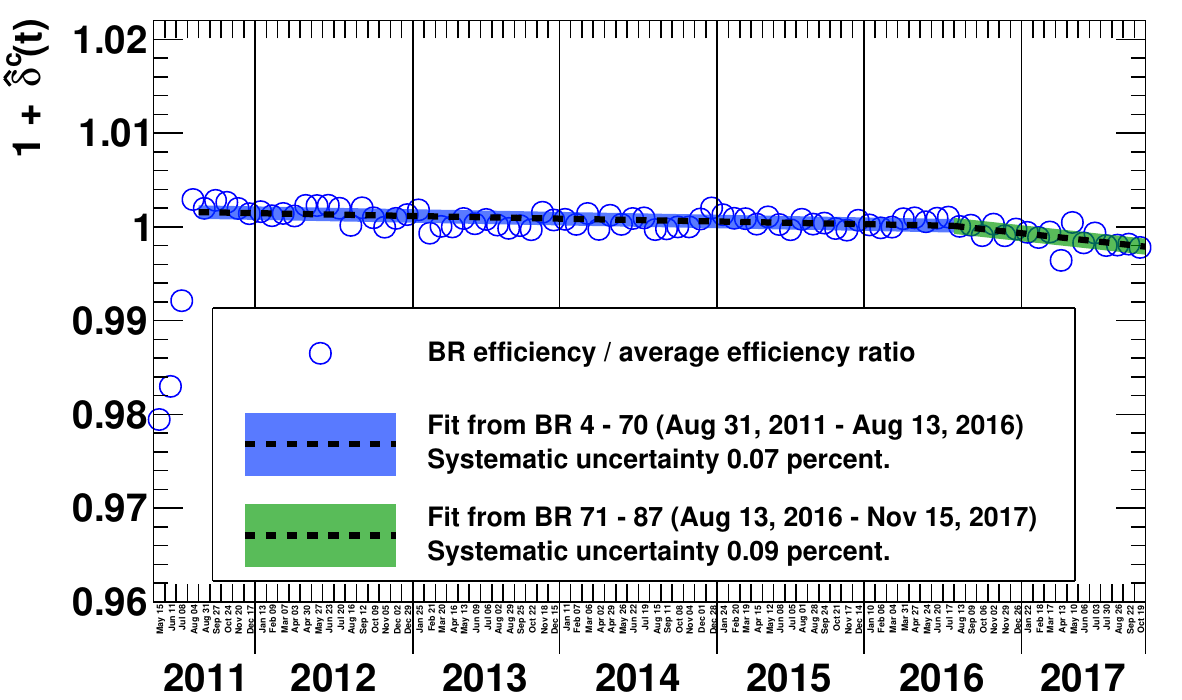}
  \caption{Determination of the time-dependent correction factor $(1 + \hat{\delta}^{c}(t))$ for the selection cut: \textbf{\enquote{Tracker hit pattern}} (\cref{sec:analysis-data-selection-selection-cuts} - \cref{enum:selection-cut-trk-pattern}). The open blue symbols show the ratio: time-dependent cut efficiency over time-averaged cut efficiency for all Bartels rotations (an example of the ratio determined in a single Bartels rotation is shown in \cref{fig:time-dependent-acceptance-tracker-pattern-br-to-average-example}). The black dashed lines show regions parameterized using straight lines. When the efficiency remains stable over an extended period of time, following a simple trend (e.g. slowly decreasing or constant) a parameterization of the time period is used instead of the individual data points to describe the time dependency. The colored bands indicate the systematic uncertainty on the time-dependent correction factor $(1 + \hat{\delta}^{c}(t))$. For individual data points, the uncertainty on the data point - from the previous ratio fit procedure - is taken.}
  \label{fig:time-dependent-acceptance-delta-tracker-pattern}
\end{figure}

\begin{figure}[H]
  \includegraphics[width=\linewidth]{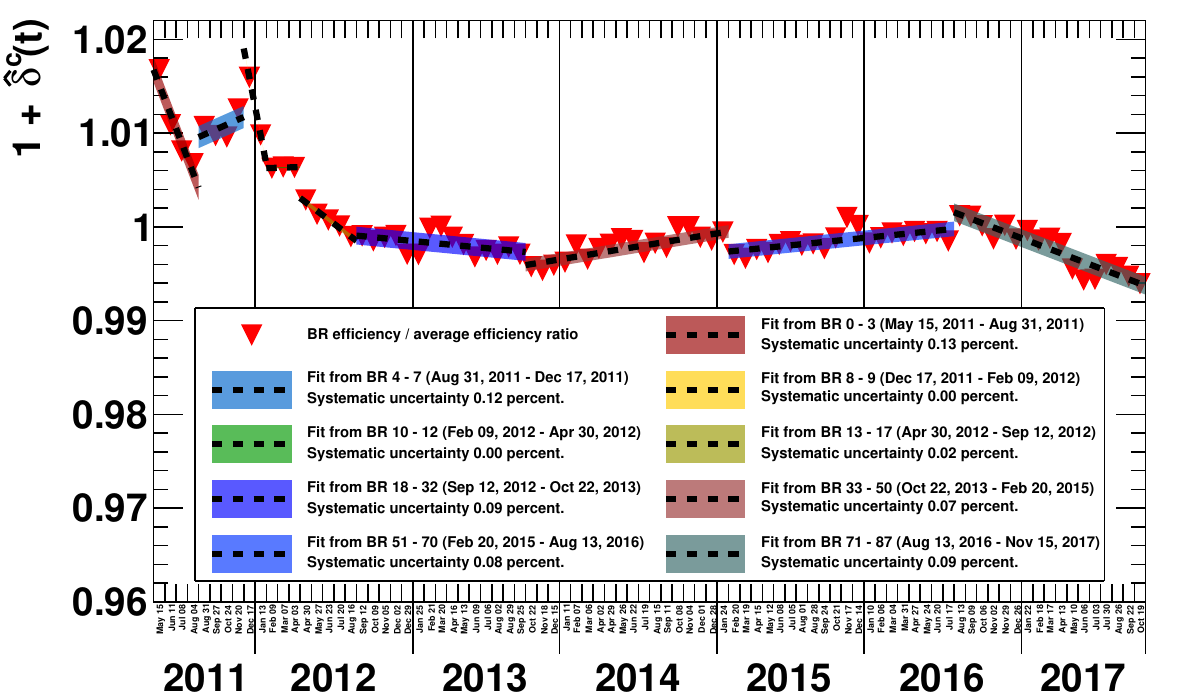}
  \caption{Determination of the time-dependent correction factor $(1 + \hat{\delta}^{c}(t))$ for the selection cut: \textbf{\enquote{TRD helium rejection}} (\cref{sec:analysis-data-selection-selection-cuts} - \cref{enum:selection-cut-trd-no-helium}). The red triangle symbols show the ratio: time-dependent cut efficiency over time-averaged cut efficiency for all Bartels rotations (an example of the ratio determined in a single Bartels rotation is shown in \cref{fig:time-dependent-acceptance-trd-no-helium-br-to-average-example}). The black dashed lines show regions parameterized using straight lines. When the efficiency remains stable over an extended period of time, following a simple trend (e.g. slowly decreasing or constant) a parameterization of the time period is used instead of the individual data points to describe the time dependency. The colored bands indicate the systematic uncertainty on the time-dependent correction factor $(1 + \hat{\delta}^{c}(t))$. For individual data points, the uncertainty on the data point - from the previous ratio fit procedure - is taken.}
  \label{fig:time-dependent-acceptance-delta-trd-no-helium}
\end{figure}

\Cref{fig:time-dependent-acceptance-delta-tracker-pattern} shows the time-dependent correction factor $(1 + \hat{\delta}^{c}(t))$
for the \enquote{Tracker hit pattern} cut and \cref{fig:time-dependent-acceptance-delta-trd-no-helium} shows the same comparison
for the \enquote{TRD helium rejection} cut.

The \enquote{Tracker hit pattern} cut is sensitive to the overall tracking efficiency of the tracker.
During the first three Bartels rotations the tracker reconstruction efficiency was lower than the average efficiency.

From the beginning of data taking up to July~\nth{24},~2011 the tracker calibration was improved (new second step calibration~\cite{BazoAlba2013})
and the reconstruction efficiency was improved by \SIapprox{2}{\percent}.

On December~\nth{1},~2011 six ladders were lost from DAQ because of a power supply malfunction in one crate, leading to a loss of \SIapprox{1.2}{\percent}
of all channels in the readout. This resulted in a slightly reduced overall reconstruction efficiency, which is partly recovered by changes in the offline
reconstruction software. From this date onwards the tracking efficiency remained stable. Since the beginning of 2017 a slight degradation is observed -
however it is on the permille level and does not significantly affect the reconstruction efficiency.

The \enquote{TRD helium rejection} cut has a complex time structure, due to recurring changes in the TRD operation (such as gas refills, high-voltage changes, etc.),
described in \cref{sec:detector-trd}. In the beginning of the data taking period the cut efficiency was higher than the
average efficiency, because the TRD was operated with a lower xenon partial pressure~\cite{Kirn2019}. This leads to a worse rejection and thus
the separation power is smaller.

Since June 2012 the TRD operation is stable and the procedure to keep the rejection at a constant level is established. This is reflected in an
extended period of time where the cut efficiency is stable and only varies at the permille level. In 2017 the overall pressure and thus the xenon
partial pressure were reduced to lower the diffusion losses. This is reflected as a slight decrease in the cut efficiency.

\begin{figure}[H]
  \centering
  \includegraphics[width=0.9\linewidth]{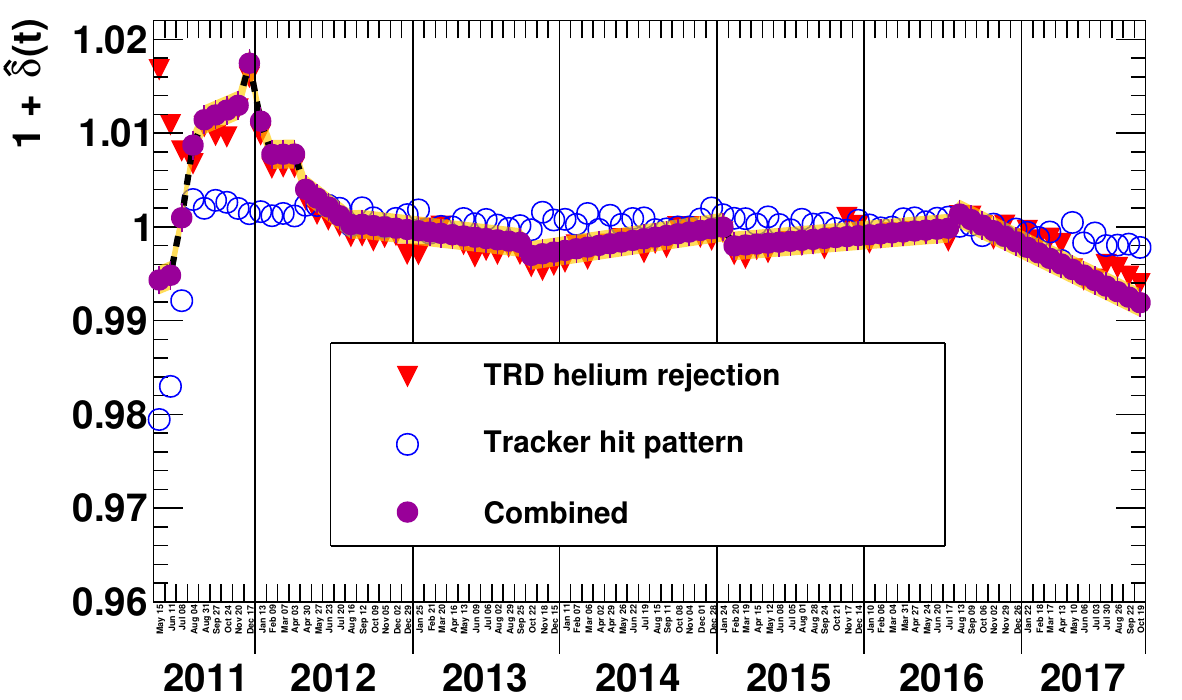}
  \caption{Determination of the time-dependent correction factor $(1 + \hat{\delta}(t))$. The red triangle symbols show the correction factor for the \textbf{\enquote{TRD helium rejection}} cut and the blue circle symbols show the correction factor for the \textbf{\enquote{Tracker hit pattern}} cut. The combined correction factor is shown as magenta symbols. The uncertainty is shown as orange band on top of the magenta symbols.}
  \label{fig:time-dependent-acceptance-delta-overall}
\end{figure}

Since the individual correction factors $(1 + \hat{\delta}^{c}(t))$ for both cuts are determined, the combined correction factor $(1 + \hat{\delta}(t))$
can be computed, as shown in \cref{fig:time-dependent-acceptance-delta-overall}.

The overall time-dependent correction factor $(1 + \hat{\delta}(t))$ varies from \SI{0.5}{\percent} to \SIapprox{2}{\percent}
for a few Bartels rotations in the beginning of the data taking period. Since the beginning of 2012 the correction is almost flat
and varies at the permille level around unity. In late 2016 the correction becomes important again as it rises up to \SIapprox{1}{\percent}
in the end of 2017.

\subsection{Charge-confusion}
\label{sec:analysis-flux-time-dependent-charge-confusion}

The charge-confusion $f_{\text{cc},\,i}(E)$ is determined individually in each Bartels rotation $i$, following the recipe given in the time-averaged flux analysis,
described in \cref{sec:analysis-lepton-counts-2d-fit}. The two-dimensional template fit procedure is executed with the charge-confusion as free fit parameter. The so-obtained
charge-confusion value for each energy bin $j$ and each Bartels rotation $j$ is then compared to the time-averaged charge-confusion value $f_{\text{cc}}(E)$.

For all energy bins the average of the charge-confusion value is compatible with the time-averaged
charge-confusion and thus $f_{\text{cc},\,i}(E) = f_{\text{cc}}(E)$ (\cref{sec:appendix-flux-time-dependent-charge-confusion}).
As consequence the charge-confusion in the time-dependent two-dimensional templates fits can be fixed to the Monte-Carlo prediction, as for the time-averaged flux analysis.

\subsection{TRD templates}
\label{sec:analysis-flux-time-dependent-trd-templates}

The procedure to extract the TRD templates from ISS data was described in \cref{sec:analysis-lepton-counts-trd-templates} for the time-averaged
analysis. These TRD templates are not applicable for each Bartels rotation, individually, due to recurring changes in the TRD operation.
The TRD operation changes are the same reason for the complex time dependence in the \enquote{TRD helium rejection} cut.

Five TRD template parameters were identified, having a non-negligible time dependence. One parameter describing the charge-confused protons:
peak position~\ccProtPeakNovo, two parameters for the electron template: peak position~\elecPeakNovo, width~\elecWidthNovo~and
two parameters concerning the proton templates: peak position~\protPeakNovo~and width~\protWidthNovo.

\begin{figure}[H]
  \includegraphics[width=\linewidth]{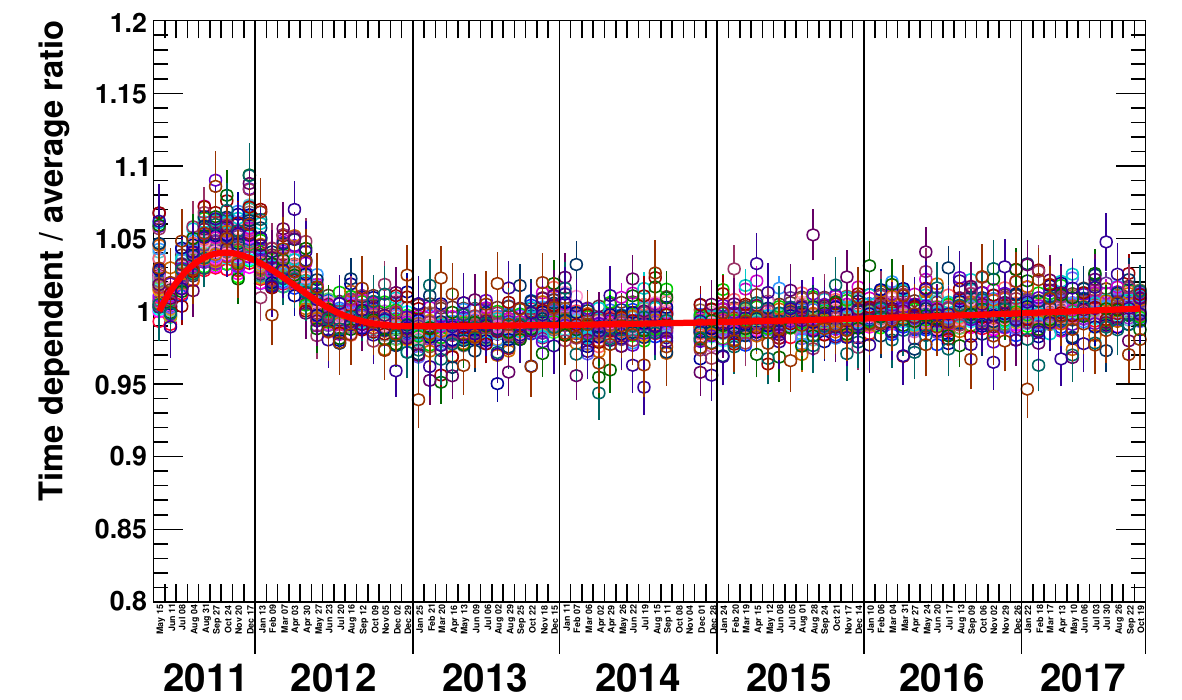}
  \caption{Time-dependence of the~\elecPeakNovo~parameter for all energy bins - \SIrange{0.65}{52.33}{\GeV} - in the time-dependent analysis for the single-track sample. Each energy bin is represented by a different color. The filled red line denote the result of a spline fit, simultaneously to all data points, describing the average time-dependence of the~\elecPeakNovo~parameter.}
  \label{fig:time-dependent-trd-templates-elec-peak-novo}
\end{figure}

\begin{figure}[H]
  \centering
  \includegraphics[width=0.9\linewidth]{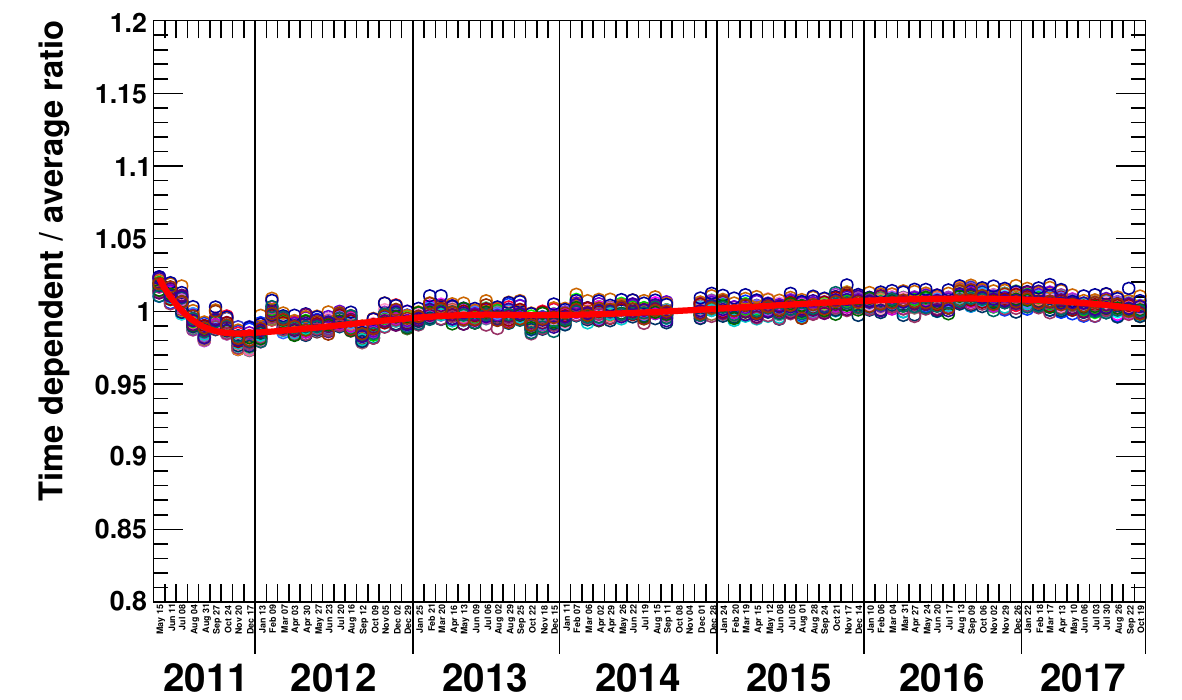}
  \caption{Time-dependence of the~\protPeakNovo~parameter for all energy bins - \SIrange{0.65}{52.33}{\GeV} - in the time-dependent analysis.}
  \label{fig:time-dependent-trd-templates-prot-peak-novo}
\end{figure}

The time-dependence of each of the TRD template parameter $c$ is examined, by repeating the TRD template extraction procedure in each
Bartels rotation $i$ for each energy bin $j$. For each TRD template parameter $c$, all energy bins that show a similar time-dependence
are grouped together. The time-dependence of all energy bins for the~\elecPeakNovo~and the~\protPeakNovo~parameters is
identical. \Cref{fig:time-dependent-trd-templates-elec-peak-novo,fig:time-dependent-trd-templates-prot-peak-novo}
show the time-dependence of the ~\elecPeakNovo~and~\protPeakNovo~parameters for all energy bins in the time-dependent analysis for the single-track sample.
Each energy $j$ bin is represented by a different color. The filled red points represent the average time-dependence of the
parameters. It was determined by executing a simultaneous fit of a spline curve to all data points of each TRD template parameter $c$.

The results for the multi-tracks samples are shown in \cref{sec:appendix-flux-time-dependent-trd-templates}.

\begin{figure}[H]
  \centering
  \includegraphics[width=0.9\linewidth]{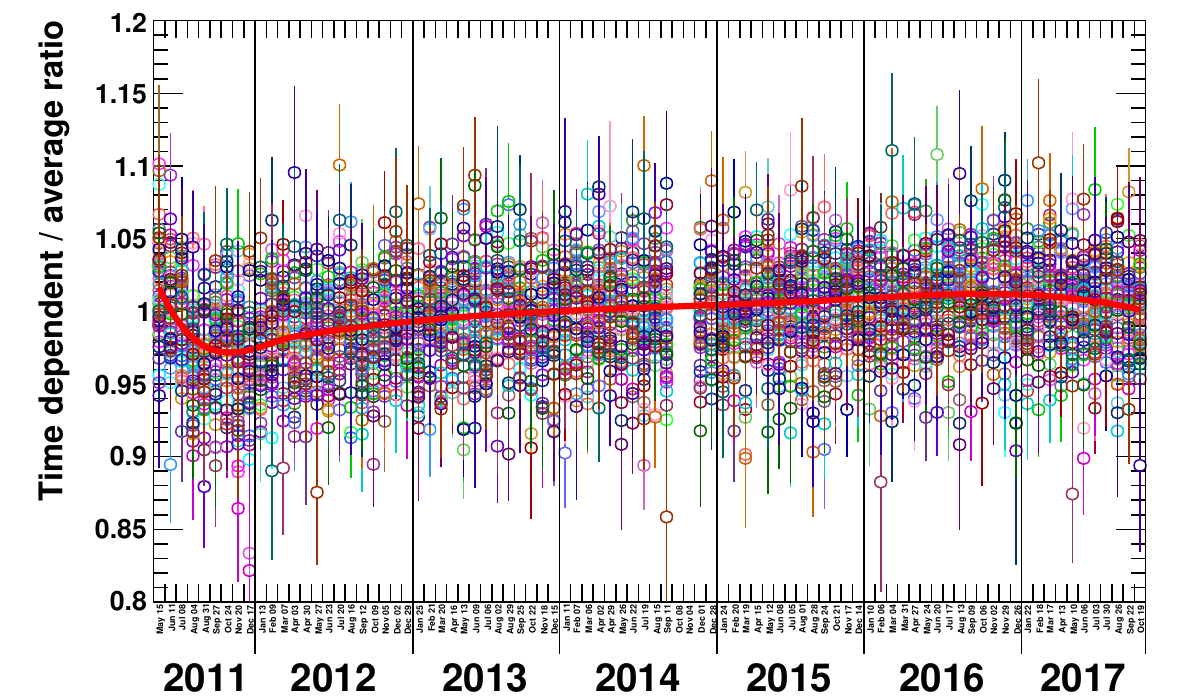}
  \caption{Time-dependence of the~\ccProtPeakNovo~parameter for selected energy bins (\SIrange{1.22}{52.33}{\GeV}), as example, in the time-dependent analysis for the single-track sample.}
  \label{fig:time-dependent-trd-templates-ccprot-peak-novo}
\end{figure}

\begin{figure}[H]
  \centering
  \includegraphics[width=0.9\linewidth]{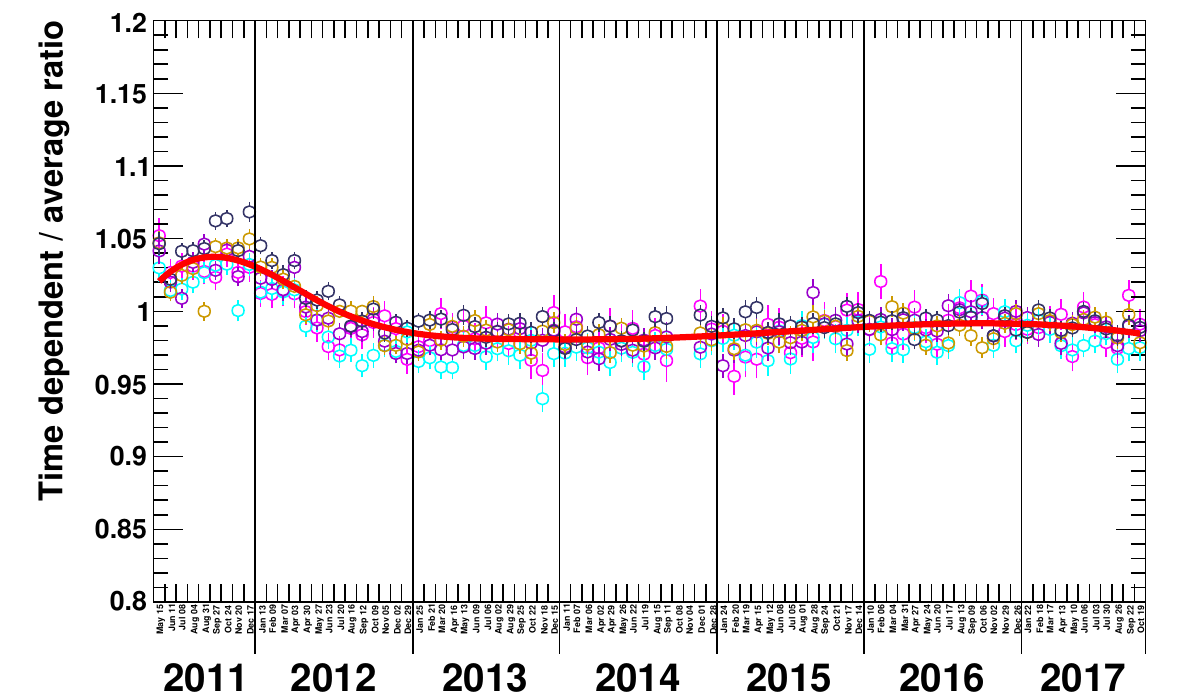}
  \caption{Time-dependence of the~\elecWidthNovo~parameter for selected energy bins (\SIrange{1.46}{3.00}{\GeV}), as example, in the time-dependent analysis for the single-track sample.}
  \label{fig:time-dependent-trd-templates-elec-width-novo}
\end{figure}

\begin{figure}[H]
  \centering
  \includegraphics[width=0.9\linewidth]{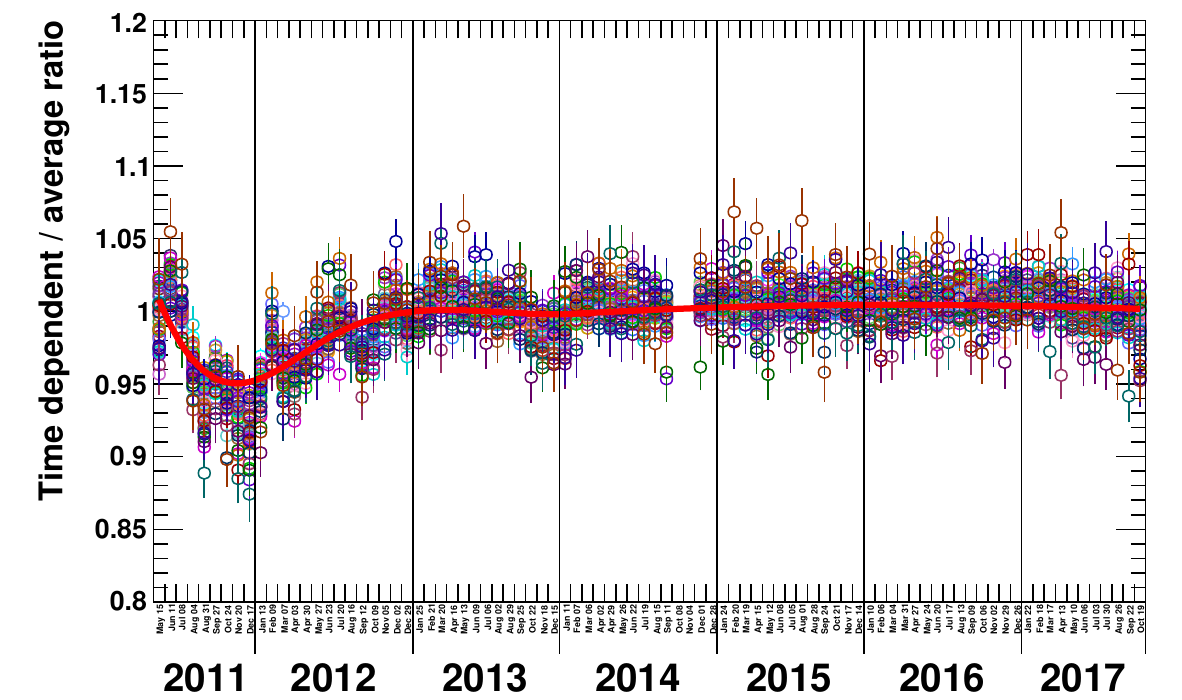}
  \caption{Time-dependence of the~\protWidthNovo~parameter for selected energy bins (\SIrange{1.72}{52.33}{\GeV}), as example, in the time-dependent analysis for the single-track sample.}
  \label{fig:time-dependent-trd-templates-prot-width-novo}
\end{figure}

The time-dependence of the three remaining parameters - \ccProtPeakNovo, \elecWidthNovo~and~\protWidthNovo, is shown in
\cref{fig:time-dependent-trd-templates-ccprot-peak-novo,fig:time-dependent-trd-templates-elec-width-novo,fig:time-dependent-trd-templates-prot-width-novo}
in selected energy bins, as example, for the single-track sample.

Since the time-dependence of all TRD template parameters needed to describe both the single-track sample and the multi-tracks sample is known,
new time-dependent TRD templates for the all-tracks sample can be constructed, following the recipe given in \cref{sec:analysis-lepton-counts-2d-fit}.

\begin{figure}[H]
  \centering
  \includegraphics[width=0.9\linewidth]{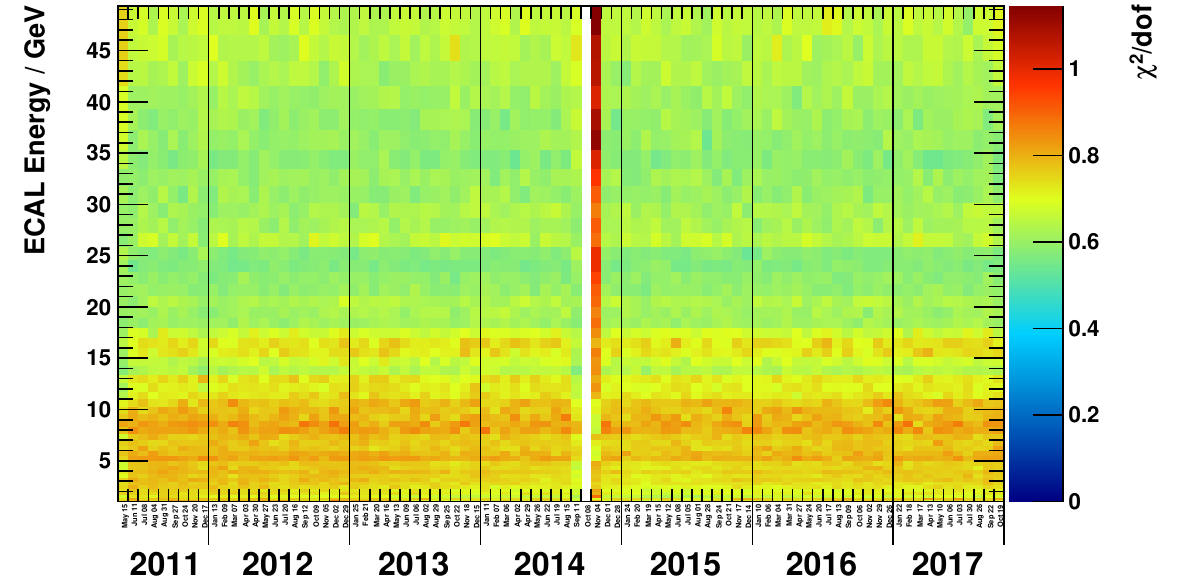}
  \caption{Goodness of fit results for the two-dimensional template fit, performed using the time-dependent TRD templates on the all-tracks sample. The color encodes the $\chi^2/\text{dof}$ value for each Bartels rotation $i$ and each energy bin $j$ of the time-dependent analysis. All $\chi^2/\text{dof}$ values are acceptable.}
  \label{fig:time-dependent-template-fit-chi2-all-tracks}
\end{figure}

To ensure the validity of the time-dependent TRD templates the two-dimensional template fit is performed for all Bartels rotations $i$ and
all energy bins $j$. \Cref{fig:time-dependent-template-fit-chi2-all-tracks} shows the $\chi^2/\text{dof}$ distribution for the all-tracks sample,
encoded as color in the plot. No particular energy or time interval shows a peculiar behaviour. The horizontal stripes appear at energies where the TRD
template parameterization changes (certain parameters become constant). The fixation of several parameters does not lead to a bias, but stabilizes
the fit procedure.

\begin{figure}[H]
  \includegraphics[width=\linewidth]{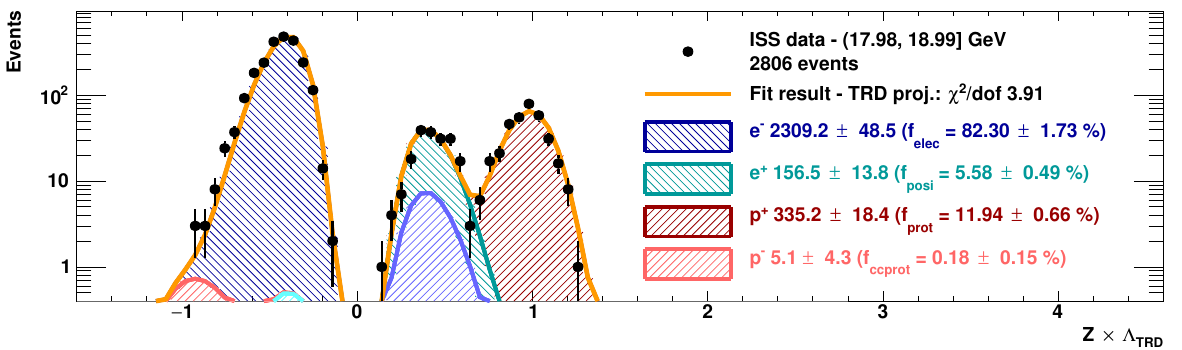}
  \caption{Results of the two-dimensional TRD / CCMVA template fit in Bartels rotation $i = 6$ for the energy bin \SIrange{17.98}{18.99}{\GeV} (all-tracks sample) using the time-averaged TRD templates.}
  \label{fig:time-dependent-template-fit-chi2-comparison-time-avg}
\end{figure}

\begin{figure}[H]
  \includegraphics[width=\linewidth]{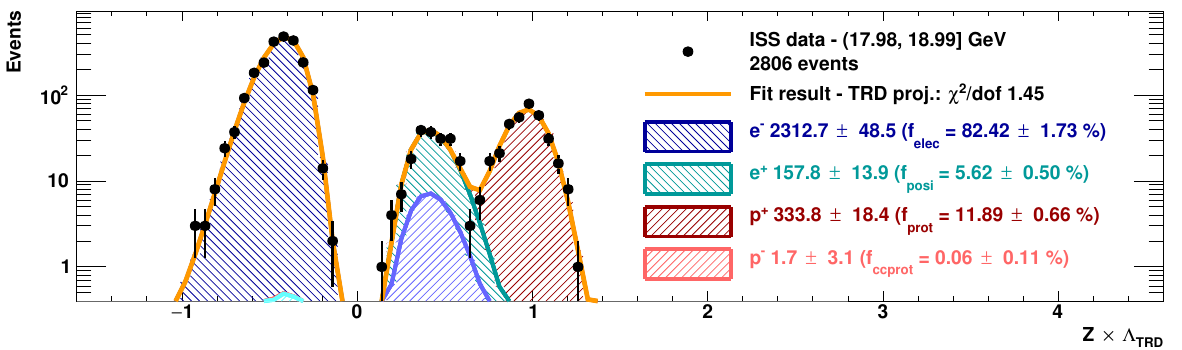}
  \caption{Results of the two-dimensional TRD / CCMVA template fit in Bartels rotation $i = 6$ for the energy bin \SIrange{17.98}{18.99}{\GeV} (all-tracks sample) using the time-dependent TRD templates.}
  \label{fig:time-dependent-template-fit-chi2-comparison-time-dep}
\end{figure}

To demonstrate the effect of the time-dependent changes of the TRD template parameters on the two-dimensional template fit, the same fit is performed twice: once using the time-averaged
templates, \cref{fig:time-dependent-template-fit-chi2-comparison-time-avg} and once using the time-dependent templates, \cref{fig:time-dependent-template-fit-chi2-comparison-time-dep},
or an example energy bin in Bartels rotation $i = 6$. It is evident that only the time-dependent templates correctly describe the data sample. When using the time-averaged templates a bias
is exposed, leading to a large $\chi^2/\text{dof}$ value.

It was verified that the systematic uncertainty $\sigma_{\text{trd}}(E) / \Phi_{e^{\pm}}(E)$, derived in \cref{sec:analysis-flux-time-averaged-sysunc-trd-estimator},
is identical when using the time-dependent TRD templates. For that reason, no extra time-dependent systematic uncertainty $\sigma_{\text{trd},\,i}(E)$ is needed.

\subsection{Energy scale stability}
\label{sec:analysis-flux-time-dependent-energy-scale-stability}

For the time-dependent analysis it is important to ensure that the energy scale (\cref{sec:analysis-event-reconstruction-ecal-shower}) is stable over time.
To quantify the stability the $E/\abs{R}$ peak position is analyzed as a function of time.

Assuming the rigidity scale is stable in time - as shown in Ref.~\cite{Berdugo2017} - the $E/\abs{R}$ peak position can be used to test if
the energy scale $E$ shows a time dependence, such as a drift.

To obtain the time-dependent $E/\abs{R}$ distribution, a negative rigidity ISS data sample is prepared in each Bartels rotation $i$,
by applying the preselection and selection cuts (described in \cref{sec:analysis-data-selection-preselection-cuts,sec:analysis-data-selection-selection-cuts}) and imposing additional cuts on the ECAL estimator ($\Lambda_{\text{ECAL}}$ > 0)
and the TRD estimator ($\Lambda_{\text{TRD}}$ > 0.7) to select clean electron samples. In these data samples the $E/\abs{R}$ distribution
is analyzed from \SIrange{2}{50}{\GeV} in six logarithmic energy bins.

For each Bartels rotation $i$, and each energy bin, the time-dependent $E/\abs{R}$ distribution is shifted along the energy axis,
to find the best matching scale factor $S_{i}$ that leads to best agreement between the resampled time-averaged $E/\abs{R}$ distribution and the
time-dependent one. The resampled time-averaged $E/\abs{R}$ distribution, consists of a randomly selected sub-sample of the time-averaged
$E/\abs{R}$ distribution, containing ten\footnote{The number needs to be sufficiently large, to smooth statistical fluctuations.} times
as much as events as in the time-dependent $E/\abs{R}$ distribution.

\begin{figure}[H]
  \begin{subfigure}{0.5\linewidth}
    \includegraphics[width=\linewidth]{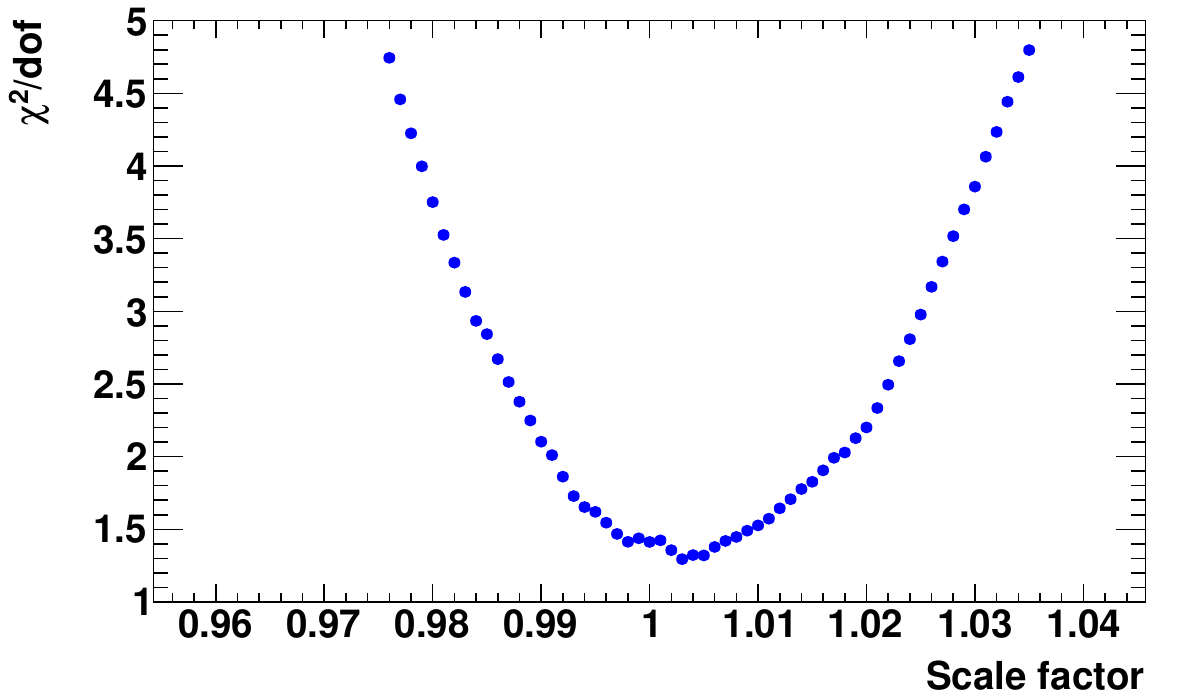}
    \caption{$\chi^2/\text{dof}$ scan}
    \label{fig:time-dependent-energy-stability-chi2-scan}
  \end{subfigure}
  \hfill
  \begin{subfigure}{0.5\linewidth}
    \includegraphics[width=\linewidth]{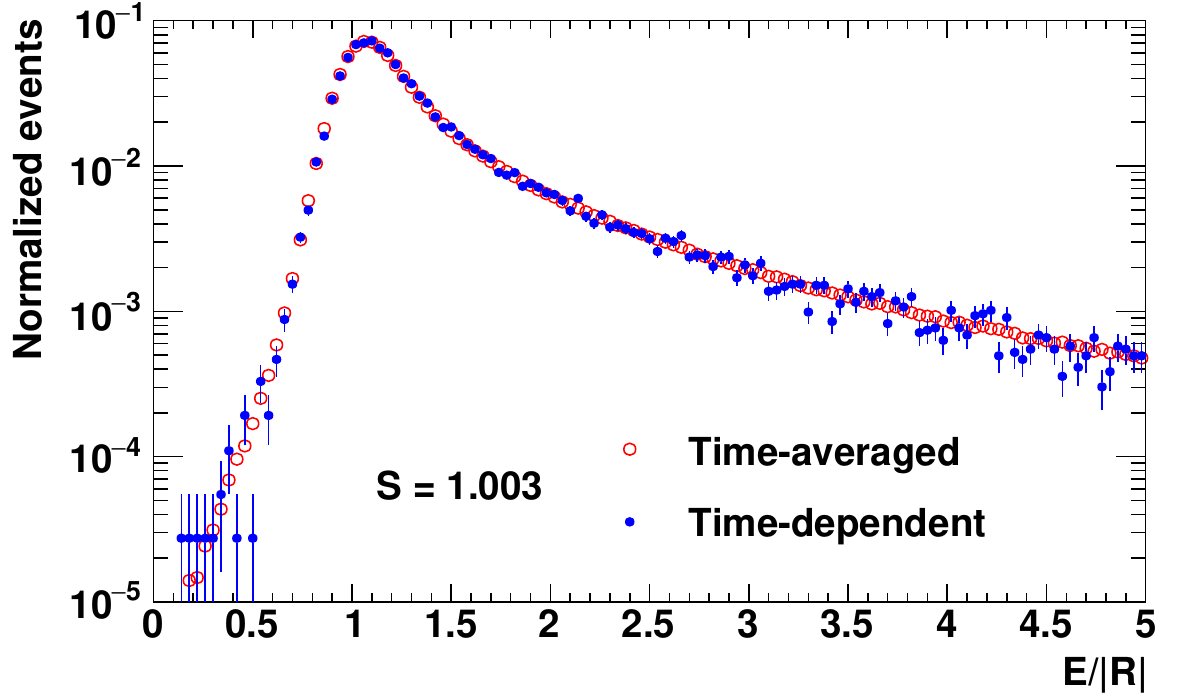}
    \caption{$E/\abs{R}$ comparison}
    \label{fig:time-dependent-energy-stability-eop-after-scan}
  \end{subfigure}
  \caption{Control plots of the $\chi^2$ scan procedure in the first $E/\abs{R}$ energy bin (\SIrange{2}{3.42}{\GeV}) to determine the best matching scale factor $S$, in an example Bartels rotation, that leads to agreement between the time-dependent and the time-averaged $E/\abs{R}$ distribution.}
  \label{fig:time-dependent-energy-stability}
\end{figure}

A $\chi^2$ scan (\cref{fig:time-dependent-energy-stability-chi2-scan}) is used to determine the best matching scale factor $S_{i}$. After shifting
the time-dependent $E/\abs{R}$ distribution along the energy axis, the time-averaged and the time-dependent distributions are on top of each other,
as shown in \cref{fig:time-dependent-energy-stability-eop-after-scan}.

\begin{figure}[H]
  \begin{subfigure}{0.5\linewidth}
    \includegraphics[width=\linewidth]{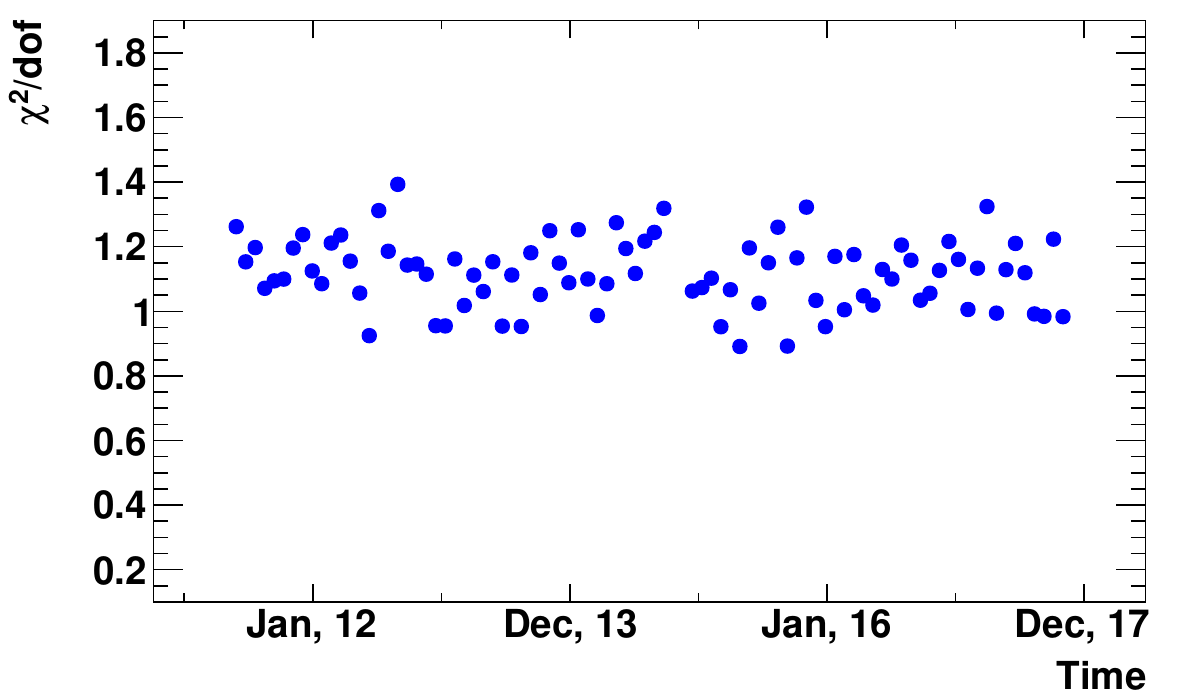}
    \caption{Minimum $\chi^2/\text{dof}$ vs.~time}
  \end{subfigure}
  \hfill
  \begin{subfigure}{0.5\linewidth}
    \includegraphics[width=\linewidth]{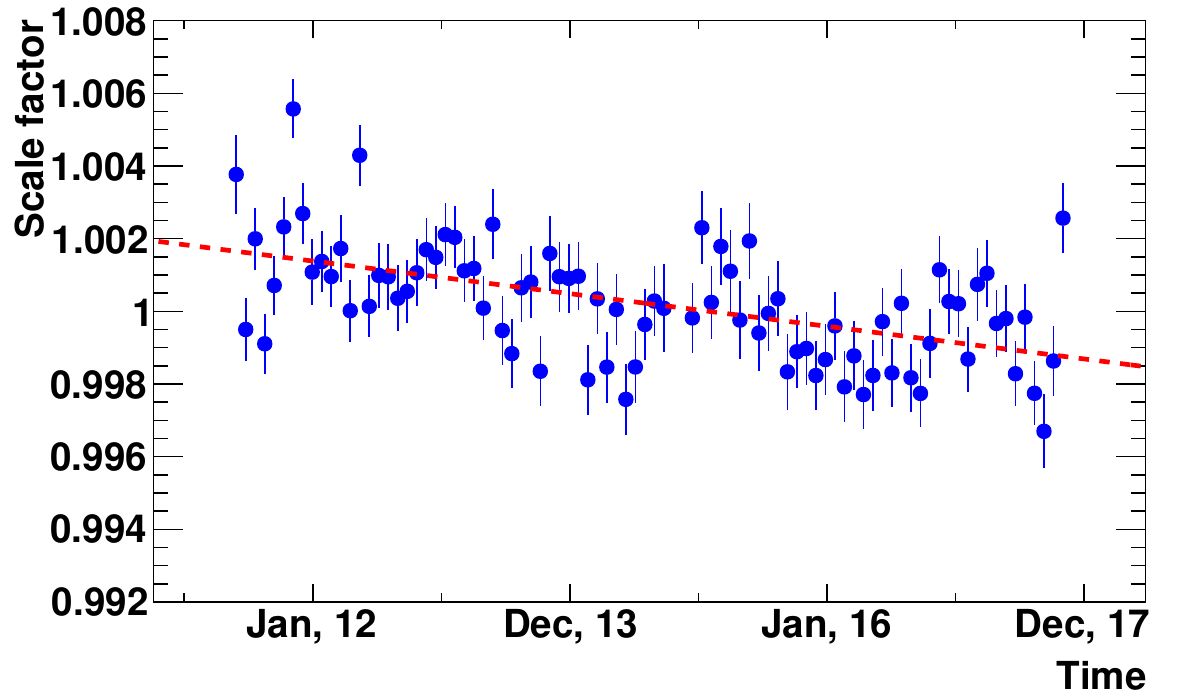}
    \caption{$E/\abs{R}$ scale factor vs.~time}
    \label{fig:time-dependent-energy-stability-scale-factor}
  \end{subfigure}
  \caption{Results of the $\chi^2$ scan procedure in the first $E/\abs{R}$ energy bin (\SIrange{2}{3.42}{\GeV}) as function of time. The left plot shows the minimum $\chi^2/\text{dof}$ as function of time and the right plot the obtained scale factor $S_{i}$ for each Bartels rotation $i$.}
  \label{fig:time-dependent-energy-stability-scale-factor-results}
\end{figure}

\Cref{fig:time-dependent-energy-stability-scale-factor-results} shows the result of the $\chi^2$ scan procedure in the first
energy bin (\SIrange{2}{3.42}{\GeV}) as function of time. The procedure is repeated for all energy bins
and yields consistent results: acceptable $\chi^2/\text{dof}$ values and a scale factor close to unity.

There appears to be a slope in the data, as exhibited in \cref{fig:time-dependent-energy-stability-scale-factor}. In a dedicated study it was determined
that the cause of the slope is a change in the rigidity reconstruction as function of time. This could either be the result of an improper
magnet temperature calibration or the observation of a slightly decreasing magnetic field over time. Both explanations are possible
and might induce a slope in the $E/\abs{R}$ distribution as function of time.

\begin{figure}[H]
  \begin{subfigure}{0.5\linewidth}
    \includegraphics[width=\linewidth]{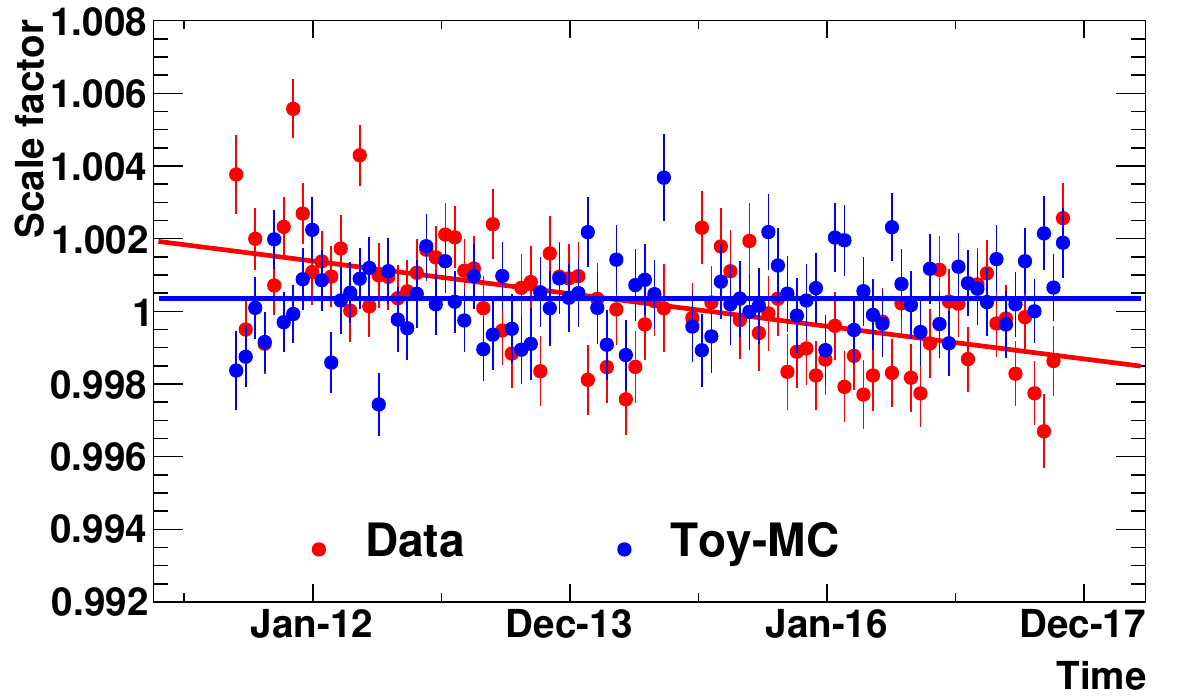}
    \caption{First $E/\abs{R}$ energy bin: \SIrange{2}{3.42}{\GeV}}
  \end{subfigure}
  \hfill
  \begin{subfigure}{0.5\linewidth}
    \includegraphics[width=\linewidth]{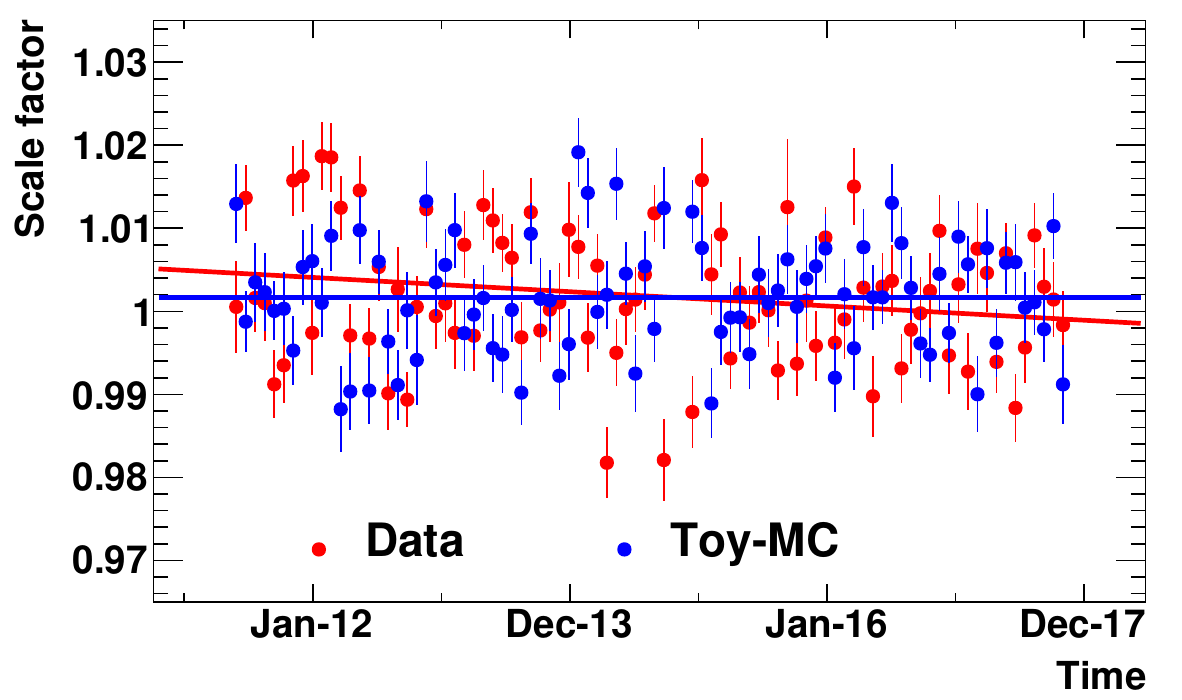}
    \caption{Last $E/\abs{R}$ energy bin: \SIrange{29.24}{50.00}{\GeV}}
  \end{subfigure}
  \caption{Scale factors $S_{i}$ of both the \textbf{ISS data sample} and the \textbf{Toy-MC sample} as function of time. The \textbf{ISS data sample} time dependence is parameterized using a straight line, and the time dependence in the \textbf{Toy-MC sample} using a constant.}
  \label{fig:time-dependent-energy-stability-scale-factor-vs-time}
\end{figure}

The $\chi^2$ scan procedure yields one scaling factor $S_{i} \pm \sigma_{S_{i}}$ per energy bin and per Bartels rotation $i$.
A model, parameterizing the time dependence, is fit to the scaling factor vs.~time data, as shown in \cref{fig:time-dependent-energy-stability-scale-factor-vs-time}
for two example energy bins. For each energy bin, the RMS $\eta_{\text{RMS}}$ is determined from the residual distribution $\eta = \text{data - model}$: it captures
the amount of fluctuations in the given data sample.

The whole $\chi^2$ scan procedure is repeated $M \approx \mathcal{O}(200)$ times, each time with a different resampled
time-averaged $E/\abs{R}$ distribution. This yields a distribution of $M$ different $\eta_{\text{RMS}}$ values. The mean of this distribution
is an estimation of the energy scale stability as function of time: $\sigma_{\text{data}}$ in the \textbf{ISS data sample}.
The estimated uncertainty $\sigma_{\text{data}}$ may overestimate the true uncertainty $\sigma_{\text{ene}}$,
because of the intrinsic statistical fluctuations in the time-dependent data samples. The intrinsic width of the $\eta$ distribution,
$\sigma_{\text{toy}}$, needs to be subtracted in order to retrieve the true uncertainty $\sigma_{\text{ene}} = \sqrt{\sigma_{\text{data}}^2 - \sigma_{\text{toy}}^2}$.

To determine $\sigma_{\text{toy}}$, the $\chi^2$ scan procedure is repeated again $M \approx \mathcal{O}(200)$ times,
but replacing the time-dependent distribution in each iteration $m$ by a randomly selected sub-sample of the time-averaged $E/\abs{R}$
distribution, containing as much events as the time-dependent $E/\abs{R}$ distribution has. This data samples is called
\textbf{Toy-MC sample}.

\begin{figure}[H]
  \begin{subfigure}{0.5\linewidth}
    \includegraphics[width=\linewidth]{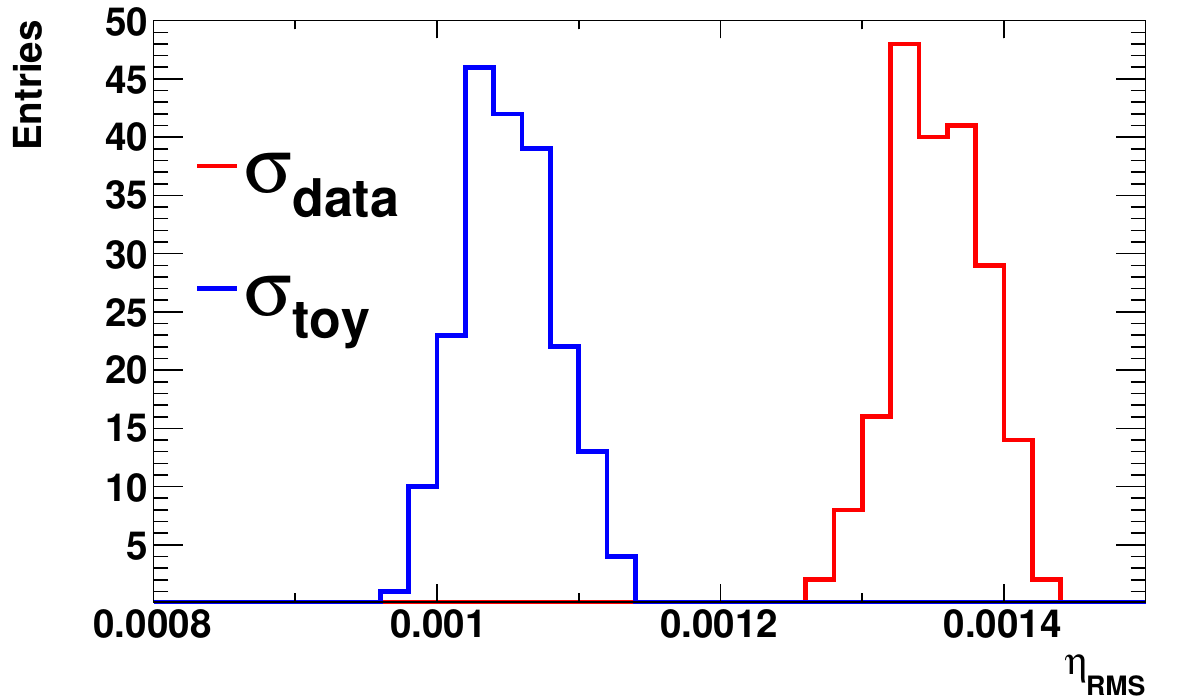}
    \caption{First $E/\abs{R}$ energy bin: \SIrange{2}{3.42}{\GeV}}
  \end{subfigure}
  \hfill
  \begin{subfigure}{0.5\linewidth}
    \includegraphics[width=\linewidth]{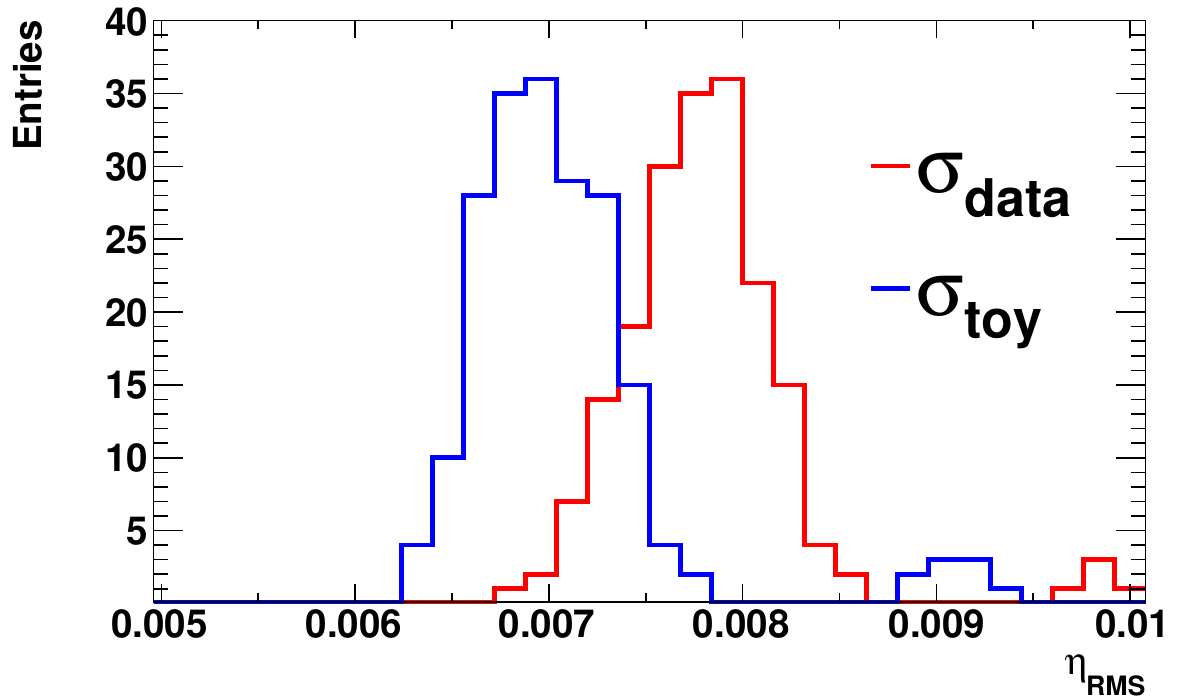}
    \caption{Last $E/\abs{R}$ energy bin: \SIrange{29.24}{50.00}{\GeV}}
  \end{subfigure}
  \caption{$\eta_{\text{RMS}}$ distribution for the \textbf{ISS data sample} and the \textbf{Toy-MC sample}. The mean of the distributions is denoted as $\sigma_{\text{data}}$ and $\sigma_{\text{toy}}$, respectively.}
  \label{fig:time-dependent-energy-stability-rms-distributions-vs-time}
\end{figure}

The scale factors determined for the both the \textbf{ISS data sample} and the \textbf{Toy-MC sample} are shown in
\cref{fig:time-dependent-energy-stability-scale-factor-vs-time}, for an example iteration $m$ in two energy bins,
and the $\eta_{\text{RMS}}$ distributions are shown in \cref{fig:time-dependent-energy-stability-rms-distributions-vs-time}
for the same example energy bins.

\begin{figure}[H]
  \centering
  \includegraphics[width=0.9\linewidth]{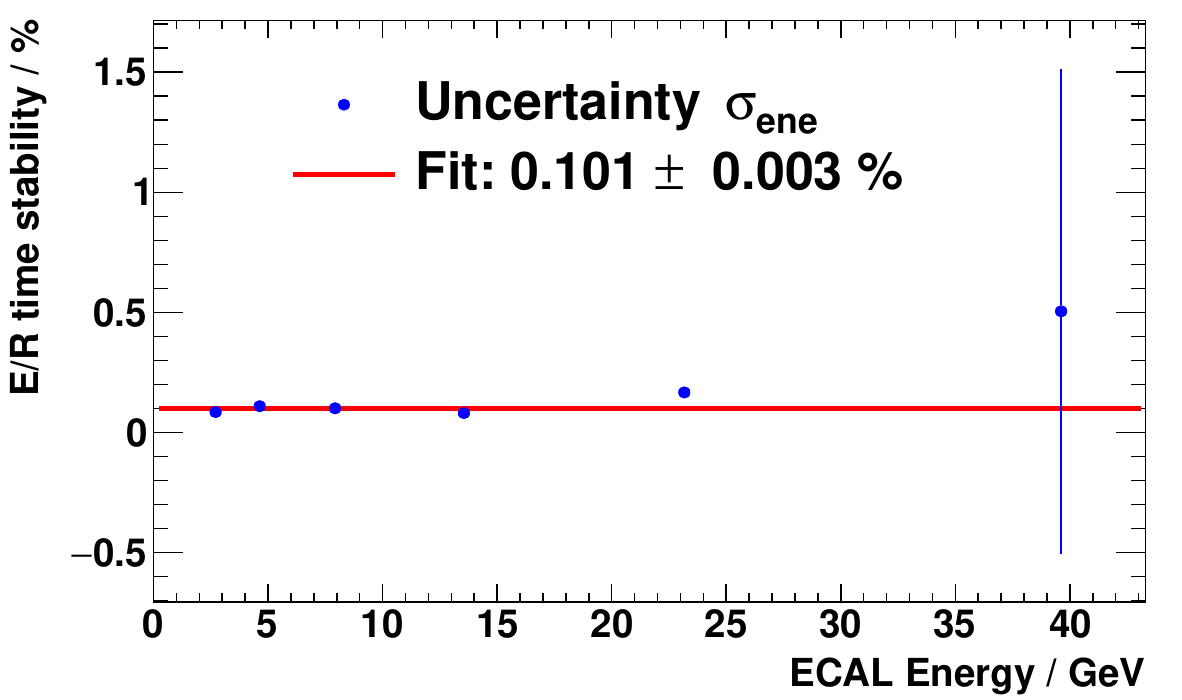}
  \caption{Relative energy scale uncertainty $\sigma_{\text{ene}} / E$ as function of energy. It is identical for all Bartels rotations $i$, since the energy scale is independent of time.}
  \label{fig:time-dependent-energy-stability-overview}
\end{figure}

\clearpage
Since the origin of the slope in the $E/\abs{R}$ distribution was attributed to the rigidity scale (which was tested in dedicated, extensive studies),
one can conclude that the energy scale is stable in time. The deviation of the scale factors $S_{i}$ from unity, indicating a perfect stable energy scale,
leads to an uncertainty of the overall energy scale - independent of energy - of $\sigma_{\text{ene}} / E = \SI{0.101(3)}{\percent}$,
as shown in \cref{fig:time-dependent-energy-stability-overview}.

The relative uncertainty $\sigma_{\text{ene}} / E$ is propagated to the fluxes as additional systematic uncertainty,
which is described in \cref{sec:analysis-flux-time-dependent-sysunc-energy-scale}. No time-dependent correction of the
energy scale is necessary. The next full reproduction of the AMS-02 data (pass7) already implements a time-dependent rigidity
correction, which will avoid the slope in the $E/\abs{R}$ distribution as function of time.

\section{Time-dependent systematic uncertainties for flux analysis}
\label{sec:analysis-flux-time-dependent-sysunc}

In this section all systematic uncertainties that contribute to the time-dependent fluxes are summarized.
The systematic uncertainties of the time-dependent flux analysis are identical to the time-averaged analysis plus two extra components,
that cover the variation with time of the acceptance and the energy scale.

\subsection{Time-dependent acceptance}
\label{sec:analysis-flux-time-dependent-sysunc-acceptance}

The derivation of the time-dependent acceptance including its uncertainty was already discussed in \cref{sec:analysis-flux-time-dependent-acceptance}.
The resulting relative systematic uncertainty $\sigma_{\text{time-acc},\,i}(E) / \Phi_{e^{\pm}}(E)$ is independent of energy and equals to \SIapprox{0.16}{\percent}
in Bartels rotation $i = 26$, as example. The magnitude of the uncertainty is almost identical in all Bartels rotations.
Furthermore the uncertainty associated with the electron flux is identical to the uncertainty associated with the positron flux.

\subsection{Time-dependent energy scale}
\label{sec:analysis-flux-time-dependent-sysunc-energy-scale}

In \cref{sec:analysis-flux-time-dependent-energy-scale-stability} it was shown how to derive $\sigma_{\text{ene}}(E) / E$ - the relative uncertainty
of the energy scale time stability as function of the energy. In the following it will be shown how to propagate the energy scale uncertainty to an uncertainty
on the fluxes.

\medskip
The flux $\Phi_{e^{\pm},\,i}(E)$ in Bartels rotation $i$ can be locally parameterized using a constant $A$ and a spectral index $\gamma_{i}^{\pm}$:
$\Phi_{e^{\pm},\,i}(E) = A \cdot E^{-\gamma_{i}^{\pm}}$. The integral of the flux in the energy interval $[E_{1},\ E_{2}]$ is given by

\begin{equation*}
  p_{i}^{\pm}(E) = \int\limits_{E_{1}}^{E_{2}} \Phi_{e^{\pm},\,i}(E) \diff E = \frac{A \cdot \left(E_{1}^{-\gamma_{i}^{\pm} + 1} - E_{2}^{-\gamma_{i}^{\pm} + 1}\right)}{\gamma_{i}^{\pm} - 1}.
\end{equation*}

When assuming that the true energy is shifted by a factor $s = \sigma_{\text{ene}}(E) / E$, the integral of the flux needs to be evaluated in the energy
interval $[s \cdot E_{1},\  s \cdot E_{2}]$

\begin{equation*}
  \hat{p}^{\pm}_{i}(E) = \int\limits_{E_{1} (1 - s)}^{E_{2} (1 - s)} \Phi_{e^{\pm},\,i}(E) \diff E.
\end{equation*}

The difference between the integral of the shifted flux and the integral of the original flux, divided by the integral of the original flux is equal
to the relative systematic uncertainty $\sigma_{\text{time-ene, }e^{\pm},\,i}(E) / \Phi_{e^{\pm},\,i}(E)$, which describes the propagation of the energy
scale uncertainty to the $e^{\pm}$ fluxes:

\begin{equation}
  \label{eq:time-ene-flux-uncertainty}
  \frac{\sigma_{\text{time-ene, }e^{\pm},\,i}(E)}{\Phi_{e^{\pm},\,i}(E)} = \frac{\hat{p}^{\pm}_{i}(E) - p_{i}^{\pm}(E)}{p_{i}^{\pm}(E)}.
\end{equation}

The difference $\hat{p}^{\pm}_{i}(E) - p_{i}^{\pm}(E)$ can be approximated using a Taylor expansion, yielding

\begin{equation}
  \label{eq:time-ene-taylor-expansion}
  \begin{aligned}
    \mathcal{T}(\hat{p}^{\pm}_{i}(E) - p_{i}^{\pm}(E))                                    &= A \cdot \left(E_{1}^{-\gamma_{i}^{\pm} + 1} - E_{2}^{-\gamma_{i}^{\pm} + 1} \right) \cdot s + \mathcal{O}(s^2) \\
    \Rightarrow \frac{\mathcal{T}(\hat{p}^{\pm}_{i}(E) - p_{i}^{\pm}(E))}{p_{i}^{\pm}(E)} &= (\gamma_{i}^{\pm} - 1) \cdot s + \mathcal{O}(s^2).
  \end{aligned}
\end{equation}

Inserting \cref{eq:time-ene-taylor-expansion} into \cref{eq:time-ene-flux-uncertainty} yields a simple expression
for the energy scale uncertainty, propagated to the $e^{\pm}$ fluxes:

\begin{equation}
  \label{eq:relsyst-unc-on-fluxes-due-to-ene-scale}
  \frac{\sigma_{\text{time-ene, }e^{\pm},\,i}(E)}{\Phi_{e^{\pm},\,i}(E)} = \abs{\gamma_{i}^{\pm} - 1} \cdot \frac{\sigma_{\text{ene}}}{E}.
\end{equation}

To evaluate the uncertainty in \cref{eq:relsyst-unc-on-fluxes-due-to-ene-scale} the spectral index $\gamma_{i}^{\pm}$ needs to be known. The initial model
to derive $\gamma_{i}^{\pm}$, described in the derivation of the $\sigma_{\text{time-ene, }e^{\pm},\,i}(E)$ uncertainty, does not hold at low energies, due to solar
modulation, which distorts the simple power law approximation. The force-field approximation~\cite{Gleeson1968} allows one to describe the low energy data very well,
by introducing another free parameter: the fisk potential $\phi_{i}$. This leads to the following analytical description of the fluxes:

\begin{equation}
  \label{eq:phi-model-with-solar-modulation}
  \Phi_{e^{\pm},\,i}(E) = A \cdot \left(\frac{E + \phi_{i}}{E_{\text{ref}}}\right)^{-\gamma_{i}^{\pm}} \cdot \frac{T \cdot (T + 2 \cdot m_{e^{\pm}})}{(T + \phi_{i}) \cdot ((T + \phi_{i}) + 2 \cdot m_{e^{\pm}})},
\end{equation}

where $m_{e^{\pm}}$ equals to the electron or positron mass, $T = E - m_{e^{\pm}}$ and $E_{\text{ref}} = \SI{5}{\GeV}$ (arbitrary choice).
After fitting the model - \cref{eq:phi-model-with-solar-modulation} - to the $e^{\pm}$ fluxes, the effective spectral index $\gamma_{i}^{\pm}(E)$
can be computed as function of energy by evaluating the spectral derivative

\begin{equation}
  \label{eq:phi-model-spectral-derivative}
  \gamma_{i}^{\pm}(E) = \diff(\log{\Phi_{e^{\pm},\,i}(E)}) / \diff(\log{E}).
\end{equation}

\begin{figure}[H]
  \begin{subfigure}{0.50\linewidth}
    \includegraphics[width=\linewidth]{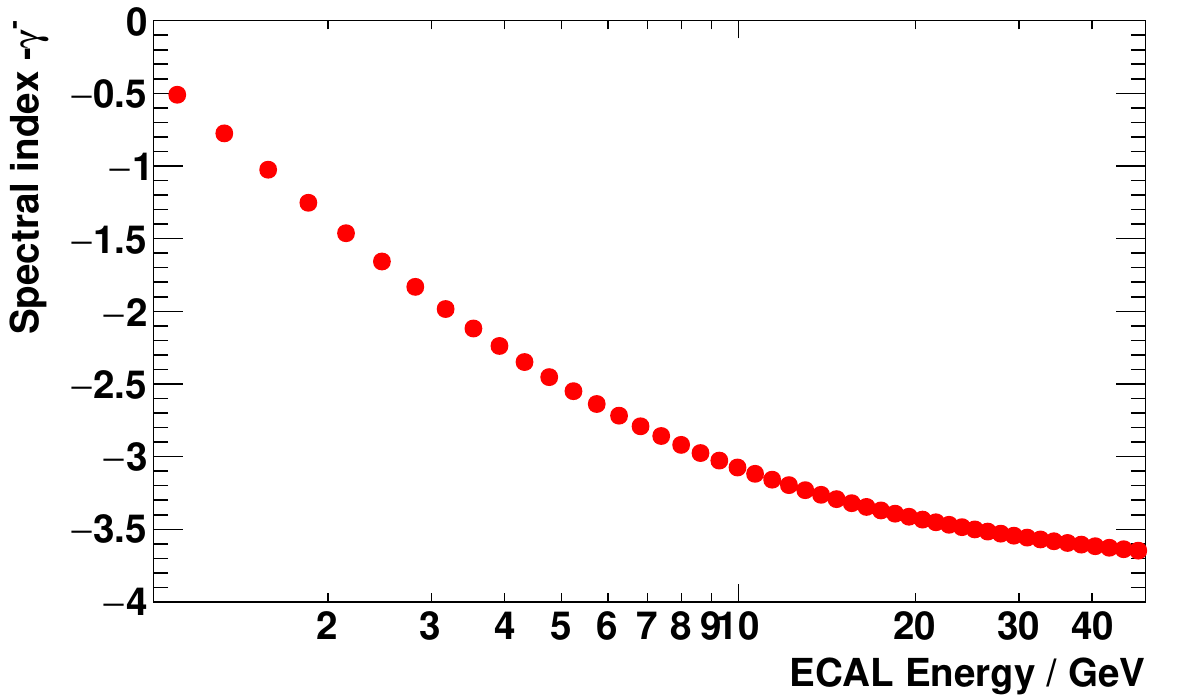}
    \caption{For electron flux}
  \end{subfigure}
  \hfill
  \begin{subfigure}{0.50\linewidth}
    \includegraphics[width=\linewidth]{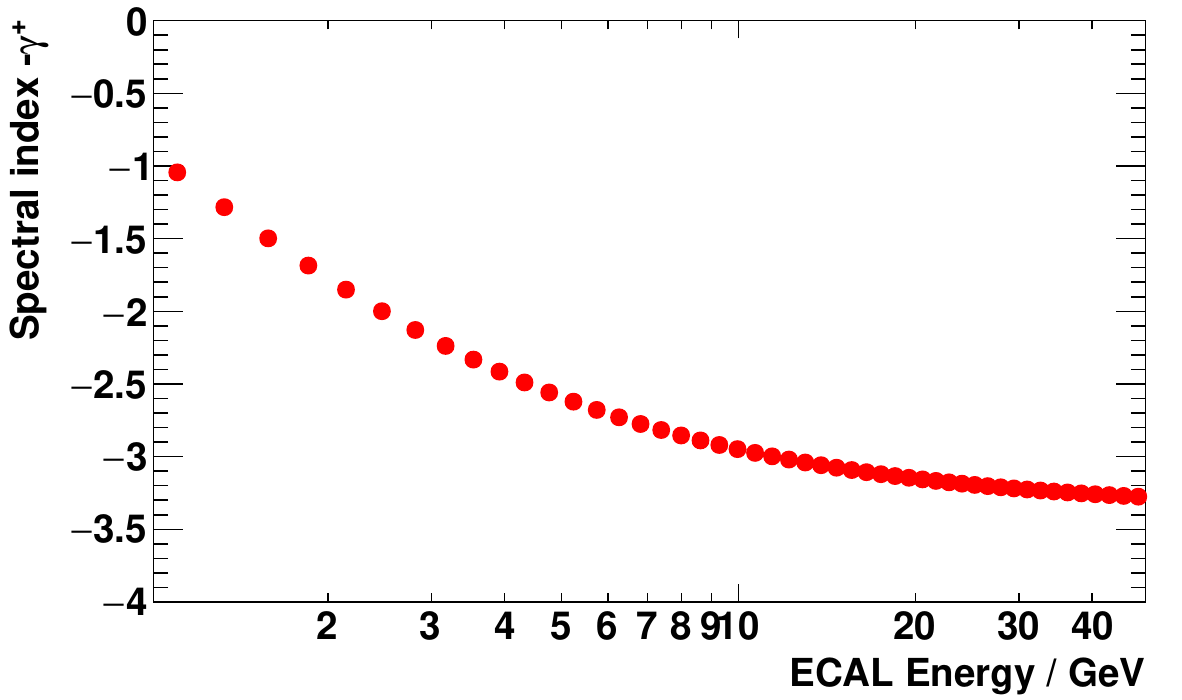}
    \caption{For positron flux}
  \end{subfigure}
  \caption{Spectral index of the electron flux (left plot) and the positron flux (right plot), determined via the computation of the spectral derivative, \cref{eq:phi-model-spectral-derivative}, using the flux model - \cref{eq:phi-model-with-solar-modulation} - for the time-dependent $e^{\pm}$ flux in Bartels rotation $i = 26$ (Apr~\nth{16},~2013 - May~\nth{13},~2013), after unfolding. Note that the y-axis shows $-\gamma_{i}^{\pm}$: by convention $\gamma_{i}^{\pm}$ is positive.}
  \label{fig:time-dependent-fluxes-spectral-index}
\end{figure}

\Cref{fig:time-dependent-fluxes-spectral-index} shows the spectral indices $\gamma_{i}^{\pm}(E)$ obtained by computing the spectral derivative
for the electron flux and the positron flux, respectively, as function of energy in an example Bartels rotation.

\begin{figure}[H]
  \begin{subfigure}{0.50\linewidth}
    \includegraphics[width=\linewidth]{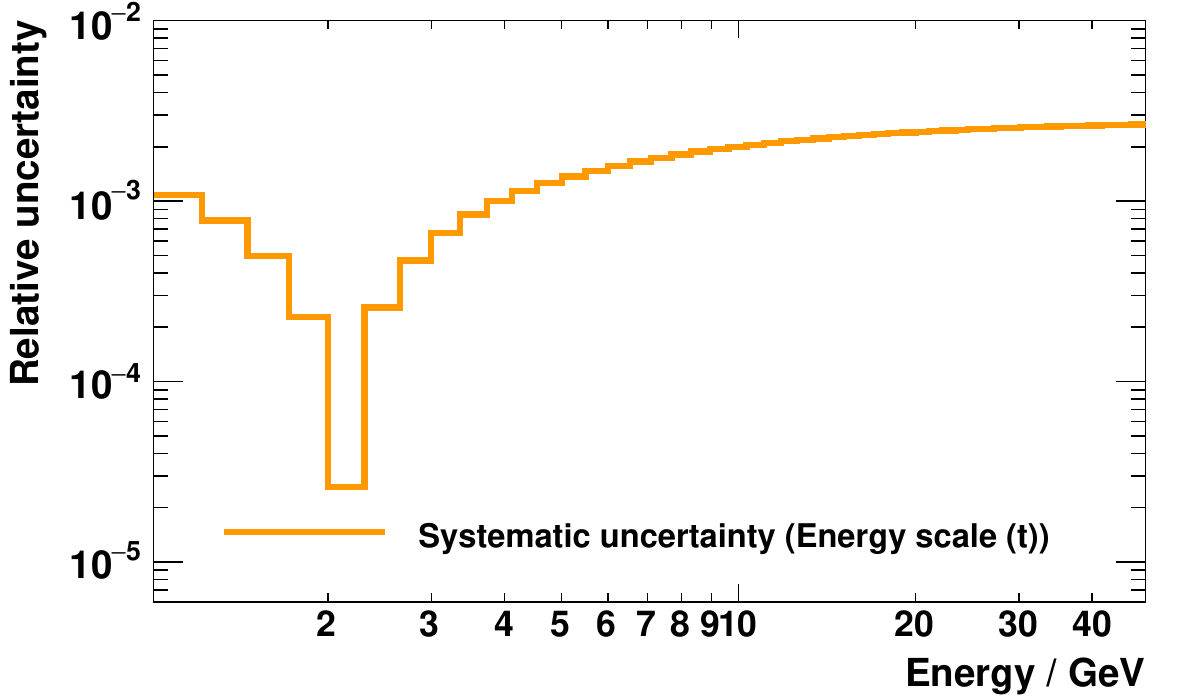}
    \caption{For electron flux}
  \end{subfigure}
  \hfill
  \begin{subfigure}{0.50\linewidth}
    \includegraphics[width=\linewidth]{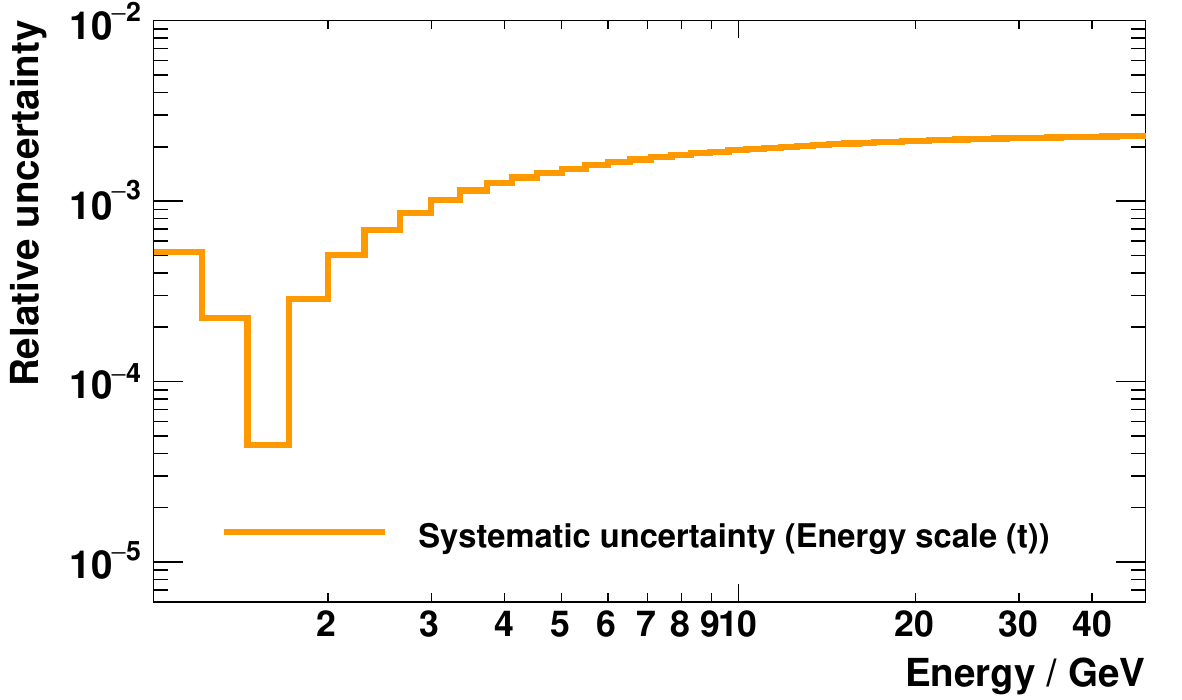}
    \caption{For positron flux}
  \end{subfigure}
  \caption{Visualization of the relative systematic uncertainty associated with the electron flux (left plot) and the positron flux (right plot) due to the time-stability of the energy scale, in an example Bartels rotation $i = 26$ (Apr~\nth{16},~2013 - May~\nth{13},~2013).}
  \label{fig:time-dependent-fluxes-syst-uncertainty-energy-scale}
\end{figure}

All ingredients are available to compute the relative systematic uncertainty $\sigma_{\text{time-ene, }e^{\pm},\,i}(E) / \Phi_{e^{\pm},\,i}(E)$.
\Cref{fig:time-dependent-fluxes-syst-uncertainty-energy-scale} shows the result as function of energy in Bartels rotation $i = 26$, as example.
The uncertainty associated with the electron flux is different from the uncertainty associated with the positron flux, due to different spectral indices, that enter
the systematic uncertainty calculation.

Between \SIrange{1}{2}{\GeV} both uncertainties show a minimum, where the spectral index $\gamma_{i}^{\pm}(E)~\approx~1$, which is consistent
with the expectation from \cref{eq:relsyst-unc-on-fluxes-due-to-ene-scale}.

\subsection{Summary}
\label{sec:analysis-flux-time-dependent-sysunc-summary}

\begin{figure}[H]
  \begin{subfigure}{\linewidth}
    \includegraphics[width=\linewidth]{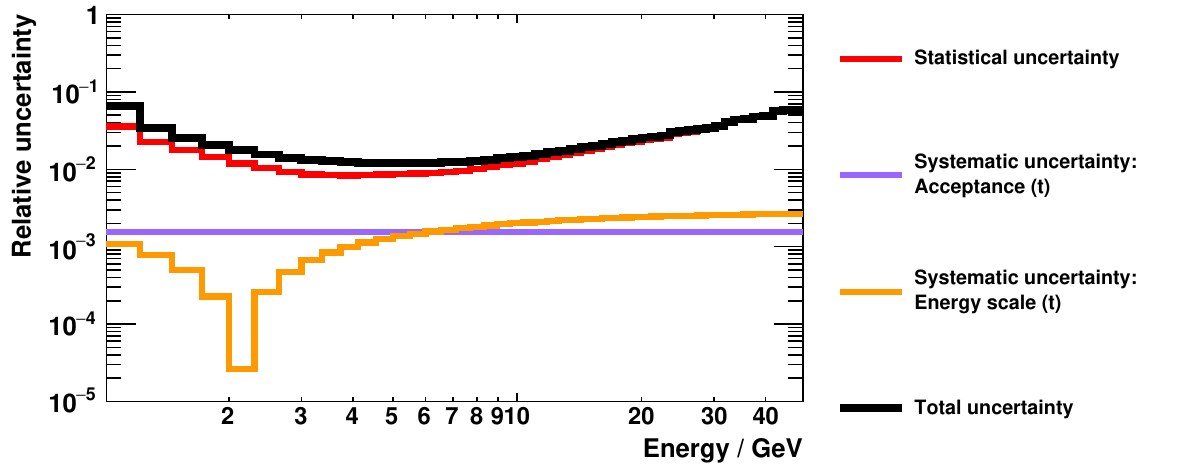}
    \caption{For electron flux}
  \end{subfigure}
  \hfill
  \begin{subfigure}{\linewidth}
    \includegraphics[width=\linewidth]{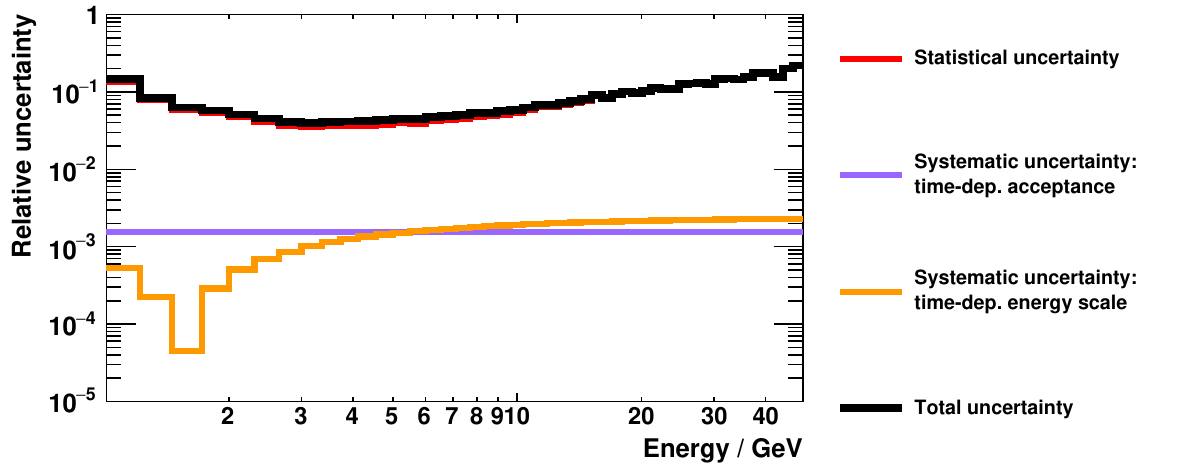}
    \caption{For positron flux}
  \end{subfigure}
  \caption{Composition of the relative systematic uncertainty associated with the electron flux (upper plot) and the positron flux (lower plot) in Bartels rotation $i = 26$ (Apr~\nth{16},~2013 - May~\nth{13},~2013), as example.}
  \label{fig:time-dependent-fluxes-syst-uncertainty-composition}
\end{figure}

\Cref{fig:time-dependent-fluxes-syst-uncertainty-composition} shows the composition of the relative systematic uncertainty $\sigma_{i}(E) / \Phi_{e^{\pm},\,i}(E)$ as function of energy
for the $e^{-}$ flux and the $e^{+}$ in Bartels rotation $i = 26$, as example.

The additional systematic uncertainties due to the time-variation of the acceptance and the energy scale are negligible
compared to the statistical uncertainty of both the electron flux and the positron flux, over all energies. The plot does not explicitly show the time-averaged systematic
uncertainty, which contributes to the total uncertainty, however it is obvious that the statistical uncertainty dominates the overall uncertainty.

\clearpage
\section{Time-dependent positron/electron ratio and positron fraction calculation}
\label{sec:analysis-ratios-time-dependent}

Following the recipe given in \cref{sec:analysis-ratios-time-averaged} for the time-averaged analysis the time-dependent ratios
for Bartels rotation $i$ are given by

\begin{equation}
  \label{eq:positron-electron-ratios-time-dependent-final}
  R_{e,\,i}(E) = \frac{N_{e^{+},\,i}(E)}{N_{e^{-},\,i}(E)}; \qquad p_{i}(E) = \frac{N_{e^{+},\,i}(E)}{N_{e^{+},\,i}(E) + N_{e^{-},\,i}(E)}.
\end{equation}

The results from the two-dimensional template fit for the \enquote{single-track sample} are used to extract the number of electrons or positrons $N_{e^{\pm},\,i}$
with the charge-confusion value fixed to the Monte-Carlo prediction, as explained in \cref{sec:analysis-ratios-time-averaged}.

\section{Time-dependent systematic uncertainties for positron/electron ratio and positron fraction}
\label{sec:analysis-ratios-time-dependent-sysunc}

In this section all systematic uncertainties that contribute to the time-dependent positron/electron ratio and positron fraction are summarized.
The systematic uncertainties of the time-dependent positron/electron ratio and positron fraction analysis are the same as for the time-averaged
analysis plus one extra component, that covers the energy scale variation with time.

\subsection{Time-dependent energy scale}
\label{sec:analysis-ratios-time-dependent-sysunc-energy-scale}

In \cref{sec:analysis-flux-time-dependent-sysunc-energy-scale} the energy scale stability was derived for the time-dependent flux analysis.
Following the same recipe the relative systematic uncertainty $\sigma_{\text{time-ene, }R_{e,\,i}}(E) / R_{e,\,i}(E)$ for the positron/electron ratio
and the relative systematic uncertainty $\sigma_{\text{time-ene, }p_{i}}(E) / p_{i}(E)$ can be derived for each Bartels rotation $i$.

\Cref{eq:relsyst-unc-on-fluxes-due-to-ene-scale} can be modified to yield the relative systematic uncertainty for the positron/electron ratio:

\begin{equation}
  \label{eq:relsyst-unc-on-elepos-due-to-ene-scale}
  \frac{\sigma_{\text{time-ene, }R_{e,\,i}}(E)}{R_{e,\,i}(E)} = \abs{\left(1 + \frac{\sigma_{\text{ene}}}{E}\right)^{\gamma_{i}^{-}(E)\ -\ \gamma_{i}^{+}(E)} - 1},
\end{equation}

and for the positron fraction:

\begin{equation}
  \label{eq:relsyst-unc-on-posfrac-due-to-ene-scale}
  \frac{\sigma_{\text{time-ene, }p_{i}(E)}}{p_{i}(E)} = \abs{\frac{1}{p_{i}(E) \cdot \left(1 + \frac{1}{R_{e,\,i}(E)} \cdot \left(1 + \frac{\sigma_{\text{ene}}}{E}\right)^{\gamma_{i}^{+}(E)\ -\  \gamma_{i}^{-}(E)}\right)} - 1}.
\end{equation}

\begin{figure}[H]
  \centering
  \includegraphics[width=0.8\linewidth]{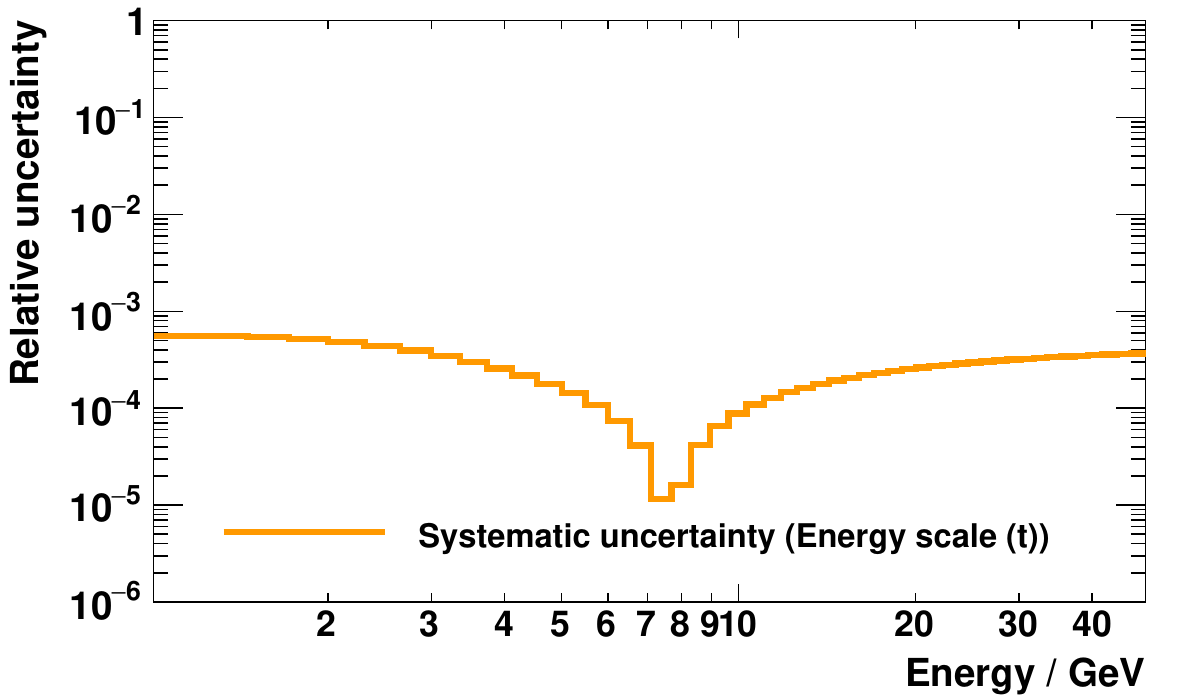}
  \caption{Visualization of the relative systematic uncertainty associated with the positron/electron ratio due to the time-stability of the energy scale, in an example Bartels rotation $i = 26$ (Apr~\nth{16},~2013 - May~\nth{13},~2013).}
  \label{fig:time-dependent-ratios-syst-uncertainty-energy-scale}
\end{figure}

\Cref{fig:time-dependent-ratios-syst-uncertainty-energy-scale} shows the result as function of energy in Bartels rotation $i = 26$, as example for the positron/electron ratio.
The uncertainty is almost identical for the positron fraction, thus the plot for the positron fraction was omitted.

Around \SIapprox{7}{\GeV} the uncertainty show a minimum, where the spectral indices of electrons and positrons $\gamma_{i}^{+}(E)~\approx~\gamma_{i}^{-}(E)$
are equal, which is consistent with the expectation from \cref{eq:relsyst-unc-on-elepos-due-to-ene-scale,eq:relsyst-unc-on-posfrac-due-to-ene-scale}.

\subsection{Summary}
\label{sec:analysis-ratios-time-dependent-sysunc-summary}

\begin{figure}[H]
  \includegraphics[width=\linewidth]{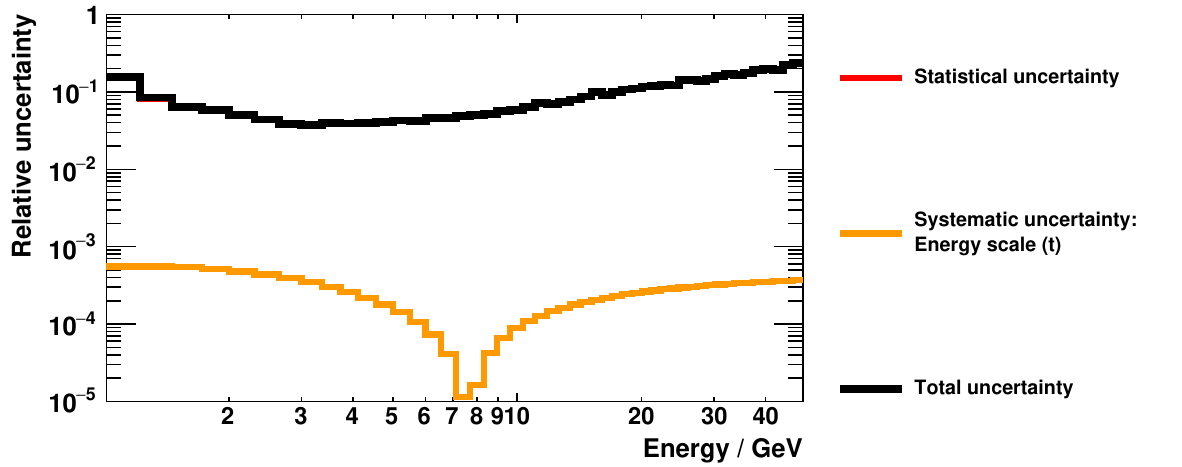}
  \caption{Composition of the relative systematic uncertainty associated with the positron/electron ratio.}
  \label{fig:time-dependent-ratios-syst-uncertainty-composition-poselectratio}
\end{figure}

The composition of the relative systematic uncertainty $\sigma_{i}(E) / R_{e,\,i}(E)$ as function of energy for the positron/electron ratio
is shown in \cref{fig:time-dependent-ratios-syst-uncertainty-composition-poselectratio}.

The composition of the relative systematic uncertainty $\sigma_{i}(E) / p_{i}(E)$ as function of energy for the positron fraction is almost identical
and thus was omitted.

The additional systematic uncertainties due to the time-variation of the energy scale is negligible compared to the statistical uncertainty of both the
positron/electron ratio and the positron fraction, over all energies. The plot does not explicitly show the time-averaged systematic
uncertainty, which contributes to the total uncertainty, however it is obvious that the statistical uncertainty dominates the overall uncertainty.

\chapter{Results}
\label{sec:results}

In this chapter the results of the time-averaged and time-dependent analyses are presented.
Comparisons against published results and other experiments are shown as well as interpretations of the results.

\section{Time-averaged results}
\label{sec:results-time-averaged}

\subsection{Electron flux}
\label{sec:results-time-averaged-electron-flux}

\begin{figure}[H]
  \centering
  \includegraphics[width=0.9\linewidth]{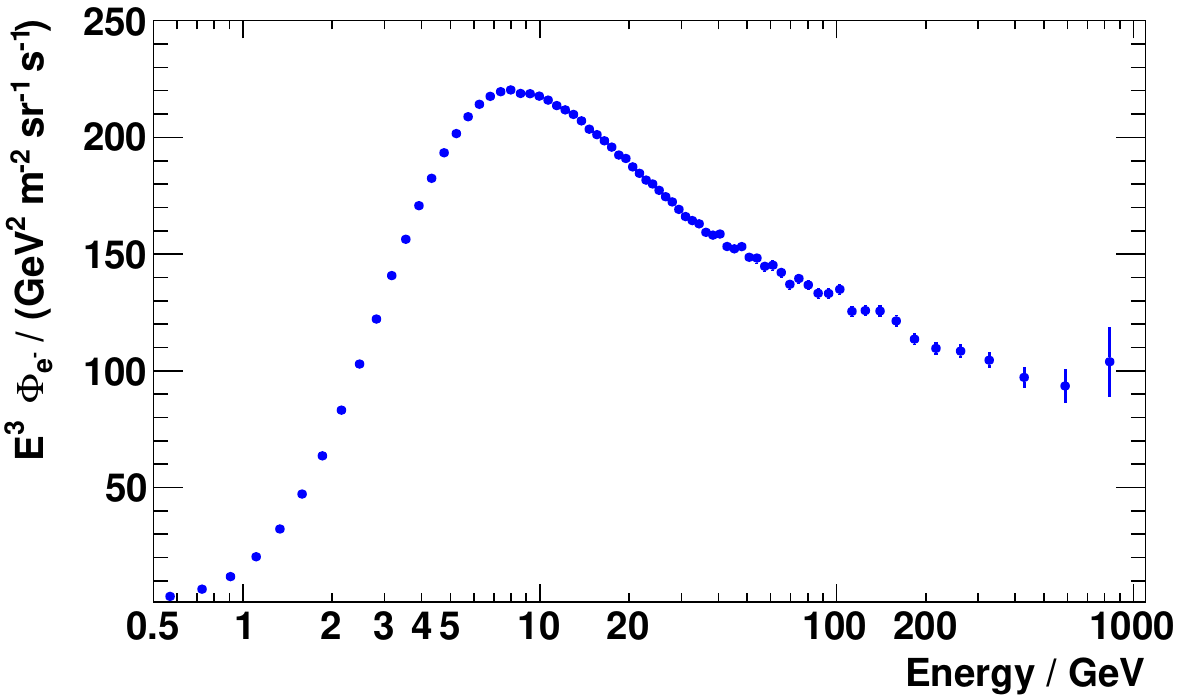}
  \caption{The electron flux measured by AMS-02 using the analysis procedure described in this work. The y-axis shows the unfolded flux, multiplied by energy $E^{3}$ for illustration purposes.}
  \label{fig:results-time-averaged-electron-flux}
\end{figure}

\Cref{fig:results-time-averaged-electron-flux} shows the time-averaged electron flux based on \textbf{28.39 million} events in the energy range \SIrange{0.5}{1000}{\GeV}, multiplied by energy $\tilde{E}^3$, where $\tilde{E}$ equals to the spectrally weighted
mean energy in each energy interval, according to Lafferty \& Wyatt~\cite{Lafferty1995}. In the following $\tilde{E}$ will be referred to as $E$ - see \cref{sec:appendix-results-energy-scale}
for a discussion of the energy scale.

\begin{figure}[H]
  \centering
  \includegraphics[width=0.85\linewidth]{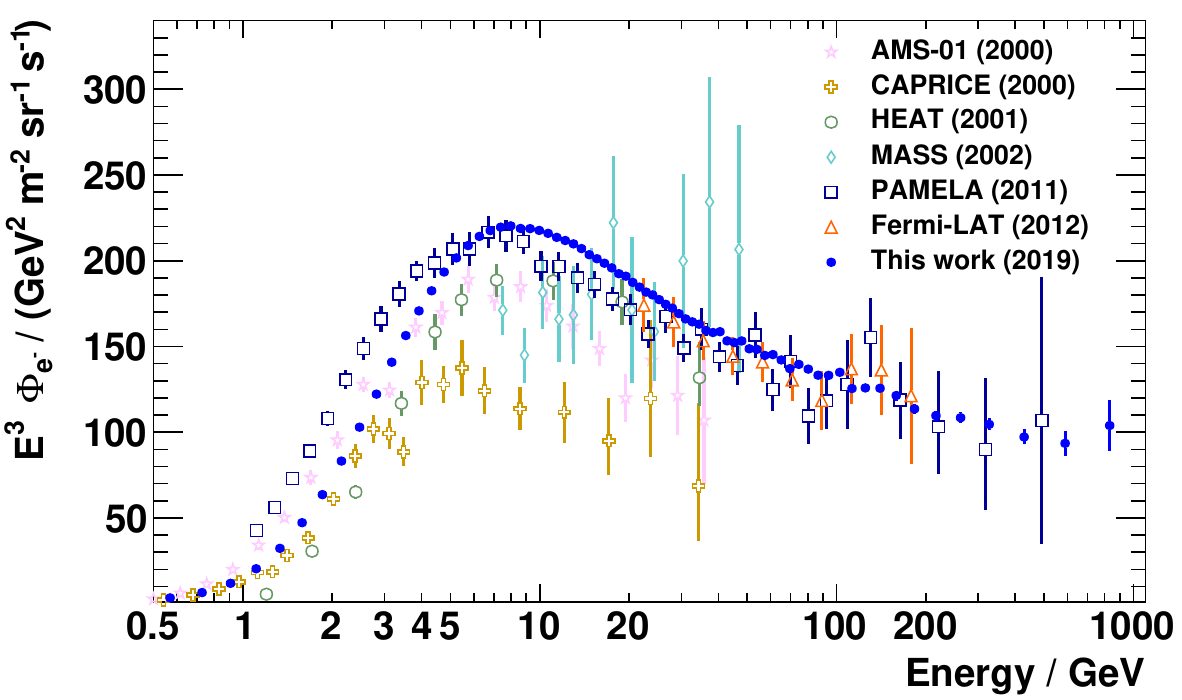}
  \caption{Comparison of the electron flux derived in this work with other experiments: AMS-01~\cite{Alcaraz2000}, CAPRICE~\cite{Boezio2000}, HEAT~\cite{DuVernois2001}, MASS~\cite{Grimani2002}, PAMELA~\cite{Adriani2011} and Fermi-LAT~\cite{Ackermann2012}.}
  \label{fig:results-time-averaged-electron-flux-aachen-others}
\end{figure}

\Cref{fig:results-time-averaged-electron-flux-aachen-others} shows a comparison of the electron flux derived in this work with previous experiments.
The electron flux was measured by AMS-02 with unprecedented accuracy, up to the \SI{}{\TeV} regime.

\begin{figure}[H]
  \centering
  \includegraphics[width=0.75\linewidth]{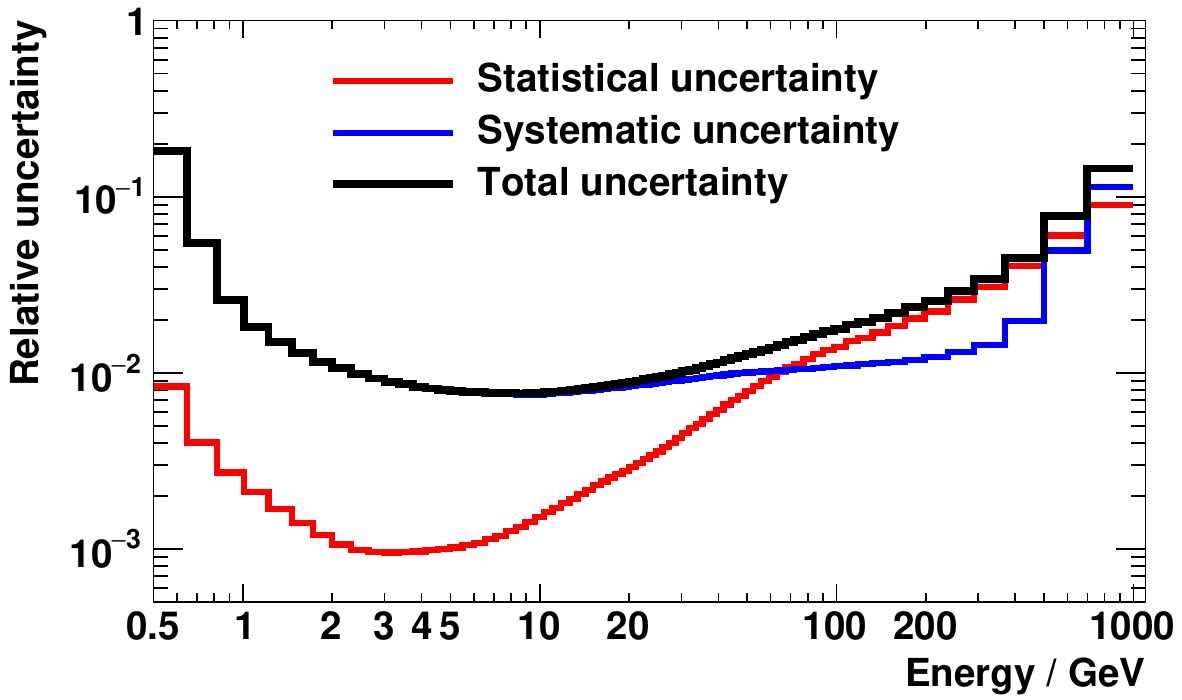}
  \caption{Breakdown of the uncertainty of the electron flux into a statistical and systematical part.}
  \label{fig:results-time-averaged-electron-flux-error-breakdown}
\end{figure}

\Cref{fig:results-time-averaged-electron-flux-error-breakdown} shows a breakdown of the total uncertainty into the statistical and systematic part.
Above \SIapprox{70}{\GeV} the statistical uncertainty dominates the uncertainty of the $e^{-}$ flux measurement. Below this energy, the systematic uncertainty always
exceeds the statistical uncertainty. See \cref{sec:analysis-flux-time-averaged-sysunc-summary} for a decomposition of the systematic uncertainties.

The electron flux derived in this work is compatible with the published electron flux~\cite{Aguilar2019a} by the AMS-02 collaboration, as shown in \cref{sec:appendix-results-comparison-published-fluxes}.

\subsubsection{Spectral analysis}
\label{sec:results-time-averaged-electron-flux-spectral-analysis}

\begin{figure}[H]
  \includegraphics[width=\linewidth]{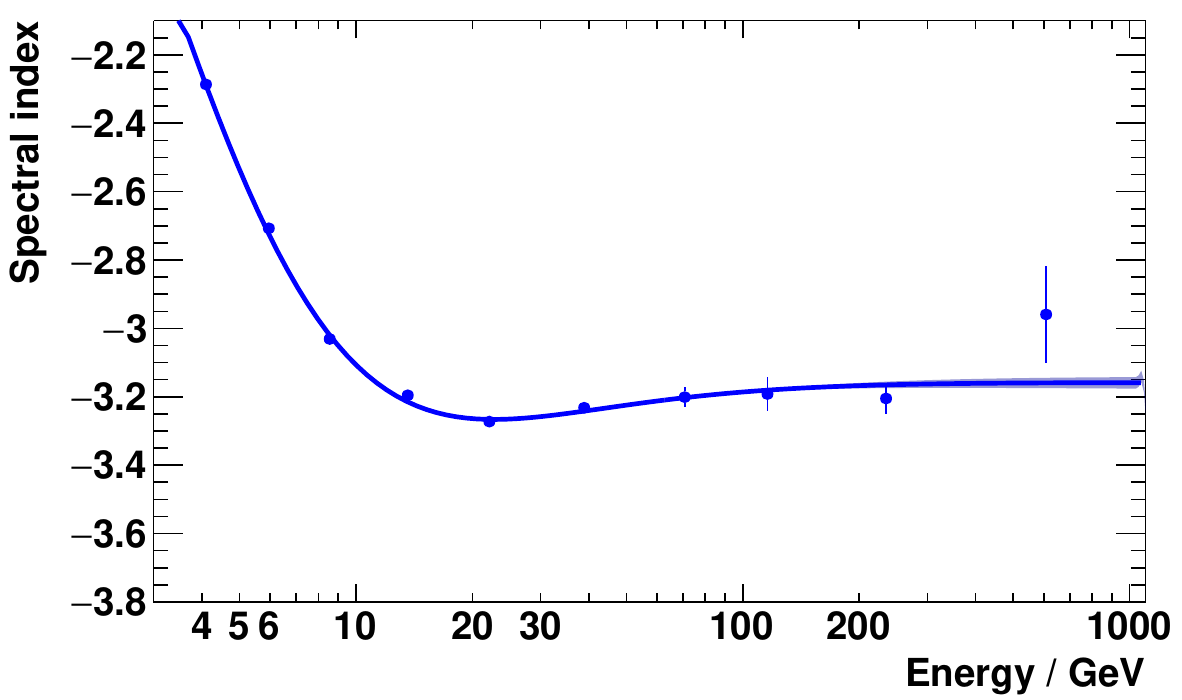}
  \caption{The spectral index of the electron flux as function of the energy in non-overlapping energy intervals shown as blue points. The blue band corresponds to the uncertainty determined from a fit of \cref{eq:results-time-averaged-electron-model} to the data.}
  \label{fig:results-time-averaged-electron-spectral-index}
\end{figure}

To examine the energy dependence of the electron flux in a model-independent way, the flux spectral index $\gamma^{-}(E)$
is calculated by evaluating $\gamma^{-}(E) = \diff(\log{\Phi_{e^{-}}(E)}) / \diff(\log{E})$ over non-overlapping energy intervals.
Following the prescription in Ref.~\cite{Aguilar2019a} the energy interval boundaries are 3.36, 5.00, 7.10, 10.32, 17.98, 27.25, 55.58, 90.19, 148.81, 370,
and \SI{1000}{\GeV}, to ensure sufficient sensitivity to the spectral index in each of the energy intervals. The results are shown in
\cref{fig:results-time-averaged-electron-spectral-index}.

The spectral index softens from $\gamma^{-}(E)~\approx~\num{-2.2}{}$ at \SI{3}{\GeV} to $\gamma^{-}(E)~\approx~\num{-3.2}{}$ above \SI{20}{\GeV}.

To determine the transition energy $E_{0}$ where the change of the electron spectral index occurs, a double power law approximation is fit to the
data\footnote{The AMS-02 electron flux publication~\cite{Aguilar2019a} uses \SI{20.04}{\GeV} as normalization constant in \cref{eq:results-time-averaged-electron-transition-fit-model}
and as beginning of the fit range for the transition fit. This work starts the fit procedure one bin later, as it decreases the uncertainty on the transition energy
and leads to smaller correlation between the $\gamma$ and $E_{0}$ parameters.} in the range \SIrange{21.13}{1000}{\GeV}:

\begin{equation}
  \label{eq:results-time-averaged-electron-transition-fit-model}
  \Phi_{e^{-}}(E) =
    \begin{cases}
       C (E / \SI{21.13}{\GeV})^{\gamma}                            & E \leq E_{0} \\
       C (E / \SI{21.13}{\GeV})^{\gamma} (E / E_{0})^{\Delta\gamma} & E > E_{0}.
    \end{cases}
\end{equation}

\begin{figure}[H]
  \centering
  \includegraphics[width=0.95\linewidth]{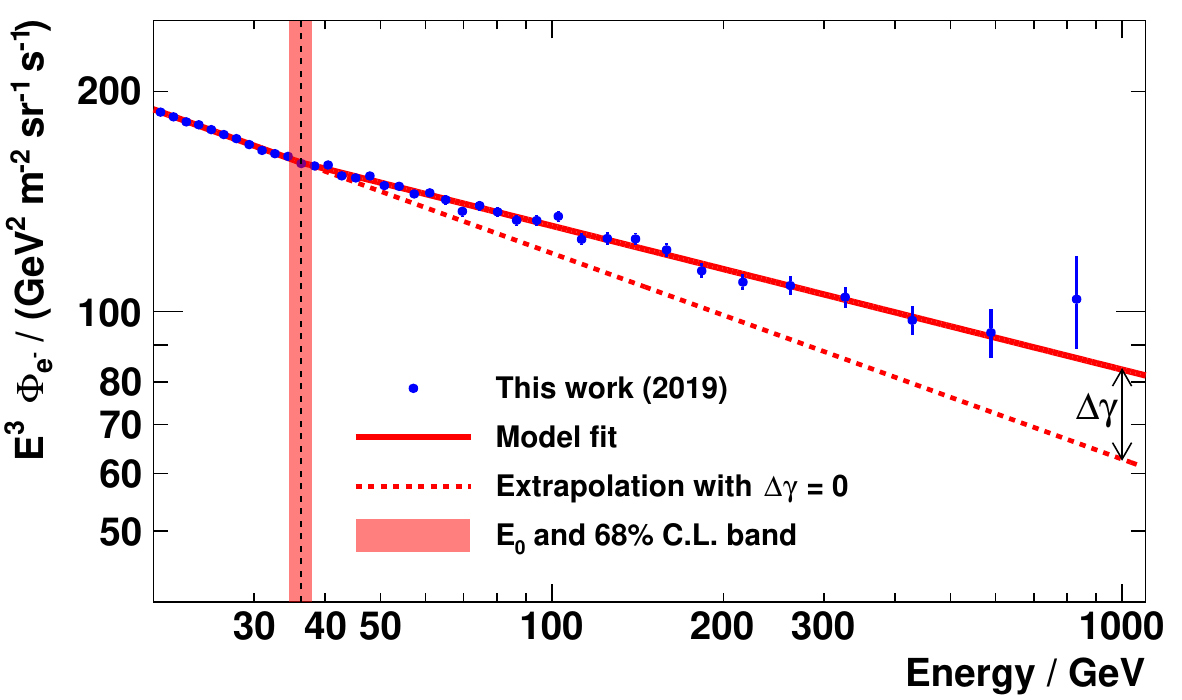}
  \caption{Fit of the double power law approximation \cref{eq:results-time-averaged-electron-transition-fit-model} to the electron flux in the range \SIrange{21.13}{1000}{\GeV}. The electron flux is shown as blue data points, scaled by $E^3$. The red line corresponds to the fit result of \cref{eq:results-time-averaged-electron-transition-fit-model}. The vertical dashed line indicates the transition energy $E_{0}$ and the red band corresponds to the uncertainty. The red dashed line corresponds to an extrapolation of the model, with $\Delta\gamma = 0$.}
  \label{fig:results-time-averaged-electron-flux-transition-fit}
\end{figure}

The fit yields $C = \SI[parse-numbers=false]{(1.973 \pm 0.006) \times 10^{-2}}{[\meter\squared~\steradian~\second~\GeV]^{-1}}$,
$\gamma = \SI[parse-numbers=false]{-3.28 \pm 0.01}{}$, $E_{0} = \SI[parse-numbers=false]{36.26 \pm 1.68}{\GeV}$
for the transition energy and $\Delta\gamma = \SI[parse-numbers=false]{0.085 \pm 0.012}{}$ for the difference in the
spectral index after the transition energy $E_{0}$ with respect to the region before the transition with $\chi^2/\text{dof} = 34.7 / 34 = 1.02$.
The results of the double power law fit are shown in \cref{fig:results-time-averaged-electron-flux-transition-fit}.

\subsubsection{Model fits}
\label{sec:results-time-averaged-electron-flux-model-fits}

As shown in Ref.~\cite{Aguilar2019a}, the whole energy range can be described by the sum of two power law components $a$ and $b$:

\begin{equation}
  \label{eq:results-time-averaged-electron-model}
  \Phi_{e^{-}}(E) = \frac{E^2}{\hat{E}^2} [1 + (\hat{E} / E_{t})^{\Delta\gamma_{t}}]^{-1} [ C_{a} (\hat{E} / E_{a})^{\gamma_{a}} + C_{b} (\hat{E} / E_{b})^{\gamma_{b}}].
\end{equation}

In order to account for the effects related to the complex spectral behavior of the electron flux at energies
below \SI{10}{\GeV}~\cite{Strong2011}, an additional transition term, $[1 + (\hat{E} / E_{t})^{\Delta\gamma_{t}}]^{-1}$, is introduced.
It is characterized by a transition energy $E_{t}$ and a spectral index $\Delta\gamma_{t}$. It has vanishing impact on the flux behavior at energies above
$E_{t}$ (e.g., \SIvarOp{}{<}{0.7}{\percent} above \SI{40}{\GeV}).

The two components, $a$ and $b$, correspond to two power law functions with corresponding normalization factors $C_{a}$ and $C_{b}$,
and spectral indices $\gamma_{a}$ and $\gamma_{b}$. To account for solar modulation effects, the force-field approximation~\cite{Gleeson1968} is used,
with the energy of particles in the interstellar space $\hat{E} = E + \phi_{e^{-}}$ and the effective fisk potential $\phi_{e^{-}}$. The constants
$E_{a} = \SI{20}{\GeV}$ and $E_{b} = \SI{300}{\GeV}$ are chosen to minimize the correlation between $C_{a}$, $C_{b}$ and $\gamma_{a}$, $\gamma_{b}$.

\begin{figure}[H]
  \includegraphics[width=\linewidth]{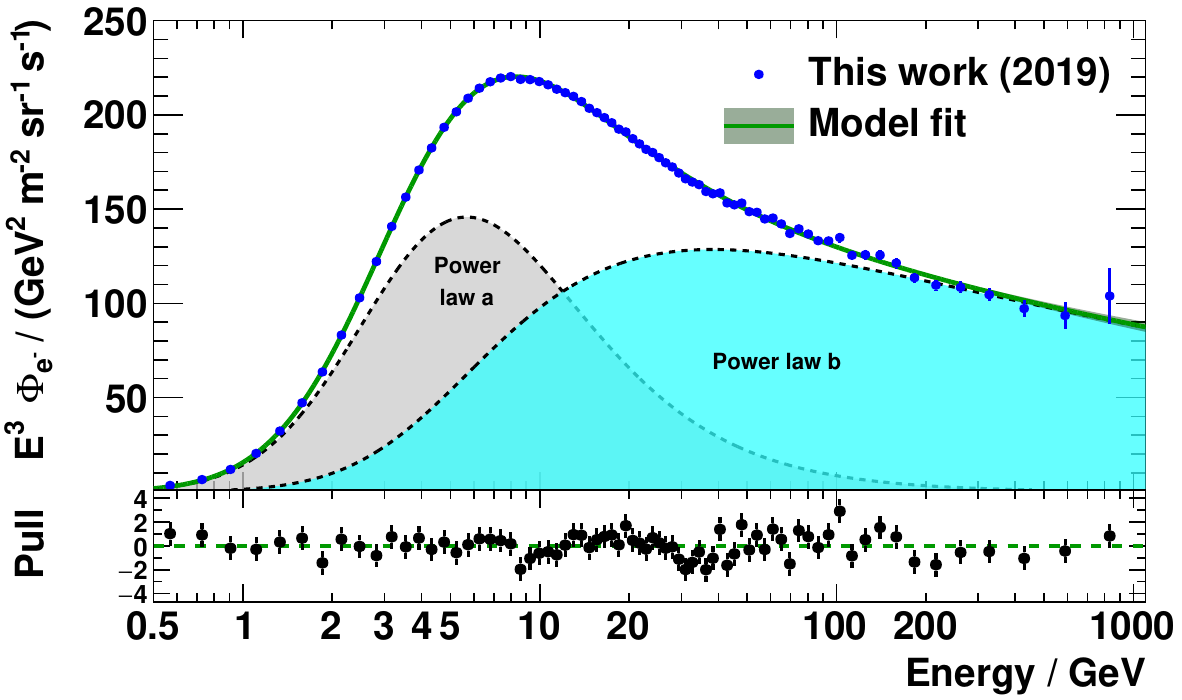}
  \caption{The electron flux, scaled with $E^3$, is shown as blue points. The result of the fit of \cref{eq:results-time-averaged-electron-model} is shown as green line, enclosed by a green band, indicating the uncertainty of the fit result. The two power law components $a$ and $b$ are indicated by the gray and turquoise areas, respectively. The lower plot shows the pull distribution of the fit (data minus model divided by data uncertainty) -- the model describes the data accurately over the whole energy range.}
  \label{fig:results-time-averaged-electron-flux-fit}
\end{figure}

A fit to the data yields, $\phi_{e^{-}} = \SI[parse-numbers=false]{1.08 \pm 0.13}{\GeV}$ for the effective
potential, $E_{t} = \SI[parse-numbers=false]{3.90 \pm 0.18}{\GeV}$ and $\Delta\gamma_{t} = \SI[parse-numbers=false]{-2.04 \pm 0.11}{}$
for the parameters of the transition term, $C_{a} = \SI[parse-numbers=false]{(1.19 \pm 0.08) \times 10^{-2}}{[\meter\squared~\steradian~\second~\GeV]^{-1}}$
and $\gamma_{a} = \SI[parse-numbers=false]{-4.46 \pm 0.12}{}$ for the power law $a$, and $C_{b} = \SI[parse-numbers=false]{(3.99 \pm 0.04) \times 10^{-6}}{[\meter\squared~\steradian~\second~\GeV]^{-1}}$,
$\gamma_{b} = \SI[parse-numbers=false]{-3.16 \pm 0.02}{}$ for the power law $b$ with $\chi^2/\text{dof} = 70.0 / 67 = 1.04$. The result of the fit
is presented in \cref{fig:results-time-averaged-electron-flux-fit}.

The origin of the low energy break at $E_{t} = \SI{3.90}{\GeV}$ is not clear. The break might already be present
in the local interstellar spectrum, or it could be a solar modulation effect. Therefore it is useful to study the low energy data in more detail,
to search for low energy breaks as function of time - which is done in the time-dependant analysis, described in \cref{sec:results-time-dependent}.

\bigskip
One can conclude that in the energy range \SIrange{0.5}{1000}{\GeV} the sum of two power law functions
with an additional transition term provides an excellent description of the data. This is consistent with the assumption
that only a few astrophysical sources of high energy electrons in the vicinity of the Solar System are present, each making a power law
contribution to the electron flux.

\clearpage
\subsection{Positron flux}
\label{sec:results-time-averaged-positron-flux}

\begin{figure}[H]
  \centering
  \includegraphics[width=0.85\linewidth]{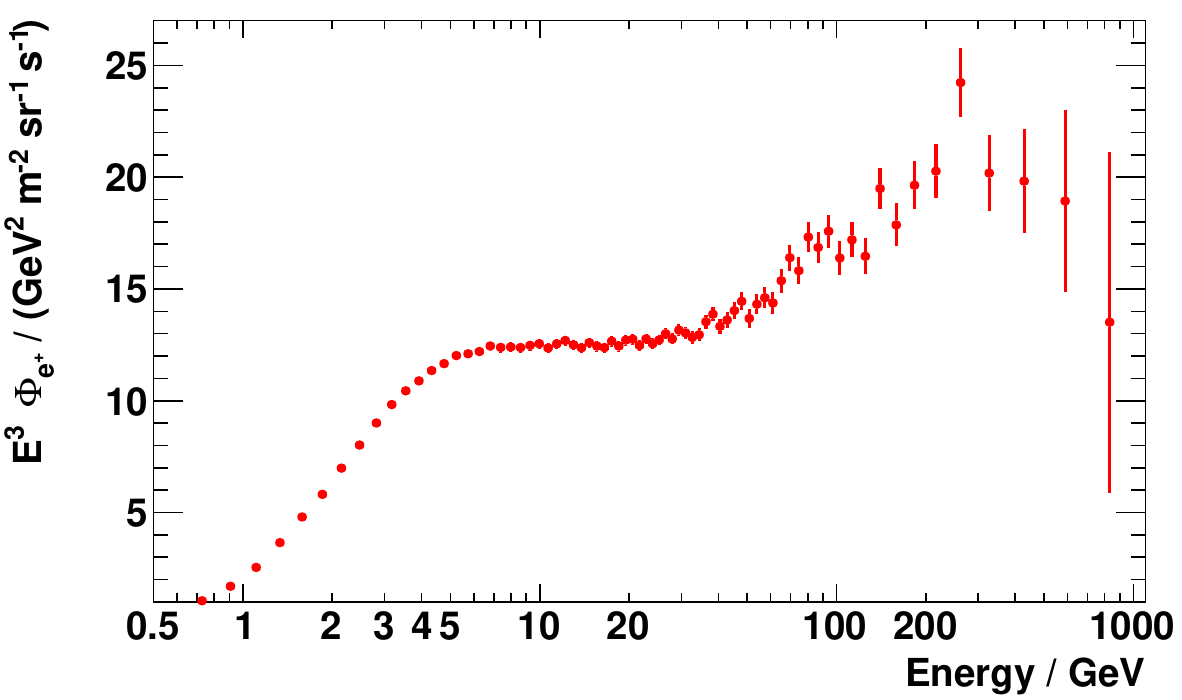}
  \caption{The positron flux measured by AMS-02 using the analysis procedure described in this work. The y-axis shows the unfolded flux, multiplied by energy $E^{3}$ for illustration purposes.}
  \label{fig:results-time-averaged-positron-flux}
\end{figure}

\Cref{fig:results-time-averaged-positron-flux} shows the time-averaged positron flux, based on \textbf{1.95 million} events in the energy range \SIrange{0.5}{1000}{\GeV}, multiplied by energy $E^{3}$.

\begin{figure}[H]
  \centering
  \includegraphics[width=0.75\linewidth]{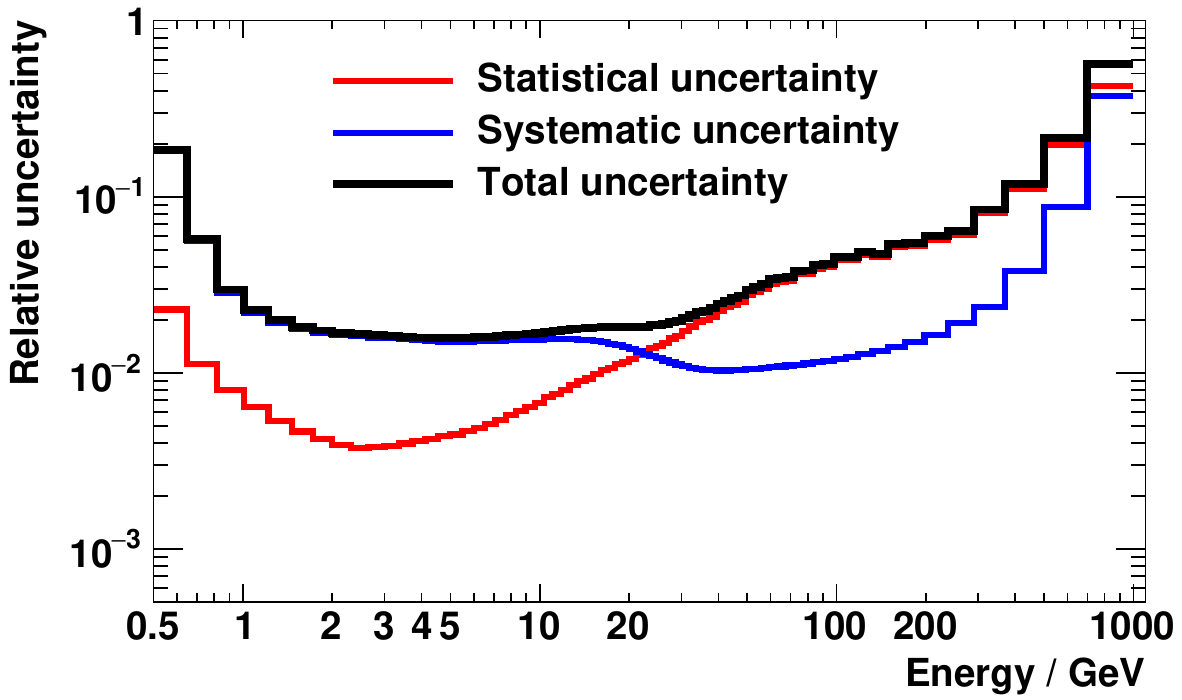}
  \caption{Breakdown of the uncertainty of the positron flux into a statistical and systematical part.}
  \label{fig:results-time-averaged-positron-flux-error-breakdown}
\end{figure}

\Cref{fig:results-time-averaged-positron-flux-error-breakdown} shows a breakdown of the total uncertainty into the statistical and systematic part.
Above \SIapprox{20}{\GeV} the statistical uncertainty dominates the uncertainty of the $e^{+}$ flux measurement. Below this energy, the systematic uncertainty always
exceeds the statistical uncertainty. See \cref{sec:analysis-flux-time-averaged-sysunc-summary} for a decomposition of the systematic uncertainties.

\begin{figure}[H]
  \includegraphics[width=\linewidth]{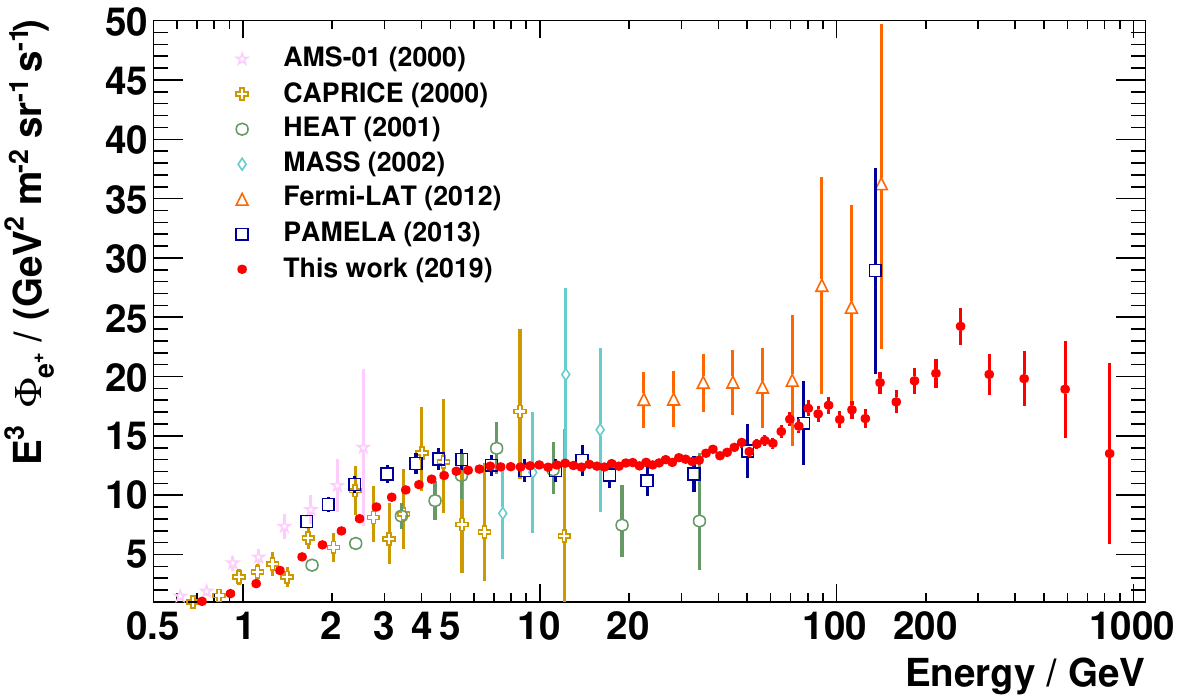}
  \caption{Comparison of the positron flux derived in this work with other experiments: AMS-01~\cite{Alcaraz2000}, CAPRICE~\cite{Boezio2000}, HEAT~\cite{DuVernois2001}, MASS~\cite{Grimani2002}, Fermi-LAT~\cite{Ackermann2012} and PAMELA~\cite{Adriani2013}.}
  \label{fig:results-time-averaged-positron-flux-aachen-others}
\end{figure}

\Cref{fig:results-time-averaged-positron-flux-aachen-others} shows a comparison of the positron flux derived in this work with previous experiments.
The positron flux was measured by AMS-02 with unprecedented accuracy, up to the \SI{}{\TeV} regime. The flux derived in this work is compatible with
the published positron flux~\cite{Aguilar2019b} by the AMS-02 collaboration, as shown in \cref{sec:appendix-results-comparison-published-fluxes}.

\subsubsection{Spectral analysis}
\label{sec:results-time-averaged-positron-flux-spectral-analysis}

\begin{figure}[H]
  \centering
  \includegraphics[width=0.8\linewidth]{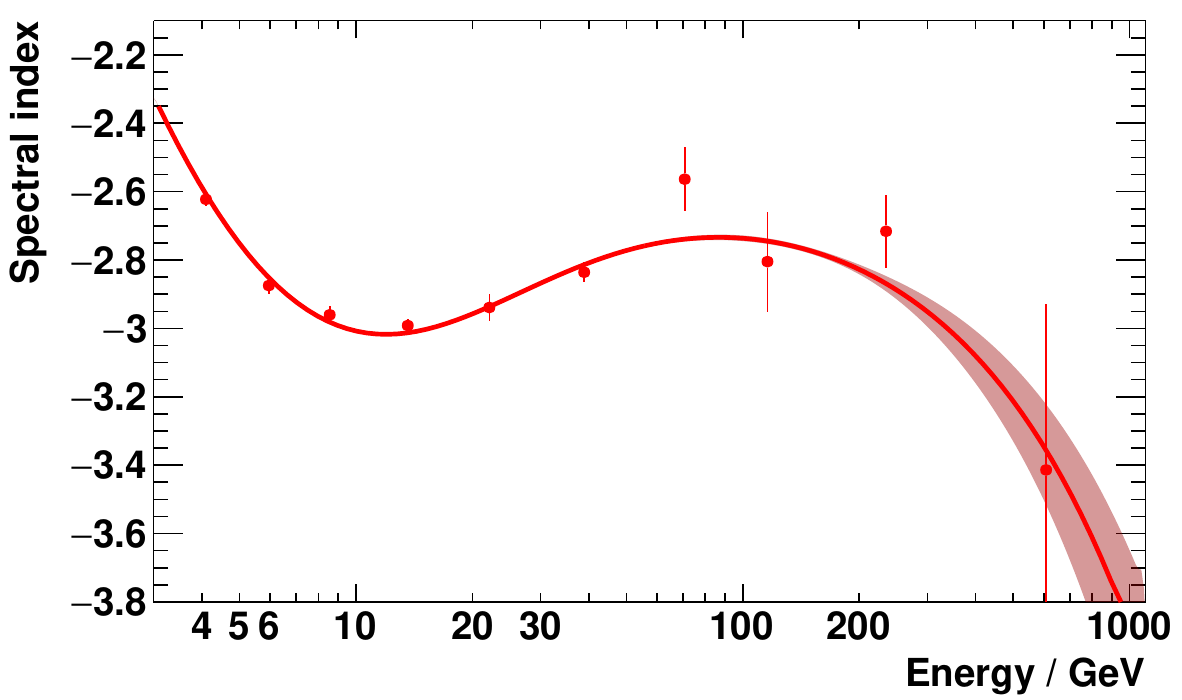}
  \caption{The spectral index of the positron flux in non-overlapping energy intervals shown as red points. The red band corresponds to the uncertainty determined from a fit of \cref{eq:results-time-averaged-positron-model} to the data.}
  \label{fig:results-time-averaged-positron-spectral-index}
\end{figure}

To examine the energy dependence of the positron flux in a model-independent way, the flux spectral index $\gamma^{+}(E)$
is calculated by evaluating $\gamma^{+}(E) = \diff(\log{\Phi_{e^{+}}(E)}) / \diff(\log{E})$, as described in \cref{sec:results-time-averaged-electron-flux}.
The results are shown in \cref{fig:results-time-averaged-positron-spectral-index}.

The spectral index softens from $\gamma^{+}(E)~\approx~\num{-2.3}{}$ at \SI{3}{\GeV} to $\gamma^{+}(E)~\approx~\num{-3.0}{}$ at \SI{10}{\GeV}.
Above that energy it gradually hardens until it reaches $\gamma^{+}(E)~\approx~\num{-2.8}{}$ at \SIapprox{80}{\GeV}. While approaching \SI{}{\TeV} energies it gradually softens again.

To determine the transition energy $E_{0}$ where the positron spectral index starts rising, a double power law approximation is fit to the
data in the range \SIrange{7.10}{55.58}{\GeV}:

\begin{equation}
  \label{eq:results-time-averaged-positron-transition-fit-model}
  \Phi_{e^{-}}(E) =
    \begin{cases}
       C (E / \SI{55.58}{\GeV})^{\gamma}                            & E \leq E_{0} \\
       C (E / \SI{55.58}{\GeV})^{\gamma} (E / E_{0})^{\Delta\gamma} & E > E_{0}.
    \end{cases}
\end{equation}

\begin{figure}[H]
  \centering
  \includegraphics[width=0.8\linewidth]{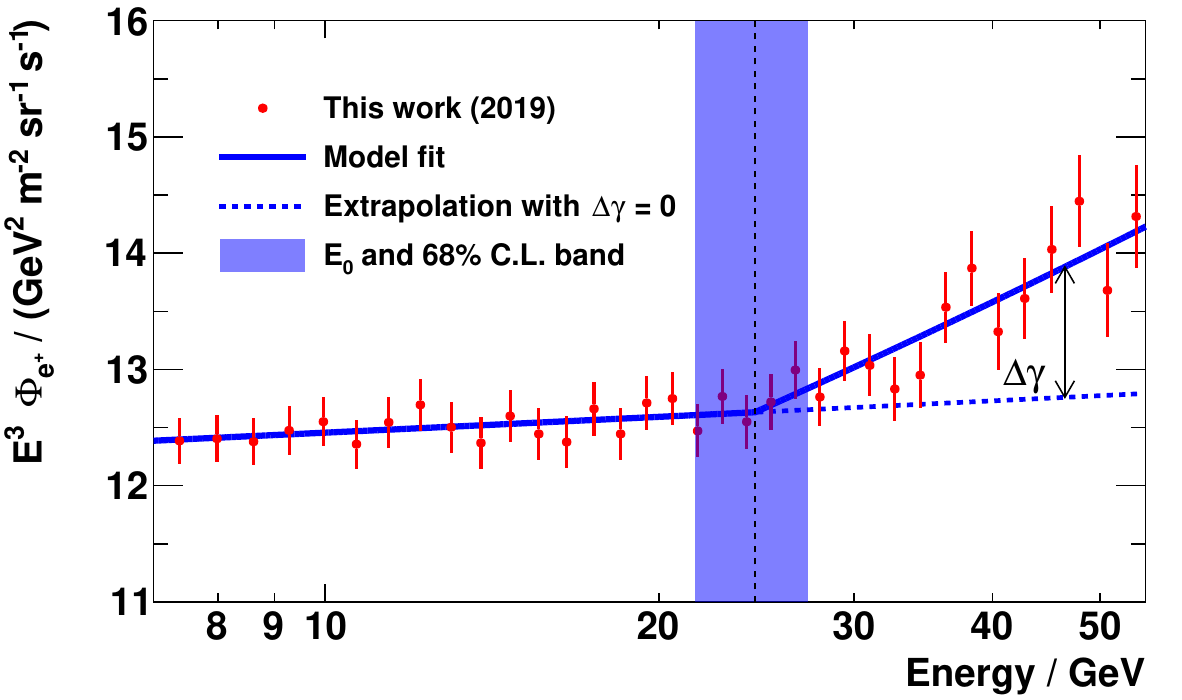}
  \caption{Fit of the double power law approximation \cref{eq:results-time-averaged-positron-transition-fit-model} to the positron flux in the range \SIrange{7.10}{55.58}{\GeV}. The positron flux is shown as red data points, scaled by $E^3$. The blue line corresponds to the fit result of \cref{eq:results-time-averaged-positron-transition-fit-model}. The vertical dashed line indicates the transition energy $E_{0}$, where the positron flux starts rising, and the blue band corresponds to the uncertainty. The blue dashed line corresponds to an extrapolation of the model, with $\Delta\gamma = 0$.}
  \label{fig:results-time-averaged-positron-flux-low-energy-transition-fit}
\end{figure}

The fit yields, $C = \SI[parse-numbers=false]{(7.45 \pm 0.11) \times 10^{-5}}{[\meter\squared~\steradian~\second~\GeV]^{-1}}$,
$\gamma = \SI[parse-numbers=false]{-2.98 \pm 0.01}{}$, $E_{0} = \SI[parse-numbers=false]{24.43 \pm 2.87}{\GeV}$ for the transition
energy and $\Delta\gamma = \SI[parse-numbers=false]{0.13 \pm 0.03}{}$ for the difference in the spectral index after the transition
energy $E_{0}$ with respect to the region before the transition. The results of the double power law fit are compatible with the
published results~\cite{Aguilar2019b} and shown in \cref{fig:results-time-averaged-positron-flux-low-energy-transition-fit}.

\begin{figure}[H]
  \includegraphics[width=\linewidth]{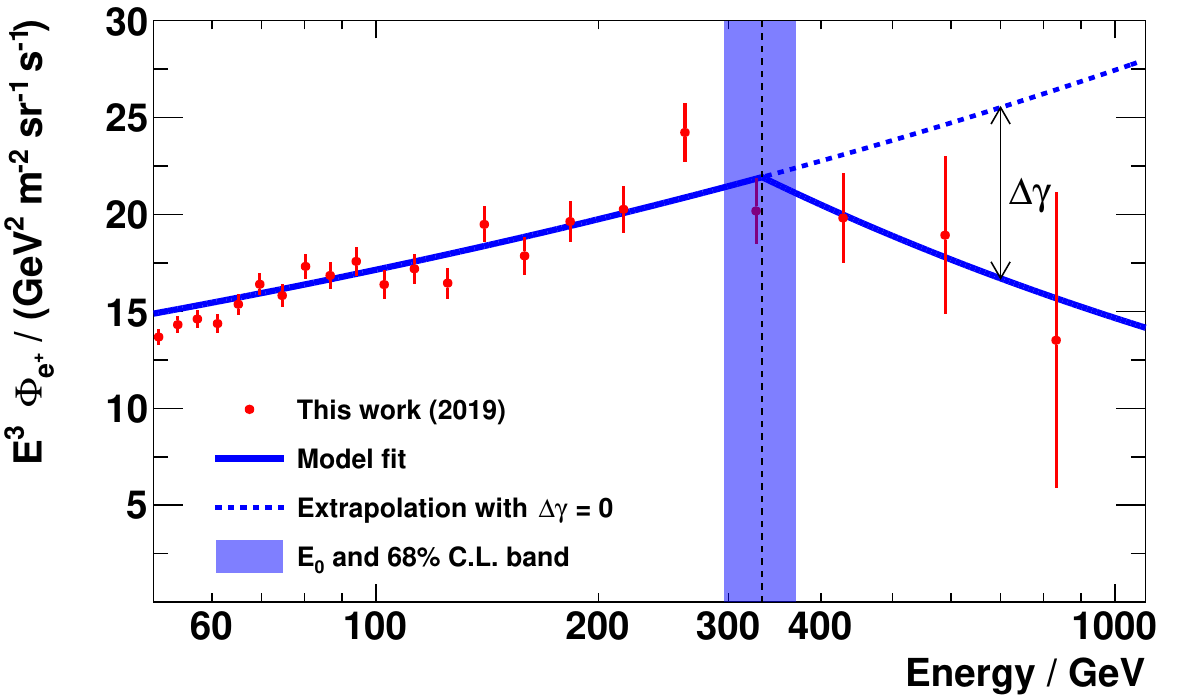}
  \caption{Fit of the double power law approximation \cref{eq:results-time-averaged-positron-transition-fit-model} to the positron flux in the range \SIrange{63.02}{1000}{\GeV}. The positron flux is shown as red data points, scaled by $E^3$. The blue line corresponds to the fit result of \cref{eq:results-time-averaged-positron-transition-fit-model}. The vertical dashed line indicates the transition energy $E_{0}$, where the positron flux starts decreasing and the blue band corresponds to the uncertainty. The blue dashed line corresponds to an extrapolation of the model, with $\Delta\gamma = 0$.}
  \label{fig:results-time-averaged-positron-flux-high-energy-transition-fit}
\end{figure}

Analogous, the energy where the positron spectral index starts decreasing can be determined, by a fit of \cref{eq:results-time-averaged-positron-transition-fit-model}
to the data\footnote{The AMS-02 positron flux publication~\cite{Aguilar2019b} uses \SI{55.58}{\GeV} as normalization constant in \cref{eq:results-time-averaged-positron-transition-fit-model}
and as beginning of the fit range for the high-energy transition fit. This work starts the fit procedure two bins later, as it improves the uncertainty on the transition energy.} in the range \SIrange{63.02}{1000}{\GeV}:

The fit yields, $C = \SI[parse-numbers=false]{(6.24 \pm 0.11) \times 10^{-5}}{[\meter\squared~\steradian~\second~\GeV]^{-1}}$,
$\gamma = \SI[parse-numbers=false]{-2.80 \pm 0.03}{}$, $E_{0} = \SI[parse-numbers=false]{333^{+61}_{-15}}{\GeV}$ for the transition energy and
$\Delta\gamma = \SI[parse-numbers=false]{-0.57 \pm 0.18}{}$ for the difference in the spectral index after
the transition energy $E_{0}$ with respect to the region before the transition. The results of the double power law fit are compatible with the
published results~\cite{Aguilar2019b} and shown in \cref{fig:results-time-averaged-positron-flux-high-energy-transition-fit}.

Note that the transition energies cannot be attributed to solar modulation, since the time variation of both the electron and positron
flux stop at above \SI{20}{\GeV} - within the measurement accuracy. This will be demonstrated in \cref{sec:results-time-dependent}.

\subsubsection{Model fits}
\label{sec:results-time-averaged-positron-flux-model-fits}

As shown in Ref.~\cite{Aguilar2019b}, the whole energy range can be described by the sum of two terms:

\begin{equation}
  \label{eq:results-time-averaged-positron-model}
  \Phi_{e^{-}}(E) = \frac{E^2}{\hat{E}^2} [ C_{d} (\hat{E} / E_{1})^{\gamma_{d}} + C_{s} (\hat{E} / E_{2})^{\gamma_{s}} \exp{(-\hat{E} / E_{s})}].
\end{equation}

The first term, called \textbf{diffuse term}, describes the low-energy part of the flux dominated by the positrons produced in the collisions of
ordinary cosmic rays with the interstellar gas. It is characterized by a normalization factor $C_{d}$ and a spectral index
$\gamma_{d}$. The second term, called \textbf{source term}, has an exponential cutoff, which describes the high-energy part of the flux dominated by
a source. It is characterized by a normalization factor $C_{s}$, a spectral index $\gamma_{s}$, and a cutoff energy $E_{s}$.

To account for solar modulation effects, the force-field approximation~\cite{Gleeson1968} is used, with the energy of particles
in the interstellar space $\hat{E} = E + \phi_{e^{+}}$ and the effective fisk potential $\phi_{e^{+}}$. The constants
$E_{1} = \SI{7}{\GeV}$ and $E_{2} = \SI{60}{\GeV}$ are chosen to minimize the correlation between $C_{s}$ and $\gamma_{s}$.

\begin{figure}[H]
  \includegraphics[width=\linewidth]{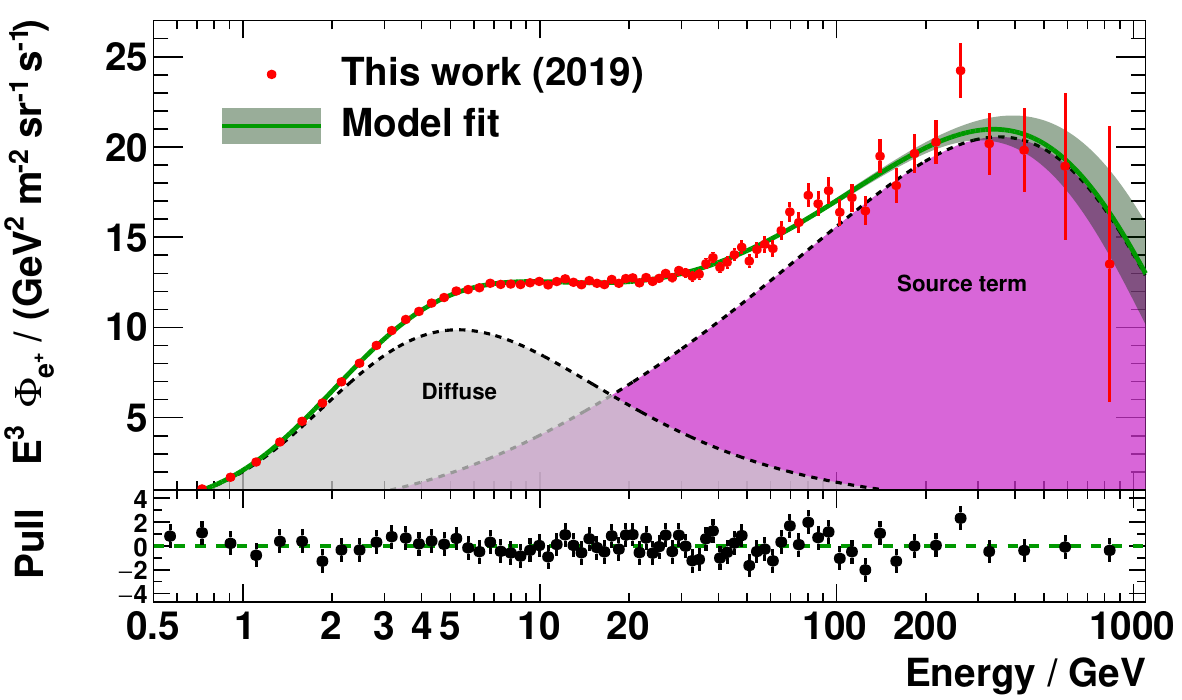}
  \caption{The positron flux, scaled with $E^3$, is shown as red points. The result of the fit of \cref{eq:results-time-averaged-positron-model} is shown as green line, enclosed by a green band, indicating the uncertainty of the fit result. The diffuse and source components are indicated by the gray and magenta areas, respectively. The lower plot shows the pull distribution of the fit (data minus model divided by data uncertainty) -- the model describes the data accurately over the whole energy range.}
  \label{fig:results-time-averaged-positron-flux-fit}
\end{figure}

A fit to the data yields $\phi_{e^{+}} = \SI[parse-numbers=false]{1.08 \pm 0.03}{\GeV}$,
$C_{d} = \SI[parse-numbers=false]{(6.61 \pm 0.14) \times 10^{-2}}{[\meter\squared~\steradian~\second~\GeV]^{-1}}$,
and $\gamma_{d} = \SI[parse-numbers=false]{-4.02 \pm 0.07}{}$ for the diffuse component, and
$C_{s} = \SI[parse-numbers=false]{(6.86 \pm 0.17) \times 10^{-5}}{[\meter\squared~\steradian~\second~\GeV]^{-1}}$,
$\gamma_{s} = \SI[parse-numbers=false]{-2.54 \pm 0.06}{}$ for the source component and
$1/E_{s} = \SI[parse-numbers=false]{1.34^{+0.39}_{-0.37}}{\TeV^{-1}}$ for the inverse cut-off energy, corresponding
to $E_{s} = \SI[parse-numbers=false]{745^{+168}_{-283}}{\GeV}$ with $\chi^2/\text{dof} = 50.7 / 68 = 0.75$.
The result of the fit is presented in \cref{fig:results-time-averaged-positron-flux-fit}.

One can conclude that the positron flux is well described by the sum of a diffuse term associated with positrons produced by collisions, which dominates at low energies, and a new source term, which dominates at high energies.

\begin{figure}[H]
  \includegraphics[width=\linewidth]{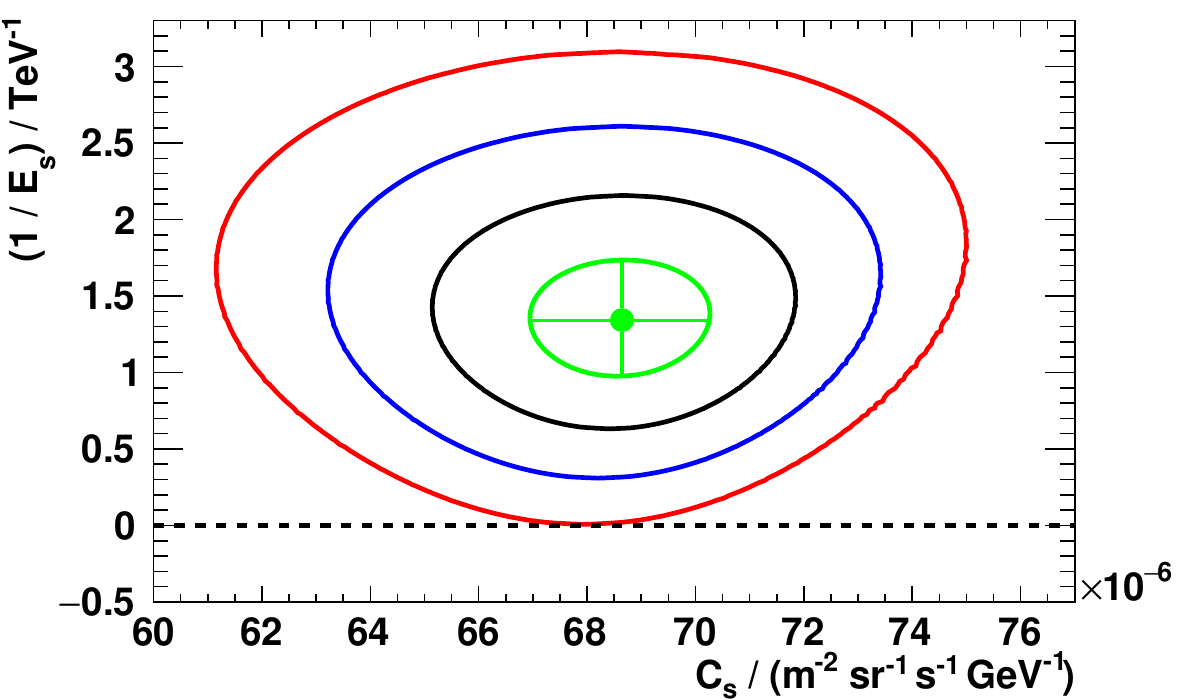}
  \caption{The contour plot between the $C_{s}$ and $1 / E_{s}$ parameters shows that the cut-off significance is established with $4\sigma$. The regions corresponding to $1\sigma$ up to $4\sigma$ are drawn in green (\SI{68.26}{\percent} C.L.), black (\SI{95.54}{\percent} C.L.), blue (\SI{99.74}{\percent} C.L.) and red (\SI{99.99}{\percent} C.L.), respectively.}
  \label{fig:results-time-averaged-positron-flux-cutoff-significance}
\end{figure}

To study the significance of the $1 / E_{s}$ measurement all six fit parameters were varied, to find regions in the parameter space corresponding to \SIrange{1}{5}{}~$\sigma$ levels.
\Cref{fig:results-time-averaged-positron-flux-cutoff-significance} shows the projection of the six dimensional parameter space to the ($1/E_{s} - C_{s}$) plane. 

A detailed analysis shows that a point where the parameter, where $1/E_{s}$ reaches 0 (cut-off at infinity) corresponds to a confidence level of \SI{4.01}{}~$\sigma$. The significance of the source
term cut-off is established at $4\sigma$: the positron flux in the energy range cannot be described by a sum of two power law functions at the \SI{99.99}{\percent} confidence level.
This is the first reported evidence for a spectral cut-off in a cosmic-ray flux at these energies.

These experimental data on cosmic-ray positrons show that, at high energies, they predominantly originate from a new source (e.g. from dark matter annihilation or from other astrophysical sources).
An important handle to disentangle whether dark matter annihilation or an astrophysical source is responsible for the source term is the \textit{anisotropy} of the arrival directions of the positrons.
An astrophysical point source - such as a pulsar - will imprint a larger anisotropy on the arrival directions of the positrons as a smooth dark mater halo. Thus measuring the dipole anisotropy
(explained in Ref.~\cite{Aguilar2013}) is important to probe the dark matter hypothesis. With the seven year dataset, an upper limit of the dipole anisotropy $\delta < 0.019$
at the \SI{95}{\percent} C.L.~for energies above \SI{16}{\GeV} is obtained~\cite{Aguilar2019b}. This does not yet rule out e.g. a pulsar origin of the high-energetic positrons: depending on the pulsar model, the expected
anisotropy is at the level of \SIapprox{0.5}{\percent}~\cite{Hooper2009}. More data is needed to be able to uncover the origin of the high-energetic positrons in future (see \cref{sec:summary}).

\clearpage
\subsection{Positron/electron ratio}
\label{sec:results-time-averaged-positron-electron-ratio}

\begin{figure}[H]
  \centering
  \includegraphics[width=0.85\linewidth]{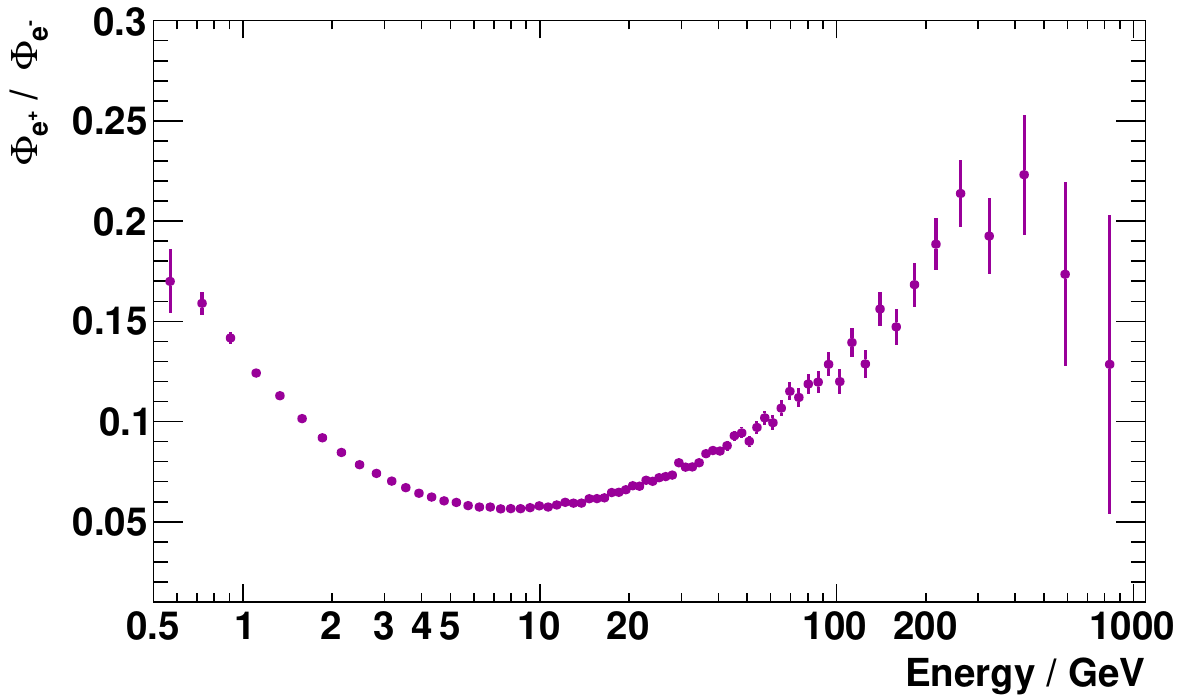}
  \caption{The positron/electron ratio measured by AMS-02.}
  \label{fig:results-time-averaged-positron-electron-ratio}
\end{figure}

\Cref{fig:results-time-averaged-positron-electron-ratio} shows the time-averaged positron/electron ratio, determined by
a dedicated analysis, using the single-track data sample, not computed from the fluxes themselves, which were derived using the all-tracks sample,
as explained in \cref{sec:analysis-lepton-counts-2d-fit}.

The positron/electron ratio is consistent between the dedicated single-track analysis and the all-tracks flux analysis,
as shown in \cref{sec:appendix-results-comparison-positron-electron-ratio-from-fluxes}.

\begin{figure}[H]
  \centering
  \includegraphics[width=0.75\linewidth]{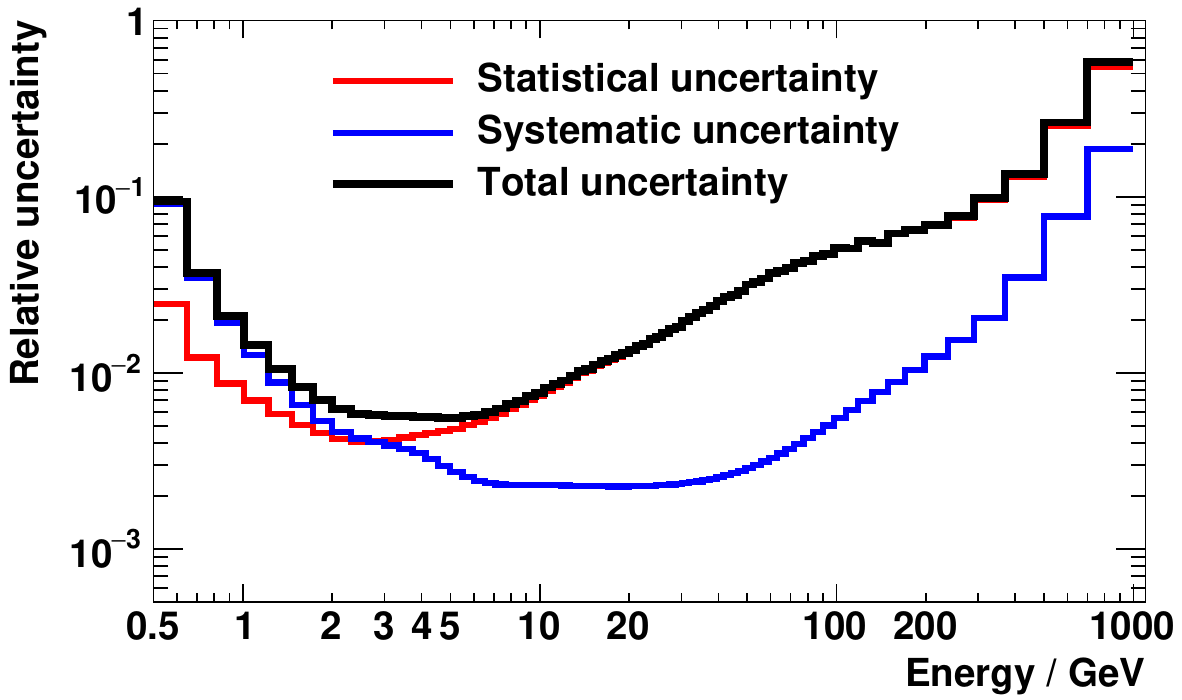}
  \caption{Breakdown of the uncertainty of the positron/electron ratio into a statistical and systematical part.}
  \label{fig:results-time-averaged-positron-electron-ratio-error-breakdown}
\end{figure}

\Cref{fig:results-time-averaged-positron-electron-ratio-error-breakdown} shows a breakdown of the total uncertainty into the statistical and systematic part.

Above \SIapprox{2}{\GeV} the statistical uncertainty dominates the uncertainty of the positron/electron ratio measurement. Below this energy, the systematic uncertainty always
exceeds the statistical uncertainty. See \cref{sec:analysis-ratios-time-averaged-sysunc-summary} for a decomposition of the systematic uncertainties.

The positron/electron ratio is an important ingredient to interpret the time-dependent flux results, and thus will be discussed in detail in
\cref{sec:results-time-dependent-positron-electron-ratio}. A related quantity, the positron fraction, is derived and discussed in \cref{sec:appendix-results-time-averaged-positron-fraction}.

\section{Time-dependent results}
\label{sec:results-time-dependent}

\subsection{Electron and positron flux}
\label{sec:results-time-dependent-electron-positron-flux}

The results of the time-dependent analysis were published in \textit{Physical Review Letters}~(Ref.~\cite{Aguilar2018}) and selected as \textbf{Editors suggestion}.
The published data is based on this work, covering the first six years of science data from AMS-02 (\textbf{May~\nth{20},~2011} until \textbf{May~\nth{12},~2017}).
The time range of the fluxes in this thesis were extended to \textbf{November~\nth{12},~2017}, covering nine more Bartels rotations.

\begin{figure}[H]
  \centering
  \includegraphics[width=0.80\linewidth]{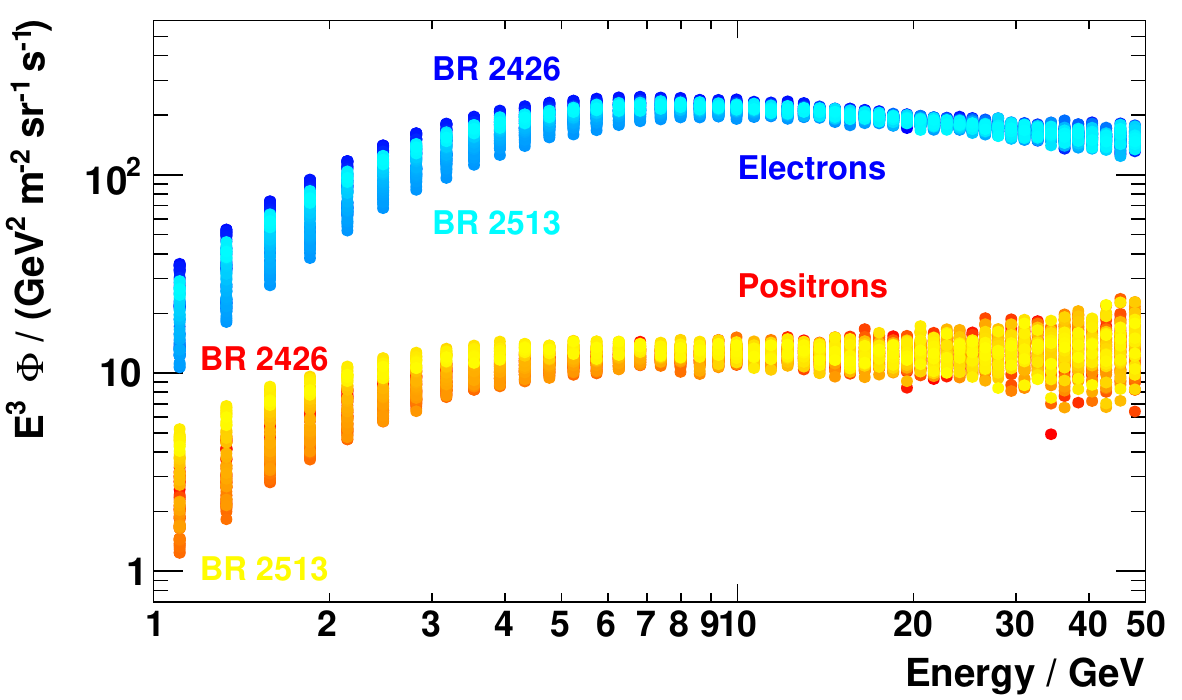}
  \caption{The electron / positron fluxes for each Bartels rotation analyzed in this work. The error bars on the individual flux points were omitted for clarity.}
  \label{fig:results-time-dependent-electron-positron-flux}
\end{figure}

\Cref{fig:results-time-dependent-electron-positron-flux} shows the extended time-dependent electron and positron flux.
Each color represent a flux from a different Bartels rotation period. At low energies a clear time order is exposed: The magnitude of the fluxes, e.g. at \SIapprox{1}{\GeV} exhibits
a maximum in the beginning of the measuring period (BR 2426), then decrease to a minimum, followed by an increase (BR 2513). The same trend is visible in both electrons and positrons.

\begin{figure}[H]
  \begin{subfigure}{0.50\linewidth}
    \includegraphics[width=\linewidth]{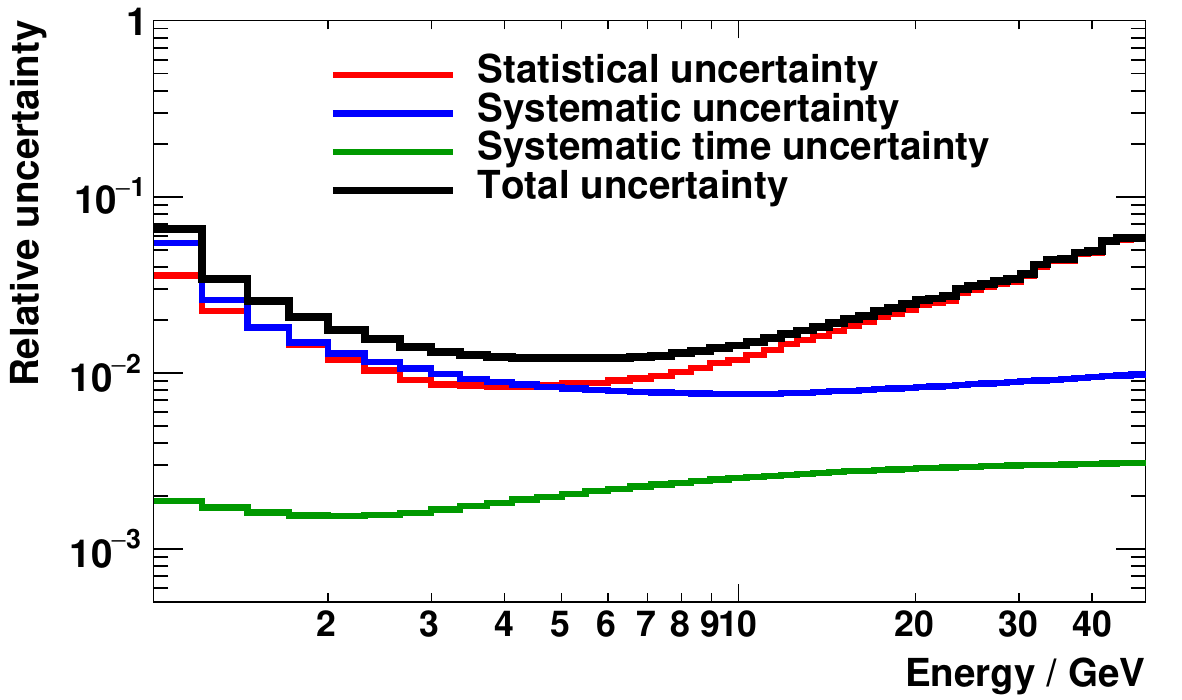}
    \caption{Electron flux}
  \end{subfigure}
  \hfill
  \begin{subfigure}{0.50\linewidth}
    \includegraphics[width=\linewidth]{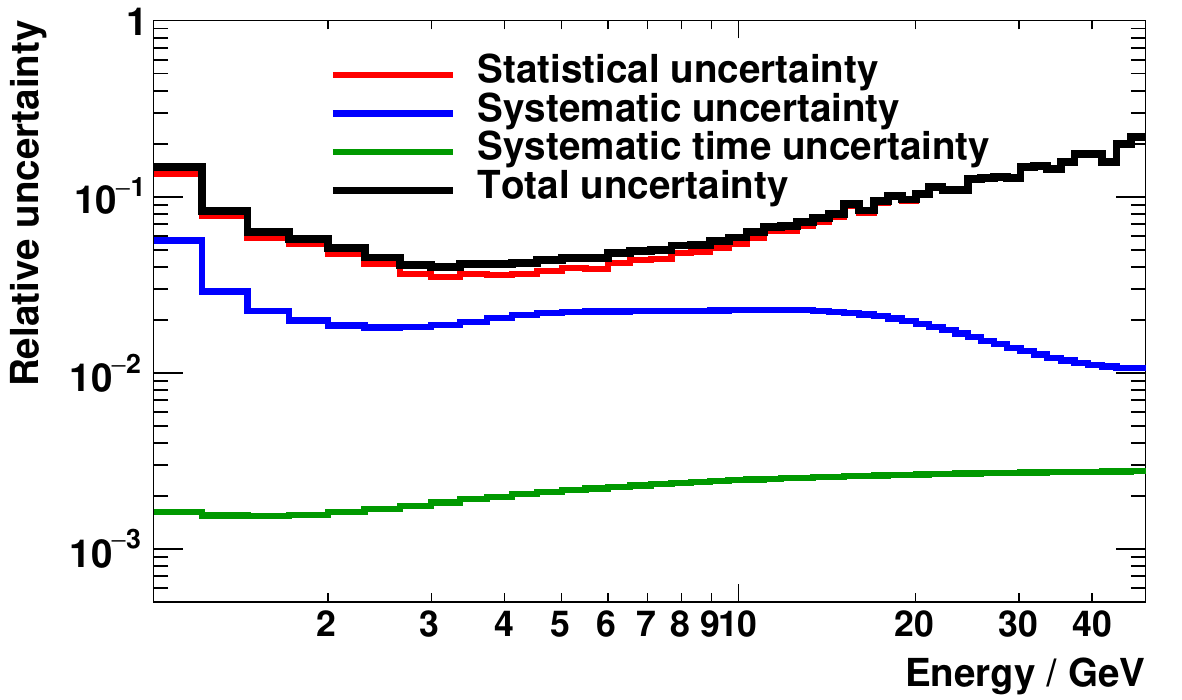}
    \caption{Positron flux}
  \end{subfigure}
  \caption{Breakdown of the uncertainty of the electron flux (left) and the positron flux (right) into a statistical and systematical part, in an example Bartels rotation $i = 26$ (Apr~\nth{16},~2013 - May~\nth{13},~2013).}
  \label{fig:results-time-dependent-electron-positron-flux-error-breakdown}
\end{figure}

\Cref{fig:results-time-dependent-electron-positron-flux-error-breakdown} shows a breakdown of the total uncertainty of the electron and the positron flux into the statistical and systematic part,
in an example Bartels rotation. The electron flux is dominated by the statistical uncertainty above \SIapprox{4}{\GeV} whereas the positron flux is dominated by the statistical uncertainty
over the whole energy range. See \cref{sec:analysis-flux-time-dependent-sysunc-summary} for a decomposition of the systematic uncertainties.

\begin{figure}[H]
  \centering
  \includegraphics[width=0.95\linewidth]{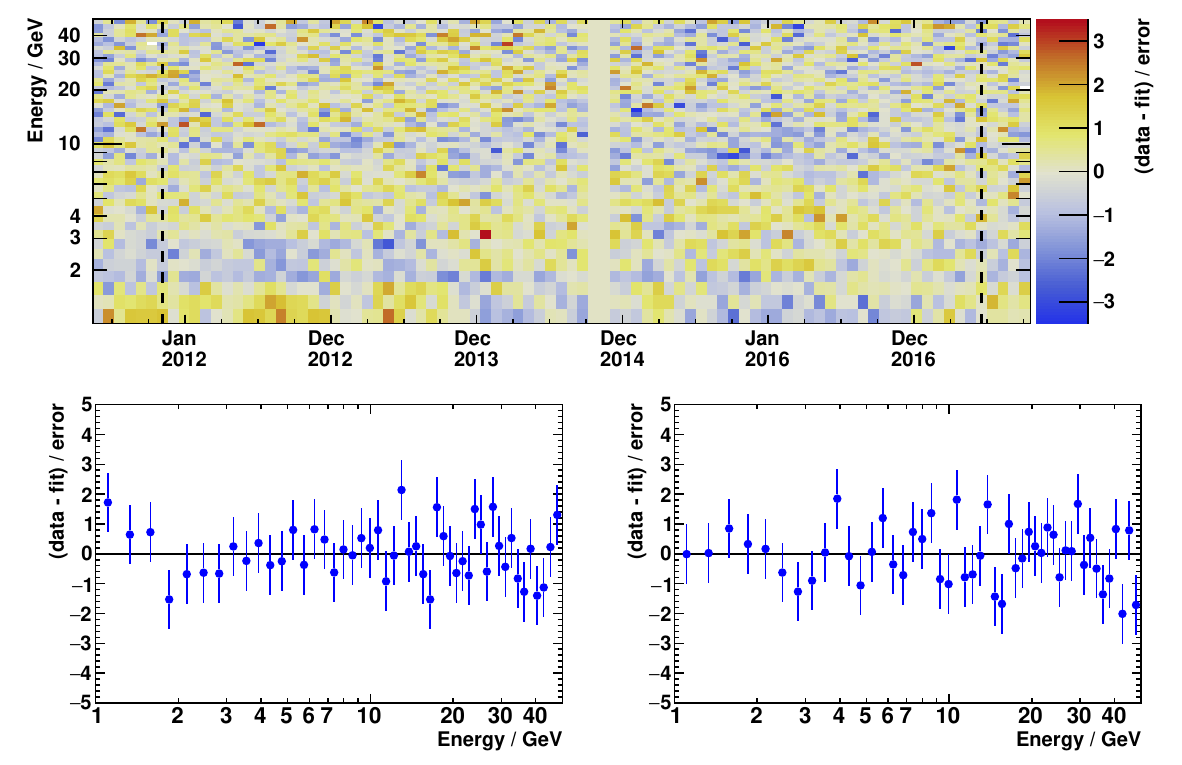}
  \caption{For the electron flux, the difference between data and model~\cite{Cavasonza2017} fits normalized to the experimental uncertainties: for each Bartels Rotation as a function of energy (upper plot). The distribution for the electrons reveals a model-dependent structure stable in time in the energy range between \SIrange{2}{3}{\GeV}, emphasized for Bartels Rotation 2432 (lower left) and Bartels Rotation 2508 (lower right). The times for these two Bartels Rotations are indicated by the vertical dashed lines in the upper graph.}
  \label{fig:results-time-dependent-electron-flux-pull}
\end{figure}

As described in Ref.~\cite{Aguilar2018}, the model given in Ref.~\cite{Cavasonza2017} was fit to the data for each Bartels rotation independently,
to search for fine structures in the energy dependence of the fluxes. The positron data show no additional structure, as shown in \cref{sec:appendix-results-absence-of-structures-in-positron-flux}.
\Cref{fig:results-time-dependent-electron-flux-pull} shows that the electron data reveal a model dependent residual structure in the energy range between \SIrange{2}{3}{\GeV}.

The residual structure is stable in time and consistent with an additional\footnote{Note that the spectral break is less pronounced in this work, than in Ref.~\cite{Aguilar2018}.
The reason is the unfolding procedure: the fluxes derived in this work are all unfolded, whereas the published fluxes in Ref.~\cite{Aguilar2018} have not been unfolded.
However the physics conclusions of \cite{Aguilar2018} remain unchanged.} smooth break~\cite{Strong2011} in the electron spectral index
$\gamma^{-}(E) = \diff(\log{\Phi_{e^{-}}(E)}) / \diff(\log{E})$ below \SI{10}{\GeV} (\cref{sec:appendix-results-energy-dependence-electron-spectral-index}),
comparable to the local interstellar electron spectrum of Ref.~\cite{Potgieter2015}.

The fits of the extended model~\cite{Cavasonza2017}, including the low energy break at \SIapprox{2}{\GeV}, to the time-dependent electron and positron fluxes
yield an average $\chi^2/\text{dof}\approx1$ for all Bartels rotations and no fine structures in the energy spectra were found.

\begin{figure}[H]
  \begin{subfigure}{0.50\linewidth}
    \includegraphics[width=\linewidth]{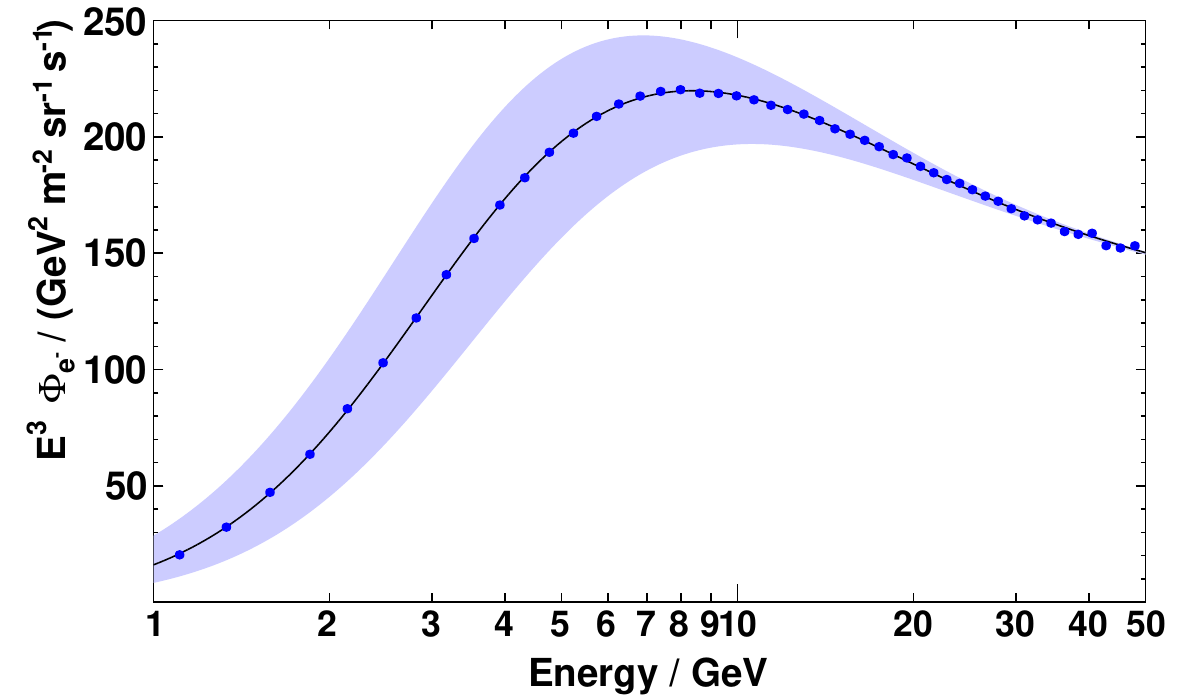}
    \caption{Electron flux}
  \end{subfigure}
  \hfill
  \begin{subfigure}{0.50\linewidth}
    \includegraphics[width=\linewidth]{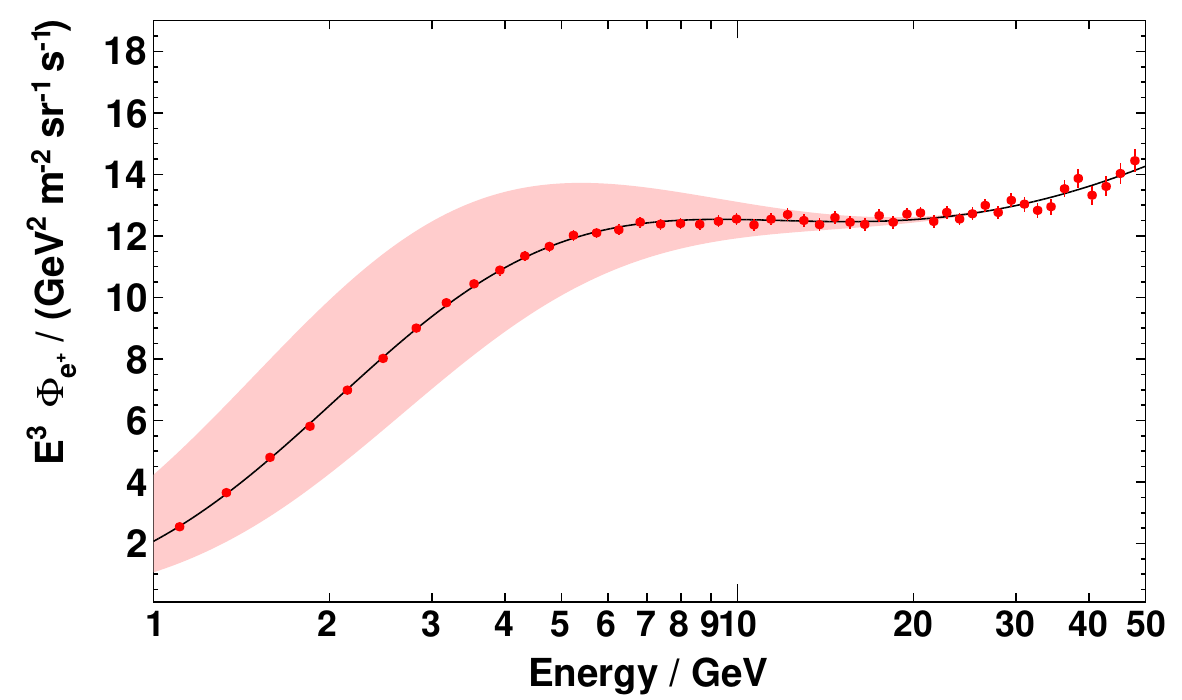}
    \caption{Positron flux}
  \end{subfigure}
  \caption{Visualization of the time-averaged electron flux (left) and positron flux (right). The time-variation range is indicated by the shaded regions, see text. The fit of the model in Ref.~\cite{Cavasonza2017} to the time-averaged data points is shown by the black curves.}
\label{fig:results-time-dependent-electron-positron-flux-time-variation}
\end{figure}

The time-averaged electron flux and positron flux are shown in \cref{fig:results-time-dependent-electron-positron-flux-time-variation}.
To visualize the magnitude of the time variations of the fluxes, the envelopes of all fitted curves are displayed as shaded regions.
The amplitude of the shaded regions decreases with increasing energy. At high energies, the statistical bin-to-bin fluctuations are larger
than the time variation.

\begin{figure}[H]
  \includegraphics[width=\linewidth]{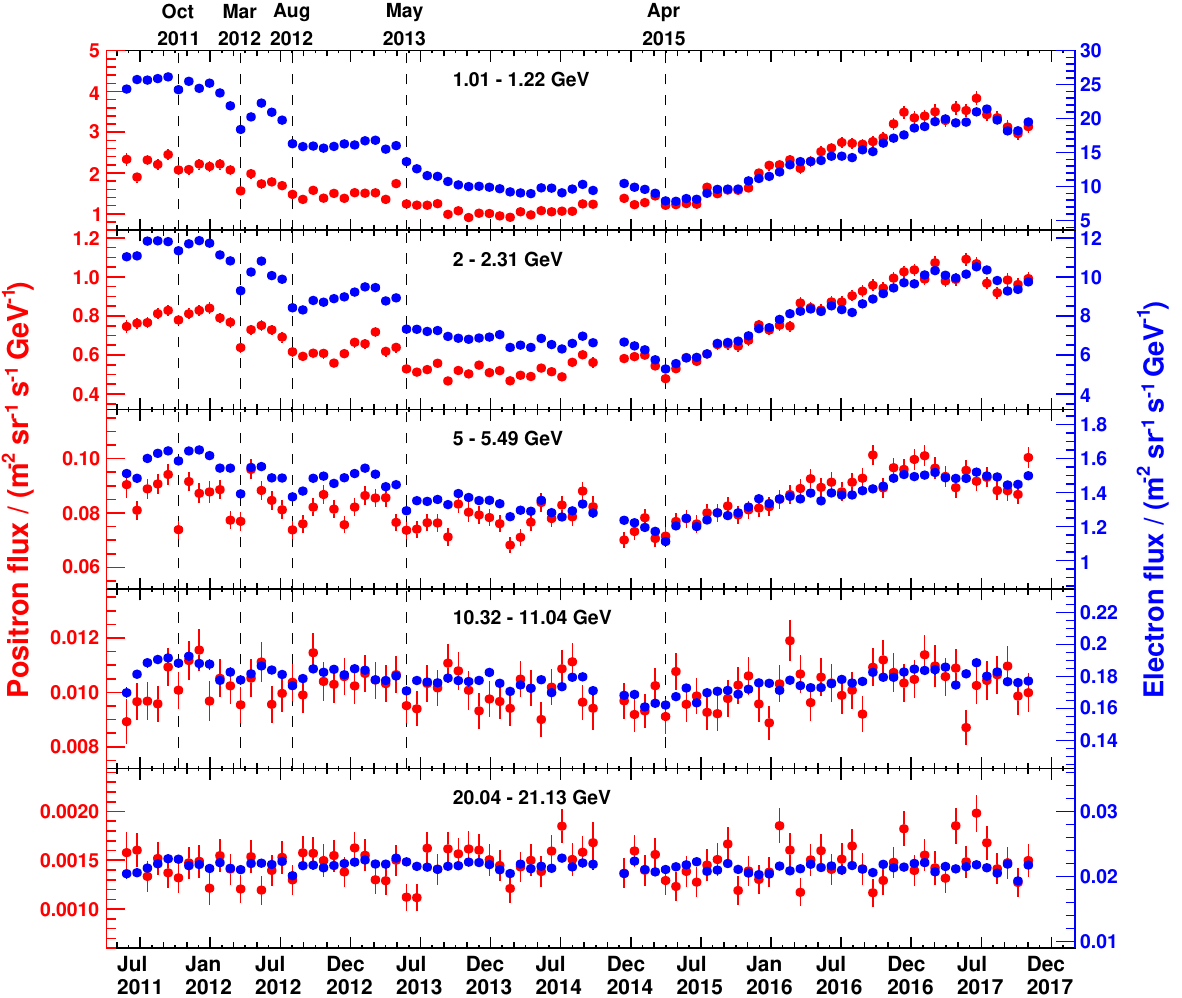}
  \caption{Fluxes of primary cosmic-ray positrons (red, left axis) and electrons (blue, right axis) as functions of time, for five of the 49 energy bins. The error bars are the statistical uncertainties. Prominent and distinct time structures visible in both the positron spectrum and the electron spectrum and at different energies are marked by dashed vertical lines.}
  \label{fig:results-time-dependent-electron-positron-flux-overview}
\end{figure}

To study the time behavior in more detail, the fluxes are shown in \cref{fig:results-time-dependent-electron-positron-flux-overview}
as a function of time for five characteristic energy bins. A clear evolution of the fluxes with time at low energies that gradually
diminishes towards high energies is exposed. At the lowest energies, the amplitudes of both the electron flux and the positron flux
change by a factor of 3. Both fluxes exhibit profound short- and long-term variations. The short-term variations occur simultaneously
in both fluxes with approximately the same relative amplitude. On the short term of Bartels rotations, several prominent and distinct
structures are observed. They are characterized by minima, visible in both the electron flux and the positron flux across the energy
range below \SIvarOp{E}{$\lesssim$}{10}{\GeV}. These are marked by dashed vertical lines in \cref{fig:results-time-dependent-electron-positron-flux-overview}.

In October 2011 and March 2012, there are sharp drops in the fluxes, followed by a quick recovery. The March
2012 event coincides with a strong Forbush decrease registered on March~\nth{8},~2012~\cite{Cheminet2013}.
Another drop occurred in August 2012; this was followed by an extended recovery period.
For \SIvarOp{E}{$\lesssim$}{10}{\GeV}, May 2013 and April 2015 mark two changes in the long-term trends of the fluxes: From May 2011
to May 2013, the fluxes of both species show a trend to decrease with time. In the period around July 2013 is the time
of the solar magnetic field reversal. From May 2013 to April 2015, the flux of electrons continues to decrease, but with
reduced slope, while the positron flux begins to increase. Then, from April 2015 until November 2017, both fluxes rise steeply.
The difference of the rate of the increase is related to the charge-sign dependent solar modulation~\cite{Potgieter2001,Heber2009}.
At energies above \SI{20}{\GeV}, neither the electron flux nor the positron flux exhibits significant time dependence.

\subsection{Positron/electron ratio}
\label{sec:results-time-dependent-positron-electron-ratio}

\begin{figure}[H]
  \centering
  \includegraphics[width=0.8\linewidth]{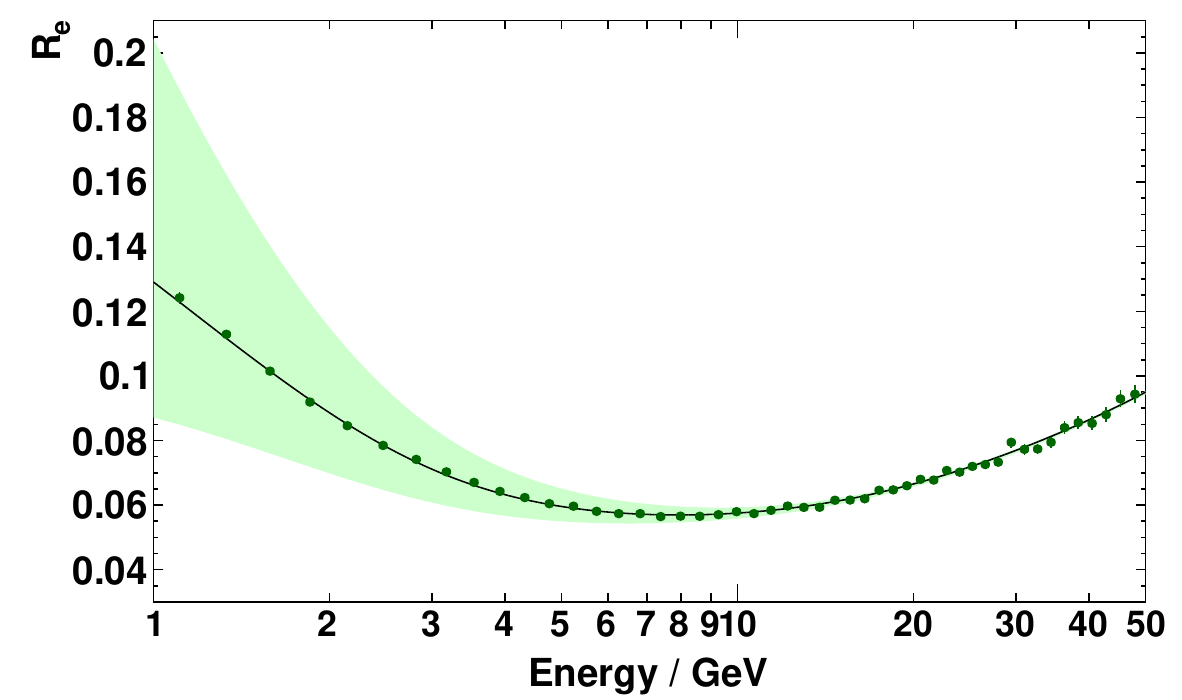}
  \caption{Visualization of the time-averaged positron/electron ratio $R_{e}$. The time-variation range is indicated by the shaded regions, see text. The fit of the model in Refs.~\cite{Corti2016,Cavasonza2017} to the time-averaged data points is shown by the black curves.}
\label{fig:results-time-dependent-positron-electron-ratio-time-variation}
\end{figure}

The time-averaged positron/electron ratio is shown in \cref{fig:results-time-dependent-positron-electron-ratio-time-variation}.
To visualize the magnitude of the time variations of the positron/electron ratio, the envelopes of all fitted curves are displayed as shaded regions.
The amplitude of the shaded regions decreases with increasing energy. At high energies, the statistical bin-to-bin fluctuations are larger
than the time variation.

\begin{figure}[H]
  \includegraphics[width=\linewidth]{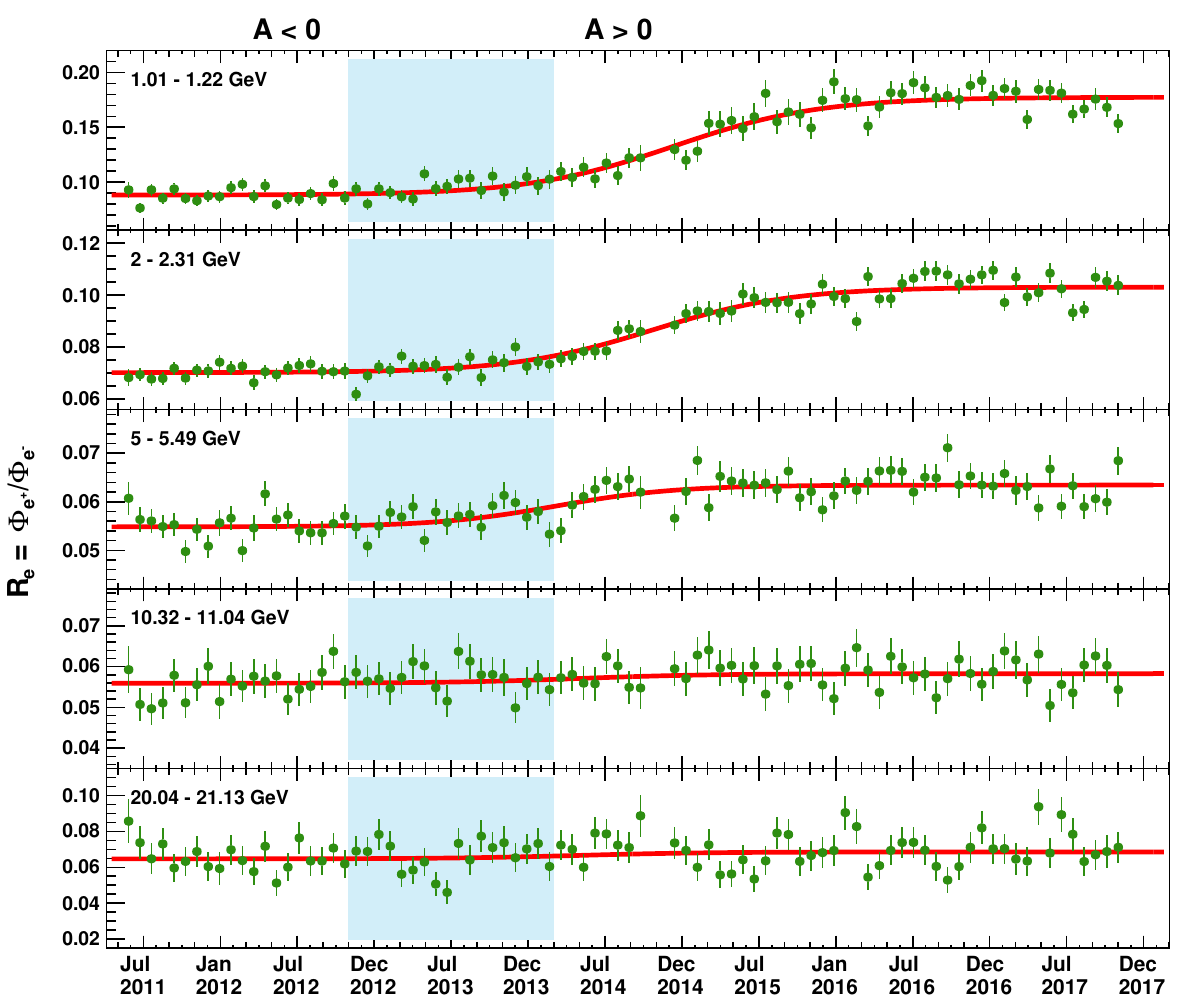}
  \caption{The ratio $R_{e}$ of the positron flux to the electron flux as a function of time. The error bars are statistical. The best-fit parametrization according to \cref{eq:re-fit-function} is shown by red curves. The polarity of the heliospheric magnetic field is denoted by A < 0 and A > 0. The period without well-defined polarity is marked by the shaded area~\cite{Sun2015}.}
  \label{fig:results-time-dependent-positron-electron-ratio-overview}
\end{figure}

The long-term time structure of the data in \cref{fig:results-time-dependent-electron-positron-flux-overview} shows that the changes
in relative amplitude are different for electrons and positrons. In the positron/electron ratio $R_{e}$, the short-term variations in the
fluxes largely cancel, and a clear overall long-term trend appears, as shown in \cref{fig:results-time-dependent-positron-electron-ratio-overview}.

At low energies, $R_{e}$ is flat at first, then smoothly increases after the time of the solar magnetic field reversal, to reach a plateau at a higher amplitude.
During the extraordinarily quiet solar minimum period from 2006 to 2011, the energy and time dependence of various cosmic-ray measurements~\cite{Potgieter2017}
are well reproduced by advanced numerical solar modulation models~\cite{Tomassetti2017}.
But for the following years covered by the new data presented in this work, important and large systematic discrepancies are observed in particular in $R_{e}$
(see SM of Ref.~\cite{Aguilar2018}), which is sensitive to charge-sign dependent effects in the solar modulation process of galactic cosmic rays.

A model-independent approach is used to extract the energy dependence of the quantities that characterize
the observed transition in $R_{e}$. With a set of four parameters, the 4312 independent $R_{e}$ measurements as a function of energy and time can be
described well with a logistic function:

\begin{equation}
  \label{eq:re-fit-function}
  R_e(t,E)=R_0(E)\left[1+\frac{C(E)}{\exp\left(-\frac{t-t_{1/2}(E)}{\Delta{}t(E)/\Delta_{80}}\right)+1}\right].
\end{equation}

\begin{figure}[H]
  \begin{subfigure}{0.32\linewidth}
    \includegraphics[width=\linewidth]{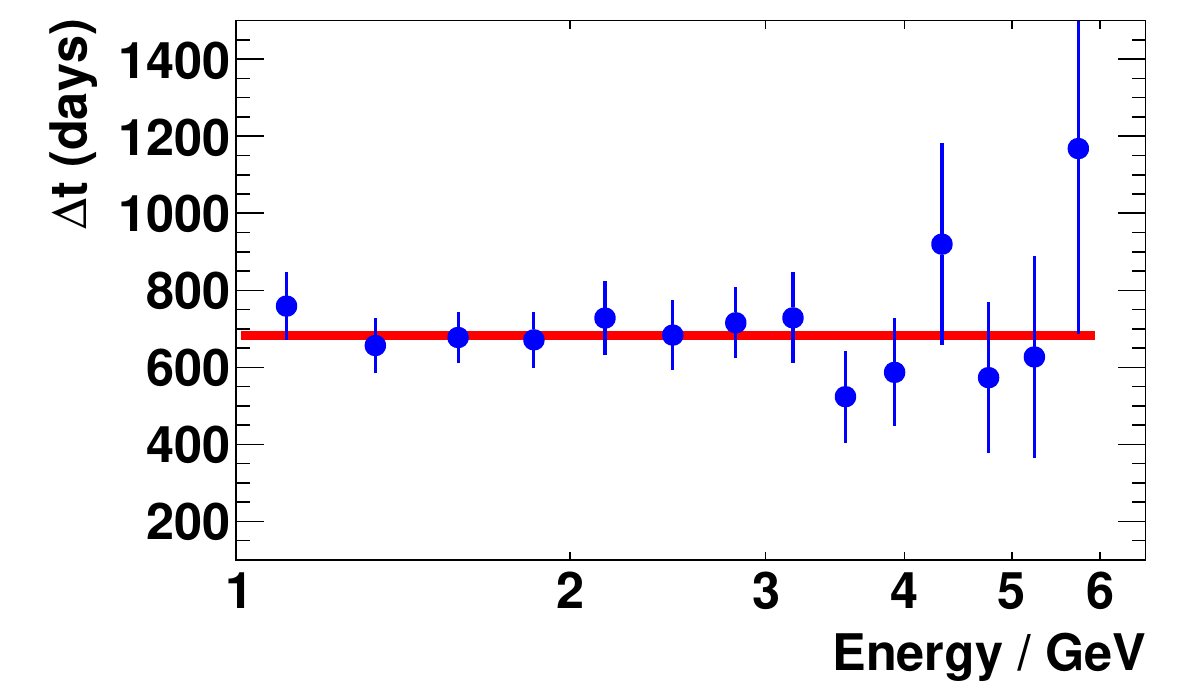}
    \caption{}
    \label{fig:results-time-dependent-re-fit-parameters-delta-t}
  \end{subfigure}
  \hfill
  \begin{subfigure}{0.32\linewidth}
    \includegraphics[width=\linewidth]{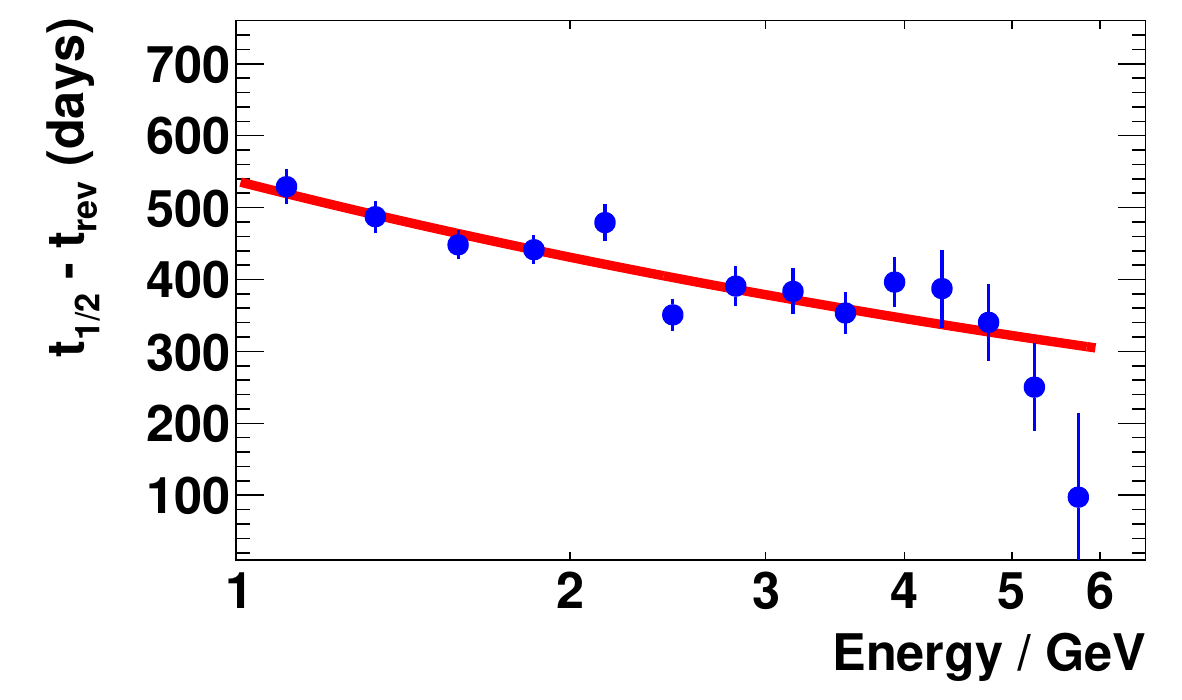}
    \caption{}
    \label{fig:results-time-dependent-re-fit-parameters-t0}
  \end{subfigure}
  \hfill
  \begin{subfigure}{0.32\linewidth}
    \includegraphics[width=\linewidth]{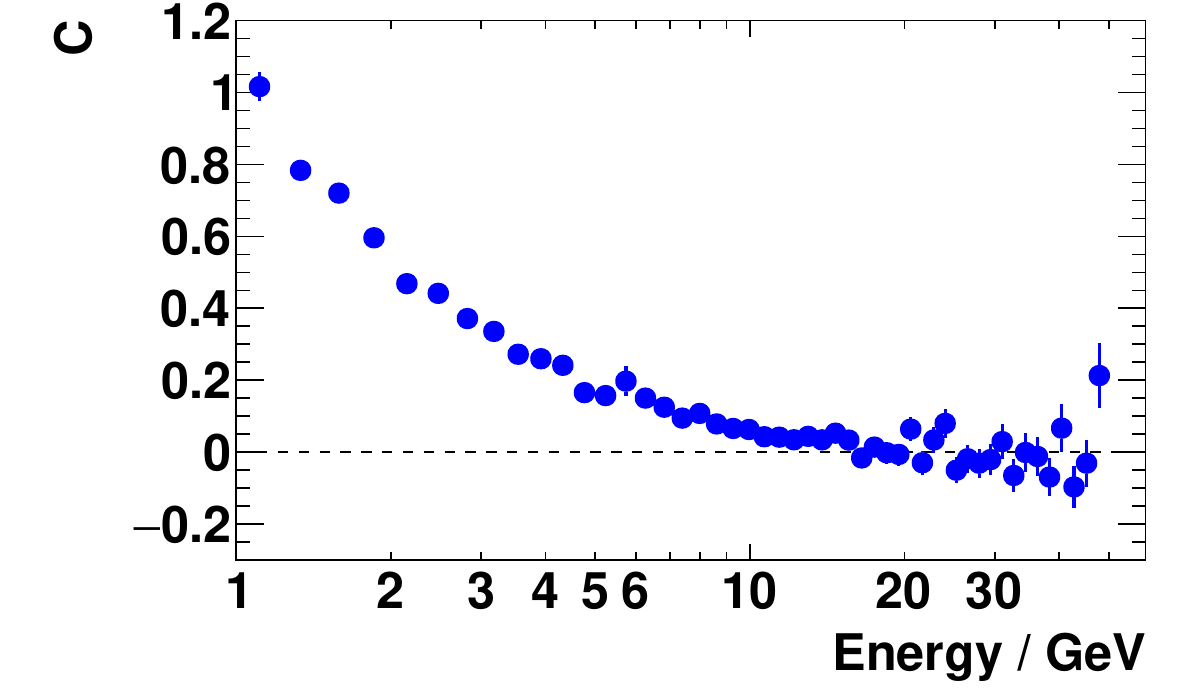}
    \caption{}
    \label{fig:results-time-dependent-re-fit-parameters-amplitude}
  \end{subfigure}
  \caption{Results of the fits of the parameterization in \cref{eq:re-fit-function} to the ratio $R_{e}$ as a function of energy (blue circles): (a) $\Delta{}t$ and the best-fit constant value of 682 days (red line), (b) $t_{1/2}-t_\mathrm{rev}$ with the parameterization according to \cref{eq:re-fit-t12} (red curve), (c) amplitude $C$ with a dashed line at zero to guide the eye.}
  \label{fig:results-time-dependent-re-fit-parameters}
\end{figure}

At a given energy $E$, the time-dependence is related to three parameters in the function: the amplitude of the transition $C$,
the midpoint of the transition $t_{1/2}$, and the duration of the transition $\Delta{}t$. $\Delta_{80}$ is set to $4.39$, such that
$\Delta{}t$ is the time it takes for the transition to proceed from \SI{10}{\percent} to \SI{90}{\percent} of the change in
magnitude. The results of fitting \cref{eq:re-fit-function} for each energy bin are shown in \cref{fig:results-time-dependent-re-fit-parameters}.
A $\chi^2/\text{dof}\approx1$ was obtained for all fits.

The parameters $t_{1/2}$ and $\Delta{}t$ can only be determined at low energies, where the amplitude of the transition is large, see
\cref{fig:results-time-dependent-positron-electron-ratio-overview}. As shown in \cref{fig:results-time-dependent-re-fit-parameters-delta-t}, the transition
duration $\Delta{}t$ is independent of energy, and a value of $682\,\pm\,27$ days was obtained.

\begin{figure}[H]
  \includegraphics[width=\linewidth]{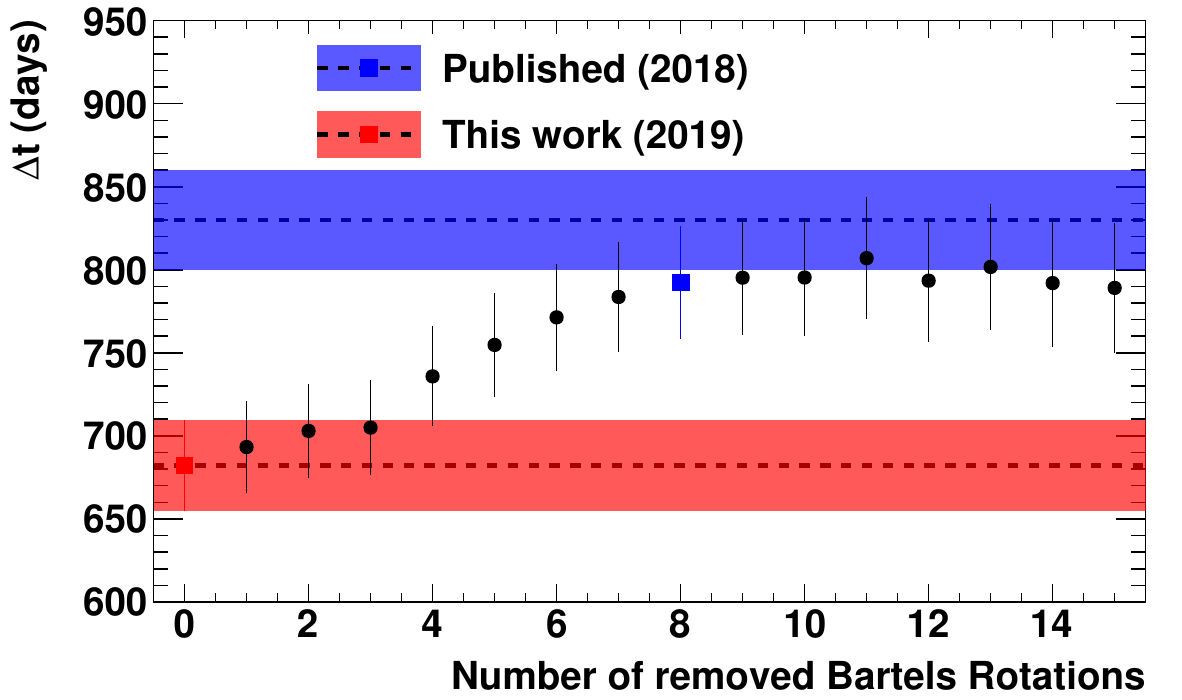}
  \caption{Dependence of the $\Delta{}t$ parameter on the amount of Bartels Rotations that are included in the fit of \cref{eq:re-fit-function} to the ratio $R_{e}$. The number of excluded Bartels Rotations at the end of the data taking period is correlated with $\Delta{}t$.}
  \label{fig:results-time-dependent-re-fit-parameter-delta-t}
\end{figure}

The value differs from the published $\Delta{}t=830\,\pm\,30$ days, presented in Ref.~\cite{Aguilar2018}. The difference arises from the additional eight
Bartels rotations from May 2017 to December 2017: when excluding the last few Bartels rotations, as shown in \cref{fig:results-time-dependent-re-fit-parameter-delta-t},
$\Delta{}t$ will increase, to a value compatible with the previously published result. The difference between the result derived in this work using the full dataset
(red point in \cref{fig:results-time-dependent-re-fit-parameter-delta-t}) and the result derived using the reduced dataset (with eight Bartels Rotations removed, blue point
in \cref{fig:results-time-dependent-re-fit-parameter-delta-t}) is used as systematic uncertainty: $\Delta{}t^{\text{syst}} = 148$ days.

Therefore the transition period is determined to be $\Delta{}t=682\,\pm\,27\,\mathrm{(stat)}\,\pm\,148\,\mathrm{(syst)}$ days.

\begin{figure}[H]
  \includegraphics[width=\linewidth]{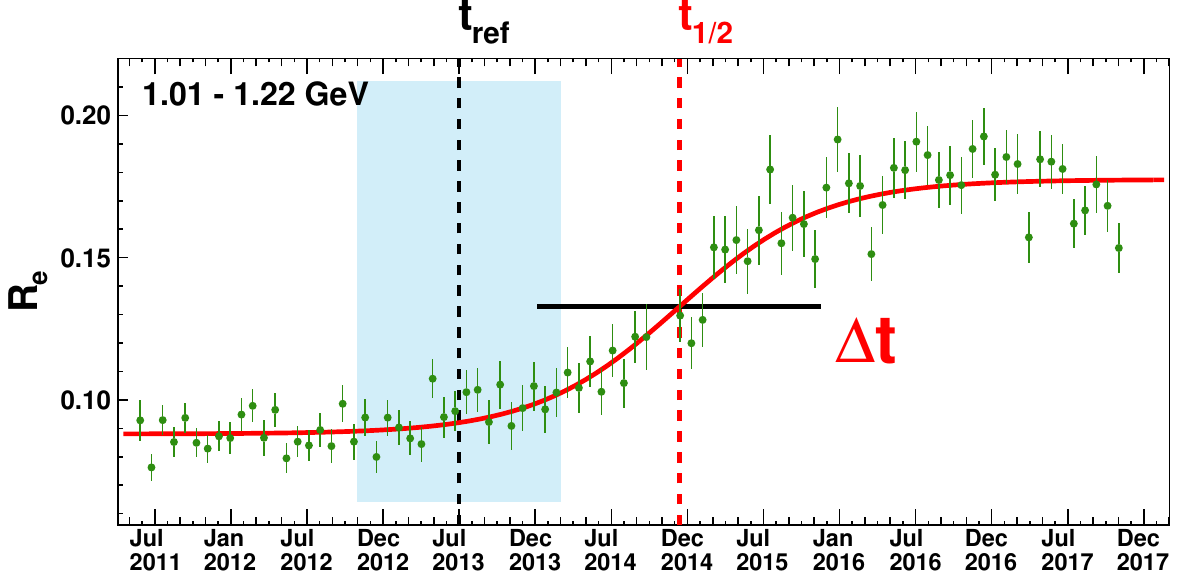}
  \caption{Illustration of the parameters in \cref{eq:re-fit-function,eq:re-fit-t12} describing the time and energy dependence of $R_{e}$, using two energy bins from \cref{fig:results-time-dependent-positron-electron-ratio-overview} as examples. The best-fit parameterizations according to \cref{eq:re-fit-function} are shown by red curves. The period without well-defined polarity is marked by the shaded area~\cite{Sun2015}. The choice for the effective time of the reversal of the solar magnetic field $t_\mathrm{rev}$ is marked by black dashed vertical lines. The fit results for the midpoint of the transition $t_{1/2}$ are marked by red dashed vertical lines. The value of $t_{1/2}$ is found to be energy dependent. The  width of the red horizontal bars indicate the duration of the transition $\Delta{}t$, which is found to be independent of energy at $682\,\pm\,27$ days. It takes time $\Delta{}t$ for the transition to proceed from \SI{10}{\percent} to \SI{90}{\percent} of the change in magnitude.}
  \label{fig:results-time-dependent-re-fit-parameters-vs-time-and-energy}
\end{figure}

\Cref{fig:results-time-dependent-re-fit-parameters-t0} shows the energy dependence of the delay $t_{1/2}$ which is well parameterized by the formula

\begin{equation}
  \label{eq:re-fit-t12}
  t_{1/2}(E)-t_\mathrm{rev}=\tau\,\cdot\,(E/\mathrm{GeV})^{\rho},
\end{equation}

where $t_\mathrm{rev}$ was fixed to be the effective time of the reversal of the solar magnetic field. For the value of $t_\mathrm{rev}$, July~\nth{1},~2013 is used,
the center of the period without well defined polarity~\cite{Sun2015}. The parameters used to describe the time and energy dependence of $R_e$ in
\cref{eq:re-fit-function,eq:re-fit-t12} are illustrated in \cref{fig:results-time-dependent-re-fit-parameters-vs-time-and-energy}.

A fit of \cref{eq:re-fit-t12} yields the parameter $\rho=-0.32\,\pm\,0.04\,\mathrm{(stat)}\,\substack{+0.08\\-0.16}\,\mathrm{(syst)}$ and the amplitude
$\tau=537\,\pm\,18\,\mathrm{(stat)}\,\pm\,136\,\mathrm{(syst)}$ days, and the value of $t_{1/2}$ changes by $233\,\pm\,31$ days from \SIrange{1}{6}{\GeV}.
The systematic uncertainties are due to the uncertainty in $t_\mathrm{rev}$. This is an important and unexpected energy dependence of $t_{1/2}$ and reflects the different response of
cosmic-ray particles and antiparticles to changing modulation conditions.

To study the amplitude $C$ in \cref{fig:results-time-dependent-re-fit-parameters-amplitude}, $\Delta{}t$ was fixed
to its average value of 682\,days and the value of $t_{1/2}$ calculated from \cref{eq:re-fit-t12} for energies
above \SI{6}{\GeV}. At high energies, the fit result for the amplitude depends only weakly on the choice of the values for $\Delta{}t$ and
$t_{1/2}$. As seen in \cref{fig:results-time-dependent-re-fit-parameters-amplitude}, the amplitude $C$ is close to 1 at \SIvarOp{E}{=}{1}{\GeV} and decreases smoothly with energy.
This is in qualitative agreement with the expectation from solar modulation models including drift effects~\cite{Potgieter1993b} and with the results
from Refs.~\cite{Heber2002,Ferreira2004,Heber2009,Heber2013}. Above \SI{20}{\GeV}, the amplitude is consistent with zero.

\chapter{Summary}
\label{sec:summary}

The presented time-averaged and time-dependent fluxes by AMS-02 are the most accurate measurements of the cosmic-ray
electron and positron flux to date. The unprecedented accuracy in the data challenges our understanding of the origin
of cosmic-ray positrons. The positron flux - at high energy - shows strong evidence for a source component
responsible for the high-energetic positrons. For the first time a cut-off in the positron flux was measured, with a
confidence of $4\sigma$. The origin of the cut-off in the positron flux could be an astrophysical source, such as a
pulsar. On the other hand the sharp drop-off of the flux could be the manifestation of a \textit{kinematic edge},
related to dark matter annihilation. The electron flux shows no hint of a cut-off: it can be described by
the sum of two power laws.

\bigskip
A key handle to differentiate between the dark matter and pulsar hypothesis is the measurement of the anisotropy
in the arrival directions. The current limits on the dipole anisotropy of $\delta < 0.019$ at the \SI{95}{\percent}
confidence level are not competitive to rule out the pulsar origin. A novel large-acceptance analysis
is under development, which will measure the fluxes as function of rigidity, not utilizing the ECAL. This allows one
to increase the acceptance by a factor $\approx{}4$.

AMS-02 will continue to measure until the end of the ISS lifetime. Improvements in the analysis techniques~\cite{Kounine2017a}
will allow us to measure the positron flux beyond the cut-off energy $E_{s} = \SI[parse-numbers=false]{745^{+168}_{-283}}{\GeV}$,
up to \SIapprox{2}{\TeV} and to determine the cut-off with more than $5\sigma$ confidence. The model independent search
for a spectral index change is currently limited by statistics. With the current dataset the break energy was determined to be
$E_{0} = \SI[parse-numbers=false]{333^{+61}_{-15}}{\GeV}$ and the change of the spectral index to $\Delta\gamma = \SI[parse-numbers=false]{-0.57 \pm 0.18}{}$.
With the extended dataset a significance of more than $5\sigma$ for the change of the spectral index and the break energy is in reach.
Furthermore the gain in statistics and the large-acceptance analysis will allow us to probe the dipole anisotropy on the sub-percent level, allowing
to detect either a signature of anisotropy or set the most stringent limits.

As already formulated in Ref.~\cite{Aguilar2018}, for the first time, the charge-sign dependent modulation during solar maximum has been investigated in detail by electrons
and positrons alone, using the time-dependent fluxes. Prominent, distinct, and coincident structures in both the positron flux
and the electron flux on a time scale of months were identified, that are not visible in the $e^{+}/e^{-}$ flux ratio. Instead
a long-term feature in the $e^{+}/e^{-}$ flux ratio was revealed: a smooth transition from one value to another, after the
polarity reversal of the solar magnetic field. The transition magnitude is decreasing as a function of energy, consistent with
expectations from solar modulation models including drift effects. This novel dataset provides accurate input to the
understanding of solar modulation. In the past, flux models were often constructed above \SIapprox{20}{\GeV}, where the
influence of solar modulation plays a minor role. Using the time-dependent precision data presented in this thesis,
sophisticated models can be developed, incorporating charge-sign dependent solar modulation, that allows one to describe
the electron and positron flux over the whole energy range from \SI{0.5}{\GeV} to \SI{1}{\TeV}.

\bigskip
Final insights on the origin of cosmic-ray positrons will be given by AMS-100~\cite{Schael2019}, which offers 1000 times
the acceptance of AMS-02, and allows for a measurement of the electron and positron flux up to \SI{10}{\TeV}. The science
program could start around the year 2040, which would mark the begin of a new era in astroparticle physics.

\appendix
\chapter{Appendix - Analysis}
\label{sec:appendix}

\section{Detector quality cuts}
\label{sec:appendix-detector-quality-cuts}

In the following all \textbf{detector quality cuts} are listed that are imposed on the ISS
data sample for each second of the data taking period.

\begin{enumerate}
  \item\textbf{Nominal reconstruction period}\hfill\\
    The ratio of reconstructed particles over the trigger rate in each second should be nominal.
    \Cref{fig:rti-particles-vs-trigger} shows a plot of the trigger rate as function of the ratio: reconstructed particles over the trigger rate.
    The red dashed line $y = \frac{1600}{0.07} \cdot x$ separates the nominal reconstruction period from the non-nominal reconstruction period, where
    the reconstruction efficiency is low. All entries left of the line are rejected for further analysis.

    \begin{minipage}{\linewidth}
      \centering
      \begin{figure}[H]
        \centering
        \includegraphics[width=0.9\linewidth]{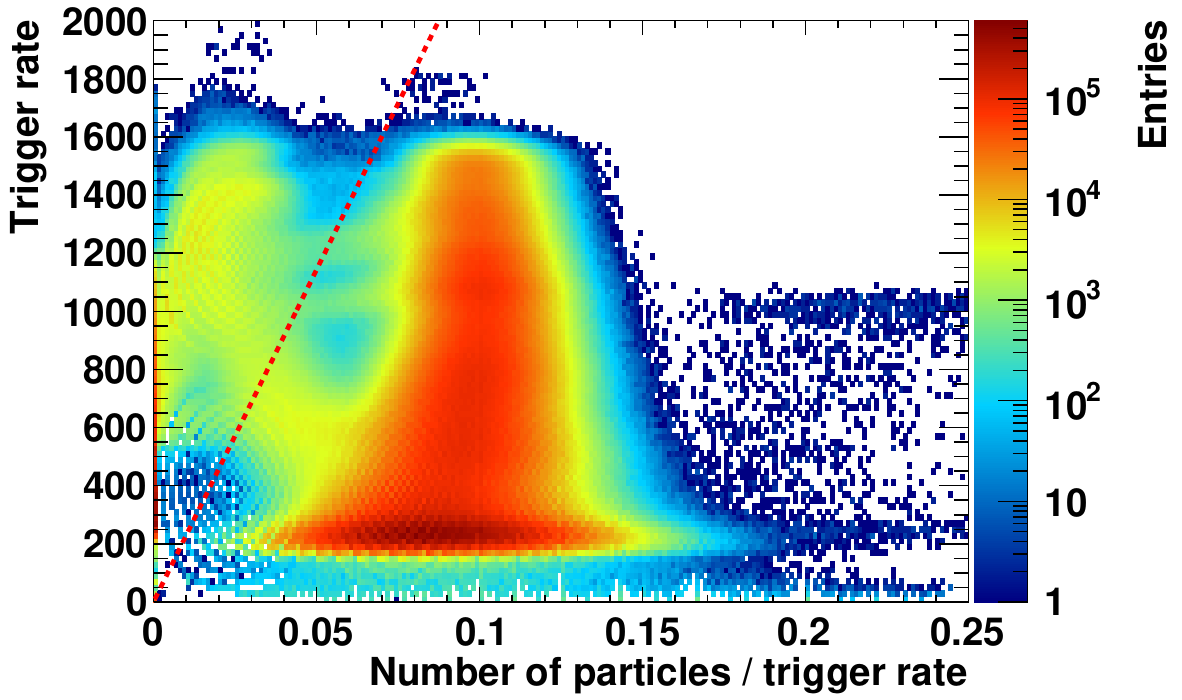}
        \caption{Plot of the trigger rate as function of the ratio reconstructed particles over trigger rate. The red dashed line illustrates the applied cut. All seconds distributed on the left of the line are rejected for further analysis, due to non-nominal reconstruction quality of that time period.}
        \label{fig:rti-particles-vs-trigger}
      \end{figure}
    \end{minipage}

  \item\textbf{Nominal data taking period}\hfill\\
    The AMS-02 collaboration keeps track of time intervals in which the detector was in an unstable condition,
    such as the first weeks of commissioning, all periods where the TRD gas system is refilled, when the DAQ
    was unavailable, etc. These so-called \enquote{bad runs} are excluded when analyzing the ISS data.

  \item\textbf{Nominal trigger performance}\hfill\\
    The amount of recorded events in each second of ISS data must be compatible with the expectation from the trigger rate:
    $f_{\text{trigger}} / N_{\text{events}} > 0.98$

  \item\textbf{Nominal ISS zenith angle}\hfill\\
    The zenith angle of the ISS must be less than \SI{40}{\degree}. Periods when the ISS was rotated must be excluded for analysis.

  \item\textbf{No missed events}\hfill\\
    If there are more than \SI{10}{\percent} of the events missing in a second, exclude the second for analysis.
    There are rare reasons that can lead to missing events, for instance transfer problems in the DAQ boards, or buffer overflows
    on the data reduction boards.

  \item\textbf{Good tracker alignment}\hfill\\
    The tracker alignment of the external tracker planes (Layer 1 and Layer 9) is performed independently by the Perugia and CIEMAT groups.
    Both alignment procedures should yield a similar set of alignment parameters for a given time period. Two important parameters are the
    shifts of the whole tracker plane with respect to the inner tracker: $\Delta\text{X}$ and $\Delta\text{Y}$. If the $\Delta\text{X}$ or $\Delta\text{Y}$ in Layer 1
    between the Perugia / CIEMAT method differs by more than \SI{70}{\micro\meter} the time period is excluded. Likewise if the $\Delta\text{X}$ or
    $\Delta\text{Y}$ in Layer 9 differs by more than \SI{100}{\micro\meter} the time period is excluded as well.

  \item\textbf{Too many events in second}\hfill\\
    If the amount of reconstructed events exceeds 1800 in a second, reject the time period. Events in these conditions are mostly taken
    near the \gls{SAA}~\cite{Kurnosova1962} or in the pole regions, where the detector is filled with low energy particles. These periods should be rejected.

  \item\textbf{Nominal live-time}\hfill\\
    If the detector was busy for more than \SI{50}{\percent} in a second, reject the time period.

  \item\textbf{Nominal DAQ condition}\hfill\\
    If hardware errors were detected in the second, reject it. Hardware errors might be bit-flips in the electronic boards, or duplicated events
    that got recorded, due to problems in the DAQ.

  \item\textbf{Nominal occupancy in TRD}\hfill\\
    Usually the mean number of hits recorded in the TRD in each second is $\approx$~60. If the mean number of hits per second exceeds 1000,
    the second will be rejected. This happens frequently at the edges of the SAA or in the pole regions.
\end{enumerate}

\section{CCMVA input variables}
\label{sec:appendix-ccmva-input-variables}

In the following the remaining nine input variables relevant for the single-track and multi-tracks sample are presented, that were omitted in
\cref{sec:analysis-event-reconstruction-construction-ccmva-estimator}.

\begin{enumerate}

  \item\textbf{Upper TOF charge}\hfill\\

    The TOF clusters belonging to the TOF $\beta$ measurement associated with the reconstructed primary particle (\cref{sec:analysis-event-reconstruction-combining-subdetectors})
    are used to estimate the charge in the upper TOF. TofUpperCharge is computed by taking the average of the available charge measurements in the TOF clusters.

    \begin{figure}[H]
      \begin{subfigure}{0.50\linewidth}
        \includegraphics[width=\linewidth]{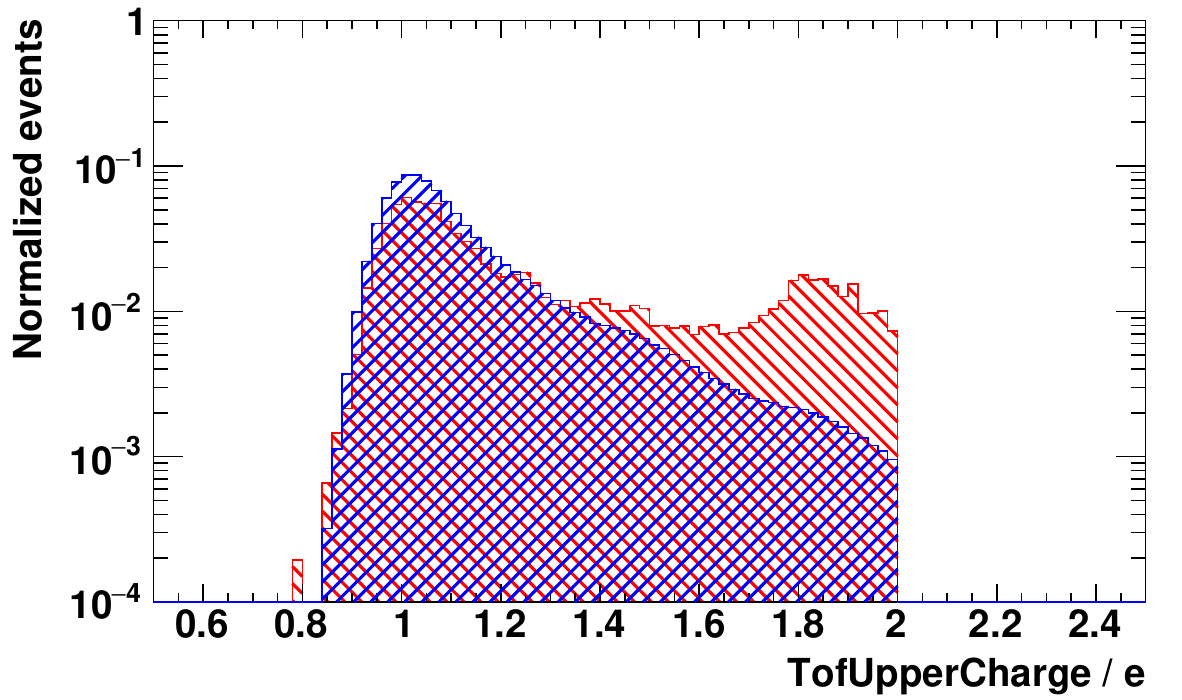}
        \caption{Single-track sample}
      \end{subfigure}
      \hfill
      \begin{subfigure}{0.50\linewidth}
        \includegraphics[width=\linewidth]{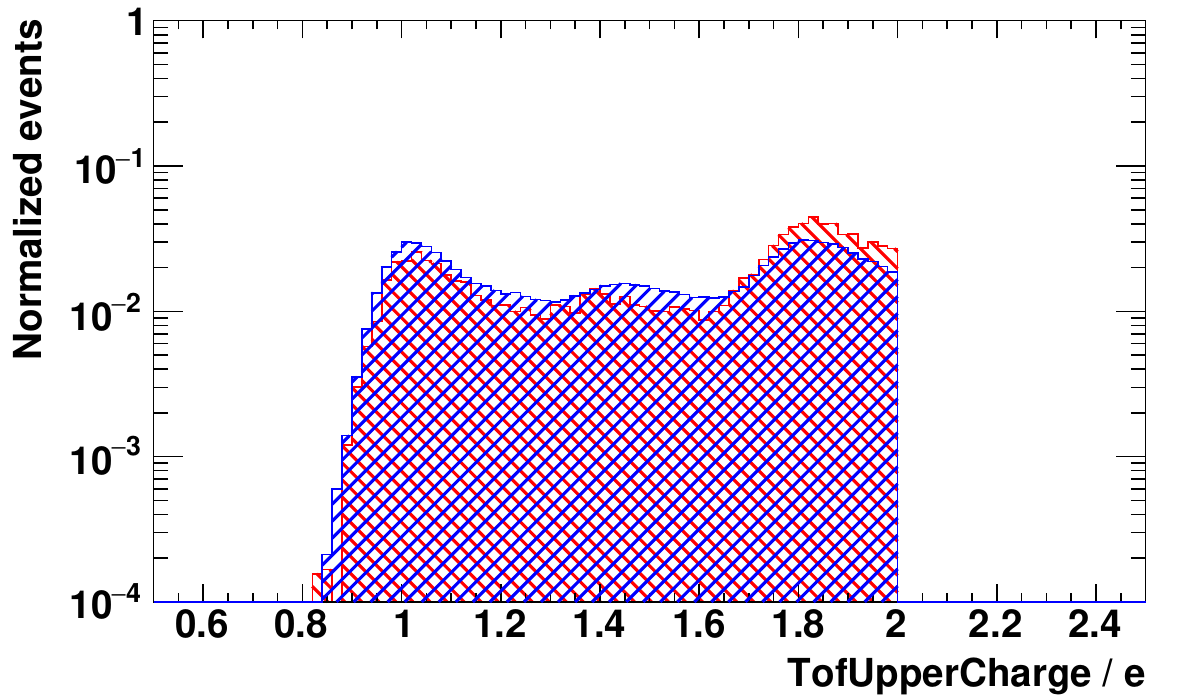}
        \caption{Multi-tracks sample}
      \end{subfigure}
      \caption{Example of TofUpperCharge distribution in the energy bin \SIrange{17.98}{18.99}{\GeV} in the electron Monte-Carlo simulation. The red histogram shows the charge-confused sample (R > 0) and the blue histogram the correct reconstructed sample (R < 0), after applying all preselection, selection and $e^{\pm}$ identification cuts.}
      \label{fig:ccmva-input-tofuppercharge-correct-wrong-sign-mc-comparison}
    \end{figure}

    \Cref{fig:ccmva-input-tofuppercharge-correct-wrong-sign-mc-comparison} shows the TofUpperCharge distribution (as defined in
    \cref{sec:analysis-data-selection-selection-cuts} - \cref{enum:selection-cut-upper-tof-charge}) in an example energy bin
    for both the correct and wrong rigidity sample in the electron Monte-Carlo simulation.

  \item\textbf{Tracker energy deposition ratio}\hfill\\

    As described in \cref{sec:analysis-event-reconstruction-tracker-track}, a maximum of one reconstructed hit per tracker layer is associated to a tracker track.
    If additional particles besides the primary particle traverse the tracker they might produce extra energy depositions in the vicinity of the reconstructed clusters, belonging to the
    primary reconstructed track. This is a useful information regarding the charge-confusion estimation, as they probability of associating the wrong hit to the primary track increases.

    Therefor it is useful to examine each reconstructed hit belonging to the selected tracker track for additional activity in the neighboring strips, adjacent to the Y clusters.
    X clusters are less interesting, as a wrong X cluster cannot change the rigidity measurement. Only Y clusters - reconstructed in the bending plane - can alter the sagitta and thus
    the rigidity measurement.

    The TrkMinSignalRatio quantity captures the information whether extra energy depositions in the vicinity of the reconstructed hits are present, in any of the tracker layers.

    The function $i_{\text{seed}}(l)$ yields the seed strip number of the reconstructed hit attached to the selected tracker track in layer $l$.
    $\text{TrkSignalRatio}(l)$ is defined as the ratio of the sum of the raw amplitudes of the strips in the Y cluster to the sum of all raw
    Y amplitudes on the same ladder in a 10 channel window, all referring to the reconstructed hit Y cluster associated with the selected tracker track in layer $l$:

    \begin{equation*}
      \text{TrkSignalRatio}(l) = \left(\sum_{s = i_{\text{seed}}(l)\ -\ \text{nStripsLeft}}^{i_{\text{seed}}(l)\ +\ \text{nStripsRight}} A(s)\right) \bigg/ \left(\sum_{s = i_{\text{seed}}(l)\ -\ 5}^{i_{\text{seed}}(l)\ +\ 5} A(s)\right).
    \end{equation*}

    TrkMinSignalRatio denotes the minimum signal ratio $\text{TrkSignalRatio}(l)$ in any of the tracker layers $l$, excluding the outer layers 1 and 9:

    \begin{equation*}
      \text{TrkMinSignalRatio} = \min_{l~\in~[2,8]} \text{TrkSignalRatio}(l).
    \end{equation*}

    \Cref{fig:ccmva-input-trkminsignalratio-correct-wrong-sign-mc-comparison} shows the TrkMinSignalRatio distribution in an example energy bin
    for both the correct and wrong rigidity sample in the electron Monte-Carlo simulation.

    \begin{figure}[H]
      \begin{subfigure}{0.50\linewidth}
        \includegraphics[width=\linewidth]{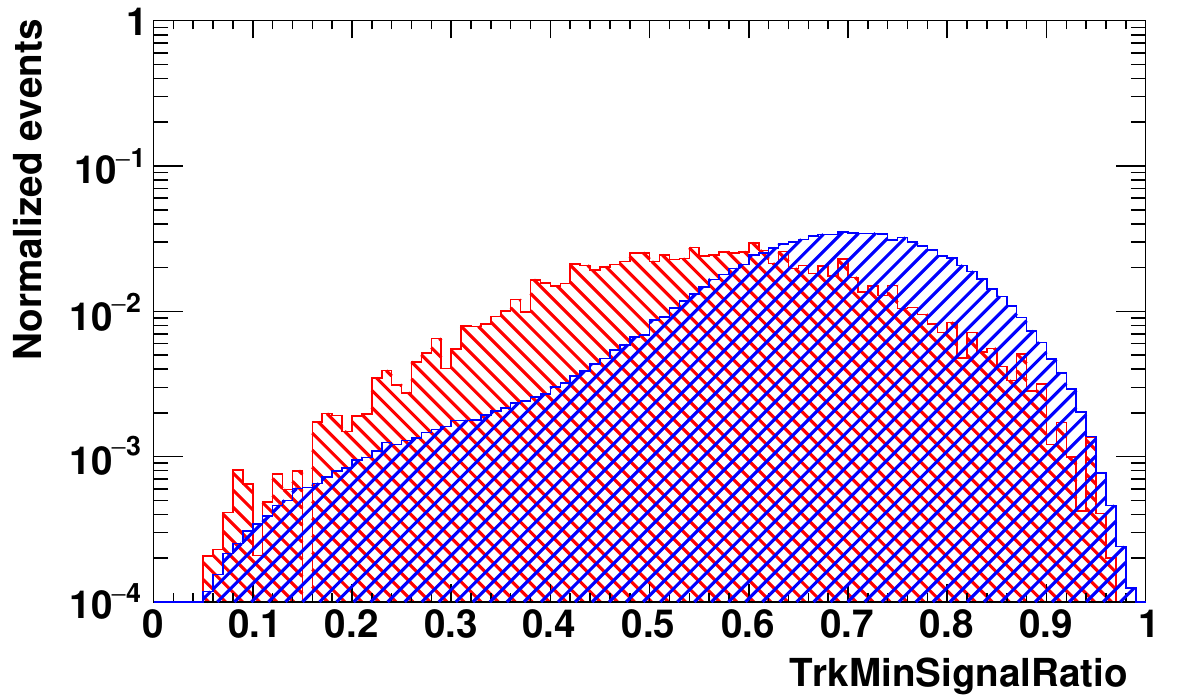}
        \caption{Single-track sample}
      \end{subfigure}
      \hfill
      \begin{subfigure}{0.50\linewidth}
        \includegraphics[width=\linewidth]{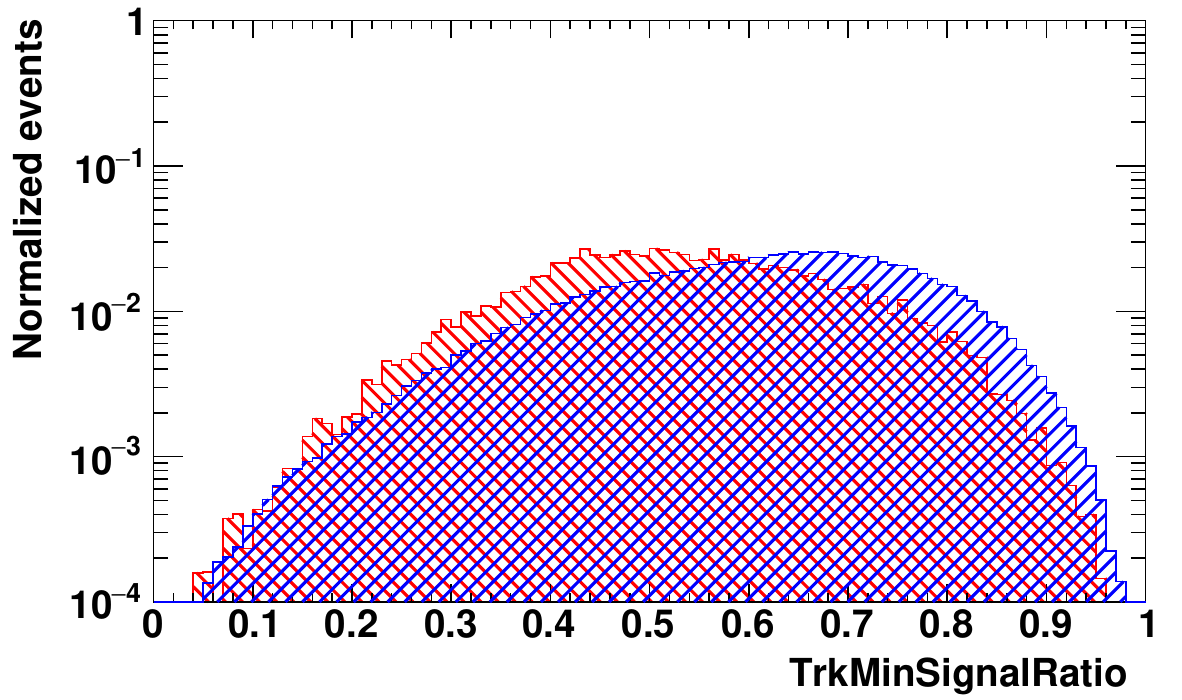}
        \caption{Multi-tracks sample}
      \end{subfigure}
      \caption{Example of TrkMinSignalRatio distribution in the energy bin \SIrange{17.98}{18.99}{\GeV} in the electron Monte-Carlo simulation.}
      \label{fig:ccmva-input-trkminsignalratio-correct-wrong-sign-mc-comparison}
    \end{figure}

  \item\textbf{Additional hits near selected tracker hits}\hfill\\

    TrkMinSignalRatio is used to quantify activity of the silicon strips near the reconstructed clusters that are used for the track fit.
    However it is also possible that there were additional reconstructed clusters nearby, which are not considered in TrkMinSignalRatio: those which are more than 5 channels away from the seed strip.
    Therefore it is necessary to inspect unused Y clusters, that are not associated to the selected tracker track but are close to the used clusters.

    TrkMissClustDist is defined as the closest distance from the cluster used for the track fit to the nearest unused one, in any of the tracker layers.
    The value is set to zero if there are no unused clusters nearby.

    \Cref{fig:ccmva-input-trkmissclustdist-correct-wrong-sign-mc-comparison} shows the TrkMissClustDist distribution in an example energy bin
    for both the correct and wrong rigidity sample in the electron Monte-Carlo simulation.

    \begin{figure}[H]
      \begin{subfigure}{0.50\linewidth}
        \includegraphics[width=\linewidth]{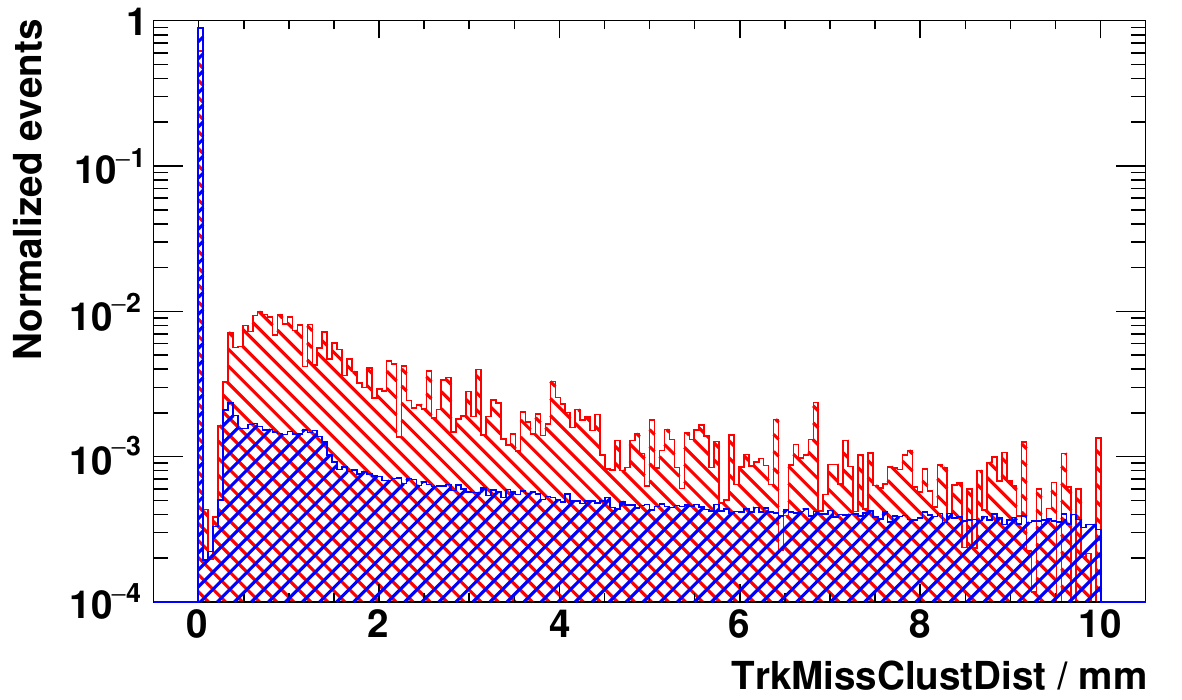}
        \caption{Single-track sample}
      \end{subfigure}
      \hfill
      \begin{subfigure}{0.50\linewidth}
        \includegraphics[width=\linewidth]{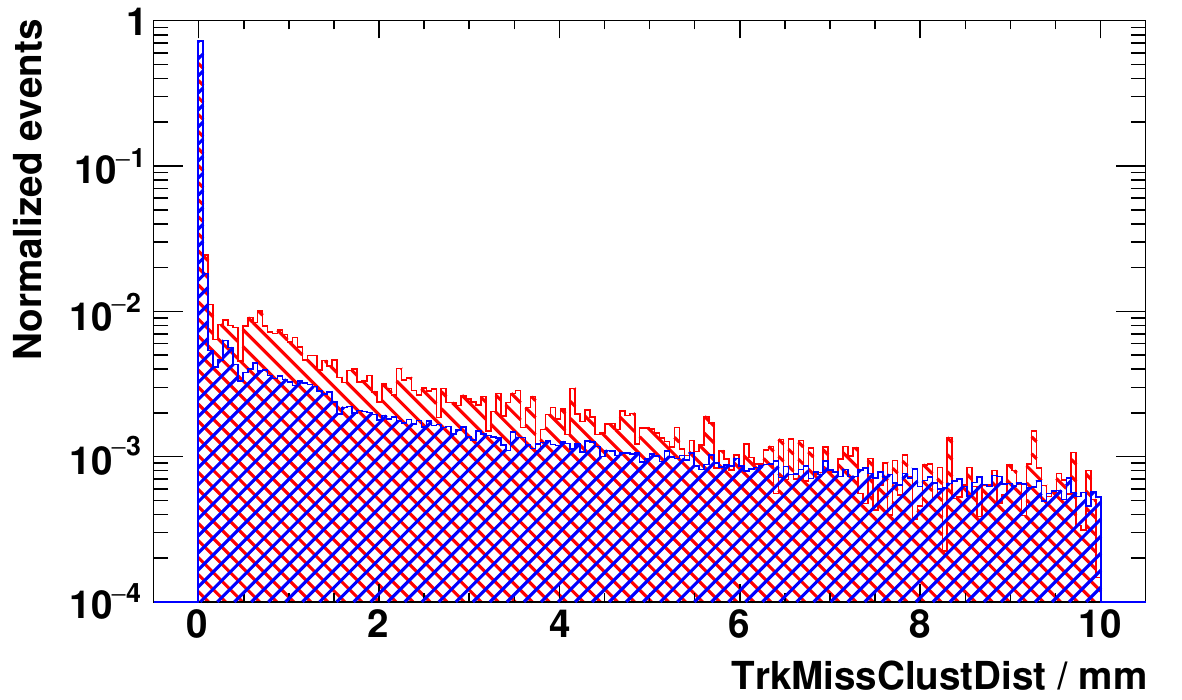}
        \caption{Multi-tracks sample}
      \end{subfigure}
      \caption{Example of TrkMissClustDist distribution in the energy bin \SIrange{17.98}{18.99}{\GeV} in the electron Monte-Carlo simulation.}
      \label{fig:ccmva-input-trkmissclustdist-correct-wrong-sign-mc-comparison}
    \end{figure}

  \item\textbf{Rigidity uncertainty}\hfill\\

    LogTrkRigRelError is defined as the logarithm of the uncertainty of the selected primary tracker track rigidity measurement.

    \begin{equation*}
      \text{LogTrkRigRelError} = \log \left(\sigma_{R_{\text{primary}}}\right)
    \end{equation*}

    \Cref{fig:ccmva-input-logtrkrigrelerror-correct-wrong-sign-mc-comparison} shows the LogTrkRigRelError distribution in an example energy bin
    for both the correct and wrong rigidity sample in the electron Monte-Carlo simulation.

    \begin{figure}[H]
      \begin{subfigure}{0.50\linewidth}
        \includegraphics[width=\linewidth]{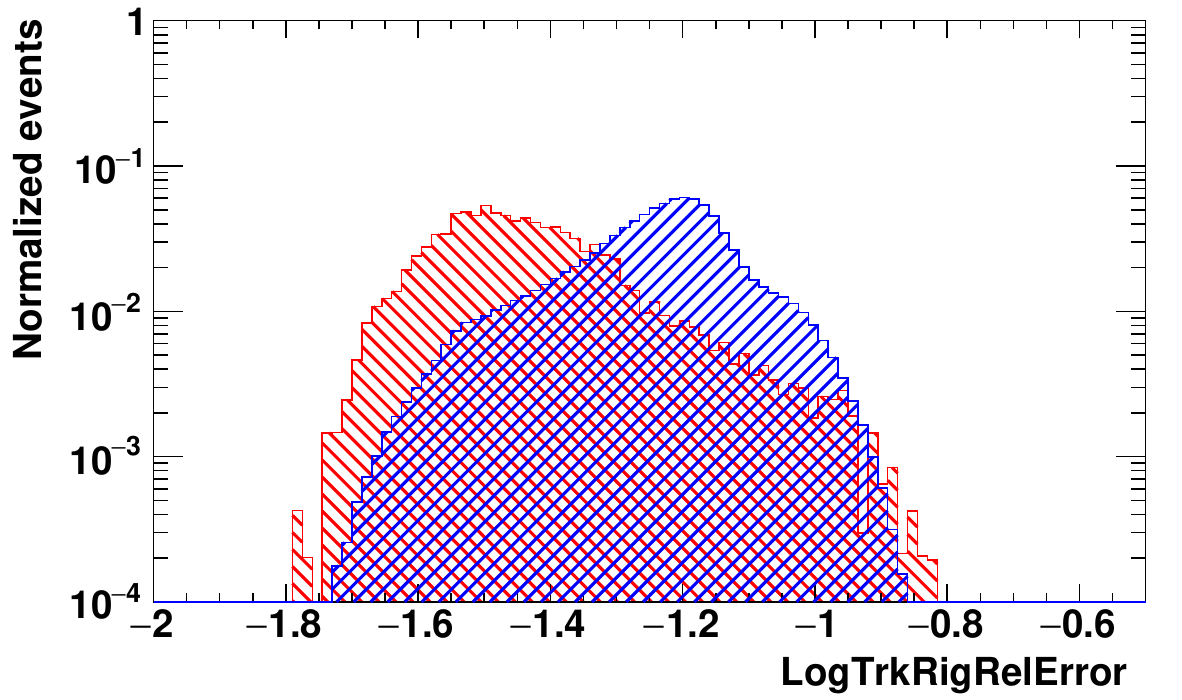}
        \caption{Single-track sample}
      \end{subfigure}
      \hfill
      \begin{subfigure}{0.50\linewidth}
        \includegraphics[width=\linewidth]{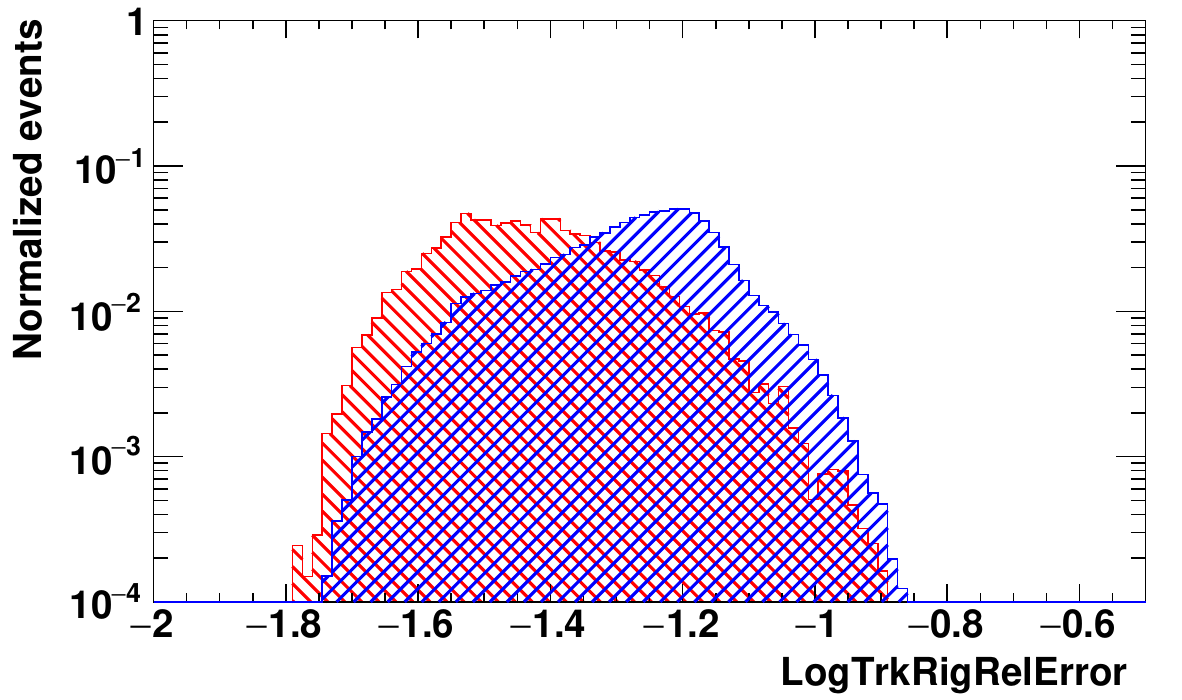}
        \caption{Multi-tracks sample}
      \end{subfigure}
      \caption{Example of LogTrkRigRelError distribution in the energy bin \SIrange{17.98}{18.99}{\GeV} in the electron Monte-Carlo simulation.}
      \label{fig:ccmva-input-logtrkrigrelerror-correct-wrong-sign-mc-comparison}
    \end{figure}

    On an event by event basis this quantity is difficult to use for discrimination in a cut based analysis, as its numerical value depends on the tracker pattern
    and thus is correlated with it. Furthermore the difference between correct and wrong reconstructed sample is not large, but noticeable. Since it offers discrimination
    power it can be included in this estimator, since correlations are properly handled during the construction of the MVA.

  \item\textbf{Tracker hit pattern}\hfill\\

    The so-called tracker pattern is a classification of the topology of the selected tracker track in disjoint categories, depending on the amount
    of outer tracker layers associated to the track. \Cref{tab:tracker-pattern} shows the numerical value of the tracker pattern and its meaning.
    The tracker patterns are sorted by \gls{MDR}. For a single-track electron the tracker pattern where all outer layers
    contribute to the rigidity measurement (L1 + L9) offers the largest detectable rigidity. The so-called full-span tracker pattern has the largest
    lever arm to detect the small displacements of the electron trajectory, due to the bending in the magnetic field, which is in the order of less than
    \SI{20}{\micro\meter} in the TeV regime, between the incident point at L1 and the exit point in L9.

    \bigskip
    \begin{minipage}{\linewidth}
      \centering
      \begin{tabular}{cccc}
        \toprule
                             Layer 1 &    Layer 2 &    Layer 9 & Tracker pattern \\
        \midrule
        \rowcolor{black!20}        x &        (x) &          x & 1 \\
                                     &          x &          x & 2 \\
        \rowcolor{black!20}        x &          x &            & 3 \\
                                     &            &          x & 4 \\
        \rowcolor{black!20}        x &            &            & 5 \\
                                     &          x &            & 6 \\
      \end{tabular}
      \captionof{table}{Tracker pattern classification. The cell entry \enquote{x} denotes that a hit is associated to the track, \enquote{(x)} that a hit might be associated to the track (optional) and an empty cell indicates no hit is associated to the track. The numerical value of the tracker pattern is sorted by \gls{MDR} - a smaller value corresponds to a higher MDR.}
      \label{tab:tracker-pattern}
    \end{minipage}

    \Cref{fig:ccmva-input-trkpattern-correct-wrong-sign-mc-comparison} shows the TrkPattern distribution in an example energy bin
    for both the correct and wrong rigidity sample in the electron Monte-Carlo simulation.

    The two tracker pattern with the smallest MDR are excluded for the analysis, indicated by the two empty bins for tracker pattern 5 and 6.

    \begin{figure}[H]
      \begin{subfigure}{0.50\linewidth}
        \includegraphics[width=\linewidth]{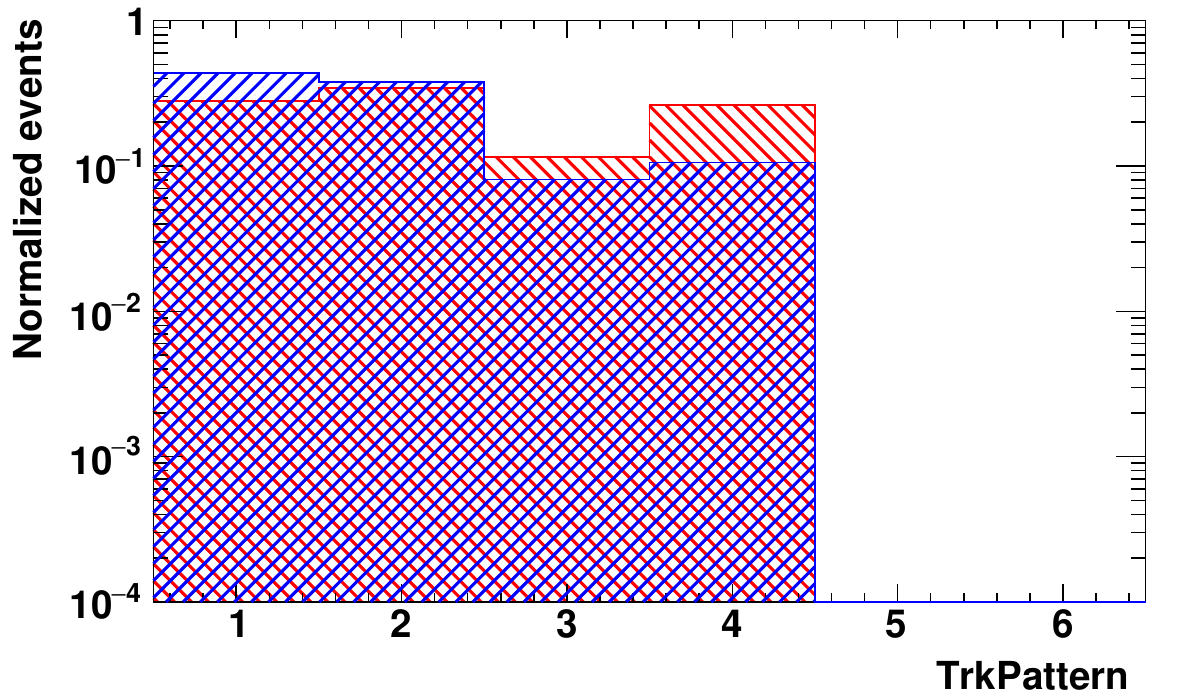}
        \caption{Single-track sample}
      \end{subfigure}
      \hfill
      \begin{subfigure}{0.50\linewidth}
        \includegraphics[width=\linewidth]{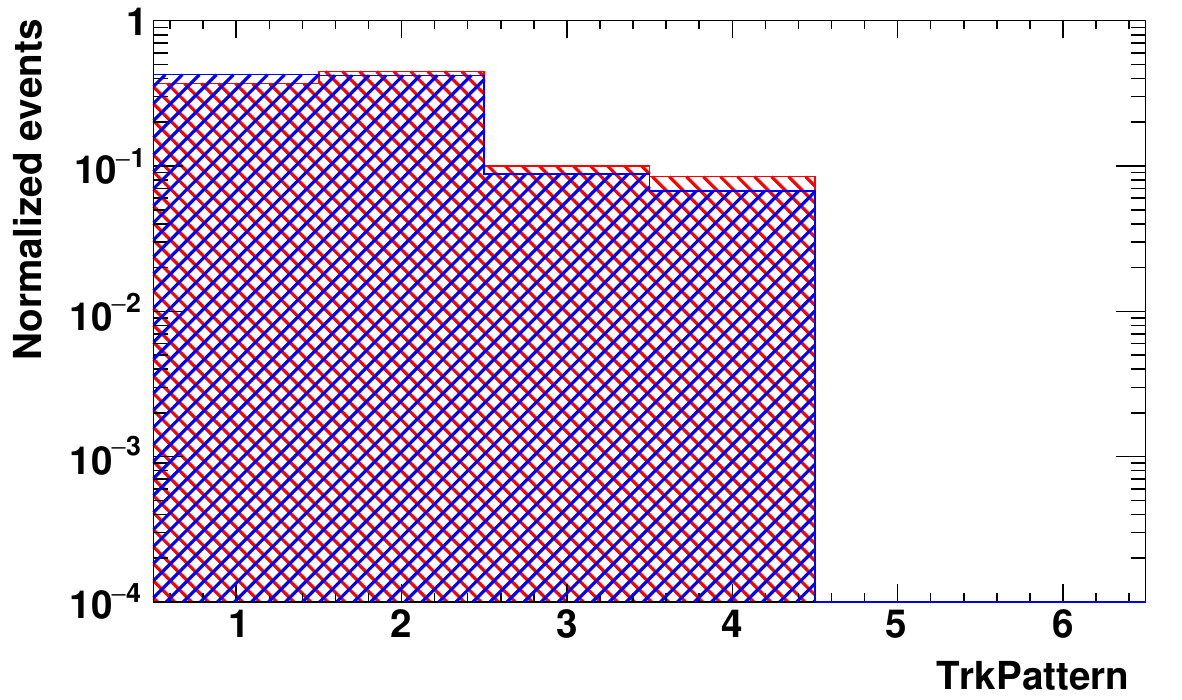}
        \caption{Multi-tracks sample}
      \end{subfigure}
      \caption{Example of TrkPattern distribution in the energy bin \SIrange{17.98}{18.99}{\GeV} in the electron Monte-Carlo simulation.}
      \label{fig:ccmva-input-trkpattern-correct-wrong-sign-mc-comparison}
    \end{figure}

  \medskip
  \item\textbf{Tracker fit algorithm compatibility}\hfill\\

    The Choutko algorithm - implemented in the AMS-02 track reconstruction - is an iterative track fit procedure, based on a fast global matrix inversion~\cite{Hart1984,Choutko2003}.
    An alternative algorithm - the Chikanian algorithm - aims to improve the multiple scattering treatment at low energies. However the Chikanian~fit~\cite{Chikanian1996}
    is also more sensitive to interactions in the detector, therefore it offers a handle to discriminate between events with interactions and those without.

    TrkChoutkoVsChikMatching is defined as asymmetry between the two track fit methods:

    \begin{equation*}
      \text{TrkChoutkoVsChikMatching} = \left(\frac{1}{R_{\text{choutko}}} - \frac{1}{R_{\text{chikanian}}}\right) \bigg/
                                        \left(\frac{1}{R_{\text{choutko}}} + \frac{1}{R_{\text{chikanian}}}\right).
    \end{equation*}

    \Cref{fig:ccmva-input-trkchoutkovschikmatching-correct-wrong-sign-mc-comparison} shows the TrkChoutkoVsChikMatching distribution in an example energy bin
    for both the correct and wrong rigidity sample in the electron Monte-Carlo simulation.

    \begin{figure}[H]
      \begin{subfigure}{0.45\linewidth}
        \includegraphics[width=\linewidth]{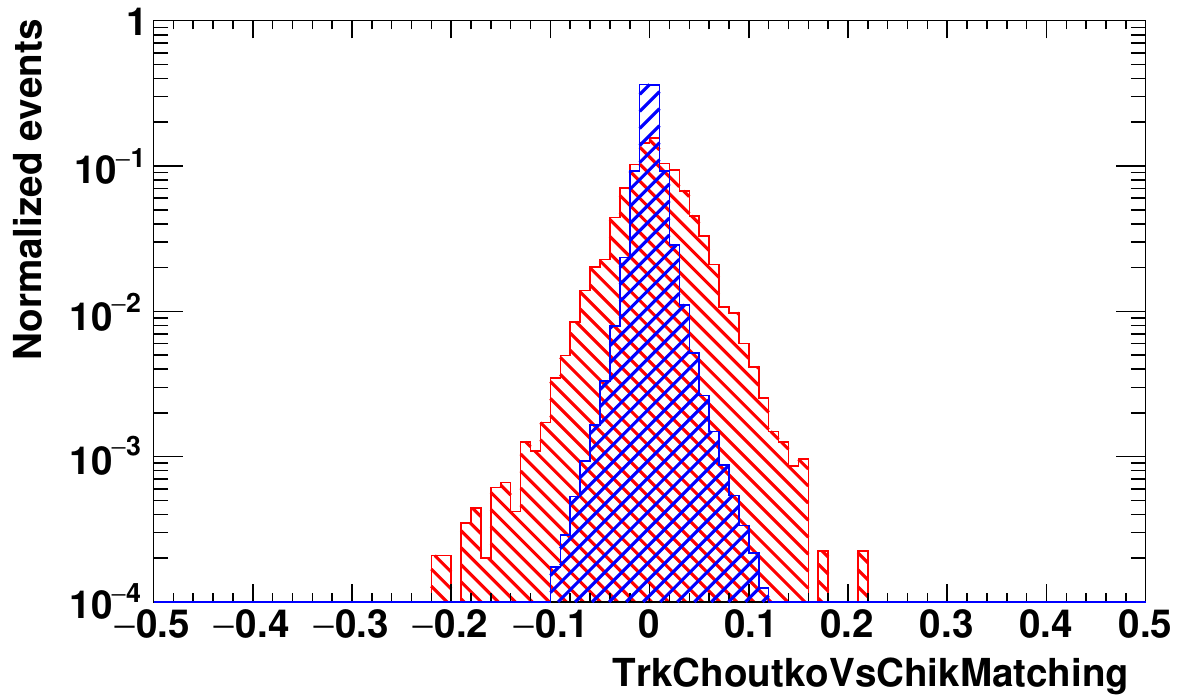}
        \caption{Single-track sample}
      \end{subfigure}
      \hfill
      \begin{subfigure}{0.45\linewidth}
        \includegraphics[width=\linewidth]{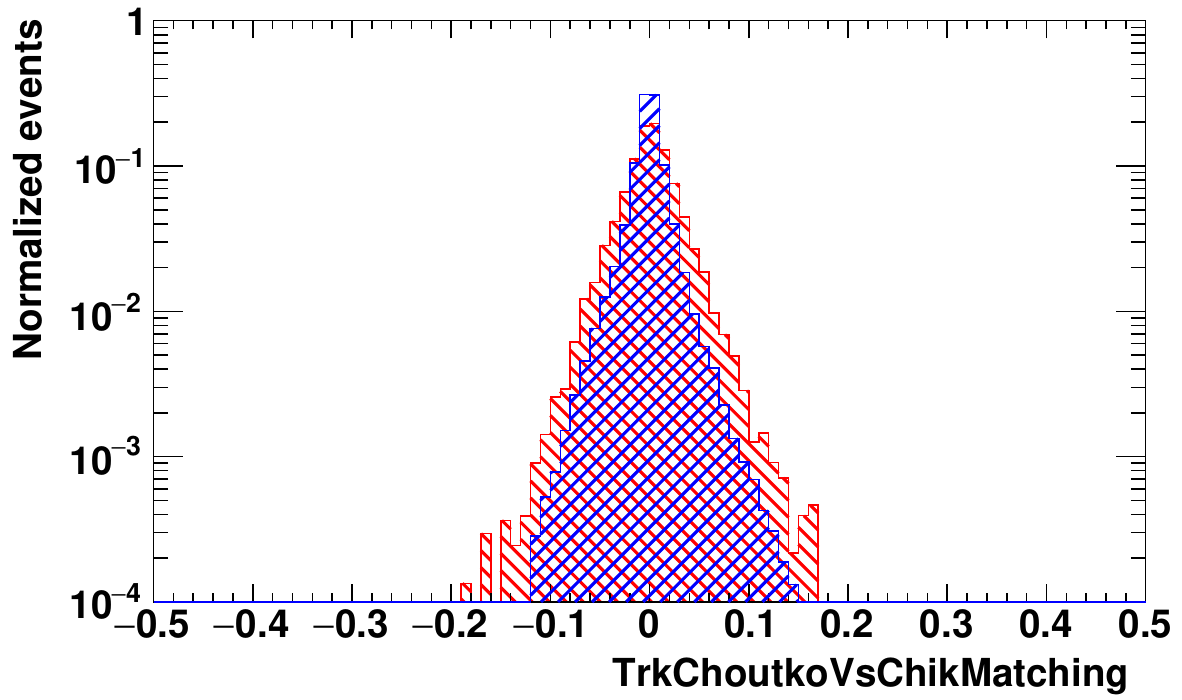}
        \caption{Multi-tracks sample}
      \end{subfigure}
      \caption{Example of TrkChoutkoVsChikMatching distribution in the energy bin \SIrange{17.98}{18.99}{\GeV} in the electron Monte-Carlo simulation.}
      \label{fig:ccmva-input-trkchoutkovschikmatching-correct-wrong-sign-mc-comparison}
    \end{figure}

  \item\textbf{Tracker upper/lower half rigidity matching}\hfill\\

    \begin{figure}[H]
      \begin{subfigure}{0.45\linewidth}
        \includegraphics[width=\linewidth]{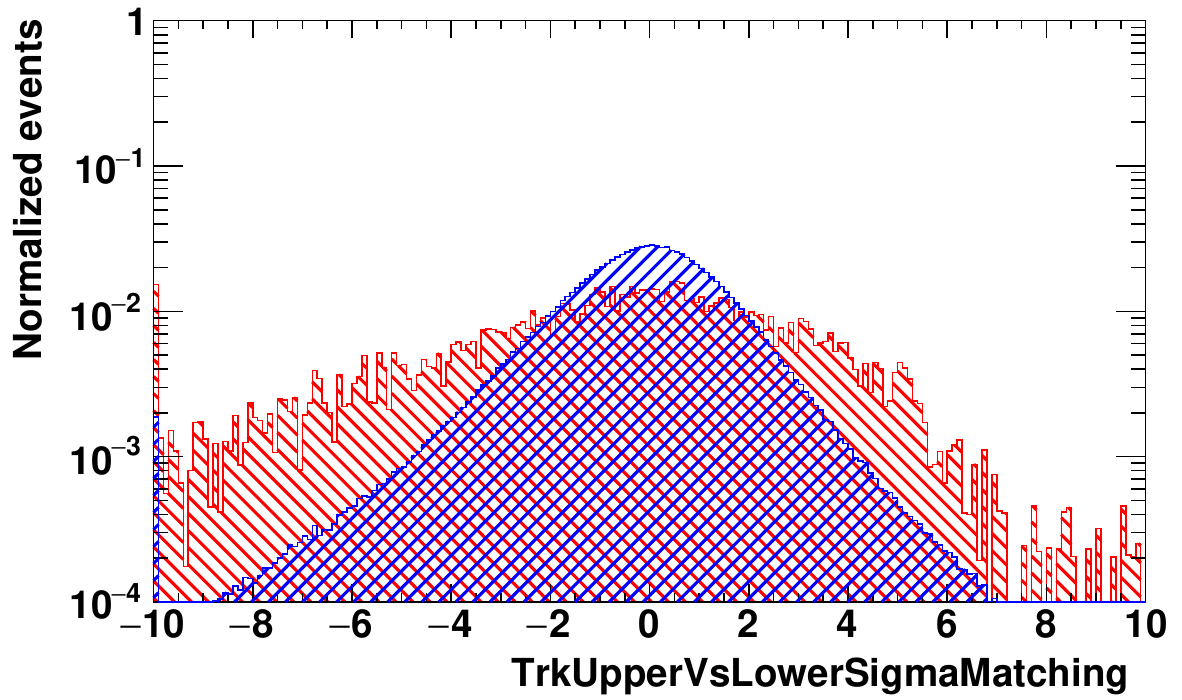}
        \caption{Single-track sample}
      \end{subfigure}
      \hfill
      \begin{subfigure}{0.45\linewidth}
        \includegraphics[width=\linewidth]{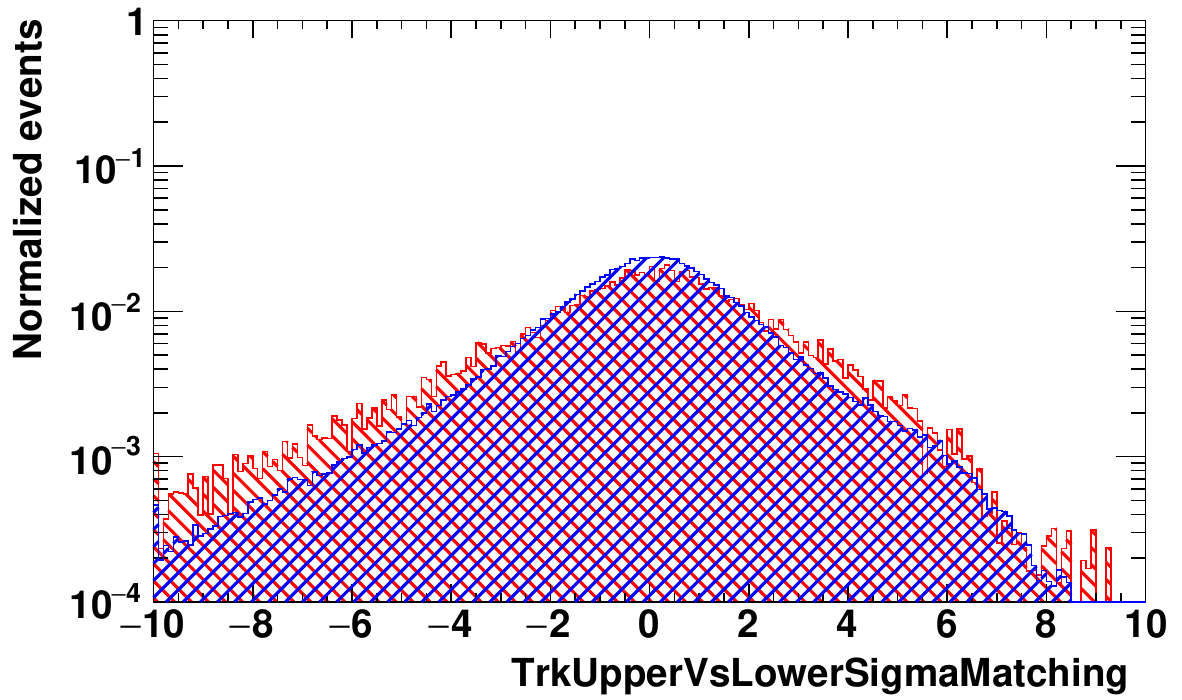}
        \caption{Multi-tracks sample}
      \end{subfigure}
      \caption{Example of TrkUpperVsLowerSigmaMatching distribution in the energy bin \SIrange{17.98}{18.99}{\GeV} in the electron Monte-Carlo simulation.}
      \label{fig:ccmva-input-trkuppervslowersigmamatching-correct-wrong-sign-mc-comparison}
    \end{figure}

    Another way to detect interactions in the inner tracker is to perform the track fit once only the upper half of the inner tracker and another time using only the lower half.
    TrkUpperVsLowerSigmaMatching is defined as the sagitta difference between these track fits, weighted by the quadratic sum of the corresponding relative uncertainties:

    \begin{equation*}
      \text{TrkUpperVsLowerSigmaMatching} = \frac{\sign(R_{\text{primary}}) \left(\frac{1}{R_{\text{upper}}} - \frac{1}{R_{\text{lower}}}\right)}
                                            {\sqrt{\left(\frac{\sigma_{R_{\text{upper}}}}{R_{\text{upper}}}\right)^2+\left(\frac{\sigma_{R_{\text{lower}}}}{R_{\text{lower}}}\right)^2}}.
    \end{equation*}

    \Cref{fig:ccmva-input-trkuppervslowersigmamatching-correct-wrong-sign-mc-comparison} shows the TrkUpperVsLowerSigmaMatching distribution in an example energy bin
    for both the correct and wrong rigidity sample in the electron Monte-Carlo simulation.

  \item\textbf{Tracker all/inner layers rigidity matching}\hfill\\

    To study the influence of the outer tracker layers to the reconstructed rigidity, the track fit is performed once only the inner tracker and once with the external layers included.
    The weighted sagitta difference is defined as TrkAllVsInnerMatching:

    \begin{equation*}
      \text{TrkAllVsInnerMatching} = \frac{\sign(R_{\text{primary}}) \left(\frac{1}{R_{\text{all}}} - \frac{1}{R_{\text{inner}}}\right)}
                                     {\frac{1}{2} \sqrt{\left(\frac{\sigma_{R_{\text{all}}}}{R_{\text{all}}}\right)^2+\left(\frac{\sigma_{R_{\text{inner}}}}{R_{\text{inner}}}\right)^2}}
    \end{equation*}

    \begin{figure}[H]
      \begin{subfigure}{0.50\linewidth}
        \includegraphics[width=\linewidth]{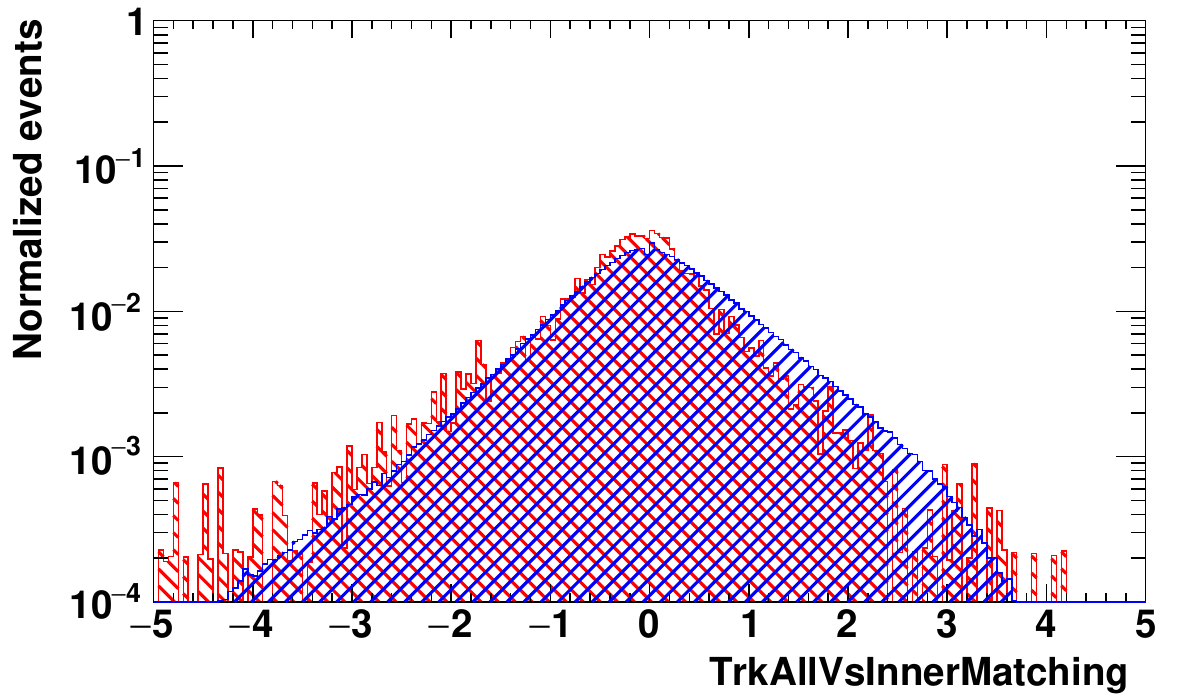}
        \caption{Single-track sample}
      \end{subfigure}
      \hfill
      \begin{subfigure}{0.50\linewidth}
        \includegraphics[width=\linewidth]{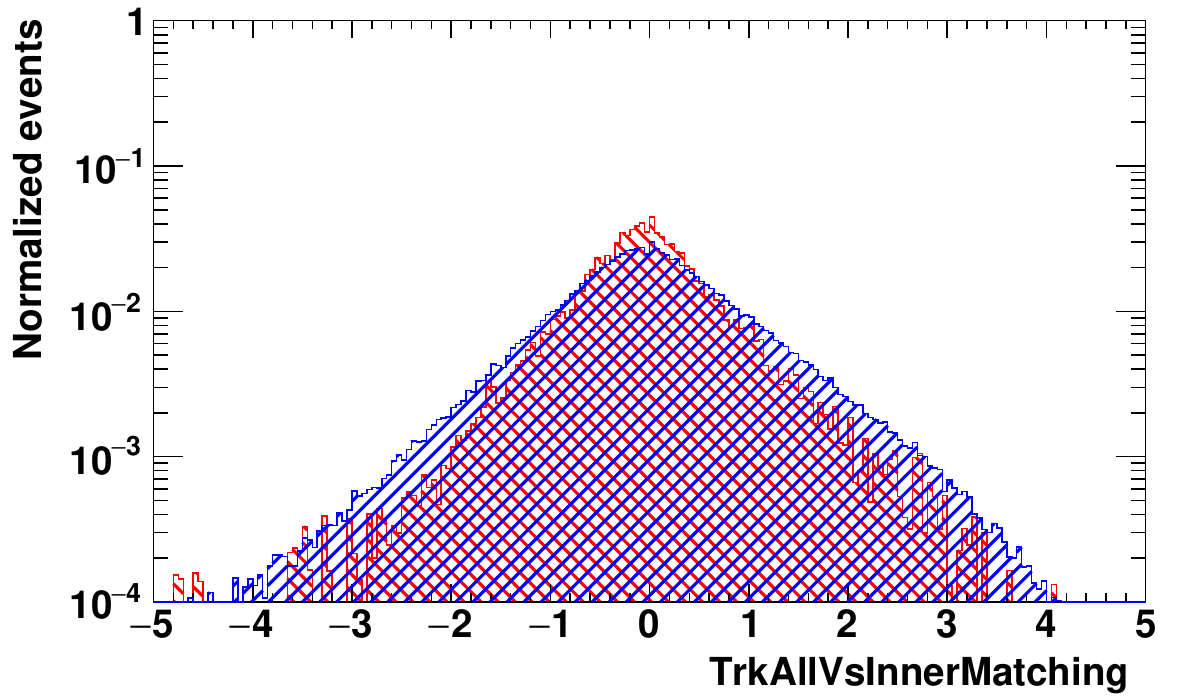}
        \caption{Multi-tracks sample}
      \end{subfigure}
      \caption{Example of TrkAllVsInnerMatching distribution in the energy bin \SIrange{17.98}{18.99}{\GeV} in the electron Monte-Carlo simulation.}
      \label{fig:ccmva-input-trkallvsinnermatching-correct-wrong-sign-mc-comparison}
    \end{figure}

    \Cref{fig:ccmva-input-trkallvsinnermatching-correct-wrong-sign-mc-comparison} shows the TrkAllVsInnerMatching distribution in an example energy bin
    for both the correct and wrong rigidity sample in the electron Monte-Carlo simulation.

  \clearpage
  \item\textbf{Number of inner tracker planes}\hfill\\

    The number of inner layers that contribute a Y hit to the track fit - TrkNLayersInnerY - enters the MVA as input quantity. It is correlated with other variables that enter
    the MVA (TrkAllVsInnerMatching, TrkUpperVsLowerSigmaMatching). Nevertheless it can be used as input variable, since the MVA can exploit this correlation to
    achieve more discrimination power.

    \Cref{fig:ccmva-input-trknlayersinnery-correct-wrong-sign-mc-comparison} shows the TrkNLayersInnerY distribution in an example energy bin
    for both the correct and wrong rigidity sample in the electron Monte-Carlo simulation.

    \begin{figure}[H]
      \begin{subfigure}{0.50\linewidth}
        \includegraphics[width=\linewidth]{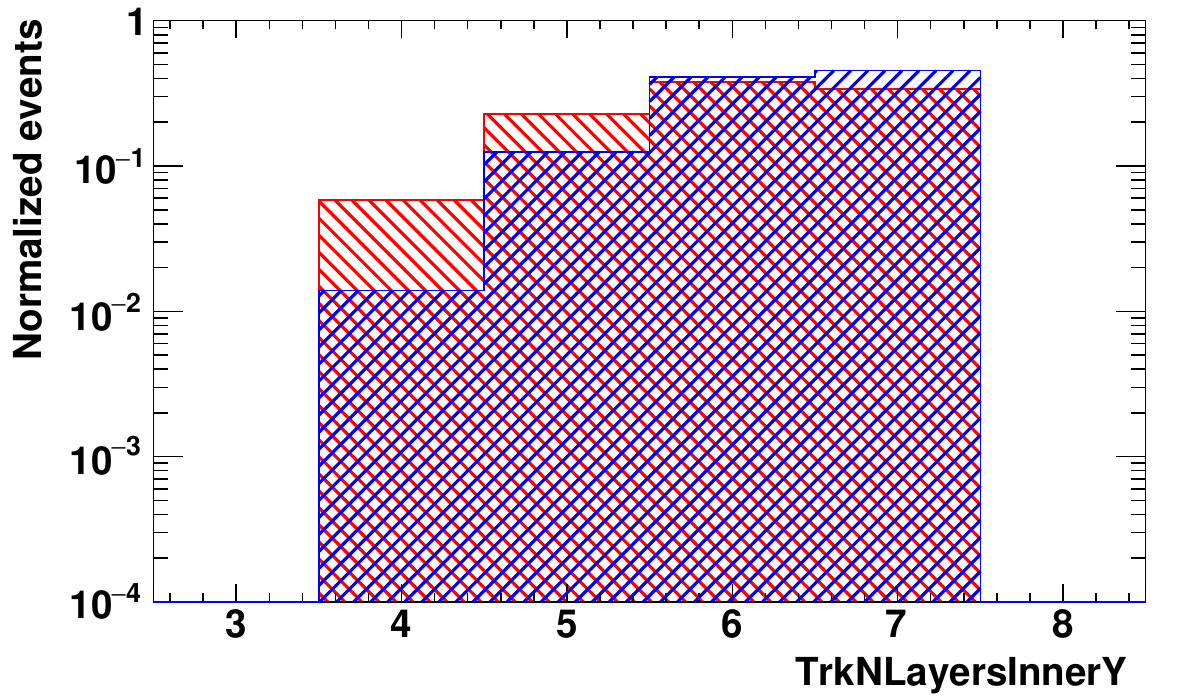}
        \caption{Single-track sample}
      \end{subfigure}
      \hfill
      \begin{subfigure}{0.50\linewidth}
        \includegraphics[width=\linewidth]{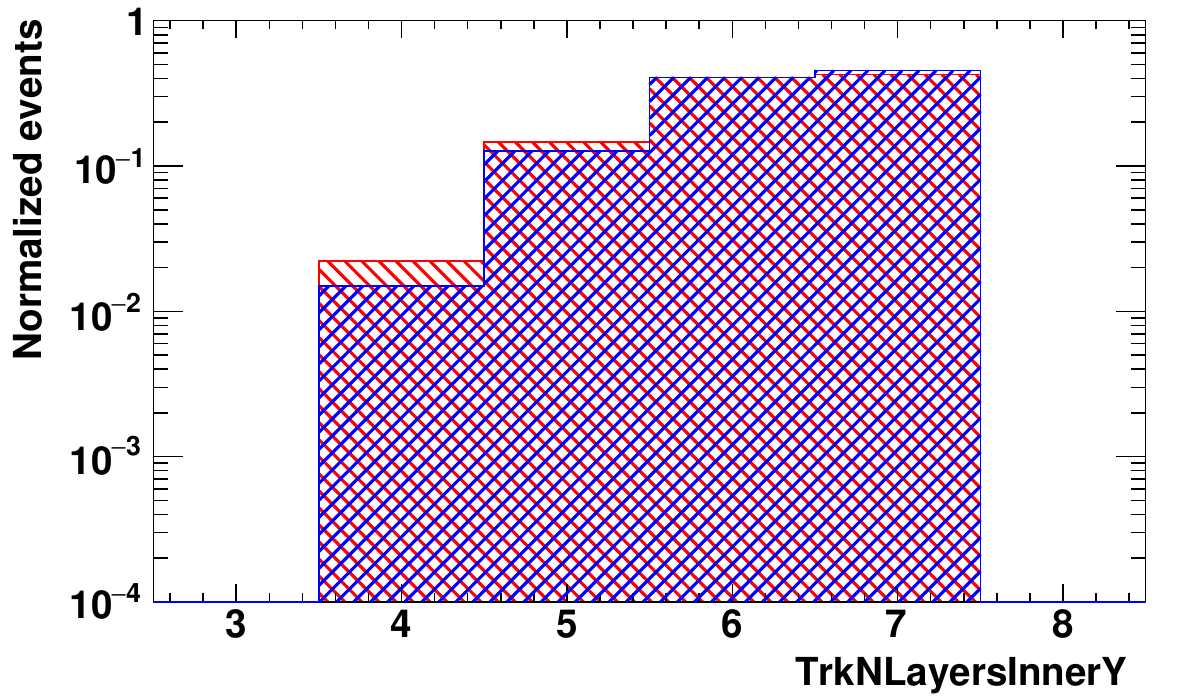}
        \caption{Multi-tracks sample}
      \end{subfigure}
      \caption{Example of TrkNLayersInnerY distribution in the energy bin \SIrange{17.98}{18.99}{\GeV} in the electron Monte-Carlo simulation.}
      \label{fig:ccmva-input-trknlayersinnery-correct-wrong-sign-mc-comparison}
    \end{figure}
\end{enumerate}

Three variables were omitted in the description in \cref{sec:analysis-event-reconstruction-construction-ccmva-estimator}, which enter the MVA estimator for the multi-tracks sample:

\begin{enumerate}
  \setcounter{enumi}{9}
  \item\textbf{Number of tracker tracks}\hfill\\

    Similar to TrkNLayersInnerY, the number of tracker tracks in the event TrkNumTracks has no discrimination power by itself. It still enters the MVA as the MVA can exploit the correlation
    between this variable and all other input quantities to achieve more discrimination power.

    \begin{figure}[H]
      \centering
      \includegraphics[width=0.50\linewidth]{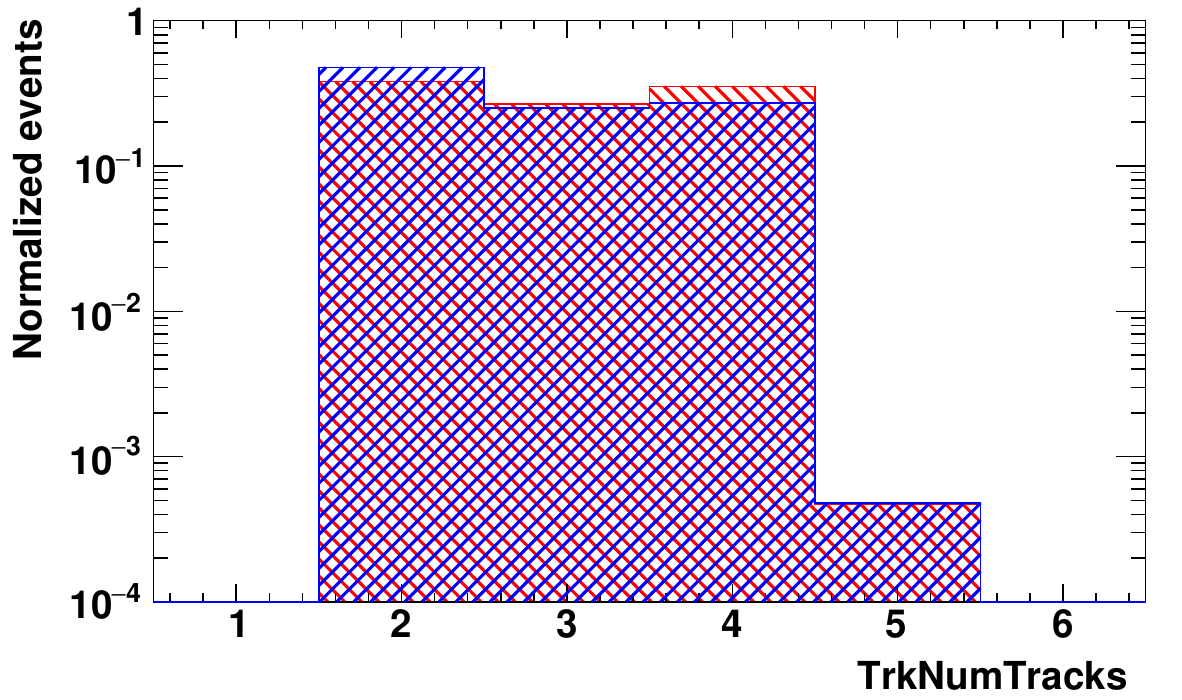}
      \caption{Example of TrkNumTracks distribution in the energy bin \SIrange{17.98}{18.99}{\GeV} in the electron Monte-Carlo simulation.}
      \label{fig:ccmva-input-trknumtracks-correct-wrong-sign-mc-comparison}
    \end{figure}

    \Cref{fig:ccmva-input-trknumtracks-correct-wrong-sign-mc-comparison} shows the TrkNumTracks distribution in an example energy bin
    for both the correct and wrong rigidity sample in the electron Monte-Carlo simulation. Note that only events with more than one reconstructed tracker track enter the distribution, explaining
    why the first bin with exactly one track is empty.

  \item\textbf{Tracker track top distance x-direction \&} \vspace{-0.75em} \item\textbf{Tracker track top distance y-direction}\hfill\\

    Using an analogous definition as TrkMinDXBottom / TrkMinDYBottom all tracker tracks are extrapolated to the TRD center position at \SIvarEquals{$z$}{113.55}{\centi\meter}.

    \begin{figure}[H]
      \centering
      \begin{subfigure}{0.62\linewidth}
        \includegraphics[width=\linewidth]{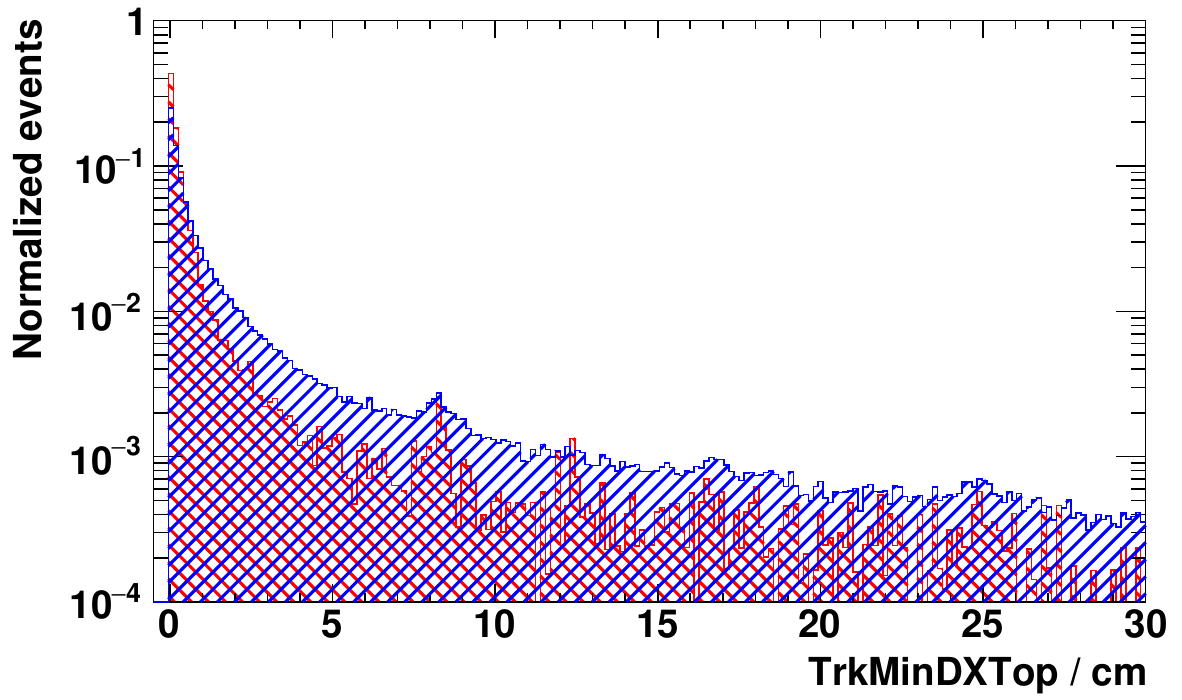}
        \caption{TrkMinDXTop}
      \end{subfigure}
      \hfill
      \begin{subfigure}{0.62\linewidth}
        \includegraphics[width=\linewidth]{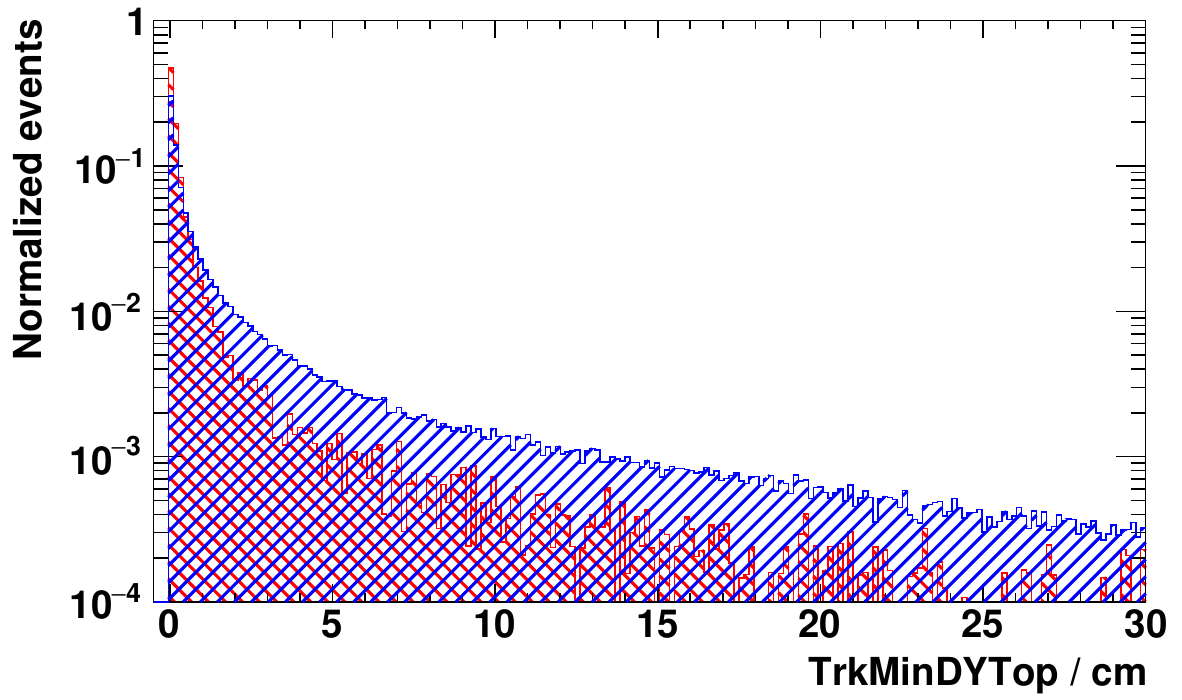}
        \caption{TrkMinDYTop}
      \end{subfigure}
      \caption{Example of TrkMinDXTop / TrkMinDYTop distributions in the energy bin \SIrange{17.98}{18.99}{\GeV} in the electron Monte-Carlo simulation.}
      \label{fig:ccmva-input-trkmindxdytop-correct-wrong-sign-mc-comparison}
    \end{figure}

    \Cref{fig:ccmva-input-trkmindxdytop-correct-wrong-sign-mc-comparison} shows the TrkMinDXTop / TrkMinDYTop distributions in an example energy bin
    for both the correct and wrong rigidity sample in the electron Monte-Carlo simulation. The separation power of this quantity is larger at low energies, but
    vanishes at medium energies above \SI{10}{\GeV}.
\end{enumerate}

\begin{figure}[H]
  \begin{subfigure}{0.32\linewidth}
    \includegraphics[width=\linewidth]{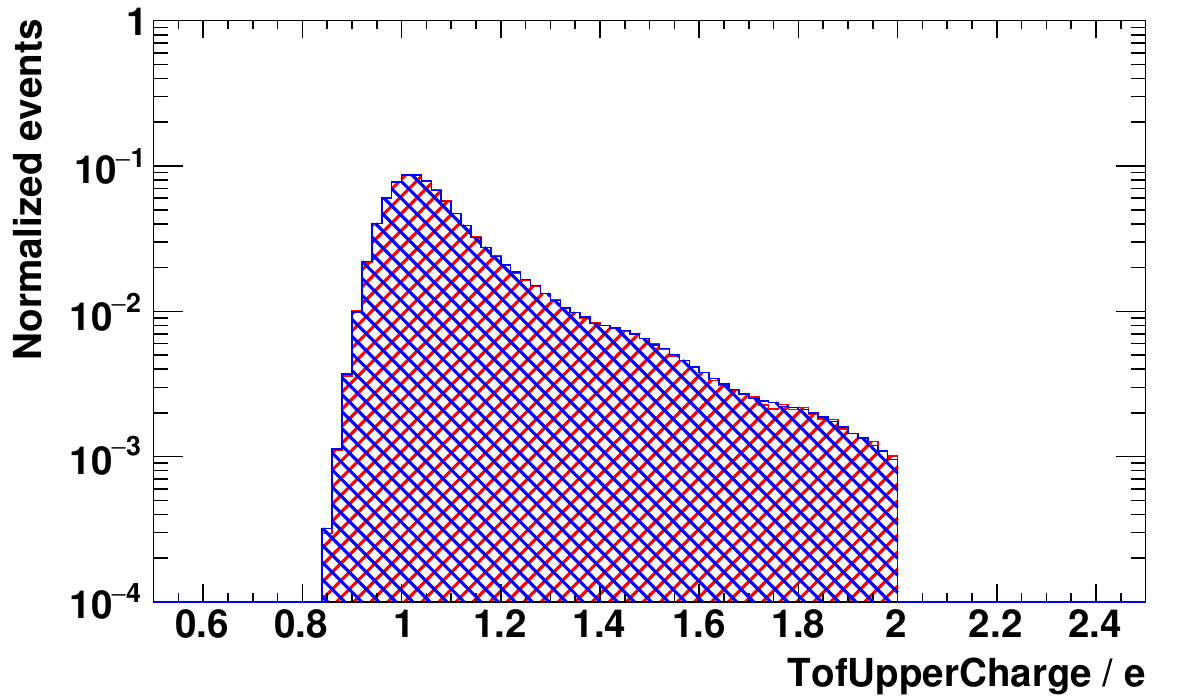}
  \end{subfigure}
  \hfill
  \begin{subfigure}{0.32\linewidth}
    \includegraphics[width=\linewidth]{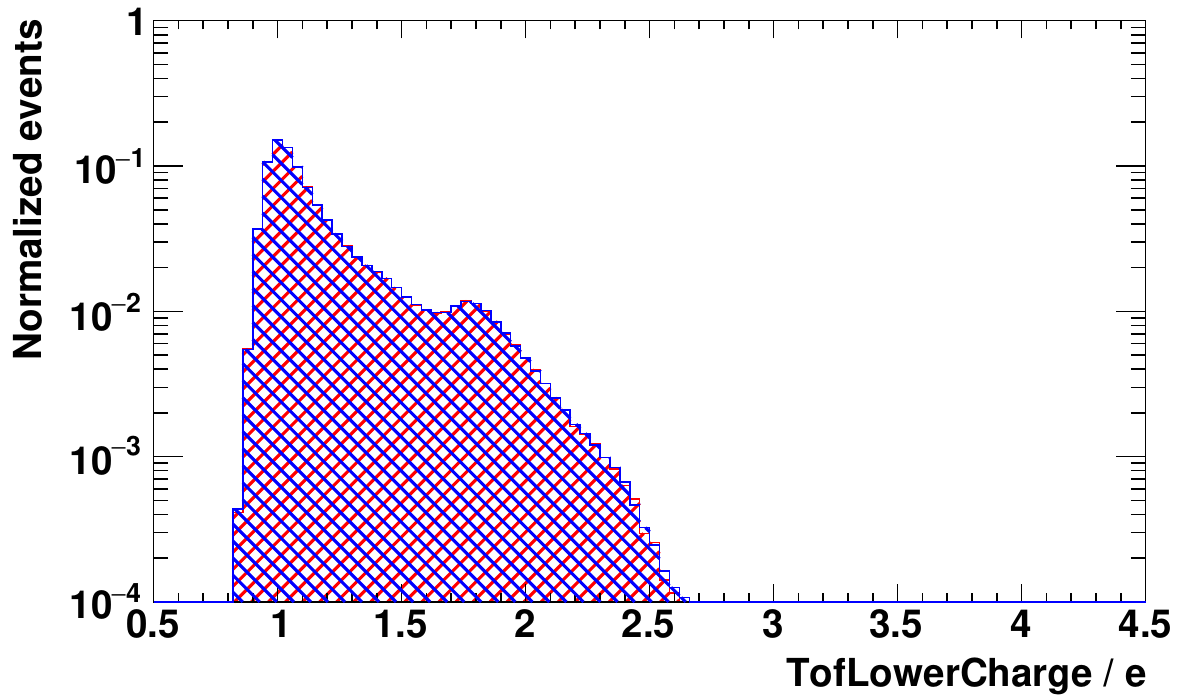}
  \end{subfigure}
  \hfill
  \begin{subfigure}{0.32\linewidth}
    \includegraphics[width=\linewidth]{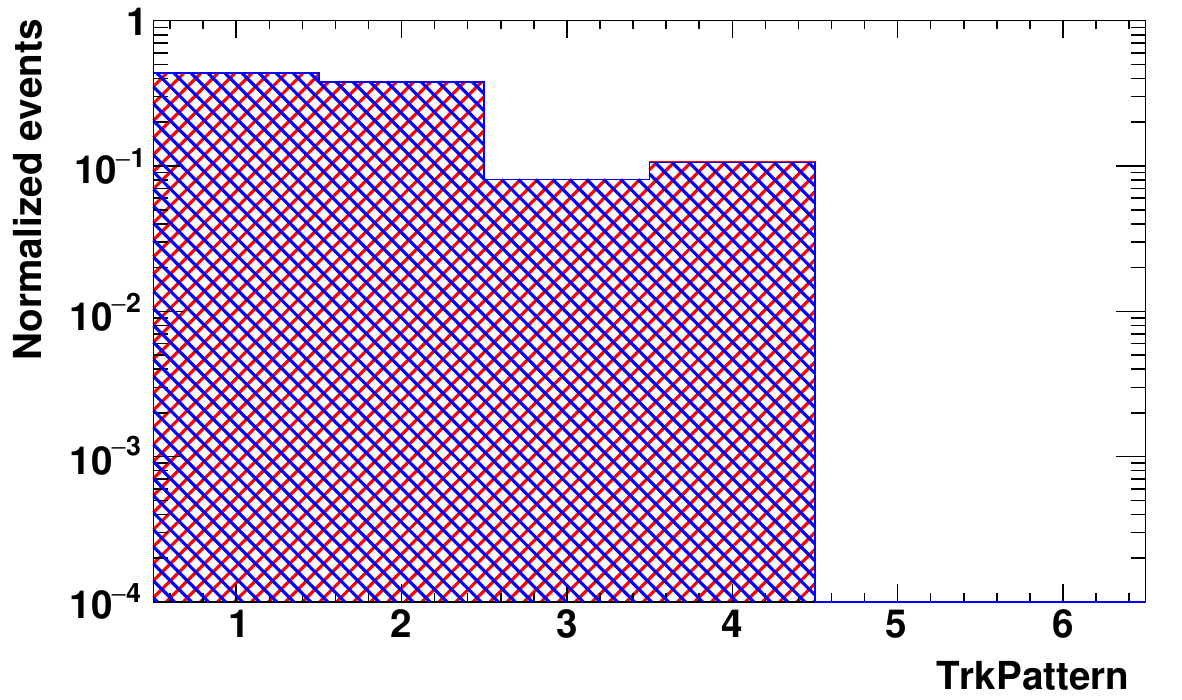}
  \end{subfigure}
  \hfill
  \begin{subfigure}{0.32\linewidth}
    \includegraphics[width=\linewidth]{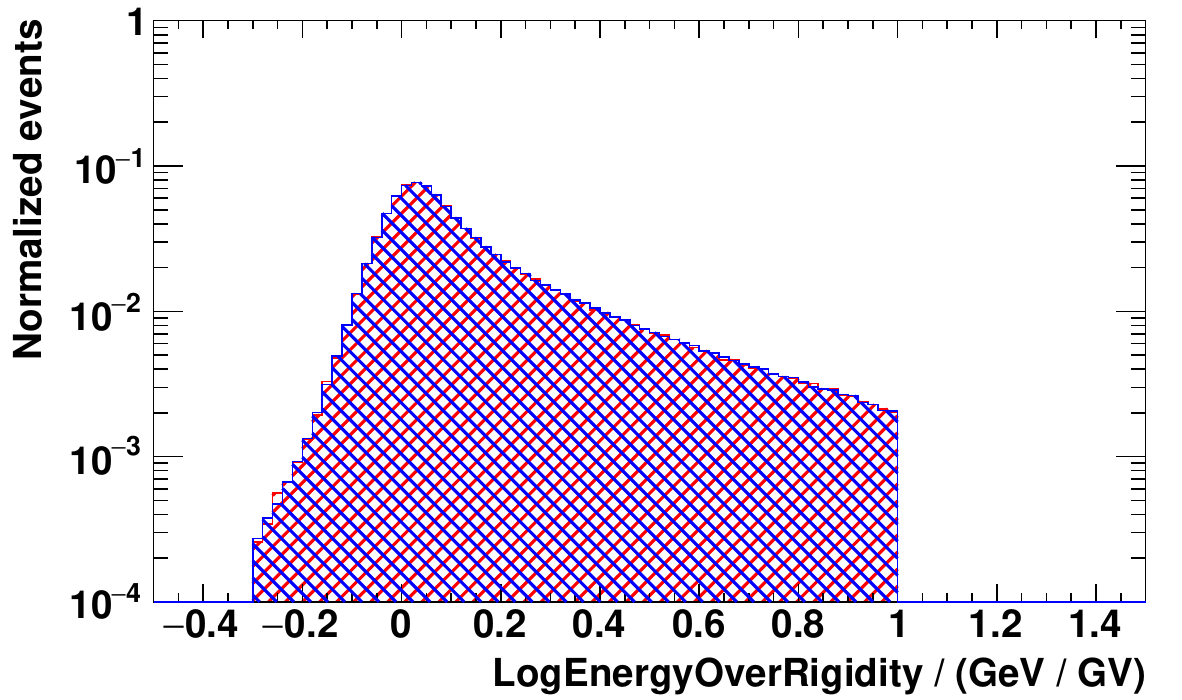}
  \end{subfigure}
  \hfill
  \begin{subfigure}{0.32\linewidth}
    \includegraphics[width=\linewidth]{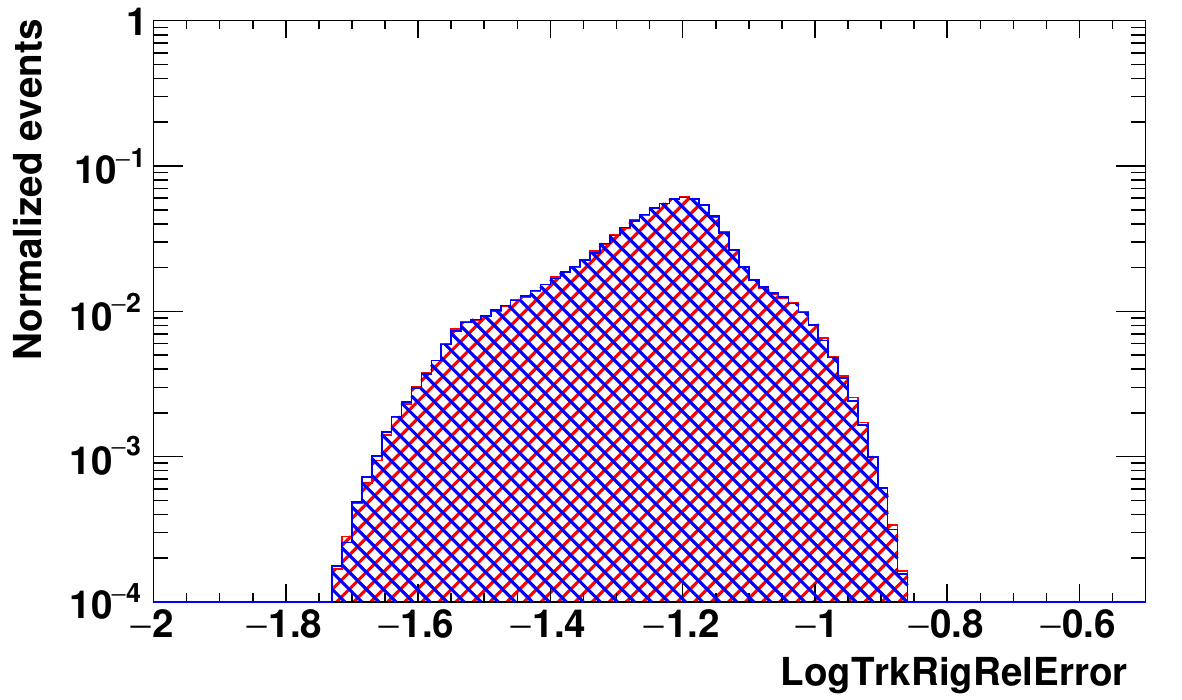}
  \end{subfigure}
  \hfill
  \begin{subfigure}{0.32\linewidth}
    \includegraphics[width=\linewidth]{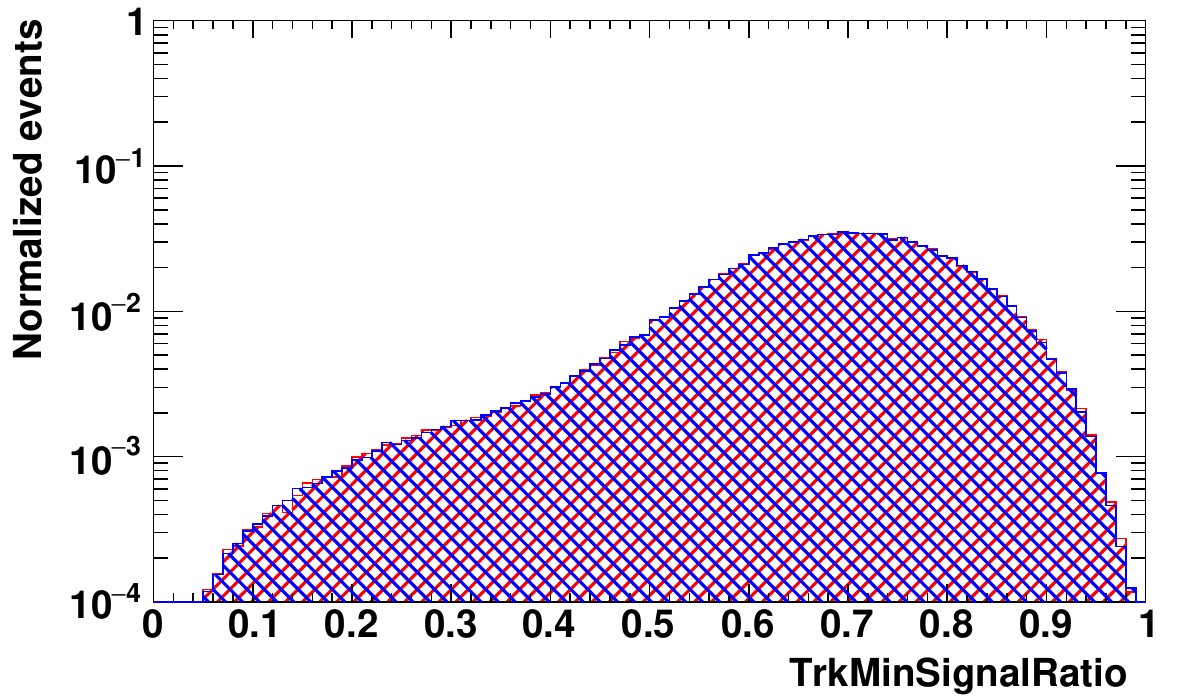}
  \end{subfigure}
  \hfill
  \begin{subfigure}{0.32\linewidth}
    \includegraphics[width=\linewidth]{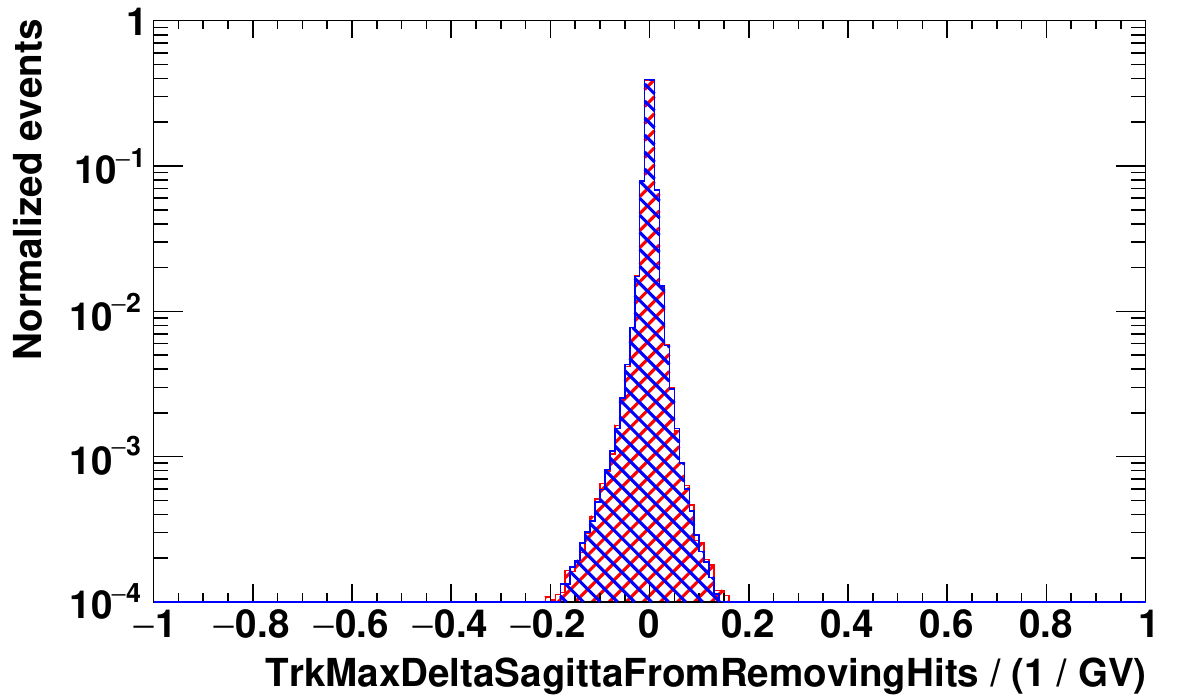}
  \end{subfigure}
  \hfill
  \begin{subfigure}{0.32\linewidth}
    \includegraphics[width=\linewidth]{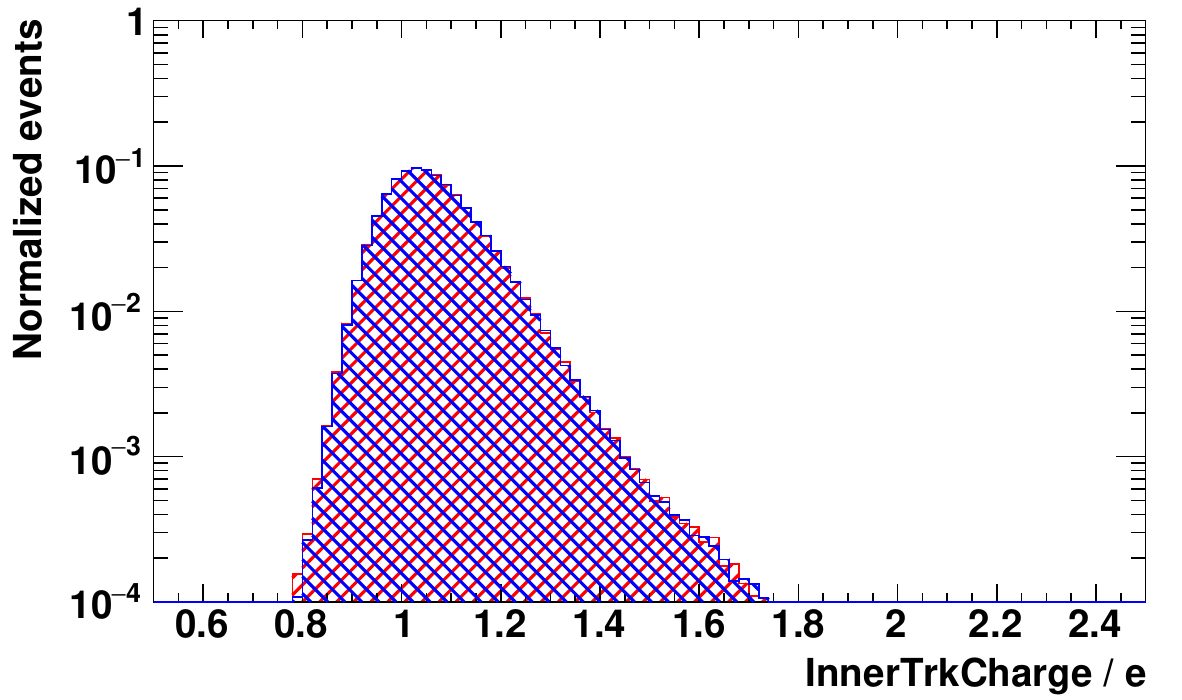}
  \end{subfigure}
  \hfill
  \begin{subfigure}{0.32\linewidth}
    \includegraphics[width=\linewidth]{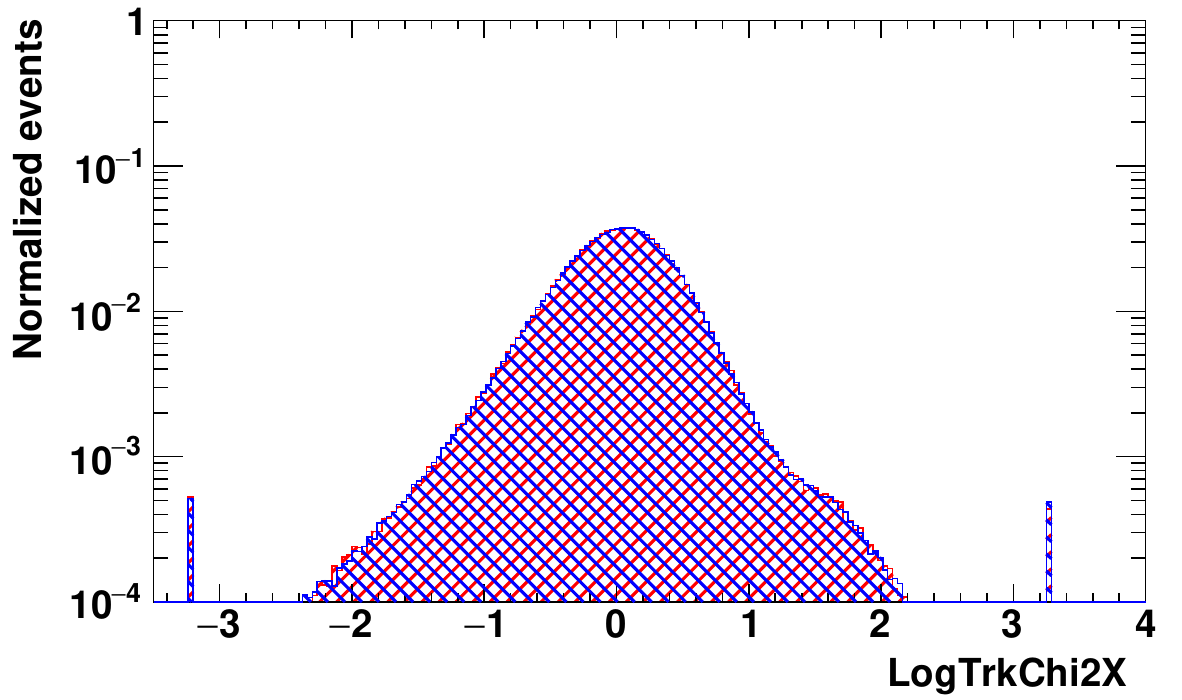}
  \end{subfigure}
  \hfill
  \begin{subfigure}{0.32\linewidth}
    \includegraphics[width=\linewidth]{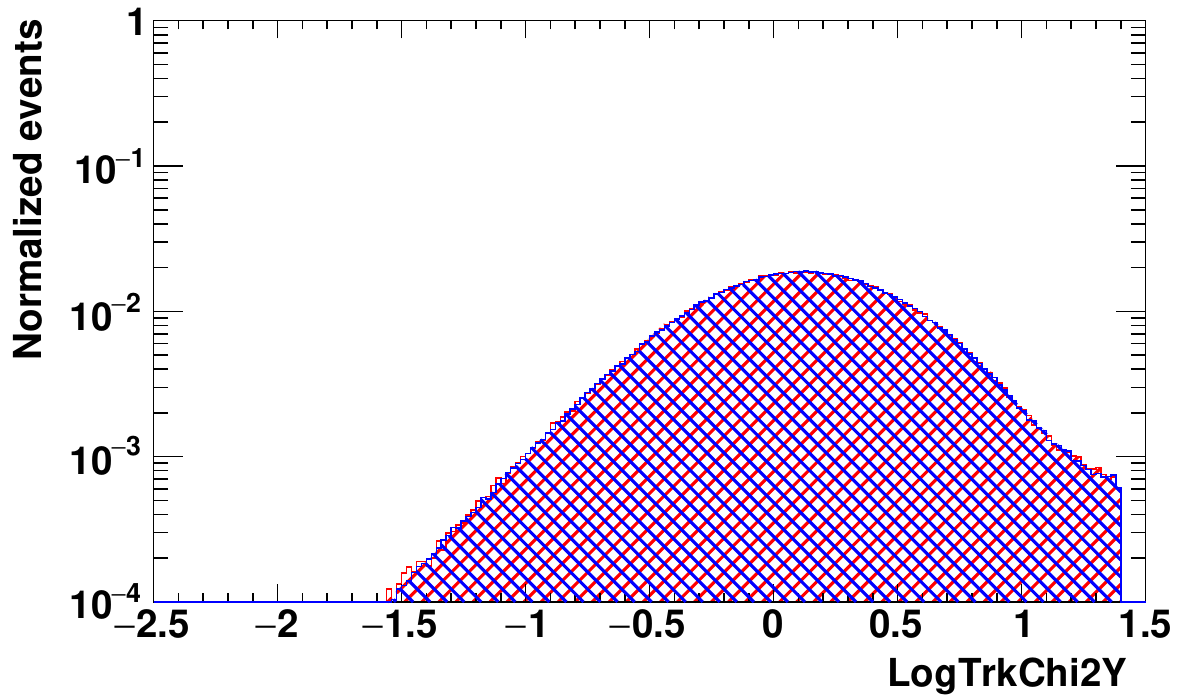}
  \end{subfigure}
  \hfill
  \begin{subfigure}{0.32\linewidth}
    \includegraphics[width=\linewidth]{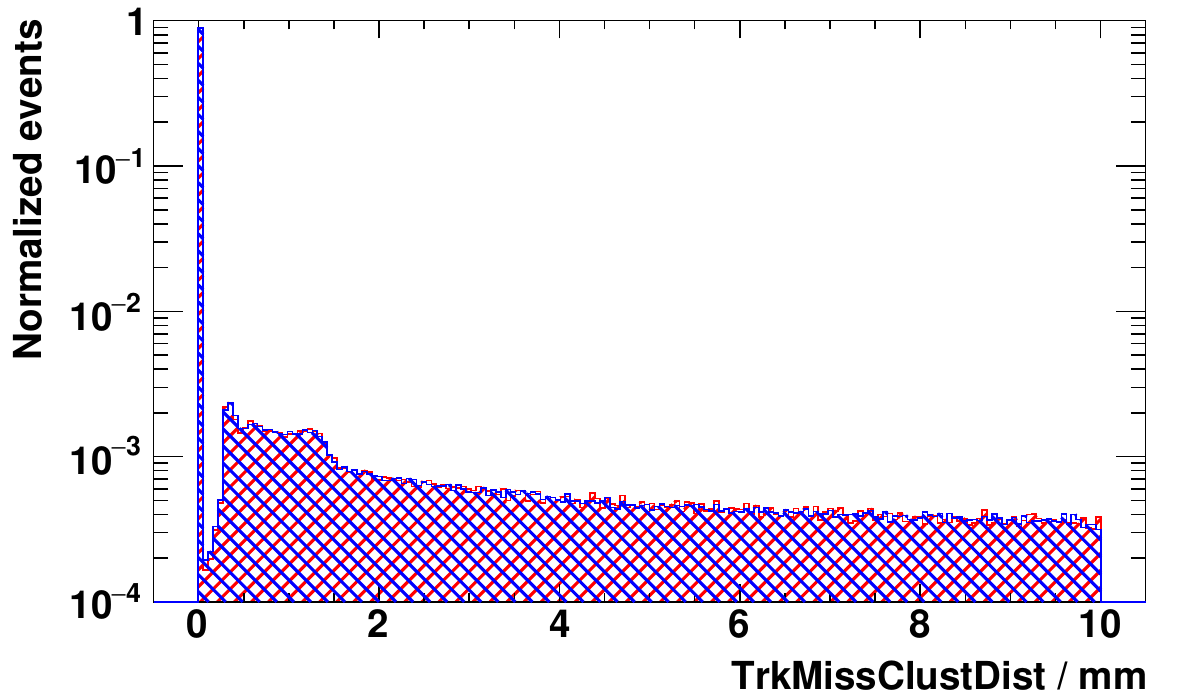}
  \end{subfigure}
  \hfill
  \begin{subfigure}{0.32\linewidth}
    \includegraphics[width=\linewidth]{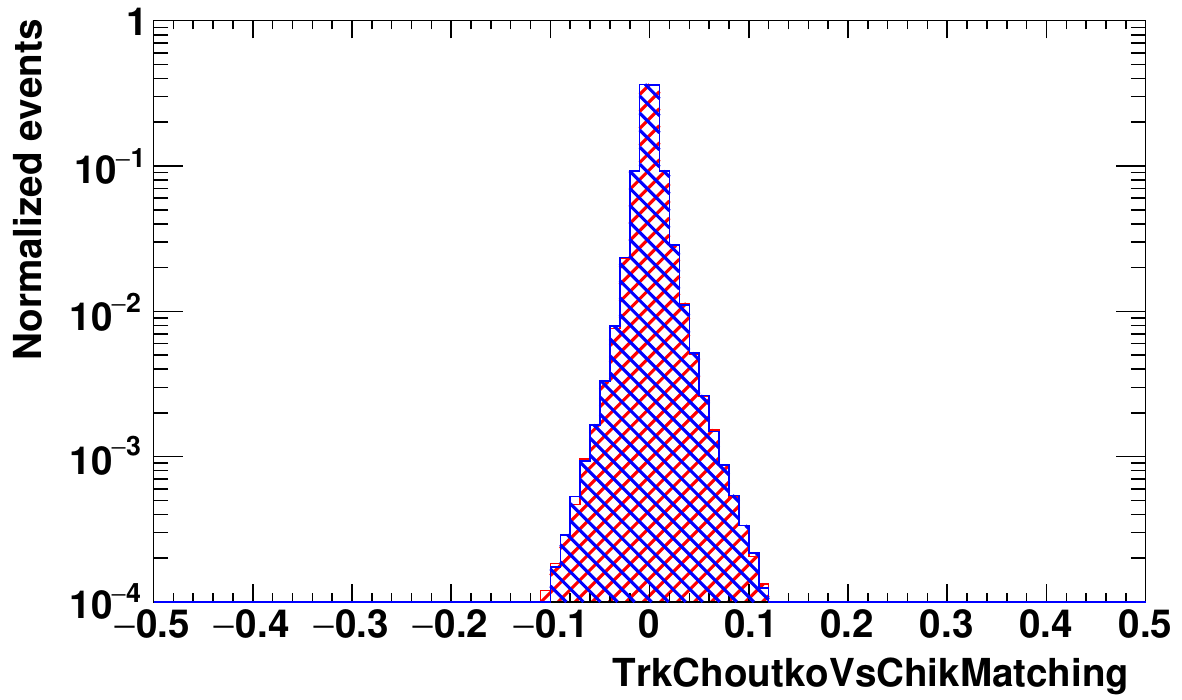}
  \end{subfigure}
  \hfill
  \begin{subfigure}{0.32\linewidth}
    \includegraphics[width=\linewidth]{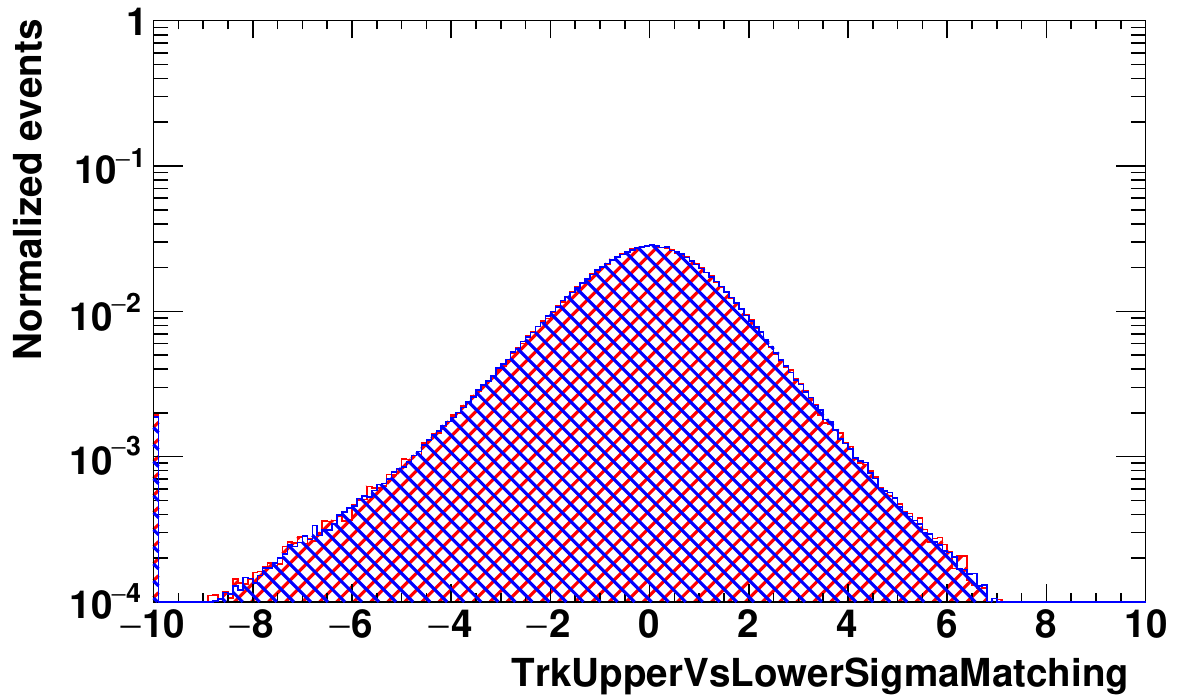}
  \end{subfigure}
  \hfill
  \begin{subfigure}{0.32\linewidth}
    \includegraphics[width=\linewidth]{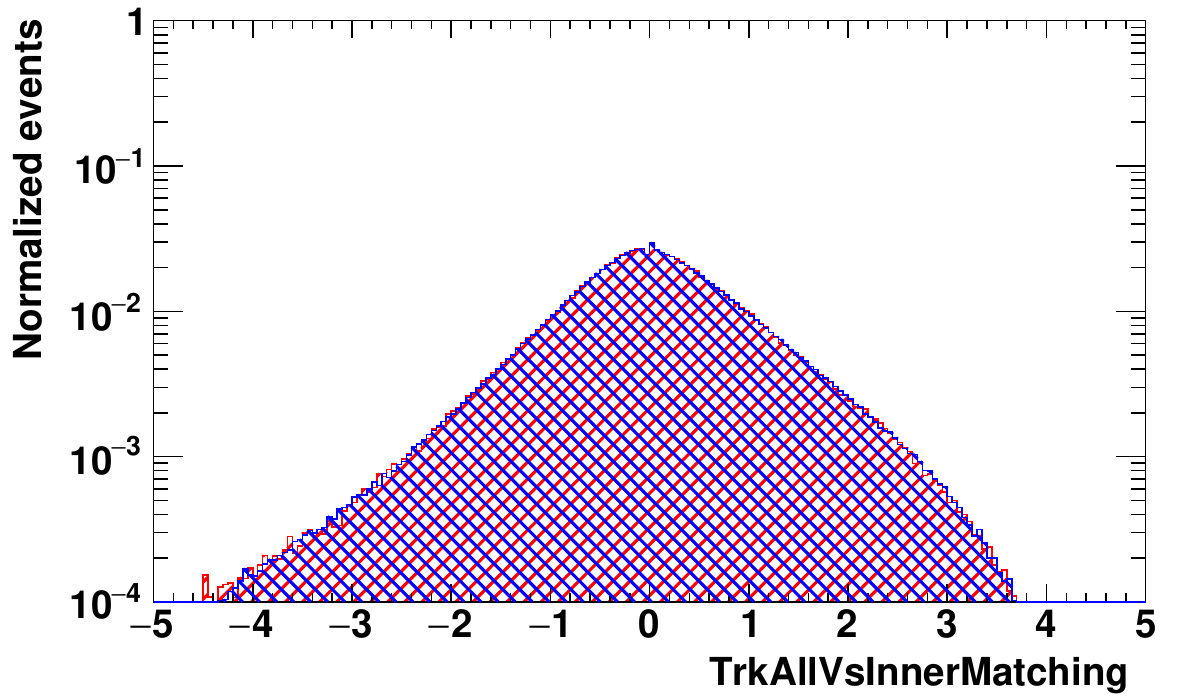}
  \end{subfigure}
  \hfill
  \begin{subfigure}{0.32\linewidth}
    \includegraphics[width=\linewidth]{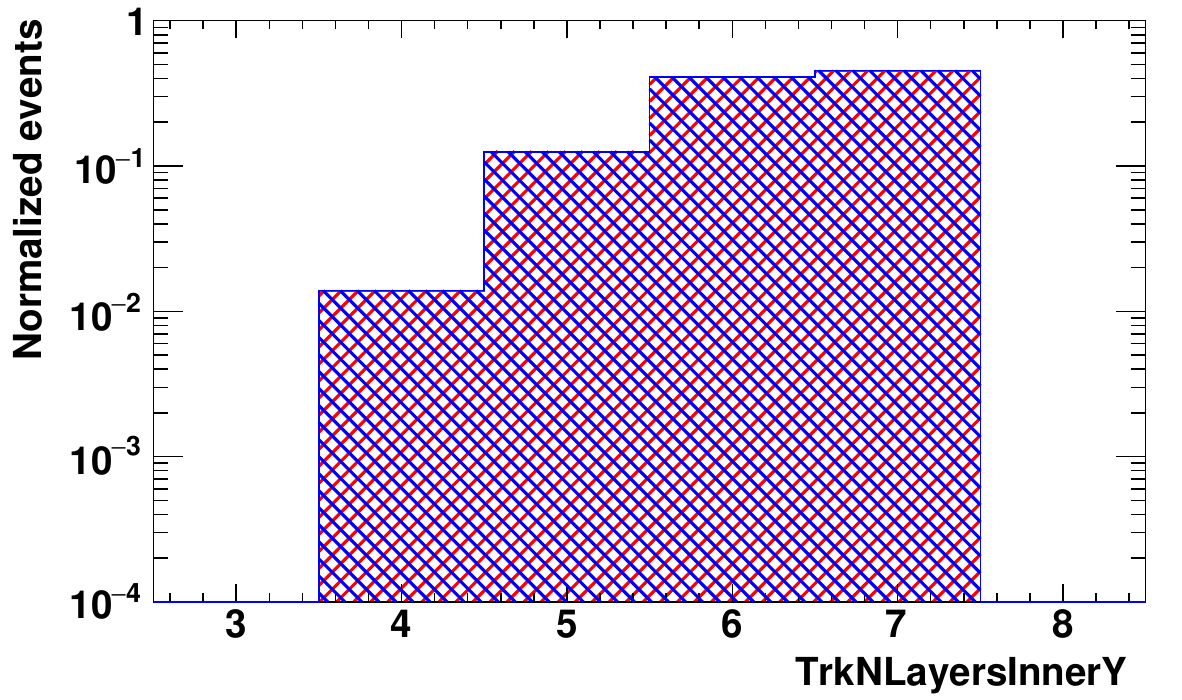}
  \end{subfigure}
  \caption{Comparison of all 15 CCMVA input variables for the single-track sample between the positron Monte-Carlo simulation (red) and the electron Monte-Carlo simulation (blue) in the energy bin \SIrange{17.98}{18.99}{\GeV}.}
  \label{fig:ccmva-input-electron-positron-single-track-mc-comparison}
\end{figure}

\begin{figure}[H]
  \begin{subfigure}{0.32\linewidth}
    \includegraphics[width=\linewidth]{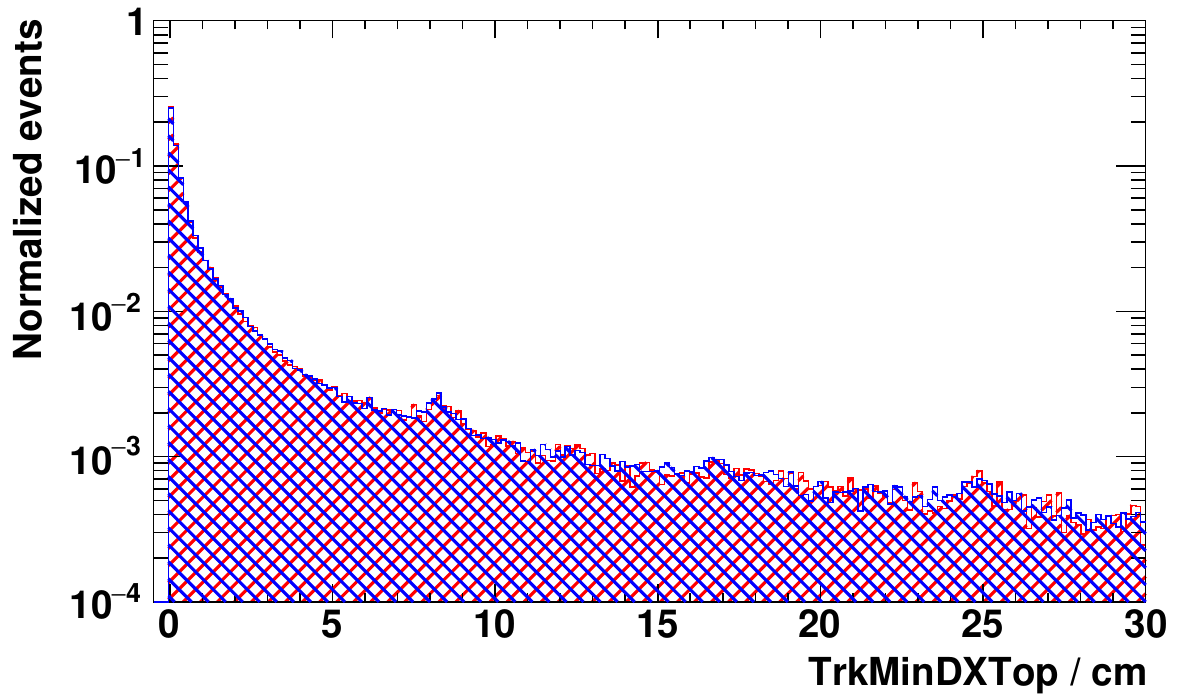}
  \end{subfigure}
  \hfill
  \begin{subfigure}{0.32\linewidth}
    \includegraphics[width=\linewidth]{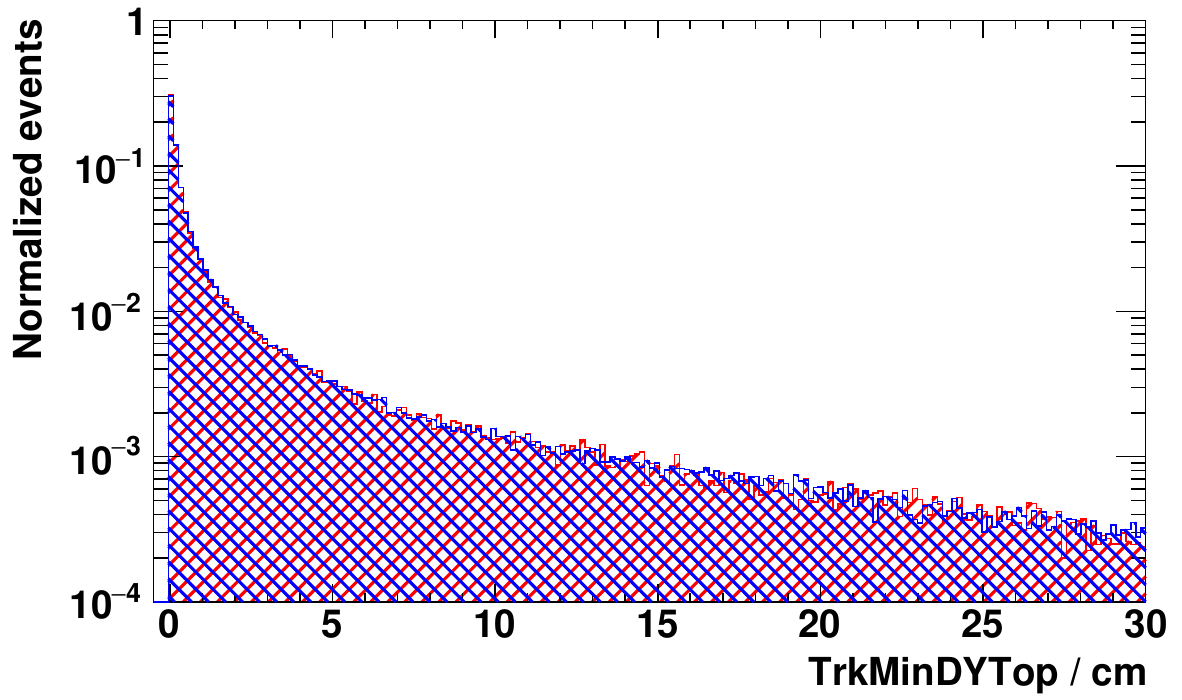}
  \end{subfigure}
  \hfill
  \begin{subfigure}{0.32\linewidth}
    \includegraphics[width=\linewidth]{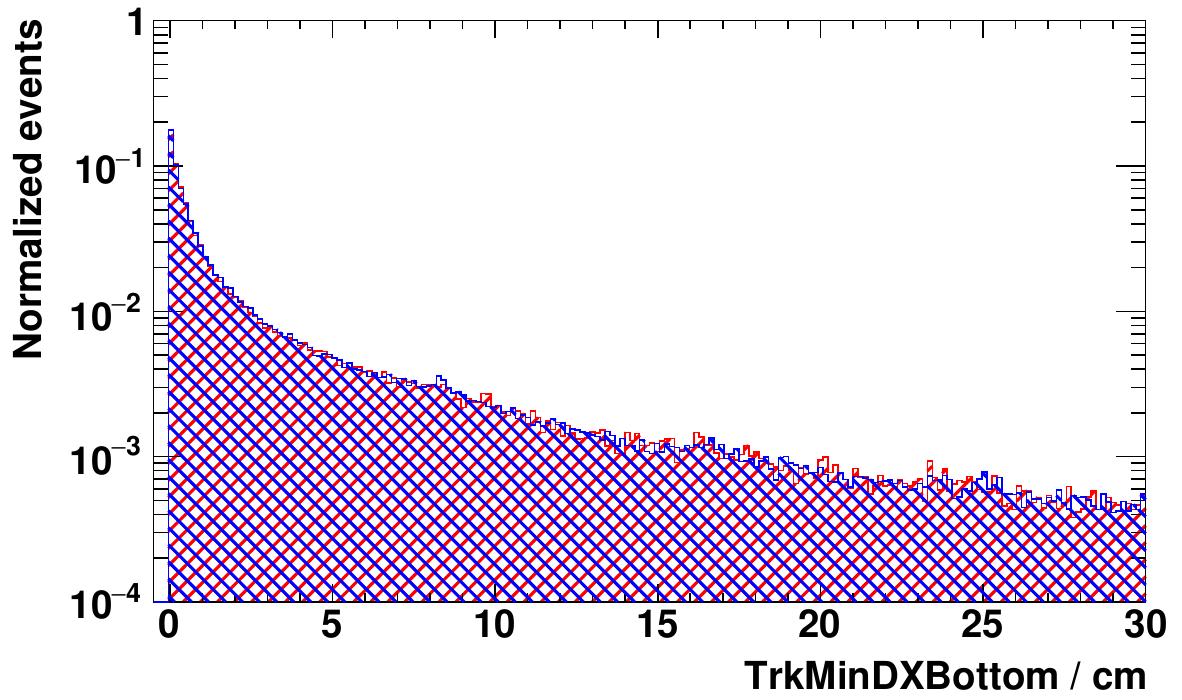}
  \end{subfigure}
  \hfill
  \begin{subfigure}{0.32\linewidth}
    \includegraphics[width=\linewidth]{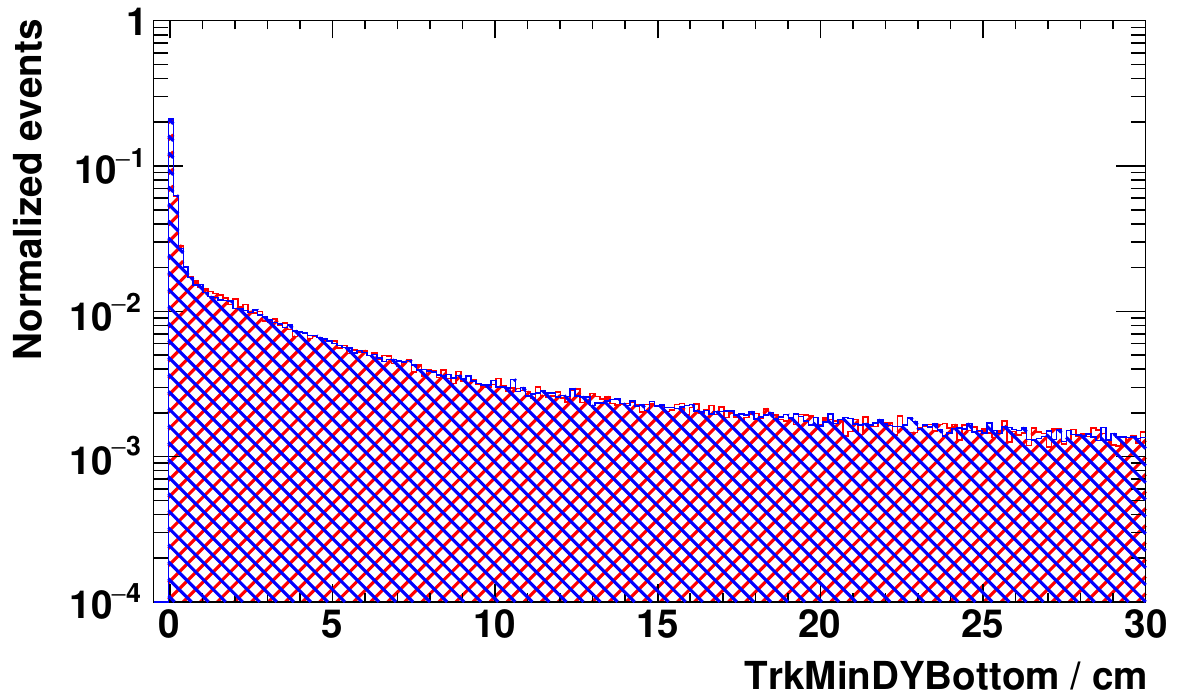}
  \end{subfigure}
  \hfill
  \begin{subfigure}{0.32\linewidth}
    \includegraphics[width=\linewidth]{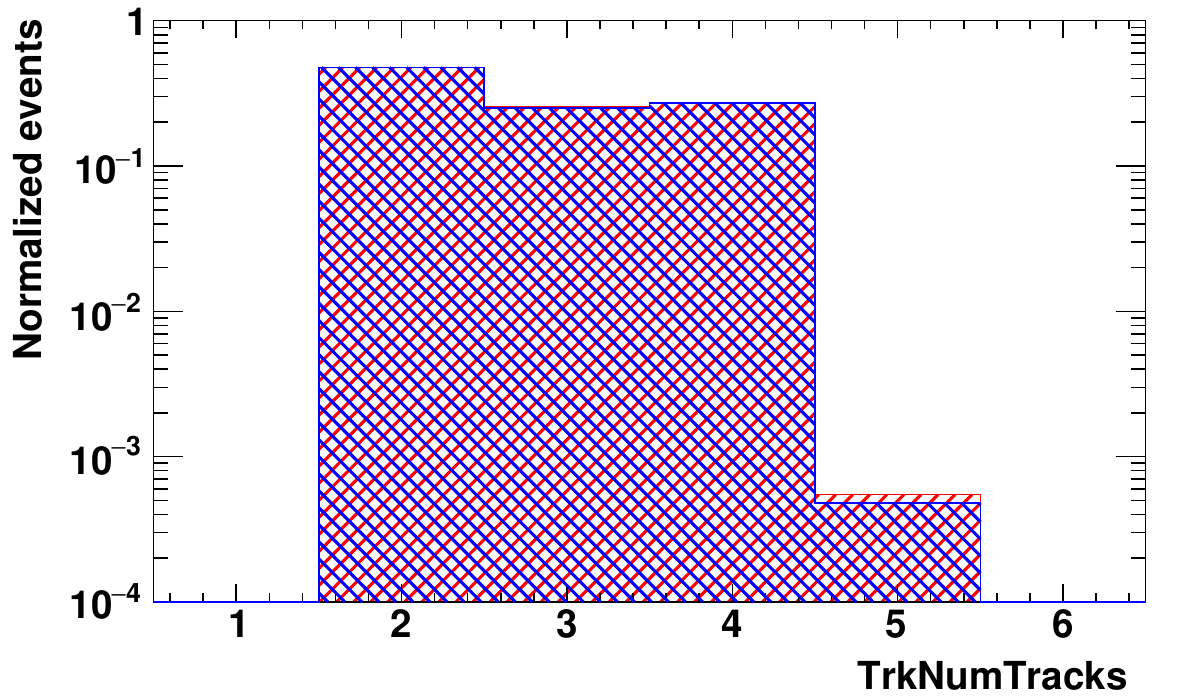}
  \end{subfigure}
  \hfill
  \begin{subfigure}{0.32\linewidth}
    \includegraphics[width=\linewidth]{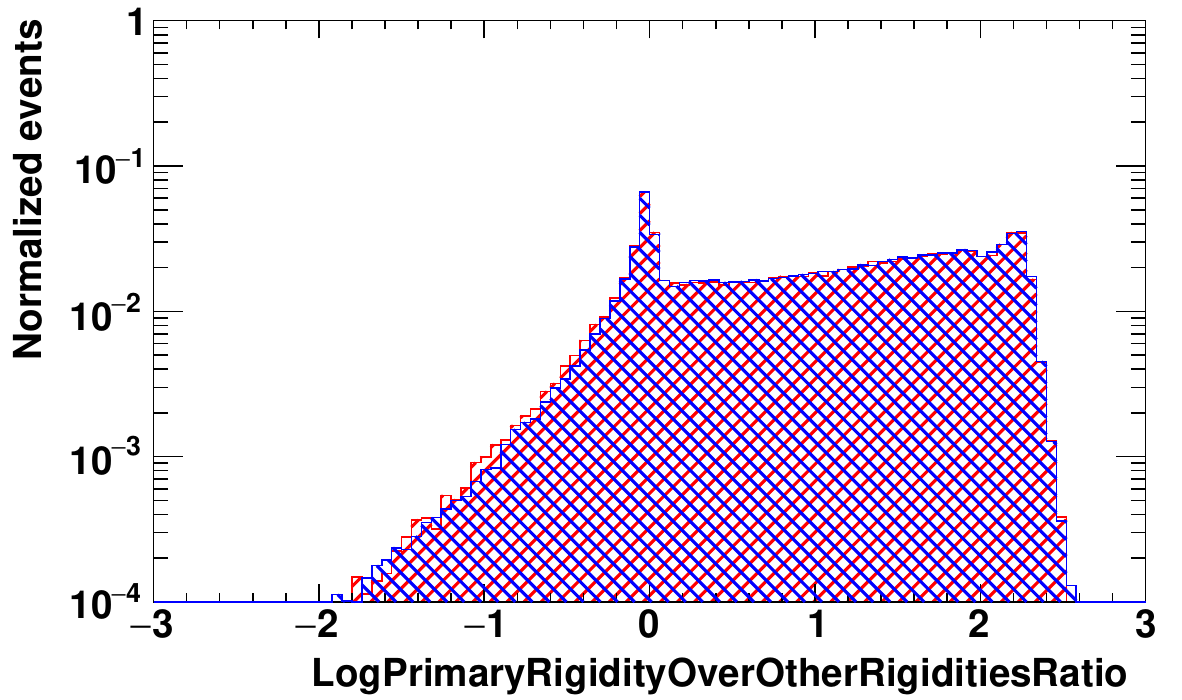}
  \end{subfigure}
  \caption{Comparison of the six multi-tracks sample specific CCMVA input variables between the positron Monte-Carlo simulation (red) and the electron Monte-Carlo simulation (blue) in the energy bin \SIrange{17.98}{18.99}{\GeV}. There are no visible differences, except for small fluctuations in the tails, proving the consistency of the input quantities between electron and positrons.}
  \label{fig:ccmva-input-electron-positron-multi-tracks-mc-comparison}
\end{figure}

All 15 input variables for the single-track sample were compared between the electron Monte-Carlo simulation and the positron Monte-Carlo, as shown in \cref{fig:ccmva-input-electron-positron-single-track-mc-comparison} for an example energy bin, after applying all preselection, selection and $e^{\pm}$ identification cuts. By construction, there is no difference between electrons and positrons in all energy bins. The same comparison was repeated for the multi-tracks sample, which also shows no differences in any of the variables. The six input variables which are specific to the multi-tracks sample are shown in \cref{fig:ccmva-input-electron-positron-multi-tracks-mc-comparison}.

\begin{figure}[H]
  \begin{subfigure}{0.32\linewidth}
    \includegraphics[width=\linewidth]{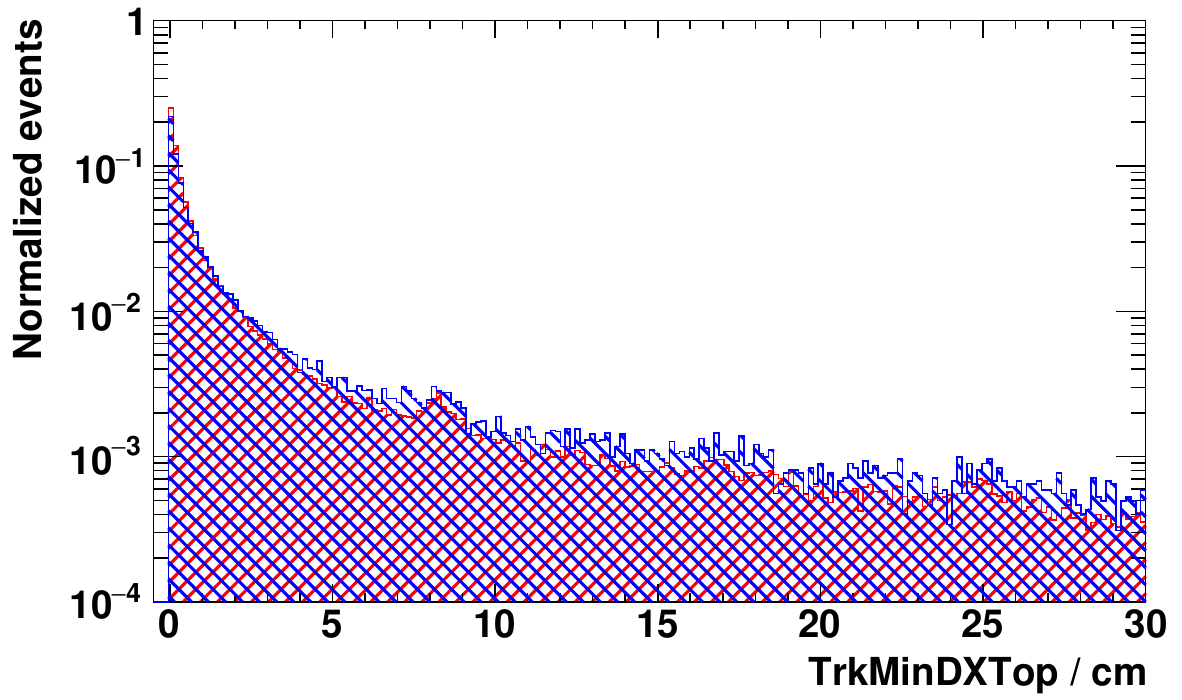}
  \end{subfigure}
  \hfill
  \begin{subfigure}{0.32\linewidth}
    \includegraphics[width=\linewidth]{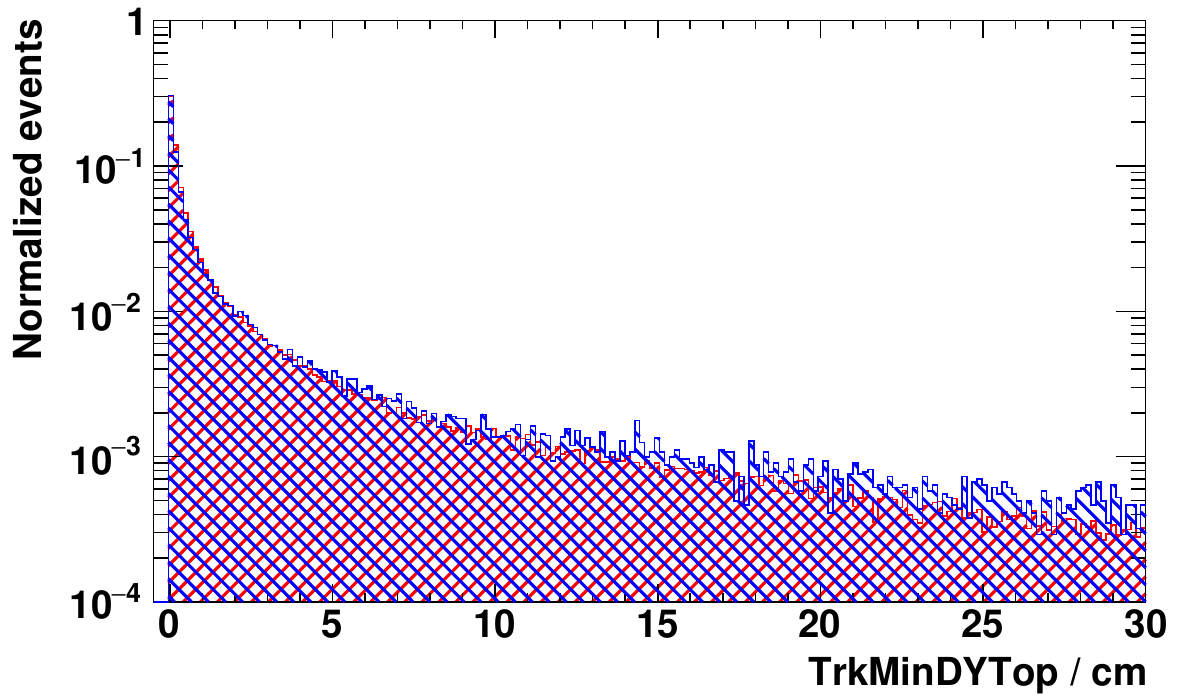}
  \end{subfigure}
  \hfill
  \begin{subfigure}{0.32\linewidth}
    \includegraphics[width=\linewidth]{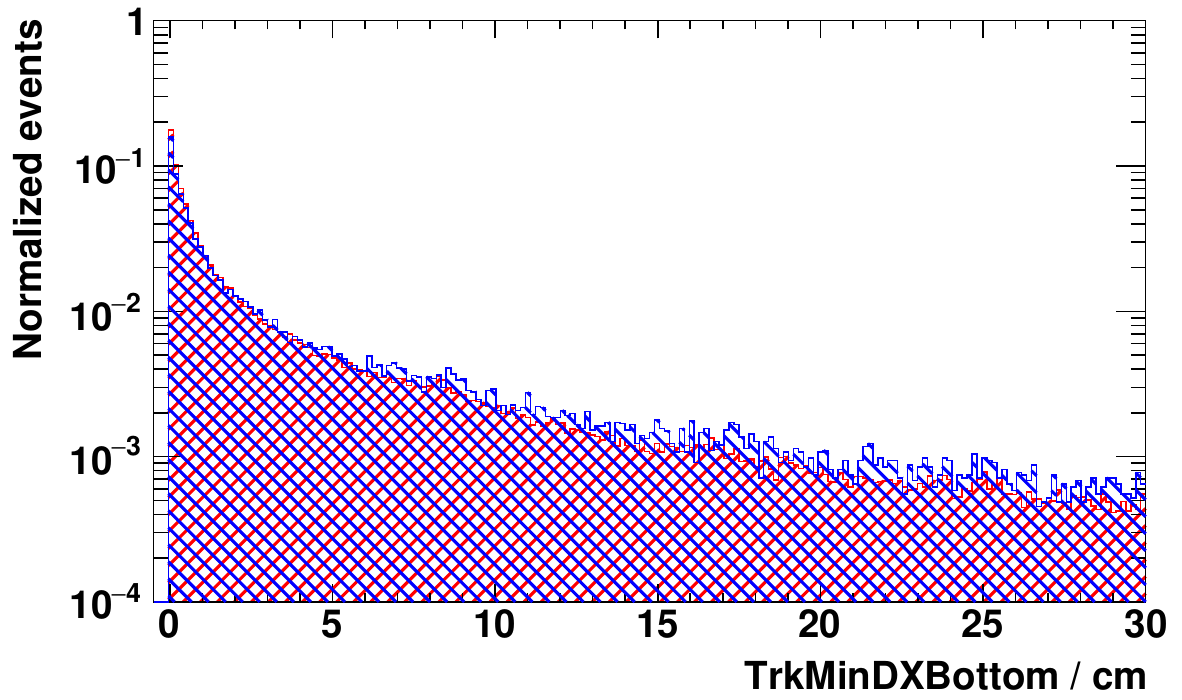}
  \end{subfigure}
  \hfill
  \begin{subfigure}{0.32\linewidth}
    \includegraphics[width=\linewidth]{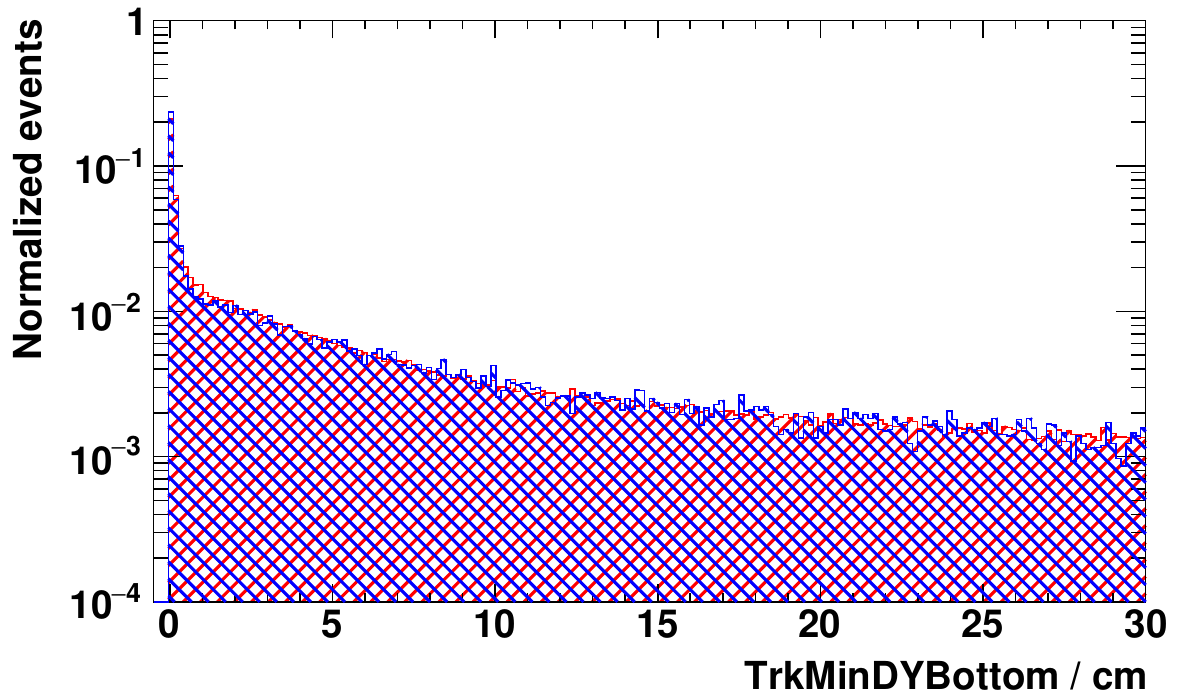}
  \end{subfigure}
  \hfill
  \begin{subfigure}{0.32\linewidth}
    \includegraphics[width=\linewidth]{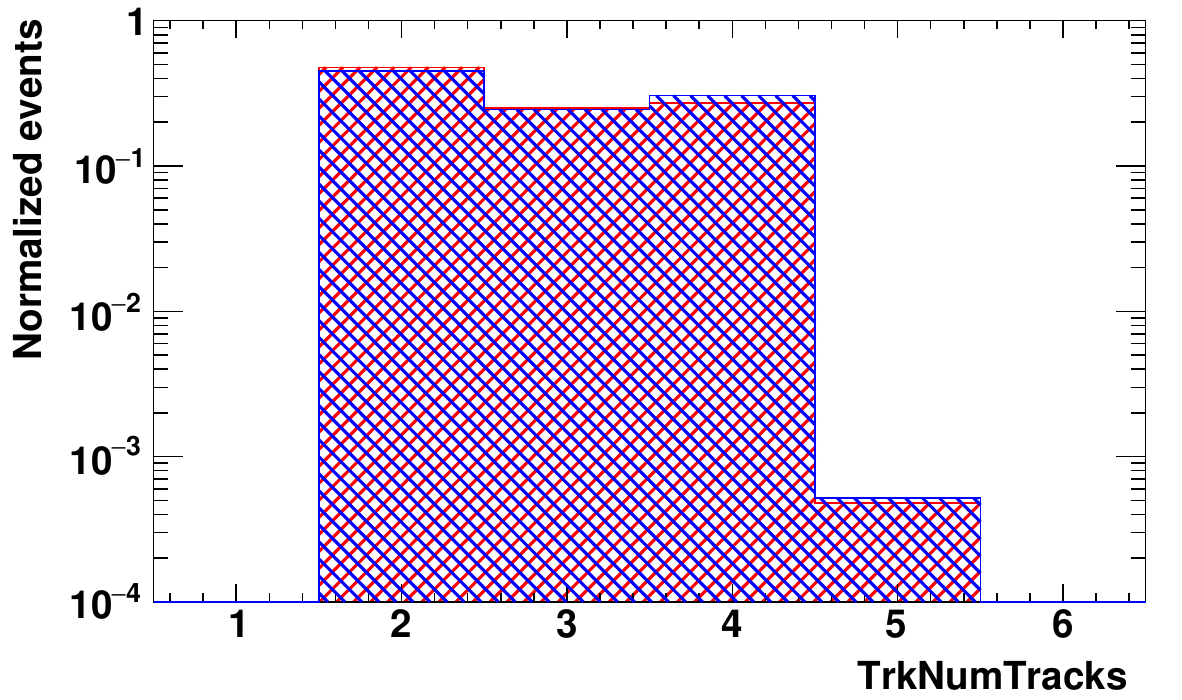}
  \end{subfigure}
  \hfill
  \begin{subfigure}{0.32\linewidth}
    \includegraphics[width=\linewidth]{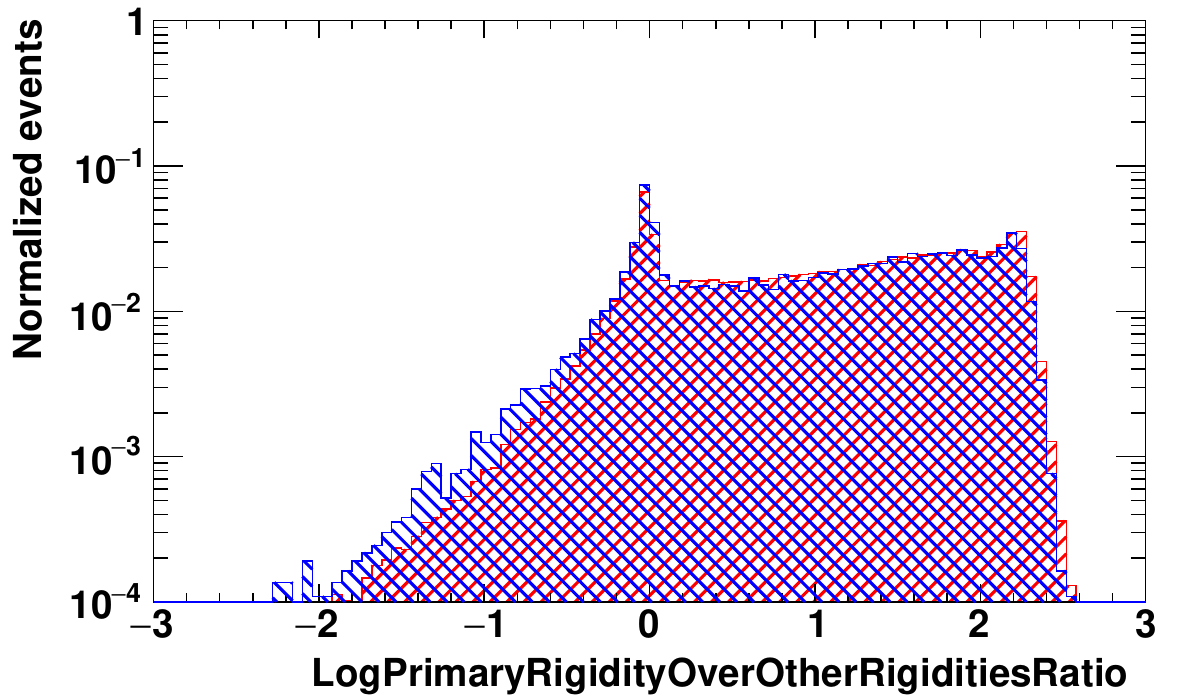}
  \end{subfigure}
  \caption{Comparison of the six multi-tracks sample specific CCMVA input variables between the ISS data (blue) and the electron Monte-Carlo (red) in the energy bin \SIrange{17.98}{18.99}{\GeV}. There are no visible differences, except for small fluctuations in the tails, proving the consistency of the input quantities between ISS data and simulation.}
  \label{fig:ccmva-input-electron-positron-multi-tracks-iss-mc-comparison}
\end{figure}

To ensure that the CCMVA is applicable on ISS data, all input variables have to match between ISS data and Monte-Carlo simulation.
\Cref{fig:ccmva-input-electron-positron-multi-tracks-iss-mc-comparison} shows a comparison in an example energy bin of all relevant variables for the multi-tracks sample
between negative rigidity ISS data and the electron Monte-Carlo simulation, after applying all preselection, selection and $e^{\pm}$ identification cuts plus additional
cuts to reduce any left-over charge-confused proton background on ISS data. All input quantities show an excellent agreement over all energies.

\begin{figure}[H]
  \begin{subfigure}{0.32\linewidth}
    \includegraphics[width=\linewidth]{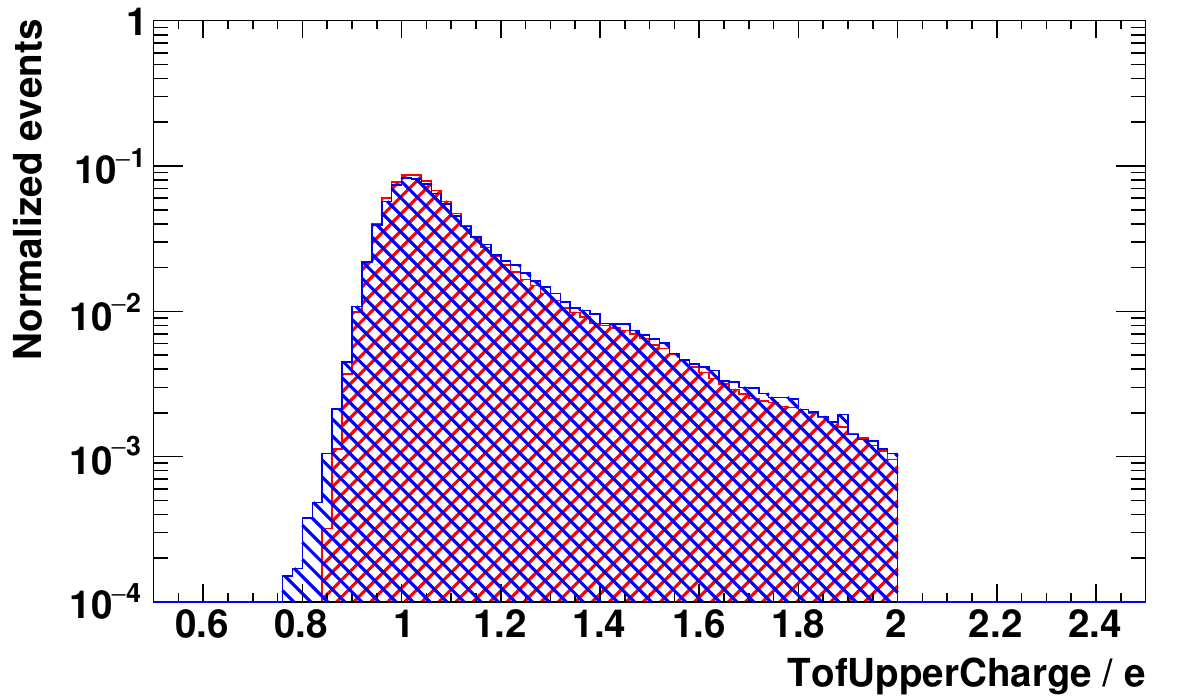}
  \end{subfigure}
  \hfill
  \begin{subfigure}{0.32\linewidth}
    \includegraphics[width=\linewidth]{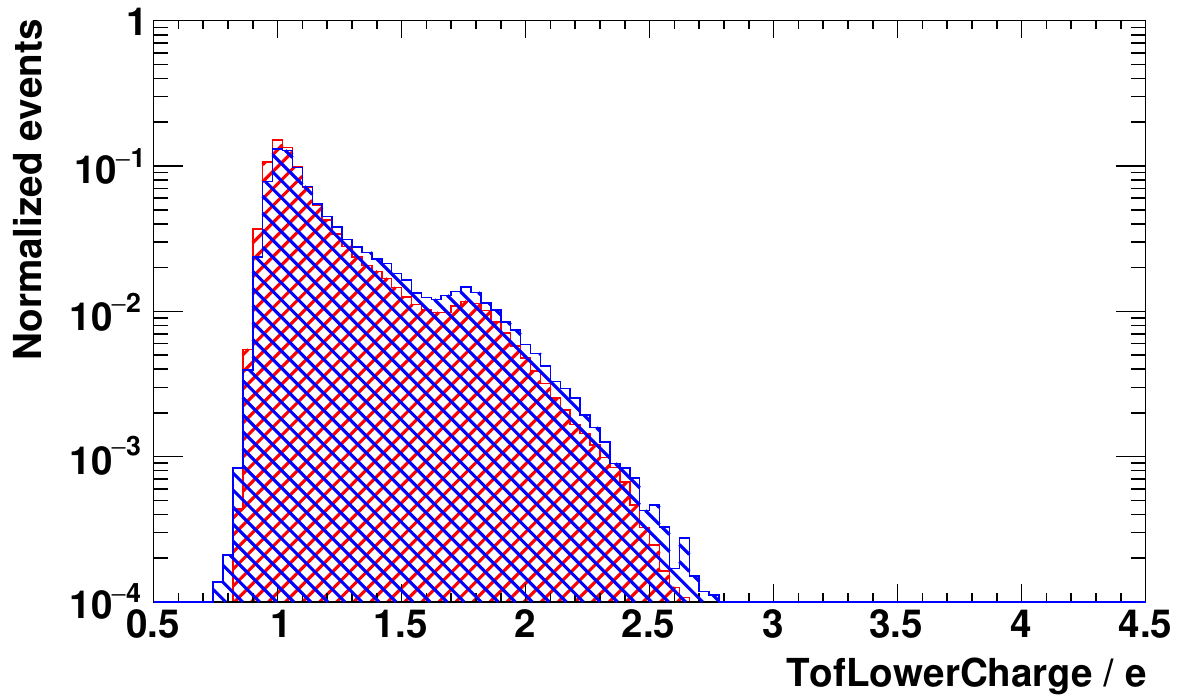}
  \end{subfigure}
  \hfill
  \begin{subfigure}{0.32\linewidth}
    \includegraphics[width=\linewidth]{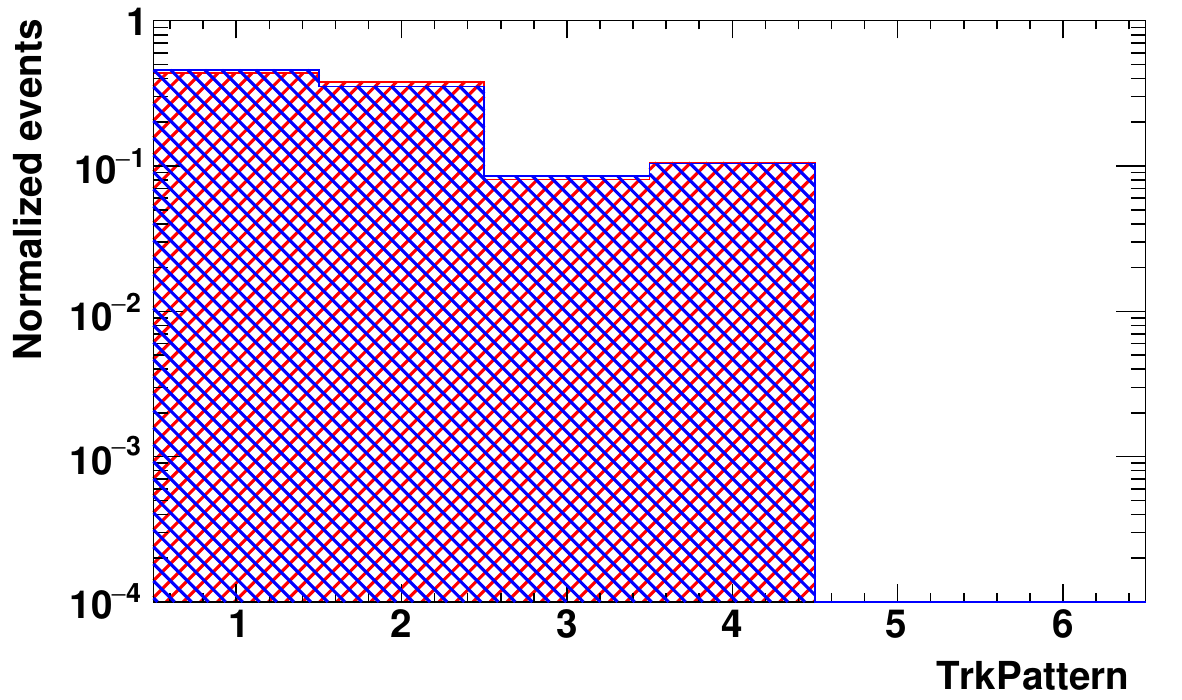}
  \end{subfigure}
  \hfill
  \begin{subfigure}{0.32\linewidth}
    \includegraphics[width=\linewidth]{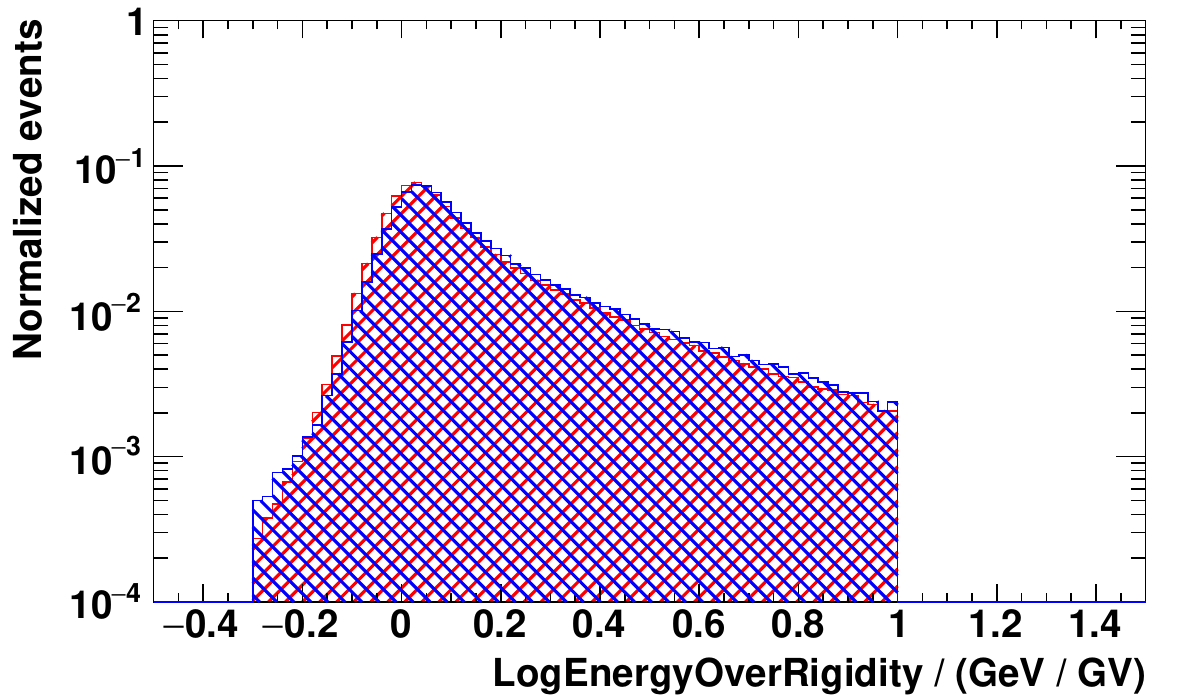}
  \end{subfigure}
  \hfill
  \begin{subfigure}{0.32\linewidth}
    \includegraphics[width=\linewidth]{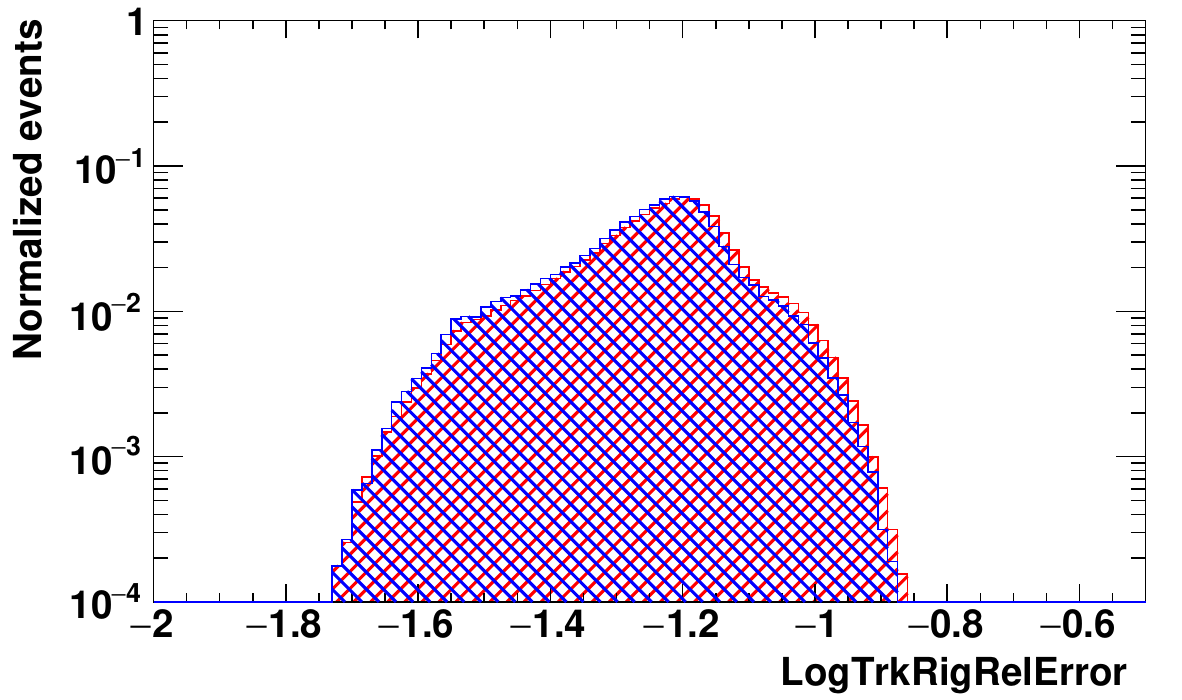}
  \end{subfigure}
  \hfill
  \begin{subfigure}{0.32\linewidth}
    \includegraphics[width=\linewidth]{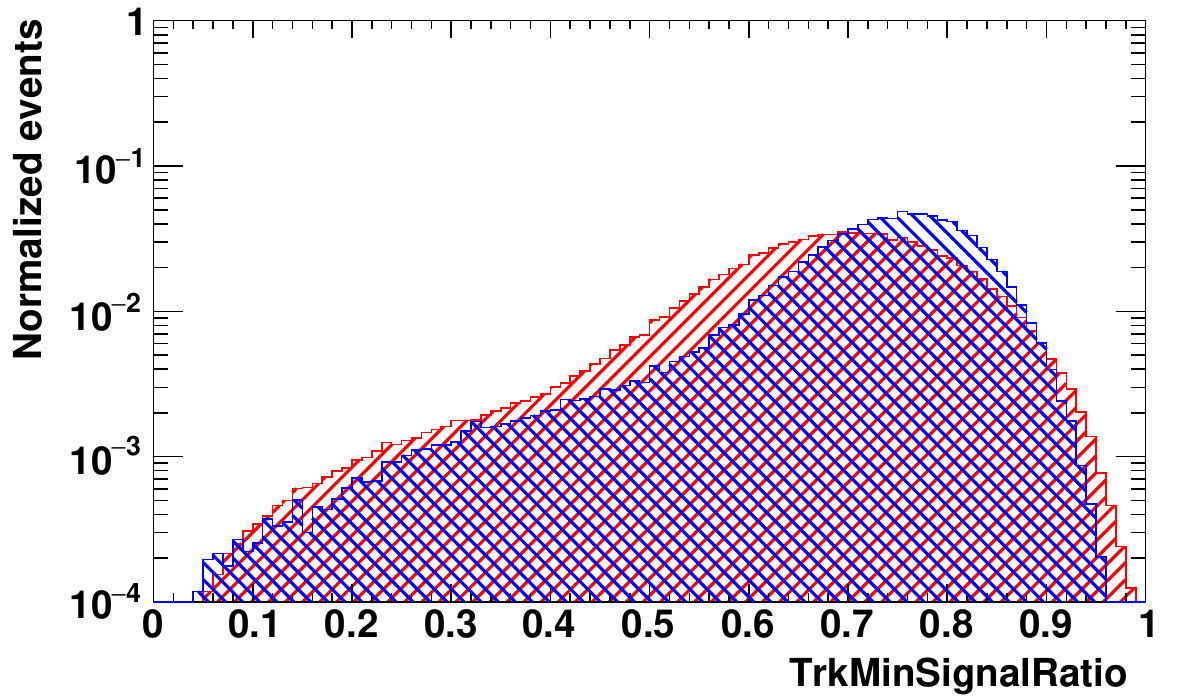}
  \end{subfigure}
  \hfill
  \begin{subfigure}{0.32\linewidth}
    \includegraphics[width=\linewidth]{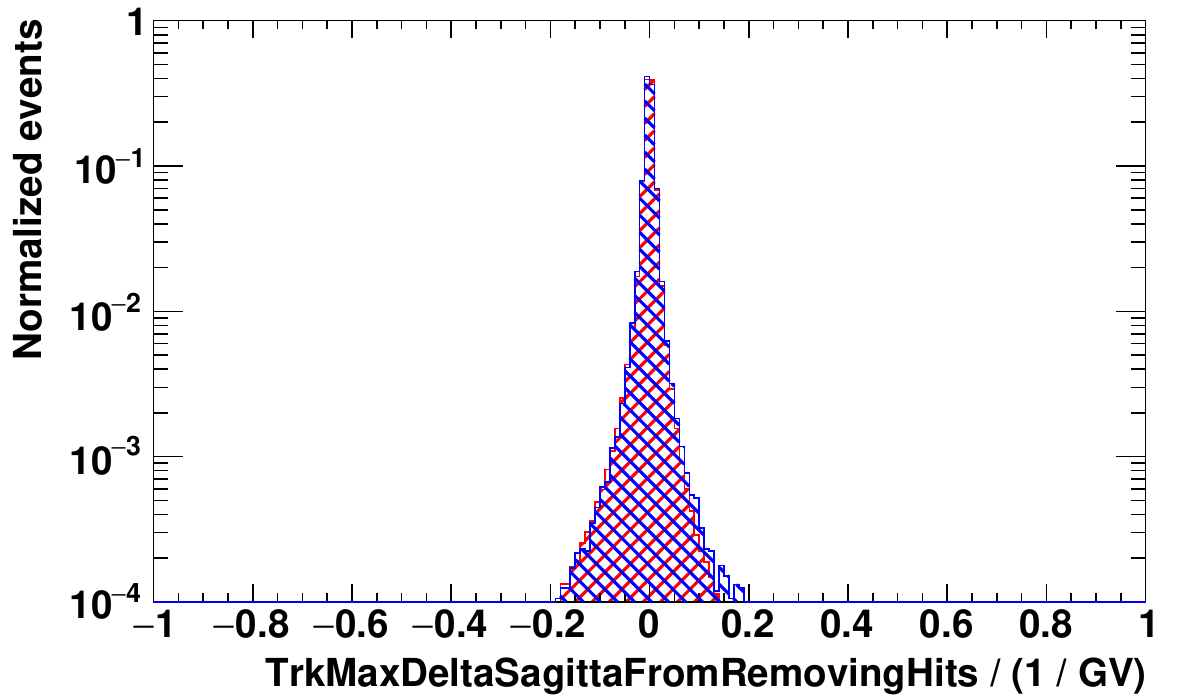}
  \end{subfigure}
  \hfill
  \begin{subfigure}{0.32\linewidth}
    \includegraphics[width=\linewidth]{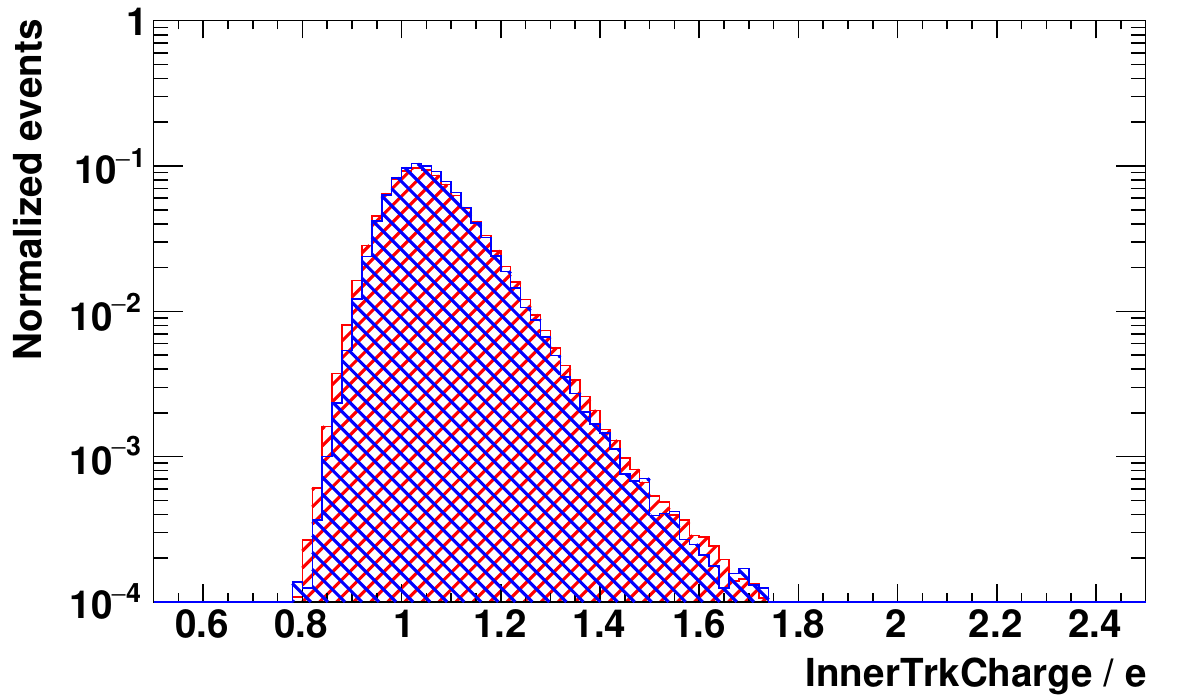}
  \end{subfigure}
  \hfill
  \begin{subfigure}{0.32\linewidth}
    \includegraphics[width=\linewidth]{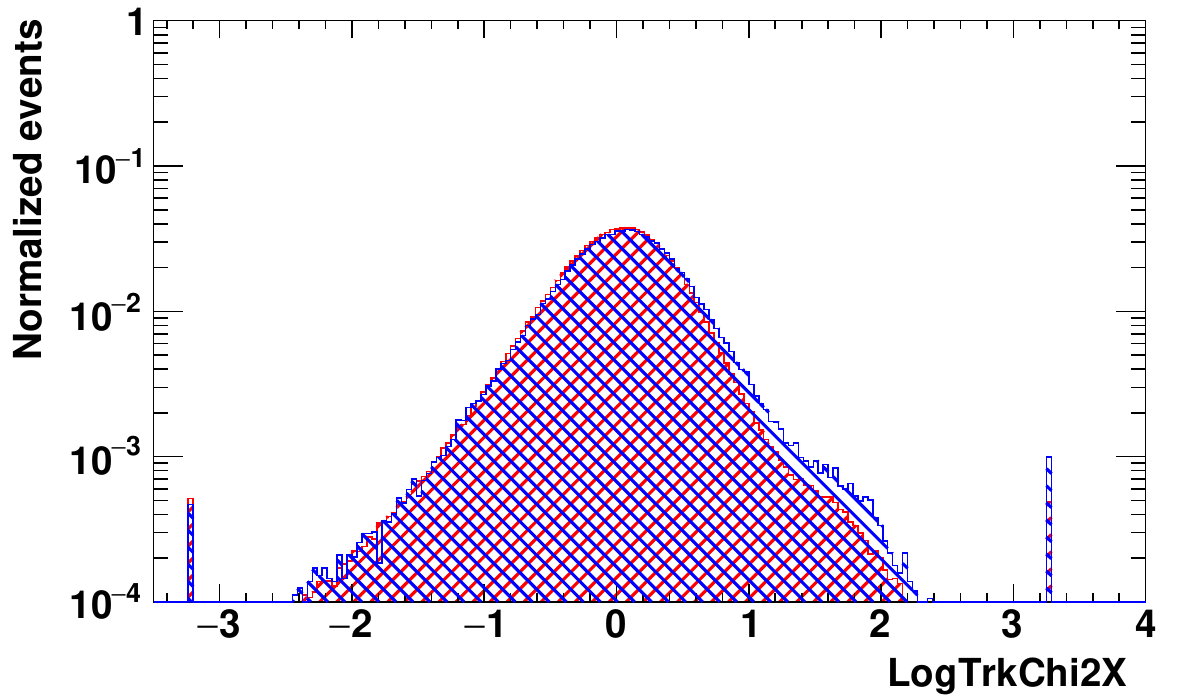}
  \end{subfigure}
  \hfill
  \begin{subfigure}{0.32\linewidth}
    \includegraphics[width=\linewidth]{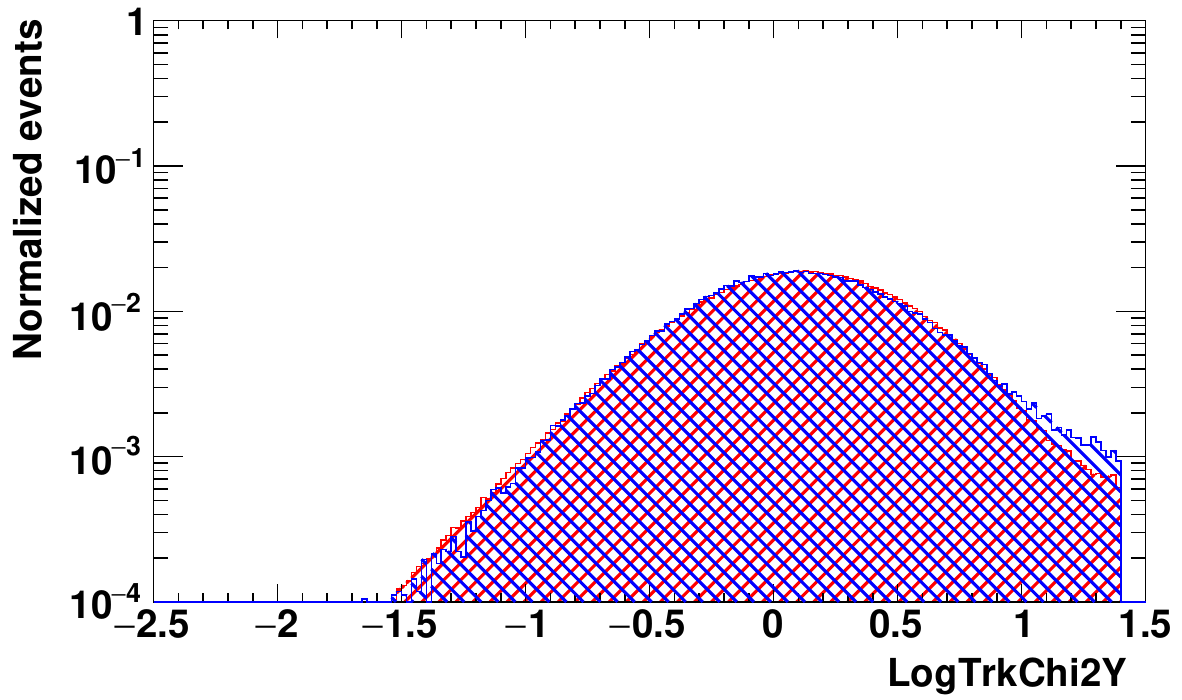}
  \end{subfigure}
  \hfill
  \begin{subfigure}{0.32\linewidth}
    \includegraphics[width=\linewidth]{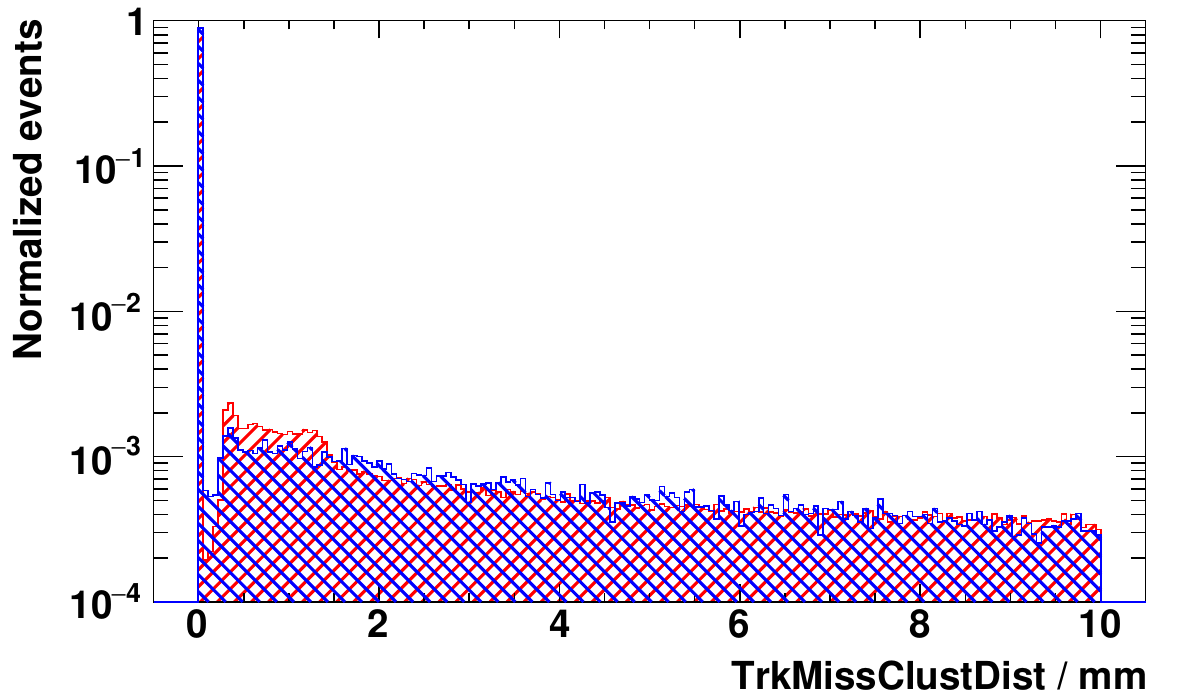}
  \end{subfigure}
  \hfill
  \begin{subfigure}{0.32\linewidth}
    \includegraphics[width=\linewidth]{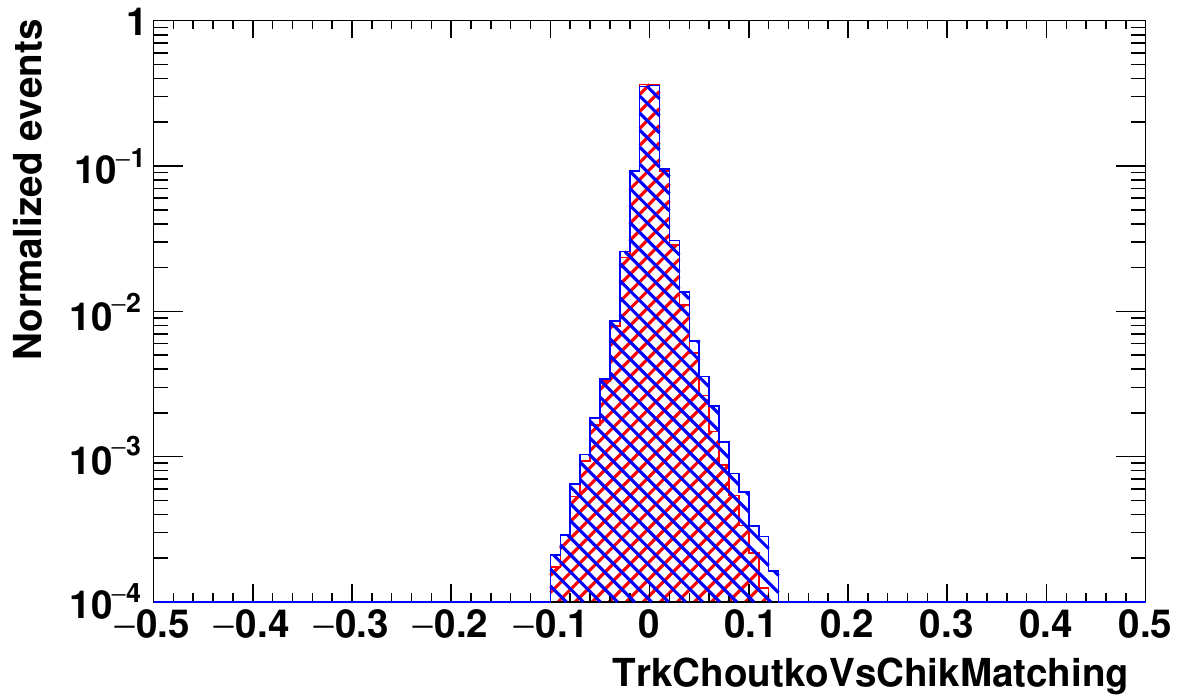}
  \end{subfigure}
  \hfill
  \begin{subfigure}{0.32\linewidth}
    \includegraphics[width=\linewidth]{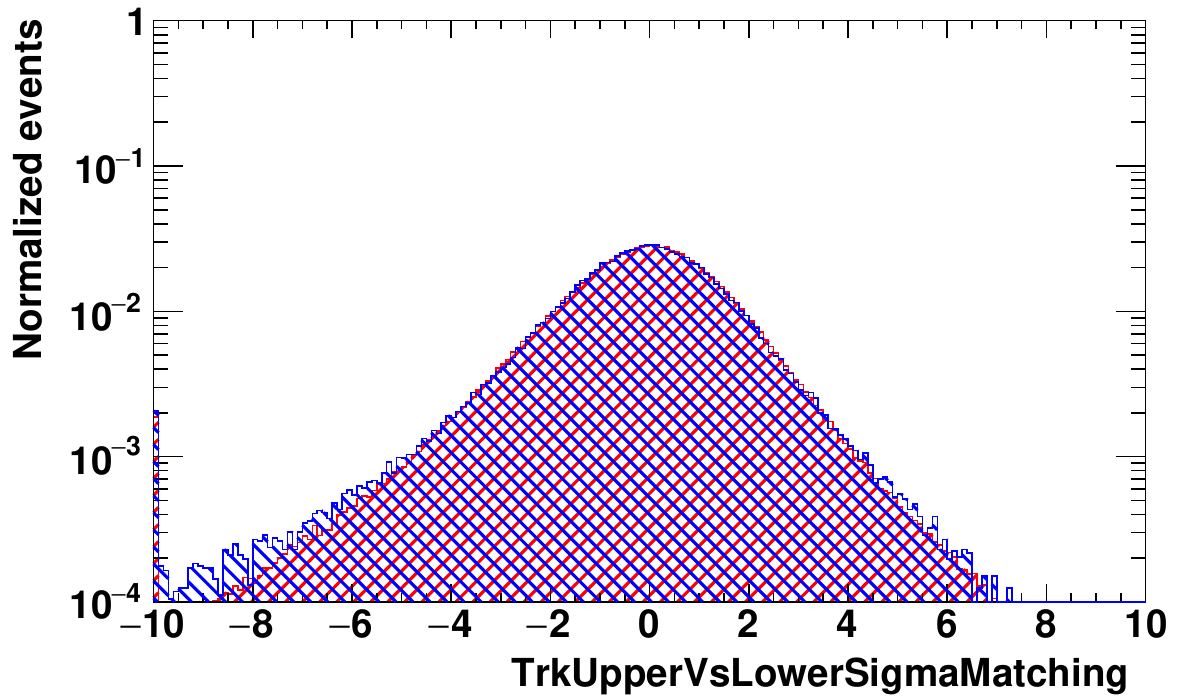}
  \end{subfigure}
  \hfill
  \begin{subfigure}{0.32\linewidth}
    \includegraphics[width=\linewidth]{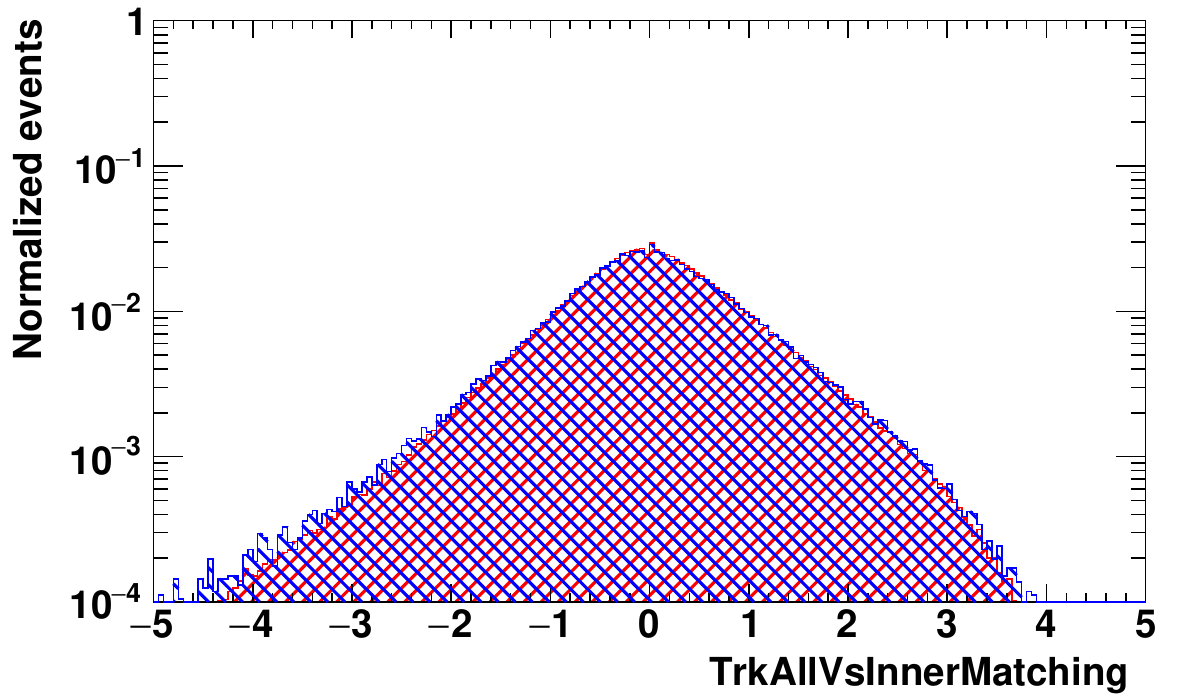}
  \end{subfigure}
  \hfill
  \begin{subfigure}{0.32\linewidth}
    \includegraphics[width=\linewidth]{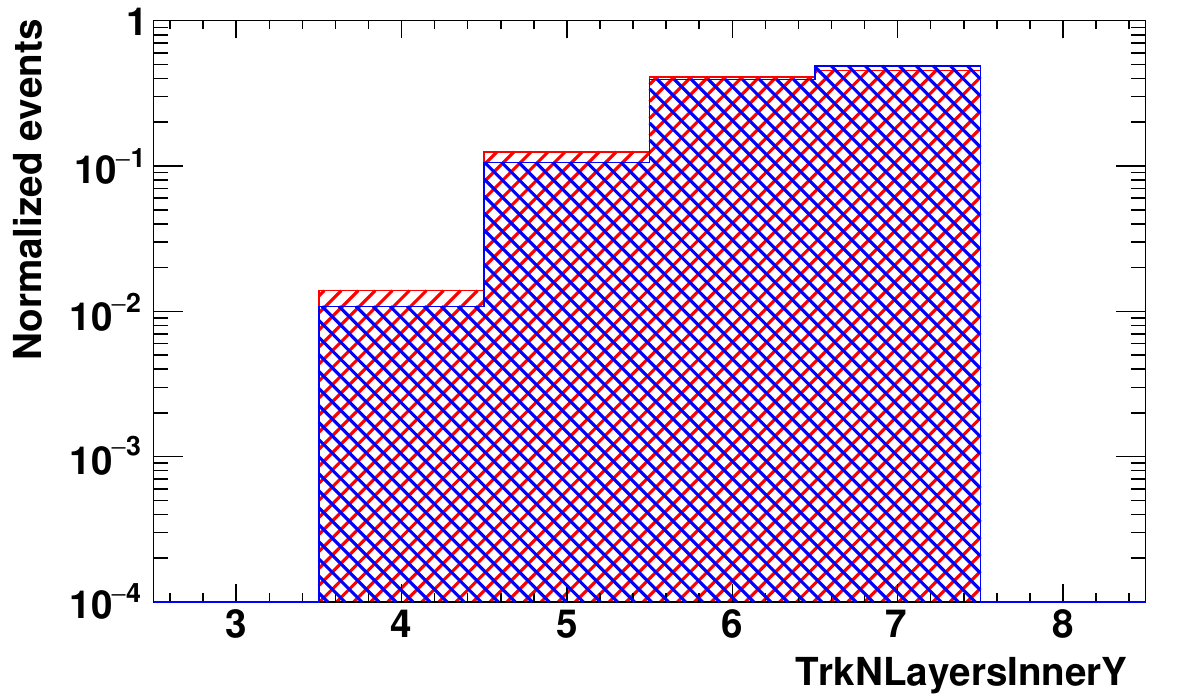}
  \end{subfigure}
  \caption{Comparison of all 15 CCMVA input variables for the single-track sample between the ISS data (blue) and the electron Monte-Carlo (red) in the energy bin \SIrange{17.98}{18.99}{\GeV}. There are no visible differences in all variables except TrkMinSignalRatio, except for small fluctuations in the tails, proving the consistency of the input quantities between ISS data and simulation. The whole comparison was performed on a negative rigidity sample, as its impossible to select pure samples of electrons with wrong reconstructed rigidity on ISS data.}
  \label{fig:ccmva-input-electron-positron-single-track-iss-mc-comparison}
\end{figure}

Furthermore the single-track sample input variables need to be verified between ISS data and Monte-Carlo simulation, as presented in
\Cref{fig:ccmva-input-electron-positron-single-track-iss-mc-comparison}. All variables except TrkMinSignalRatio show excellent agreement
over all energies. The TrkMinSignalRatio quantity requires an accurate simulation of the energy depositions in the silicon tracker on the strip
level. The ISS TrkMinSignalRatio distribution favors slightly higher values and a narrower peak than the Monte-Carlo simulation, which means
that the energy is deposited in a wider area in the Monte-Carlo simulation, covering more strips than on ISS data. However this has no impact
on the applicability of the MVA on ISS data, as the TrkMinSignalRatio observable is of less importance compared to others and thus does not dominate
the discrimination power of the MVA. If it would be an important observable the bias would have an impact on the output shape of the MVA, leading to
ISS / Monte-Carlo differences.

\section{TRD templates (multi-tracks sample)}
\label{sec:appendix-trd-templates}

In \cref{sec:analysis-lepton-counts-trd-templates} the energy dependence of all template parameters was presented for the single-track sample.
In this section the TRD template parameters as function of energy are shown for the multi-tracks sample.

\begin{figure}[H]
  \begin{subfigure}{0.48\linewidth}
    \includegraphics[width=\linewidth]{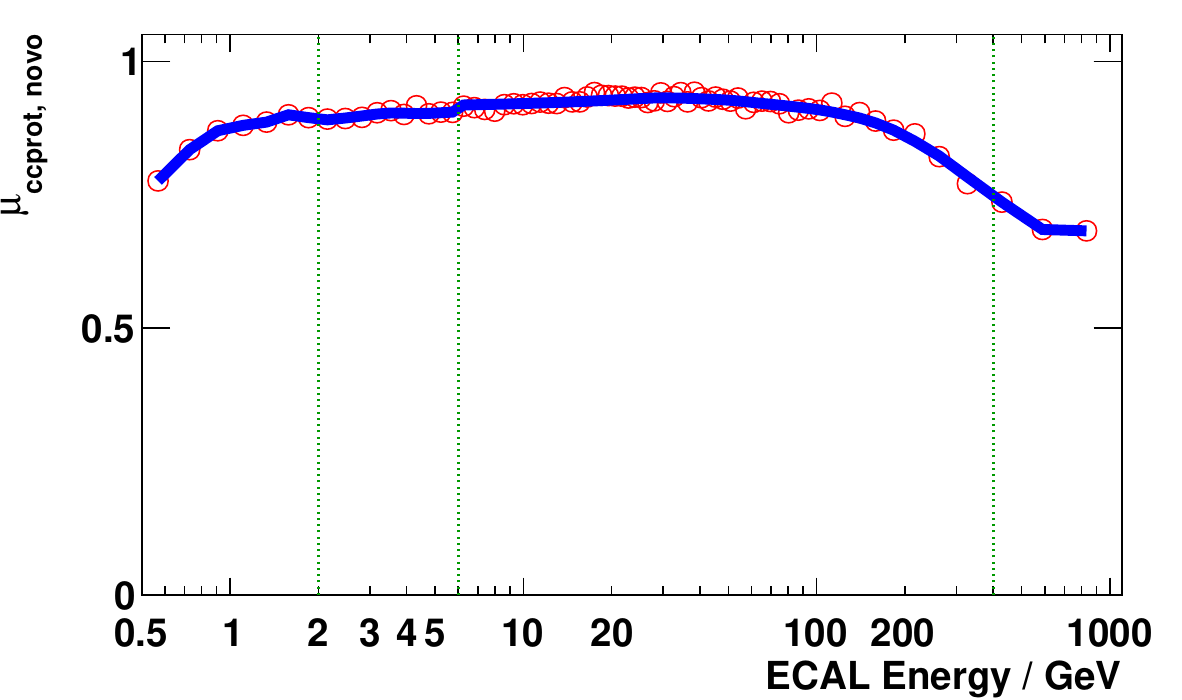}
    \caption{Template parameter~\ccProtPeakNovo}
  \end{subfigure}
  \hfill
  \begin{subfigure}{0.48\linewidth}
    \includegraphics[width=\linewidth]{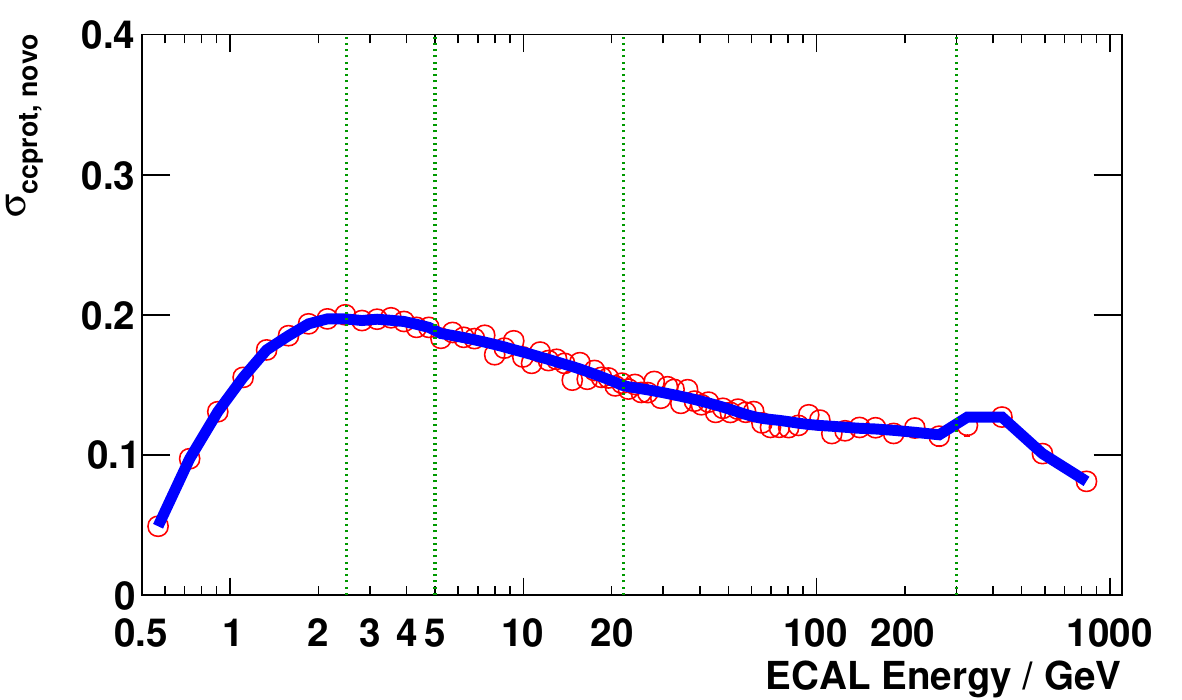}
    \caption{Template parameter~\ccProtWidthNovo}
  \end{subfigure}
  \hfill
  \begin{subfigure}{0.48\linewidth}
    \includegraphics[width=\linewidth]{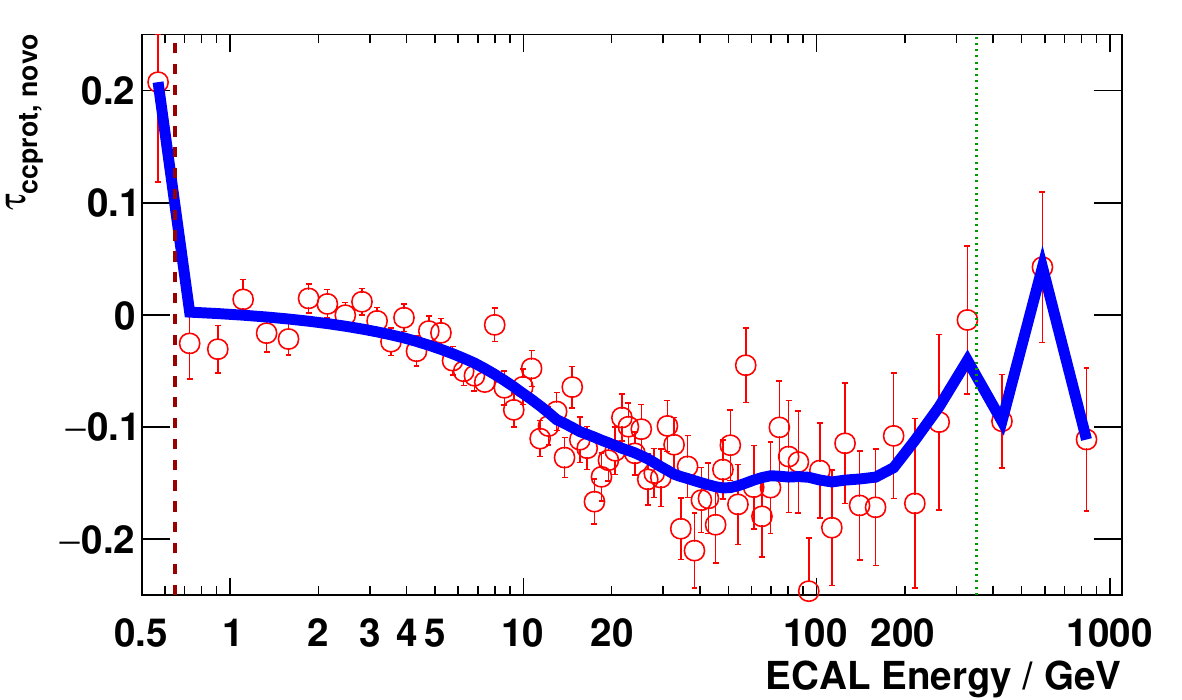}
    \caption{Template parameter~\ccProtTailNovo}
  \end{subfigure}
  \hfill
  \begin{subfigure}{0.48\linewidth}
    \includegraphics[width=\linewidth]{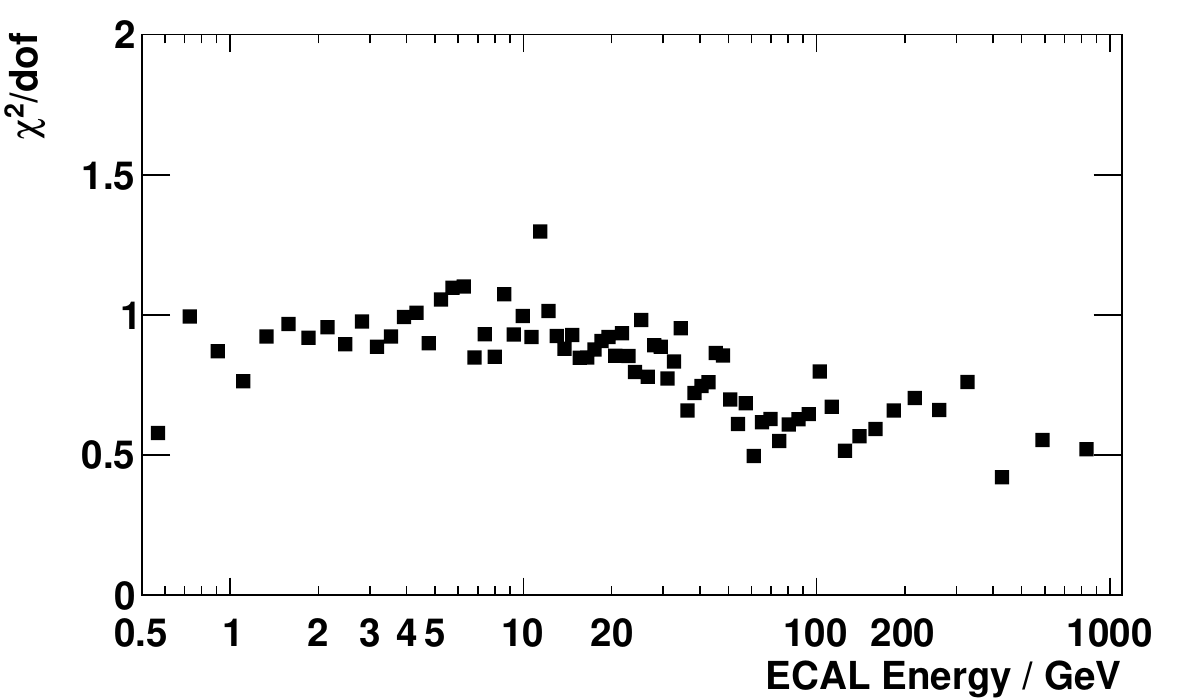}
    \caption{Goodness-of-fit}
    \label{fig:trd-template-parameters-ccprotons-chisquare-multi-tracks-vs-energy}
  \end{subfigure}
  \caption{Analytical TRD template parameters describing the evolution of the charge-confused proton template as function of energy for the multi-tracks sample. In each energy bin the analytical function -~\cref{eq:trd-model-ccprotons}~- is fit to the template data sample, yielding the red points. A smoothing procedure yields the blue curves, which are used as template parameters for the analysis. The vertical dashed green lines mark areas where different isolated smoothing procedures are applied. The goodness-of-fit is indicated by the black points in the lower right plot.}
  \label{fig:trd-template-parameters-ccprotons-multi-tracks-vs-energy}
\end{figure}

\Cref{fig:trd-template-parameters-ccprotons-multi-tracks-vs-energy} shows the charge-confused proton template parameters
as function of energy for the multi-tracks sample.

\begin{figure}[H]
  \begin{subfigure}{0.47\linewidth}
    \includegraphics[width=\linewidth]{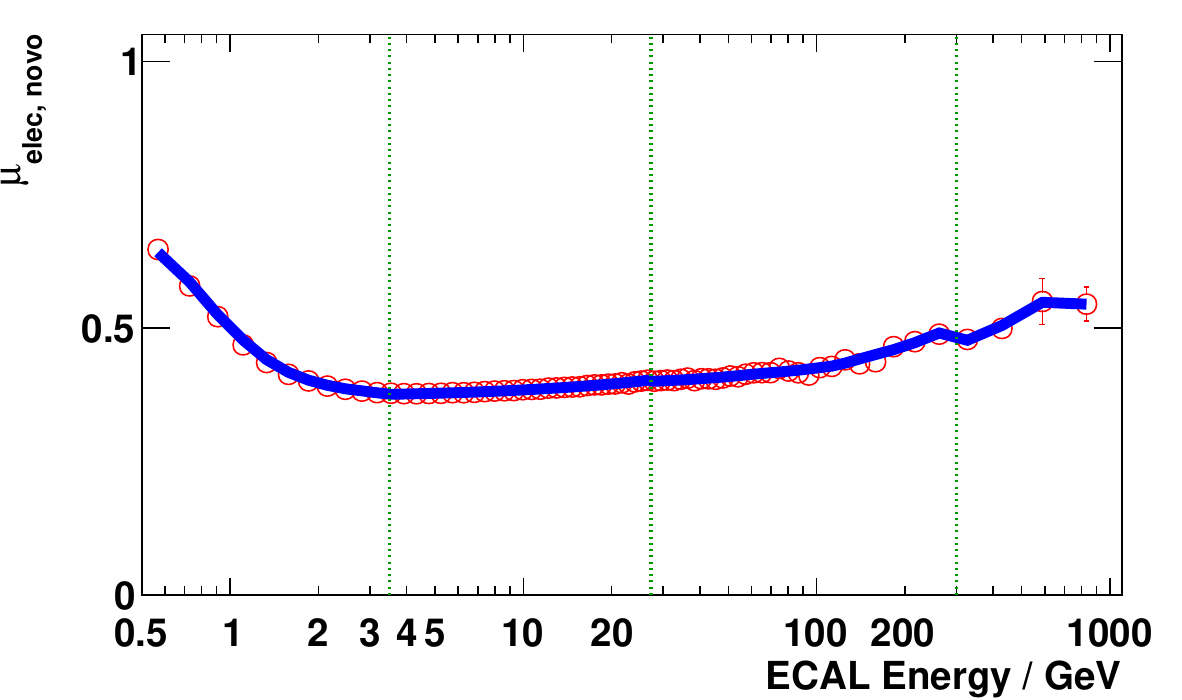}
    \caption{Template parameter~\elecPeakNovo}
  \end{subfigure}
  \hfill
  \begin{subfigure}{0.47\linewidth}
    \includegraphics[width=\linewidth]{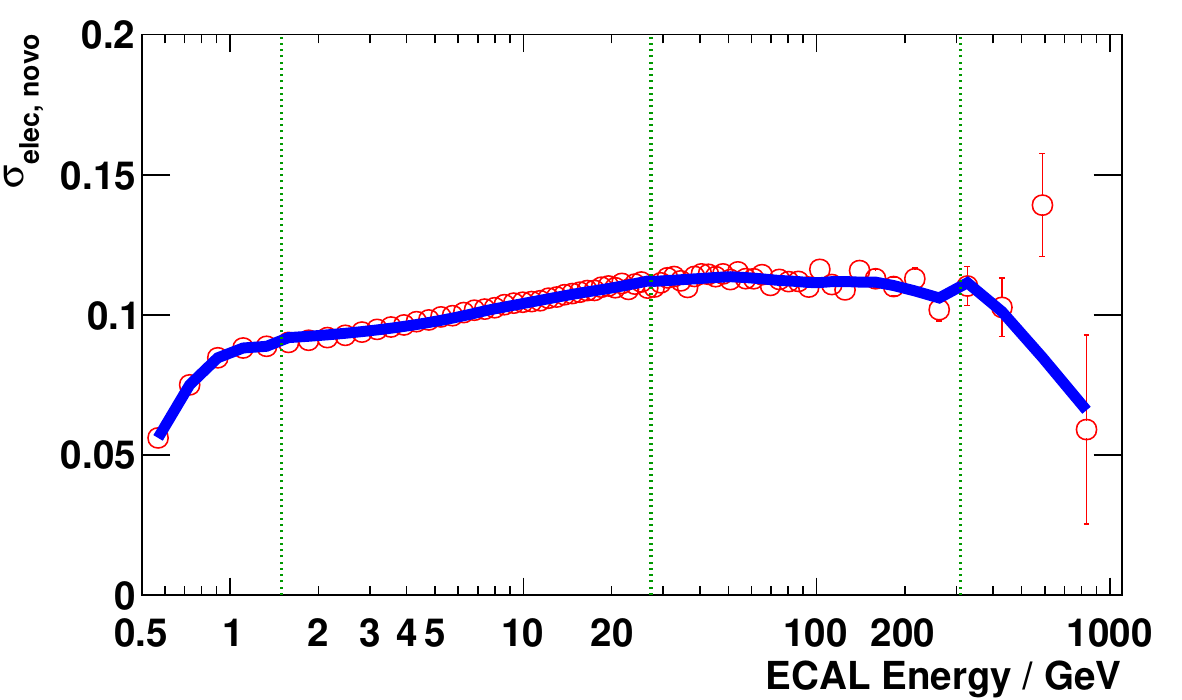}
    \caption{Template parameter~\elecWidthNovo}
  \end{subfigure}
  \hfill
  \begin{subfigure}{0.47\linewidth}
    \includegraphics[width=\linewidth]{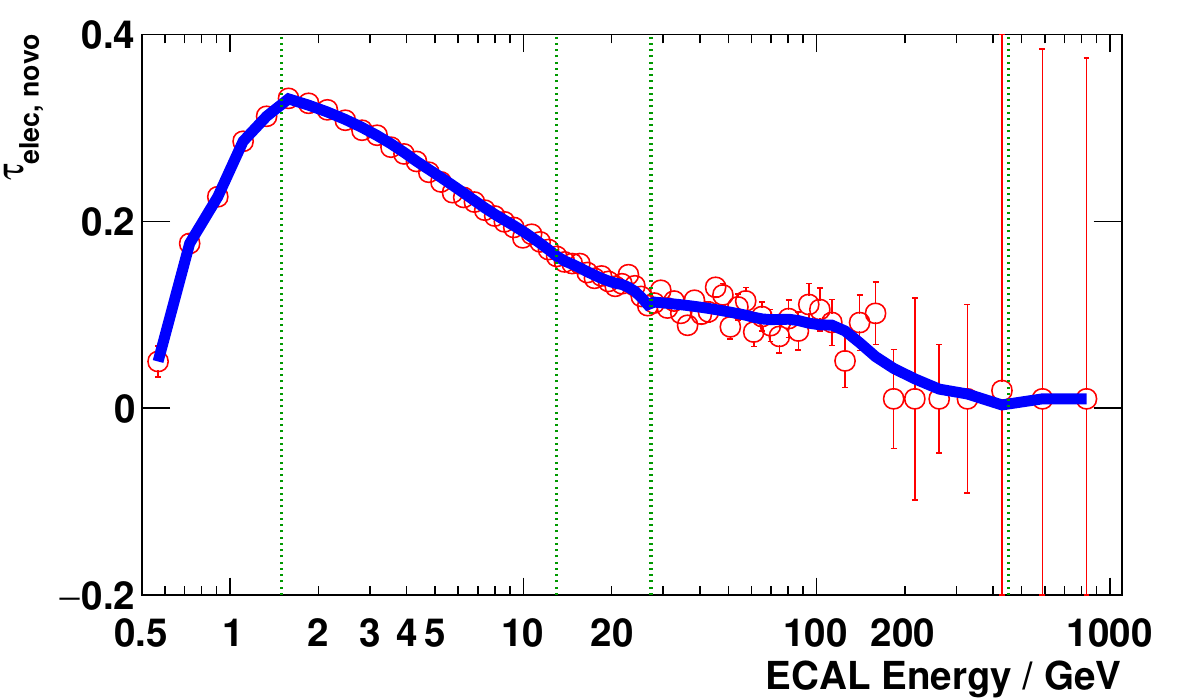}
    \caption{Template parameter~\elecTailNovo}
  \end{subfigure}
  \hfill
  \begin{subfigure}{0.47\linewidth}
    \includegraphics[width=\linewidth]{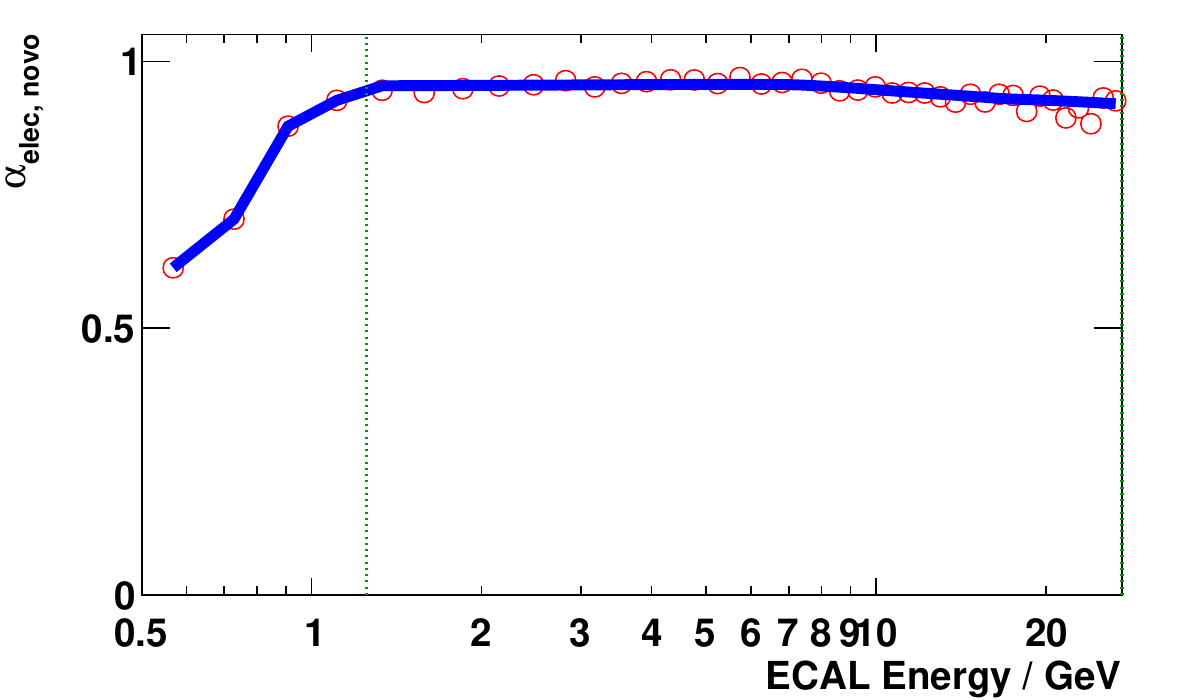}
    \caption{Template parameter~\elecFractionNovo}
    \label{fig:trd-template-parameters-electrons-elecfractionnovo-multi-tracks-vs-energy}
  \end{subfigure}
  \hfill
  \begin{subfigure}{0.47\linewidth}
    \includegraphics[width=\linewidth]{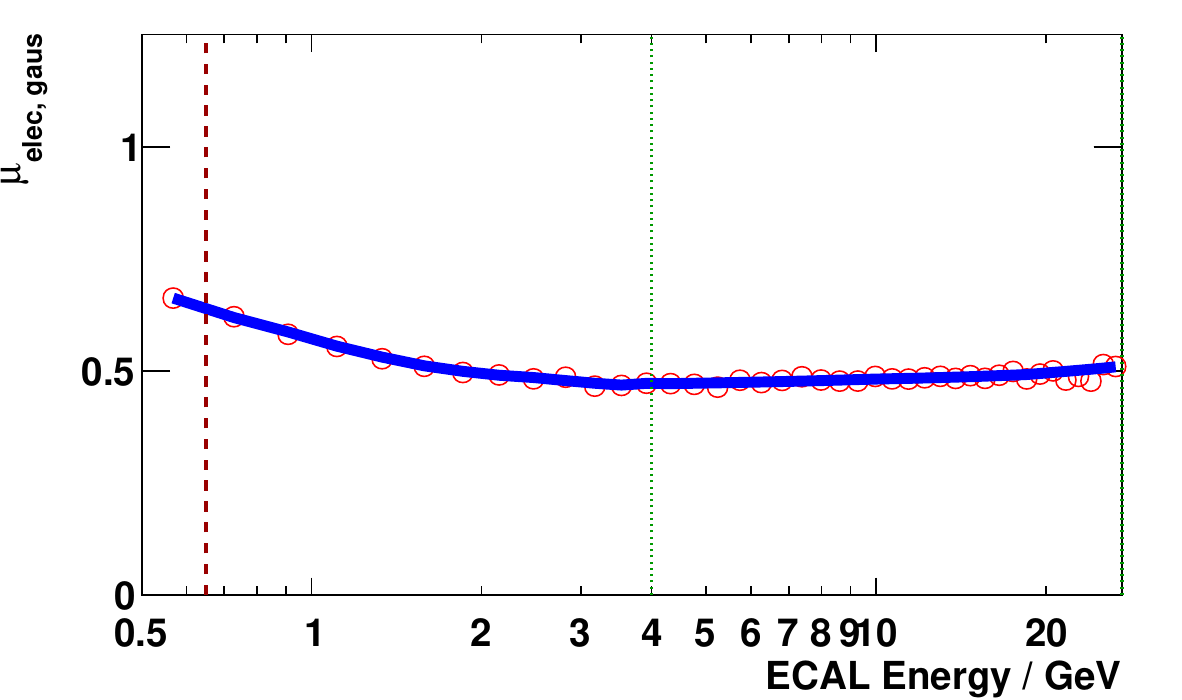}
    \caption{Template parameter~\elecPeakGaus}
    \label{fig:trd-template-parameters-electrons-elecpeakgaus-multi-tracks-vs-energy}
  \end{subfigure}
  \hfill
  \begin{subfigure}{0.47\linewidth}
    \includegraphics[width=\linewidth]{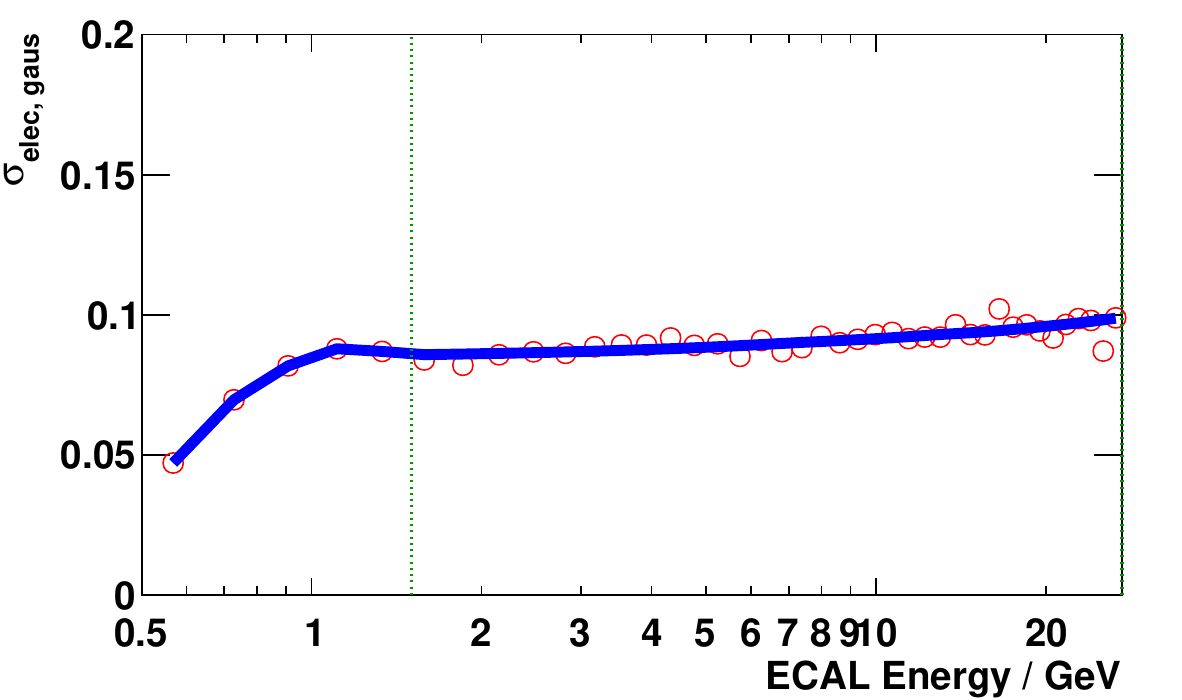}
    \caption{Template parameter \elecWidthGaus}
    \label{fig:trd-template-parameters-electrons-elecwidthgaus-multi-tracks-vs-energy}
  \end{subfigure}
  \hfill
  \begin{subfigure}{0.47\linewidth}
    \includegraphics[width=\linewidth]{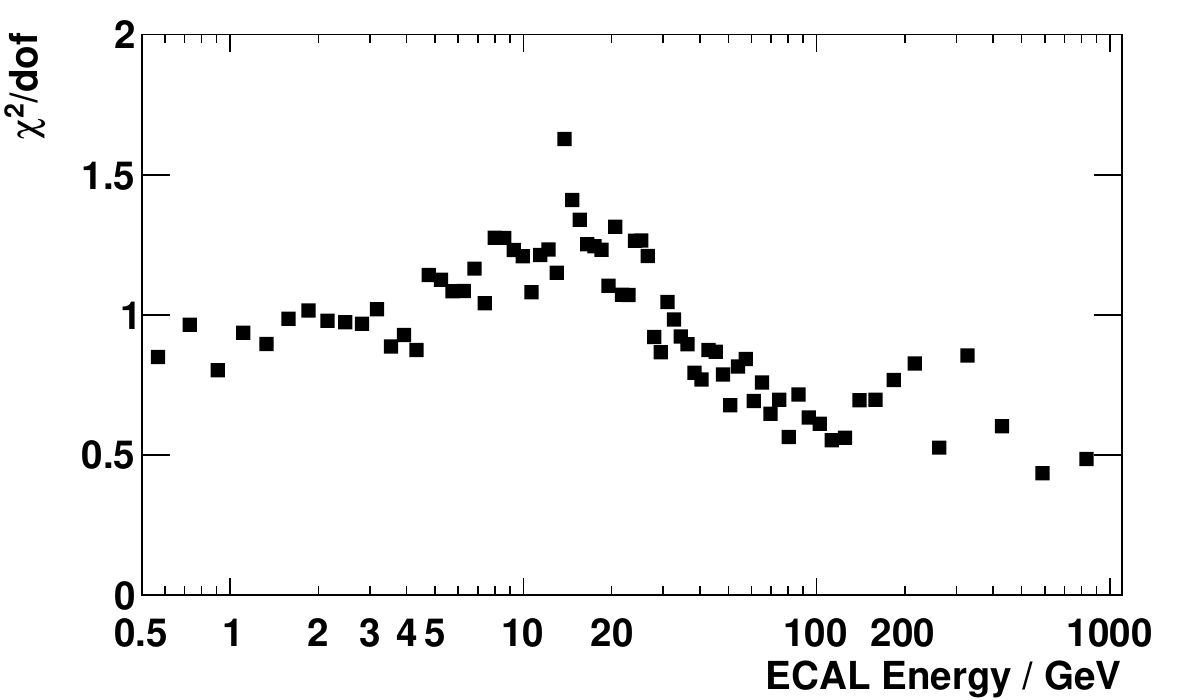}
    \caption{Goodness-of-fit}
    \label{fig:trd-template-parameters-electrons-chisquare-multi-tracks-vs-energy}
  \end{subfigure}
  \caption{Analytical TRD template parameters describing the evolution of the electron template as function of energy for the multi-tracks sample. The meaning of the red points, the blue curve and the dashed lines is given in the caption of \cref{fig:trd-template-parameters-ccprotons-multi-tracks-vs-energy}.}
  \label{fig:trd-template-parameters-electrons-multi-tracks-vs-energy}
\end{figure}

\Cref{fig:trd-template-parameters-electrons-multi-tracks-vs-energy} shows the electron template parameters
as function of energy for the multi-tracks sample.

\begin{figure}[H]
  \begin{subfigure}{0.48\linewidth}
    \includegraphics[width=\linewidth]{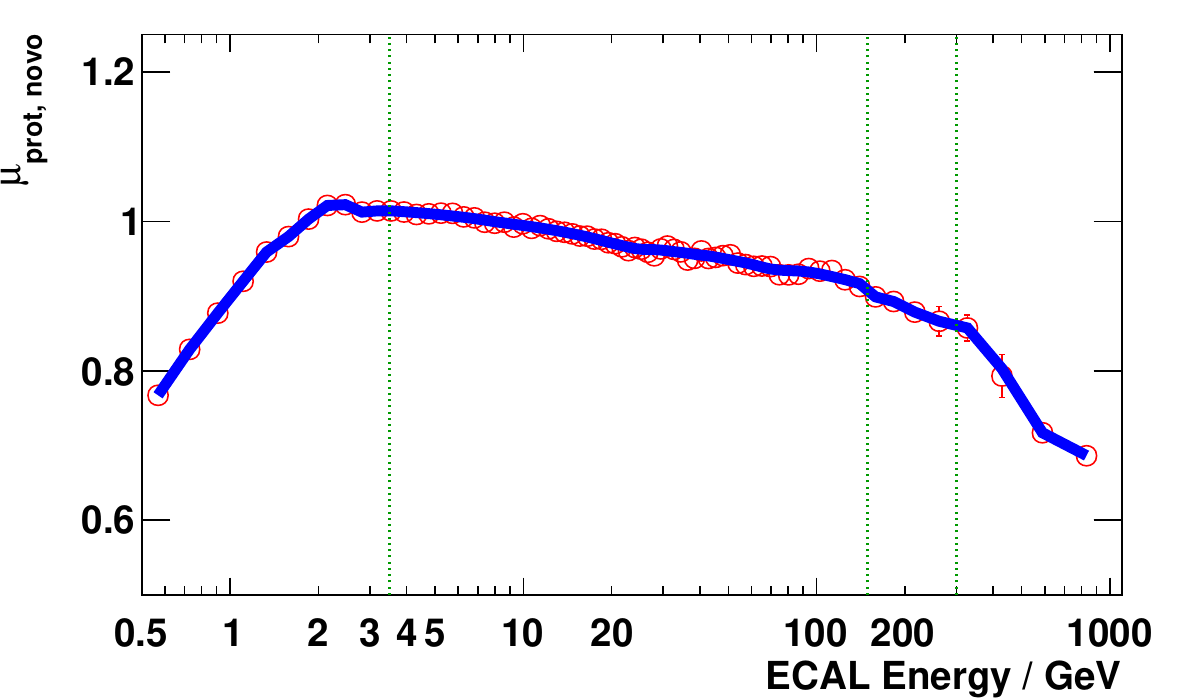}
    \caption{Template parameter~\protPeakNovo}
  \end{subfigure}
  \hfill
  \begin{subfigure}{0.48\linewidth}
    \includegraphics[width=\linewidth]{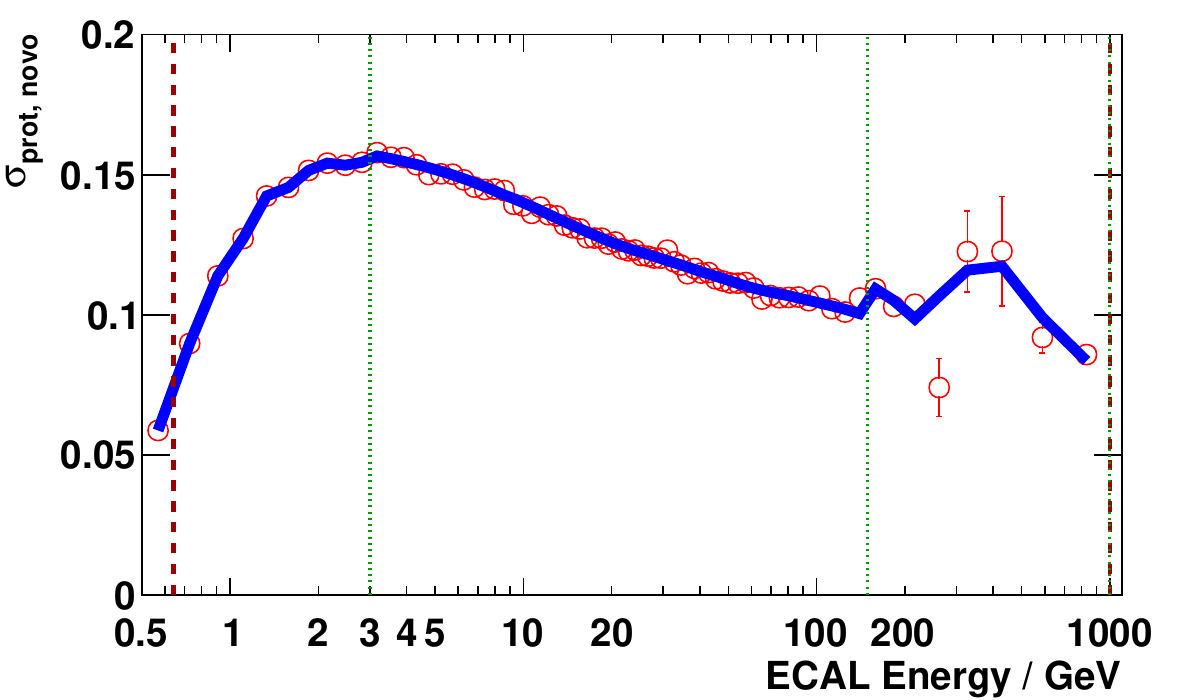}
    \caption{Template parameter~\protWidthNovo}
  \end{subfigure}
  \hfill
  \begin{subfigure}{0.48\linewidth}
    \includegraphics[width=\linewidth]{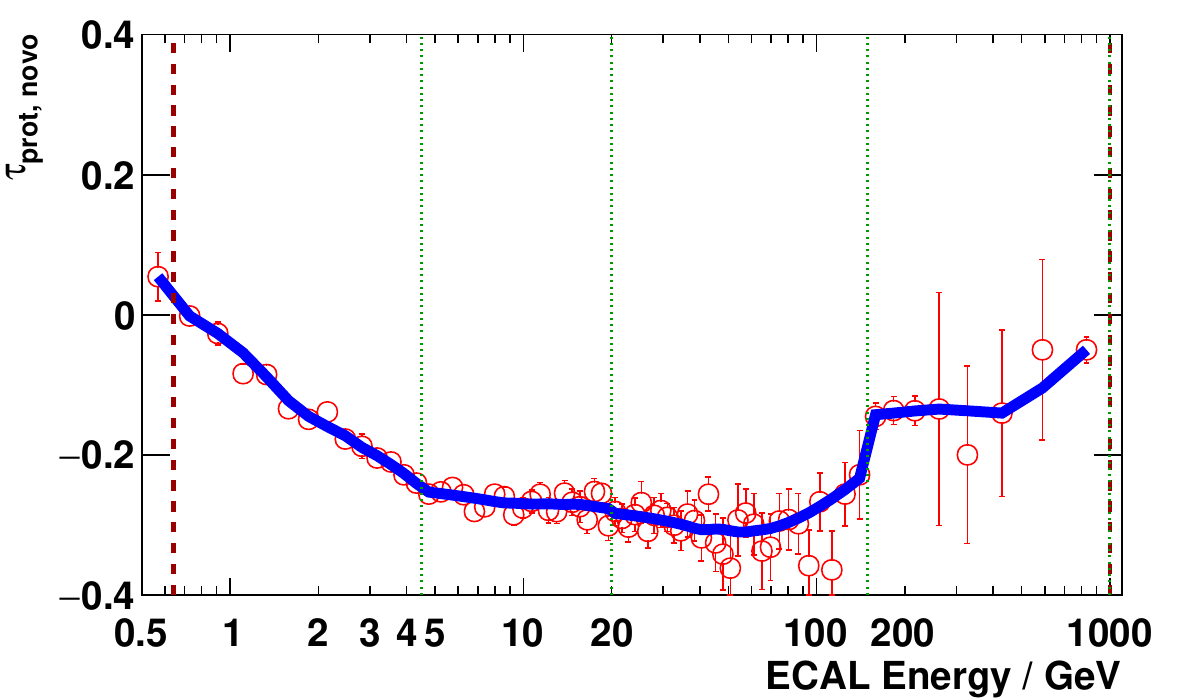}
    \caption{Template parameter~\protTailNovo}
  \end{subfigure}
  \hfill
  \begin{subfigure}{0.48\linewidth}
    \includegraphics[width=\linewidth]{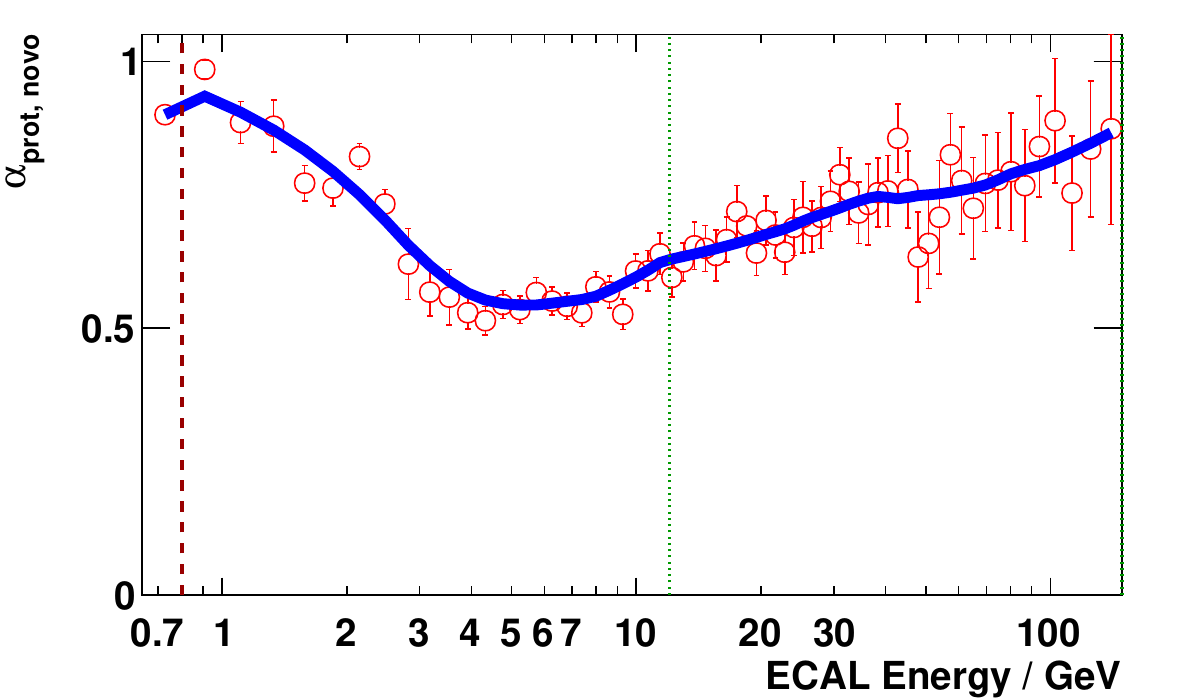}
    \caption{Template parameter~\protFractionNovo}
    \label{fig:trd-template-parameters-protons-protfractionnovo-multi-tracks-vs-energy}
  \end{subfigure}
  \hfill
  \begin{subfigure}{0.48\linewidth}
    \includegraphics[width=\linewidth]{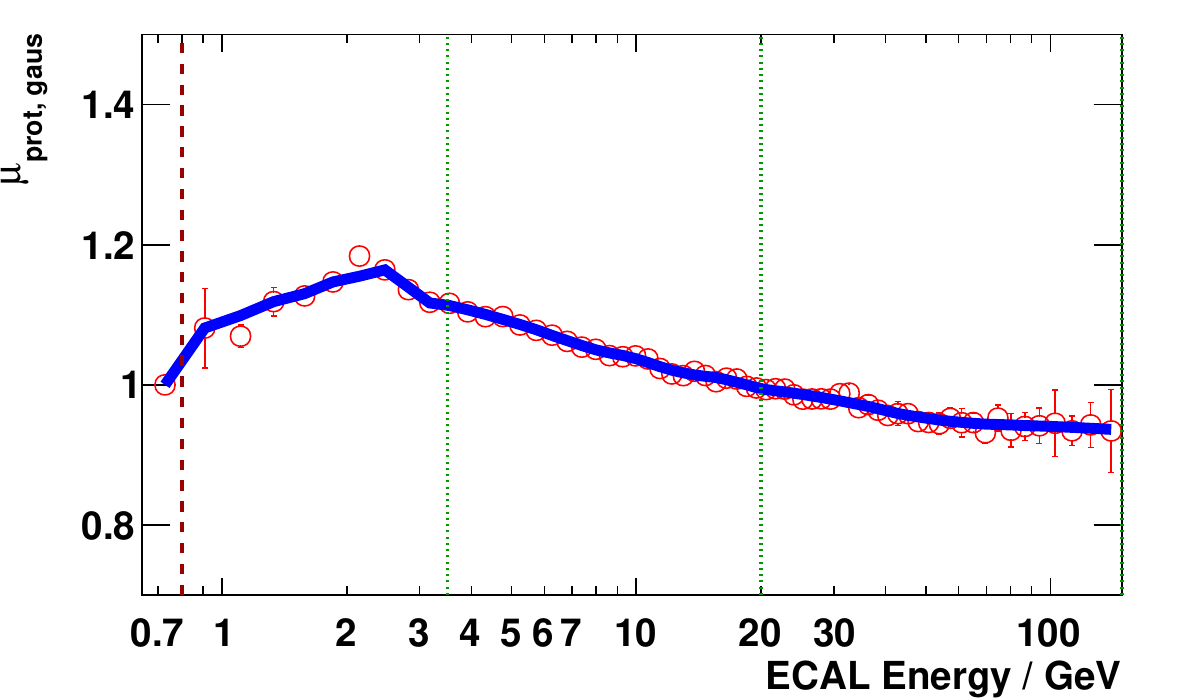}
    \caption{Template parameter~\protPeakGaus}
    \label{fig:trd-template-parameters-protons-protpeakgaus-multi-tracks-vs-energy}
  \end{subfigure}
  \hfill
  \begin{subfigure}{0.50\linewidth}
    \includegraphics[width=\linewidth]{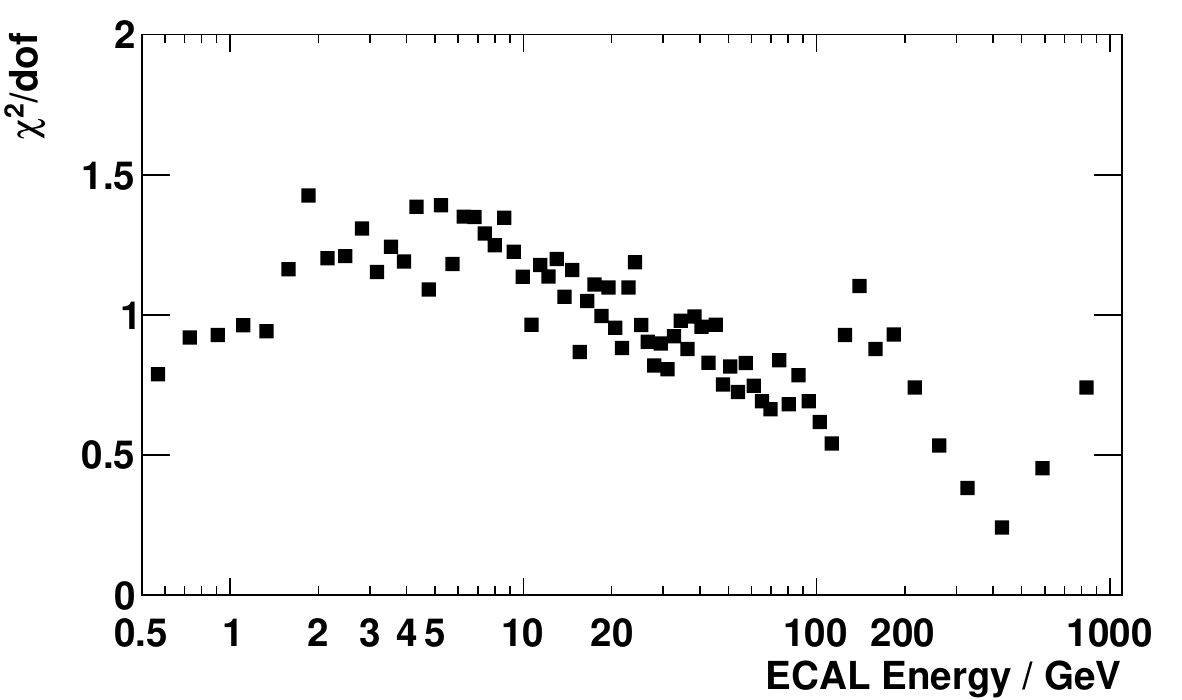}
    \caption{Goodness-of-fit}
    \label{fig:trd-template-parameters-protons-chisquare-multi-tracks-vs-energy}
  \end{subfigure}
  \caption{Analytical TRD template parameters describing the evolution of the proton template as function of energy for the multi-tracks sample.}
  \label{fig:trd-template-parameters-protons-multi-tracks-vs-energy}
\end{figure}

\Cref{fig:trd-template-parameters-electrons-chisquare-multi-tracks-vs-energy} shows that between \SIrange{6}{20}{\GeV} a tendency towards higher $\chi^2/\text{dof}$ is revealed. A closer inspection shows that the template data sample contains an additional component, that is not respected in the simultaneous fit procedure: fragmenting helium. In \cref{sec:analysis-lepton-counts-2d-fit} it was shown that this bias is not a practical problem when applying these templates to extract electron and positron counts, as the fragmenting helium component is not present in the final data samples, from which the number of electrons and positrons are extracted. \Cref{fig:trd-template-parameters-protons-multi-tracks-vs-energy} shows the proton template parameters
as function of energy for the multi-tracks sample.

For the multi-tracks sample, \cref{fig:trd-template-parameters-protons-chisquare-multi-tracks-vs-energy}, a tendency towards larger $\chi^2/\text{dof}$ values approaching lower energies is visible, due to the absence of the~\protWidthGausDelta~parameter. It was shown in \cref{sec:analysis-lepton-counts-2d-fit} that this bias is gone when applying these templates to the multi-tracks data sample, when extracting the final electron and positron counts. The systematic effect at low energies can be safely ignored.

\section{Construction of two-dimensional TRD / CCMVA templates (multi-tracks sample)}
\label{sec:appendix-2d-trd-ccmva-templates}

\begin{figure}[H]
  \begin{subfigure}{0.50\linewidth}
    \includegraphics[width=\linewidth]{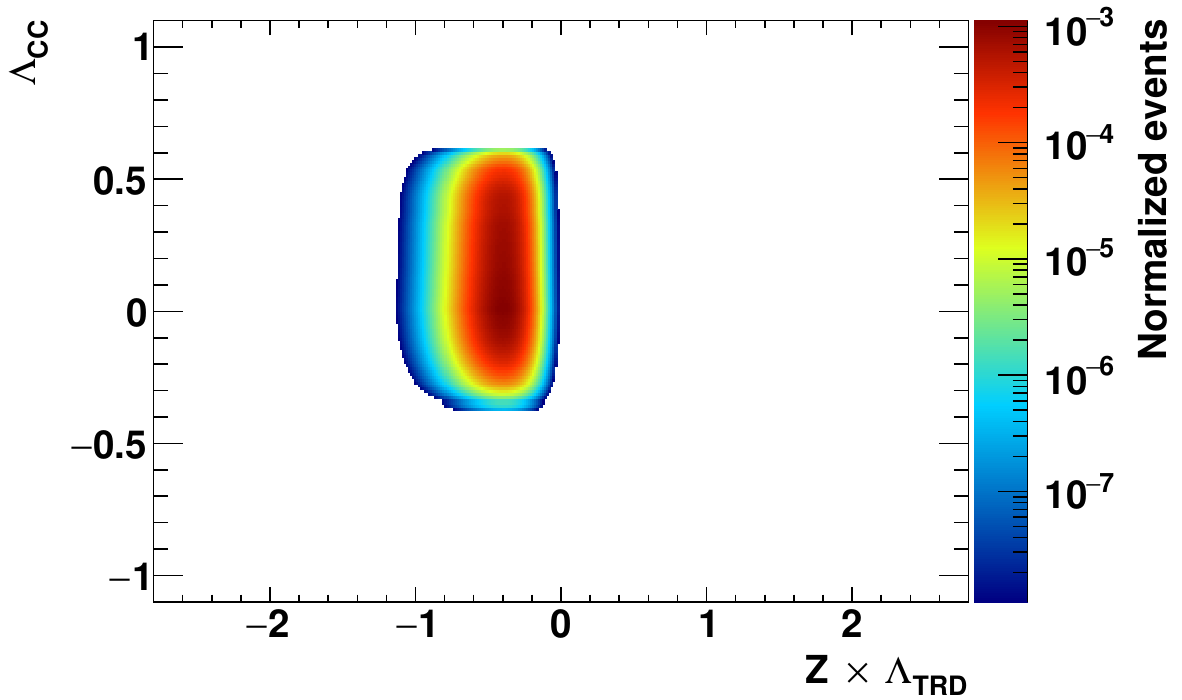}
    \caption{Electrons}
  \end{subfigure}
  \hfill
  \begin{subfigure}{0.50\linewidth}
    \includegraphics[width=\linewidth]{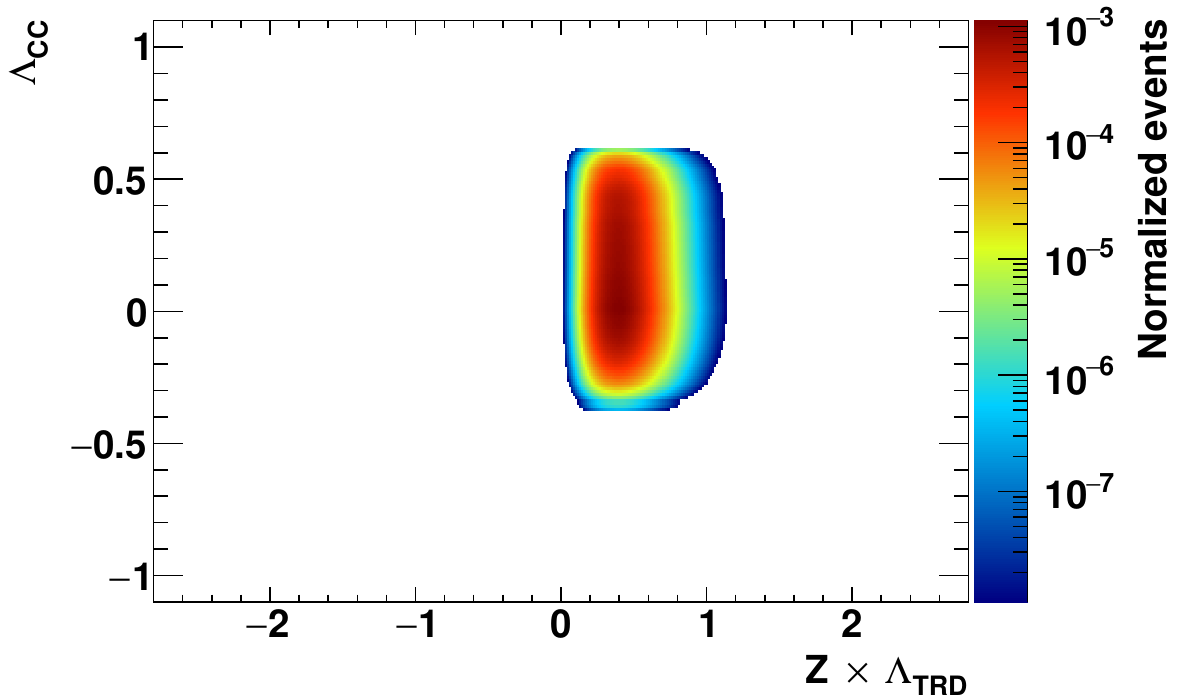}
    \caption{Positrons}
  \end{subfigure}
  \hfill
  \begin{subfigure}{0.50\linewidth}
    \includegraphics[width=\linewidth]{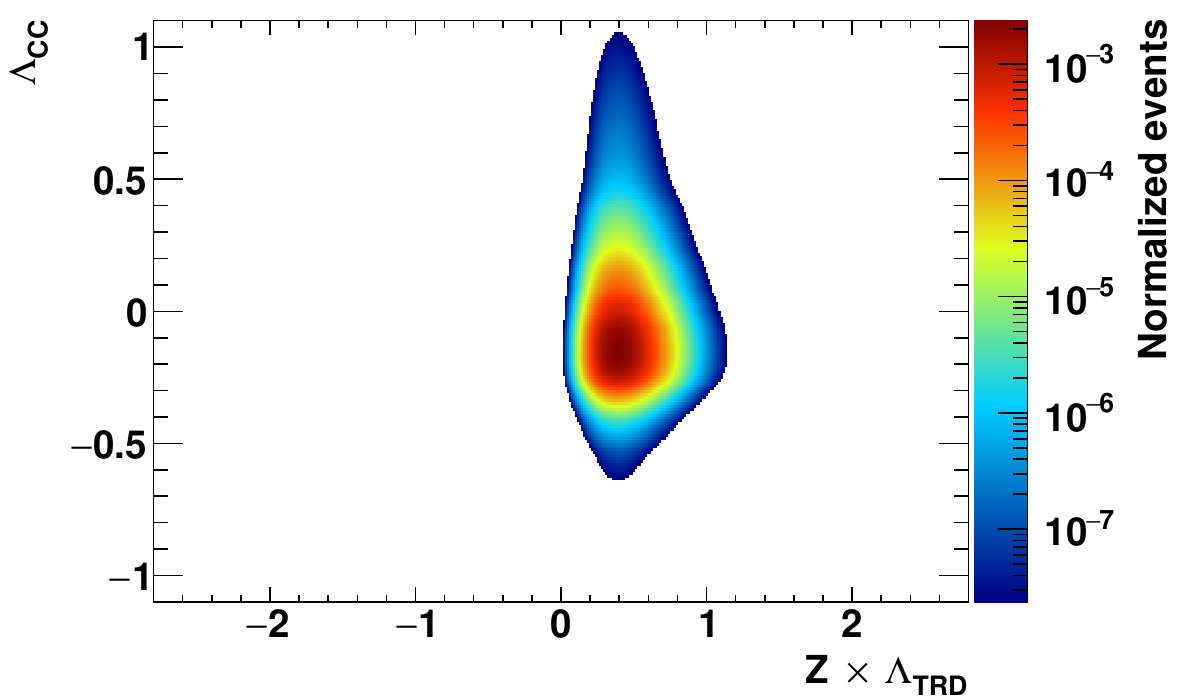}
    \caption{Charge-confused electrons}
  \end{subfigure}
  \hfill
  \begin{subfigure}{0.50\linewidth}
    \includegraphics[width=\linewidth]{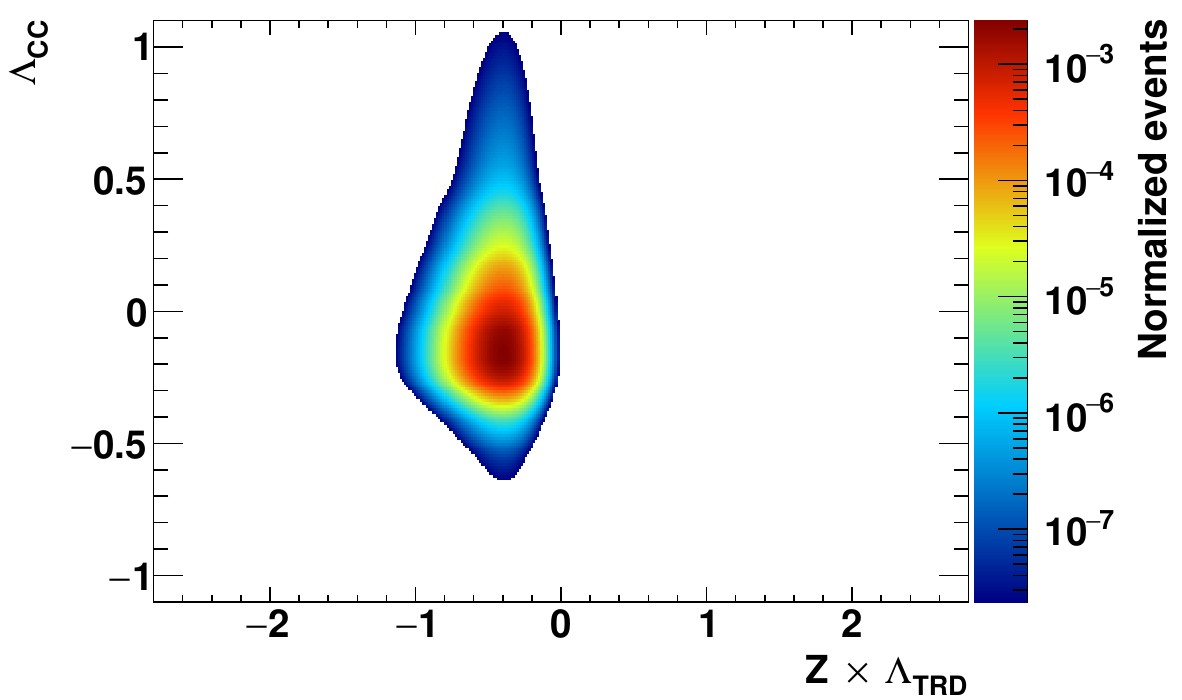}
    \caption{Charge-confused positrons}
  \end{subfigure}
  \hfill
  \begin{subfigure}{0.50\linewidth}
    \includegraphics[width=\linewidth]{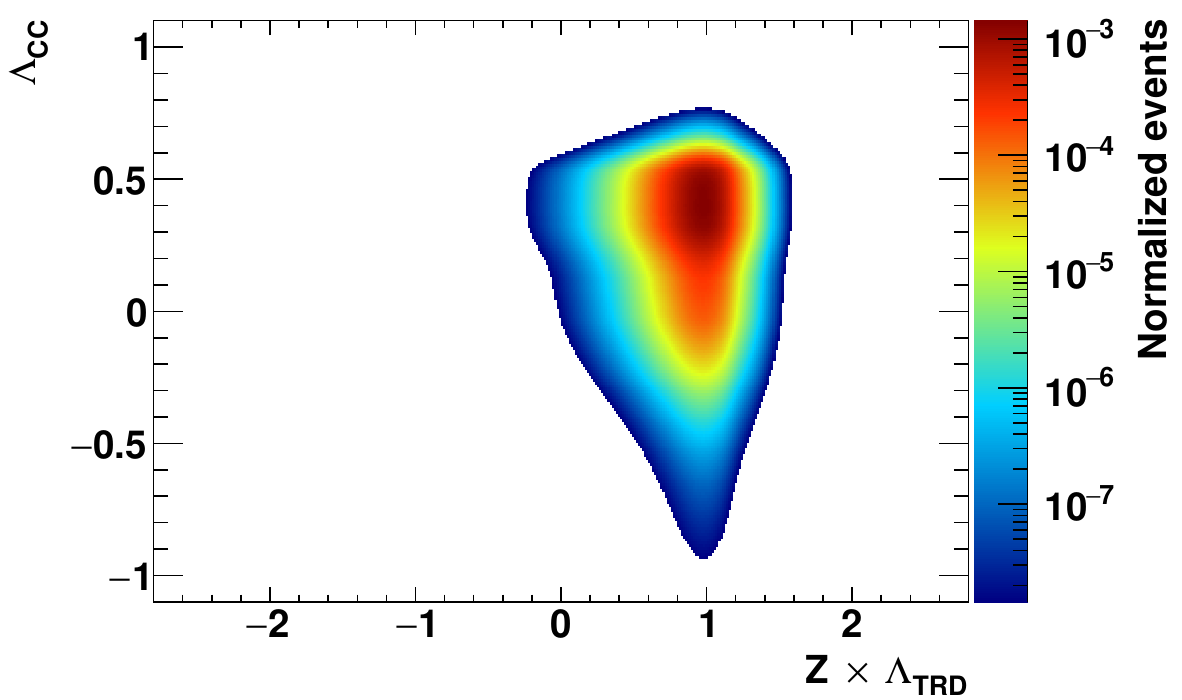}
    \caption{Protons}
  \end{subfigure}
  \hfill
  \begin{subfigure}{0.50\linewidth}
    \includegraphics[width=\linewidth]{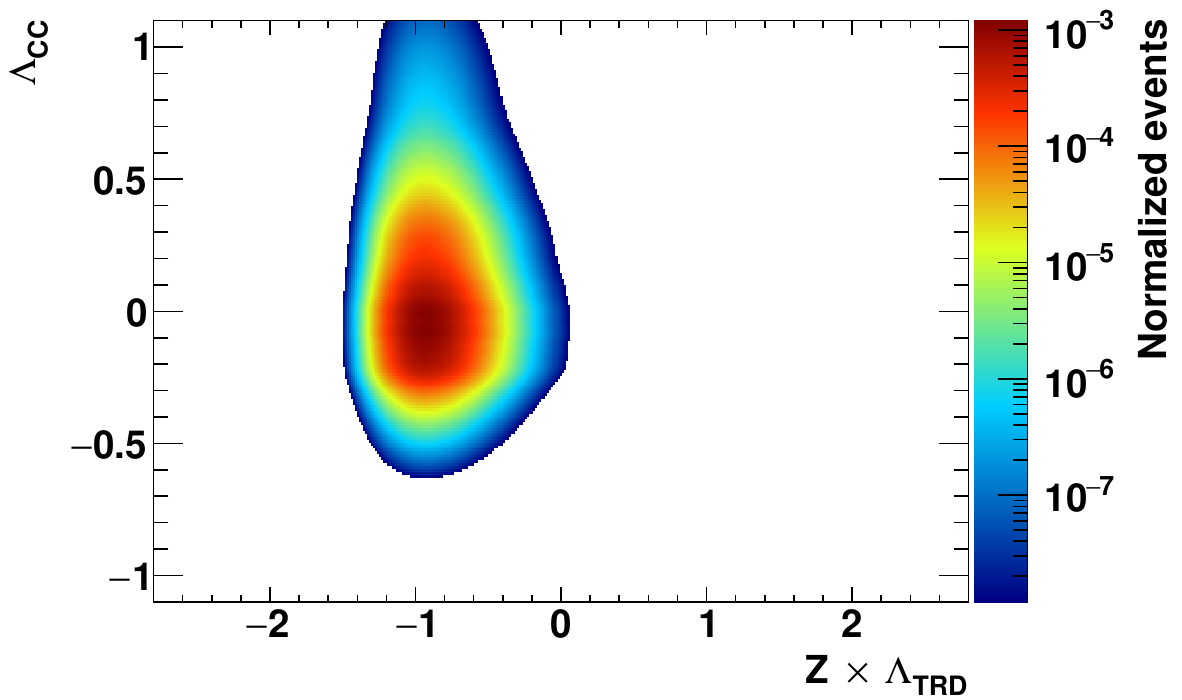}
    \caption{Charge-confused protons}
  \end{subfigure}
  \caption{Two-dimensional TRD / CCMVA templates in the energy bin \SIrange{17.98}{18.99}{\GeV} for the multi-tracks sample.}
  \label{fig:2d-templates-multi-tracks}
\end{figure}

\Cref{fig:2d-templates-multi-tracks} shows the two-dimensional templates for the multi-tracks sample in an example energy bin,
which were omitted in the introduction in \cref{sec:analysis-lepton-counts-ccmva-templates}.

\section{Tag cut definitions}
\label{sec:appendix-tag-and-probe-cuts}

In the following a list of all tag sample selection cuts that are used in the analysis
is presented. The quantities that appear in the cut conditions refer to branch names in the ROOT~\cite{Brun1997}
analysis tree used in this work to perform the data analysis.

\begin{enumerate}
  \item\label{enum:appendix-tag-and-probe-cuts-NegativeRigidityTag}\textbf{NegativeRigidityTag}\hfill
    \begin{itemize}
      \item R < 0\hfill\\
        The reconstructed rigidity of the selected tracker track must be negative.
    \end{itemize}

  \item\label{enum:appendix-tag-and-probe-cuts-HasTrackerTag}\textbf{HasTrackerTag}\hfill
    \begin{itemize}
      \item TrkNumTracks = 1\hfill\\
        There must be only one reconstructed tracker track in the event.
      \item R $\neq$ 0\hfill\\
        The reconstructed rigidity must be non-zero.
      \item TrkPattern = 1 or TrkPattern = 2\hfill\\
        The tracker track must have a hit in Layer 1 and Layer 9, or the tracker track must have a hit in Layer 2 and Layer 9.
      \item \SI{0.5}{\elementarycharge} < TrkCharge < \SI{1.8}{\elementarycharge}\hfill\\
        The tracker charge must be compatible with a $Z = 1$ particle.
      \item 0.01 < TrackerChiSquareY < 25\hfill\\
        The tracker track goodness-of-fit in Y projection must be in a reasonable range, to filter out wrongly reconstructed tracks.
      \item Tracker track must point to the ECAL\hfill\\
        The tracker track extrapolation needs to go through the ECAL.
    \end{itemize}

  \item\label{enum:appendix-tag-and-probe-cuts-HasTofTag}\textbf{HasTofTag}\hfill
    \begin{itemize}
      \item \SI{0.5}{\elementarycharge} < TofUpperCharge < \SI{1.8}{\elementarycharge}\hfill\\
        The upper TOF charge must be compatible with a $Z=1$ particle.
      \item 0.8 < TofBeta < 1.25\hfill\\
        The reconstructed TOF velocity must be compatible with a relativistic particle.
    \end{itemize}

  \item\label{enum:appendix-tag-and-probe-cuts-HasTrdTag}\textbf{HasTrdTag}\hfill
    \begin{itemize}
      \item TrdActiveLayers > 15\hfill\\
        At least 15 layers in the TRD must have a hit.
      \item $\mathcal{L}_{e^{-}/\ce{He}}$ > 0.8 \hfill\\
        Helium identified using the TRD must be rejected.
    \end{itemize}

  \item\label{enum:appendix-tag-and-probe-cuts-HasEcalTag}\textbf{HasEcalTag}\hfill
    \begin{itemize}
      \item $\Lambda_{\text{ECAL}}$ > -1\hfill\\
        The ECAL estimator measurement must be available.
    \end{itemize}

  \bigskip
  \item\label{enum:appendix-tag-and-probe-cuts-EcalElectronTag}\textbf{EcalElectronTag}\hfill
    \begin{itemize}
      \item E/R > 0.5\hfill\\
        If both a rigidity and ECAL energy measurement is available, it must be larger than 0.5.
      \item EcalShowerDirectionZ > 0\hfill\\
        The ECAL shower must be down-going.
      \item $\Lambda_{\text{ECAL}}$ > 0\hfill\\
        The ECAL estimator measurement must be electron like.
    \end{itemize}

  \item\label{enum:appendix-tag-and-probe-cuts-TofElectronTag}\textbf{TofElectronTag}\hfill
    \begin{itemize}
      \item $\abs{1 - 1 / \text{TofBeta}} < 0.15$\hfill\\
        The reconstructed TOF velocity must be compatible with a relativistic $e^{\pm}$.
    \end{itemize}

  \item\label{enum:appendix-tag-and-probe-cuts-TrdElectronTag}\textbf{TrdElectronTag}\hfill
    \begin{itemize}
      \item $\Lambda_{\text{TRD}}$ < 0.75\hfill\\
        TRD selection of electrons using the log-likelihood ratio estimator.
    \end{itemize}

  \item\label{enum:appendix-tag-and-probe-cuts-TrdHasUsefulSegmentsInBothProjectionsTag}\textbf{TrdHasUsefulSegmentsInBothProjectionsTag}\hfill\\
    Note: A TRD segment refers to a partially reconstructed TRD track in two dimensions, utilizing only hits of one projection.

    \begin{itemize}
      \item Any of the segments in X-Z projection must span at least 6 layers.\hfill
      \item Any of the segments in Y-Z projection must span at least 4 layers.\hfill
      \item The first hit of any of the segments in Y-Z projection must be in the upper half of the TRD.\hfill
      \item The last hit of any of the segments in Y-Z projection must be in the lower half of the TRD.\hfill
    \end{itemize}

  \item\label{enum:appendix-tag-and-probe-cuts-TrdHasUsefulTrackTag}\textbf{TrdHasUsefulTrackTag}\hfill\\
    Note: A TRD track is composed from a single X-Z segment and a single Y-Z segment.

    \begin{itemize}
      \item The TRD track in X-Z projection must span at least 6 layers.\hfill
      \item The TRD track in Y-Z projection must span at least 4 layers.\hfill
      \item The first hit of the TRD track must be in the upper half of the TRD.\hfill
      \item The last hit of the TRD track must be in the lower half of the TRD.\hfill
    \end{itemize}

  \item\label{enum:appendix-tag-and-probe-cuts-TrdActiveLayersTag}\textbf{TrdActiveLayersTag}\hfill
    \begin{itemize}
      \item Energy-dependent TrdActiveLayers cut\hfill\\
        Same definition as in the corresponding selection cut (\cref{sec:analysis-data-selection-selection-cuts} - \cref{enum:selection-cut-trd-active-layers}).
    \end{itemize}

  \item\label{enum:appendix-tag-and-probe-cuts-TrdNoHeliumTag}\textbf{TrdNoHeliumTag}\hfill
    \begin{itemize}
      \item $\mathcal{L}_{e^{-}/\ce{He}}$ > 0.8 \hfill\\
        Helium identified using the TRD must be rejected.
    \end{itemize}

  \item\label{enum:appendix-tag-and-probe-cuts-TofBetaTag}\textbf{TofBetaTag}\hfill
    \begin{itemize}
      \item 0.8 < TofBeta < 1.25\hfill\\
        The reconstructed TOF velocity must be compatible with a relativistic particle.
    \end{itemize}

  \item\label{enum:appendix-tag-and-probe-cuts-TofNumberOfLayersTag}\textbf{TofNumberOfLayersTag}\hfill
    \begin{itemize}
      \item TofNumberOfLayers > 2\hfill\\
        At least 3 out of the 4 TOF layers must have a reconstructed cluster.
    \end{itemize}

  \item\label{enum:appendix-tag-and-probe-cuts-TofTimeDifferenceTag}\textbf{TofTimeDifferenceTag}\hfill
    \begin{itemize}
      \item \SI{0}{\nano\second} < TofDeltaT < \SI{20}{\nano\second}\hfill\\
        The particle passage from upper to lower TOF must not last longer than \SI{20}{\nano\second}. A relativistic
        $e^{\pm}$ takes on average \SI{4.3}{\nano\second} for the passage. This cut only rejects wrongly associated clusters,
        which might lead to a wrong time difference measurement.
    \end{itemize}

  \item\label{enum:appendix-tag-and-probe-cuts-TrackerChargeTag}\textbf{TrackerChargeTag}\hfill
    \begin{itemize}
      \item \SI{0.5}{\elementarycharge} < TrkCharge < \SI{1.8}{\elementarycharge}\hfill\\
        The tracker charge must be compatible with a $Z = 1$ particle.
    \end{itemize}

  \item\label{enum:appendix-tag-and-probe-cuts-TrackerPatternTag}\textbf{TrackerPatternTag}\hfill
    \begin{itemize}
      \item 1 $\leq$ TrkPattern $\leq$ 4\hfill\\
        Same definition as in the corresponding selection cut (\cref{sec:analysis-data-selection-selection-cuts} - \cref{enum:selection-cut-trk-pattern}).
    \end{itemize}

  \item\label{enum:appendix-tag-and-probe-cuts-ElectronUpperEnergyOverRigidityTag}\textbf{ElectronUpperEnergyOverRigidityTag}\hfill
    \begin{itemize}
      \item E/R < 10\hfill\\
        This matches the upper cut value of the corresponding $e^{\pm}$ identification cut.
    \end{itemize}
  \item\label{enum:appendix-tag-and-probe-cuts-EcalPreselectionTag}\textbf{EcalPreselectionTag}\hfill
    \begin{itemize}
      \item $\abs{\text{EcalCentreOfGravityX}}$ < \SI{32}{\centi\meter} and $\abs{\text{EcalCentreOfGravityY}}$ < \SI{32}{\centi\meter}\hfill\\
        The centre-of-gravity of the reconstructed shower is at least 1.5 Moli\`{e}re radii away from the borders of the ECAL.
    \end{itemize}

  \item\label{enum:appendix-tag-and-probe-cuts-EcalTrdPreselectionTag}\textbf{EcalTrdPreselectionTag}\hfill
    \begin{itemize}
      \item All cuts from \textbf{TrdHasUsefulTrackTag} / \textbf{EcalPreselectionTag} (\cref{enum:appendix-tag-and-probe-cuts-TrdHasUsefulTrackTag,enum:appendix-tag-and-probe-cuts-EcalPreselectionTag})\hfill
      \item TRD track / ECAL shower matching\hfill\\
        The TRD track extrapolated to the ECAL Z position must be spatially compatible with the reconstructed ECAL shower centre-of-gravity in X-Z and Y-Z projection.
    \end{itemize}
\end{enumerate}

\section{Tag sample selection cuts}
\label{sec:appendix-tag-and-probe-samples}

\subsection{Tag cuts used for the \enquote{At least one useful TRD track} cut}
\label{sec:appendix-tag-and-probe-samples-at-least-one-useful-trd-track}

The tag sample for the \textbf{\enquote{At least one useful TRD track}} cut (\cref{sec:analysis-data-selection-preselection-cuts} - \cref{enum:preselection-cut-at-least-one-useful-trd-track}) is prepared using following cuts:

\begin{itemize}
  \item\textbf{All detector quality cuts}~(\cref{sec:analysis-data-selection-detector-quality-cuts})
  \item\textbf{NegativeRigidityTag}~(\cref{sec:appendix-tag-and-probe-cuts} - \cref{enum:appendix-tag-and-probe-cuts-NegativeRigidityTag})
  \item\textbf{HasTrackerTag}~(\cref{sec:appendix-tag-and-probe-cuts} - \cref{enum:appendix-tag-and-probe-cuts-HasTrackerTag})
  \item\textbf{HasTofTag}~(\cref{sec:appendix-tag-and-probe-cuts} - \cref{enum:appendix-tag-and-probe-cuts-HasTofTag})
  \item\textbf{HasEcalTag}~(\cref{sec:appendix-tag-and-probe-cuts} - \cref{enum:appendix-tag-and-probe-cuts-HasEcalTag})
  \item\textbf{EcalElectronTag}~(\cref{sec:appendix-tag-and-probe-cuts} - \cref{enum:appendix-tag-and-probe-cuts-EcalElectronTag})
  \item\textbf{TofElectronTag}~(\cref{sec:appendix-tag-and-probe-cuts} - \cref{enum:appendix-tag-and-probe-cuts-TofElectronTag})
  \item\textbf{TrdHasUsefulSegmentsInBothProjectionsTag}~(\cref{sec:appendix-tag-and-probe-cuts} - \cref{enum:appendix-tag-and-probe-cuts-TrdHasUsefulSegmentsInBothProjectionsTag})
  \item\textbf{EcalPreselectionTag}~(\cref{sec:appendix-tag-and-probe-cuts} - \cref{enum:appendix-tag-and-probe-cuts-EcalPreselectionTag})
\end{itemize}

\subsection{Tag cuts used for the \enquote{At least one useful TOF cluster combination} cut}
\label{sec:appendix-tag-and-probe-samples-at-least-one-useful-tof-track}

The tag sample for the \textbf{\enquote{At least one useful TOF cluster combination}} cut (\cref{sec:analysis-data-selection-preselection-cuts} - \cref{enum:preselection-cut-at-least-one-useful-tof-track}) is prepared using following cuts:

\begin{itemize}
  \item\textbf{All detector quality cuts}~(\cref{sec:analysis-data-selection-detector-quality-cuts})
  \item\textbf{NegativeRigidityTag}~(\cref{sec:appendix-tag-and-probe-cuts} - \cref{enum:appendix-tag-and-probe-cuts-NegativeRigidityTag})
  \item\textbf{HasTrackerTag}~(\cref{sec:appendix-tag-and-probe-cuts} - \cref{enum:appendix-tag-and-probe-cuts-HasTrackerTag})
  \item\textbf{HasTrdTag}~(\cref{sec:appendix-tag-and-probe-cuts} - \cref{enum:appendix-tag-and-probe-cuts-HasTrdTag})
  \item\textbf{HasEcalTag}~(\cref{sec:appendix-tag-and-probe-cuts} - \cref{enum:appendix-tag-and-probe-cuts-HasEcalTag})
  \item\textbf{EcalElectronTag}~(\cref{sec:appendix-tag-and-probe-cuts} - \cref{enum:appendix-tag-and-probe-cuts-EcalElectronTag})
  \item\textbf{TrdElectronTag}~(\cref{sec:appendix-tag-and-probe-cuts} - \cref{enum:appendix-tag-and-probe-cuts-TrdElectronTag})
  \item\textbf{TofNumberOfLayersTag}~(\cref{sec:appendix-tag-and-probe-cuts} - \cref{enum:appendix-tag-and-probe-cuts-TofNumberOfLayersTag})
  \item\textbf{TofTimeDifferenceTag}~(\cref{sec:appendix-tag-and-probe-cuts} - \cref{enum:appendix-tag-and-probe-cuts-TofTimeDifferenceTag})
  \item\textbf{EcalTrdPreselectionTag}~(\cref{sec:appendix-tag-and-probe-cuts} - \cref{enum:appendix-tag-and-probe-cuts-EcalTrdPreselectionTag})
\end{itemize}

\subsection{Tag cuts used for the \enquote{Upper TOF charge} cut}
\label{sec:appendix-tag-and-probe-samples-upper-tof-charge}

The tag sample for the \textbf{\enquote{Upper TOF charge}} cut (\cref{sec:analysis-data-selection-selection-cuts} - \cref{enum:selection-cut-upper-tof-charge}) is prepared using following cuts:

\begin{itemize}
  \item\textbf{All detector quality cuts}~(\cref{sec:analysis-data-selection-detector-quality-cuts})
  \item\textbf{All preselection cuts}~(\cref{sec:analysis-data-selection-preselection-cuts})
  \item\textbf{NegativeRigidityTag}~(\cref{sec:appendix-tag-and-probe-cuts} - \cref{enum:appendix-tag-and-probe-cuts-NegativeRigidityTag})
  \item\textbf{EcalElectronTag}~(\cref{sec:appendix-tag-and-probe-cuts} - \cref{enum:appendix-tag-and-probe-cuts-EcalElectronTag})
  \item\textbf{TrdElectronTag}~(\cref{sec:appendix-tag-and-probe-cuts} - \cref{enum:appendix-tag-and-probe-cuts-TrdElectronTag})
  \item\textbf{TrdNoHeliumTag}~(\cref{sec:appendix-tag-and-probe-cuts} - \cref{enum:appendix-tag-and-probe-cuts-TrdNoHeliumTag})
  \item\textbf{TofBetaTag}~(\cref{sec:appendix-tag-and-probe-cuts} - \cref{enum:appendix-tag-and-probe-cuts-TofBetaTag})
  \item\textbf{TrackerChargeTag}~(\cref{sec:appendix-tag-and-probe-cuts} - \cref{enum:appendix-tag-and-probe-cuts-TrackerChargeTag})
\end{itemize}

\subsection{Tag cuts used for the \enquote{Enough active layers in TRD} cut}
\label{sec:appendix-tag-and-probe-samples-trd-active-layers}

The tag sample for the \textbf{\enquote{Enough active layers in TRD}} cut (\cref{sec:analysis-data-selection-selection-cuts} - \cref{enum:selection-cut-trd-active-layers}) is prepared using following cuts:

\begin{itemize}
  \item\textbf{All detector quality cuts}~(\cref{sec:analysis-data-selection-detector-quality-cuts})
  \item\textbf{All preselection cuts}~(\cref{sec:analysis-data-selection-preselection-cuts})
  \item\textbf{NegativeRigidityTag}~(\cref{sec:appendix-tag-and-probe-cuts} - \cref{enum:appendix-tag-and-probe-cuts-NegativeRigidityTag})
  \item\textbf{HasTofTag}~(\cref{sec:appendix-tag-and-probe-cuts} - \cref{enum:appendix-tag-and-probe-cuts-HasTofTag})
  \item\textbf{EcalElectronTag}~(\cref{sec:appendix-tag-and-probe-cuts} - \cref{enum:appendix-tag-and-probe-cuts-EcalElectronTag})
  \item\textbf{TrdElectronTag}~(\cref{sec:appendix-tag-and-probe-cuts} - \cref{enum:appendix-tag-and-probe-cuts-TrdElectronTag})
  \item\textbf{TrackerChargeTag}~(\cref{sec:appendix-tag-and-probe-cuts} - \cref{enum:appendix-tag-and-probe-cuts-TrackerChargeTag})
\end{itemize}

\subsection{Tag cuts used for the \enquote{Tracker track goodness-of-fit in Y-projection} cut}
\label{sec:appendix-tag-and-probe-samples-tracker-chi-square-y}

The tag sample for the \textbf{\enquote{Tracker track goodness-of-fit in Y-projection}} cut (\cref{sec:analysis-data-selection-selection-cuts} - \cref{enum:selection-cut-trk-chi-square-y}) is prepared using following cuts:

\begin{itemize}
  \item\textbf{All detector quality cuts}~(\cref{sec:analysis-data-selection-detector-quality-cuts})
  \item\textbf{All preselection cuts}~(\cref{sec:analysis-data-selection-preselection-cuts})
  \item\textbf{NegativeRigidityTag}~(\cref{sec:appendix-tag-and-probe-cuts} - \cref{enum:appendix-tag-and-probe-cuts-NegativeRigidityTag})
  \item\textbf{HasTofTag}~(\cref{sec:appendix-tag-and-probe-cuts} - \cref{enum:appendix-tag-and-probe-cuts-HasTofTag})
  \item\textbf{EcalElectronTag}~(\cref{sec:appendix-tag-and-probe-cuts} - \cref{enum:appendix-tag-and-probe-cuts-EcalElectronTag})
  \item\textbf{TrdElectronTag}~(\cref{sec:appendix-tag-and-probe-cuts} - \cref{enum:appendix-tag-and-probe-cuts-TrdElectronTag})
  \item\textbf{TrdActiveLayersTag}~(\cref{sec:appendix-tag-and-probe-cuts} - \cref{enum:appendix-tag-and-probe-cuts-TrdActiveLayersTag})
  \item\textbf{TrackerChargeTag}~(\cref{sec:appendix-tag-and-probe-cuts} - \cref{enum:appendix-tag-and-probe-cuts-TrackerChargeTag})
  \item\textbf{TrackerPatternTag}~(\cref{sec:appendix-tag-and-probe-cuts} - \cref{enum:appendix-tag-and-probe-cuts-TrackerPatternTag})
\end{itemize}

\subsection{Tag cuts used for the \enquote{Energy $\leftrightarrow$ rigidity matching} cut}
\label{sec:appendix-tag-and-probe-samples-energy-rigidity-matching}

The tag sample for the \textbf{\enquote{Energy $\leftrightarrow$ rigidity matching}} cut (\cref{sec:analysis-data-selection-electron-positron-identification-cuts} - \cref{enum:electron-positron-identification-cut-energy-rigidity-matching}) is prepared using following cuts:

\begin{itemize}
  \item\textbf{All detector quality cuts}~(\cref{sec:analysis-data-selection-detector-quality-cuts})
  \item\textbf{All preselection cuts}~(\cref{sec:analysis-data-selection-preselection-cuts})
  \item\textbf{All selection cuts}~(\cref{sec:analysis-data-selection-selection-cuts})
  \item\textbf{NegativeRigidityTag}~(\cref{sec:appendix-tag-and-probe-cuts} - \cref{enum:appendix-tag-and-probe-cuts-NegativeRigidityTag})
  \item\textbf{EcalElectronTag}~(\cref{sec:appendix-tag-and-probe-cuts} - \cref{enum:appendix-tag-and-probe-cuts-EcalElectronTag})
  \item\textbf{TrdElectronTag}~(\cref{sec:appendix-tag-and-probe-cuts} - \cref{enum:appendix-tag-and-probe-cuts-TrdElectronTag})
\end{itemize}

\subsection{Tag cuts used for the \enquote{Tracker $\leftrightarrow$ ECAL matching in X-projection} cut}
\label{sec:appendix-tag-and-probe-samples-tracker-track-ecal-cog-delta-x}

The tag sample for the \textbf{\enquote{Tracker $\leftrightarrow$ ECAL matching in X-projection}} cut (\cref{sec:analysis-data-selection-electron-positron-identification-cuts} - \cref{enum:electron-positron-identification-cut-tracker-track-ecal-cog-delta-x}) is prepared using following cuts:

\begin{itemize}
  \item\textbf{All detector quality cuts}~(\cref{sec:analysis-data-selection-detector-quality-cuts})
  \item\textbf{All preselection cuts}~(\cref{sec:analysis-data-selection-preselection-cuts})
  \item\textbf{All selection cuts}~(\cref{sec:analysis-data-selection-selection-cuts})
  \item\textbf{NegativeRigidityTag}~(\cref{sec:appendix-tag-and-probe-cuts} - \cref{enum:appendix-tag-and-probe-cuts-NegativeRigidityTag})
  \item\textbf{EcalElectronTag}~(\cref{sec:appendix-tag-and-probe-cuts} - \cref{enum:appendix-tag-and-probe-cuts-EcalElectronTag})
  \item\textbf{TrdElectronTag}~(\cref{sec:appendix-tag-and-probe-cuts} - \cref{enum:appendix-tag-and-probe-cuts-TrdElectronTag})
  \item\textbf{ElectronUpperEnergyOverRigidityTag}~(\cref{sec:appendix-tag-and-probe-cuts} - \cref{enum:appendix-tag-and-probe-cuts-ElectronUpperEnergyOverRigidityTag})
\end{itemize}

\section{Iterative Bayesian unfolding method}
\label{sec:appendix-unfolding-iterative-bayesian-method}

Mathematically the unfolding process can be written as

\begin{equation}
  \label{eq:unfolding-general}
  \mathbf{\hat{n}} = U(\mathbf{n}; \boldsymbol{\theta}),
\end{equation}

where $\mathbf{n}$ denotes the observed distribution in data, $\boldsymbol{\theta}$ the unfolding parameters and $\mathbf{\hat{n}}$ the unfolded, true distribution.
The goal of the unfolding procedure is to find a suitable function $U$ and its parameters. Different unfolding methods represent different choices of the $U$ function.
Most methods utilize a matrix to represent $U$ and construct $U$ from the migration matrix $\mathbf{M}$. The unfolding parameters $\boldsymbol{\theta}$ have to be estimated (e.g. regularization parameter,
number of iterations, etc.).

In this work, after comparing different unfolding methods, the \textbf{\enquote{iterative Bayesian unfolding}}~\cite{DAgostini1995} proposed by D’Agostini - implemented in RooUnfold~\cite{Adye2011} - was chosen, as it delivers stable results and a reasonable statistical uncertainty after unfolding.

The fundamental idea behind the iterative Bayesian unfolding method is based on the \textbf{Bayes theorem}~\cite{Bayes1763}, which represents the relation of conditional probabilities:

\begin{equation}
  \label{eq:bayes-theorem}
  P(A \given B) = \frac{P(B \given A) \cdot P(A)}{P(B)},
\end{equation}

where $P(A)$ and $P(B)$ denote the probabilities of observing event A and B independently of each other, $P(A \given B$) the likelihood of event A occurring given that B is true and $P(B \given A)$ the likelihood of event B occurring given that A is true.

D’Agostini starts introducing his method by defining the terms \enquote{causes} and \enquote{effect}: There are a $n_{C}$ causes $C_{i}$ which can produce \textit{one} effect $E$. If we know the initial probability of the causes $P(C_{i})$ and the conditional probability of the $i^{\text{th}}$ cause to produce the effect $P(E \given C_{i})$ the Bayes formula can be written as

\begin{equation}
  \label{eq:bayes-theorem-generalized-one-effect}
  \begin{aligned}
    P(C_{i} \given E) &= \frac{P(E \given C_{i}) \cdot P(C_{i})}{P(E)}
                      &= \frac{P(E \given C_{i}) \cdot P(C_{i})}{\sum^{n_{C}}_{k = 1} P(E \given C_{k}) \cdot P(C_{k})}.
  \end{aligned}
\end{equation}

$P(E)$ can be decomposed into conditional probabilities, as there are $n_{C}$ disjoint causes that can lead to the same effect $E$, according to the \enquote{Law of total probability}~\cite{Kokoska2000}.

If $n(E)$ events for effect $E$ are observed, the expected number of events assignable to each of the causes $C_{i}\ (i = 1,\ \ldots, n_{C})$ is

\begin{equation}
  \label{eq:bayes-expected-events-single-effect}
  \hat{n}(C_{i}) = P(C_{i} \given E) \cdot n(E).
\end{equation}

The outcome of a measurement may have multiple effects $E_{j}\ (j = 1,\ \ldots, n_{E})$ instead of a single effect $E$. For each of the effects the~\cref{eq:bayes-theorem-generalized-one-effect} holds. Therefore it can be generalized to $n_{E}$ effects:

\begin{equation}
  \label{eq:bayes-theorem-generalized-multiple-effects}
  \begin{aligned}
    P(C_{i} \given E_{j}) &= \frac{P(E_{j} \given C_{i}) \cdot P_{0}(C_{i})}{\sum^{n_{C}}_{k = 1} P(E_{j} \given C_{k}) \cdot P_{0}(C_{k})},
  \end{aligned}
\end{equation}

\noindent
where $P_{0}(C_{i})$ denote the \textit{initial} probabilities of the causes $C_{i}$.

After $N_{\text{obs}}$ observations a distribution of frequencies is obtained: $\mathbf{n}(E) = \{n(E_{1}),\ \ldots, n(E_{n_{E}})\}$. The expected
number of events to be assigned to each of the causes and only due to the observed events can be calculated, by applying \cref{eq:bayes-expected-events-single-effect}
to each of the effects:

\begin{equation}
  \label{eq:bayes-expected-events-multiple-effects}
  \left.\hat{n}(C_{i})\right\lvert_{\text{obs}} = \sum_{j = 1}^{n_{E}} P(C_{i} \given E_{j}) \cdot n(E_{j}).
\end{equation}

One needs to take into account that there is no need for each cause to produce at least one effect. The efficiency of detecting the cause $C_{i}$ in any of the possible effects is given by:

\begin{equation}
  \label{eq:bayes-epsilon}
  \epsilon_{i} = \sum_{j = 1}^{N_{E}} P(E_{j} \given C_{i}).
\end{equation}

Thus the best estimate of the true number of events is:

\begin{equation}
  \label{eq:bayes-expected-events-multiple-effects-corrected}
  \hat{n}(C_{i}) = \frac{1}{\epsilon_{i}} \cdot \sum_{j = 1}^{n_{E}} P(C_{i} \given E_{j}) \cdot n(E_{j}).
\end{equation}

Finally the true total number of events and the probabilities of the causes can be extracted:

\begin{equation}
  \label{eq:bayes-total-events-and-final-probability}
  \hat{N}_{\text{true}} = \sum_{i = 1}^{n_{C}} \hat{n}(C_{i}),\qquad\hat{P}(C_{i}) = \frac{\hat{n}(C_{i})}{\hat{N}_{\text{true}}}
\end{equation}

From \cref{eq:bayes-theorem-generalized-multiple-effects} it is evident, that the conditional probability $P(C_{i} \given E_{j})$ depends on the choice of the \textit{initial} probabilities of the causes $P_{0}(C_{i})$. Thus $\hat{n}(C_{i})$ and $\hat{P}(C_{i})$ also depend on the \textit{initial} probabilities.

If $n_{0}(C_{i}) = P_{0}(C_{i}) \cdot N_{\text{obs}}$ is not compatible with $\hat{n}(C_{i})$, the choice of the \textit{initial} probabilities $P_{0}(C_{i})$ was incorrect. This leads to an \textbf{iterative} procedure: $P_{0}(C_{i})$ needs to be refined, by replacing it with $\hat{P}(C_{i})$ from \cref{eq:bayes-total-events-and-final-probability} and the whole procedure needs to be repeated, until $\hat{n}(C_{i})$ agrees with $n_{0}(C_{i})$. If that is the case the true distribution $\hat{n}(C_{i}) $ was found and the unfolding process is finished.

\bigskip
Coming back to \cref{eq:unfolding-general}, the unfolding function $U$ - suitable of transforming the measured distribution into the true distribution - can be derived.
By reformulating \cref{eq:bayes-expected-events-multiple-effects-corrected} the unfolding function $U$ can be identified as matrix $\mathbf{U}$:

\begin{equation}
  \label{eq:unfolding-matrix-definition}
  \begin{aligned}
    \hat{n}(C_{i}) &= \frac{1}{\epsilon_{i}} \cdot \sum_{j = 1}^{n_{E}} P(C_{i} \given E_{j}) \cdot n(E_{j}) \\
      &\leftstackrel{\eqref{eq:bayes-epsilon}}{=} \sum_{j = 1}^{n_{E}} \frac{P(C_{i} \given E_{j})}{\sum_{k = 1}^{N_{E}} P(E_{k} \given C_{i})} \cdot n(E_{j}) \\
      &\leftstackrel{\eqref{eq:bayes-theorem-generalized-multiple-effects}}{=} \sum_{j = 1}^{n_{E}} \underbrace{\frac{P(E_{j} \given C_{i}) \cdot P_{0}(C_{i})}{\left(\sum_{k = 1}^{N_{E}} P(E_{k} \given C_{i})\right) \cdot \left(\sum^{n_{C}}_{k = 1} P(E_{j} \given C_{k}) \cdot P_{0}(C_{k})\right)}}_{\vcentcolon= \mathbf{U}_{ij}} \cdot n(E_{j}).
  \end{aligned}
\end{equation}

This leads to a simple formulation in matrix style:

\begin{equation*}
  \mathbf{\hat{n}} = \mathbf{U} \cdot \mathbf{n}.
\end{equation*}

There is only one unfolding parameter $\boldsymbol{\theta}$ for the iterative Bayesian unfolding: the number of iterations $N_{\text{iter}}$ needed until convergence is reached.
There is no a-priori information how many iterations are sufficient -- it depends on the problem. Thus the number of iterations needs to be varied and after each iteration the
results should be carefully checked. A high number of iterations might lead to an unfolded distribution that heavily fluctuates around the true distribution, which is not the case
for the unfolding of the event counts in this work.

\clearpage
\section{Time-dependent charge-confusion}
\label{sec:appendix-flux-time-dependent-charge-confusion}

\begin{figure}[H]
  \includegraphics[width=\linewidth]{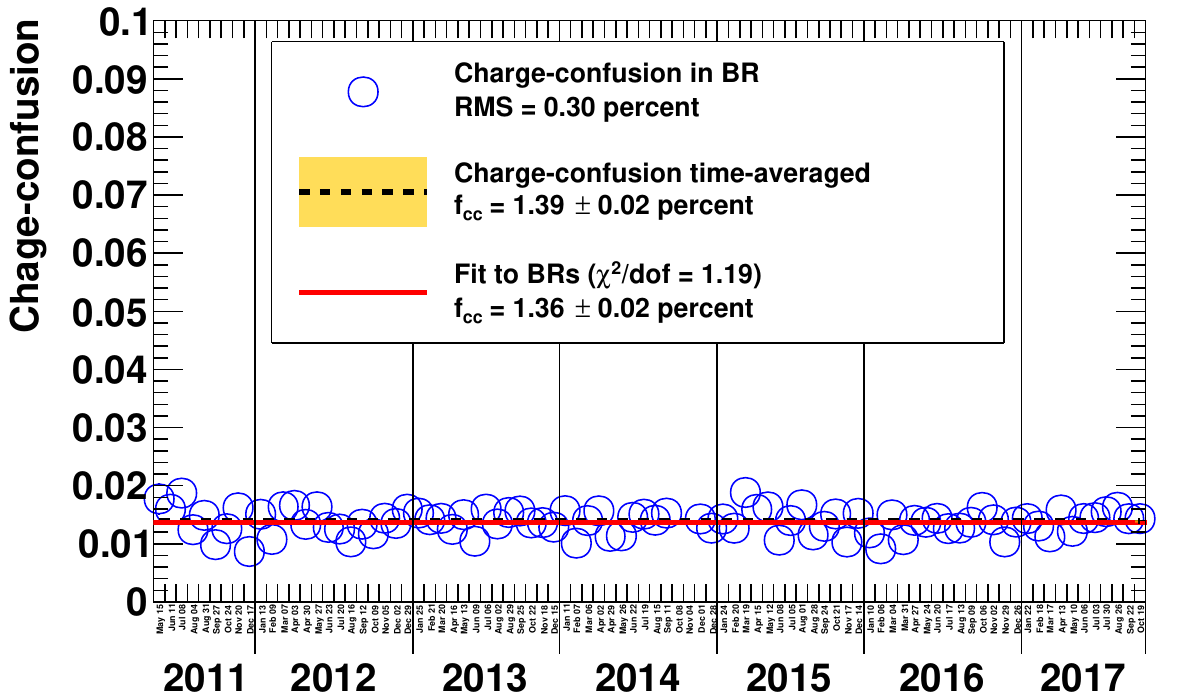}
  \caption{Time-dependent charge-confusion for the energy bin \SIrange{9.62}{10.32}{\GeV}. The x-axis spans 88 Bartels rotations and each bin label denotes the date when the Bartels rotation started (in UTC). The blue symbols show the charge-confusion for the all-tracks sample in the specific energy bin as function of time. The black dashed line shows the time-averaged charge-confusion for comparison. The red line shows a fit of a constant to the red symbols, yielding an acceptable $\chi^2/\text{dof} = 1.19$ value.}
  \label{fig:time-dependent-charge-confusion}
\end{figure}

\Cref{fig:time-dependent-charge-confusion} shows the charge-confusion as function of time for an example energy bin of the time-dependent analysis
for the all-tracks sample. For all energy bins the average of the charge-confusion value is compatible with the time-averaged
charge-confusion and thus $f_{\text{cc},\,i}(E) = f_{\text{cc}}(E)$.

\section{Time-dependent TRD templates (multi-tracks sample)}
\label{sec:appendix-flux-time-dependent-trd-templates}

\begin{figure}[H]
  \centering
  \includegraphics[width=0.95\linewidth]{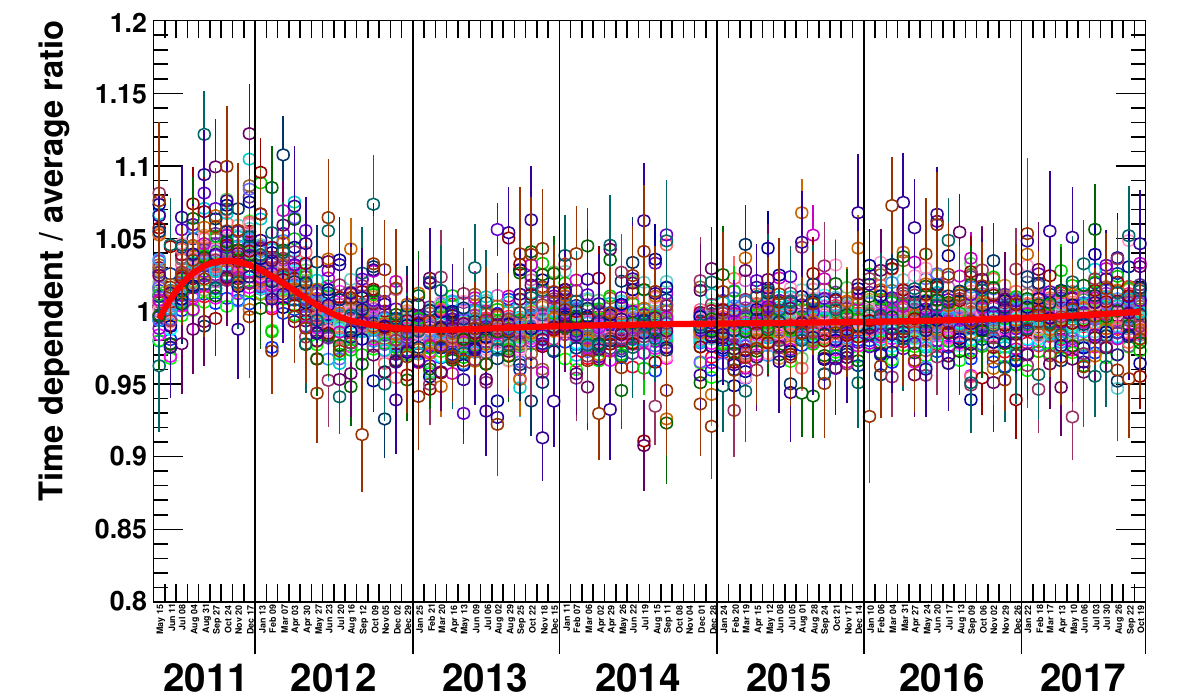}
  \caption{Time-dependence of the~\elecPeakNovo~parameter for all energy bins - \SIrange{0.65}{52.33}{\GeV} - in the time-dependent analysis for the multi-tracks sample. Each energy bin is represented by a different color. The filled red line denote the result of a spline fit, simultaneously to all data points, describing the average time-dependence of the~\elecPeakNovo~parameter.}
  \label{fig:time-dependent-trd-templates-elec-peak-novo-multi-tracks}
\end{figure}

\begin{figure}[H]
  \centering
  \includegraphics[width=0.95\linewidth]{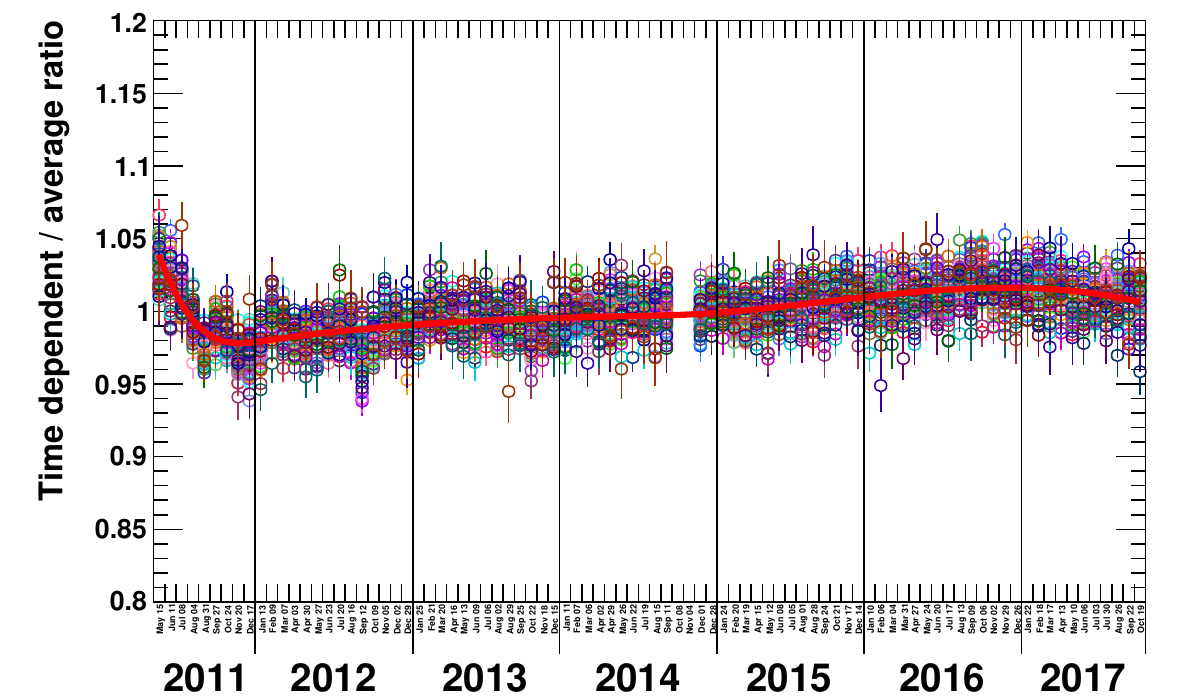}
  \caption{Time-dependence of the~\protPeakNovo~parameter for all energy bins - \SIrange{0.65}{52.33}{\GeV} - in the time-dependent analysis for the multi-tracks sample.}
  \label{fig:time-dependent-trd-templates-prot-peak-novo-multi-tracks}
\end{figure}

In \cref{sec:analysis-flux-time-dependent-trd-templates} the time dependence of the TRD templates was shown for the single-track sample.
The results for the multi-tracks samples is shown in the following: \Cref{fig:time-dependent-trd-templates-elec-peak-novo-multi-tracks} shows
the~\elecPeakNovo~parameter, \cref{fig:time-dependent-trd-templates-prot-peak-novo-multi-tracks} for the~\protPeakNovo~parameter.

\begin{figure}[H]
  \centering
  \includegraphics[width=0.90\linewidth]{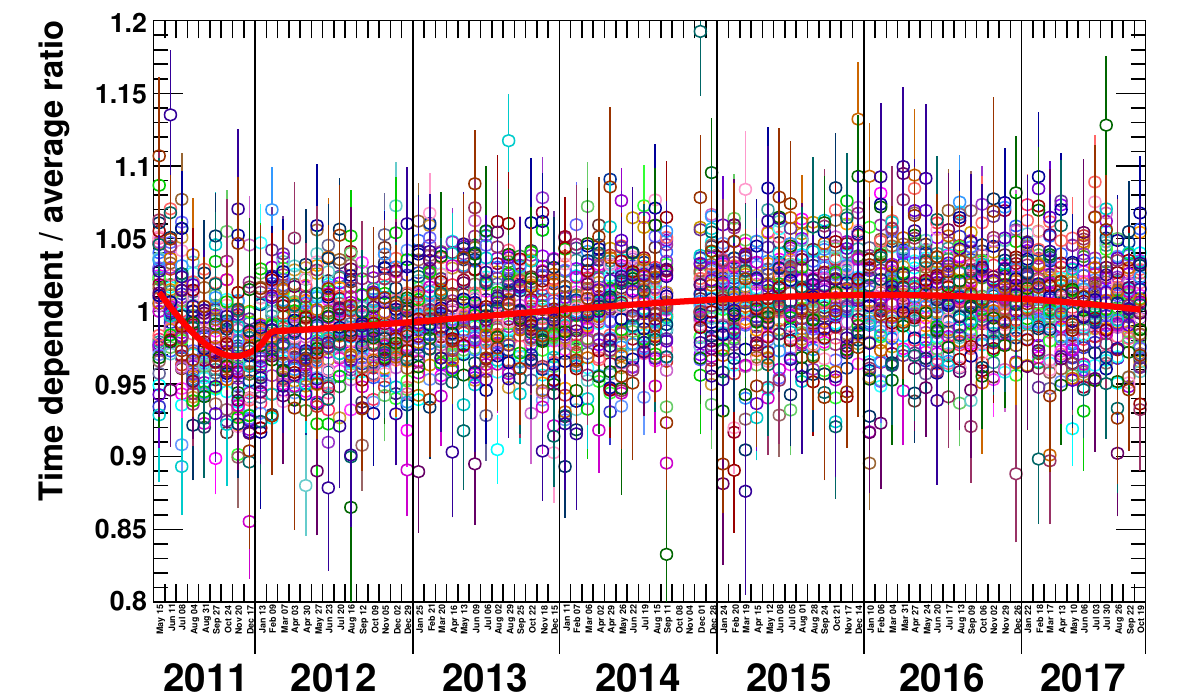}
  \caption{Time-dependence of the~\ccProtPeakNovo~parameter for selected energy bins (\SIrange{1.22}{52.33}{\GeV}), as example, in the time-dependent analysis for the multi-tracks sample.}
  \label{fig:time-dependent-trd-templates-ccprot-peak-novo-multi-tracks}
\end{figure}

\begin{figure}[H]
  \centering
  \includegraphics[width=0.90\linewidth]{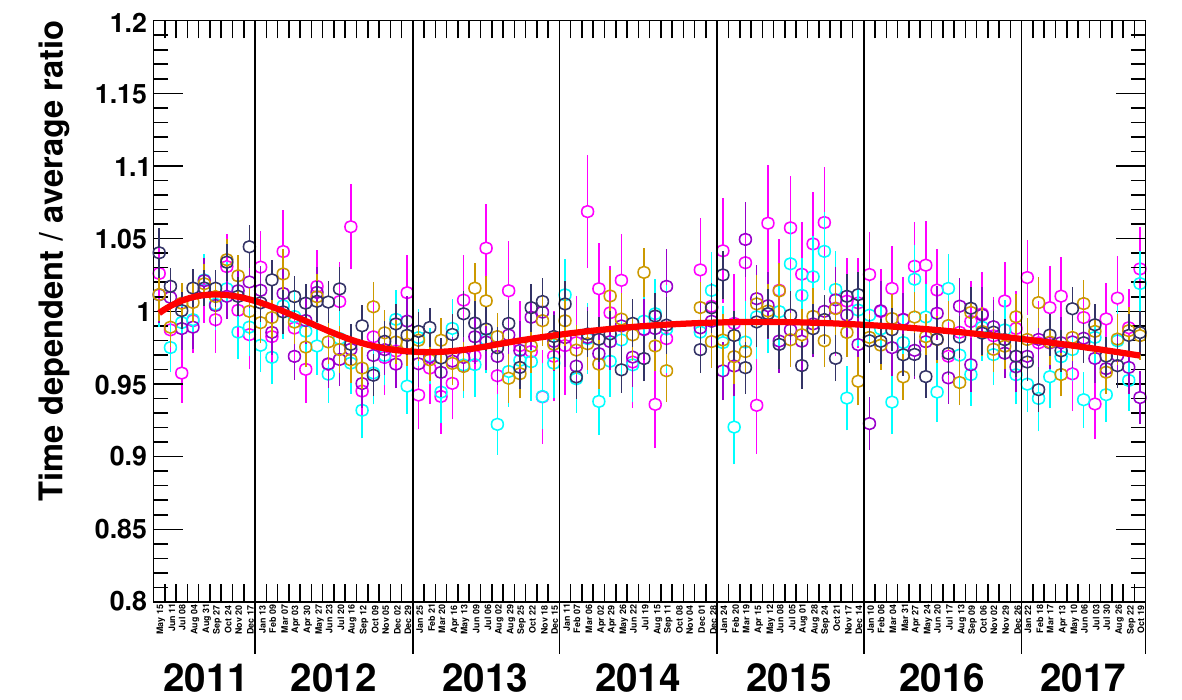}
  \caption{Time-dependence of the~\elecWidthNovo~parameter for selected energy bins (\SIrange{1.46}{3.00}{\GeV}), as example, in the time-dependent analysis for the multi-tracks sample.}
  \label{fig:time-dependent-trd-templates-elec-width-novo-multi-tracks}
\end{figure}

\begin{figure}[H]
  \centering
  \includegraphics[width=0.95\linewidth]{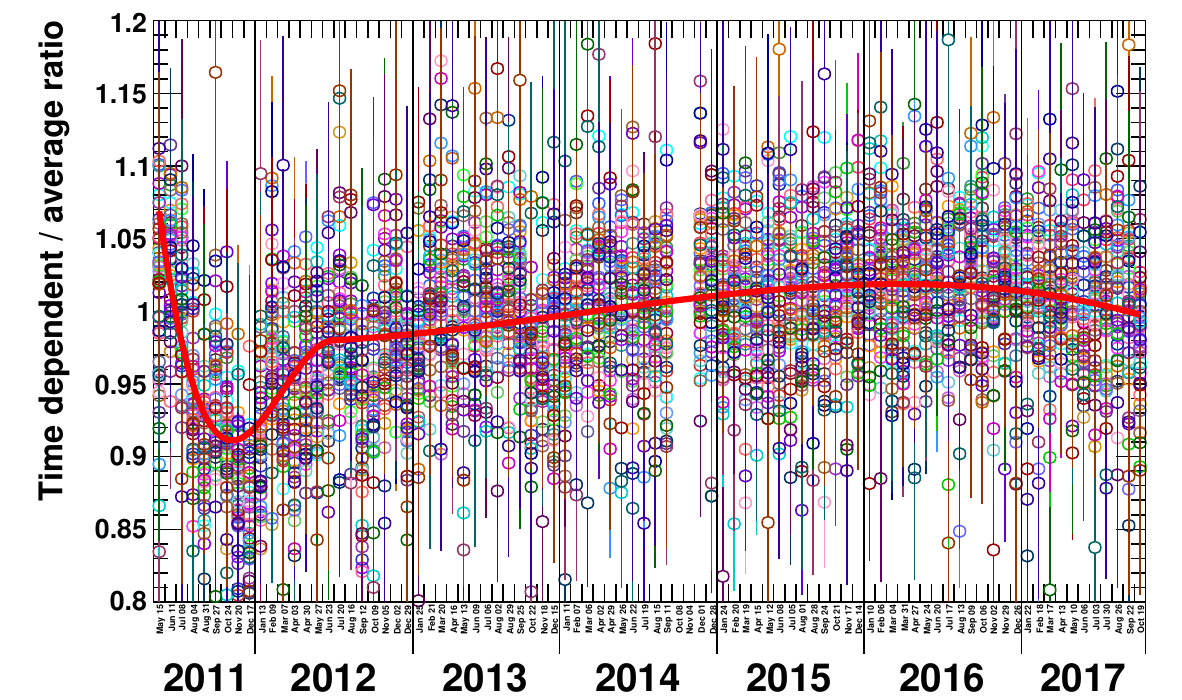}
  \caption{Time-dependence of the~\protWidthNovo~parameter for selected energy bins (\SIrange{1.72}{52.33}{\GeV}), as example, in the time-dependent analysis for the multi-tracks sample.}
  \label{fig:time-dependent-trd-templates-prot-width-novo-multi-tracks}
\end{figure}

The time-dependence of the three remaining parameters - \ccProtPeakNovo, \elecWidthNovo~and~\protWidthNovo, is shown in
\cref{fig:time-dependent-trd-templates-ccprot-peak-novo-multi-tracks,fig:time-dependent-trd-templates-elec-width-novo-multi-tracks,fig:time-dependent-trd-templates-prot-width-novo-multi-tracks}
in selected energy bins, as example, for the multi-tracks sample.

\chapter{Appendix - Results}
\label{sec:appendix-results}

\section{Energy scale}
\label{sec:appendix-results-energy-scale}

It is important to note that the ECAL shower reconstruction used in this work is different compared
to the latest AMS-02 publications covering the high-energy electrons~\cite{Aguilar2019a} and positrons~\cite{Aguilar2019b}.
In this work the conventional ECAL shower reconstruction was used, as for the first electron and positron
flux publications~\cite{Aguilar2014a}, but with an improved Top-Of-Instrument correction derived in a dedicated study.
This energy scale will be referred to as \enquote{Aachen energy scale} in the following.

The improved ECAL shower reconstruction is described in detail in~\cite{Kounine2017a} and allows to exploit
even higher energies, due to a better positron/proton separation and an improved energy resolution. In the following
it will be referred to as \enquote{MIT energy scale}.

Both ECAL shower reconstruction methods yield slightly different reconstructed energies for the same events,
compatible with the quoted absolute energy scale uncertainty (see~Ref.~\cite{Aguilar2014a}). For a fair comparison
of this work with the recent published results, the reconstructed energy scale in this work has to match - on average -
the results of improved ECAL shower reconstruction. To achieve this goal the MIT and the Aachen energy scale were
compared on the electron Monte-Carlo simulation and migration matrices were obtained using the MIT energy scale and
the Aachen energy scale sharing the same true energy axis. By studying the differences between both migration matrices,
an effective migration matrix can be deduced, assuring that the unfolded flux is identical when using either the Aachen
energy scale and the effective migration matrix or the MIT energy scale and the unmodified migration matrix.

For technical reasons the MIT energy scale and the Aachen energy scale could only be compared on the electron Monte-Carlo
simulation and not on the whole time period for the ISS data. As the MIT energy scale is unavailable for the whole ISS time
period in this work, it could not be used for the analysis. However it will be shown that the fluxes derived with the
Aachen energy scale, unfolded with the effective migration matrix, are fully compatible with the published results, that
exclusively use the MIT energy scale. Thus using the Aachen energy scale is acceptable, with the
only drawback that energies higher than \SI{1}{\TeV} cannot be exploited.

\begin{figure}[H]
  \centering
  \includegraphics[width=0.7\linewidth]{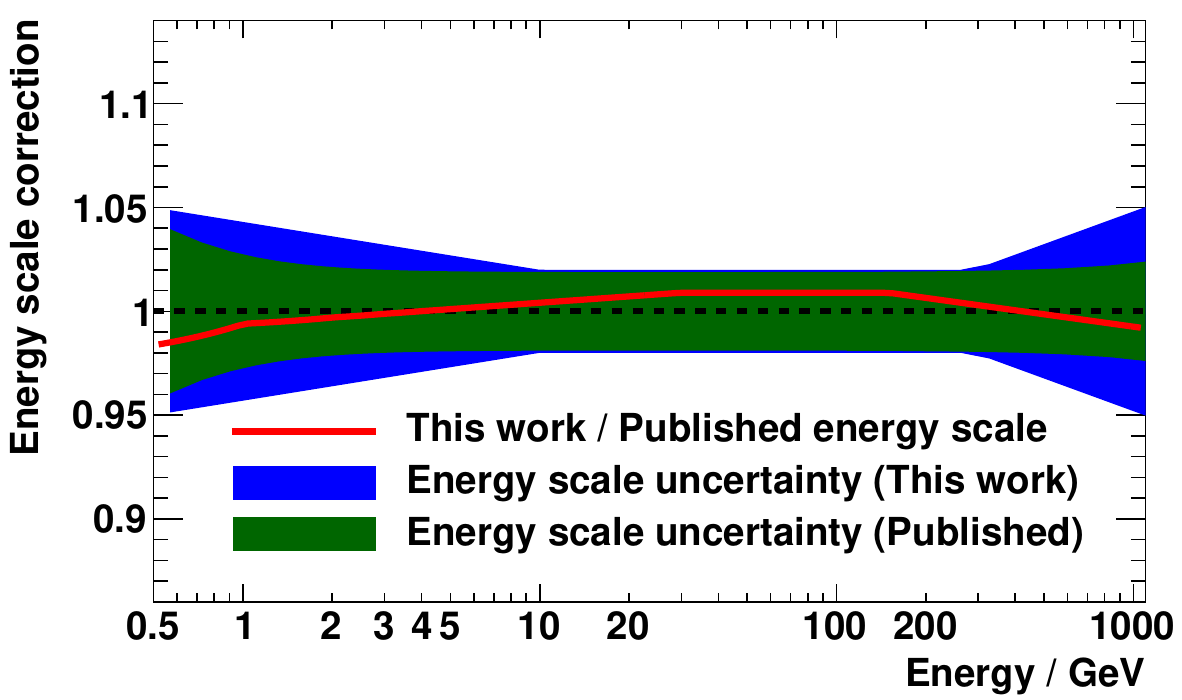}
  \caption{Correction applied to the Aachen energy scale used in this work to match the MIT energy scale used in recent AMS-02 electron and positron flux publications~\cite{Aguilar2019a,Aguilar2019b}. The red line shows the necessary energy scale correction and the colored bands the energy scale uncertainties used for this work (blue band) and for recent publications (green band).}
  \label{fig:results-energy-scale-correction}
\end{figure}

\Cref{fig:results-energy-scale-correction} shows the average difference of both reconstructed energy scales as
function of the true energy. The obtained average correction (red line) is fully compatible with the quoted
absolute energy scale uncertainties, limited by the knowledge of AMS-02 beam test.

\section{Time-averaged results}
\label{sec:appendix-results-time-averaged}

\subsection{Comparison with published electron and positron fluxes}
\label{sec:appendix-results-comparison-published-fluxes}

\begin{figure}[H]
  \begin{subfigure}{0.50\linewidth}
    \includegraphics[width=\linewidth]{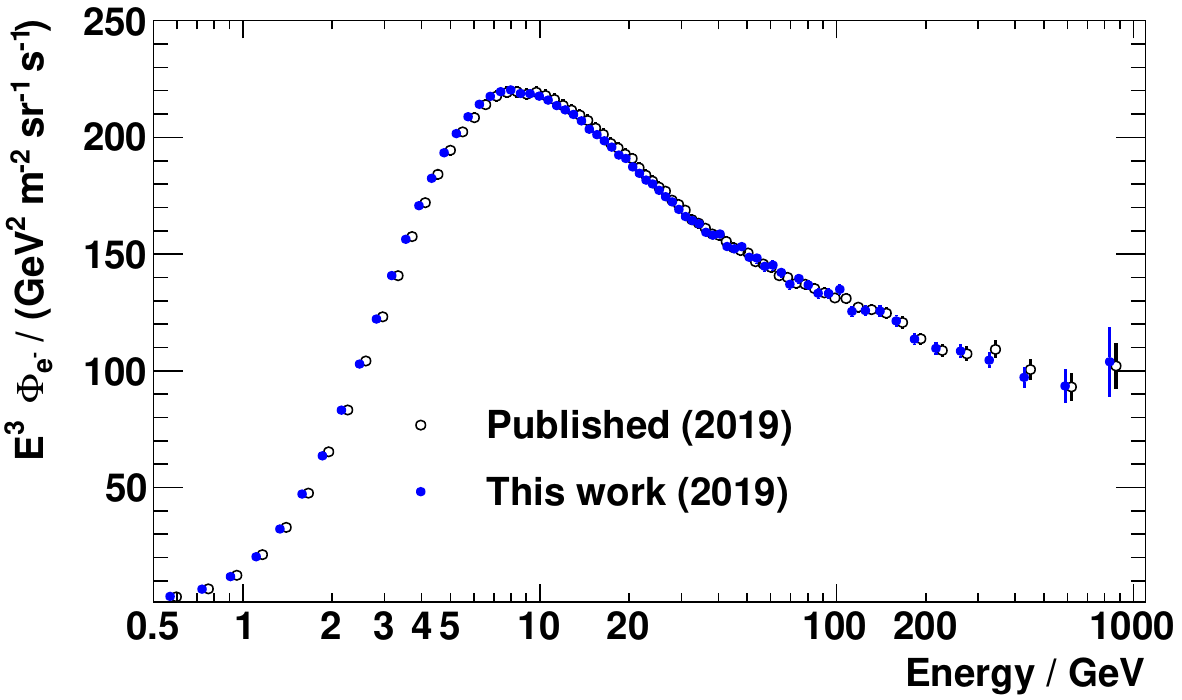}
  \end{subfigure}
  \begin{subfigure}{0.50\linewidth}
    \includegraphics[width=\linewidth]{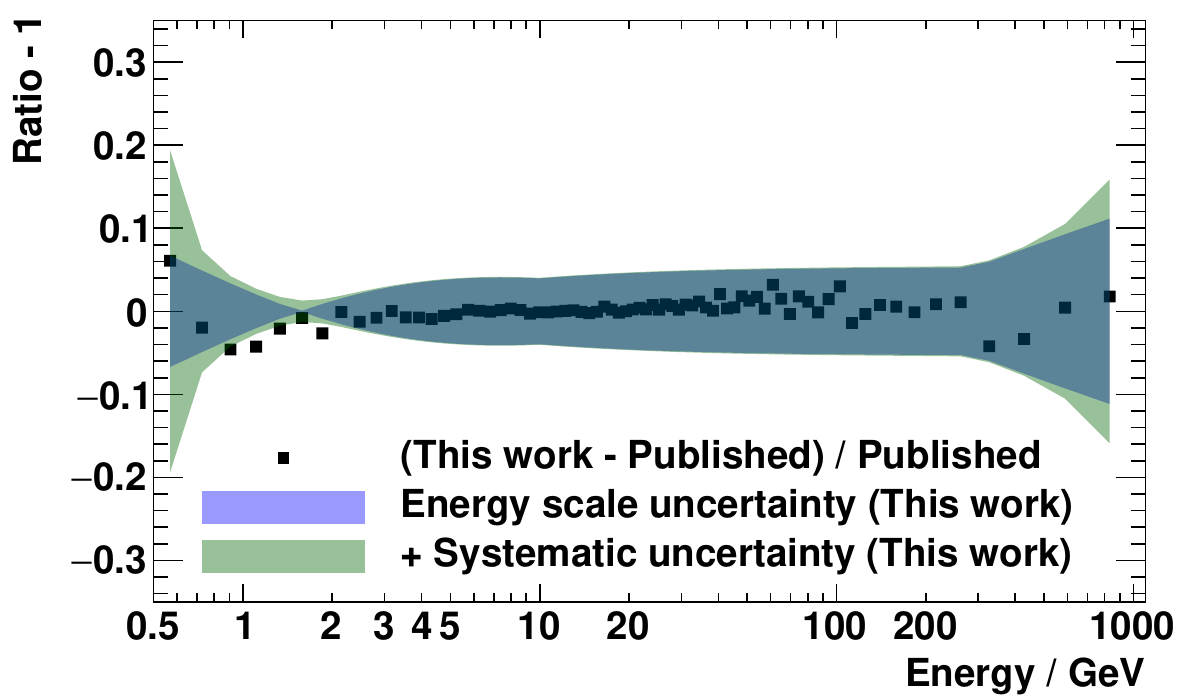}
  \end{subfigure}
  \caption{Comparison of the electron flux derived in this work with the recent AMS-02 electron flux publication~\cite{Aguilar2019a}. To ease the comparison the published data points are shifted horizontally by \SI{5}{\percent}. Within the quoted systematic uncertainties the results are compatible.}
  \label{fig:results-time-averaged-electron-flux-aachen-mit}
\end{figure}

The electron flux derived in this work is compatible with the recently published AMS-02 electron flux~\cite{Aguilar2019a}, as shown in \cref{fig:results-time-averaged-electron-flux-aachen-mit}.

\begin{figure}[H]
  \centering
  \includegraphics[width=0.75\linewidth]{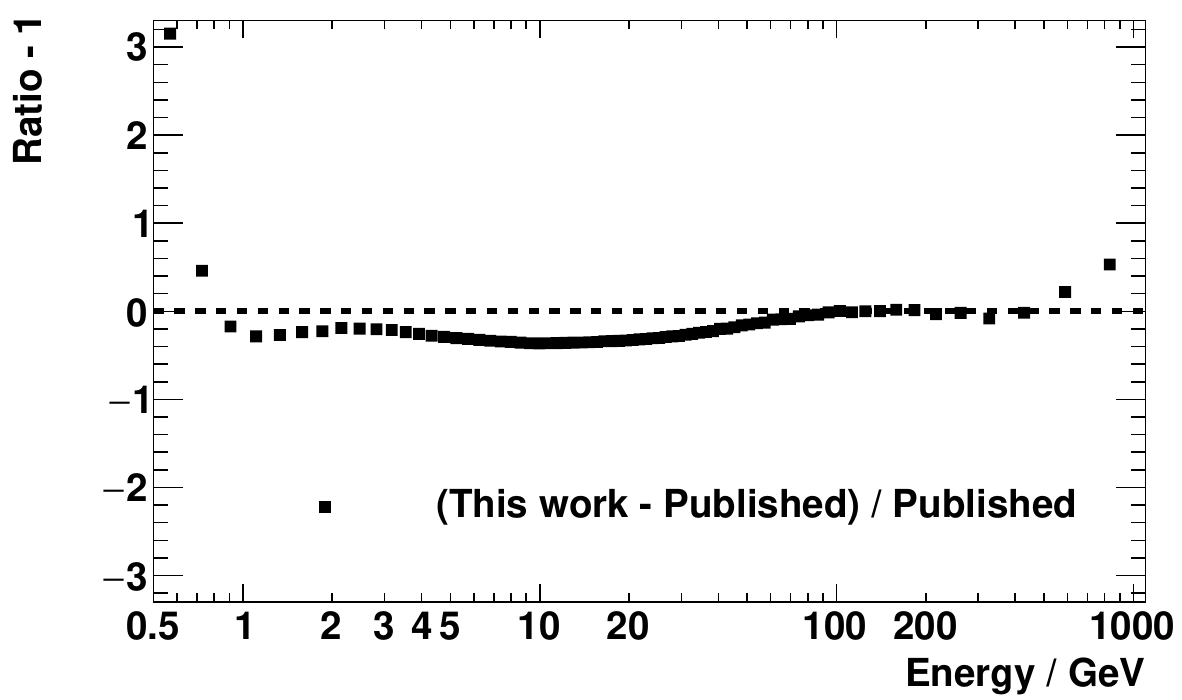}
  \caption{Ratio of the total uncertainty of the electron flux derived in this work over the uncertainty of the recently published AMS-02 electron flux~\cite{Aguilar2019a}.}
  \label{fig:results-time-averaged-electron-flux-aachen-mit-uncertainty}
\end{figure}

\Cref{fig:results-time-averaged-electron-flux-aachen-mit-uncertainty} shows a comparison of the total uncertainty of the electron flux derived in this
work and the recently published AMS-02 electron flux~\cite{Aguilar2019a}.

From \SIrange{0.7}{500}{\GeV} the uncertainty of the electron flux derived in this work is smaller than the published result. The recent publication was
optimized for the high-energy part and offers a smaller uncertainty in the high-energy regime than the flux derived in this work.

\begin{figure}[H]
  \begin{subfigure}{0.50\linewidth}
    \includegraphics[width=\linewidth]{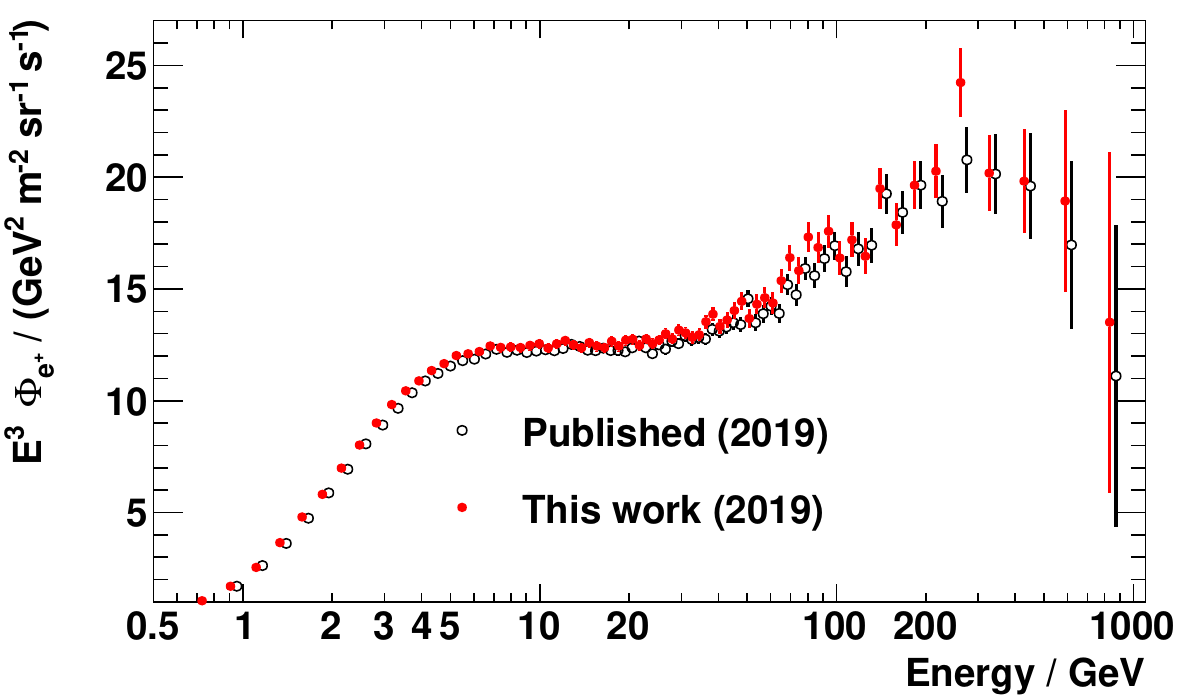}
  \end{subfigure}
  \begin{subfigure}{0.50\linewidth}
    \includegraphics[width=\linewidth]{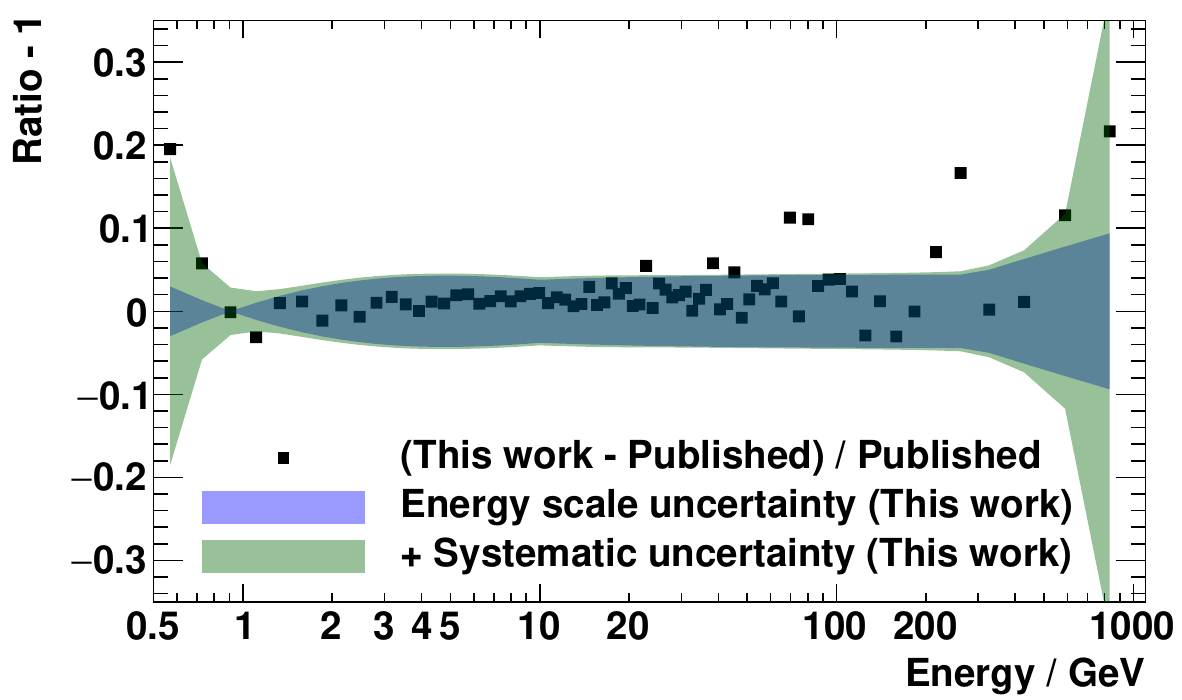}
  \end{subfigure}
  \caption{Comparison of the positron flux derived in this work with the recent AMS-02 positron flux publication~\cite{Aguilar2019b}. To ease the comparison the published data points are shifted horizontally by \SI{5}{\percent}. Within the quoted systematic uncertainties the results are compatible.}
  \label{fig:results-time-averaged-positron-flux-aachen-mit}
\end{figure}

The positron flux derived in this work is compatible with the recently published AMS-02 positron flux~\cite{Aguilar2019b}, as shown in \cref{fig:results-time-averaged-positron-flux-aachen-mit}.

\begin{figure}[H]
  \centering
  \includegraphics[width=0.75\linewidth]{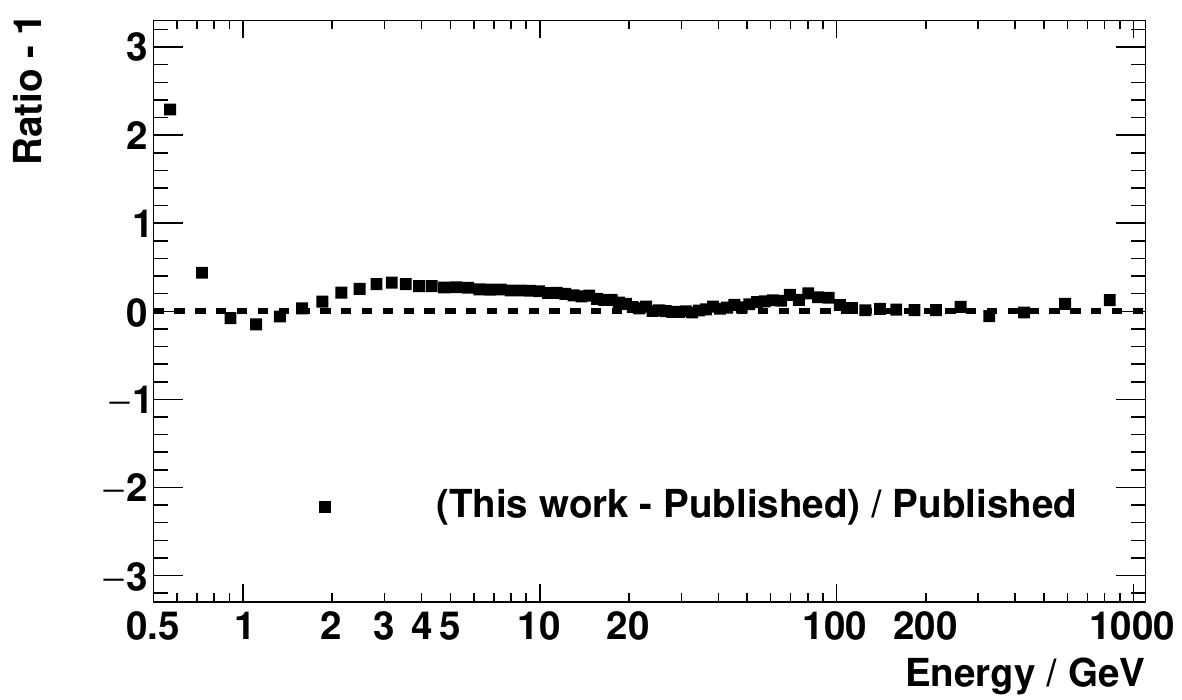}
  \caption{Ratio of the total uncertainty of the positron flux derived in this work over the uncertainty of the recently published AMS-02 positron flux~\cite{Aguilar2019b}.}
  \label{fig:results-time-averaged-positron-flux-aachen-mit-uncertainty}
\end{figure}

\Cref{fig:results-time-averaged-positron-flux-aachen-mit-uncertainty} shows a comparison of the total uncertainty of the positron flux derived in this
work and the recently published AMS-02 positron flux~\cite{Aguilar2019b}.

The uncertainty of the positron flux derived in this work is comparable to the published result above \SI{20}{\GeV}. Below that energy, the uncertainty on the published
result is slightly smaller.

\subsection{Positron fraction}
\label{sec:appendix-results-time-averaged-positron-fraction}

\begin{figure}[H]
  \includegraphics[width=\linewidth]{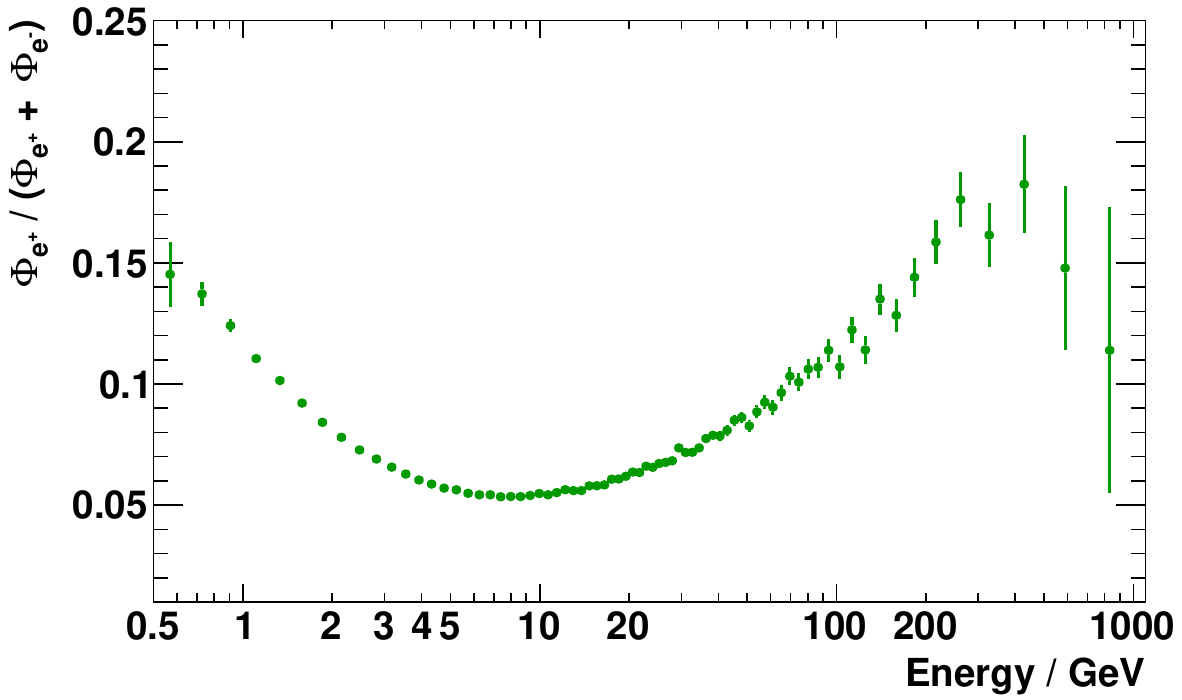}
  \caption{The positron fraction measured by AMS-02.}
  \label{fig:results-time-averaged-positron-fraction}
\end{figure}

\Cref{fig:results-time-averaged-positron-fraction} shows the time-averaged positron fraction, determined by
a dedicated analysis, using the single-track data sample, not computed from the fluxes themselves, which were derived using the all-tracks sample,
as explained in \cref{sec:analysis-lepton-counts-2d-fit}.

The positron fraction derived in this work is compatible with the published positron fraction~\cite{Aguilar2019b} by the AMS-02 collaboration,
as shown in \cref{sec:appendix-results-comparison-published-positron-fraction}. Furthermore the positron fraction is consistent between the dedicated
single-track analysis and the all-tracks flux analysis, as shown in \cref{sec:appendix-results-comparison-positron-fraction-from-fluxes}.

\begin{figure}[H]
  \centering
  \includegraphics[width=0.8\linewidth]{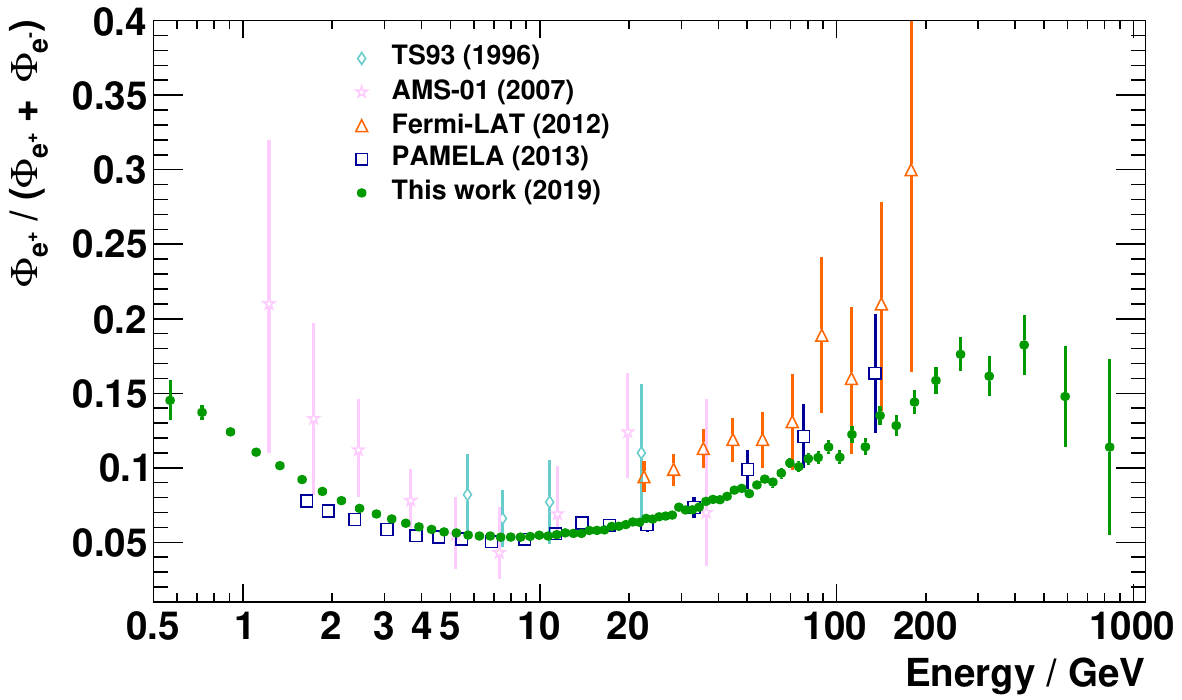}
  \caption{Comparison of the positron fraction derived in this work with other experiments: TS-93~\cite{Golden1996}, AMS-01~\cite{Alcaraz2000}, Fermi-LAT~\cite{Ackermann2012} and PAMELA~\cite{Adriani2013}.}
  \label{fig:results-time-averaged-positron-fraction-aachen-others}
\end{figure}

\Cref{fig:results-time-averaged-positron-fraction-aachen-others} shows a comparison of the positron fraction derived in this work with previous experiments.
The positron fraction was measured by AMS-02 with unprecedented accuracy, up to the \SI{}{\TeV} regime.

\begin{figure}[H]
  \centering
  \includegraphics[width=0.75\linewidth]{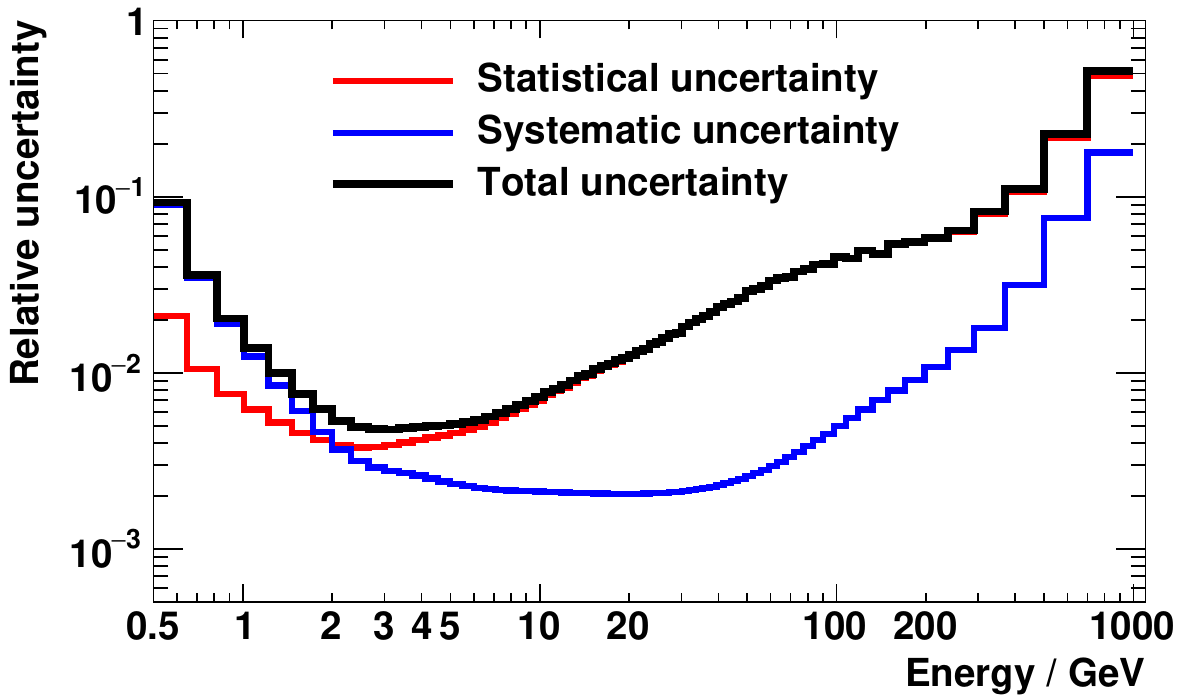}
  \caption{Breakdown of the uncertainty of the positron fraction into a statistical and systematical part.}
  \label{fig:results-time-averaged-positron-fraction-error-breakdown}
\end{figure}

\Cref{fig:results-time-averaged-positron-fraction-error-breakdown} shows a breakdown of the total uncertainty into the statistical and systematic part.

Above \SIapprox{2}{\GeV} the statistical uncertainty dominates the uncertainty of the positron fraction measurement. Below this energy, the systematic uncertainty always
exceeds the statistical uncertainty. See \cref{sec:analysis-ratios-time-averaged-sysunc-summary} for a decomposition of the systematic uncertainties.

\subsection{Comparison with published positron fraction}
\label{sec:appendix-results-comparison-published-positron-fraction}

\begin{figure}[H]
  \begin{subfigure}{0.50\linewidth}
    \includegraphics[width=\linewidth]{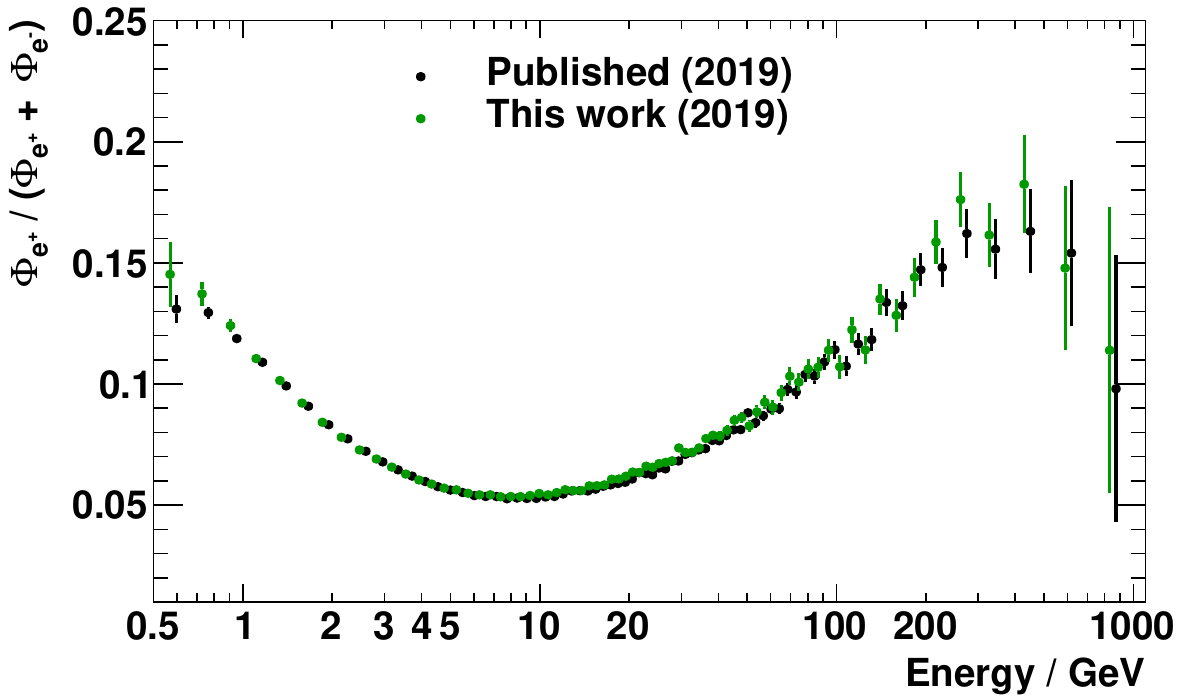}
  \end{subfigure}
  \begin{subfigure}{0.50\linewidth}
    \includegraphics[width=\linewidth]{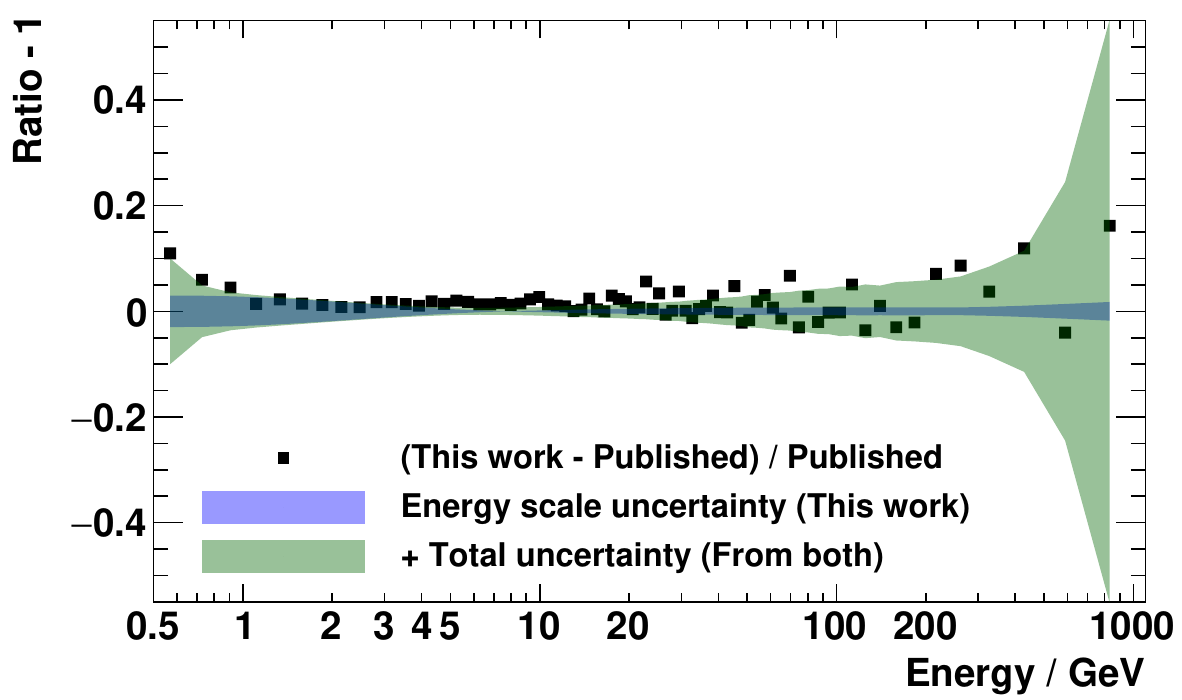}
  \end{subfigure}
  \caption{Comparison of the positron fraction derived in this work with the recent AMS-02 positron fraction publication~\cite{Aguilar2019b}. To ease the comparison the published data points are shifted horizontally by \SI{5}{\percent}. Within the quoted systematic uncertainties the results are compatible.}
  \label{fig:results-time-averaged-positron-fraction-aachen-mit}
\end{figure}

The positron fraction derived in this work is compatible with the recently published AMS-02 positron fraction~\cite{Aguilar2019b}, as shown in \cref{fig:results-time-averaged-positron-fraction-aachen-mit}.

Over the whole energy range the uncertainty of the positron fraction derived in this work is larger than the published result\footnote{The analysis in this work was not tuned specifically for the positron fraction analysis.}.

\begin{figure}[H]
  \centering
  \includegraphics[width=0.8\linewidth]{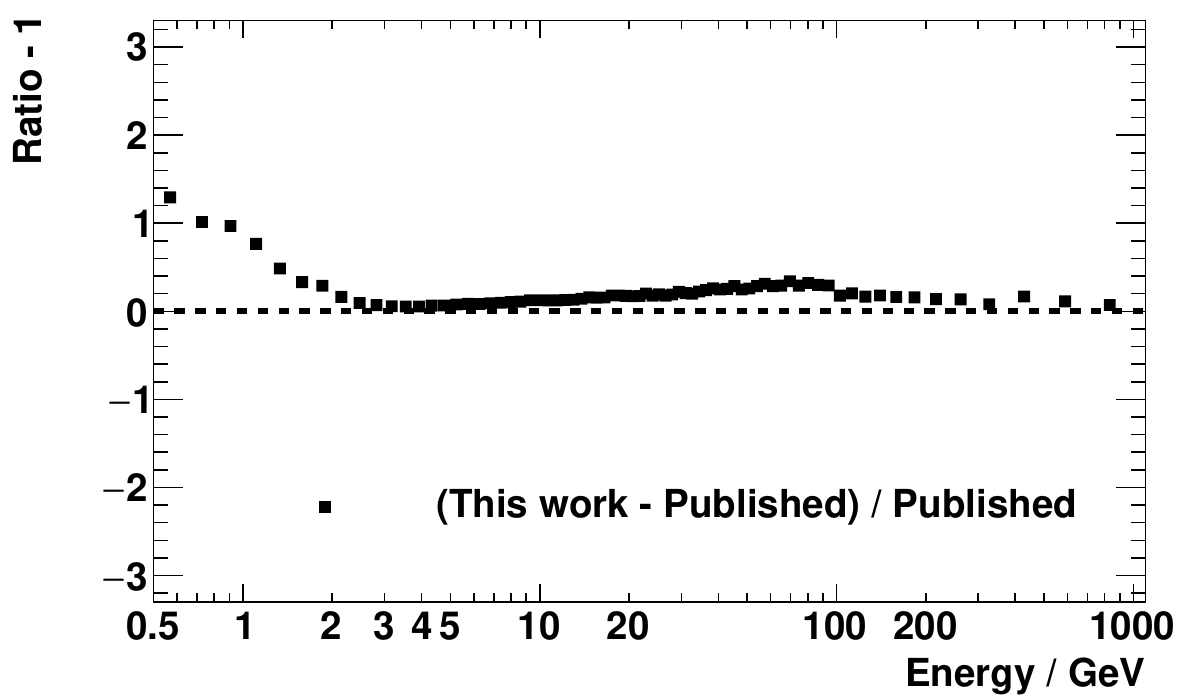}
  \caption{Ratio of the total uncertainty of the positron fraction derived in this work over the uncertainty of the recently published AMS-02 positron fraction~\cite{Aguilar2019b}.}
  \label{fig:results-time-averaged-positron-fraction-aachen-mit-uncertainty}
\end{figure}

\subsection{Cross-check of the positron/electron ratio between flux / dedicated analysis}
\label{sec:appendix-results-comparison-positron-electron-ratio-from-fluxes}

\begin{figure}[H]
  \begin{subfigure}{0.50\linewidth}
    \includegraphics[width=\linewidth]{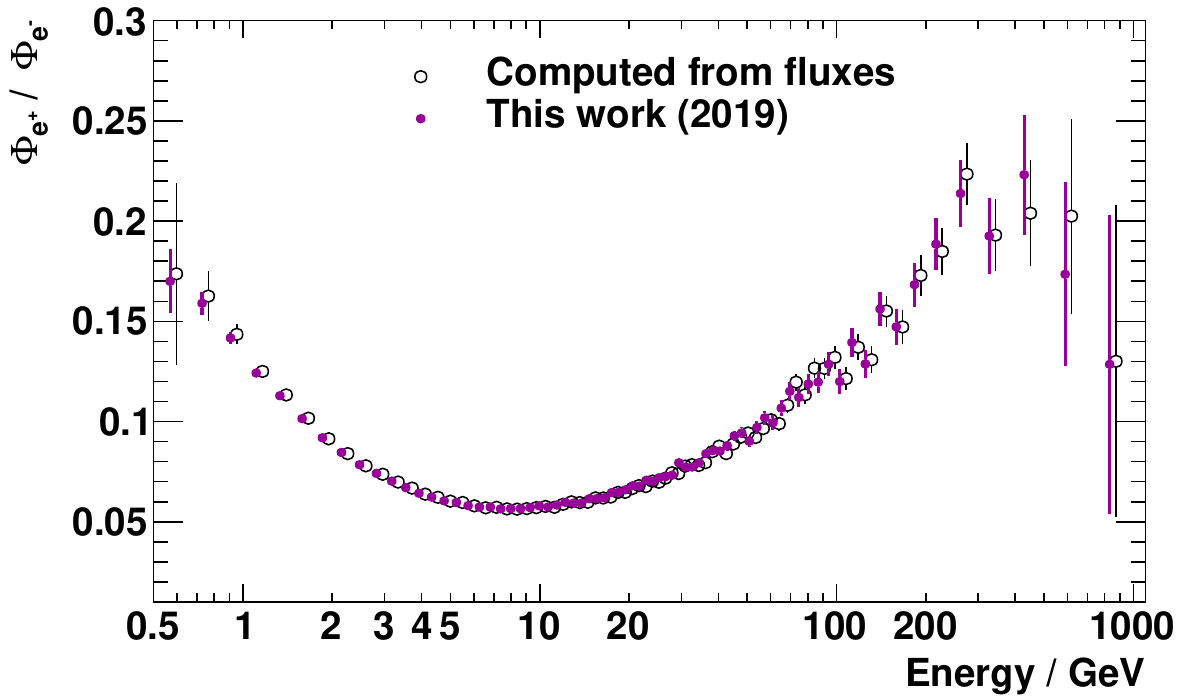}
  \end{subfigure}
  \begin{subfigure}{0.50\linewidth}
    \includegraphics[width=\linewidth]{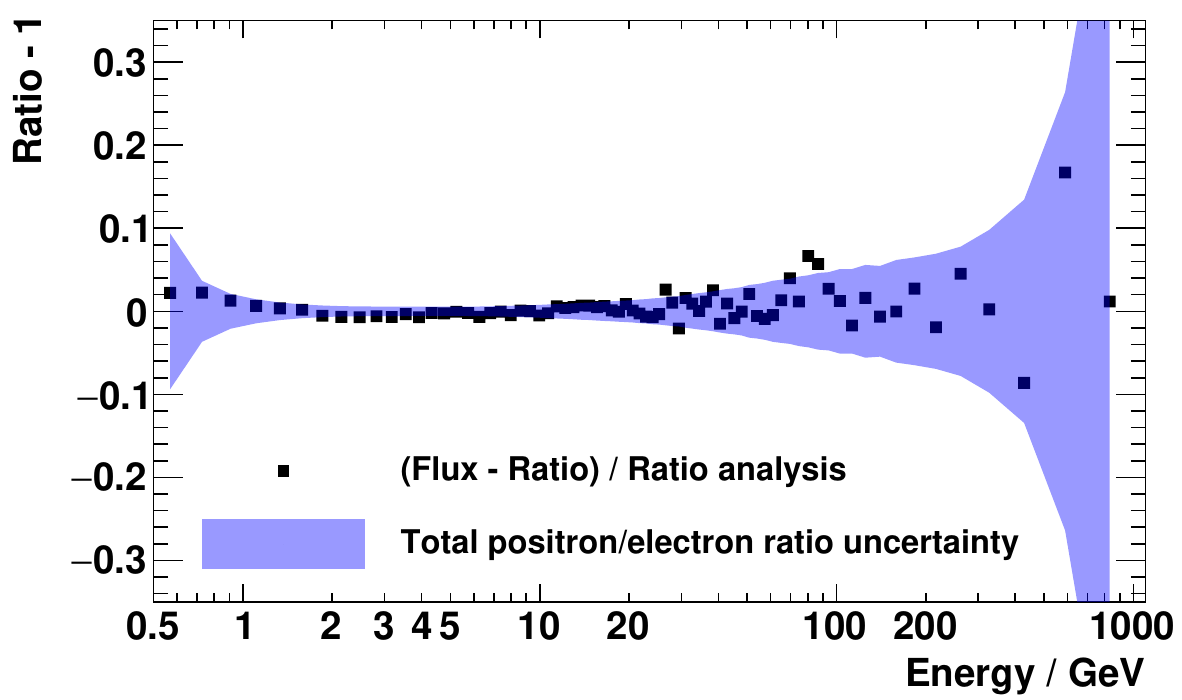}
  \end{subfigure}
  \caption{Comparison of the positron/electron ratio derived using the single-track analysis with the ratio computed from the fluxes, that were derived from the all-track analysis. To ease the comparison the all-track analysis data points are shifted horizontally by \SI{5}{\percent}. Within the quoted systematic uncertainties the results are compatible.}
  \label{fig:results-time-averaged-positron-electron-ratio-aachen-from-fluxes-aachen}
\end{figure}

As cross-check the positron/electron ratio is also computed from the fluxes, to verify the consistency between the single-track
and the all-tracks analyses. As shown in \cref{fig:results-time-averaged-positron-electron-ratio-aachen-from-fluxes-aachen} the agreement is good.

\begin{figure}[H]
  \centering
  \includegraphics[width=0.75\linewidth]{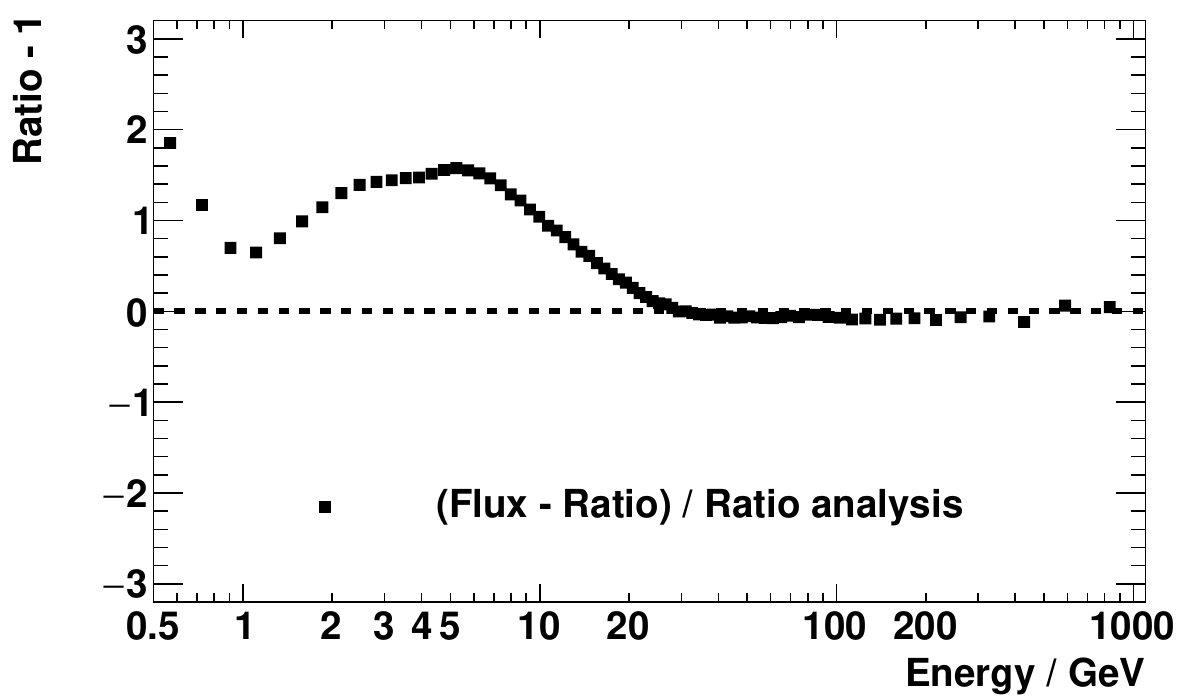}
  \caption{Ratio of the total uncertainty of the positron/electron ratio derived using the single-track analysis over the positron/electron ratio derived from the all-track analysis.}
  \label{fig:results-time-averaged-positron-electron-ratio-aachen-from-fluxes-aachen-uncertainty}
\end{figure}

\Cref{fig:results-time-averaged-positron-electron-ratio-aachen-from-fluxes-aachen-uncertainty} shows the advantage of the single-track analysis
for the positron/electron ratio derivation in the low-energy part. Above \SIapprox{20}{\GeV} the flux analysis yields smaller uncertainties.

\subsection{Cross-check of the positron fraction between flux / dedicated analysis}
\label{sec:appendix-results-comparison-positron-fraction-from-fluxes}

\begin{figure}[H]
  \begin{subfigure}{0.50\linewidth}
    \includegraphics[width=\linewidth]{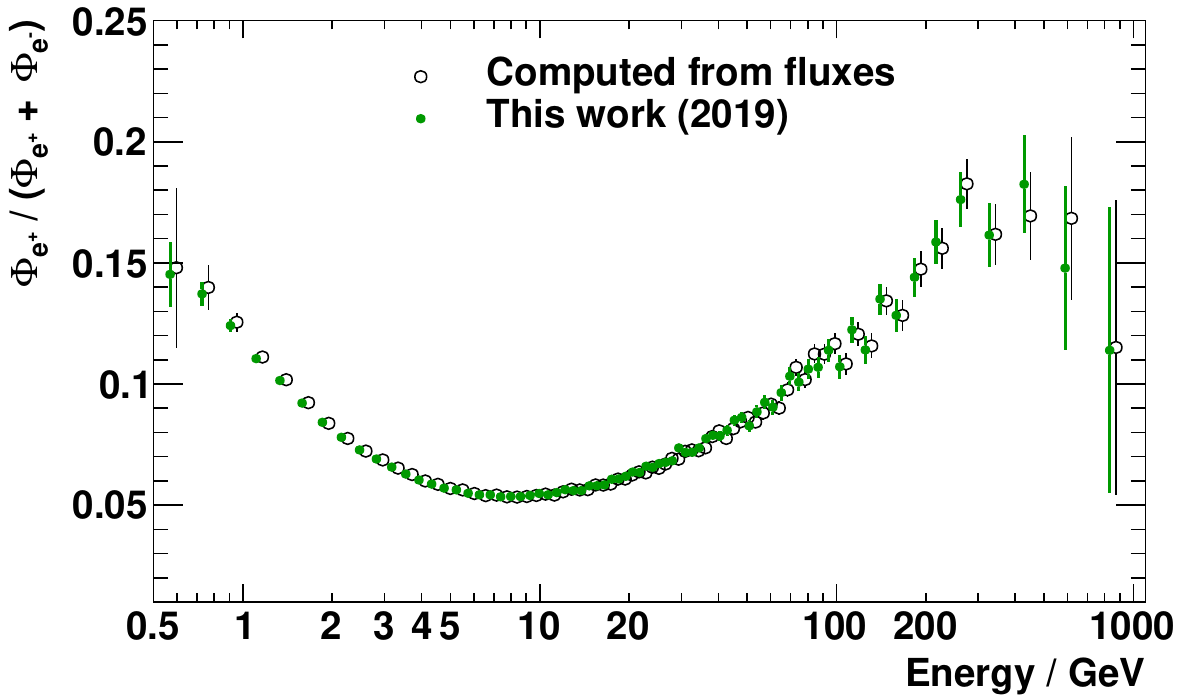}
  \end{subfigure}
  \begin{subfigure}{0.50\linewidth}
    \includegraphics[width=\linewidth]{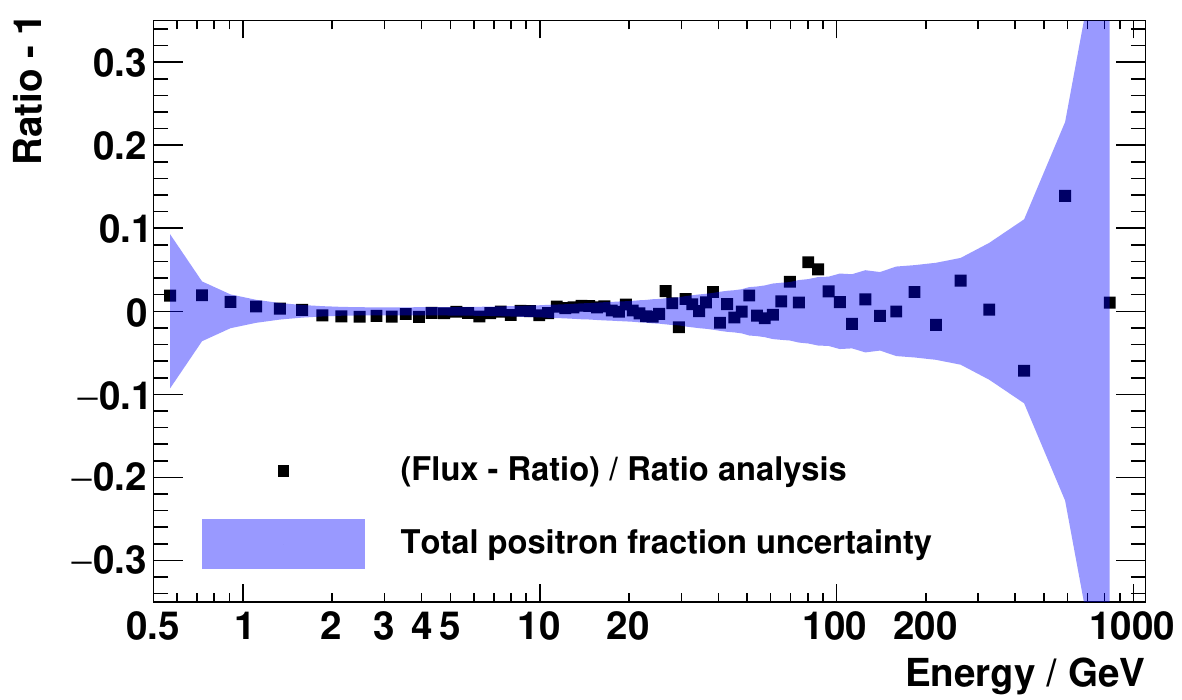}
  \end{subfigure}
  \caption{Comparison of the positron fraction derived using the single-track analysis with the ratio computed from the fluxes, that were derived from the all-track analysis. To ease the comparison the all-track analysis data points are shifted horizontally by \SI{5}{\percent}. Within the quoted systematic uncertainties the results are compatible.}
  \label{fig:results-time-averaged-positron-fraction-aachen-from-fluxes-aachen}
\end{figure}

As cross-check the positron fraction is also computed from the fluxes, to verify the consistency between the single-track
and the all-tracks analyses. As shown in \cref{fig:results-time-averaged-positron-fraction-aachen-from-fluxes-aachen} the agreement is good.

\begin{figure}[H]
  \includegraphics[width=\linewidth]{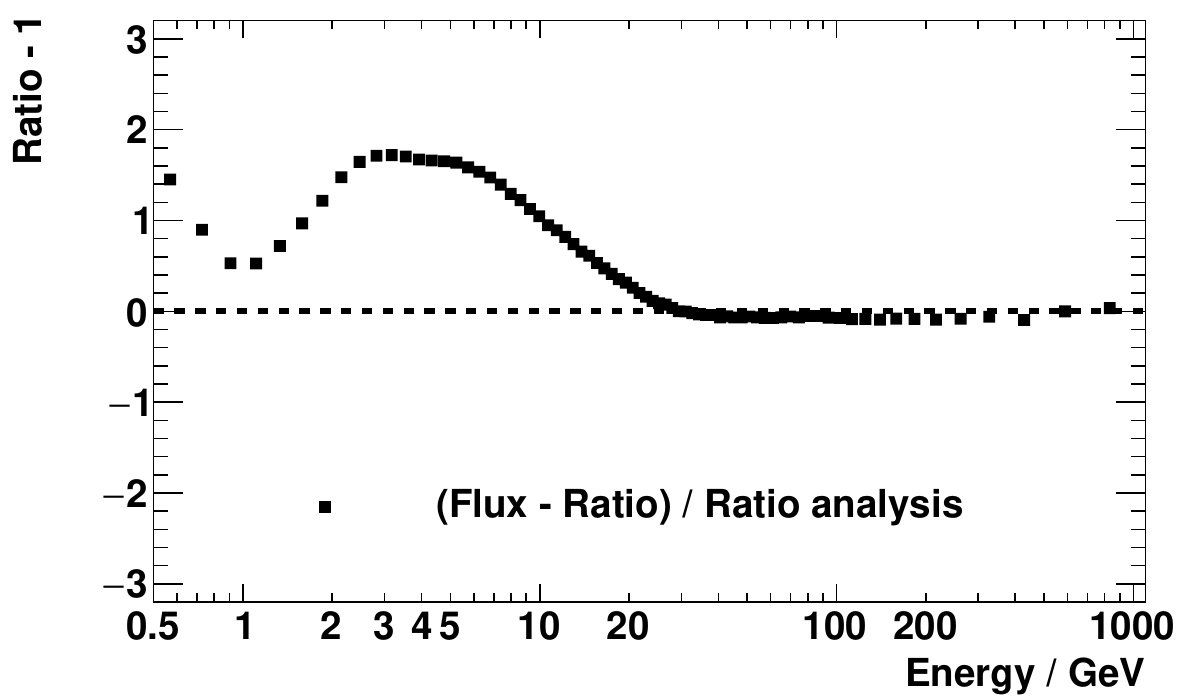}
  \caption{Ratio of the total uncertainty of the positron fraction derived using the single-track analysis over the positron fraction derived from the all-track analysis.}
  \label{fig:results-time-averaged-positron-fraction-aachen-from-fluxes-aachen-uncertainty}
\end{figure}

\Cref{fig:results-time-averaged-positron-fraction-aachen-from-fluxes-aachen-uncertainty} shows the advantage of the single-track analysis
for the positron fraction derivation in the low-energy part. Above \SIapprox{20}{\GeV} the flux analysis yields smaller uncertainties.

\clearpage
\section{Time-dependent results}
\label{sec:appendix-results-time-dependent}

\subsection{Energy dependence of the electron spectral index}
\label{sec:appendix-results-energy-dependence-electron-spectral-index}

\begin{figure}[H]
  \includegraphics[width=\linewidth]{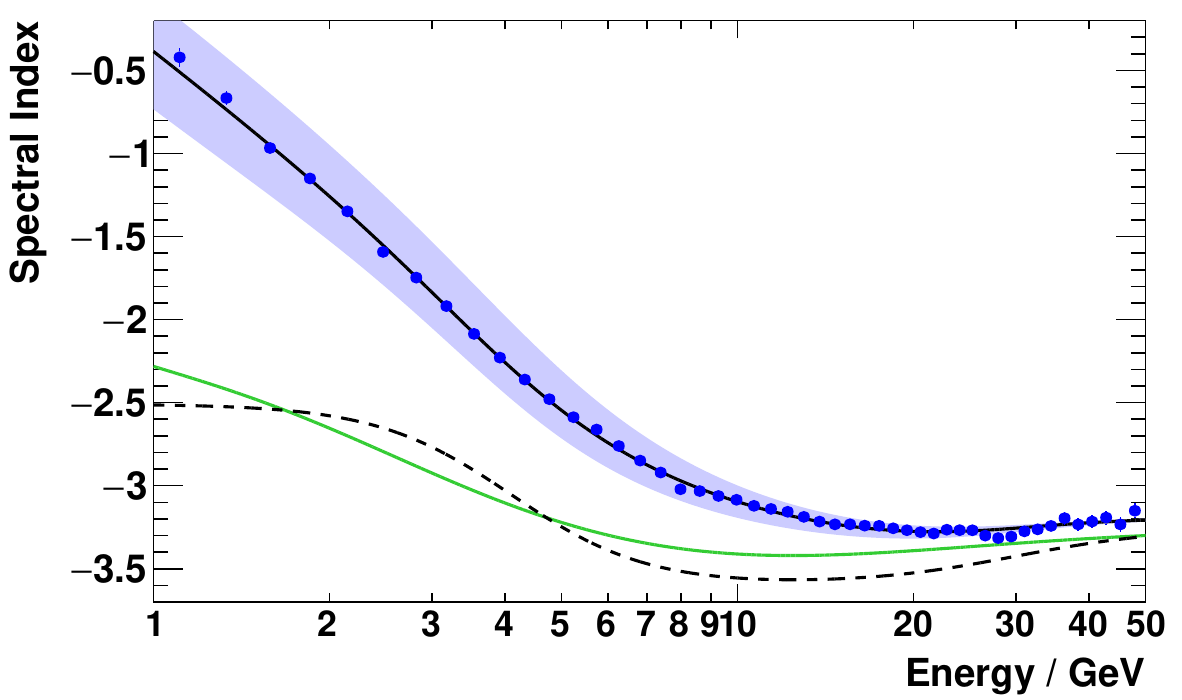}
  \caption{Energy dependence of the electron spectral index $\gamma^{-}(E) = \diff(\log{\Phi_{e^{-}}(E)}) / \diff(\log{E})$ obtained in a model independent way (see \cref{sec:results-time-averaged-electron-flux}) from the time averaged data and the spectral index obtained from the model described in Ref.~\cite{Cavasonza2017} fitted to the time-averaged electron flux data (solid black curve). The shaded band indicates the time-variation. The spectral index from this model without solar modulation (dashed black curve) clearly shows a break in the spectral index between \SIrange{2}{10}{\GeV}. A recent model describing the local interstellar electron spectrum~\cite{Potgieter2015} is also shown (green curve).}
  \label{fig:results-time-dependent-electron-spectral-index-lis}
\end{figure}

The energy dependence of the electron spectral index is shown in \cref{fig:results-time-dependent-electron-spectral-index-lis}.

\clearpage
\subsection{Absence of structures in the positron flux}
\label{sec:appendix-results-absence-of-structures-in-positron-flux}

As described in Ref.~\cite{Aguilar2018}, the model given in Ref.~\cite{Cavasonza2017} was fit to the data for each Bartels rotation independently,
to search for fine structures in the energy dependence of the fluxes. The positron data show no additional structure, as shown in \cref{sec:appendix-results-absence-of-structures-in-positron-flux}.

\begin{figure}[H]
  \includegraphics[width=\linewidth]{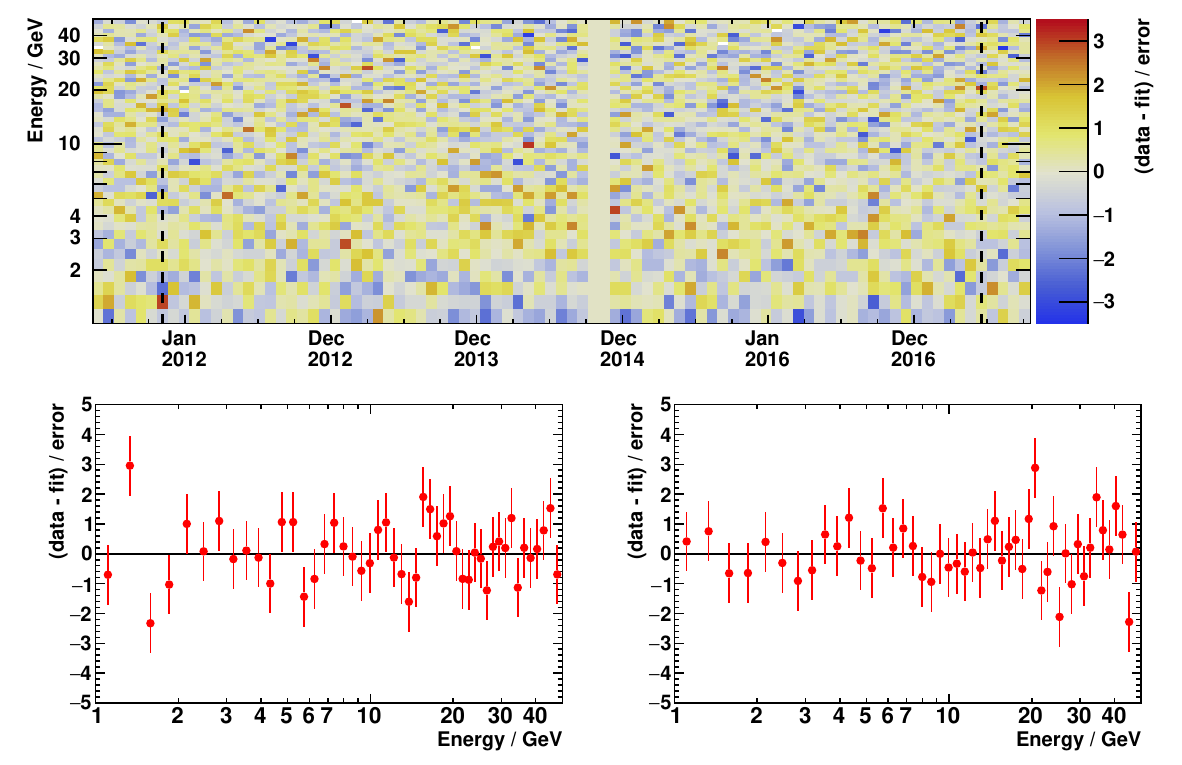}
  \caption{For the positron flux, the difference between data and model~\cite{Cavasonza2017} fits normalized to the experimental uncertainties: for each Bartels Rotation as a function of energy (upper plot). The distribution for the positrons reveals no visible structures, emphasized for Bartels Rotation 2432 (lower left) and Bartels Rotation 2508 (lower right). The times for these two Bartels Rotations are indicated by the vertical dashed lines.}
  \label{fig:results-time-dependent-positron-flux-pull}
\end{figure}

\printbibliography[heading=bibintoc]

\backmatter
\printglossary

%
\chapter{Acknowledgements}
\label{sec:ack}

I am very grateful to many people who made the creation of this dissertation possible.

\medskip
\noindent
First of all I would like to thanks my advisor Prof. Dr. Schael, who offered me the
PhD position after completing my Diploma in his institute. During the past 13 years
I learned a lot about physics and data analysis, the interpretation of the results
and many soft skills from Prof. Schael. I am sincerely thankful for his continuous
support and encouragements in my decisions throughout the thesis. The enthusiasm for
astroparticle physics is truly inspiring and made working in his institute a pleasure.

\medskip
\noindent
Furthermore, I thankfully acknowledge Prof. Dr. Christopher Wiebusch who kindly agreed to be
my second supervisor.

\medskip
\noindent
My special thanks goes to my colleagues and friends from the Aachen group with whom
I had many fruitful discussions throughout my time as PhD student: Dr. Henning Gast,
Dr. Thorsten Siedenburg, Andreas Bachlechner, Bastian Beischer, Fabian Machate and
Dr. Leila Ali Cavasonza. Many things became clear after discussions with all of you.

\medskip
\noindent
In the first year Dr. Thorsten Siedenburg answered numerous of questions regarding
the detector, the reconstruction algorithms and physics in general. Each week he took
time for me and patiently explained all things I wanted to know. The \textit{ACQt} data
format was inspired by his ideas - thanks for all the time you devoted to me.

\medskip
\noindent
Many colleagues and friends shared an office with me over the years: Sarah Beranek,
Bastian Beischer, Tobias Verlage, Nikolay Nikonov, Leila Ali Cavasonza, Sichen Li
and Robin Sonnabend. Many thanks to all of you for making our office a nice play to stay,
discuss and drink coffee. Special thanks go to Bastian and Henning: it was a pleasure to work with you,
and design the \textit{ACsoft} software package, the \textit{ACQt} file format and all the
surrounding software infrastructure. Discussions with both of you always led to the best possible,
most flexible, future-proof solution. Henning patiently took time to read through my thesis and
gave excellent comments, which helped me a lot. Many thanks for your time.

\medskip
\noindent
I would like to express my deep gratitude to the AMS-02 collaboration, who supported my
analysis and encouraged me to go further. Special thanks go to Prof. Dr. Samuel Ting, Dr. Andrei
Kounine, Dr. Vitaly Choutko, Dr. Zhili Weng and Dr. Marco Incagli. Many discussions with Andrei
and Vitaly helped in understanding the detector, the inner workings of the reconstruction software
and the utilized algorithms. The analysis described in this dissertation would not be finished today,
without many productive discussion and debates during various analysis meeting, the ideas and
suggestions from the collaboration.

\medskip
\noindent
Special thanks go to Dr. Marco Incagli who patiently walked me through the whole ECAL calibration sequence,
which I carefully rechecked during 2017 (not described in this thesis). I gained many insights during
this time. Long and fruitful discussions with Marco at CERN and via numerous mails helped a lot. Marco
always took the time to provide all necessary information necessary to complete the ECAL calibration,
thank you very much for your time and help.

\medskip
\noindent
The acknowledgements would be incomplete without a big thank you for my parents. During the past
three decades you continuously supported me and my decisions and helped me to find my way in life.
I am thankful that you bought me a computer when I was 12 years old. This changed my life and
the focus for the following decades. It also paved the way for studying physics: to understand
the inner workings of such a machine, on the lowest level.

\medskip
\noindent
Most importantly, I thank my wife Stefanie, who gave birth to our wonderful sons - Hanno \& Bene.
There are no words that can describe what I feel for you: without your continuous support this thesis
would not have been possible. You always supported me throughout this entire process and made
countless sacrifices to help me get to this point. I am thankful to have a women like you by my side.
Thanks for sharing your life with me since almost 10 years!

\end{document}